\documentclass[12pt]{article}
\makeatletter \@addtoreset{equation}{section} \makeatother
\renewcommand{\theequation}{\thesection.\arabic{equation}}
\newcommand{\ba}{\begin{array}}
\newcommand{\ea}{\end{array}}
\newcommand{\beq}{\begin{equation}}
\newcommand{\eeq}{\end{equation}}
\newcommand{\bea}{\begin{eqnarray}}
\newcommand{\eea}{\end{eqnarray}}

\def\bce{\begin{center}}
\def\ece{\end{center}}

\def\nonu{\nonumber}

\def\pa{\partial}

\def\be{\beta}

\def\eps6{{\displaystyle \mathop{\epsilon}^{6}}{}}
\def\g6{{\displaystyle \mathop{g}^{6}}{}}

\def\nab6{{\displaystyle \mathop{\nabla}^{6}}{}}

\def\0{{\sst{(0)}}}
\def\1{{\sst{(1)}}}
\def\2{{\sst{(2)}}}
\def\3{{\sst{(3)}}}
\def\4{{\sst{(4)}}}
\def\5{{\sst{(5)}}}
\def\6{{\sst{(6)}}}
\def\7{{\sst{(7)}}}
\def\8{{\sst{(8)}}}

\def\ba{\begin{array}}
\def\ea{\end{array}}
\def\beq{\begin{equation}}
\def\eeq{\end{equation}}
\def\be{\begin{equation}}
\def\ee{\end{equation}}

\def\eps{\epsilon}

\def\ba{\begin{array}}
\def\ea{\end{array}}
\def\beq{\begin{equation}}
\def\eeq{\end{equation}}
\def\be{\begin{equation}}
\def\ee{\end{equation}}

\def\eps{\epsilon}

\def\eps6{{\displaystyle \mathop{\epsilon}^{6}}{}}

\def\nab6{{\displaystyle \mathop{\nabla}^{6}}{}}

\newcommand{\bean}{\begin{eqnarray*}}
\newcommand{\eean}{\end{eqnarray*}}

\begin{document}
\thispagestyle{empty} \addtocounter{page}{-1}
   \begin{flushright}
\end{flushright}

\vspace*{1.3cm}
 
\centerline{ \Large \bf   
Higher Spin Currents in Wolf Space: Part II }
\vspace*{0.5cm}
\centerline{{\bf Changhyun Ahn 
}} 
\vspace*{1.0cm} 
\centerline{\it 
Department of Physics, Kyungpook National University, Taegu
702-701, Korea} 
\vspace*{0.8cm} 
\centerline{\tt ahn@knu.ac.kr 
} 
\vskip2cm

\centerline{\bf Abstract}
\vspace*{0.5cm}
The $16$ lowest higher spin currents of spins 
$(1, \frac{3}{2}, \frac{3}{2}, 2)$, 
$(\frac{3}{2}, 2, 2, \frac{5}{2} )$, $(\frac{3}{2}, 2, 2, \frac{5}{2})$ 
and $(2, \frac{5}{2}, \frac{5}{2}, 3)$ 
in terms of ${\cal N}=2$ WZW affine currents
were obtained in the 
${\cal N}=4$ superconformal Wolf space coset  
$\frac{SU(5)}{SU(3) \times SU(2) \times U(1)}$ previously.
By calculating the operator product expansions (OPEs) 
between the above higher spin currents 
which are contained in 
an extension of large ${\cal N}=4$ nonlinear superconformal algebra, 
the next $16$ 
higher spin currents of spins
$(2, \frac{5}{2}, \frac{5}{2}, 3)$, 
$(\frac{5}{2}, 3, 3, \frac{7}{2} )$, $(\frac{5}{2}, 3, 3, \frac{7}{2})$ 
and $(3, \frac{7}{2}, \frac{7}{2}, 4)$ are determined
from the right hand sides of these OPEs.  
Moreover,
the composite fields consisting of both the $11$ currents in the large 
${\cal N}=4$ nonlinear superconformal algebra and the above 
$16$ lowest higher spin currents also occur in the right hand sides of these
OPEs. The latter appears quadratically (and linearly) in the fusion rules 
together with large ${\cal N}=4$
nonlinear superconformal family of the identity operator.  

\baselineskip=18pt
\newpage
\renewcommand{\theequation}
{\arabic{section}\mbox{.}\arabic{equation}}

\section{Introduction}

For the ${\cal N}=4$ superconformal coset theory (in two dimensions) 
described by 
the Wolf space $\frac{SU(N+2)}{ SU(N) \times SU(2) \times U(1)}$ 
with $N=3$,  
the ${\cal N}=2$ WZW affine 
current algebra with  constraints was obtained in \cite{Ahn1311}. 
The $16$ generators 
of the large ${\cal N}=4$ (linear) superconformal algebra were described by 
these ${\cal N}=2$ WZW affine currents explicitly. 
By factoring out both four spin-$\frac{1}{2}$ currents 
and the spin-$1$ current from these $16$ generators, the remaining $11$ 
generators (i.e., spin-$2$ current, four spin-$\frac{3}{2}$ currents, 
and six spin-$1$ currents)  
leading to 
the large ${\cal N}=4$ (nonlinear) superconformal algebra were obtained.    
Based on the large ${\cal N}=4$ holography \cite{GG1305},
the extra $16$ higher spin currents, with spin contents 
\bea
\left(1,\frac{3}{2},  \frac{3}{2}, 2 \right), \,\, 
\left(\frac{3}{2}, 2, 2, \frac{5}{2} \right), \,\, 
\left(\frac{3}{2}, 2, 2, \frac{5}{2} \right), \,\,  
\left(2, \frac{5}{2}, 
\frac{5}{2}, 3 \right)
\label{lowest}
\eea
 described in terms of 
four ${\cal N}=2$ multiplets, 
were found and realized by the above ${\cal N}=2$ WZW affine currents. 
Then the
 operator product expansions (OPEs) between the above $11$ currents and 
these extra $16$ higher spin currents (\ref{lowest}), which are a part of 
${\cal N}=4$ ${\cal W}_{\infty}$ algebra, 
were described explicitly. 
It turned out that the composite fields with definite 
$U(1)$ charges, made of above $(11+16)$ generating currents commuting  with 
the Wolf space subgroup,
have occurred in the right hand sides of these OPEs. 

As suggested in \cite{Ahn1311}, it is natural to
calculate the OPEs between the higher spin currents in (\ref{lowest})
and themselves in order to determine next $16$ higher spin currents. 
The spin contents are  characterized 
by \cite{GG1305}
\bea
\left(2,\frac{5}{2},  \frac{5}{2}, 3 \right), \,\, 
\left(\frac{5}{2}, 3, 3, \frac{7}{2} \right), \,\, 
\left(\frac{5}{2}, 3, 3, \frac{7}{2} \right), \,\,  
\left(3, \frac{7}{2}, 
\frac{7}{2}, 4 \right).
\label{nextlowest}
\eea
How does one determine these $16$ higher spin currents (\ref{nextlowest})
in terms of ${\cal N}=2 $ WZW affine currents? 
First of all, by recalling that the four 
higher spin-$\frac{3}{2}$ currents in (\ref{lowest}) 
were obtained in \cite{Ahn1311} 
from the OPEs between the 
higher spin-$1$ current and four spin-$\frac{3}{2}$ currents in the 
large ${\cal N}=4$ nonlinear algebra, 
the four higher spin-$\frac{5}{2}$ currents in (\ref{nextlowest})
can be determined by the OPEs between the higher spin-$1$ current in 
(\ref{lowest}) and the four higher spin-$\frac{5}{2}$ currents 
in (\ref{lowest}).
The next step is to determine the six higher spin-$3$ currents in 
(\ref{nextlowest}).
One way to obtain these currents is to use the spin-$\frac{3}{2}$ currents
in the large ${\cal N}=4$ nonlinear superconformal algebra appropriately
and calculate the OPEs between them and the above 
four higher spin-$\frac{5}{2}$ currents 
in (\ref{nextlowest}). 
This motivation comes from the observations of how to determine 
the six higher spin-$2$ currents in (\ref{lowest}) in \cite{Ahn1311}.
During this calculation, the presence of the higher spin-$2$ current 
in (\ref{nextlowest}) is observed also.
For the last higher spin-$3$ current locating at the last ${\cal N}=2$
multiplet in (\ref{nextlowest}), one can take, for example,  the correponding 
higher spin-$\frac{5}{2}$ current is the one in the second ${\cal N}=2$ 
multiplet.  
What about four higher spin-$\frac{7}{2}$ currents in (\ref{nextlowest})? 
Once again,  
by recalling how one has determined four higher spin-$\frac{5}{2}$ currents
in (\ref{lowest}) \cite{Ahn1311},
the four higher spin-$\frac{7}{2}$ currents in (\ref{nextlowest})
can be determined by acting spin-$\frac{3}{2}$ currents of the 
large  ${\cal N}=4$ nonlinear superconformal algebra  on the above higher
spin-$3$ currents living in the last three ${\cal N}=2$ multiplets 
of (\ref{nextlowest}).
Finally, the higher spin-$4$ current in (\ref{nextlowest})
can be fixed by calculating the OPE between one of the spin-$\frac{3}{2}$
currents and one of the higher spin-$\frac{7}{2}$ currents living in 
the last ${\cal N}=2$ multiplet of (\ref{nextlowest}).

In section $2$, 
the new 16 higher spin currents in (\ref{nextlowest}) 
are obtained explicitly.

In section $3$, the OPEs between the higher spin-1 current in (\ref{lowest})
and the remaining higher spin currents are obtained. 
Then all the other OPEs between  the higher spin currents in 
(\ref{lowest}) are obtained similarly.

In section $4$, we summarize the main results of this paper with the fusion 
rules and comment on the 
future works.

In Appendices $A$-$L$, 
all the OPEs between the higher spin currents in 
(\ref{lowest}) and themselves are written explicitly \footnote{ The asymptotic 
symmetry of the $AdS_3$ higher spin theory based on super higher spin algebra
$shs_2[\lambda]$ has been studied further and it matches with 
those of the two-dimensional CFT Wolf space coset in the 't Hooft 
limit \cite{GP}. The extension of large ${\cal N}=4$ superconformal 
algebra which contains one ${\cal N}=4$ multiplet for each integer spin $s=1,
2, \cdots$ as well as the currents of large ${\cal N}=4$ superconformal 
algebra has been studied in \cite{BCG}. 
The first two ${\cal N}=4$  multiplets with $s=1, 2$ correspond to 
(\ref{lowest}) and (\ref{nextlowest}) respectively.
For the particular level at 
the Kazama-Suzuki model, the ${\cal N}=3$ (enhanced) supersymmetry 
is observed in \cite{CHR}.  The full spectrum of the tensionless 
string theory (in $AdS_3 \times {\bf S}^3 \times {\bf T}^4$) can be 
reorganized in terms of representations of the ${\cal N}=4$ super 
${\cal W}_{\infty}$ algebra where the large level limit is taken 
\cite{GG1406}.   }.

\section{The sixteen second  lowest 
higher spin currents in the Wolf space coset    }

Let us recall that the $16$ lowest higher spin currents  
are described by \cite{Ahn1311}
\bea
\left(1, \frac{3}{2}, \frac{3}{2}, 2 \right)
& : & (T^{(1)}, T_{+}^{(\frac{3}{2})}, T_{-}^{(\frac{3}{2})}, T^{(2)}), 
\nonu \\
 \left(\frac{3}{2}, 2, 2, \frac{5}{2} \right) & : & 
(U^{(\frac{3}{2})}, U_{+}^{(2)}, U_{-}^{(2)}, U^{(\frac{5}{2})} ), \nonu \\
\left(\frac{3}{2}, 2, 2, \frac{5}{2} \right) & : & 
(V^{(\frac{3}{2})}, V^{(2)}_{+}, V^{(2)}_{-}, V^{(\frac{5}{2})}),  \nonu \\
\left(2, \frac{5}{2}, \frac{5}{2}, 3 \right) & : &
 (W^{(2)}, W_{+}^{(\frac{5}{2})}, W_{-}^{(\frac{5}{2})}, W^{(3)}),
\label{lowesthigher}
\eea
where the spins are specified in each primary field.
By adding these (\ref{lowesthigher}) to spin with one unit,
one obtains the following $16$ next higher spin currents  \cite{GG1305}
\bea
\left(2, \frac{5}{2}, \frac{5}{2}, 3 \right) & : & 
( P^{(2)}, P_{+}^{(\frac{5}{2})}, P_{-}^{(\frac{5}{2})}, P^{(3)}),
\nonu \\
\left( \frac{5}{2}, 3, 3, \frac{7}{2} \right) & : & 
( Q^{(\frac{5}{2})}, Q_{+}^{(3)}, Q_{-}^{(3)}, Q^{(\frac{7}{2})}),
\nonu \\
\left( \frac{5}{2}, 3, 3, \frac{7}{2}  \right) & : & 
( R^{(\frac{5}{2})}, R_{+}^{(3)}, R_{-}^{(3)}, R^{(\frac{7}{2})}),
\nonu \\
\left( 3, \frac{7}{2}, \frac{7}{2}, 4 \right)  & : & 
( S^{(3)}, S_{+}^{(\frac{7}{2})}, S_{-}^{(\frac{7}{2})}, S^{(4)}).
\label{nexthigher}
\eea
It turned out that the $U(1)$ charge (defined in $(3.19)$ of \cite{Ahn1311}) 
of each field in (\ref{nexthigher})
is exactly the same as the one of each field in (\ref{lowesthigher}).


We would like to construct the $16$ 
next lowest higher spin currents of spins 
in (\ref{nexthigher}) \footnote{
It is also useful to write down the spin-$1$ affine current 
in terms of purely bosonic spin-$1$ current and spin-$\frac{1}{2}$ current. 
One has the following expressions (in the 
reference $[34]$
of \cite{Ahn1311})
\bea
\overline{D} K^m|_{\theta=\bar{\theta}=0}(z) & = & 
V^m(z) -\frac{1}{(5+k)} \left( f_{p \bar{q}}^{\,\,\,\, m} K^p K^{\bar{q}}
+ f_{a \bar{b}}^{\,\,\,\, m}  J^a J^{\bar{b}} \right)(z),
\nonu \\
D K^{\bar{m}}|_{\theta=\bar{\theta}=0}(z) & = & 
V^{\bar{m}}(z) -\frac{1}{(5+k)} \left( f_{\bar{p} q}^{\,\,\,\, \bar{m}} K^{\bar{p}} K^{q}
+ f_{\bar{a} b}^{\,\,\,\, \bar{m}}  J^{\bar{a}} J^{b} \right)(z),
\nonu \\
\overline{D} J^a|_{\theta=\bar{\theta}=0}(z) & = & 
V^a(z) -\frac{1}{(5+k)}  f_{b \bar{m}}^{\,\,\,\, a} J^b K^{\bar{m}}(z),
\nonu \\
D J^{\bar{a}}|_{\theta=\bar{\theta}=0}(z) & = & 
V^{\bar{a}}(z) -\frac{1}{(5+k)}  f_{\bar{b} m}^{\,\,\,\, \bar{a}} J^{\bar{b}} K^{m}(z).
\nonu
\eea
The following OPEs between the purely bosonic spin-$1$ currents satisfy
\bea
V^A(z) \, V^{B}(w) & = & 
 -\frac{1}{(z-w)} \, f_{\bar{A} \bar{B}}^{\,\,\,\,\,\,\bar{C}} V^C(w)
+\cdots,
\nonu \\
V^{\bar{A}}(z) \, V^{\bar{B}}(w) & = & 
 -\frac{1}{(z-w)} \, f_{A B}^{\,\,\,\,\,\,C} V^{\bar{C}}(w)
+\cdots,
\nonu \\
V^A(z) \, V^{\bar{B}}(w) & = & \frac{1}{(z-w)^2} \, k \, \delta^{A\bar{B}} 
 -\frac{1}{(z-w)} \, \left( 
f_{\bar{A} B}^{\,\,\,\,\,\,\bar{C}} V^C
+  f_{\bar{A} B}^{\,\,\,\,\,\,C} V^{\bar{C}}
\right)(w)
+\cdots,
\nonu
\eea
where the indices $A, B, \cdots$ and $\bar{A}, \bar{B}, \cdots $
stand for the indices of the group $SU(N+2=5)$ in the complex basis
and furthermore the following regularity holds
\bea
V^{A}(z) \, K^m(w) & = & V^{A}(z) \, K^{\bar{m}}(w) =
V^{A}(z) \, J^a(w) = V^{A}(z) \, J^{\bar{a}}(w) = \cdots,
\nonu \\
V^{\bar{A}}(z) \, K^m(w) & = &  V^{\bar{A}}(z) \, K^{\bar{m}}(w) =
V^{\bar{A}}(z) \, J^a(w) = V^{\bar{A}}(z) \, J^{\bar{a}}(w) = \cdots.
\nonu 
\eea
The OPEs between the spin-$\frac{1}{2}$ currents are given in the 
reference in \cite{Ahn1311} (with the footnote $11$).
The reason why one should express all the higher spin currents in this basis
is due to the fact that the number of independent affine currents for given
higher spin current is 
significantly reduced. 
For example, the number of independent terms in the higher spin-$3$ 
current $W^{(3)}(z)$ in the basis of \cite{Ahn1311} is given by $3438$
while those in the present basis is given by $681$. 
}.  
Because there are $11$ currents from the large ${\cal N}=4$ nonlinear algebra
and $16$ lowest higher spin currents given by (\ref{lowesthigher}),
one expects that the next lowest 
higher spin currents (\ref{nexthigher}) will occur 
in the various OPEs between these $27(=11+16)$ currents.
In particular, the spin-$\frac{3}{2}$ currents, $\hat{G}_{21}(z)$ and 
$\hat{G}_{12}(z)$, which are the generators of large ${\cal N}=4$ nonlinear 
algebra, and the higher spin-$1$ current, $T^{(1)}(z)$, 
which is one of  the generators in 
(\ref{lowesthigher}), will play an important role in this construction
\footnote{Note that 
in the convention of \cite{BCG}, the last component of any
${\cal N}=2$ multiplet is not a primary but quasiprimary field under the 
stress energy tensor. In our case,  all the fields,
$T^{(2)}(w), U^{(\frac{5}{2})}(w), V^{(\frac{5}{2})}(w), W^{(3)}(w), P^{(3)}(w), 
Q^{(\frac{7}{2})}(w), R^{(\frac{7}{2})}(w)$ and $S^{(4)}(w)$ are primary fields 
under the $\hat{T}(z)$. Furthermore in \cite{GP}, some higher spin currents 
in the 
extension of large ${\cal N}=4$ superconformal algebra in terms of
affine currents were found for arbitrary
$N$ but only up to  higher spin-$2$ currents. For example, the higher 
spin-$\frac{5}{2}, \cdots, \frac{7}{2}, 4 $ currents  
are not known for general $N$ so far.  }.  
Of course, when the higher spin-$3$ current $S^{(3)}(w)$ is determined, 
other spin-$\frac{3}{2}$ current $\hat{G}_{22}(z)$ is also useful.

\subsection{Construction of four higher spin-$\frac{5}{2}$ currents:
$P_{\pm}^{(\frac{5}{2})}(z)$, $Q^{(\frac{5}{2})}(z)$ and $R^{(\frac{5}{2})}(z)$}

Let us construct the four lowest fermionic higher spin-$\frac{5}{2}$ 
currents contained in (\ref{nexthigher}). 
Recall that the four lowest fermionic higher spin-$\frac{3}{2}$ 
currents in (\ref{lowesthigher}) were obtained from the OPEs between 
the higher spin-$1$ current, $T^{(1)}(z)$, 
and four spin-$\frac{3}{2}$ currents, $\hat{G}_a(w)$ 
where $a=11,12,21,22$, via Appendices $(C.1)$ and $(C.2)$ of \cite{Ahn1311}.
Then it is natural to consider the OPEs between the higher spin-$1$ current,
 $T^{(1)}(z)$,
and the four fermionic higher spin-$\frac{5}{2}$ currents in
(\ref{lowesthigher}).  

It turns out that the OPEs between the higher spin-$1$ current, $T^{(1)}(z)$,
and the spin-$\frac{5}{2}$ currents living in the last components of
the second and third 
${\cal N}=2$ multiplets of (\ref{lowesthigher}) respectively, from $(4.7), 
(4.32)$ and $(4.46)$ of \cite{Ahn1311}, 
are summarized by
\bea
T^{(1)}(z) \, 
\left(
\begin{array}{c}
U^{(\frac{5}{2})} \\
V^{(\frac{5}{2})} \\
\end{array} \right) (w) & = & \frac{1}{(z-w)^2} \,
\frac{4}{3(5+k)} 
\left[ \mp(-3+k) \left( 
\begin{array}{c} 
\hat{G}_{11} \\
\hat{G}_{22} \\
\end{array} \right)  +  \left(
\begin{array}{c} 
U^{(\frac{3}{2})} \\
-V^{(\frac{3}{2})} \\
\end{array} \right) \right](w)   
\nonu \\
& + & \frac{1}{(z-w)} \, \left[  
\left(
\begin{array}{c} 
U^{(\frac{5}{2})} \\
-V^{(\frac{5}{2})} \\
\end{array} \right)  + 
\left(
\begin{array}{c}
 Q^{(\frac{5}{2})} \\
 R^{(\frac{5}{2})} \\
\end{array} \right)  \right](w)  + \cdots.
\label{t1uv}
\eea
From the $U(1)$ charge in the Table $2$ of \cite{Ahn1311},
one can easily see that the right hand side of (\ref{t1uv}) preserves
the $U(1)$ charges, $\pm \frac{(-3+k)}{(5+k)}$, 
of the left hand side of (\ref{t1uv}). 
The $U(1)$ charge of $Q^{(\frac{5}{2})}(z)[R^{(\frac{5}{2})}(z)]$ 
is the same as the one of 
$U^{(\frac{5}{2})}(z)[V^{(\frac{5}{2})}(z)]$ \footnote{One can explicitly check that 
the OPE between the $U(1)$ current given in $(5.2)$ of \cite{Ahn1311} and 
 $Q^{(\frac{5}{2})}(w)[R^{(\frac{5}{2})}(w)]$ provides these $U(1)$ charges. }.
Note that the factor $(k-3)$ appears in the first term of 
the second order pole in (\ref{t1uv}) which will vanish at $k=3$.

Similarly, 
 the OPEs between the higher spin-$1$ current, $T^{(1)}(z)$,
and the higher spin-$\frac{5}{2}$ currents living in the second and third 
components of last
${\cal N}=2$ multiplet of (\ref{lowesthigher}) 
are described by
\bea
T^{(1)}(z) \, W_{\pm}^{(\frac{5}{2})}(w) & = & \frac{1}{(z-w)^2}
\, \frac{4}{3(5+k)} 
\left[ \mp (-4+k) \left( 
\begin{array}{c} 
\hat{G}_{21} \\
\hat{G}_{12} \\
\end{array} \right) +  2 T_{\pm}^{(\frac{3}{2})}\right](w)   
\nonu \\
& + & \frac{1}{(z-w)} \, 
\left[ \pm W_{\pm}^{(\frac{5}{2})} + P_{\pm}^{(\frac{5}{2})} 
\right](w)  +
\cdots,
\label{t1w+w-}
\eea
where the equations $(4.7), (4.53) $ and $(4.56)$ of \cite{Ahn1311} are used.
The $U(1)$ charges, $\pm \frac{(3+k)}{(5+k)}$, are preserved in 
(\ref{t1w+w-}) respectively. 
The $U(1)$ charge of $P_{\pm}^{(\frac{5}{2})}(z)$ 
is the same as the one of 
$W_{\pm}^{(\frac{5}{2})}(z)$.

The OPEs (\ref{t1uv}) and (\ref{t1w+w-})
imply that the higher spin-$1$ current, which is the lowest component of 
${\cal N}=4$  multiplet in (\ref{lowesthigher}),  
generates the higher spin-$\frac{5}{2}$ currents 
in other ${\cal N}=4$ multiplet 
(\ref{nexthigher}) by acting with those currents with same spins 
in (different) ${\cal N}=4$  multiplet (\ref{lowesthigher}). 
Note that the four higher spin-$\frac{5}{2}$ currents, 
$P_{+}^{(\frac{5}{2})}(z)[W_{+}^{(\frac{5}{2})}(z)]$, 
$P_{-}^{(\frac{5}{2})}(z)[W_{-}^{(\frac{5}{2})}(z)]$, 
$Q^{(\frac{5}{2})}(z)[U^{(\frac{5}{2})}(z)]$ and 
$R^{(\frac{5}{2})}(z)[V^{(\frac{5}{2})}(z)]$, transform as 
$({\bf 2}, {\bf 2})$ under the $SU(2) \times SU(2)$ \cite{GG1305} 
\footnote{
Similarly, the four spin-$\frac{3}{2}$ currents, 
$\hat{G}_{21}(z)[T_{+}^{(\frac{3}{2})}(z)]$, 
$\hat{G}_{12}(z)[T_{-}^{(\frac{3}{2})}(z)]$, 
$\hat{G}_{11}(z)[U^{(\frac{3}{2})}(z)]$ and 
$\hat{G}_{22}(z)[V^{(\frac{3}{2})}(z)]$, transform as 
$({\bf 2}, {\bf 2})$ under the $SU(2) \times SU(2)$.}.
The descendant fields of the spin-$\frac{3}{2}$ currents 
in the second-order pole in (\ref{t1uv}) and (\ref{t1w+w-})
do not occur in the first-order pole because the difference of spins
in the left hand side, $1-\frac{5}{2} =-\frac{3}{2}$, is 
equal to the minus spin for the primary current appearing 
in the second-order pole of the right hand side. 
See the footnote $51$ of \cite{Ahn1311}.
Furthermore, one can imagine the ${\cal N}=2$ fusion rule from the two 
results (\ref{t1uv}) and (\ref{t1w+w-}). In an appropriate basis,
for example,
the OPE between the first ${\cal N}=2$ multiplet
and the second ${\cal N}=2$ multiplet of (\ref{lowesthigher}) will produce
the second ${\cal N}=2$ multiplet  
of (\ref{nexthigher}) in the right hand side of the 
OPE.   
We will come back this issue in next section $4$.

\subsection{Construction of six higher spin-$3$ currents
and one higher
spin-$2$ current: $P^{(3)}(z)$, 
$Q_{\pm}^{(3)}(z)$, $R_{\pm}^{(3)}(z)$, $S^{(3)}(z)$ and  $P^{(2)}(z)$}

Based on the previous four higher 
spin-$\frac{5}{2}$ currents, we would like to 
construct six higher spin-$3$ currents in (\ref{nexthigher}).
As a byproduct, one also determines the higher spin-$2$ current, which is the 
lowest component of ${\cal N}=4$ multiplet (\ref{nexthigher}).

Recall that the OPE between the second component of ${\cal N}=2$ super stress
energy tensor, corresponding to $\hat{G}_{21}(z)$ (after factoring out 
four spin-$\frac{1}{2}$ and one spin-$1$ currents), and 
the third component of  any ${\cal N}=2$ super primary
current  provides its first and fourth components  
\footnote{Similarly,  
the OPE between the third component of ${\cal N}=2$ super stress
energy tensor, corresponding to $\hat{G}_{12}(z)$, and 
the second component of  any ${\cal N}=2$ super primary
current  provides its first and fourth components. }. 
One calculates the following OPE between $\hat{G}_{21}(z)$ and 
$ P_{-}^{(\frac{5}{2})}(w)$ (whose explicit expression was 
obtained in previous subsection via (\ref{t1w+w-}))
and the right hand side can be 
summarized as follows: 
\bea
\hat{G}_{21}(z) \, P_{-}^{(\frac{5}{2})}(w) & = & \frac{1}{(z-w)^3}
\left[ \frac{16(-4+k)}{(5+k)^2} i \hat{A}_3 + \frac{16k(8+k)}{3(5+k)^2} 
i \hat{B}_3 -\frac{8(-4+k)}{3(5+k)}  T^{(1)}\right](w) 
\nonu \\
& + & 
\frac{1}{(z-w)^2}
\, P^{(2)}(w) \nonu \\
& + & \frac{1}{(z-w)} \, 
\left[\frac{1}{4} \pa P^{(2)} +\frac{8(-4+k)}{(19+23k)} 
\left( \hat{T} T^{(1)} -\frac{1}{2} \pa^2 T^{(1)}  \right) \right.
\nonu \\
&+ &  \frac{48(-20+k+k^2)}{(5+k)^2(19+23k)} i  
\left( \hat{T} \hat{A}_3 -\frac{1}{2} \pa^2 \hat{A}_3  \right)
\nonu \\
& + & \left. \frac{16k(8+k)}{(5+k)(19+23k)} i
\left( \hat{T} \hat{B}_3 -\frac{1}{2} \pa^2 \hat{B}_3  \right)
+ P^{(3)} 
\right](w)  +\cdots,
\label{g21p5half-}
\eea
where the equation $(3.17)$ of \cite{Ahn1311} is used.
The second-order pole in (\ref{g21p5half-})
can give the lowest component of the first ${\cal N}=2$ multiplet in 
(\ref{nexthigher})
while the first-order pole 
provides the higher spin-$3$ current of same ${\cal N}=2$ multiplet.
No descendant fields for the spin-$1$ current appearing in the 
third-order pole arise in the second-order pole.
The higher spin-$3$ current can be extracted from the first-order pole 
after subtracting the descendant field for the spin-$2$ current with 
correct coefficient and the three quasi-primary fields of spin $3$
\footnote{This higher spin-$3$ current 
is not to be confused with the expression
$P^{(3)}(w)$ in Appendix $(C.26)$ of \cite{Ahn1311}.}.
Because there exist three spin-$1$ currents with vanishing $U(1)$ charges,
$T^{(1)}(w), \hat{A}_3(w)$ and $\hat{B}_3(w)$,
according to the Table $2$ of \cite{Ahn1311},
one obtains
three quasi primary  fields as in (\ref{g21p5half-}).
The numerical factor $\frac{1}{2}$ is fixed by 
the property of quasi primary field.
In different basis, one can find the higher spin-$\frac{5}{2}$ current 
by adding the extra field contents and requiring that 
third order pole in (\ref{g21p5half-})
should vanish (and there should be no quasi primary fields in the 
first order pole). 
From the $U(1)$ charge in the Table $2$ in \cite{Ahn1311},
both sides of (\ref{g21p5half-}) preserve
the vanishing $U(1)$ charge. The $U(1)$ charge of $P_{-}^{(\frac{5}{2})}(z)$
is equal to $-\frac{(3+k)}{(5+k)}$ as explained before. 
Then the explicit form for the first ${\cal N}=2$ 
multiplet in (\ref{nexthigher})
is determined completely in terms of ${\cal N}=2$ WZW affine currents
\footnote{
\label{exp1}
One has the following OPEs 
\bea
\left(
\begin{array}{c}
\hat{G}_{21} \\
\hat{G}_{12} \\
\end{array} \right) (z) \, 
P^{(2)}(w) & = & \mp\frac{1}{(z-w)^2} \,
\left[ \frac{8(11+k)}{3(5+k)^2}
T_{\pm}^{(\frac{3}{2})} \right] (w) \nonu \\
& + &  \frac{1}{(z-w)} \left[ \mp 4 
 P_{\pm}^{(\frac{5}{2})} -\frac{4}{(5+k)}  W_{\pm}^{(\frac{5}{2})} 
-\frac{8(19+2k)}{
3(5+k)^2} i  \hat{A}_{\mp} \left(
\begin{array}{c} 
U^{(\frac{3}{2})} \\
V^{(\frac{3}{2})} \\
\end{array} \right) \right. \nonu \\
&\mp &   \frac{24}{(5+k)^2} i \hat{A}_3 \left( 
\begin{array}{c} \hat{G}_{21} \\
\hat{G}_{12} \\
\end{array} \right) \mp \frac{8}{3(5+k)} i
\hat{B}_{\mp} \left( 
\begin{array}{c}  
\hat{G}_{22} \\
\hat{G}_{11} \\
\end{array} \right)  \nonu \\
& - & \frac{8(19+2k)}{3(5+k)^2} i \hat{B}_{\mp} 
\left( 
\begin{array}{c} 
V^{(\frac{3}{2})} \\
U^{(\frac{3}{2})} \\
\end{array} \right)  
\pm \frac{8(8+k)}{(5+k)^2} i \hat{B}_3 \left(
\begin{array}{c} 
\hat{G}_{21} \\
\hat{G}_{12} \\
\end{array} \right) +   
\frac{16}{(5+k)}  i \hat{B}_3 T_{\pm}^{(\frac{3}{2})} 
\nonu \\
& \mp & \left. 
\frac{4}{(5+k)}  T^{(1)} \left(
\begin{array}{c} 
\hat{G}_{21} \\
\hat{G}_{12} \\
\end{array} \right)  + \frac{8(8+k)}{3(5+k)^2}  \pa 
\left(
\begin{array}{c}
\hat{G}_{21} \\
\hat{G}_{12} \\
\end{array} \right) \pm \frac{8(9+k)}{3(5+k)^2} \pa T_{\pm}^{(\frac{3}{2})} 
\right](w) \nonu \\
& + & \cdots.
\label{g2112p2}
\eea
Now the first-order pole of these OPEs give 
the second[third] component of the first ${\cal N}=2$ multiplet 
in (\ref{nexthigher}).
The second order pole of this OPE can be 
removed by introducing other higher spin-$2$ current $\hat{P}^{(2)}(z)$ 
correctly.
Let us write down  each primary field in terms of ${\cal N}=2$ WZW affine 
currents as follows:
\bea
P^{(2)}(z) & = & \frac{16(8+k)}{3(5+k)^2} \left( V^1 V^{\bar{1}}
+ V^2 V^{\bar{2}} + V^3 V^{\bar{3}} \right)(z) + \mbox{other 650 terms},
\nonu \\
P_{+}^{(\frac{5}{2})}(z) & = &  \frac{4\sqrt{2}}{(5+k)^2} \left( K^1 V^4 V^{\bar{2}}
+ K^1 V^5 V^{\bar{3}} \right) +   \mbox{other 33 terms},
\nonu \\
P_{-}^{(\frac{5}{2})}(z) & = &  \frac{4 i \sqrt{2}}{(5+k)^3} \left( K^{\bar{1}} V^1 
V^{7}
- K^1 V^1 V^{\bar{7}} \right) +   \mbox{other 33 terms},
\nonu \\
P^{(3)}(z) & = & \frac{8}{(5+k)^2} \left( V^1 V^4 V^{\bar{2}}
+ V^1 V^5 V^{\bar{3}}  \right)(z) + \mbox{other 75 terms}.
\nonu
\eea
}.

Recall that the OPE between the second[third] 
component of ${\cal N}=2$ super stress
energy tensor, corresponding to $\hat{G}_{21}(z)[\hat{G}_{12}(z)]$, and 
the first component of  any ${\cal N}=2$ super primary
current  provides its second[third] component.  
Because the first components of the second and third  ${\cal N}=2$ 
multiplets
in (\ref{nexthigher}) are determined, 
one can calculate the following OPEs in order to obtain 
other kind of higher spin-$3$ currents appearing in those 
${\cal N}=2$ multiplets.
It turns out that
\bea
\left(
\begin{array}{c}
\hat{G}_{21} \\
\hat{G}_{12} \\
\end{array} \right) (z) \, \left(
\begin{array}{c} 
Q^{(\frac{5}{2})} \\
R^{(\frac{5}{2})} \\
\end{array} \right) (w) & = & \frac{1}{(z-w)^3} 
\, \left[ \frac{16k(3+k)}{3(5+k)^2} i \hat{B}_{\mp} \right](w) \nonu \\
& + & \frac{1}{(z-w)^2} \, \left[ 
\frac{4(9+4k)}{3(5+k)} 
\left(
\begin{array}{c}
U_{+}^{(2)}  \\
V_{-}^{(2)} \\
\end{array} \right)  
\pm \frac{8(9+2k)}{3(5+k)^2} \hat{A}_3 \hat{B}_{\mp} 
\right. \nonu \\
& \mp & \left.  \frac{8k}{(5+k)^2}  \hat{B}_{\mp} 
\hat{B}_3 
-  \frac{4k}{(5+k)^2}  i \pa  \hat{B}_{\mp}
\mp \frac{4}{(5+k)} i T^{(1)} \hat{B}_{\mp} \right](w) 
\nonu \\
&+ &\frac{1}{(z-w)}  \, \left[ 
 \frac{1}{4} \pa  \left( 
\begin{array}{c} 
\{ \hat{G}_{21} \, Q^{(\frac{5}{2})} \}_{-2} \\
\{ \hat{G}_{12} \, R^{(\frac{5}{2})} \}_{-2} \\
\end{array} \right)
\right. 
\label{g2112qr} \\
&+& \left. \frac{16k(3+k)}{(5+k)(19+23k)} 
i \left( \hat{T}  \hat{B}_{\mp} 
-\frac{1}{2} \pa^2 \hat{B}_{\mp} \right)  
 + \left(
\begin{array}{c} 
Q_{+}^{(3)} \\
R_{-}^{(3)} \\
\end{array} \right) 
\right](w)
+\cdots,
\nonu
\eea
where  the explicit form for $Q^{(\frac{5}{2})}(w)$
and $R^{(\frac{5}{2})}(w)$ via (\ref{t1uv}) is used again. 
Once again, the higher spin-$3$ current occurs 
in the first-order pole in 
(\ref{g2112qr}) after subtracting the descendant field 
with coefficient $\frac{1}{4}$ for the spin-$2$ field
appearing 
in the second-order pole and the quasi-primary field 
which is the only unique quasi primary field for spin-$3$
with given $U(1)$ charge  \footnote{For the 
$n$-th order pole in the OPE $\Phi_i(z) \, \Phi_j(w)$, one denotes by
$\{ \Phi_i \, \Phi_j \}_{-n}(w)$.}.
 The $U(1)$ charge of $Q_{+}^{(3)}(z)$
is equal to $\frac{2k}{(5+k)}$ while 
the one of $R_{-}^{(3)}(z)$
is equal to $-\frac{2k}{(5+k)}$.
No descendant field for the spin-$1$ in the third order pole
because the spin difference of left hand side  is
given by $\frac{3}{2}-\frac{5}{2} = -1$ 
which is equal to the minus spin for this spin-$1$ current. 
Note that the nonlinear terms arise in the second-order pole of 
(\ref{g2112qr}) and all the possible five composite fields of spin-$2$ in Table 
$3$ of \cite{Ahn1311} with given $U(1)$ charge occur.
In this case, the third order pole can be removed in an appropriate basis
by changing the higher spin-$\frac{5}{2}$ current.

Now one calculates the different combination of OPEs between 
the spin-$\frac{3}{2}$ currents and the  higher 
spin-$\frac{5}{2}$ currents 
with different total $U(1)$ charges as follows:
\bea
\left(
\begin{array}{c}
\hat{G}_{12} \\
\hat{G}_{21} \\
\end{array} \right)(z) \, 
\left(
\begin{array}{c}
Q^{(\frac{5}{2})} \\
R^{(\frac{5}{2})} \\
\end{array} \right)(w) & = & \frac{1}{(z-w)^3} 
\, \left[ \frac{8(9+k)}{(5+k)^2} i \hat{A}_{\pm} \right] (w) 
\nonu \\
& + & \frac{1}{(z-w)^2} \left[ -\frac{4(15+2k)}{3(5+k)} 
\left(
\begin{array}{c} 
U_{-}^{(2)} \\
V_{+}^{(2)} \\
\end{array} \right) 
\pm \frac{24}{(5+k)^2} \hat{A}_{\pm} \hat{A}_3 
\right. \nonu \\
& \mp & \left. 
 \frac{8(12+k)}{3(5+k)^2} \hat{A}_{\pm} \hat{B}_3
-  \frac{12}{(5+k)^2} i \pa \hat{A}_{\pm}  \mp 
\frac{4}{(5+k)} i
T^{(1)} \hat{A}_{\pm} \right](w)
\label{g1221qr}
\\ 
& + &  \frac{1}{(z-w)} \left[ 
\frac{1}{4} \pa  \left(
\begin{array}{c} 
\{ \hat{G}_{12} \, Q^{(\frac{5}{2})} \}_{-2} \\
\{ \hat{G}_{21} \, R^{(\frac{5}{2})} \}_{-2} \\
\end{array} \right)
\right. \nonu \\
& + & \left. 
 \frac{24(9+k)}{(5+k)(19+23k)} 
i \left( \hat{T}  \hat{A}_{\pm} 
-\frac{1}{2} \pa^2 \hat{A}_{\pm} \right) 
+\left(
\begin{array}{c} 
Q_{-}^{(3)} \\
R_{+}^{(3)} \\
\end{array} \right) 
\right](w) +\cdots,
\nonu
\eea
where  the explicit form for $Q^{(\frac{5}{2})}(w)$
and $R^{(\frac{5}{2})}(w)$ via (\ref{t1uv}) is used again. 
Compared to the previous case (\ref{g2112qr}), 
the right hand side of (\ref{g1221qr}) 
contains $\hat{A}_{\pm}(w)$ dependent terms corresponding to
$\hat{B}_{\mp}(w)$ in (\ref{g2112qr}).
 The $U(1)$ charge of $Q_{-}^{(3)}(z)$
is equal to $-\frac{6}{(5+k)}$ and
the one of $R_{+}^{(3)}(z)$
is equal to $\frac{6}{(5+k)}$.
The fusion rule from the three 
results (\ref{g21p5half-}), (\ref{g2112qr}) 
and (\ref{g1221qr}) can be described as follows. 
The OPE between the first ${\cal N}=2$ 
multiplet, $(2.44)$ 
of \cite{Ahn1311}, of large ${\cal N}=4$ nonlinear 
algebra 
and the first[second](third) ${\cal N}=2$ multiplet of (\ref{nexthigher}) will produce
the sum of the first[second](third) ${\cal N }=2 $ 
multiplet in (\ref{lowesthigher}) 
and the first[second](third) 
multiplet of (\ref{nexthigher}) as well as other ${\cal N}=2$
multiplets of large ${\cal N}=4$ nonlinear superconformal algebra 
respectively 
in the right hand side of the 
OPE.   

How does one determine the last undetermined higher 
spin-$3$ current appearing in the fourth ${\cal N}=2$ multiplet 
in (\ref{nexthigher})?
Recall that the higher spin-$2$ current $W^{(2)}(w)$  in (\ref{lowesthigher}) 
was obtained from the 
OPE between $\hat{G}_{22}(z)$ and $U^{(\frac{3}{2})}(w)$ via $(4.48)$ 
of \cite{Ahn1311}.
Then one expects that the undetermined higher 
spin-$3$ current can be obtained from 
the OPE between $\hat{G}_{22}(z)$ and $Q^{(\frac{5}{2})}(w)$
and it turns out that
\bea
\hat{G}_{22}(z) \, Q^{(\frac{5}{2})}(w) & = &
\frac{1}{(z-w)^3} \left[
\frac{16(3+2k)}{(5+k)^2} i \hat{A}_3 +\frac{16k(6+k)}{3(5+k)^2} i \hat{B}_3
+ \frac{8(-3+k)}{3(5+k)} T^{(1)}  \right](w)
\nonu \\
& + & \frac{1}{(z-w)^2} \left[ - P^{(2)} -\frac{8(-3+32k+6k^2)}{
3(3+7k)(5+k)} \hat{T} -\frac{4(13+2k)}{3(5+k)} T^{(2)} +
\frac{4(15+2k)}{3(5+k)} W^{(2)} \right. \nonu \\
&+ &  \frac{8}{3(5+k)^2} \hat{A}_1 \hat{A}_1  +
\frac{8}{3(5+k)^2} \hat{A}_2 \hat{A}_2  +
\frac{152}{3(5+k)^2} \hat{A}_3 \hat{A}_3 -
\frac{16(7+k)}{3(5+k)^2} \hat{A}_3 \hat{B}_3 \nonu \\
& + & \left. \frac{8(1+3k)}{3(5+k)^2} \hat{B}_1 \hat{B}_1 
+ \frac{8(1+3k)}{3(5+k)^2} \hat{B}_2 \hat{B}_2 
+\frac{8}{3(5+k)^2} \hat{B}_3 \hat{B}_3
-\frac{8}{(5+k)} i  T^{(1)} \hat{A}_3 \right](w) \nonu \\
& + & \frac{1}{(z-w)} \left[  \frac{1}{4} \pa 
\{ \hat{G}_{22} \, Q^{(\frac{5}{2})}\}_{-2} + 
\frac{48(3+2k)}{(5+k)(19+23k)}  
i \left( \hat{T} \hat{A}_3 -\frac{1}{2} 
\pa^2 \hat{A}_3\right) \right. \nonu \\
& + &  \frac{16k(6+k)}{(5+k)(19+23k)} i  
\left( \hat{T} \hat{B}_3 -\frac{1}{2} 
\pa^2 \hat{B}_3\right) + \frac{8(-3+k)}{(19+23k)}  \left( \hat{T} T^{(1)} 
-\frac{1}{2} 
\pa^2 T^{(1)} \right) 
\nonu \\
& + & \left.  S^{(3)} 
\right](w) + \cdots,
\label{g22q5half}
\eea
where  the explicit form for $Q^{(\frac{5}{2})}(w)$
can be found in  (\ref{t1uv}) and the equation $(3.15)$ of
\cite{Ahn1311} is used. 
Compared to the case (\ref{g21p5half-}), 
the same 
three quasi-primary fields in the first-order pole of (\ref{g22q5half}) 
occur but the second-order pole contains other spin-$2$ currents as well as 
$P^{(2)}(w)$ itself. The spin-$3$ current $S^{(3)}(w)$ has vanishing $U(1)$ 
charge. One can rewrite the quadratic expression appearing in the 
second order pole $ (\hat{A}_1 \hat{A}_1+
\hat{A}_2 \hat{A}_2)(w) = (\hat{A}_{+} \hat{A}_{-} + i \pa \hat{A}_3)(w)$
and similarly  one has the relation $ (\hat{B}_1 \hat{B}_1+
\hat{B}_2 \hat{B}_2)(w) = (\hat{B}_{+} \hat{B}_{-} + i \pa \hat{B}_3)(w)$ 
with definite
$U(1)$ charge.
Note the appearance of the factor $(k-3)$ in the last term of 
third order pole (and the term containing $T^{(1)}(w)$ 
in the first order pole) in (\ref{g22q5half}).

Then the explicit forms for the higher 
spin-$3$ currents 
in (\ref{nexthigher}), from (\ref{g21p5half-}),  
(\ref{g2112qr}), (\ref{g1221qr}) and 
(\ref{g22q5half}),
are determined completely in terms of ${\cal N}=2$ WZW affine currents
\footnote{
Note that the six higher spin-$3$ currents, 
$P^{(3)}(z)[T^{(2)}(z)]$, 
$Q_{+}^{(3)}(z)[U_{+}^{(2)}(z)]$, 
$Q_{-}^{(3)}(z)[U_{-}^{(2)}(z)]$,  
$R_{+}^{(3)}(z)[V_{+}^{(2)}(z)]$, 
$R_{-}^{(3)}(z)[V_{-}^{(2)}(z)]$,  
 and $S^{(3)}(z)[W^{(2)}(z)]$, 
transform as 
$({\bf 3}, {\bf 1}) \oplus ({\bf 1}, {\bf 3} )$ 
under the $SU(2) \times SU(2)$ \cite{GG1305}.}.

\subsection{Construction of four higher spin-$\frac{7}{2}$ currents:
$Q^{(\frac{7}{2})}(z)$, $R^{(\frac{7}{2})}(z)$ and $S_{\pm}^{(\frac{7}{2})}(z)$}

Based on the previous six higher spin-$3$ currents, we would like to 
construct four higher spin-$\frac{7}{2}$ currents in ${\cal N}=4$ 
multiplet in (\ref{nexthigher}).

As done in (\ref{g21p5half-}) for the first ${\cal N}=2$ 
multiplet in (\ref{nexthigher}),
one can calculate, from $(3.17)$ of \cite{Ahn1311} and (\ref{g1221qr}), 
the following OPE
\bea
\hat{G}_{21}(z) \, Q_{-}^{(3)}(w) & = & 
-\frac{1}{(z-w)^3} \left[ \frac{8(-174+114k+187k^2+23k^3)}{(5+k)^2} 
\hat{G}_{11} + \frac{20(-1+k)}{(5+k)^2} U^{(\frac{3}{2})} \right](w)
\nonu \\
& + & \frac{1}{(z-w)^2} \left[ \frac{6(3+k)}{(5+k)} Q^{(\frac{5}{2})}
+ \frac{(17+6k)}{(5+k)} U^{(\frac{5}{2})} -
\frac{2(924+409k+41k^2)}{(5+k)^2(19+23k)} i  \hat{A}_{+} \hat{G}_{21} \right.
\nonu \\
&+ &  \frac{2(9+5k)}{(5+k)^2} i \hat{A}_{+} 
T_{+}^{(\frac{3}{2})} + \frac{18}{(5+k)^2} i \hat{A}_3 \hat{G}_{11} +
\frac{16(3+k)}{(5+k)^2} i \hat{A}_3 U^{(\frac{3}{2})}  
\nonu \\
& + &  \frac{8(-1+k)}{(5+k)^2} i \hat{B}_{-} \hat{G}_{12} -
\frac{8(-1+k)}{(5+k)^2} i \hat{B}_{-} T_{-}^{(\frac{3}{2})} 
-\frac{2(12+k)}{(5+k)^2}  i \hat{B}_3 \hat{G}_{11} \nonu \\
&- &  \frac{16}{(5+k)} i \hat{B}_3 U^{(\frac{3}{2})}  + \frac{3}{(5+k)}
T^{(1)} \hat{G}_{11} + \frac{2(867+359k+64k^2)}{3(5+k)^2(19+23k)} 
\pa \hat{G}_{11} \nonu \\
& - & \left. \frac{4(11+7k)}{3(5+k)^2} \pa 
U^{(\frac{3}{2})} \right](w) \nonu \\
& + & \frac{1}{(z-w)} \left[ \frac{1}{5} \pa 
\{ \hat{G}_{21} \, Q_{-}^{(3)}\}_{-2} \right. \nonu \\
&  - &  \frac{32(-870+396k+1049k^2+302k^3 +23k^4)}{
(5+k)^2(19+23k)(47+35k)} \left( \hat{T} \hat{G}_{11} -\frac{3}{8} \pa^2 
\hat{G}_{11}  \right) \nonu \\
& - & \left.  \frac{80(-1+k)}{(5+k)(47+35k)} 
\left( \hat{T} U^{(\frac{3}{2})} -\frac{3}{8} \pa^2 
U^{(\frac{3}{2})}  \right)  + 
Q^{(\frac{7}{2})}
\right](w) +\cdots.
\label{g21q3-}
\eea
All  the composite fields with spin-$\frac{5}{2}$ except 
$T^{(1)} U^{(\frac{3}{2})}(w)$ appearing in 
Table 4 of \cite{Ahn1311} arise in the second order pole in (\ref{g21q3-}). 
In other words, this implies that the OPEs between the $11$ currents 
and the second lowest higher spin currents will do not contain any 
quadratic expressions in the lowest higher spin currents. 
As explained before, the first and last components of the second 
${\cal N}=2$ multiplet in (\ref{nexthigher}) occur in the second order 
and first order poles in (\ref{g21q3-}) respectively. 
The higher spin-$\frac{7}{2}$ current occurs 
in the first-order pole in 
(\ref{g21q3-}) after subtracting the descendant field 
with coefficient $\frac{1}{5}$ for the spin-$\frac{5}{2}$ field
appearing 
in the second-order pole and the two quasi-primary fields where the numerical
factor $\frac{3}{8}$ is fixed. 
Recall that the two structure constants appearing in two quasi primary fields 
can be fixed by using the property that the above reduced first order pole 
subtracted by 
descendant terms should transform as a primary field.  

Similarly, 
 the first and last components of the third 
${\cal N}=2$ multiplet in (\ref{nexthigher}) can be obtained from the 
following OPE 
\bea
\hat{G}_{21}(z) \, R_{-}^{(3)}(w) & = & 
\frac{1}{(z-w)^3} \left[ -\frac{8(684+1035k+361k^2+14k^3)}{3(5+k)^2(19+23k)} 
\hat{G}_{22} + \frac{8(-1+2k)}{(5+k)^2} V^{(\frac{3}{2})} \right](w)
\nonu \\
& + & \frac{1}{(z-w)^2} \left[ \frac{4(6+k)}{(5+k)} R^{(\frac{5}{2})}
- \frac{(23+4k)}{(5+k)} V^{(\frac{5}{2})}  \right.
\nonu \\
&+ &  \frac{16}{(5+k)^2} i \hat{A}_{-} 
T_{-}^{(\frac{3}{2})} - \frac{2(9+2k)}{(5+k)^2} i \hat{A}_3 \hat{G}_{22} +
\frac{16}{(5+k)} i \hat{A}_3 V^{(\frac{3}{2})}  
\nonu \\
& - &  \frac{6(152+263k+55k^2+4k^3)}{(5+k)^2(19+23k)} 
i \hat{B}_{+} \hat{G}_{21} -
\frac{2(21+k)}{(5+k)^2} i \hat{B}_{+} T_{+}^{(\frac{3}{2})} 
\nonu \\
& + & \frac{6k}{(5+k)^2}  i \hat{B}_3 \hat{G}_{22} 
-   \frac{16(3+k)}{(5+k)^2} i \hat{B}_3 V^{(\frac{3}{2})}  
- 
\frac{3}{(5+k)}
T^{(1)} \hat{G}_{22} 
\nonu \\
& + & \left.  \frac{2(114+337k+119k^2+12k^3)}{3(5+k)^2(19+23k)} 
\pa \hat{G}_{22} 
 +  \frac{4(23+3k)}{3(5+k)^2} \pa 
V^{(\frac{3}{2})} \right](w) \nonu \\
& + & \frac{1}{(z-w)} \left[ \frac{1}{5} \pa 
\{ \hat{G}_{21} \, R_{-}^{(3)}\}_{-2} \right. \nonu \\
&  - &  \frac{32(684+1035k+361k^2+14k^3)}{
3(5+k)(19+23k)(47+35k)} \left( \hat{T} \hat{G}_{22} -\frac{3}{8} \pa^2 
\hat{G}_{22}  \right) \nonu \\
& + & \left.  \frac{32(-1+2k)}{(5+k)(47+35k)} 
\left( \hat{T} V^{(\frac{3}{2})} -\frac{3}{8} \pa^2 
V^{(\frac{3}{2})}  \right)  + 
R^{(\frac{7}{2})}
\right](w) +\cdots.
\label{g21r3-}
\eea
In this case,
all  the composite fields with spin-$\frac{5}{2}$ except 
$\hat{A}_{-} \hat{G}_{12}(w)$ 
and $T^{(1)} V^{(\frac{3}{2})}(w)$ appearing in 
Table 4 of \cite{Ahn1311} arise in the second order pole in (\ref{g21r3-}). 
As observed in previous case, in the OPE of (\ref{g21r3-}), the only linear 
terms in the lowest higher spin currents can occur.
The two structure constants appearing in two quasi primary fields 
can be fixed by using the fact that the first order pole subtracted by  
two descendant terms containing $\hat{G}_{22}(w)$ and $V^{(\frac{3}{2})}(w)$ 
terms should behave as a primary field under the 
stress energy tensor $\hat{T}(z)$.

Motivated by (\ref{g2112p2}) or the equation $(4.52)$ of \cite{Ahn1311},
one calculates the following OPE
\bea
\hat{G}_{21}(z) \, S^{(3)}(w) & = &
\frac{1}{(z-w)^3} \left[ \frac{4(1809+1382k-63k^2+28k^3)}{
3(5+k)^2(19+23k)} \hat{G}_{21} \right. \nonu \\
&- & \left. \frac{4(-2433-964k+731k^2+110k^3)}{
3(5+k)^2(19+23k)} T_{+}^{(\frac{3}{2})} 
\right](w) \nonu \\
& + & \frac{1}{(z-w)^2} \left[ \frac{(3-k)}{(5+k)} 
P_{+}^{(\frac{5}{2})} - \frac{6(3+k)}{(5+k)}  
W_{+}^{(\frac{5}{2})}  -\frac{4(-2+k)}{(5+k)^2} 
i \hat{A}_{-} \hat{G}_{11} \right. \nonu \\
& - &   \frac{4(7+4k)}{(5+k)^2}  i  \hat{A}_{-} 
U^{(\frac{3}{2})} -\frac{4(327+322k+59k^2)}{(5+k)^2(19+23k)} i \hat{A}_3 
\hat{G}_{21} + \frac{4}{(5+k)} i \hat{A}_3 
T_{+}^{(\frac{3}{2})} \nonu \\
& - &  \frac{12}{(5+k)^2} 
i \hat{B}_{-} \hat{G}_{22}  + \frac{4(17+2k)}{(5+k)^2} i \hat{B}_{-} 
V^{(\frac{3}{2})} \nonu \\
& - & \frac{4(171+368k+43k^2+6k^3)}{(5+k)^2(19+23k)} i
\hat{B}_3 \hat{G}_{21} \nonu \\
& +&  \frac{20(1+k)}{(5+k)^2} i \hat{B}_3 
T_{+}^{(\frac{3}{2})} -\frac{4(-26+29k+3k^2)}{(5+k)(19+23k)} 
T^{(1)} \hat{G}_{21}    \nonu \\
& -& \left. \frac{4(423+851k+118k^2+6k^3)}{3(5+k)^2(19+23k)} 
\pa \hat{G}_{21} -\frac{4(-55+86k-9k^2+2k^3)}{(5+k)^2(19+23k)} \pa
T_{+}^{(\frac{3}{2})}
\right](w) \nonu \\
&+ &  \frac{1}{(z-w)} \left[ \frac{1}{5} \pa \{ \hat{G}_{21} \, S^{(3)} 
\}_{-2} \right. \nonu \\
& + &  \frac{16(1809+1382k-63k^2+28k^3)}{3(5+k)
(19+23k)(47+35k)} \left( \hat{T} \hat{G}_{21} -\frac{3}{8}
\pa^2 \hat{G}_{21} \right) \nonu \\
&- & \left.   \frac{16(-2433-964k+731k^2+110k^3)}{3(5+k)(19+23k)(47+35k)}  
\left( \hat{T} T_{+}^{(\frac{3}{2})} -\frac{3}{8}
\pa^2  T_{+}^{(\frac{3}{2})}  \right) 
 +
 S_{+}^{(\frac{7}{2})}
\right](w) 
\nonu \\
& + & \cdots.
\label{g21s3}
\eea
All  the composite fields with spin-$\frac{5}{2}$ except 
$T^{(1)} T_{+}^{(\frac{3}{2})}(w)$ appearing in 
Table 4 of \cite{Ahn1311} arise in the second order pole in (\ref{g21s3}). 
The linear terms in the lowest higher spin currents arise in (\ref{g21s3}).  
The factor $(k-3)$ appears in the $P_{+}^{(\frac{5}{2})}(w)$ term in the second 
order pole of (\ref{g21s3}).
The analysis for the two quasi primary fields in the first order pole
can be done as before.

Similarly,
by taking other spin-$\frac{3}{2}$ current with same higher spin current
\bea
\hat{G}_{12}(z) \, S^{(3)}(w) & = &
-\frac{1}{(z-w)^3} \left[ \frac{4(1695+1586k+351k^2+28k^3)}{
3(5+k)^2(19+23k)} \hat{G}_{12} \right. \nonu \\
&- & \left. \frac{4(-2205-1372k-97k^2+110k^3)}{
3(5+k)^2(19+23k)} T_{-}^{(\frac{3}{2})} 
\right](w) 
\nonu \\
 \nonu \\
& + & \frac{1}{(z-w)^2} \left[ \frac{(3-k)}{(5+k)} 
P_{-}^{(\frac{5}{2})} - \frac{6(3+k)}{(5+k)}  
W_{-}^{(\frac{5}{2})}  -\frac{4k}{(5+k)^2} 
i \hat{A}_{+} \hat{G}_{22} \right. \nonu \\
& + &   \frac{4(11+4k)}{(5+k)^2}  i  \hat{A}_{+} 
V^{(\frac{3}{2})} -\frac{4(593+682k+105k^2)}{(5+k)^2(19+23k)} i \hat{A}_3 
\hat{G}_{12} + \frac{4(17+k)}{(5+k)^2} i \hat{A}_3 
T_{-}^{(\frac{3}{2})} \nonu \\
& - &  \frac{4}{(5+k)^2} 
i \hat{B}_{+} \hat{G}_{11}  - \frac{4(13+2k)}{(5+k)^2} i \hat{B}_{+} 
U^{(\frac{3}{2})} \nonu \\
& - & \frac{4(209+452k+89k^2+6k^3)}{(5+k)^2(19+23k)} i
\hat{B}_3 \hat{G}_{12} \nonu \\
& +&  \frac{4}{(5+k)} i \hat{B}_3 
T_{-}^{(\frac{3}{2})} -\frac{4(-64-17k+3k^2)}{(5+k)(19+23k)} 
T^{(1)} \hat{G}_{12}    \nonu \\
& +& \left. \frac{4(309+713k+118k^2+6k^3)}{3(5+k)^2(19+23k)} 
\pa \hat{G}_{12} -\frac{4(-131-6k-9k^2+2k^3)}{(5+k)^2(19+23k)} \pa
T_{-}^{(\frac{3}{2})}
\right](w) 
\nonu \\
&+ &  \frac{1}{(z-w)} \left[ \frac{1}{5} \pa \{ \hat{G}_{12} \, S^{(3)} 
\}_{-2} \right. \nonu \\
& - &  \frac{16(1695+1586k+351k^2+28k^3)}{3(5+k)
(19+23k)(47+35k)} \left( \hat{T} \hat{G}_{12} -\frac{3}{8}
\pa^2 \hat{G}_{12} \right) \nonu \\
&- & \left.   \frac{16(-2205-1372k-97k^2+110k^3)}{3(5+k)(19+23k)(47+35k)}  
\left( \hat{T} T_{-}^{(\frac{3}{2})} -\frac{3}{8}
\pa^2  T_{-}^{(\frac{3}{2})}  \right)  +
 S_{-}^{(\frac{7}{2})} \right](w) \nonu \\
& + & \cdots.
\label{g12s3}
\eea
The composite field with spin-$\frac{5}{2}$, 
$T^{(1)} T_{-}^{(\frac{3}{2})}(w)$, appearing in 
Table 4 of \cite{Ahn1311} does not 
arise in the second order pole in (\ref{g12s3}). 
The factor $(k-3)$ appears in the $P_{-}^{(\frac{5}{2})}(w)$ term in the second 
order pole of (\ref{g12s3}).
The structure constants appearing in the two quasi primary fields 
in the first order pole can be determined as before.
No quadratic terms in the lowest higher spin current in (\ref{g12s3}) appear.

The first order poles in (\ref{g21q3-}), (\ref{g21r3-}), (\ref{g21s3}) and 
(\ref{g12s3}) have common feature in the sense that the two 
quasi primary fields for fixed $U(1)$ charge occur 
\footnote{
\label{exp2}
One has the following expressions for the second and third ${\cal N}=2$ 
multiplets in (\ref{nexthigher})
\bea
Q^{(\frac{5}{2})}(z) & = & -\frac{4 i \sqrt{2}}{(5+k)^2} \left( K^1 V^4 V^{11}
+ K^1 V^5 V^{12}  \right)(z) + \mbox{other 27 terms},
\nonu \\
Q_{+}^{(3)}(z) & = &  -\frac{16 ik(3+k)}{(5+k)^3(19+23k)} \left( K^1 
J^{10} V^1 V^{\bar{1}}
+ K^1 J^{10} V^2 V^{\bar{2}} \right)(z) +   \mbox{other 307 terms},
\nonu \\
Q_{-}^{(3)}(z) & = &  \frac{8 i}{(5+k)^3} \left( J^{10} J^{\bar{10}} V^1 
V^{10}
+ J^{10} J^{\bar{10}} V^2 V^{11} \right)(z) +   \mbox{other 143 terms},
\nonu \\
Q^{(\frac{7}{2})}(z) & = & \frac{8i\sqrt{2}(763 + 1582 k + 623 k^2 + 92 k^3)}{
(5+k)^3(19 + 23 k) (47 + 35 k)} \left( J^{10} V^1 V^1 V^{\bar{1}}
+ J^{10} V^1 V^2 V^{\bar{2}}  \right)(z) + \mbox{other 850 terms},
\nonu \\
R^{(\frac{5}{2})}(z) & = & -\frac{4 \sqrt{2}}{(5+k)^2} \left( K^{\bar{1}} V^7 V^{\bar{10}}
+ i K^{\bar{1}} V^{\bar{4}} V^{\bar{11}}  \right)(z) + \mbox{other 27 terms},
\nonu \\
R_{+}^{(3)}(z) & = &  -\frac{24 i(9+k)}{(5+k)^2(19+23k)} \left( V^1 V^{\bar{1}} 
V^{\bar{9}}
+ V^2 V^{\bar{2}} V^{\bar{9}} \right)(z) +   \mbox{other 143 terms},
\nonu \\
R_{-}^{(3)}(z) & = & - \frac{16 ik(3+k)}{(5+k)^3(19+23k)} 
\left( K^{\bar{1}} J^{\bar{10}} V^1 
V^{\bar{1}}
+ K^{\bar{1}} J^{\bar{10}} V^2 V^{\bar{2}} \right)(z) +   \mbox{other 307 terms},
\nonu \\
R^{(\frac{7}{2})}(z) & = & \frac{16i\sqrt{2}(1482 + 2403 k + 902 k^2 + 133 k^3)}{
3(5+k)^3(19 + 23 k) (47 + 35 k)} \left( J^{\bar{10}} V^1 V^{\bar{1}} V^{\bar{1}}
+ J^{\bar{10}} V^2 V^{\bar{1}} V^{\bar{2}}  \right)(z) + \mbox{other 859 terms}.
\nonu
\eea
The four higher spin-$\frac{7}{2}$ currents, $Q^{(\frac{7}{2})}(z)$, 
$R^{(\frac{7}{2})}(z)$, and $S_{\pm}^{(\frac{7}{2})}(z)$, 
transform as $({\bf 2},{\bf 2})$
under the $SU(2) \times SU(2)$ as before \cite{GG1305}.}.



\subsection{Construction of one higher spin-$4$ current: $S^{(4)}(z)$}

How does one determine the highest component of 
the last ${\cal N}=2$ multiplet in (\ref{nexthigher})?
One way to obtain the highest spin-$4$ current is to consider the 
following OPE
\bea
\hat{G}_{21}(z) \, S_{-}^{(\frac{7}{2})}(w)  & = & \frac{1}{(z-w)^4} \,
\left[-\frac{32(-24198 - 29457 k + 10669 k^2 + 23557 k^3 + 3085 k^4)}{
5(5+k)^3(19+23k)(47+35k)} i \hat{A}_3 \right. \nonu \\
&+ &   \frac{32(120642 + 228177 k + 152639 k^2 + 36707 k^3 + 3635 k^4)}{
15(5+k)^3(19+23k)(47+35k)} i \hat{B}_3  \nonu \\
&- & \left.  \frac{16(-5859 - 84466 k - 83502 k^2 - 22818 k^3 + 1885 k^4)}{
15(5+k)^3(19+23k)(47+35k)} T^{(1)}\right](w) \nonu \\
& + & \frac{1}{(z-w)^3} \, \left[ 
\frac{12(-3+k)}{5(5+k)} P^{(2)}
-\frac{8(-57 + 11 k + 6 k^2)}{5(5+k)^3} W^{(2)}
\right. \nonu \\
& - & \frac{32(-145224 - 597681 k - 737779 k^2 - 289037 k^3 + 22531 k^4 + 21350 k^5)}{15(5+k)^3(3+7k)(19+23k)(47+35k)} \hat{T}
 \nonu \\
&-& \frac{8(-9327 + 13091 k + 84711 k^2 + 37305 k^3 + 4900 k^4)}{
15(5+k)^3(19+23k)(47+35k)} T^{(2)} 
\nonu \\
& + & \frac{32(-40371 - 116053 k - 94358 k^2 - 24257 k^3 + 35 k^4)}{
15(5+k)^3(19+23k)(47+35k)} (\hat{A}_1 \hat{A}_1 +
 \hat{A}_2 \hat{A}_2) 
\nonu \\
&+ & \frac{16(-136578 - 409469 k - 456070 k^2 - 172309 k^3 + 70 k^4)}{
15(5+k)^3(19+23k)(47+35k)} \hat{A}_3 \hat{A}_3
\nonu \\
& - & \frac{32(-51273 + 40960 k + 212891 k^2 + 122858 k^3 + 8680 k^4)}{
15(5+k)^3(19+23k)(47+35k)} \hat{A}_3 \hat{B}_3
\nonu \\
&- & \frac{32(48408 + 104977 k + 57260 k^2 + 15821 k^3 + 7210 k^4)}{
15(5+k)^3(19+23k)(47+35k)} 
(\hat{B}_1 \hat{B}_1 +\hat{B}_2 \hat{B}_2)
\nonu \\
&- & \frac{16(96816 + 217991 k + 148000 k^2 + 45067 k^3 + 10298 k^4 + 840 k^5)}
{15(5+k)^3(19+23k)(47+35k)} \hat{B}_3 \hat{B}_3
\nonu \\
& + & \frac{16(-24198 - 29457 k + 10669 k^2 + 23557 k^3 + 3085 k^4)}{
5(5+k)^3(19+23k)(47+35k)} i \pa \hat{A}_3
\nonu \\
& - & \frac{16k(120642 + 228177 k + 152639 k^2 + 36707 k^3 + 3635 k^4)}{
15(5+k)^3(19+23k)(47+35k)} i \pa \hat{B}_3
\nonu \\
& + & \frac{8(-5859 - 84466 k - 83502 k^2 - 22818 k^3 + 1885 k^4)}{
15(5+k)^2(19+23k)(47+35k)} \pa T^{(1)}
\nonu \\
&+& \left. 
\frac{16(-389 - 39 k + 242 k^2)}{5(5+k)^2(19+23k)} i T^{(1)} \hat{A}_3
+\frac{16(19 - 113 k - 84 k^2 + 4 k^3)}{
5(5+k)^2(19+23k)} i T^{(1)} \hat{B}_3 \right](w)
 \nonu \\
& + &  \frac{1}{(z-w)^2} \, \left[
\frac{4}{(5+k)} i P^{(2)} \hat{B}_3 -
\frac{4}{(5+k)} i P^{(2)} \hat{A}_3  +\frac{2(13+3k)}{(5+k)} 
S^{(3)}  \right. \nonu \\
&+&  \frac{6(-3+k)}{5(5+k)} P^{(3)}  -\frac{4(22+9k)}{5(5+k)} W^{(3)}
-\frac{256(-201 - 613 k - 199 k^2 + 5 k^3)}{15(5+k)(19+23k)(47+35k)} 
T^{(1)} \hat{T}
\nonu \\
& -& \frac{4(1493 + 1115 k + 98 k^2)}{
5(5+k)^2(19+23k)} i T^{(1)} \pa \hat{A}_3 
+\frac{4(1732 + 2121 k + 365 k^2 + 12 k^3)}{
5(5+k)^2(19+23k)} i T^{(1)} \pa \hat{B}_3 
\nonu \\
& -& \frac{8(8+7k)}{5(5+k)^2} i \hat{A}_{+} V_{+}^{(2)} 
-\frac{32}{(5+k)^2} i \hat{A}_{-} U_{-}^{(2)} 
-\frac{16(59+7k)}{15(5+k)^2} i \hat{A}_{3} T^{(2)} 
\nonu \\
& + & \frac{4(-131009 - 210397 k + 30389 k^2 + 85501 k^3 + 4620 k^4)}{
15(5+k)^3(19+23k)(47+35k)} \hat{A}_{+} \pa \hat{A}_{-}
\nonu \\
& + & 
\frac{4(535451 + 1125793 k + 640141 k^2 + 73811 k^3 + 4620 k^4)}{
15(5+k)^3(19+23k)(47+35k)} \hat{A}_{-} \pa \hat{A}_{+}
\nonu \\
&-& \frac{32(-213447 - 561982 k - 242654 k^2 + 62770 k^3 + 30905 k^4)}{
15(5+k)^2(3+7k)(19+23k)(47+35k)} i \hat{A}_3 \hat{T}
\nonu \\
&+& \frac{8(43643 + 135346 k + 182215 k^2 + 78956 k^3 + 4620 k^4)}{
15(5+k)^3(19+23k)(47+35k)} i \hat{A}_3 \pa \hat{A}_3
\nonu \\
& + & \frac{16}{(5+k)} i \hat{A}_3 W^{(2)} 
\nonu \\
& + &  \frac{8(-392947 - 520298 k + 64729 k^2 + 233948 k^3 + 39900 k^4)}{
15(5+k)^3(19+23k)(47+35k)} i \hat{A}_3 \pa \hat{B}_3
\nonu \\
&+&  \frac{4(1091 + 709 k + 102 k^2)}{
5(5+k)^2(19+23k)} i \hat{A}_3 \pa T^{(1)}
+  \frac{8(24 + k)}{
5(5+k)^2} i \hat{B}_{+}  U_{+}^{(2)} 
+ \frac{8(-1 + k)}{
(5+k)^2} i \hat{B}_{-}  V_{-}^{(2)} 
\nonu \\
&-&  \frac{4(-152213 - 467881 k - 461995 k^2 - 135887 k^3 + 4200 k^4)}{
15(5+k)^3(19+23k)(47+35k)}  \hat{B}_{+} \pa \hat{B}_{-}
\nonu \\
&+&  \frac{4(252229 + 447515 k + 208535 k^2 + 23425 k^3 + 13440 k^4)}{
15(5+k)^3(19+23k)(47+35k)}  \hat{B}_{-} \pa \hat{B}_{+}
\nonu \\
&+&  \frac{32(18753 + 292055 k + 584767 k^2 + 370207 k^3 + 77464 k^4 + 
3010 k^5)}{
15(5+k)^2(3+7k)(19+23k)(47+35k)}  i \hat{B}_{3} \hat{T}
\nonu \\
&+&  \frac{64(20 + 7k)}{
15(5+k)^2} i \hat{B}_{3}  T^{(2)}  
- \frac{16(3 + k)}{
(5+k)^2} i \hat{B}_{3}  W^{(2)}  
\nonu \\
& - & \frac{12(570 + 725 k + 123 k^2 + 4 k^3)}{
5(5+k)^2(19+23k)} i \hat{B}_{3}  \pa T^{(1)}  
\nonu \\
& - &  \frac{8(184361 + 676672 k + 781117 k^2 + 342086 k^3 + 48720 k^4)}{
15(5+k)^3(19+23k)(47+35k)}   \hat{B}_{3} \pa \hat{A}_3
\nonu \\
&+&  \frac{8(227225 + 470020 k + 300589 k^2 + 56762 k^3 + 4200 k^4)}{
15(5+k)^3(19+23k)(47+35k)}   \hat{B}_{3} \pa \hat{B}_3
\nonu \\
&+&  \frac{4(202221 + 1319940 k + 2127646 k^2 + 1300340 k^3 + 238973 k^4 + 
 10000 k^5)}{
5(5+k)^2(3+7k)(19+23k)(47+35k)} \pa \hat{T}
\nonu \\
&-&  \frac{2(-130131 - 241473 k - 146133 k^2 - 16163 k^3 + 260 k^4)}{
5(5+k)^2(19+23k)(47+35k)} \pa T^{(2)}
\nonu \\
&+&  \frac{2(143+37k)}{
5(5+k)^2} \pa W^{(2)}
-\frac{8(-13+3k)}{5(5+k)^2} \hat{G}_{11} \hat{G}_{22}
-\frac{8(37+3k)}{5(5+k)^2} \hat{G}_{11} V^{(\frac{3}{2})}
\nonu \\
&+&  \frac{16(232555 + 471215 k + 264827 k^2 + 41029 k^3 + 2190 k^4)}{
15(5+k)^3(19+23k)(47+35k)} i \pa^2 \hat{A}_3
\nonu \\
&+ &  \frac{32(-6251 + 55591 k + 124930 k^2 + 64288 k^3 + 4989 k^4 + 365 k^5)}{
15(5+k)^3(19+23k)(47+35k)} i \pa^2 \hat{B}_3
\nonu \\
&+&  \frac{8(66831 + 131711 k + 62169 k^2 - 2127 k^3 + 40 k^4)}{
15(5+k)^2(19+23k)(47+35k)} i \pa^2 T^{(1)}
\nonu \\
&-&  \frac{16(26898 + 57784 k + 36995 k^2 + 6638 k^3 + 385 k^4)}{
5(5+k)^2(19+23k)(47+35k)} \hat{G}_{12} \hat{G}_{21}
\nonu \\
&-&  \frac{8(-700 - 704 k - 117 k^2 + 3 k^3)}{
5(5+k)^2(19+23k)} \hat{G}_{12} T_{+}^{(\frac{3}{2})}
\nonu \\
& + & \frac{8(-36184 - 71452 k - 47423 k^2 - 4530 k^3 + 25 k^4)}{
5(5+k)^2(19+23k)(47+35k)} \hat{G}_{21} T_{-}^{(\frac{3}{2})}
\nonu \\
&-&  \frac{16(2014 + 2485 k - 95 k^2 + 6 k^3)}{
15(5+k)^3(19+23k)} i \hat{A}_{+} \hat{A}_{-} \hat{B}_3
+ \frac{8(11 +9 k)}{
5(5+k)^2} \hat{G}_{22} U^{(\frac{3}{2})}
\nonu \\
& + &  \frac{32(2126 + 2639 k + 5 k^2)}{
15(5+k)^3(19+23k)} i \hat{A}_{3} \hat{A}_{3} \hat{A}_3
+ \frac{16(1687 + 2173 k + 10 k^2)}{
15(5+k)^3(19+23k)} i \hat{A}_{+} \hat{A}_{-} \hat{A}_3
\nonu \\
&- &  \frac{32(1183 + 1683 k + 229 k^2 + k^3)}{
5(5+k)^3(19+23k)} i \hat{A}_{3} \hat{A}_{3} \hat{B}_3
+  \frac{16(889 + 1207 k + 10 k^2)}{
15(5+k)^3(19+23k)} i \hat{A}_{3} \hat{B}_{+} \hat{B}_{-}
\nonu \\
&+ & \frac{32(-330 - 153 k + 223 k^2 + 2 k^3)}{
5(5+k)^3(19+23k)} i \hat{A}_{3} \hat{B}_{3} \hat{B}_3
- \frac{16(-243 - 274 k + k^2)}{
5(5+k)^2(19+23k)} T^{(1)} \hat{A}_{3} \hat{A}_3
\nonu \\
&- &  \frac{16(-266 + 67 k + 319 k^2 + 6 k^3)}{
15(5+k)^3(19+23k)} i \hat{B}_{-} \hat{B}_{+} \hat{B}_3
+ \frac{32(-148 - 159 k + k^2)}{
5(5+k)^2(19+23k)} T^{(1)} \hat{A}_{3} \hat{B}_3
\nonu \\
& - &   \frac{32(-133 - 109 k - 13 k^2 + 3 k^3)}{
15(5+k)^3(19+23k)} i \hat{B}_{3} \hat{B}_{3} \hat{B}_3
- \frac{8(-11 + 27 k + 2 k^2)}{
5(5+k)^2(19+23k)} T^{(1)} \hat{B}_{+} \hat{B}_{-}
\nonu \\
&- & \left.
\frac{16(-53 - 44 k + k^2)}{
5(5+k)^2(19+23k)} T^{(1)} \hat{B}_{3} \hat{B}_{3}
-\frac{8(-201 - 203 k + 2 k^2)}{
5(5+k)^2(19+23k)} \hat{A}_{-} T^{(1)} \hat{A}_{+}
\right](w)
\nonu \\
& + &  \frac{1}{(z-w)} \, \left[ \frac{1}{6}
 \pa \{ \hat{G}_{21} \, 
 S_{-}^{(\frac{7}{2})} \}_{-2} 
-\frac{24 (-57+11 k+6 k^2))}{(5+k) (265+149 k)}
\left( \hat{T} W^{(2)} -\frac{3}{10} \pa^2 W^{(2)}\right)
\right. \nonu \\
& - & 
\frac{8 (-9327+13091 k+84711 k^2+37305 k^3+4900 k^4)}{
(5+k) (19+23 k) (47+35 k) (265+149 k)}
\left( \hat{T} T^{(2)} -\frac{3}{10} \pa^2 T^{(2)}\right)
\nonu \\
&+ & \frac{36 (-3+k)}{(265+149 k)}
\left( \hat{T} P^{(2)} -\frac{3}{10} \pa^2 P^{(2)}\right)
\nonu \\
&+ & 
\frac{32 
}{
(5+k) (3+7 k) (19+23 k) (29+25 k) (47+35 k) (155+127 k) (265+149 k)}
\nonu \\
& \times & (-705062520-4044532095 k-8746728739 k^2
-9146242164 k^3\nonu \\
& - & 4619929142 k^4
 -  717097199 k^5+225222129 k^6+67792130 k^7)
\nonu \\
&\times & 
 \left( \hat{T} \hat{T} -\frac{3}{10} \pa^2 \hat{T} \right)
\nonu \\
&+ & \frac{16 (-136578-409469 k-456070 k^2-172309 k^3+70 k^4)}{
(5+k)^2 (19+23 k) (47+35 k) (265+149 k)}
\nonu \\
&\times &
\left( \hat{T} \hat{A}_3 \hat{A}_3 -\frac{3}{10} \pa^2 (\hat{A}_3 \hat{A}_3) 
\right)
\nonu \\ 
&-  & 
\frac{32 (-51273+40960 k+212891 k^2+122858 k^3+8680 k^4)}{
(5+k)^2 (19+23 k) (47+35 k) (265+149 k)}
\nonu \\
& \times & 
\left( \hat{T} \hat{A}_3 \hat{B}_3 -\frac{3}{10} \pa^2 
(\hat{A}_3 \hat{B}_3) 
\right)
\nonu \\
&+ & 
\frac{48  (-389-39 k+242 k^2)}{(5+k) (19+23 k) (265+149 k)}
i \left( \hat{T} T^{(1)} \hat{A}_3  -\frac{3}{10} \pa^2 
(T^{(1)} \hat{A}_3) 
\right)
\nonu \\
&- & 
\frac{16 (96816+217991 k+148000 k^2+45067 k^3+10298 k^4+840 k^5)}{
(5+k)^2 (19+23 k) (47+35 k) (265+149 k)}
 \nonu \\
& \times & \left( \hat{T} \hat{B}_3 \hat{B}_3 -\frac{3}{10} \pa^2 
(\hat{B}_3 \hat{B}_3) 
\right)
\nonu \\
& + & 
\frac{48  (19-113 k-84 k^2+4 k^3)}{(5+k) (19+23 k) (265+149 k)}
i \left( \hat{T} T^{(1)} \hat{B}_3  -\frac{3}{10} \pa^2 
(T^{(1)} \hat{B}_3) 
\right)
\nonu \\
&+ & 
\frac{8
}
{15 (5+k)^2 (1+5 k) (19+23 k) (29+25 k) (47+35 k)}
  (-3993162-11110911 k
\nonu \\
& - & 10230596 k^2
-  1203174 k^3+1936950 k^4+387725 k^5)
 i 
\nonu \\
& \times & \left( \hat{T} \pa \hat{A}_3  -\frac{1}{2} \pa 
\hat{T} \hat{A}_3 -\frac{1}{4} \pa^3 \hat{A}_3 
\right)
\nonu \\
&- &  
\frac{8
}{45 (5+k)^2 (1+5 k) (19+23 k) (29+25 k) (47+35 k)}
 \nonu \\
& \times &  
(1742688+29985702 k+69123135 k^2+61531136 k^3+27509730 k^4
\nonu \\
&& +6413250 k^5+454375 k^6)
i \left( \hat{T} \pa \hat{B}_3  -\frac{1}{2} \pa 
\hat{T} \hat{B}_3 -\frac{1}{4} \pa^3 \hat{B}_3 
\right)
\nonu \\
&+ &  
\frac{4 (-5859-84466 k-83502 k^2-22818 k^3+1885 k^4)}{
9 (5+k) (1+5 k) (19+23 k) (47+35 k)}
\nonu \\
& \times & \left( \hat{T} \pa T^{(1)}  -\frac{1}{2} \pa 
\hat{T} T^{(1)} -\frac{1}{4} \pa^3 T^{(1)}
\right)
\nonu \\
&+ & 
\frac{32 (-40371-116053 k-94358 k^2-24257 k^3+35 k^4)}{
5 (5+k)^2 (19+23 k) (29+25 k) (47+35 k)}
\nonu \\
& \times &  \left( \hat{T} \hat{A}_{+} \hat{A}_{-}  -\frac{1}{2} \pa^2 
\hat{A}_{+} \hat{A}_{-} -\frac{1}{2} \hat{A}_{+} \pa^2 \hat{A}_{-}
-\frac{1}{12} i \pa^3 \hat{A}_3 +\frac{1}{2} i \pa \hat{T} \hat{A}_3
\right)
\nonu \\
&- & \frac{32 (48408+104977 k+57260 k^2+15821 k^3+7210 k^4)}{
5 (5+k)^2 (19+23 k) (29+25 k) (47+35 k)}
\nonu \\
& \times &
 \left( \hat{T} \hat{B}_{+} \hat{B}_{-}  -\frac{1}{2} \pa^2 
\hat{B}_{+} \hat{B}_{-} -\frac{1}{2} \hat{B}_{+} \pa^2 \hat{B}_{-}
-\frac{1}{12} i \pa^3 \hat{B}_3 +\frac{1}{2} i \pa \hat{T} \hat{B}_3
\right)
\nonu \\
&+ & \left.
S^{(4)} \right](w) +  \cdots.
\label{g21s7half-}
\eea
All  the composite fields with spin-$2$ except 
$T^{(1)} T^{(1)}(w)$ appearing in 
Table $3$ of \cite{Ahn1311} arise 
in the third order pole in (\ref{g21s7half-}).
No quadratic terms in the lowest higher spin current in (\ref{g21s7half-}) 
appear.
The $(k-3)$ factor appears in $P^{(2)}(w)$ term of third order pole, 
 $P^{(3)}(w)$ term of second order pole, and 
$( \hat{T} P^{(2)} -\frac{3}{10} \pa^2 P^{(2)})(w)$ term of the first order pole.
One can reexpress the third order pole in terms of 
the descendant field of spin-$1$ current appearing in the fourth order pole
plus other (quasi) primary fields as describe before.
There is no descendant field for spin-$2$ fields appearing 
in the third order pole in the second order pole
because the difference between the two fields in the left hand side
is equal to $\frac{3}{2}-\frac{7}{2} =-2$. 
Furthermore, there is no descendant field in the second order pole
for spin-$1$ fields in the fourth order pole.
Note the presence of higher spin-$3$ current, $S^{(3)}(w)$ in the second order 
pole.
In the first order pole, all the expressions except the first term and last
term are expressed in terms of $14$ quasi primary fields. 
The first four quasi primary fields 
have standard quadratic expressions with the coefficient $\frac{3}{10}$.
The next five quasi primary fields are cubic in their expressions.
This feature is rather special because all the previous constructions on 
the quasi primary fields are restricted to the quadratic case.
See for example the reference $[54]$ of \cite{Ahn1311}.

One can easily check that $\hat{A}_3 \hat{A}_3(w)[\hat{B}_3 \hat{B}_3(w)]$ 
is a quasi primary field.
The next three quasi primary fields are again quadratic in their
expressions.
The last two quasi primary fields are cubic in their expressions.
Let us emphasize that  
when one calculates the OPE between the stress energy tensor $\hat{T}(z)$
and the first order pole of (\ref{g21s7half-}) subtracted by 
$ \frac{1}{6}
 \pa \{ \hat{G}_{21} \, 
 S_{-}^{(\frac{7}{2})} \}_{-2}(w)$,
the fourth order pole of this OPE 
contains nontrivial expressions with spin-$2$ with vanishing $U(1)$ charge.
One can rewrite these expressions in terms of $14$ independent terms 
appearing in the third order pole of (\ref{g21s7half-}). 
This implies that the quasi primary fields look like as 
$\hat{T} \Phi(w)$ plus other terms where $\Phi(w)$ is above $14$ independent 
terms: the field contents with vanishing $U(1)$ charge in Table $3$ of
\cite{Ahn1311} where $T^{(1)} T^{(1)}(w)$ is replaced by $P^{(2)}(w)$.

Therefore, one has the complete expressions for the last ${\cal N}=2$ 
multiplet in (\ref{nexthigher})
\footnote{
\label{exp3}
Also the last ${\cal N}=2$ multiplet in (\ref{nexthigher})
has the following expressions
\bea
S^{(3)}(z) & = & \frac{8}{(5+k)^2} \left(  V^4 V^{11} V^{\bar{10}}
+ V^5 V^{12} V^{\bar{10}}  \right)(z) + \mbox{other 633 terms},
\nonu \\
S_{+}^{(\frac{7}{2})}(z) & = &  \frac{ 8 \sqrt{2}}{(5+k)^3} \left(
J^{10} V^4 V^{\bar{4}} V^{\bar{10}}
+ J^{10} V^5 V^{\bar{5}} V^{\bar{10}} \right)(z) +   \mbox{other 879 terms},
\nonu \\
S_{-}^{(\frac{7}{2})}(z) & = &  \frac{8 \sqrt{2}}{(5+k)^3} \left( J^{\bar{10}} 
V^{5} V^{10} 
V^{\bar{5}}
+ J^{\bar{10}} V^5 V^{11} V^{\bar{6}} \right)(z) +   \mbox{other 876 terms},
\nonu \\
S^{(4)}(z) & = & \frac{16
}{
3(5+k)^3(19 + 23 k)(29+25k) (47 + 35 k)}
(5849400525 + 21901659454 k + 31822883478 k^2 
\nonu \\
&+ & 
22570705436 k^3 + 
 8016661577 k^4 + 1327087470 k^5 + 93733660 k^6) 
\left( V^1 V^1 V^{\bar{1}} V^{\bar{1}}
+ 2 V^1 V^2 V^{\bar{1}} V^{\bar{2}}  \right)(z) 
\nonu \\
& + & \mbox{other 4926 terms}.
\nonu
\eea
Note that the  higher spin-$2$ current, $P^{(2)}(z)$, 
and the higher spin-$4$ current, $S^{(4)}(z)$, 
transform as $({\bf 1},{\bf 1})$ under the $SU(2) \times US(2)$ 
respectively \cite{GG1305}.}.

\section{The OPEs between the sixteen 
 lowest higher spin currents in the Wolf space coset    }

We would like to construct the OPEs between the ${\cal N}=4$ multiplet
in (\ref{lowesthigher}) and itself and to express them in terms of
the $11$ generators of large ${\cal N}=4$ nonlinear superconformal algebra,
$16$ generators of  the ${\cal N}=4$ multiplet
in (\ref{lowesthigher}) and $16$
generators of
 the other ${\cal N}=4$ multiplet
in (\ref{nexthigher}) (and their composite operators with possible 
derivatives).

Because the pole structures are written in terms of ${\cal N}=2$ WZW affine
currents, the main step is to write them in terms of
above $(11+16+16)$ generators.
The $(11+16)$ generators in terms of ${\cal N}=2$ affine currents are
given in \cite{Ahn1311} and the other $16$ generators are given in the 
footnotes \ref{exp1}, \ref{exp2} and \ref{exp3}.
Basically we would like to calculate the following OPEs
\bea
\left(
\begin{array}{cccc}
T^{(1)} & T_{+}^{(\frac{3}{2})} & T_{-}^{(\frac{3}{2})} & 
T^{(2)} \\
U^{(\frac{3}{2})} & U_{+}^{(2)} & U_{-}^{(2)} &  U^{(\frac{5}{2})} 
 \\
V^{(\frac{3}{2})} & V_{+}^{(2)} & 
V_{-}^{(2)} &  V^{(\frac{5}{2})} 
\\
W^{(2)} & W_{+}^{(\frac{5}{2})} &
W_{-}^{(\frac{5}{2})} & W^{(3)} \\
\end{array}
\right)(z)  \,
\left(
\begin{array}{cccc} 
T^{(1)} & T_{+}^{(\frac{3}{2})} & T_{-}^{(\frac{3}{2})} & 
T^{(2)} \\
U^{(\frac{3}{2})} & U_{+}^{(2)} & U_{-}^{(2)} & U^{(\frac{5}{2})} 
 \\
V^{(\frac{3}{2})} & V_{+}^{(2)} & 
V_{-}^{(2)} & V^{(\frac{5}{2})} 
\\
W^{(2)} & W_{+}^{(\frac{5}{2})} & 
W_{-}^{(\frac{5}{2})} & W^{(3)} \\
\end{array} \right)(w).  
\label{opediagram}
\eea
In the Tables $2, 3, \cdots, 7$ in \cite{Ahn1311}, 
the composite fields with spins, $s=1, \frac{3}{2}, 2, \frac{5}{2}, 3, 
\frac{7}{2}$ are listed according to their $U(1)$ charges. 
In the OPEs (\ref{opediagram}), 
the composite fields
with maximum spin $5$ appear in the first order pole in the OPE 
$W^{(3)}(z) \, W^{(3)}(w)$. 
Then one should find the composite fields with spins 
$s=4, \frac{9}{2}$ and $5$.
Furthermore, by including the extra $16$ generators in (\ref{nexthigher}),
the Tables of \cite{Ahn1311} have more composite fields and the extra Tables 
with above spins $s=4, \frac{9}{2}$ and $5$ can be obtained. 
We will present the OPEs (\ref{opediagram}) completely except 
the first order pole of $W^{(3)}(z) \, W^{(3)}(w)$ 
\footnote{The first order pole of this OPE is very special 
in the sense that there is no new primary field in this singular term.
Suppose that there exists such a singular term.
After reversing the arguments $z$ and $w$ and expanding around $w$, 
the same term will appear with opposite sign. This implies 
that it is identically zero. See also Appendix in the reference 
$[63]$ of \cite{Ahn1311}. }.

One can calculate the OPEs (\ref{opediagram}) step by step.
Let us take $T^{(1)}(z)$ in the first operator
and take $16$ operators for the second operator.
The nontrivial four OPEs are given in (\ref{t1uv}) and (\ref{t1w+w-}).
Now let us compute the remaining nontrivial OPEs as follows.
 
$\bullet$ The OPEs between the higher spin-$1$ current and the
first ${\cal N}=2$ multiplet in (\ref{lowest})

We use the equations $(4.7), (4.10), (4.14)$, and $(4.17)$ of \cite{Ahn1311}
where the expressions for 
the ${\cal N}=2$ WZW affine currents are known. 
It turns out that
\bea
T^{(1)}(z) \, T^{(1)}(w)  & = & \frac{1}{(z-w)^2} \,
\left[\frac{6k}{(5+k)} \right] +\cdots,
\nonu \\
T^{(1)}(z) \, T_{\pm}^{(\frac{3}{2})}(w) & = & \pm \frac{1}{(z-w)}  \,
T_{\pm}^{(\frac{3}{2})}(w)  +\cdots,
\nonu \\
T^{(1)}(z) \, T^{(2)}(w) & = & \frac{1}{(z-w)^2}  \,
\left[ -\frac{6i}{(5+k)} \hat{A}_3 -\frac{2ik}{(5+k)} \hat{B}_3 +
\frac{(3+k)}{(3+7k)} T^{(1)}\right](w)  +\cdots.
\label{t1first}
\eea

$\bullet$ The OPEs between the higher spin-$1$ current and some components in
the
second and third ${\cal N}=2$ multiplets in (\ref{lowesthigher})

We use the equations $(4.21), (4.24), (4.28)$, and $(4.32)$ 
and  $(4.35), (4.38), (4.42)$, and $(4.46)$ 
of \cite{Ahn1311}. The results are as follows:
\bea
T^{(1)}(z) \, 
\left(
\begin{array}{c}
U^{(\frac{3}{2})} \\
V^{(\frac{3}{2})} \\
\end{array} \right) (w) 
& = & \frac{1}{(z-w)}  
\, 
\left(
\begin{array}{c}
U^{(\frac{3}{2})} \\
-V^{(\frac{3}{2})} \\
\end{array} \right) (w) 
  +\cdots,
\nonu \\
T^{(1)}(z) \, 
\left(
\begin{array}{c} 
U_{+}^{(2)} \\
V_{-}^{(2)} \\
\end{array} \right) (w) & = & \pm \frac{1}{(z-w)^2}\,
\left[ \frac{2k}{(5+k)} i \hat{B}_{\mp} \right] (w)   +\cdots,
\nonu \\
T^{(1)}(z) \, 
\left(
\begin{array}{c} 
U_{-}^{(2)} \\
V_{+}^{(2)} \\
\end{array} \right) (w) & = & \mp \frac{1}{(z-w)^2}\,
\left[ \frac{6}{(5+k)} i \hat{A}_{\pm} \right] (w)   +\cdots.
\nonu 
\eea

$\bullet$ The OPEs between the higher spin-$1$ current and the
fourth ${\cal N}=2$ multiplet in (\ref{lowesthigher}) 

We use the equations   $(4.49), (4.53), (4.56)$, and $(4.60)$ 
of \cite{Ahn1311}.
One has the following OPEs
\bea
T^{(1)}(z) \, W^{(2)}(w) & = & \frac{1}{(z-w)^2} \,
\left[  \frac{6i}{(5+k)} \hat{A}_3  -\frac{2ik}{(5+k)} 
\hat{B}_3 +   T^{(1)} \right](w)    +
\cdots,
\nonu \\
T^{(1)}(z) \, W^{(3)}(w) & = & \frac{1}{(z-w)^3} \, \frac{i}{(5+k)^2} 
\left[ 12(1+k) \hat{A}_3 -16 k \hat{B}_3\right](w) 
\nonu \\
& + & \frac{1}{(z-w)^2}  \, \left[-  P^{(2)} -\frac{4(741+2530k+1873k^2+
436k^3)}{3(3+7k)(19+23k)(5+k)} 
\hat{T} +\frac{4(-1+k)}{3(5+k)} T^{(2)} \right.
\nonu \\
&-&  \frac{8(2+k)}{3(5+k)^2} \hat{A}_1 \hat{A}_1 -
   \frac{8(2+k)}{3(5+k)^2} \hat{A}_2 \hat{A}_2-
 \frac{4(-5+2k)}{3(5+k)^2} \hat{A}_3 \hat{A}_3 -
 \frac{8(-4+k)}{3(5+k)^2} \hat{A}_3 \hat{B}_3 \nonu \\
&+ &  \frac{4(-13+k)}{3(5+k)^2} \hat{B}_1 \hat{B}_1 +
   \frac{4(-13+k)}{3(5+k)^2} \hat{B}_2 \hat{B}_2+
 \frac{4(-13+4k)}{3(5+k)^2} \hat{B}_3 \hat{B}_3 
 \nonu \\
&-&   \frac{6i(1+k)}{(5+k)^2} \pa \hat{A}_3 +
\frac{8ik}{(5+k)^2} \pa \hat{B}_3  -\frac{4i(25+41k)}{(5+k)(19+23k)}
T^{(1)}  \hat{A}_3
\nonu \\
&-& \left.  \frac{8(-3+k)}{(19+23k)} T^{(1)} T^{(1)} 
+ \frac{4i(19+3k)}{(5+k)(19+23k)} T^{(1)} 
\hat{B}_3 +\frac{4(4+k)}{(5+k)} W^{(2)}
\right](w)
\nonu \\
& + & \frac{1}{(z-w)} \, \frac{1}{(5+k)}\left[  
\frac{24k}{(3+7k)} \pa \hat{T} + 4 \pa T^{(2)} 
+ 4 \hat{G}_{12} T_{+}^{(\frac{3}{2})} - 4 \hat{G}_{21}  
T_{-}^{(\frac{3}{2})} \right](w)  
\nonu \\
& + &
\cdots.
\label{t1w3}
\eea
Compared to the previous expression for the third order pole in 
(\ref{g21s7half-}), 
the composite fields with spin $2$ are the same as those in the third order
pole in (\ref{g21s7half-}) except $\pa T^{(1)}(w)$. Also the quadratic
expression $T^{(1)} T^{(1)}(w)$ occurs in (\ref{t1w3}). 
In the first line of the second order pole in (\ref{t1w3}),
each term is a primary field. In the second line, the first three terms is
a quasi primary field and the last term is a primary field. 
In the third line, the whole expression is a quasi primary field.
In the fourth line, the first two terms are the descendant fields
with relative coefficient $-\frac{1}{2}$ 
coming from  the third order pole in (\ref{t1w3}) and the last term is a 
primary field.
In the last line of the second order pole in (\ref{t1w3}), 
the first term is a quasi primary field and each field in 
the last two terms is a primary 
field.  One can check that 
the first order pole in (\ref{t1w3}) is a primary field. Note that 
there is no descendant field coming from the second order pole in (\ref{t1w3})
because the spin difference of the left hand side of this OPE is given by
$1-3=-2$. 

Now one can move to other OPEs. Let us take $T_{+}^{(\frac{3}{2})}(z)$ and 
$15$ remaining higher spin currents except the higher spin-$1$ current 
from (\ref{lowesthigher}). Note that the OPE 
$T_{+}^{(\frac{3}{2})}(z) \, T^{(1)}(w)$ can be read off from the second equation in
(\ref{t1first}). The full expression is given by Appendix $A$.
Next, one can calculate the OPEs between $T_{-}^{(\frac{3}{2})}(z)$ and other 
remaining higher spin currents and they are given also in Appendix $A$.
Furthermore, the OPEs between the higher spin-$2$ current $T^{(2)}(z)$ 
and other $13$ remaining higher spin currents are given in Appendix $B$.
Then at the moment, the OPEs between the first ${\cal N}=2$ multiplet 
and other four ${\cal N}=2$ multiplets in (\ref{lowesthigher})
are listed in Appendices $A$ and $B$ as well as  the subsection $2.1$.  

Let us move the second ${\cal N}=2$ multiplet in (\ref{lowesthigher}).
The OPEs between the higher spin-$\frac{3}{2}$ current $U^{(\frac{3}{2})}(z)$
and other $12$ higher spin currents can be found in Appendix $C$.
Similarly, the OPEs between the higher spin-$2$ current $U_{+}^{(2)}(z)$
and others are in Appendix $D$. Furthermore, 
the remaining Appendices $E, F$ have the OPEs corresponding to 
the higher spin-$2$ current $U_{-}^{(2)}(z)$ 
and the higher spin-$\frac{5}{2}$ current  $U^{(\frac{5}{2})}(z)$ respectively.
Then   the OPEs between the second ${\cal N}=2$ multiplet 
and other three ${\cal N}=2$ multiplets in (\ref{lowesthigher})
are listed in Appendices $C, D, E$ and $F$.

Appendices $G, H, I$ and $J$ contain the OPEs between 
 the third ${\cal N}=2$ multiplet 
and other two ${\cal N}=2$ multiplets in (\ref{lowesthigher}).

Appendix $K$ lists   the OPEs between 
 the fourth ${\cal N}=2$ multiplet 
and itself in (\ref{lowesthigher}). 

The final Appendix $L$ contains the composite spin-$\frac{9}{2}$ fields
in the first order poles of the OPEs, $U^{(\frac{5}{2})}(z) \, W^{(3)}(w),
V^{(\frac{5}{2})}(z) \, W^{(3)}(w),  W_{+}^{(\frac{5}{2})}(z) \, W^{(3)}(w)$
and $W_{-}^{(\frac{5}{2})}(z) \, W^{(3)}(w)$.
In next section, we will summarize the main results with the fusion rules.

\section{Conclusions and outlook }

Let us describe the fusion rules from the main results of this paper.
Among the OPEs in (\ref{opediagram}),
we would like to focus on the right hand sides of these OPEs which 
contain any components appearing in the second ${\cal N}=4$ multiplet in 
(\ref{nexthigher}). There are four ${\cal N}=2$ multiplets in 
(\ref{lowesthigher}). Then there exist $10(=4+3+2+1)$ 
${\cal N}=2$ superfusion   
rules.
From the sections $2, 3$, and Appendices $A, B$,
one has the following fusion rules between the first and second ${\cal N}=2$ 
multiplets in (\ref{lowesthigher}) in component approach  \footnote{There are
no second higher spin currents in the OPE between the 
first ${\cal N}=2$ multiplet in (\ref{lowesthigher}) by collecting the 
results of (\ref{t1first}) and Appendices $A$ and $B$. Then the total number of
${\cal N}=2$ superfusion rule is given by $9$. They will appear in the 
equations (\ref{fusion1})-(\ref{fusion9}) below.}
\bea
\left[T^{(1)}\right] \cdot \left[U^{(\frac{5}{2})}\right]  & = 
& \left[Q^{(\frac{5}{2})}\right] 
 +\cdots,
\qquad
\left[T_{\pm}^{(\frac{3}{2})}\right] \cdot \left[U_{\mp}^{(2)}\right]   = 
 \left[Q^{(\frac{5}{2})}\right] 
 +\cdots,
\nonu \\
\left[T_{\pm}^{(\frac{3}{2})}\right] \cdot \left[U^{(\frac{5}{2})}\right]  & = 
& \left[Q_{\pm}^{(3)}\right] 
 +\cdots,
\qquad
\left[T^{(2)}\right] \cdot \left[U^{(\frac{3}{2})}\right]   = 
 \left[Q^{(\frac{5}{2})}\right] 
 +\cdots,
\nonu \\
\left[T^{(2)}\right] \cdot \left[U_{\pm}^{(2)}\right]   & = 
& \left[Q_{\pm}^{(3)}\right] 
 +\cdots,
\nonu \\
\left[T^{(2)}\right] \cdot \left[U^{(\frac{5}{2})}\right] &  =  &
 \left[P_{-}^{(\frac{5}{2})}\right] + \left[Q^{(\frac{5}{2})}\right]  
+\left[Q^{(\frac{7}{2})}\right] 
 +\cdots.
\label{fusion1}
\eea
One denotes  
the large ${\cal N}=4$ nonlinear superconformal family of
the identity operator by $\left[ I \right]$.
In the first OPE of (\ref{fusion1}), there exist other  
terms, $\left[ U^{(\frac{3}{2})}\right]$, $\left[U^{(\frac{5}{2})} \right]$ and 
$\left[ I \right]$ (we ignore) by looking at the equation (\ref{t1uv}).
The abbreviated part in the upper case  in the second OPE
of (\ref{fusion1}) 
contains   $\left[ U^{(\frac{3}{2})}\right]$ and $\left[U^{(\frac{5}{2})} \right]$  
and those in the lower case has 
 $\left[ U^{(\frac{3}{2})}\right]$ and 
$\left[ I \right]$.
One has  $\left[ U_{+}^{(2)}\right]$, $\left[ T^{(1)} \right]$ and 
$\left[ I \right]$ in the upper case of the third OPE.
For the lower case, there are  
$\left[ U_{-}^{(2)}\right]$, $\left[T^{(1)} \right]$, $\left[ W^{(2)}\right]$, 
$\left[ T_{-}^{(\frac{3}{2})}\right]$, $\left[ T_{-}^{(\frac{3}{2})} U^{(\frac{3}{2})} 
\right]$, $\left[ U^{(\frac{3}{2})} \right]$ and 
$\left[ I \right]$ if one specifies the complete form.
Then the quadratic fields appear in this OPE.
Similarly, the fourth OPE contains  $\left[ U^{(\frac{3}{2})}\right]$ and 
$ \left[U^{(\frac{5}{2})} \right]$.
The fifth OPE with upper case has  
$\left[ U_{+}^{(2)}\right], \left[ T^{(1)} \right], \left[ T^{(2)} \right], 
\left[ T_{+}^{(\frac{3}{2})} \right], \left[ T_{+}^{(\frac{3}{2})} U^{(\frac{3}{2})} 
\right]$ and 
$\left[ I \right]$. For the lower case, one has 
the following extra terms
$\left[ U_{-}^{(2)}\right], \left[ T^{(1)} \right], \left[ W^{(2)} \right], 
\left[ U^{(\frac{3}{2})}\right], \left[ T_{-}^{(\frac{3}{2})} \right], 
\left[ T_{-}^{(\frac{3}{2})} U^{(\frac{3}{2})} 
\right]$ and 
$\left[ I \right]$ if one describes the full result.
For the final OPE of (\ref{fusion1}), one also has
$\left[ U^{(\frac{3}{2})}\right]$, 
$\left[ U^{(\frac{5}{2})} \right]$, $\left[ T_{-}^{(\frac{3}{2})} \right]$, 
$\left[ T^{(1)}\right]$, $ \left[T^{(1)} U^{(\frac{3}{2})} \right]$, 
$\left[ V^{(\frac{3}{2})} \right]$, 
$\left[ W_{-}^{(\frac{5}{2})}  
\right]$, $\left[ W_{+}^{(\frac{5}{2})} \right]$, $\left[ T_{+}^{(\frac{3}{2})} 
\right]$, $\left[ U_{+}^{(2)}\right]$, 
$\left[ T_{-}^{(\frac{3}{2})} U_{+}^{(2)} \right]$, 
$\left[ U^{(\frac{3}{2})}
W^{(2)} \right]$ and 
$\left[ I \right]$ if one writes down the full expression.
Then one realizes that the quadratic terms (as well as linear ones)
in the lowest higher spin currents
can arise in the fusion rules. 

It would be interesting to see whether the above fusion rules can be 
written in terms of ${\cal N}=2$ superspace by adding these six equations.
It is obvious to see that there is a second ${\cal N}=2$ multiplet in 
(\ref{nexthigher}). Moreover, there is also  
$\left[ P_{-}^{(\frac{5}{2})}\right]$ coming from the last OPE of (\ref{fusion1}).
It is an open problem whether one can remove this unwanted term (or
one can generate other three partners, $\left[P^{(2)} \right]$, 
$\left[P_{+}^{(\frac{5}{2})} \right]$ 
and $\left[P^{(3)} \right]$ in the above fusion rules) 
by redefining the primary fields
correctly.
Note that according to $U(1)$ charge conservation in (\ref{fusion1}),
the left hand side of last fusion rule 
has $\frac{(-3+k)}{(5+k)}$ while the $U(1)$ charge of
$P_{-}^{(\frac{5}{2})}(w)$ is given by $-\frac{(3+k)}{(5+k)}$.  
In the notation of $\left[ P_{-}^{(\frac{5}{2})}\right]$, this is actually 
given by $P_{-}^{(\frac{5}{2})} \hat{B}_{-}(w)$ in Appendix $B$ which gives the 
correct $U(1)$ charge. 

Now one can summarize the following fusion rules 
between the first and third ${\cal N}=2$ multiplets in (\ref{lowesthigher})
\bea
\left[T^{(1)}\right] \cdot \left[V^{(\frac{5}{2})}\right]  & = 
& \left[R^{(\frac{5}{2})}\right] 
 +\cdots,
\qquad
\left[T_{\pm}^{(\frac{3}{2})}\right] \cdot \left[V_{\mp}^{(2)}\right]   = 
 \left[R^{(\frac{5}{2})}\right] 
 +\cdots,
\nonu \\
\left[T_{\pm}^{(\frac{3}{2})}\right] \cdot \left[V^{(\frac{5}{2})}\right] &  = 
& \left[R_{\pm}^{(3)}\right] 
 +\cdots,
\qquad
\left[T^{(2)}\right] \cdot \left[V^{(\frac{3}{2})}\right]   = 
 \left[R^{(\frac{5}{2})}\right] 
 +\cdots,
\nonu \\
\left[T^{(2)}\right] \cdot \left[V_{\pm}^{(2)}\right]   & = 
& \left[R_{\pm}^{(3)}\right] 
 +\cdots,
\nonu \\
\left[T^{(2)}\right] \cdot \left[V^{(\frac{5}{2})}\right] &  = &
 \left[P_{-}^{(\frac{5}{2})}\right] + \left[R^{(\frac{5}{2})}\right]  
+\left[R^{(\frac{7}{2})}\right] 
 +\cdots.
\label{fusion2} 
\eea
These are
very similar to the previous fusion rules (\ref{fusion1}).
We can see the extra structures we ignored in (\ref{fusion2}) from sections
$2$ and $3$ and Appendices $A$ and $B$ as done in previous considerations on 
(\ref{fusion1}). 
Simply adding these fusion rules gives the third ${\cal N}=2$ multiplet 
in (\ref{nexthigher}) and there exists a $\left[ P_{-}^{(\frac{5}{2})}\right]$ 
which will disappear in an appropriate basis.

Next one can select, from Appendices $A$ and $B$, 
the following fusion rules between the first and fourth ${\cal N}=2$ multiplets 
in (\ref{lowesthigher})
\bea
\left[T^{(1)}\right] \cdot \left[W_{\pm}^{(\frac{5}{2})}\right]  & = 
& \left[P_{\pm}^{(\frac{5}{2})}\right] 
 +\cdots,
\qquad
\left[T^{(1)} \right] \cdot \left[W^{(3)}\right]  =  \left[P^{(2)}\right]
 +\cdots,
\nonu \\
\left[T_{\pm}^{(\frac{3}{2})}\right] \cdot \left[W^{(2)}\right]  & = 
& \left[P_{\pm}^{(\frac{5}{2})}\right] 
 +\cdots,
\qquad
\left[T_{\pm}^{(\frac{3}{2})}\right] \cdot \left[W_{\mp}^{(\frac{5}{2})}\right]   = 
 \left[P^{(2)}\right] + \left[P^{(3)}\right] 
 +\cdots,
\nonu \\
\left[T_{+}^{(\frac{3}{2})}\right] \cdot \left[W^{(3)}\right]  & = 
& \left[P_{+}^{(\frac{5}{2})}\right] +   \left[Q^{(\frac{5}{2})}\right] +
 \left[R^{(\frac{5}{2})}\right]
 +\cdots,
\qquad
\left[T_{-}^{(\frac{3}{2})}\right] \cdot \left[W^{(3)}\right]   = 
 \left[P_{-}^{(\frac{5}{2})}\right] 
 +\cdots,
\nonu \\
\left[T^{(2)}\right] \cdot \left[W^{(2)}\right]  & = 
& \left[P^{(2)}\right] 
 +\cdots,
\qquad
\left[T^{(2)}\right] \cdot \left[W_{\pm}^{(\frac{5}{2})}\right]   = 
 \left[P_{\pm}^{(\frac{5}{2})}\right]
 +\cdots,
\nonu \\
\left[T^{(2)}\right] \cdot \left[W^{(3)}\right] &  = 
& \left[P^{(2)}\right] + \left[P^{(3)}\right] +
\left[Q_{+}^{(3)}\right] + \left[Q_{-}^{(3)}\right]
+\left[R_{-}^{(3)}\right] +
 \left[S^{(3)}\right]
 +\cdots.
\label{fusion3}
\eea 
One can see that by adding these fusion rules (\ref{fusion3})
there is no $\left[ Q^{(\frac{7}{2})}\right]$ dependence we want to include 
and there exists 
$\left[ S^{(3)}\right]$ we want to remove in the context of ${\cal N}=2$
supersymmetric fusion rule.
The extra structures we ignored can be found in Appendices $A$ and $B$.

For the fusion rules between the second ${\cal N}=2$ multiplet in 
(\ref{lowesthigher}), one obtains, from Appendices $D$ and $F$, 
\bea
\left[U_{+}^{(2)}\right] \cdot \left[U^{(\frac{5}{2})}\right]  & = 
& \left[Q^{(\frac{5}{2})}\right]  +\cdots,
\qquad
\left[U^{(\frac{5}{2})}\right] \cdot \left[U^{(\frac{5}{2})}\right]   = 
 \left[Q_{-}^{(3)}\right]
+\cdots.
\label{fusion4}
\eea
In the first equation of (\ref{fusion4}), 
there exist $\left[U^{(\frac{3}{2})}\right]$, $\left[U_{+}^{(2)}\right]$, 
and $\left[I \right]$.
If one specifies the complete expression, 
the extra structures  $\left[U^{(\frac{3}{2})} U^{(\frac{5}{2})}\right]$, 
$\left[U_{+}^{(2)} U_{-}^{(2)}\right]$, $\left[U_{+}^{(2)} \right]$, 
$\left[W^{(2)} \right]$, $\left[ U_{-}^{(2)} \right]$, 
$\left[ T^{(1)} \right]$, $\left[ T_{-}^{(\frac{3}{2})} U^{(\frac{3}{2})} 
\right]$, $\left[ T_{-}^{(\frac{3}{2})}\right]$, $\left[U^{(\frac{3}{2})}\right]$ 
and $\left[I \right]$ appear in the second equation of (\ref{fusion4}).
Again,
 the quadratic terms (as well as linear ones)
in the lowest higher spin currents
occur in these fusion rules. 

From Appendices $C$, $D$, $E$ and $F$, one obtains 
the following fusion rules between the second and third ${\cal N}=2$ 
multiplets in (\ref{lowesthigher}) 
\bea
\left[U^{(\frac{3}{2})}\right] \cdot \left[V_{\pm}^{(2)}\right]  & = 
& \left[P_{\pm}^{(\frac{5}{2})}\right] +\cdots,
\qquad
\left[U^{(\frac{3}{2})}\right] \cdot \left[V^{(\frac{5}{2})}\right]   = 
 \left[P^{(2)}\right] + \left[S^{(3)}\right]  +\cdots,
\nonu \\
\left[U_{\pm}^{(2)}\right] \cdot \left[V^{(\frac{3}{2})}\right]  & = 
& \left[P_{\pm}^{(\frac{5}{2})}\right]  +\cdots,
\qquad
\left[U_{\pm}^{(2)}\right] \cdot \left[V_{\mp}^{(2)}\right]   = 
 \left[P^{(2)}\right] + \left[P^{(3)}\right]
+   \left[S^{(3)}\right]
+\cdots,
\nonu \\
\left[U_{+}^{(2)}\right] \cdot \left[V^{(\frac{5}{2})}\right]  & = 
& \left[P_{+}^{(\frac{5}{2})}\right] + \left[Q^{(\frac{5}{2})}\right]
+   \left[S_{+}^{(\frac{7}{2})}\right]
+\cdots,
\nonu \\
\left[U_{-}^{(2)}\right] \cdot \left[V^{(\frac{5}{2})}\right]  & = 
& \left[P_{-}^{(\frac{5}{2})}\right]
+   \left[S_{-}^{(\frac{7}{2})}\right]
+\cdots,
\qquad
\left[U^{(\frac{5}{2})}\right] \cdot \left[V^{(\frac{3}{2})}\right]   = 
 \left[P^{(2)}\right]
+\cdots,
\nonu \\
\left[U^{(\frac{5}{2})}\right] \cdot \left[V_{+}^{(2)}\right]  & = 
& \left[P_{+}^{(\frac{5}{2})}\right] 
+  \left[R^{(\frac{5}{2})}\right]
+   \left[S_{+}^{(\frac{7}{2})}\right] 
+\cdots,
\nonu \\
\left[U^{(\frac{5}{2})}\right] \cdot \left[V_{-}^{(2)}\right]  & = 
& \left[P_{-}^{(\frac{5}{2})}\right] 
+   \left[S_{-}^{(\frac{7}{2})}\right] 
+\cdots,
\nonu \\
\left[U^{(\frac{5}{2})}\right] \cdot \left[V^{(\frac{5}{2})}\right]  & = 
& \left[P^{(2)}\right]
+ \left[P^{(3)}\right]
+ \left[S^{(3)}\right]
+ \left[S^{(4)}\right]
+\cdots.
\label{fusion5}
\eea
One expects that the right hand side of ${\cal N}=2$ superfusion rule  
contains the first and fourth 
${\cal N}=2$ multiplets
in (\ref{nexthigher}) by adding (\ref{fusion5}) each other.
It is nontrivial to find how one can remove the unwanted terms 
$\left[ Q^{(\frac{5}{2})} \right]$ and $\left[ R^{(\frac{5}{2})}\right]$.

Once again, 
one obtains 
the following fusion rules between the second and fourth ${\cal N}=2$ 
multiplets in (\ref{lowesthigher}), from Appendices $C$, $D$, $E$ and $F$,
\bea
\left[U^{(\frac{3}{2})}\right] \cdot \left[W_{\pm}^{(\frac{5}{2})}\right]  & = 
& \left[Q_{\pm}^{(3)}\right]  +\cdots,
\qquad
\left[U^{(\frac{3}{2})}\right] \cdot \left[W^{(3)}\right]   = 
 \left[P_{-}^{(\frac{5}{2})}\right] + \left[Q^{(\frac{5}{2})}\right]
 +\cdots,
\nonu \\
\left[U_{\pm}^{(2)}\right] \cdot \left[W^{(2)}\right]  & = 
& \left[Q_{\pm}^{(3)}\right]
+\cdots,
\qquad
\left[U_{+}^{(2)}\right] \cdot \left[W_{-}^{(\frac{5}{2})}\right]   = 
 \left[P_{-}^{(\frac{5}{2})}\right] + \left[Q^{(\frac{5}{2})}\right]
+   \left[Q^{(\frac{7}{2})}\right]
+\cdots,
\nonu \\
\left[U_{+}^{(2)}\right] \cdot \left[W^{(3)}\right]  & = 
& \left[P^{(2)}\right] + \left[P^{(3)}\right]
+  \left[Q_{+}^{(3)}\right]
+   \left[S^{(3)}\right]
+\cdots,
\nonu \\
\left[U_{-}^{(2)}\right] \cdot \left[W_{+}^{(\frac{5}{2})}\right]  & = 
& 
\left[P_{-}^{(\frac{5}{2})}\right] + \left[Q^{(\frac{5}{2})}\right]
+   \left[Q^{(\frac{7}{2})}\right]
+\cdots,
\qquad
\left[U_{-}^{(2)}\right] \cdot \left[W^{(3)}\right]   = 
   \left[Q_{-}^{(3)}\right]
+\cdots,
\nonu \\
\left[U^{(\frac{5}{2})}\right] \cdot \left[W^{(2)}\right]  & = 
& \left[P_{-}^{(\frac{5}{2})}\right] 
+  \left[Q^{(\frac{5}{2})}\right]
+\cdots,
\nonu \\
\left[U^{(\frac{5}{2})}\right] \cdot \left[W_{+}^{(\frac{5}{2})}\right]  & = 
& \left[P^{(2)}\right]
+ \left[P^{(3)}\right]
+\left[Q_{+}^{(3)}\right]
+ \left[S^{(3)}\right]
+\cdots,
\nonu \\
\left[U^{(\frac{5}{2})}\right] \cdot \left[W_{-}^{(\frac{5}{2})}\right]  & = 
& \left[Q_{-}^{(3)}\right]
+\cdots,
\nonu \\
\left[U^{(\frac{5}{2})}\right] \cdot \left[W^{(3)}\right]  & = 
& 
 \left[P^{(2)}\right]+ 
\left[P_{+}^{(\frac{5}{2})}\right] +
\left[P_{-}^{(\frac{5}{2})}\right] 
+  \left[P^{(3)}\right]
+  \left[Q^{(\frac{5}{2})}\right]
+  \left[Q_{+}^{(3)}\right]
+  \left[Q^{(\frac{7}{2})}\right]
\nonu \\
& + & \left[R^{(\frac{5}{2})}\right]
+  \left[S^{(3)}\right]
+ \left[S_{-}^{(\frac{7}{2})}\right]
+\cdots.
\label{fusion6}
\eea
By adding the fusion rules (\ref{fusion6}), 
one sees the first and second ${\cal N}=2$ multiplets in (\ref{lowesthigher}).
It would be interesting to see whether 
the other ${\cal N}=2$ multiplets can appear in ${\cal N}=2$ superfusion
rules. In the right hand side of last fusion rule of (\ref{fusion6}), 
the fermion currents are multiplied by the bosonic currents, $P^{(2)}(w)$ and 
so on. 

For the fusion rules between the third ${\cal N}=2$ multiplet in 
(\ref{lowesthigher}), one obtains, from Appendices $H$ and $J$, 
\bea
\left[V_{+}^{(2)}\right] \cdot \left[V^{(\frac{5}{2})}\right]  & = 
&  \left[R^{(\frac{5}{2})}\right]
+\cdots,
\qquad
\left[V^{(\frac{5}{2})}\right] \cdot \left[V^{(\frac{5}{2})}\right]   =  
 \left[R_{-}^{(3)}\right]
+\cdots.
\label{fusion7}
\eea
In the first equation of (\ref{fusion7}), 
there exist $\left[V^{(\frac{3}{2})}\right]$, $\left[T^{(1)}\right]$, $\left[ 
V_{+}^{(2)}\right]$ 
and $\left[I \right]$.
The extra structures, if one writes down them explicitly,  
$\left[V^{(\frac{3}{2})} V^{(\frac{5}{2})}\right]$, 
$\left[V_{+}^{(2)} V_{-}^{(2)}\right]$, $\left[V_{-}^{(2)} \right]$, 
$\left[W^{(2)} \right]$, $\left[ V_{+}^{(2)} \right]$, 
$\left[ T^{(1)} \right]$, $\left[ T_{-}^{(\frac{3}{2})} V^{(\frac{3}{2})} 
\right]$, $\left[ T_{-}^{(\frac{3}{2})}\right]$, $\left[V^{(\frac{3}{2})}\right]$ 
and $\left[I \right]$ appear in the second equation of (\ref{fusion4}).
The quadratic terms (as well as linear ones)
in the lowest higher spin currents
can be seen from these fusion rules. 

One obtains 
the following fusion rules between the third and fourth ${\cal N}=2$ 
multiplets in (\ref{lowesthigher}) from Appendices $G$, $H$, $I$ and $J$
\bea
\left[V^{(\frac{3}{2})}\right] \cdot \left[W_{\pm}^{(\frac{5}{2})}\right]  & = 
& \left[R_{\pm}^{(3)}\right]  +\cdots,
\qquad
\left[V^{(\frac{3}{2})}\right] \cdot \left[W^{(3)}\right]   = 
 \left[P_{-}^{(\frac{5}{2})}\right] + \left[R^{(\frac{5}{2})}\right]
 +\cdots,
\nonu \\
\left[V_{\pm}^{(2)}\right] \cdot \left[W^{(2)}\right]  & = 
& \left[R_{\pm}^{(3)}\right]
+\cdots,
\qquad
\left[V_{+}^{(2)}\right] \cdot \left[W_{-}^{(\frac{5}{2})}\right]   =  
\left[P_{-}^{(\frac{5}{2})}\right] + \left[R^{(\frac{5}{2})}\right]
+   \left[R^{(\frac{7}{2})}\right]
+\cdots,
\nonu \\
\left[V_{+}^{(2)}\right] \cdot \left[W^{(3)}\right]  & = 
& \left[P^{(2)}\right] + \left[P^{(3)}\right]
+  \left[R_{+}^{(3)}\right]
+   \left[S^{(3)}\right]
+\cdots,
\nonu \\
\left[V_{-}^{(2)}\right] \cdot \left[W_{+}^{(\frac{5}{2})}\right]  & = 
& 
\left[P_{-}^{(\frac{5}{2})}\right] + \left[R^{(\frac{5}{2})}\right]
+   \left[R^{(\frac{7}{2})}\right]
+\cdots,
\qquad
\left[V_{-}^{(2)}\right] \cdot \left[W^{(3)}\right]   = 
   \left[R_{-}^{(3)}\right]
+\cdots,
\nonu \\
\left[V^{(\frac{5}{2})}\right] \cdot \left[W^{(2)}\right]  & = 
& \left[P_{-}^{(\frac{5}{2})}\right] 
+  \left[R^{(\frac{5}{2})}\right]
+\cdots,
\nonu \\
\left[V^{(\frac{5}{2})}\right] \cdot \left[W_{+}^{(\frac{5}{2})}\right]  & = 
& \left[P^{(2)}\right]
+ \left[P^{(3)}\right]
+\left[R_{+}^{(3)}\right]
+\cdots,
\nonu \\
\left[V^{(\frac{5}{2})}\right] \cdot \left[W_{-}^{(\frac{5}{2})}\right]  & = 
& 
 \left[P^{(2)}\right]
+
\left[R_{-}^{(3)}\right]
+\cdots,
\nonu \\
\left[V^{(\frac{5}{2})}\right] \cdot \left[W^{(3)}\right]  & = 
& 
\left[P_{+}^{(\frac{5}{2})}\right] +
\left[P_{-}^{(\frac{5}{2})}\right] 
+  \left[R^{(\frac{5}{2})}\right]
+  \left[R^{(\frac{7}{2})}\right]
+ \left[S_{-}^{(\frac{7}{2})}\right]
+\cdots.
\label{fusion8}
\eea
By adding the fusion rules (\ref{fusion8}), 
one sees the first and third ${\cal N}=2$ multiplets in (\ref{lowesthigher}).
It is an open problem to see whether 
the other ${\cal N}=2$ multiplets can appear in ${\cal N}=2$ superfusion
rules.

Finally, from Appendix $K$, one obtains the following 
fusion rules between the last ${\cal N}=2$ multiplet in (\ref{lowesthigher})
\bea
\left[W^{(2)}\right] \cdot \left[W_{\pm}^{(\frac{5}{2})}\right]  & = 
&  \left[P_{\pm}^{(\frac{5}{2})}\right]
+  \left[S_{\pm}^{(\frac{7}{2})}\right]
+\cdots,
\nonu \\
\left[W^{(2)}\right] \cdot \left[W^{(3)}\right]  & = 
&  \left[P^{(2)}\right]
+  \left[P^{(3)}\right]
+  \left[Q_{+}^{(3)}\right]
+  \left[Q_{-}^{(3)}\right]
+  \left[S^{(3)}\right]
+\cdots,
\nonu \\
\left[W_{+}^{(\frac{5}{2})}\right] \cdot \left[W_{-}^{(\frac{5}{2})}\right]  & = 
&  \left[P^{(2)}\right]
+ \left[P^{(3)}\right]
+  \left[Q_{+}^{(3)}\right]
+  \left[Q_{-}^{(3)}\right]
+  \left[R_{-}^{(3)}\right]
+  \left[S^{(3)}\right]
+\left[S^{(4)}\right]
+\cdots,
\nonu \\
\left[W_{+}^{(\frac{5}{2})}\right] \cdot \left[W^{(3)}\right]  & = 
&  
 \left[P_{+}^{(\frac{5}{2})}\right] 
+  \left[Q^{(\frac{5}{2})}\right]
+  \left[Q^{(\frac{7}{2})}\right]
+  \left[R^{(\frac{5}{2})}\right]
+ \left[R^{(\frac{7}{2})}\right]
+ \left[S_{+}^{(\frac{7}{2})}\right]
+\cdots,
\nonu \\
\left[W_{-}^{(\frac{5}{2})}\right] \cdot \left[W^{(3)}\right]  & = 
&  
 \left[P_{-}^{(\frac{5}{2})}\right] 
+ \left[S_{-}^{(\frac{7}{2})}\right]
+\cdots,
\nonu \\
\left[W^{(3)}\right] \cdot \left[W^{(3)}\right]  & = 
&  \left[P^{(2)}\right]
+  \left[P^{(3)}\right]
+  \left[Q_{-}^{(3)}\right]
+  \left[R_{-}^{(3)}\right]
+  \left[S^{(3)}\right]
+  \left[S^{(4)}\right]
+\cdots.
\label{fusion9}
\eea
By adding the fusion rules (\ref{fusion9}), 
one sees the first, second and fourth 
${\cal N}=2$ multiplets in (\ref{lowesthigher}) in the 
${\cal N}=2$ supersymmetric fusion rule.
It would be interesting to see whether the other remaining 
third ${\cal N}=2$ multiplet can join this ${\cal N}=2$ superfusion rule.

Then one has the following fusion rules from (\ref{fusion1})-(\ref{fusion9})
\bea
\left[ \Phi_i \right] \cdot \left[ \Phi_j \right] =
\left[ I \right] + \sum_{k} C_{ijk} \left[\Phi_k \right] + \sum_{l,m} 
C_{ijlm}\left[\Phi_l \, \Phi_m \right] +
\sum_a C_{ija} \left[ \Psi_a \right],
\qquad
i,j,k,l,m,a = 1,2, \cdots, 16,
\nonu
\eea
where $\Phi_i$ stands for any $16$ components in ${\cal N}=4$ multiplet in 
(\ref{lowesthigher}) and 
$\Psi_a$  stands for any $16$ components in ${\cal N}=4$ multiplet in 
(\ref{nexthigher}). All the structure constants,
$C_{ijk}$, $C_{ijlm}$ and $C_{ija}$, are known \footnote{ In \cite{Ahn1311},
the main results in the fusion rules can be written in terms of
$\left[ I \right] \cdot \left[ I \right] =
\left[ I \right]$ 
and $\left[ I \right] \cdot \left[ \Phi_i \right] =
\left[ I \right] + \sum_{j} C_{ij} \left[\Phi_j \right]$. }.

Let us comment future directions as follows.

$\bullet$ The complete OPEs between the $11$ nonlinear currents and 
the second higher spin ${\cal N}=4$ multiplet

In (\ref{g21p5half-})-(\ref{g21s7half-}), 
we have seen some of these OPEs before.
For the first ${\cal N}=4$ multiplet, the complete OPEs  with $11$ currents 
were given and 
the right hand sides of these OPEs consist of above $11$ currents and
the first ${\cal N}=4$ multiplet (and their composite fields).
One expects that 
the right hand sides of the complete  OPEs will 
consist of above $11$ currents and
the first and second ${\cal N}=4$ multiplets (and their composite fields).

$\bullet$ The OPEs between the first higher spin ${\cal N}=4$ multiplet and 
the second higher spin ${\cal N}=4$ multiplet 

We have seen some of these OPEs in (\ref{u3halfp2}), (\ref{v3halfp2})
and (\ref{t1r5half}). 
In order to find the complete algebra, it is necessary to calculate 
these OPEs.  

$\bullet$ The OPEs between  
the second higher spin ${\cal N}=4$ multiplet and itself

Furthermore, it is natural to complete these OPEs and 
to check whether there exist new primary fields in the right hand sides 
of these OPEs.

$\bullet$ The ${\cal N}=2$ and ${\cal N}=4$ superfusion rules

In (\ref{fusion1})-(\ref{fusion9}) we presented the fusion rules 
in the component approach.
It is nontrivial to write them in terms of ${\cal N}=2$ supercurrents
and furthermore, it would be interesting to see whether these ${\cal N}=2$
superspace approach can be generalized to ${\cal N}=4$ superspace approach.  
For example, one expects the following ${\cal N}=4$ superfusion rules
$\left[ {\cal U} \right] \cdot \left[  {\cal U} \right] 
= \left[ {\cal I} \right] + 
\left[{\cal U}\right] + \left[ {\cal U} {\cal U} \right] + \left[ {\cal V} 
\right]$
where ${\cal U}$ stands for the whole 
$16$ higher spin currents in (\ref{lowesthigher})
and  ${\cal V}$ stands for the whole 
$16$ higher spin currents in (\ref{nexthigher}).

$\bullet$ An extension of small ${\cal N}=4$ superconformal algebra

It is well-known that the small (or regular) ${\cal N}=4$ superconformal algebra
can be obtained by taking one of the level as zero and the other level as 
an infinity in the large ${\cal N}=4$ linear superconformal algebra.  
Therefore, an extension of small ${\cal N}=4$ superconformal algebra,
which will be useful in \cite{GG1406},
can be obtained by taking these limits to the results of this paper as well as
the one in \cite{Ahn1311}.

\vspace{.7cm}

\centerline{\bf Acknowledgments}

CA would like to thank 
H. Kim for discussions. 
This work was supported by the Mid-career Researcher Program through
the National Research Foundation of Korea (NRF) grant 
funded by the Korean government (MEST) (No. 2012-045385/2013-056327).
CA acknowledges warm hospitality from 
the School of  Liberal Arts (and Institute of Convergence Fundamental
Studies), Seoul National University of Science and Technology.

\newpage

\appendix

\renewcommand{\thesection}{\large \bf \mbox{Appendix~}\Alph{section}}
\renewcommand{\theequation}{\Alph{section}\mbox{.}\arabic{equation}}


\section{The nontrivial 
OPEs between  the higher spin-$\frac{3}{2}$ currents, 
$T_{\pm}^{(\frac{3}{2})}(z)$, and 
other $15$ higher spin currents}

We list the main OPE results case by case. 

$\bullet$ The OPEs between the higher spin-$\frac{3}{2}$ currents 
and the
first ${\cal N}=2$ multiplet of (\ref{lowesthigher})

\bea
T_{+}^{(\frac{3}{2})}(z) \, 
T_{-}^{(\frac{3}{2})}(w)  & = & -\frac{1}{(z-w)^3} \, \left[
\frac{6k}{(5+k)} \right]
\nonu \\
& + & \frac{1}{(z-w)^2} \, \left[ \frac{6i}{(5+k)}\hat{A}_3 +\frac{2ik}{(5+k)}
\hat{B}_3 
-  T^{((1)} \right](w) \nonu \\
& + & \frac{1}{(z-w)} \, \left[ 
\frac{1}{2} \pa \{ T_{+}^{(\frac{3}{2})} \, 
T_{-}^{(\frac{3}{2})} \}_{-2}
-T^{(2)} -
\frac{6k}{(3+7k)} \hat{T}  \right](w) \nonu \\
& + & \cdots,
\nonu \\
T_{\pm}^{(\frac{3}{2})}(z) \, 
T^{(2)}(w) & = & \frac{1}{(z-w)^2} \, 
\left[ \frac{12(1+k)(3+k)}{(5+k)(3+7k)}
T_{\pm}^{(\frac{3}{2})} \right](w) +\frac{1}{(z-w)} \,
\frac{1}{3} \pa \{ T_{\pm}^{(\frac{3}{2})} \, 
T^{(2)} \}_{-2}(w)  \nonu \\
& + & \cdots.
\nonu 
\eea
Note that the relative coefficients, $\frac{1}{2}$ and $\frac{1}{3}$,
can be easily determined by counting the spins in  both the  left and right 
hand sides. See also the footnote $51$ of \cite{Ahn1311}. 

$\bullet$ The OPEs between the higher spin-$\frac{3}{2}$ currents and the
second and third ${\cal N}=2$ multiplets of (\ref{lowesthigher})

\bea
T_{\pm}^{(\frac{3}{2})}(z) \, 
\left(
\begin{array}{c}
V^{(\frac{3}{2})} \\
U^{(\frac{3}{2})} \\
\end{array} \right) (w) & = & \mp \frac{1}{(z-w)^2} \, 
\left[ \frac{6}{(5+k)} 
i \hat{A}_{\mp} \right] (w) \nonu \\
& + & 
\frac{1}{(z-w)} \, \left[ \left(
\begin{array}{c} 
- V_{+}^{(2)} \\
U_{-}^{(2)} \\
\end{array} \right)   
+\frac{1}{2} \pa \{ T_{\pm}^{(\frac{3}{2})} \, 
\left(
\begin{array}{c}
V^{(\frac{3}{2})} \\
U^{(\frac{3}{2})} \\
\end{array} \right)\}_{-2}
\right](w) +\cdots,
\nonu \\
T_{\pm}^{(\frac{3}{2})}(z) \, 
\left(
\begin{array}{c}
U_{+}^{(2)} \\
V_{-}^{(2)} \\
\end{array} \right)(w) & = & \frac{1}{(z-w)} \,
\left[ \frac{2}{(5+k)}i \hat{B}_{\mp}
T_{\pm}^{(\frac{3}{2})} \right] (w) +\cdots,
\nonu \\
T_{+}^{(\frac{3}{2})}(z) \,
U_{-}^{(2)}(w) & = & -\frac{1}{(z-w)^2} \,
\left[ \frac{(5+2k)}{(5+k)} U^{(\frac{3}{2})} \right](w) 
\nonu \\
& + & \frac{1}{(z-w)} \, \left[ -\frac{1}{2} {\bf Q^{(\frac{5}{2})}} 
-U^{(\frac{5}{2})}
+\frac{1}{3} \pa \{  T_{+}^{(\frac{3}{2})} \,
U_{-}^{(2)} \}_{-2}
\right](w) 
+  \cdots,
\nonu \\
T_{-}^{(\frac{3}{2})}(z) \, V_{+}^{(2)}(w) & = &
\frac{1}{(z-w)^2} \,
\left[ \frac{(5+2k)}{(5+k)} V^{(\frac{3}{2})} \right](w)
\nonu \\
& + & \frac{1}{(z-w)} \left[ 
\frac{1}{2} {\bf R^{(\frac{5}{2})}} 
-V^{(\frac{5}{2})} 
+\frac{2}{(5+k)} i \hat{A}_{-} T_{-}^{(\frac{3}{2})}
+ \frac{4}{(5+k)} i \hat{A}_3 V^{(\frac{3}{2})}  \right.
\nonu \\
&- &  
\frac{2}{(5+k)} i \hat{B}_{+} \hat{G}_{21}
-\frac{2}{(5+k)} i \hat{B}_{+} T_{+}^{(\frac{3}{2})}  
-\frac{4}{(5+k)} i \hat{B}_3  V^{(\frac{3}{2})}
\nonu \\
& + & \left.  \frac{(13+2k)}{3(5+k)} \pa  V^{(\frac{3}{2})}
\right](w) +  \cdots, 
\nonu \\
T_{\pm}^{(\frac{3}{2})}(z) \, 
\left(
\begin{array}{c}
V_{-}^{(2)} \\
U_{+}^{(2)} \\
\end{array} \right) (w) & = & \frac{1}{(z-w)^2} \, \frac{(6+k)}{(5+k)} 
\left[ \left(
\begin{array}{c}
 \hat{G}_{22} \\
\hat{G}_{11} \\
\end{array} \right) +\left(
\begin{array}{c}
-  V^{(\frac{3}{2})} \\
U^{(\frac{3}{2})} \\
\end{array} \right)  \right](w)
\nonu \\
& + & \frac{1}{(z-w)} \, \left[ \frac{1}{2} \left(
\begin{array}{c} 
-{\bf R^{(\frac{5}{2})}} \\
{\bf Q^{(\frac{5}{2})}} \\
\end{array} \right)
+\frac{1}{3} \pa \{ T_{\pm}^{(\frac{3}{2})} \, 
\left(
\begin{array}{c}
V_{-}^{(2)} \\
U_{+}^{(2)} \\
\end{array} \right) \}_{-2}
\right](w)
\nonu \\
& + & \cdots,
\nonu \\
T_{+}^{(\frac{3}{2})}(z) \,
U^{(\frac{5}{2})}(w) & = & \frac{1}{(z-w)^3} \,
\left[ \frac{8k(3+k)}{3(5+k)^2} i \hat{B}_{-} \right](w) \nonu \\
& + & \frac{1}{(z-w)^2} \, \left[ \frac{2(9+4k)}{3(5+k)}
U_{+}^{(2)}  +\frac{4(9+2k)}{3(5+k)^2} \hat{A}_3 \hat{B}_{-} -
\frac{4k}{(5+k)^2} \hat{B}_{-} \hat{B}_3 \right.
\nonu \\
&-& \left.  \frac{2k}{(5+k)^2} i \pa \hat{B}_{-}
-\frac{2}{(5+k)} T^{(1)} i \hat{B}_{-} \right](w) \nonu \\
&+ &\frac{1}{(z-w)}  \,  \left[ 
 \frac{1}{4} \pa \{  T_{+}^{(\frac{3}{2})} \,
U^{(\frac{5}{2})} \}_{-2}
+ \frac{1}{2} {\bf Q_{+}^{(3)}} \right. \nonu \\
& + &  \left. \frac{8k(3+k)}{(5+k)(19+23k)} 
i \left( \hat{T}  \hat{B}_{-} 
-\frac{1}{2} \pa^2 \hat{B}_{-} \right)  \right](w)
+\cdots,
\nonu \\
T_{+}^{(\frac{3}{2})}(z) \, 
V^{(\frac{5}{2})}(w) & = & \frac{1}{(z-w)^3} \, 
\left[ \frac{4}{(5+k)} i  
\hat{A}_{-} \right](w)
\nonu \\
& + & 
\frac{1}{(z-w)^2} \, \left[ \frac{2(13+2k)}{3(5+k)} V_{+}^{(2)}+
\frac{12}{(5+k)^2} \hat{A}_{-} \hat{A}_3-\frac{4(8+k)}{(5+k)^2}
\hat{A}_{-} \hat{B}_3
\right. \nonu \\
&+ & \left. \frac{6}{(5+k)^2} i  \pa \hat{A}_{-}
-\frac{2}{(5+k)} i T^{(1)} \hat{A}_{-}
\right](w) \nonu \\
& + & \frac{1}{(z-w)} \, \left[ \frac{1}{2} {\bf R_{+}^{(3)}} -
\frac{3}{2(5+k)} i T^{(1)} \pa
\hat{A}_{-}  \right. \nonu \\
& + & \frac{4(138+343k+113k^2)}{
(5+k)(3+7k)(19+23k)} i \hat{A}_{-} \hat{T}  
-  \frac{2}{(5+k)} i  \hat{A}_{-} T^{(2)} \nonu \\
& - & \frac{2}{(5+k)} 
i \hat{A}_{-} W^{(2)} -\frac{15}{(5+k)^2} \hat{A}_{-} 
\pa \hat{A}_3 -\frac{(16+k)}{(5+k)^2} \hat{A}_{-} \pa \hat{B}_3 
\nonu \\
&+&  \frac{1}{2(5+k)} i \hat{A}_{-} \pa T^{(1)}  +
\frac{9}{(5+k)^2} \hat{A}_3 \pa \hat{A}_{-} + \frac{(13+2k)}{6(5+k)} \pa 
V_{+}^{(2)}  \nonu \\
&+&  \frac{4}{(5+k)^2} i \hat{A}_{+} \hat{A}_{-} \hat{A}_{-}
  + \frac{4}{(5+k)^2} i \hat{A}_{-} \hat{A}_3 \hat{A}_3 -
\frac{(12+k)}{(5+k)^2} i \hat{A}_{-} \hat{A}_3 \hat{B}_3 
\nonu \\
&+&  \frac{8}{(5+k)^2} i \hat{A}_{-} \hat{B}_{+} \hat{B}_{-}
+ \frac{4}{(5+k)^2} i \hat{A}_{-} \hat{B}_3 \hat{B}_3  +
\frac{(4+k)}{(5+k)^2} i \hat{A}_3 \hat{A}_{-} \hat{B}_3 
\nonu \\
&-& \left.  \frac{8}{(5+k)^2} i \pa^2 \hat{A}_{-}
\right](w)
+\cdots,
\nonu \\
T_{-}^{(\frac{3}{2})}(z) \, U^{(\frac{5}{2})}(w) & = & \frac{1}{(z-w)^3}
\, \left[ \frac{8(5+2k)}{(5+k)^2} i \hat{A}_{+} \right] (w) 
\nonu \\
& + & \frac{1}{(z-w)^2} \,
\left[ \frac{4(7+k)}{3(5+k)} U_{-}^{(2)} \right](w)
\nonu \\
&+& \frac{1}{(z-w)} \left[ 
\frac{1}{2} {\bf Q_{-}^{(3)}}
+\frac{3}{2(5+k)} i  T^{(1)} \pa \hat{A}_{+}
 + \frac{8(23+13k)}{(5+k)(19+23k)} i \hat{A}_{+} \hat{T} 
\right. \nonu \\
& - & \frac{2}{(5+k)} i \hat{A}_{+} W^{(2)} 
+ \frac{9}{(5+k)^2} \hat{A}_{+} \pa \hat{A}_3 
-\frac{(-8+k)}{(5+k)^2} \hat{A}_{+} \pa \hat{B}_3 
\nonu \\
& - & \frac{3}{2(5+k)} i \hat{A}_{+} \pa T^{(1)}
- \frac{4}{(5+k)} i \hat{A}_3 U_{-}^{(2)}
+ \frac{3}{(5+k)^2} \hat{A}_3 \pa \hat{A}_{+}
\nonu \\
& + & \frac{4}{(5+k)} i \hat{B}_3  U_{-}^{(2)} +
\frac{(-8+k)}{(5+k)^2} \hat{B}_3 \pa \hat{A}_{+}
\nonu \\
& + & \frac{(29+2k)}{6(5+k)} \pa  U_{-}^{(2)}
-\frac{2}{(5+k)} \hat{G}_{11} \hat{G}_{12} +
\frac{2}{(5+k)} \hat{G}_{11} T_{-}^{(\frac{3}{2})}
\nonu \\
& - & 
\frac{2}{(5+k)} T_{-}^{(\frac{3}{2})}  U^{(\frac{3}{2})}
+\frac{4}{(5+k)^2} i \hat{A}_{+} \hat{A}_3 \hat{A}_3 
-\frac{8}{(5+k)^2} i \hat{A}_{+} \hat{A}_3 \hat{B}_3 
\nonu \\
& + &  \frac{4}{(5+k)^2} i \hat{A}_{+} \hat{B}_{+} \hat{B}_{-} 
+\frac{2}{(5+k)} \hat{G}_{12} U^{(\frac{3}{2})} 
\nonu \\
& + & \left.
\frac{4}{(5+k)^2} i \hat{A}_{+} \hat{B}_3 \hat{B}_3
+\frac{4}{(5+k)^2} i \hat{A}_{-} \hat{A}_{+} \hat{A}_{+}
\right](w) +\cdots,
\nonu \\
T_{-}^{(\frac{3}{2})}(z) \, V^{(\frac{5}{2})}(w) & = &
\frac{1}{(z-w)^3} \,
\left[ \frac{8k(3+k)}{3(5+k)^2} i \hat{B}_{+} \right] (w)
\nonu \\
& + & \frac{1}{(z-w)^2} \left[ \frac{2(9+4k)}{3(5+k)} V_{-}^{(2)} -
\frac{4(9+2k)}{3(5+k)^2} \hat{A}_3 \hat{B}_{+} +\frac{4k}{(5+k)^2} \hat{B}_{+} 
\hat{B}_3 \right. 
\nonu \\
&- & \left. \frac{2k}{(5+k)^2} i \pa \hat{B}_{+}
+ \frac{2}{(5+k)} i T^{(1)} \hat{B}_{+}
\right](w) \nonu \\
& + & 
\frac{1}{(z-w)} \left[ 
\frac{1}{4} \pa \{ T_{-}^{(\frac{3}{2})} \, V^{(\frac{5}{2})} \}_{-2}
+\frac{1}{2} {\bf R_{-}^{(3)}} + \frac{8k(3+k)}{(5+k)(19+23k)}  
i \hat{B}_{+} \hat{T} 
\right. \nonu \\
&-& \left. \frac{2}{(5+k)} T_{-}^{(\frac{3}{2})} V^{(\frac{3}{2})} 
\right](w) +\cdots.
\label{opeabove}  
\eea
The composite fields $\hat{B}_{\mp} T_{\pm}^{(\frac{3}{2})}(w)$ appearing in the 
second OPE of  (\ref{opeabove}) are primary
fields.
In the third order OPE of (\ref{opeabove}), there exists a field 
with boldface \footnote{In Appendices we use a boldface notation 
for the second ${\cal N}=4$ multiplet in (\ref{nexthigher}), ${\bf P^{(2)}}$,
${\bf P_{+}^{(\frac{5}{2})}}$, $\cdots$, ${\bf S^{(4)}}$,  in order to
see them 
more clearly in the right hand side of all the OPEs. They will appear 
at the beginning of each singular term. }. 
In the fourth OPE of (\ref{opeabove}),
the last term in the first order pole can be written in terms of
two terms with the coefficient
$\frac{(13+2k)}{3(5+k)} = \frac{(5+2k)}{3(5+k)} +\frac{8}{3(5+k)}$.
Then the first factor $\frac{(5+2k)}{3(5+k)}$ comes from the 
the higher spin-$\frac{3}{2}$ current living in the second order pole.
One can easily check that  the five composite (quadratic) 
fields in the first 
order pole plus the derivative term with above coefficient $\frac{8}{3(5+k)}$
is a primary field. 
In the sixth OPE, the second order pole subtracted by $U_{+}^{(2)}(w)$ term 
is a primary field.
Similarly,  the second order pole subtracted by $V_{+}^{(2)}(w)$ term 
in the seventh OPE is a primary field.
Furthermore, one can check that the first order pole, subtracted 
by $R_{+}^{(3)}(w)$
term and the descendant terms coming from the primary spin-$2$ field
located at the second singular term,   
behaves as a quasi primary field.
In the eighth OPE,  the first order pole, subtracted by $Q_{-}^{(3)}(w)$
term and the descendant terms coming from the primary spin-$2$ field
located at the second singular term,   
is a quasi primary field.
In the ninth OPE, the second order pole is a primary field
and the last two terms in the first order pole is a quasi primary field.

$\bullet$ The OPEs between the higher spin-$\frac{3}{2}$ currents and the
fourth ${\cal N}=2$ multiplet of (\ref{lowesthigher})

\bea
T_{\pm}^{(\frac{3}{2})}(z) \, 
W^{(2)}(w) & = & \frac{1}{(z-w)^2} \, 
\left[ \frac{(7+k)}{(5+k)} 
T_{\pm}^{(\frac{3}{2})} \right] (w) \nonu \\
& + & 
\frac{1}{(z-w)} \, \left[  -\frac{1}{2} {\bf P_{\pm}^{(\frac{5}{2})}} \mp
W_{\pm}^{(\frac{5}{2})}  
+\frac{1}{3} \pa \{ T_{\pm}^{(\frac{3}{2})} \, 
W^{(2)} \}_{-2}
\right](w)
+  \cdots,
\nonu \\
T_{\pm}^{(\frac{3}{2})}(z) \, 
W_{\pm}^{(\frac{5}{2})}(w) & = & \pm 
\frac{1}{(z-w)^2} \, \left[ \frac{4(11+k)}{3(5+k)^2} 
 \hat{A}_{\mp} \hat{B}_{\mp} \right](w)
\nonu \\
&+ &\frac{1}{(z-w)} \, \left[ \pm \frac{2(8+k)}{3(5+k)}  
\left(
\begin{array}{c}
\hat{G}_{21} \hat{G}_{21} \\
\hat{G}_{12} \hat{G}_{12} \\
\end{array} \right)
\mp \frac{2}{(5+k)} i \hat{A}_{\mp} \left(
\begin{array}{c} 
U_{+}^{(2)} \\
V_{-}^{(2)} \\
\end{array} \right)  \right. \nonu \\
& \mp & \left.
\frac{2(6+k)}{(5+k)^2} \hat{A}_{\mp} \pa \hat{B}_{\mp}
 + \frac{2}{(5+k)} \left( 
\begin{array}{c} 
\hat{G}_{21} \\
\hat{G}_{12} \\
\end{array} \right)  T_{\pm}^{(\frac{3}{2})} 
\right](w) +\cdots,
\nonu \\
T_{+}^{(\frac{3}{2})}(z) \, 
W_{-}^{(\frac{5}{2})}(w) & = & 
\frac{1}{(z-w)^3} \, \left[ -\frac{8(5+2k)}{(5+k)^2} i \hat{A}_3+
\frac{8k(-2+k)}{3(5+k)^2} i  \hat{B}_3 -\frac{4(-2+k)}{3(5+k)} T^{(1)} \right](w)
\nonu \\
& + & \frac{1}{(z-w)^2} 
\, \left[ 
\frac{1}{2} {\bf P^{(2)}}
+\frac{4(15+65k+22k^2)}{3(3+7k)(5+k)} \hat{T}
-\frac{4(-3+k)}{3(5+k)} T^{(2)} \right. \nonu \\
&-&   \frac{4(4+k)}{(5+k)}
W^{(2)}
+\frac{16(-1+k)}{3(5+k)^2}
 \hat{A}_1 \hat{A}_1
+\frac{16(-1+k)}{3(5+k)^2}
\hat{A}_2 \hat{A}_2 +\frac{4(-13+4k)}{3(5+k)^2} \hat{A}_3 \hat{A}_3
\nonu \\
& - & \frac{8(-4+k)}{3(5+k)^2} \hat{A}_3 \hat{B}_3   
 + \frac{4}{3(5+k)} \hat{B}_1 \hat{B}_1 +
\frac{4}{3(5+k)} \hat{B}_2 \hat{B}_2 -\frac{4(-5+2k)}{3(5+k)^2}
\hat{B}_3 \hat{B}_3  \nonu \\
& + & \left. \frac{4}{(5+k)} i T^{(1)} \hat{A}_3 -
\frac{4}{(5+k)} i T^{(1)} \hat{B}_3 \right](w)
\nonu \\
& + &  
\frac{1}{(z-w)} \, \left[ \frac{1}{4} \pa \{ T_{+}^{(\frac{3}{2})} \, 
W_{-}^{(\frac{5}{2})} \}_{-2}  + \frac{1}{2} {\bf P^{(3)}} - W^{(3)} 
-\frac{12k}{(5+k)(3+7k)} \pa \hat{T}
\right. \nonu \\
&- &  \frac{4(-2+k)}{(19+23k)} \left( \hat{T} T^{(1)} -\frac{1}{2} \pa^2 
T^{(1)} \right) -\frac{24(25+15k+2k^2)}{(5+k)^2(19+23k)} i \left( \hat{T} 
\hat{A}_3 -\frac{1}{2} \pa^2 
\hat{A}_3 \right) \nonu \\
& + & 
\frac{8k(-2+k)}{(5+k)(19+23k)} i \left( \hat{T} \hat{B}_3 -\frac{1}{2} \pa^2 
\hat{B}_3 \right) -\frac{2}{(5+k)} \pa T^{(2)} \nonu \\
&-& \left. \frac{2}{(5+k)} \hat{G}_{12} T_{+}^{(\frac{3}{2})} +
\frac{2}{(5+k)} \hat{G}_{21} T_{-}^{(\frac{3}{2})} \right](w) +\cdots,
\nonu \\
T_{-}^{(\frac{3}{2})}(z) \, W_{+}^{(\frac{5}{2})}(w) & = &
\frac{1}{(z-w)^3} \, \left[ -\frac{8(5+2k)}{(5+k)^2} i \hat{A}_3+
\frac{8k(-2+k)}{3(5+k)^2} i  \hat{B}_3 -\frac{4(-2+k)}{3(5+k)} 
T^{(1)} \right](w)
\nonu \\
& + & \frac{1}{(z-w)^2}
\left[-\frac{1}{2} {\bf P^{(2)}} 
-\frac{4(75+340k+175k^2+22k^3)}{3(5+k)^2(3+7k)} \hat{T}
\right. \nonu \\
&+ & \frac{4(-3+k)}{3(5+k)} T^{(2)} +
\frac{4(4+k)}{(5+k)} W^{(2)}
-\frac{16(-1+k)}{3(5+k)^2} \hat{A}_{+} \hat{A}_{-}
-\frac{4(-13+4k)}{3(5+k)^2} \hat{A}_3 \hat{A}_3
\nonu \\
&+ & \frac{8(-4+k)}{3(5+k)^2} \hat{A}_3 \hat{B}_3
-\frac{4}{3(5+k)} \hat{B}_{+} \hat{B}_{-}
+\frac{4(-5+2k)}{3(5+k)^2} \hat{B}_3 \hat{B}_3
-\frac{16(-1+k)}{3(5+k)^2} i \pa \hat{A}_3 
\nonu \\
&-& \left. \frac{4}{3(5+k)} i \pa \hat{B}_3
-\frac{4}{(5+k)} i T^{(1)} \hat{A}_3 + \frac{4}{(5+k)} i
T^{(1)} \hat{B}_3 
\right](w) \nonu \\
& + & \frac{1}{(z-w)} \left[ \frac{1}{2} {\bf P^{(3)}} -\frac{1}{8} 
{\bf \pa P^{(2)} }
-W^{(3)} -\frac{4(-2+k)}{(19+23k)} T^{(1)} \hat{T} -\frac{3}{(5+k)} i
T^{(1)} \pa \hat{A}_3
\right. \nonu \\
&- & \frac{2}{(5+k)} i \hat{A}_{+} V_{+}^{(2)}
-\frac{4(-1+k)}{3(5+k)^2} \hat{A}_{+} \pa \hat{A}_{-}
-\frac{2}{(5+k)} i \hat{A}_{-} U_{-}^{(2)}  
\nonu \\
&- & \frac{24(25+15k+2k^2)}{(5+k)^2(19+23k)} i \hat{A}_3 \hat{T} 
+\frac{2(13-4k)}{3(5+k)^2} \hat{A}_3 \pa \hat{A}_3
+\frac{2(-4+7k)}{3(5+k)^2} \hat{A}_3 \pa \hat{B}_3
\nonu \\
& + & \frac{1}{(5+k)} i \hat{A}_3 \pa T^{(1)}
-\frac{1}{3(5+k)} \hat{B}_{+} \pa \hat{B}_{-}
-\frac{1}{3(5+k)} \hat{B}_{-} \pa \hat{B}_{+}
\nonu \\
&+ & \frac{8k(-2+k)}{(5+k)(19+23k)} i \hat{B}_3 \hat{T}
-\frac{2(4+5k)}{3(5+k)^2} \hat{B}_3 \pa \hat{A}_3 
+\frac{2(-5+2k)}{3(5+k)^2} \hat{B}_3 \pa \hat{B}_3
\nonu \\
& + & \frac{1}{(5+k)} i \hat{B}_3 \pa T^{(1)}
-\frac{(15+83k+22k^2)}{3(5+k)(3+7k)} \pa \hat{T}
+\frac{(-6+k)}{3(5+k)} \pa T^{(2)} +
\pa W^{(2)}
\nonu \\
&-& \left. \frac{1}{2(5+k)} T^{(1)} \hat{B}_{+} \hat{B}_{-} 
+\frac{1}{2(5+k)} \hat{B}_{-} T^{(1)} \hat{B}_{+}
-\frac{4(-1+k)}{3(5+k)^2} \hat{A}_{-} \pa \hat{A}_{+}
\right](w) \cdots,
\nonu \\
T_{+}^{(\frac{3}{2})}(z) \, 
W^{(3)}(w) & = & \frac{1}{(z-w)^3} \, \left[ \frac{4(-1+3k)}{(5+k)^2} 
\hat{G}_{21}  + \frac{4(-1170-918k+119k^2 + 55k^3)}{
3(5+k)^2(19+23k)} T_{+}^{(\frac{3}{2})} \right](w)
\nonu \\
& + & \frac{1}{(z-w)^2} \, \left[ \frac{(24+5k)}{2(5+k)} 
{\bf P_{+}^{(\frac{5}{2})}} + \frac{(24+5k)}{(5+k)} W_{+}^{(\frac{5}{2})} 
-\frac{4}{(5+k)^2} i \hat{A}_{-} \hat{G}_{11}
\right. \nonu \\
&+ &  \frac{4(11+3k)}{(5+k)^2} 
i \hat{A}_{-} U^{(\frac{3}{2})} 
-\frac{12}{(5+k)^2} i \hat{A}_3 \hat{G}_{21}
\nonu \\
& - &  \frac{2(413+774k+169k^2)}{(5+k)^2(19+23k)} i \hat{A}_3 
T_{+}^{(\frac{3}{2})} +\frac{(1+k)}{(5+k)^2} i \hat{B}_{-} 
\hat{G}_{22} \nonu \\
& -&   \frac{1}{(5+k)} 
i \hat{B}_{-} V^{(\frac{3}{2})} 
-\frac{4k}{(5+k)^2} i \hat{B}_3 \hat{G}_{21}
 -   \frac{2(57+350k+37k^2)}{(5+k)^2(19+23k)} i \hat{B}_3 
T_{+}^{(\frac{3}{2})}
\nonu \\
&+&  \frac{2}{(5+k)} T^{(1)} \hat{G}_{21} -
\frac{12(-3+k)}{(19+23k)} T^{(1)}  T_{+}^{(\frac{3}{2})} 
+\frac{4}{(5+k)^2} \pa \hat{G}_{21} \nonu \\
& + & \left.
\frac{(-1731-1754k-283k^2+12k^3)}{3(5+k)^2(19+23k)} \pa T_{+}^{(\frac{3}{2})}
\right](w) \nonu \\
& + & \frac{1}{(z-w)} \, 
\left[ -\frac{1}{(5+k)} i {\bf R^{(\frac{5}{2})}} \hat{B}_{-} 
+\frac{(8+k)}{2(5+k)} 
{\bf \pa P_{+}^{(\frac{5}{2})}} + \frac{1}{(5+k)} i {\bf Q^{(\frac{5}{2})}} 
\hat{A}_{-} \right. \nonu \\
&+& \frac{(37+4k)}{3(5+k)^2} \hat{A}_{-} \hat{A}_{+} \hat{G}_{21} -
\frac{(9+k)}{(5+k)^2} \hat{A}_{-} \hat{A}_{+} T_{+}^{(\frac{3}{2})} +
\frac{2}{(5+k)} i \hat{A}_{-} U^{(\frac{5}{2})} 
\nonu \\
& - & \frac{4(7+k)}{(5+k)^2}
\hat{A}_{-} \hat{A}_3 U^{(\frac{3}{2})}
-\frac{4}{(5+k)^2} \hat{A}_{-} \hat{B}_{-} \hat{G}_{12}
+\frac{8}{(5+k)^2} \hat{A}_{-} \hat{B}_3 U^{(\frac{3}{2})} 
\nonu \\ 
&+& \frac{4}{(5+k)^2} i \hat{A}_{-} \pa \hat{G}_{11} 
+\frac{4(4+k)}{3(5+k)^2} i \hat{A}_{-} \pa U^{(\frac{3}{2})} 
+\frac{k}{(5+k)^2} \hat{B}_{+} \hat{B}_{-} \hat{G}_{21}
\nonu \\
& + & \frac{(1+k)}{(5+k)^2} \hat{B}_{+} \hat{B}_{-} 
T_{+}^{(\frac{3}{2})}
-\frac{k}{(5+k)^2} \hat{B}_{-} \hat{B}_{+} \hat{G}_{21}
-\frac{(1+k)}{(5+k)^2} \hat{B}_{-} \hat{B}_{+} T_{+}^{(\frac{3}{2})}
\nonu \\
& - & \frac{1}{(5+k)} \hat{B}_{-} \hat{B}_3 \hat{G}_{22}
+\frac{1}{(5+k)} \hat{B}_{-} \hat{B}_3 V^{(\frac{3}{2})}
-\frac{(21+k)}{3(5+k)^2} i \hat{B}_{-} \hat{B}_3 \pa \hat{G}_{22} 
\nonu \\
&
+ & 
\frac{(33+k)}{3(5+k)^2} i \hat{B}_{-}  \pa V^{(\frac{3}{2})} 
-\frac{(37+4k)}{3(5+k)^2} \hat{A}_{+} \hat{A}_{-} \hat{G}_{21}
+\frac{(9+k)}{(5+k)^2} \hat{A}_{+} \hat{A}_{-} T^{(\frac{3}{2})}
\nonu \\
&+ & \frac{4}{(5+k)} \hat{A}_3 \hat{A}_{-} U^{(\frac{3}{2})} 
+\frac{4}{(5+k)^2} \hat{A}_3 \hat{A}_3 T_{+}^{(\frac{3}{2})}
-\frac{8}{(5+k)^2} \hat{A}_3 \hat{B}_3 T_{+}^{(\frac{3}{2})}
\nonu \\
& - &
\frac{2(49+138k+41k^2)}{(5+k)^2(19+23k)} i \hat{A}_3 \pa  T_{+}^{(\frac{3}{2})}
+\frac{1}{(5+k)} \hat{B}_3 \hat{B}_{-} \hat{G}_{22}
-\frac{1}{(5+k)} \hat{B}_3 \hat{B}_{-} V^{(\frac{3}{2})}
\nonu \\
&+& \frac{4}{(5+k)^2} \hat{B}_3 \hat{B}_3 T_{+}^{(\frac{3}{2})}
+\frac{2(19-58k+3k^2)}{(5+k)^2(19+23k)} i \hat{B}_3 \pa T_{+}^{(\frac{3}{2})}
-\frac{4(-3+k)}{(19+23k)} T^{(1)} \pa T_{+}^{(\frac{3}{2})}
\nonu \\
&+ &  \frac{(6+k)}{(5+k)} \pa W_{+}^{(\frac{5}{2})}
+
\frac{1}{(5+k)} \pa T^{(1)} \hat{G}_{21}
-\frac{2(25+4k)}{3(5+k)^2} \pa^2 T_{+}^{(\frac{3}{2})}
\nonu \\
& + & \frac{12k}{(5+k)(3+7k)} \hat{G}_{21} \hat{T}
+\frac{2}{(5+k)} \hat{G}_{21} T^{(2)} -\frac{4(23+2k)}{3(5+k)^2} i \hat{G}_{21}
\pa \hat{A}_3
\nonu \\
&+& \left. \frac{8(-7+4k+k^2)}{(5+k)(19+23k)} T_{+}^{(\frac{3}{2})} \hat{T}
\right](w) +\cdots,
\nonu \\
T_{-}^{(\frac{3}{2})}(z) \, W^{(3)}(w) & = &
\frac{1}{(z-w)^3}
 \, \left[ \frac{4(-1+3k)}{(5+k)^2} 
\hat{G}_{12}  - \frac{4(-1170-918k+119k^2 + 55k^3)}{
3(5+k)^2(19+23k)} T_{-}^{(\frac{3}{2})} \right](w)
\nonu \\
& + & \frac{1}{(z-w)^2} \, \left[ -\frac{(22+5k)}{2(5+k)} 
{\bf P_{-}^{(\frac{5}{2})}} + \frac{(22+5k)}{(5+k)} W_{-}^{(\frac{5}{2})} 
+\frac{4}{(5+k)^2} i \hat{A}_{+} \hat{G}_{22}
\right. \nonu \\
&+ &  \frac{8(4+k)}{(5+k)^2} 
i \hat{A}_{+} V^{(\frac{3}{2})} 
+\frac{12}{(5+k)^2} i \hat{A}_3 \hat{G}_{12}
\nonu \\
& - &  \frac{6(25+41k)}{(5+k)(19+23k)} i \hat{A}_3 
T_{-}^{(\frac{3}{2})} +\frac{(13+k)}{(5+k)^2} i \hat{B}_{+} 
\hat{G}_{11} \nonu \\
& +&   \frac{(9+k)}{(5+k)^2} 
i \hat{B}_{+} U^{(\frac{3}{2})} 
+\frac{4k}{(5+k)^2} i \hat{B}_3 \hat{G}_{12}
 -   \frac{2(19+342k+83k^2)}{(5+k)^2(19+23k)} i \hat{B}_3 
T_{-}^{(\frac{3}{2})}
\nonu \\
&-&  \frac{2}{(5+k)} T^{(1)} \hat{G}_{12} -
\frac{12(-3+k)}{(19+23k)} T^{(1)}  T_{-}^{(\frac{3}{2})} 
+\frac{4}{(5+k)^2} \pa \hat{G}_{12} \nonu \\
& - & \left.
\frac{(-1693-1670k-237k^2+12k^3)}{3(5+k)^2(19+23k)} \pa T_{-}^{(\frac{3}{2})}
\right](w) \nonu \\
& + & \frac{1}{(z-w)} \left[ 
-\frac{(6+k)}{2(5+k)} {\bf \pa P_{-}^{(\frac{5}{2})}} 
-\frac{k}{(5+k)^2} \hat{A}_{-} \hat{A}_{+} \hat{G}_{12} -
\frac{(-15+k)}{3(5+k)^2} \hat{A}_{-} \hat{A}_{+} T_{-}^{(\frac{3}{2})}
\right. \nonu \\
&- & \frac{4}{(5+k)^2} \hat{B}_{+} \hat{A}_{+} \hat{G}_{21} 
-\frac{k}{(5+k)^2} \hat{B}_{+} \hat{B}_{-} \hat{G}_{12} 
- \frac{(1+k)}{(5+k)^2} \hat{B}_{+} \hat{B}_{-} T_{-}^{(\frac{3}{2})}
\nonu \\
& - & \frac{(7+k)}{(5+k)^2} \hat{B}_{+} \hat{B}_3 \hat{G}_{11} 
\nonu \\
& - &  \frac{(11+k)}{(5+k)^2} \hat{B}_{+} \hat{B}_3 U^{(\frac{3}{2})} 
+ \frac{(9+k)}{(5+k)^2} i \hat{B}_{+} \pa \hat{G}_{11} 
+ \frac{(9+k)}{(5+k)^2} i \hat{B}_{+} \pa U^{(\frac{3}{2})}
\nonu \\
&+ &  \frac{k}{(5+k)^2} \hat{B}_{-} \hat{B}_{+} \hat{G}_{12}
+\frac{(-3+k)}{(5+k)^2} \hat{B}_{-} \hat{B}_{+} T_{-}^{(\frac{3}{2})}
+\frac{k}{(5+k)^2} \hat{A}_{+} \hat{A}_{-} \hat{G}_{12}
\nonu \\
& + & \frac{(-15+k)}{3(5+k)^2} \hat{A}_{+} \hat{A}_{-} T_{-}^{(\frac{3}{2})}
+\frac{4}{(5+k)^2} \hat{A}_{+} \hat{A}_3 \hat{G}_{22}
+\frac{2(11+2k)}{(5+k)^2} \hat{A}_{+} \hat{A}_3 V^{(\frac{3}{2})}
\nonu \\
&- & \frac{8}{(5+k)^2} \hat{A}_{+} \hat{B}_3 V^{(\frac{3}{2})}
-\frac{4}{(5+k)^2} i \hat{A}_{+} \pa V^{(\frac{3}{2})}
-\frac{4}{(5+k)^2} \hat{A}_3 \hat{A}_{+} \hat{G}_{22} 
\nonu \\
& - & \frac{2(7+2k)}{(5+k)^2} 
\hat{A}_3 \hat{A}_{+} V^{(\frac{3}{2})}
-\frac{4}{(5+k)^2} \hat{A}_3 \hat{A}_3 T_{-}^{(\frac{3}{2})}
+\frac{8}{(5+k)^2} \hat{A}_3 \hat{B}_3  T_{-}^{(\frac{3}{2})}
\nonu \\
& + & 
\frac{2(-11-54k+5k^2)}{(5+k)(95+134k+23k^2)} i \hat{A}_3 \pa 
T_{-}^{(\frac{3}{2})}
+\frac{(7+k)}{(5+k)^2} \hat{B}_3 \hat{B}_{+} \hat{G}_{11} 
\nonu \\
& - & \frac{4}{(5+k)^2} \hat{B}_3 \hat{B}_3 T_{-}^{(\frac{3}{2})}
-\frac{2(19+142k+43k^2)}{(5+k)(95+134k+23k^2)} i \hat{B}_3 \pa 
T_{-}^{(\frac{3}{2})}
\nonu \\
& + & \frac{(11+k)}{(5+k)^2} 
\hat{B}_3 \hat{B}_{+} U^{(\frac{3}{2})}
-\frac{4(-3+k)}{(19+23k)} T^{(1)} \pa T_{-}^{(\frac{3}{2})}
+\frac{12k}{(5+k)(3+7k)} \hat{G}_{12} \hat{T}
\nonu \\
& + & \frac{2}{(5+k)} \hat{G}_{12} T^{(2)} + \frac{2(3+k)}{(5+k)^2} 
i \hat{G}_{12} \pa \hat{A}_3 -\frac{1}{(5+k)} \hat{G}_{12} \pa T^{(1)}
\nonu \\
& - & \frac{8(-7+4k+k^2)}{(5+k)(19+23k)} T_{-}^{(\frac{3}{2})} \hat{T}
-\frac{16(6+k)}{3(5+k)^2} i  T_{-}^{(\frac{3}{2})} \pa \hat{A}_3
+\frac{2}{(5+k)} \hat{G}_{11} V_{-}^{(2)}
\nonu \\
&+& \left. \frac{2}{(5+k)} U^{(\frac{3}{2})} V_{-}^{(2)} -
\frac{2}{(5+k)} V^{(\frac{3}{2})} U_{-}^{(2)}  + \frac{(6+k)}{(5+k)}
\pa W_{-}^{(\frac{5}{2})} \right](w) + \cdots.
\label{opeopeabove}
\eea
In the second OPE of (\ref{opeopeabove}), the first order pole subtracted 
by the correct descendant field leads to a primary field.
Note that one has a relation $\hat{G}_{21} \hat{G}_{21}(w) = \frac{2}{(5+k)}
\pa (\hat{A}_{-} \hat{B}_{-} )(w)$.
In the third OPE, the second order pole containing the nonlinear terms 
is a quasi primary field.
The first order term consisting of last three terms with $\pa \hat{T}(w)$
term leads to a primary field.
In the fourth OPE, the second order pole starting from $\hat{A}_{+} 
\hat{A}_{-}(w)$ gives a quasi primary field.
The first order term starting from $T^{(1)} \hat{T}(w)$ term
by subtracting the correct descendant field  provides 
a quasi primary field.  
In the fifth OPE, the second order pole is a primary field and 
 the first order pole is a quasi primary field after subtracting the 
descendant field.
In the sixth OPE, the second order pole is a primary field. 

\section{The nontrivial OPEs between  the higher spin-$2$ current, $T^{(2)}(z)$, 
and 
other $13$ higher currents}

We present the OPEs between the last component 
in the first ${\cal N}=2$ multiplet and the remaining higher spin currents.

$\bullet$ The OPEs between the higher spin-$2$ current and the
first ${\cal N}=2$ multiplet of (\ref{lowesthigher})

\bea
T^{(2)}(z) \, T^{(2)}(w) & = & \frac{1}{(z-w)^4} \, 
\left[ \frac{72k(3+4k+k^2)}{(5+k)^2(3+7k)} \right] \nonu \\
& + & 
\frac{1}{(z-w)^2} \, \left[ \frac{96k(1+k)(3+k)}{(5+k)(3+7k)^2} \hat{T}+
\frac{4(12+k+k^2)}{(5+k)(3+7k)} T^{(2)}\right](w) \nonu \\ 
& + & 
\frac{1}{(z-w)}  \, \frac{1}{2} \pa \{ T^{(2)} \, T^{(2)} \}_{-2}(w) 
+\cdots.
\nonu
\eea

$\bullet$ The OPEs between the higher spin-$2$ current and the
second and third ${\cal N}=2$ multiplet of (\ref{lowesthigher})

\bea
T^{(2)}(z) \, U^{(\frac{3}{2})}(w) & = & 
-\frac{1}{(z-w)^2}\, 
\left[ \frac{(-21-7k+2k^2)}{(5+k)(3+7k)} U^{(\frac{3}{2})} \right] (w) 
\nonu \\
& + & \frac{1}{(z-w)} \, \left[ \frac{2}{3} \pa \{ T^{(2)} \, 
U^{(\frac{3}{2})}\}_{-2} +\frac{1}{2} {\bf Q^{(\frac{5}{2})}} +U^{(\frac{5}{2})} 
\right](w) +\cdots,
\nonu \\
T^{(2)}(z) \, V^{(\frac{3}{2})}(w) & = & 
-\frac{1}{(z-w)^2}\, 
\left[ \frac{(-21-7k+2k^2)}{(5+k)(3+7k)} V^{(\frac{3}{2})} \right] (w) 
\nonu \\
& + & \frac{1}{(z-w)} \, \left[ \frac{1}{2} {\bf R^{(\frac{5}{2})}} -
V^{(\frac{5}{2})} 
+\frac{2}{(5+k)} i \hat{A}_{-} T_{-}^{(\frac{3}{2})} +
\frac{4}{(5+k)} i \hat{A}_3 V^{(\frac{3}{2})} \right. 
\nonu \\
& -&  \frac{2}{(5+k)} i \hat{B}_{+} T_{+}^{(\frac{3}{2})} 
-\frac{4}{(5+k)} i \hat{B}_3 V^{(\frac{3}{2})} -\frac{2(-33-35k+2k^2)}{
3(5+k)(3+7k)} \pa V^{(\frac{3}{2})}
\nonu \\
&-& \left.  \frac{2}{(5+k)} i \hat{B}_{+}
\hat{G}_{21}
\right](w) +\cdots,
\nonu \\
T^{(2)}(z) \, U_{+}^{(2)}(w) & = & \frac{1}{(z-w)^3}
\, \left[ \frac{2k(6+k)}{(5+k)^2} i \hat{B}_{-} \right](w) \nonu \\
& + & 
\frac{1}{(z-w)^2} \, \left[ \frac{2(12+k+k^2)}{(5+k)(3+7k)}
U_{+}^{(2)} + \frac{4}{(5+k)} \hat{A}_3 \hat{B}_{-} -\frac{4k}{(5+k)^2}
\hat{B}_{-} \hat{B}_3  \right. \nonu \\
&+ & \left. \frac{k(4+k)}{(5+k)^2} i \pa \hat{B}_{-} -\frac{2}{(5+k)} i 
T^{(1)} \hat{B}_{-}\right](w) 
\nonu \\
& + & \frac{1}{(z-w)} \, \left[ 
\frac{1}{2} {\bf Q_{+}^{(3)}}
-\frac{1}{2(5+k)} i T^{(1)} \pa \hat{B}_{-}
+\frac{(11+2k)}{(5+k)^2} \hat{A}_3 \pa \hat{B}_{-}
\right. \nonu \\
&+ & \frac{4k(75+117k+14k^2)}{(5+k)(3+7k)(19+23k)} i \hat{B}_{-} \hat{T}
+\frac{2}{(5+k)} i \hat{B}_{-} T^{(2)}
+\frac{(5+2k)}{(5+k)^2} \hat{B}_{-} \pa \hat{A}_3
\nonu \\
& - & \frac{3k}{(5+k)^2} \hat{B}_{-} \pa \hat{B}_3
-\frac{3}{2(5+k)} i \hat{B}_{-} \pa T^{(1)}
-\frac{k}{(5+k)^2} \hat{B}_3 \pa \hat{B}_{-}
\nonu \\
& + & \frac{(21-5k+2k^2)}{2(5+k)(3+7k)} \pa U_{+}^{(2)}
+\frac{k(6+k)}{3(5+k)^2} i \pa^2 \hat{B}_{-}
\nonu \\
&-& \left. \frac{2}{(5+k)} \hat{G}_{11} T_{+}^{(\frac{3}{2})}
+ \frac{2}{(5+k)} T_{+}^{(\frac{3}{2})} 
U^{(\frac{3}{2})} \right](w) +\cdots,
\nonu \\
T^{(2)}(z) \, V_{-}^{(2)}(w) & = & \frac{1}{(z-w)^3}
\, \left[ \frac{2k(6+k)}{(5+k)^2} i \hat{B}_{+} \right](w) \nonu \\
& + & 
\frac{1}{(z-w)^2} \, \left[ \frac{2(12+k+k^2)}{(5+k)(3+7k)}
V_{-}^{(2)} - \frac{4}{(5+k)} \hat{A}_3 \hat{B}_{+} +\frac{4k}{(5+k)^2}
\hat{B}_{+} \hat{B}_3  \right. \nonu \\
&+ & \left. \frac{k(4+k)}{(5+k)^2} i \pa \hat{B}_{+} +\frac{2}{(5+k)} i 
T^{(1)} \hat{B}_{+}\right](w) 
\nonu \\
& + & \frac{1}{(z-w)} \, \left[ 
\frac{1}{2} {\bf R_{-}^{(3)}}
+\frac{1}{2(5+k)} i T^{(1)} \pa \hat{B}_{+}
+\frac{(11+2k)}{(5+k)^2} i \hat{A}_3 \hat{B}_{+} \hat{B}_3
\right. \nonu \\
&+ & \frac{4k(75+117k+14k^2)}{(5+k)(3+7k)(19+23k)} i \hat{B}_{+} \hat{T}
+\frac{2}{(5+k)} i \hat{B}_{+} T^{(2)}
-\frac{k(6+k)}{3(5+k)^2} i \hat{B}_{+} \hat{B}_{+} \hat{B}_{-} 
\nonu \\
& + & \frac{k(33+4k)}{3(5+k)^2} \hat{B}_{+} \pa \hat{B}_3
+\frac{3}{2(5+k)} i \hat{B}_{+} \pa T^{(1)}
+\frac{k}{(5+k)^2} \hat{B}_3 \pa \hat{B}_{+}
\nonu \\
& + & \frac{(21-5k+2k^2)}{2(5+k)(3+7k)} \pa V_{-}^{(2)}
-\frac{(5+2k)}{2(5+k)^2} i \hat{A}_{+} \hat{A}_{-} \hat{B}_{+}
+\frac{(5+2k)}{2(5+k)^2} i  \hat{A}_{-} \hat{A}_{+} \hat{B}_{+}
\nonu \\
&-&  \frac{2}{(5+k)} \hat{G}_{22} T_{-}^{(\frac{3}{2})}
- \frac{2}{(5+k)} T_{-}^{(\frac{3}{2})} 
V^{(\frac{3}{2})} +\frac{k(6+k)}{3(5+k)^2} i 
\hat{B}_{-} \hat{B}_{+} \hat{B}_{+}
\nonu \\
& -& \left. \frac{(11+2k)}{(5+k)^2} i \hat{B}_3 \hat{A}_3 \hat{B}_{+}
\right](w) +\cdots,
\nonu \\
T^{(2)}(z) \, U_{-}^{(2)}(w) & = & \frac{1}{(z-w)^3}
\, \left[ \frac{6(5+2k)}{(5+k)^2} i \hat{A}_{+} \right](w) \nonu \\
& + & 
\frac{1}{(z-w)^2} \, \left[ \frac{2(12+k+k^2)}{(5+k)(3+7k)}
U_{-}^{(2)} + \frac{3(5+2k)}{(5+k)^2} i \pa \hat{A}_{+} \right](w) 
\nonu \\
& + & \frac{1}{(z-w)} \, \left[ 
\frac{1}{2} {\bf Q_{-}^{(3)} }
+\frac{3}{2(5+k)} i T^{(1)} \pa \hat{A}_{+}
-\frac{4}{(5+k)} i \hat{A}_3 \pa U_{-}^{(2)}
\right. \nonu \\
&+ & \frac{8(23+13k)}{(5+k)(19+23k)} i \hat{A}_{+} \hat{T}
-\frac{2}{(5+k)} i \hat{A}_{+} W^{(2)}
+\frac{(29+8k)}{(5+k)^2} \hat{A}_{+} \pa \hat{A}_3
\nonu \\
& - & \frac{(-8+k)}{(5+k)^2} \hat{A}_{+} \pa \hat{B}_3
-\frac{3}{2(5+k)} i \hat{A}_{+} \pa T^{(1)}
+\frac{3}{(5+k)^2} \hat{A}_3 \pa \hat{A}_{+}
+\frac{4}{(5+k)} i  \hat{B}_3 U_{-}^{(2)}
\nonu \\
& + & \frac{(39+37k+2k^2)}{2(5+k)(3+7k)} \pa U_{-}^{(2)}
-\frac{2}{(5+k)} \hat{G}_{11} \hat{G}_{12}
+\frac{2}{(5+k)} \hat{G}_{12} U^{(\frac{3}{2})}
\nonu \\
&+&  \frac{2}{(5+k)} \hat{G}_{11} T_{-}^{(\frac{3}{2})}
- \frac{2}{(5+k)} T_{-}^{(\frac{3}{2})} 
U^{(\frac{3}{2})} -\frac{(5+2k)}{(5+k)^2}  i \hat{A}_{+} \hat{A}_{+}
\hat{A}_{-} \nonu \\
&+ & \frac{4}{(5+k)^2}  i \hat{A}_{+} \hat{A}_3 \hat{A}_3 
- \frac{k}{(5+k)^2} i \hat{A}_{+} \hat{A}_3 \hat{B}_3 
+ \frac{4}{(5+k)^2} 
i \hat{A}_{+} \hat{B}_{+} \hat{B}_{-} \nonu \\
& + &  \left. \frac{4}{(5+k)^2} i 
\hat{A}_{+} \hat{B}_3 \hat{B}_3 
+ \frac{(9+2k)}{(5+k)^2} i \hat{A}_{-} \hat{A}_{+} \hat{A}_{+}
+\frac{(-8+k)}{(5+k)^2} i \hat{A}_3 \hat{A}_{+} \hat{B}_3
\right](w) +\cdots,
\nonu \\
T^{(2)}(z) \, V_{+}^{(2)}(w) & = & \frac{1}{(z-w)^3}
\, \left[ 
\frac{6(5+2k)}{(5+k)^2} i \hat{A}_{-} \right] (w) \nonu \\
& + & 
\frac{1}{(z-w)^2} \, \left[ \frac{2(12+k+k^2)}{(5+k)(3+7k)}
V_{+}^{(2)} + \frac{3(5+2k)}{(5+k)^2} i \pa \hat{A}_{-} \right](w) 
\nonu \\
& + & \frac{1}{(z-w)} \, \left[ 
\frac{1}{2} {\bf R_{+}^{(3)} }
-\frac{3}{2(5+k)} i T^{(1)} \pa \hat{A}_{-}
+\frac{4}{(5+k)} i \hat{A}_3  V_{+}^{(2)}
\right. \nonu \\
&+ & \frac{8(23+13k)}{(5+k)(19+23k)} i \hat{A}_{-} \hat{T}
-\frac{2}{(5+k)} i \hat{A}_{-} W^{(2)}
-\frac{9}{(5+k)^2} \hat{A}_{-} \pa \hat{A}_3
\nonu \\
& + & \frac{(-8+k)}{(5+k)^2} \hat{A}_{-} \pa \hat{B}_3
+\frac{3}{2(5+k)} i \hat{A}_{-} \pa T^{(1)}
-\frac{3}{(5+k)^2} \hat{A}_3 \pa \hat{A}_{-}
\nonu \\
& + & \frac{(39+37k+2k^2)}{2(5+k)(3+7k)} \pa V_{+}^{(2)}
-\frac{2}{(5+k)} \hat{G}_{21} \hat{G}_{22}
+\frac{2}{(5+k)} \hat{G}_{21} V^{(\frac{3}{2})}
\nonu \\
&+&  \frac{2}{(5+k)} \hat{G}_{22} T_{+}^{(\frac{3}{2})}
+ \frac{2}{(5+k)} T_{+}^{(\frac{3}{2})} 
V^{(\frac{3}{2})} +\frac{4}{(5+k)^2}  i \hat{A}_{+} \hat{A}_{-}
\hat{A}_{-} 
+\frac{(5+2k)}{(5+k)^2} i \pa^2 \hat{A}_{-}
\nonu \\
&+ & \frac{4}{(5+k)^2}  i \hat{A}_{-} \hat{A}_3 \hat{A}_3 
- \frac{8}{(5+k)^2} i \hat{A}_{-} \hat{A}_3 \hat{B}_3 
+ \frac{4}{(5+k)^2} 
i \hat{A}_{-} \hat{B}_{+} \hat{B}_{-} \nonu \\
& + &  \left. \frac{4}{(5+k)^2} i 
\hat{A}_{-} \hat{B}_3 \hat{B}_3 
-\frac{(-16+k)}{(5+k)^2} \hat{B}_3 \pa \hat{A}_{-}
-\frac{4}{(5+k)} i  \hat{B}_3 V_{+}^{(2)}
\right](w) +\cdots,
\nonu \\
T^{(2)}(z) \, U^{(\frac{5}{2})}(w) & = & \frac{1}{(z-w)^3}
\, \left[ -\frac{8(12+8k+k^2)}{3(5+k)^2} \hat{G}_{11} + 
\frac{4(11+3k)}{3(5+k)^2} U^{(\frac{3}{2})} \right](w)
\nonu \\
& + & \frac{1}{(z-w)^2} \, \left[ -\frac{(-1+k)}{6(5+k)} 
{\bf Q^{(\frac{5}{2})}} 
+\frac{(93+13k+4k^2)}{3(5+k)(3+7k)} U^{(\frac{5}{2})}
+\frac{4(9+2k)}{3(5+k)^2} i \hat{A}_3 \hat{G}_{11}
\right. \nonu \\
& + & \frac{4(9+2k)}{3(5+k)^2} i \hat{A}_3 U^{(\frac{3}{2})}
+\frac{4(3+2k)}{3(5+k)^2} i \hat{B}_{-} T_{-}^{(\frac{3}{2})}
-\frac{4k}{(5+k)^2} i \hat{B}_3 \hat{G}_{11}
\nonu \\
&- &  \frac{4k}{(5+k)^2} i \hat{B}_3  U^{(\frac{3}{2})} 
-\frac{4(18+15k+2k^2)}{9(5+k)^2} \pa \hat{G}_{11}
+\frac{4(17+4k)}{9(5+k)^2} \pa U^{(\frac{3}{2})}
\nonu \\
&+ & \left. \frac{2}{(5+k)} T^{(1)} \hat{G}_{11} +
\frac{2}{(5+k)} T^{(1)} U^{(\frac{3}{2})} \right](w) 
\nonu \\
& + & 
\frac{1}{(z-w)} \, \left[ 
-\frac{(-7+k)}{15(5+k)} {\bf \pa Q^{(\frac{5}{2})}}
+\frac{1}{(5+k)} i {\bf P_{-}^{(\frac{5}{2})}} \hat{B}_{-}
+\frac{1}{2} {\bf Q^{(\frac{7}{2})} }
 \right. \nonu \\
&+ & \frac{4(9+5k)}{15(5+k)^2} \hat{A}_{-} \hat{A}_{+} \hat{G}_{11}
+\frac{(171+32k)}{15(5+k)^2} \hat{A}_{-} \hat{A}_{+} U^{(\frac{3}{2})}
+ \frac{8(-3+k)}{5(5+k)^2} \hat{B}_{+} \hat{B}_{-} \hat{G}_{11}
\nonu \\
& + & \frac{4}{(5+k)^2} \hat{B}_{+} \hat{B}_{-} U^{(\frac{3}{2})}
 -\frac{4}{(5+k)^2} \hat{B}_{-} \hat{A}_{+} \hat{G}_{22}
+\frac{4}{(5+k)^2} \hat{B}_{-} \hat{A}_{+} V^{(\frac{3}{2})}
\nonu \\
& - & \frac{2}{(5+k)} i \hat{B}_{-} W_{-}^{(\frac{5}{2})}
-\frac{8}{(5+k)^2} \hat{B}_{-} \hat{A}_3 \hat{G}_{12}
+ \frac{8}{(5+k)^2} \hat{B}_{-} \hat{A}_3 T_{-}^{(\frac{3}{2})}
\nonu \\
& - & \frac{8(-3+k)}{5(5+k)^2} \hat{B}_{-} \hat{B}_{+} \hat{G}_{11}
+\frac{4(4+k)}{5(5+k)^2} i \hat{B}_{-} \pa \hat{G}_{12}
-\frac{4(22+3k)}{15(5+k)^2} i \hat{B}_{-} \pa T_{-}^{(\frac{3}{2})}
\nonu \\
& - & \frac{2}{(5+k)} i \hat{A}_{+} W_{+}^{(\frac{5}{2})}
-\frac{4(9+5k)}{15(5+k)^2} \hat{A}_{+} \hat{A}_{-} \hat{G}_{11}
-\frac{(51+32k)}{15(5+k)^2} \hat{A}_{+} \hat{A}_{-} U^{(\frac{3}{2})}
\nonu \\
&+ & \frac{2(13+418k+37k^2)}{(5+k)^2(95+115k)} \hat{A}_{+} \hat{A}_3
\hat{G}_{21} 
-\frac{4(13+418k+37k^2)}{(5+k)^2(285+345k)} 
i \hat{A}_{+} \pa \hat{G}_{21}
\nonu \\
& - & 
\frac{8(6+5k)}{15(5+k)^2} i \hat{A}_{+} \pa T_{+}^{(\frac{3}{2})} 
-\frac{2(13+418k+37k^2)}{
(5+k)^2(95+115k)} \hat{A}_3 \hat{A}_{+} \hat{G}_{21}
-\frac{4}{(5+k)} i \hat{A}_3 U^{(\frac{5}{2})}
\nonu \\
& + & \frac{4}{(5+k)^2} \hat{A}_3 \hat{A}_3 U^{(\frac{3}{2})}
-\frac{8}{(5+k)^2} \hat{A}_3 \hat{B}_3  U^{(\frac{3}{2})}
+\frac{24}{5(5+k)^2} i \hat{A}_3 \pa \hat{G}_{11} 
\nonu \\
&+ & \frac{4}{(5+k)} i \hat{B}_3  U^{(\frac{5}{2})}
+\frac{4}{(5+k)^2} \hat{B}_3 \hat{B}_3  U^{(\frac{3}{2})}
-\frac{8(12+k)}{15(5+k)^2} i \hat{B}_3 \pa \hat{G}_{11} 
\nonu \\
&- & \frac{4(20+k)}{15(5+k)^2} i \hat{B}_3 \pa  U^{(\frac{3}{2})}
+ \frac{4}{5(5+k)} T^{(1)} \pa \hat{G}_{11}
+\frac{2}{(5+k)} T^{(1)} \pa  U^{(\frac{3}{2})}
\nonu \\
&+& \frac{4(21+5k)}{5(5+k)^2} i \pa \hat{A}_{+} T_{+}^{(\frac{3}{2})}
-\frac{2(-8+3k)}{5(5+k)^2} i \pa \hat{B}_{-} \hat{G}_{12}
+\frac{2(36+29k)}{15(5+k)^2} i \pa \hat{B}_{-}  T_{-}^{(\frac{3}{2})}
\nonu \\
&+& \frac{4}{5(5+k)} \pa T^{(1)} \hat{G}_{11}
+\frac{4(57+25k)}{(5+k)(47+35k)}  U^{(\frac{3}{2})} \hat{T}
\nonu \\
&+ & \frac{(-56159-38721k+170k^2)}{90(5+k)(235+222k+35k^2)}
\pa^2 U^{(\frac{3}{2})}
-\frac{4(-23+4k)}{15(5+k)^2} i \hat{A}_3 \pa  U^{(\frac{3}{2})}
\nonu \\
&- & \frac{2}{(5+k)} \hat{G}_{12} U_{+}^{(2)} 
+\frac{2}{(5+k)}  T_{-}^{(\frac{3}{2})}  U_{+}^{(2)} 
+\frac{2(147+139k+4k^2)}{15(5+k)(3+7k)} \pa U^{(\frac{5}{2})}
\nonu \\
& -& 
\frac{4(2865+8283k+5807k^2+1313k^3+92k^4)}{(5+k)^2(19+23k)(47+35k)}
\hat{G}_{11} \hat{T}
\nonu \\
&-&  \frac{2}{(5+k)}  U^{(\frac{3}{2})} 
W^{(2)} -\frac{2(10+9k)}{5(5+k)^2} i U^{(\frac{3}{2})} \pa \hat{B}_3
-\frac{1}{(5+k)}
 U^{(\frac{3}{2})} \pa T^{(1)}
\nonu \\
&-& \left.
\frac{(36423+46821k+71201k^2+43935k^3+5980k^4)}{45(5+k)^2(19+23k)(47+35k)}
\pa^2 \hat{G}_{11}
\right](w) + \cdots,
\nonu  \\
T^{(2)}(z) \, V^{(\frac{5}{2})}(w) & = & \frac{1}{(z-w)^3}
\, \left[ -\frac{8(2+k)(6+k)}{3(5+k)^2} \hat{G}_{22} + 
\frac{4(11+k)}{3(5+k)^2} V^{(\frac{3}{2})} \right](w)
\nonu \\
& + & \frac{1}{(z-w)^2} \, \left[ \frac{(-5+k)}{6(5+k)} 
{\bf R^{(\frac{5}{2})} } 
+\frac{(105+41k+4k^2)}{3(5+k)(3+7k)} V^{(\frac{5}{2})}
-\frac{4(9+2k)}{3(5+k)^2} i \hat{A}_3 \hat{G}_{22}
\right. \nonu \\
& - & \frac{4(5+2k)}{3(5+k)^2} i \hat{A}_3 V^{(\frac{3}{2})}
+\frac{4(19+4k)}{3(5+k)^2} i \hat{B}_{+} T_{+}^{(\frac{3}{2})}
+\frac{4k}{(5+k)^2} i \hat{B}_3 \hat{G}_{22}
\nonu \\
&+&  \frac{4(8+k)}{3(5+k)^2} i \hat{B}_3  V^{(\frac{3}{2})} 
-\frac{4(18+15k+2k^2)}{9(5+k)^2} \pa \hat{G}_{22}
-\frac{4(14+5k)}{9(5+k)^2} \pa V^{(\frac{3}{2})}
\nonu \\
&- &  \frac{2}{(5+k)} T^{(1)} \hat{G}_{22} +
\frac{2}{(5+k)} T^{(1)} V^{(\frac{3}{2})} +\frac{4(8+k)}{3(5+k)^2} i 
\hat{A}_{-}  T_{-}^{(\frac{3}{2})}
\nonu \\
& + & \left. \frac{8(8+k)}{3(5+k)^2} i \hat{B}_{+} \hat{G}_{21} \right](w) 
\nonu \\
& + & 
\frac{1}{(z-w)} \, \left[ 
\frac{(1+k)}{15(5+k)} {\bf \pa R^{(\frac{5}{2})}}
-\frac{1}{(5+k)} i {\bf P_{-}^{(\frac{5}{2})}} \hat{A}_{-}
+\frac{1}{2} {\bf R^{(\frac{7}{2})} }
 \right. \nonu \\
&- & 
\frac{8}{(5+k)^2} \hat{A}_{-} \hat{A}_{+} V^{(\frac{3}{2})}
+\frac{2}{(5+k)} i \hat{A}_{-} W_{-}^{(\frac{5}{2})}
- \frac{2(47+10k)}{15(5+k)^2} \hat{A}_{-} \hat{A}_3 T_{-}^{(\frac{3}{2})}
\nonu \\
& - & \frac{8}{(5+k)^2} \hat{A}_{-} \hat{B}_{+} \hat{G}_{11}
 -\frac{8}{(5+k)^2} \hat{A}_{-} \hat{B}_{+} U^{(\frac{3}{2})}
-\frac{4}{(5+k)^2} i \hat{A}_{-} \pa \hat{G}_{12}
\nonu \\
& + & \frac{164}{15(5+k)^2} i \hat{A}_{-} \pa T_{-}^{(\frac{3}{2})}
-\frac{(134+13k)}{15(5+k)^2} \hat{B}_{+} \hat{B}_{-} V^{(\frac{3}{2})}
\nonu \\
& + &  \frac{4(323+557k+140k^2+6k^3)}{(5+k)^2(285+345k)} 
i \hat{B}_{+} \pa \hat{G}_{21}
\nonu \\
& + & \frac{32(1+k)}{15(5+k)^2}  i \hat{B}_{+} \pa T_{+}^{(\frac{3}{2})}
+\frac{(14+13k)}{15(5+k)^2}  \hat{B}_{-} \hat{B}_{+} V^{(\frac{3}{2})}
\nonu \\
& + & \frac{2(47+10k)}{15(5+k)^2}  \hat{A}_3 \hat{A}_{-} T_{-}^{(\frac{3}{2})}
-\frac{4}{(5+k)^2} \hat{A}_3 \hat{A}_3 V^{(\frac{3}{2})}
+\frac{8}{(5+k)^2} \hat{A}_3 \hat{B}_3 V^{(\frac{3}{2})}
\nonu \\
&- & \frac{8(9+2k)}{15(5+k)^2} i \hat{A}_3 \pa \hat{G}_{22} 
-\frac{4(-5+4k)}{15(5+k)^2} i \hat{A}_3 \pa V^{(\frac{3}{2})}
\nonu \\
& - & 
\frac{4}{(5+k)^2}  \hat{B}_3 \hat{B}_3 V^{(\frac{3}{2})} 
+\frac{8k}{
5(5+k)^2} i \hat{B}_3 \pa \hat{G}_{22}
-\frac{4(-22+k)}{15(5+k)^2} i \hat{B}_3\pa V^{(\frac{3}{2})}
\nonu \\
& - & \frac{4}{5(5+k)} T^{(1)} \pa \hat{G}_{22}
+\frac{2}{(5+k)} T^{(1)} \pa  V^{(\frac{3}{2})}
+\frac{8(33+26k+k^2)}{15(5+k)(3+7k)}  \pa V^{(\frac{5}{2})} 
\nonu \\
&- & \frac{4}{(5+k)^2} i \pa \hat{A}_{-}  \hat{G}_{12}
-\frac{8(9+2k)}{15(5+k)^2} i \pa \hat{A}_3   \hat{G}_{22}
-\frac{2(65+8k)}{15(5+k)^2} i \pa \hat{A}_3 V^{(\frac{3}{2})} 
\nonu \\
&- & \frac{4(-893-1022k-20k^2+9k^3)}{(5+k)^2(285+345k)} i \pa
\hat{B}_{+} \hat{G}_{21}
\nonu \\
& - &  \frac{4(-836+749k+685k^2)}{
45(5+k)(235+222k+35k^2)} i \pa \hat{B}_{+} T_{+}^{(\frac{3}{2})}
+\frac{8k}{5(5+k)^2} i \pa \hat{B}_3  \hat{G}_{22}
\nonu \\
&-& \frac{4}{5(5+k)}  \pa T^{(1)} \hat{G}_{22}
-\frac{1}{(5+k)}  \pa T^{(1)} V^{(\frac{3}{2})}
\nonu \\
&-& \frac{2(53371+170323k+163134k^2+48359k^3+3185k^4)}{
45(5+k)^2(893+1746k+805k^2)} \pa^2  \hat{G}_{22}
\nonu \\
& + & \frac{4(4522+5867k+1525k^2)}{45(5+k)(235+222k+35k^2)} i  
T_{+}^{(\frac{3}{2})} \pa \hat{B}_{+}
\nonu \\
&+ & \frac{2}{(5+k)}
T_{-}^{(\frac{3}{2})} V_{+}^{(2)}
-\frac{8(1368+2493k+1178k^2+133k^3)}{
3(5+k)(19+23k)(47+35k)}  \hat{G}_{22} \hat{T}
\nonu \\
&- & \left.
\frac{12(17+9k)}{(5+k)(47+35k)} V^{(\frac{3}{2})} \hat{T} 
+\frac{2}{(5+k)}  V^{(\frac{3}{2})}  W^{(2)} 
\right](w) + \cdots.
\nonu 
\eea

$\bullet$ The OPEs between the higher spin-$2$ current and the
fourth ${\cal N}=2$ multiplet of (\ref{lowesthigher})

\bea
T^{(2)}(z) \, W^{(2)}(w) & = & \frac{1}{(z-w)^2}
\, \left[ -\frac{1}{2} {\bf P^{(2)}} -\frac{4(15+23k+16k^2)}{
3(5+k)(3+7k)} \hat{T}  + \frac{8(2+k)}{3(5+k)} T^{(2)} \right.
\nonu \\
& + & \frac{16(3+4k+k^2)}{(5+k)(3+7k)} W^{(2)}
-\frac{16(-1+k)}{3(5+k)^2} \hat{A}_{+} \hat{A}_{-}-\frac{4(-13+4k)}{
3(5+k)^2} \hat{A}_3 \hat{A}_3 \nonu \\
& + & \frac{8(-4+k)}{3(5+k)^2} \hat{A}_3 \hat{B}_3 
-\frac{4}{3(5+k)} \hat{B}_{+} \hat{B}_{-} +
\frac{4(-5+2k)}{3(5+k)^2} \hat{B}_3 \hat{B}_3 \nonu \\
& - & \left.
\frac{16(-1+k)}{3(5+k)^2} i \pa \hat{A}_3 -\frac{4}{3(5+k)} i 
\pa \hat{B}_3 -\frac{4}{(5+k)} i T^{(1)} \hat{A}_3
+\frac{4}{(5+k)} i T^{(1)} \hat{B}_3
  \right](w) 
\nonu \\
& + & \frac{1}{(z-w)} \, \left[ -\frac{1}{4} {\bf \pa P^{(2)}}
-\frac{4}{(5+k)} i T^{(1)} \pa \hat{A}_3 +\frac{2}{(5+k)} i 
T^{(1)} \pa \hat{B}_3
\right. \nonu \\
&- & \frac{2}{(5+k)} i \hat{A}_{+} V_{+}^{(2)}
-\frac{8(-1+k)}{3(5+k)^2} \hat{A}_{+} \pa \hat{A}_{-}
-\frac{2}{(5+k)} i \hat{A}_{-} U_{-}^{(2)}
\nonu \\
& -& \frac{4(-13+4k)}{3(5+k)^2} \hat{A}_3 \pa \hat{A}_3
+\frac{16(-1+k)}{3(5+k)^2} \hat{A}_3 \pa \hat{B}_3
-\frac{2}{3(5+k)} \hat{B}_{+} \pa \hat{B}_{-}
\nonu \\
& - & \frac{2}{3(5+k)} \hat{B}_{-} \pa \hat{B}_{+}
-\frac{8(2+k)}{3(5+k)^2} \hat{B}_3 \pa \hat{A}_3
+\frac{4(-5+2k)}{3(5+k)^2} \hat{B}_3 \pa \hat{B}_3
\nonu \\
& -& \frac{2(15+32k+16k^2)}{3(5+k)(3+7k)} \pa \hat{T}
-\frac{8(-1+k)}{3(5+k)^2} \hat{A}_{-} \pa \hat{A}_{+}
+\frac{2}{(5+k)} i \hat{B}_3 \pa T^{(1)} 
\nonu \\
&+ & \left. \frac{(5+4k)}{3(5+k)} \pa T^{(2)}
+\frac{(27+39k+8k^2)}{(5+k)(3+7k)} \pa W^{(2)}
\right](w) +\cdots,
\nonu \\
T^{(2)}(z) \, W_{+}^{(\frac{5}{2})}(w) & = & -\frac{1}{(z-w)^3}
\, \left[ \frac{4(11+k)}{3(5+k)^2} \hat{G}_{21} +
\frac{4(11+k)}{3(5+k)^2} T_{+}^{(\frac{3}{2})} \right](w)
\nonu \\
& + & \frac{1}{(z-w)^2} \, \left[ \frac{(21+5k)}{2(5+k)} 
{\bf P_{+}^{(\frac{5}{2})}} 
+\frac{2(33+47k+10k^2)}{(5+k)(3+7k)} W_{+}^{(\frac{5}{2})}
+\frac{4(8+k)}{3(5+k)^2} i \hat{A}_{-} \hat{G}_{11}
\right. \nonu \\
& + & \frac{4(23+7k)}{3(5+k)^2} i \hat{A}_{-} U^{(\frac{3}{2})}
-\frac{12}{(5+k)^2} i \hat{A}_3 \hat{G}_{21} -
\frac{8(4+k)}{(5+k)^2} i \hat{A}_3 T_{+}^{(\frac{3}{2})}
\nonu \\
& + & \frac{4(8+k)}{3(5+k)^2} i \hat{B}_{-} V^{(\frac{3}{2})}
 -  \frac{4k}{(5+k)^2} i \hat{B}_3 \hat{G}_{21}
+\frac{2}{(5+k)} T^{(1)} \hat{G}_{21}
\nonu \\
&-& \left.  \frac{4(8+k)}{3(5+k)^2} \pa \hat{G}_{21} -
\frac{4(13+3k)}{3(5+k)^2} \pa T_{+}^{(\frac{3}{2})} \right](w)
\nonu \\
& + & \frac{1}{(z-w)} \, \left[ 
{\bf  \pa P_{+}^{\frac{5}{2}}} 
- \frac{6}{(5+k)^2} \hat{A}_{-} \hat{A}_{+} \hat{G}_{21}
 - \frac{4(4+k)}{(5+k)^2} \hat{A}_{-} \hat{A}_{+} T_{+}^{(\frac{3}{2})}
\right. \nonu \\
&   - &  \frac{2}{(5+k)} i \hat{A}_{-} U^{(\frac{5}{2})}
 - \frac{2(25+8k)}{3(5+k)^2} \hat{A}_{-} \hat{A}_3 U^{(\frac{3}{2})}
 - \frac{4}{(5+k)^2} \hat{A}_{-} \hat{B}_{-} T_{-}^{(\frac{3}{2})}
\nonu \\
&  + &  \frac{8}{3(5+k)^2} i \hat{A}_{-} \pa \hat{G}_{11}
 + \frac{4(7+2k)}{3(5+k)^2} i \hat{A}_{-} \pa U^{(\frac{3}{2})}
+\frac{2k}{(5+k)^2} \hat{B}_{+} \hat{B}_{-} \hat{G}_{21} 
\nonu \\
&- &  \frac{2k}{(5+k)^2} \hat{B}_{-} \hat{B}_{+} \hat{G}_{21}
+\frac{4(8+k)}{3(5+k)^2}  i \hat{B}_{-} \pa V^{(\frac{3}{2})}
+ \frac{6}{(5+k)^2} \hat{A}_{+} \hat{A}_{-} \hat{G}_{21}
\nonu \\
& + & \frac{4(4+k)}{(5+k)^2}   \hat{A}_{+} \hat{A}_{-} T_{+}^{(\frac{3}{2})}
+\frac{2(25+8k)}{3(5+k)^2} \hat{A}_3 \hat{A}_{-} U^{(\frac{3}{2})}
+\frac{2(3+4k)}{(3+7k)} \pa W_{+}^{(\frac{5}{2})}
\nonu \\
& - & \frac{2(8+k)}{3(5+k)^2} i \pa \hat{B}_{-} V^{(\frac{3}{2})}
+\frac{2}{(5+k)} \pa T^{(1)} \hat{G}_{21}
+\frac{(2+k)}{3(5+k)^2} \pa^2 \hat{G}_{21}
\nonu \\
& - & \frac{(13+3k)}{3(5+k)^2} \pa^2 T_{+}^{(\frac{3}{2})}
+\frac{4}{3(5+k)} i \hat{G}_{11} \pa \hat{A}_{-}
\nonu \\
&- & \left. \frac{2}{(5+k)}U^{(\frac{3}{2})} V_{+}^{(2)} 
+ \frac{2}{(5+k)}  V^{(\frac{3}{2})} U_{+}^{(2)}
\right](w) +\cdots,
\nonu \\
T^{(2)}(z) \, W_{-}^{(\frac{5}{2})}(w) & = & \frac{1}{(z-w)^3}
\, \left[ - \frac{4(11+k)}{3(5+k)^2} \hat{G}_{12} +
\frac{4(11+k)}{3(5+k)^2} T_{-}^{(\frac{3}{2})} \right](w)
\nonu \\
& + & \frac{1}{(z-w)^2} \, \left[ -\frac{(21+5k)}{2(5+k)} 
{\bf P_{-}^{(\frac{5}{2})}} 
+\frac{2(33+47k+10k^2)}{(5+k)(3+7k)} W_{-}^{(\frac{5}{2})}
\right. \nonu \\
& + & \frac{4(23+7k)}{3(5+k)^2} i \hat{A}_{+} V^{(\frac{3}{2})}
+\frac{12}{(5+k)^2} i \hat{A}_3 \hat{G}_{12} -
\frac{8(4+k)}{(5+k)^2} i \hat{A}_3 T_{-}^{(\frac{3}{2})}
\nonu \\
& + & \frac{4(8+k)}{3(5+k)^2} i \hat{B}_{+} U^{(\frac{3}{2})}
+  \frac{4k}{(5+k)^2} i \hat{B}_3 \hat{G}_{12}
-\frac{2}{(5+k)} T^{(1)} \hat{G}_{12}
\nonu \\
&-& \left.  \frac{4(8+k)}{3(5+k)^2} \pa \hat{G}_{12} +
\frac{4(13+3k)}{3(5+k)^2} \pa T_{-}^{(\frac{3}{2})} 
-\frac{4(8+k)}{3(5+k)^2} i \hat{A}_{+} \hat{G}_{22}
\right](w)
\nonu \\
& + & \frac{1}{(z-w)} \, \left[ -\pa {\bf P_{-}^{(\frac{5}{2})}} 
+\frac{6}{(5+k)^2} \hat{A}_{-} \hat{A}_{+} \hat{G}_{12} 
+\frac{2(27+4k)}{3(5+k)^2} \hat{A}_{-} \hat{A}_{+} T_{-}^{(\frac{3}{2})}
\right. \nonu \\
&- & \frac{4}{(5+k)^2} \hat{B}_{+} \hat{A}_{+} \hat{G}_{21}
-\frac{2k}{(5+k)^2} \hat{B}_{+} \hat{B}_{-} \hat{G}_{12}
+\frac{4(8+k)}{3(5+k)^2} i \hat{B}_{+} \pa U^{(\frac{3}{2})}
\nonu \\
&+ & \frac{2k}{(5+k)^2} \hat{B}_{-} \hat{B}_{+} \hat{G}_{12}
+\frac{2}{(5+k)} i \hat{A}_{+} V^{(\frac{5}{2})}
-\frac{6}{(5+k)^2} \hat{A}_{+} \hat{A}_{-} \hat{G}_{12}
\nonu \\
& - & \frac{2(21+4k)}{3(5+k)^2} \hat{A}_{+} \hat{A}_{-} T_{-}^{(\frac{3}{2})}
- \frac{4}{3(5+k)} \hat{A}_{+} \hat{A}_3 \hat{G}_{22} 
 +\frac{2(37+8k)}{3(5+k)^2} \hat{A}_{+} \hat{A}_3 V^{(\frac{3}{2})}
\nonu \\
&- & \frac{8}{(5+k)^2} \hat{A}_{+} \hat{B}_3 V^{(\frac{3}{2})}
 - \frac{8}{3(5+k)^2} i \hat{A}_{+} \pa \hat{G}_{22}
+ \frac{4(3+2k)}{3(5+k)^2}  i \hat{A}_{+} \pa V^{(\frac{3}{2})}
\nonu \\
&+ & \frac{4}{3(5+k)} \hat{A}_3 \hat{A}_{+} \hat{G}_{22} 
-\frac{2(25+8k)}{3(5+k)^2} \hat{A}_3 \hat{A}_{+} V^{(\frac{3}{2})}
-\frac{2}{(5+k)} \hat{G}_{12} \pa T^{(1)}
\nonu \\
&- & \frac{4(51+10k)}{3(5+k)^2} i T_{-}^{(\frac{3}{2})} \pa \hat{A}_3
+\frac{2}{(5+k)} U^{(\frac{3}{2})} V_{-}^{(2)}
-\frac{2(8+k)}{3(5+k)^2} i \hat{U}^{(\frac{3}{2})} \pa \hat{B}_{+} 
\nonu \\
&+ & \left. \frac{2(3+4k)}{(3+7k)} 
\pa W_{-}^{(\frac{5}{2})}  
-\frac{(31+5k)}{3(5+k)^2} \pa^2 \hat{G}_{12} 
-\frac{2}{(5+k)} V^{(\frac{3}{2})} U_{-}^{(2)}
\right](w) +\cdots,
\nonu \\
T^{(2)}(z) \, W^{(3)}(w) & = & \frac{1}{(z-w)^4}
\, \left[  \frac{24(24+83k+21k^2)}{(5+k)^2(19+23k)} i \hat{A}_3 -
\frac{8k(32+435k+160k^2+21k^3)}{
(5+k)^3(19+23k)} i \hat{B}_3  \right. \nonu \\
&+& \left. \frac{4(-82+117k+34k^2+3k^3)}{
(5+k)^2(19+23k)} T^{(1)} \right](w)
\nonu \\
& + & \frac{1}{(z-w)^3} \, \left[ 
\frac{1}{(5+k)} {\bf P^{(2)}}
+\frac{4(-33-11k+20k^2)}{3(5+k)^2(3+7k)} 
\hat{T} -\frac{2(-17+5k)}{3(5+k)^2} T^{(2)}
\right. \nonu \\
&-&  \frac{2(3+k)}{(5+k)^2} W^{(2)} +\frac{4(25+23k)}{3(5+k)^3}
\hat{A}_{+} \hat{A}_{-} +\frac{4(-29+23k)}{3(5+k)^3} \hat{A}_3 
\hat{A}_3 \nonu \\
& + &  \frac{8(-4+k)(1+3k)}{3(5+k)^3} \hat{A}_3 \hat{B}_3
\nonu \\
& - & \frac{4(11+22k+3k^2)}{3(5+k)^3}
\hat{B}_{+} \hat{B}_{-} -\frac{4(11+k+6k^2)}{3(5+k)^3} \hat{B}_3 
\hat{B}_3 
 +   \frac{4(25+23k)}{3(5+k)^3} i \pa \hat{A}_3 
\nonu \\
&-& \left. \frac{4(11+22k+3k^2)}{3(5+k)^3} i \pa \hat{B}_3
+\frac{4(3+k)}{(5+k)^2} i T^{(1)} \hat{A}_3 -\frac{4(1+k)}{(5+k)^2} i
T^{(1)} \hat{B}_3
\right](w) 
\nonu \\
& + & \frac{1}{(z-w)^2} \, \left[-\frac{(13+3k)}{(5+k)} {\bf P^{(3)}} +
\frac{1}{4(5+k)}  {\bf \pa P^{(2)}}  +\frac{4(21+31k+6k^2)}{(5+k)(3+7k)} 
W^{(3) }
\right. \nonu \\
&+ & \frac{8(5+k)(-3+2k)}{(3+7k)(19+23k)} T^{(1)} \hat{T} 
-\frac{16(-3+k)}{(19+23k)} T^{(1)} T^{(2)}
+\frac{2(15+2k)}{(5+k)^2} i T^{(1)} \pa \hat{A}_3
\nonu \\
& + & \frac{2(1+k)}{(5+k)^2} i T^{(1)} \pa \hat{B}_3 
+\frac{2(22+5k)}{(5+k)^2} i \hat{A}_{+} V_{+}^{(2)}
-\frac{4(17+k)}{3(5+k)^3} \hat{A}_{+} \pa \hat{A}_{-}
\nonu \\
& - & \frac{2(6+k)}{(5+k)^2} i \hat{A}_{-} U_{-}^{(2)}
+\frac{32(237+604k+179k^2+12k^3)}{(5+k)^2(3+7k)(19+23k)} 
i \hat{A}_3 \hat{T}
\nonu \\
& - & \frac{8(-7+k)}{3(5+k)^3} \hat{A}_{-} \pa \hat{A}_{+}
+\frac{28(-1+k)}{3(5+k)^3} \hat{A}_3 \pa \hat{A}_3 
-\frac{4k}{(5+k)^2} i \hat{B}_3 \pa T^{(1)}
\nonu \\
& - & \frac{8(87+184k+41k^2)}{(5+k)(95+134k+23k^2)} i
\hat{A}_3 T^{(2)}
+\frac{8}{(5+k)^2} i \hat{A}_3 W^{(2)}
\nonu \\
& - & \frac{2(12+k)}{(5+k)^2} i \hat{A}_3 \pa T^{(1)}
+\frac{(9+k)}{(5+k)^2} i \hat{B}_{+} U_{+}^{(2)}
-\frac{(88+107k+15k^2)}{3(5+k)^3} \hat{B}_{+} \pa \hat{B}_{-}
\nonu \\
&+ & \frac{5(1+k)}{(5+k)^2} i \hat{B}_{-} V_{-}^{(2)} 
+\frac{3k}{(5+k)^2} \hat{B}_{-} \pa \hat{B}_{+}
\nonu \\
&- &  \frac{8(19+123k+20k^2)}{(5+k)^2(19+23k)} i \hat{B}_3 T^{(2)}
+\frac{4(10+17k+6k^2)}{3(5+k)^3} \hat{B}_3 \pa \hat{A}_3
\nonu \\
& + & \frac{2(-66+203k+64k^2)}{3(5+k)^2(3+7k)} \pa \hat{T}
+\frac{(121+26k)}{3(5+k)^2} \pa T^{(2)}
+\frac{5(4+k)}{(5+k)^2} \pa W^{(2)}
\nonu \\
& + & \frac{4(20+13k)}{3(5+k)^3} i \pa^2 \hat{A}_3
-\frac{2(22+77k+13k^2)}{3(5+k)^3} i \pa^2 \hat{B}_3
+\frac{2(17+3k)}{3(5+k)^2} \pa^2 T^{(1)}
\nonu \\
&+ & \frac{4}{(5+k)^2} \hat{G}_{11} \hat{G}_{22}
-\frac{4}{(5+k)^2} \hat{G}_{11} V^{(\frac{3}{2})} 
+\frac{2(11+3k)}{(5+k)^2} \hat{G}_{12} T_{+}^{(\frac{3}{2})}
\nonu \\
&- & \frac{6}{(5+k)} \hat{G}_{21} T_{-}^{(\frac{3}{2})}
-\frac{4}{(5+k)^2} \hat{G}_{22} U^{(\frac{3}{2})}
+\frac{4(13+3k)}{(5+k)^2} U^{(\frac{3}{2})} V^{(\frac{3}{2})}
\nonu \\
& - & \frac{64}{(5+k)^3} i \hat{A}_{+} \hat{A}_{-} \hat{B}_3
+ \frac{32}{(5+k)^3} i \hat{A}_{-} \hat{A}_{+} \hat{A}_3
+\frac{8}{(5+k)^3} i \hat{A}_3 \hat{A}_3 \hat{A}_3   
\nonu \\
&+ & \frac{8(-2+k)}{(5+k)^3} i \hat{A}_3 \hat{A}_3 \hat{B}_3 
-\frac{2(-61+19k+3k^2)}{3(5+k)^3} 
i \hat{A}_3  \hat{B}_{+} \hat{B}_{-}
\nonu \\
&+ & \frac{2(-25+19k+3k^2)}{3(5+k)^3} i \hat{B}_{-} \hat{A}_3 \hat{B}_{+}
+\frac{2(22+11k+3k^2)}{3(5+k)^3} i \hat{B}_{-} \hat{B}_{+} \hat{B}_3
\nonu \\
& - & \frac{2(22-k+3k^2)}{3(5+k)^3} i \hat{B}_3 \hat{B}_{+} \hat{B}_{-}
+  \frac{8k}{(5+k)^3} i \hat{B}_3 \hat{B}_3 \hat{B}_3
-\frac{4}{(5+k)^2} T^{(1)} \hat{A}_3 \hat{A}_3
\nonu \\
&+ &  \frac{8}{(5+k)^2} T^{(1)} \hat{A}_3 \hat{B}_3
-\frac{4}{(5+k)^2} T^{(1)} \hat{B}_3 \hat{B}_3
- \frac{4}{(5+k)^2} \hat{B}_{-} T^{(1)} \hat{B}_{+} 
\nonu \\
&- & \left.
\frac{16k(-30+175k+87k^2+14k^3)}{(5+k)^2(3+7k)(19+23k)}  
i \hat{B}_3 \hat{T}
-\frac{8(-1+2k)}{(5+k)^3} i \hat{A}_3 \hat{B}_3 \hat{B}_3
\right](w) 
\nonu \\
& + &  \frac{1}{(z-w)} \, \left[ 
-\frac{1}{(5+k)} i \hat{A}_{-} {\bf Q_{-}^{(3)}}
-\frac{4}{(5+k)} i \hat{B}_3 {\bf S^{(3)}}
-\frac{4}{(5+k)} i  \hat{B}_3 {\bf P^{(3)}}
\right. \nonu \\
& + & \frac{2}{(5+k)} \hat{B}_3 \hat{B}_3 {\bf P^{(2)}}
+\frac{3}{(5+k)} i \hat{B}_{-} {\bf R_{-}^{(3)}} 
+\frac{2}{(5+k)} i \hat{B}_{+} {\bf Q_{+}^{(3)}}
\nonu \\
& - & \frac{1}{(5+k)} \hat{B}_{+} \hat{B}_{-} {\bf P^{(2)}}
-  \frac{1}{(5+k)} i \pa \hat{B}_3  {\bf P^{(2)}}
-\frac{(3+k)}{(5+k)} \pa {\bf P^{(3)}}
-\frac{1}{4(5+k)} \pa^2 {\bf P^{(2)}}
\nonu \\
& - & \frac{2}{(5+k)} W^{(2)} W^{(2)}
+\frac{4(3+k)}{(5+k)^2} U^{(\frac{3}{2})} \pa  V^{(\frac{3}{2})}
-\frac{2}{(5+k)}  U_{-}^{(2)}  V_{+}^{(2)}
 \nonu \\
&-& \frac{2}{(5+k)}  U_{+}^{(2)}  V_{-}^{(2)}
-\frac{4}{(5+k)} T^{(2)} W^{(2)}
-\frac{2}{(5+k)} T^{(2)} T^{(2)}
\nonu \\
& + & \frac{8}{(5+k)^2} \hat{A}_3 \hat{A}_3  W^{(2)}
\nonu \\
&+ &  \frac{8}{(5+k)^2} \hat{A}_3 \hat{A}_3  T^{(2)}
- \frac{8}{(5+k)^3} \hat{A}_3 \hat{A}_3 \hat{A}_3 \hat{A}_3 
+ \frac{16}{(5+k)^2} i \hat{A}_3 \hat{A}_3 \hat{B}_3 T^{(1)}
\nonu \\
&- & \frac{24}{(5+k)^3}  \hat{A}_3 \hat{A}_3 \hat{B}_{+} \hat{B}_{-} 
+\frac{8(-3+k)}{(5+k)^3} i \hat{A}_3 \hat{A}_3 \pa \hat{B}_3
-\frac{4}{(5+k)^2}  \hat{A}_3 \hat{A}_3 \pa T^{(1)}
\nonu \\
&- & \frac{32(-4+k)}{3(5+k)^3}  \hat{A}_3 \hat{B}_3 \hat{B}_3 \hat{B}_3  
+ \frac{16}{(5+k)^2} i \hat{A}_3 \hat{B}_3 \hat{B}_3 T^{(1)}
+ \frac{8}{(5+k)^2} \hat{A}_3 \hat{B}_3 \pa T^{(1)}
\nonu \\
&+ & \frac{8}{(5+k)^2} \hat{A}_3 \hat{B}_{+}  U_{+}^{(2)} 
+\frac{16(-18+k)}{(5+k)^3} \hat{A}_3 \hat{B}_{+} \hat{B}_{-} \hat{B}_3
-\frac{24}{(5+k)^2} i  \hat{A}_3 \hat{B}_{+} \hat{B}_{-} T^{(1)}
\nonu \\
&+ & \frac{4(-59+4k)}{(5+k)^3} i \hat{A}_3 \hat{B}_{+} \pa \hat{B}_{-} 
- \frac{16(21+2k)}{(5+k)^3} i \hat{A}_3 \pa \hat{B}_{3}  \hat{B}_{3} 
+\frac{24}{(5+k)^2} \hat{A}_3 \pa \hat{B}_3 T^{(1)}
\nonu \\
&- & \frac{2(-107+10k)}{(5+k)^3} i \hat{A}_3 \pa \hat{B}_{+} \hat{B}_{-}
-\frac{24(1+3k)}{(5+k)(19+23k)} i \hat{A}_3 \pa T^{(2)}
\nonu \\
& + & \frac{2(233-51k+184k^2)}{(5+k)^3(19+23k)} \hat{A}_3 \pa^2 \hat{B}_3
-\frac{6}{(5+k)^2} i \hat{A}_3 \pa^2 T^{(1)} 
+\frac{4}{(5+k)^2} i \hat{A}_{-}  T_{-}^{(\frac{3}{2})}  U^{(\frac{3}{2})}
\nonu \\
& -& \frac{4}{(5+k)^2} i \hat{A}_{-} \hat{G}_{11}  T_{-}^{(\frac{3}{2})} 
-\frac{7}{(5+k)^2} i \hat{A}_{-} \pa U_{-}^{(2)}
+\frac{8}{(5+k)^2} \hat{A}_{+} \hat{A}_3 V_{+}^{(2)}
\nonu \\
&+ & \frac{4}{(5+k)^2} \hat{A}_{+} \hat{A}_{-} W^{(2)} 
+\frac{8}{(5+k)^2} \hat{A}_{+} \hat{A}_{-} T^{(2)}
-\frac{8}{(5+k)^3} \hat{A}_{+} \hat{A}_{-} \hat{A}_3 \hat{A}_3
\nonu \\
&- & \frac{16}{(5+k)^3} \hat{A}_{+} \hat{A}_{-} \hat{A}_3 \hat{B}_3 
+ \frac{120}{(5+k)^3} \hat{A}_{+} \hat{A}_{-} \hat{B}_3 \hat{B}_3 
+  \frac{16}{(5+k)^2} i \hat{A}_{+} \hat{A}_{-} \hat{B}_3 T^{(1)} 
\nonu \\
&- & \frac{88}{(5+k)^3} \hat{A}_{+} \hat{A}_{-} \hat{B}_{+} \hat{B}_{-} 
-\frac{2}{(5+k)^3} i \hat{A}_{+} \hat{A}_{-} \pa \hat{A}_{3}
+\frac{2(-56+k)}{(5+k)^3} i  \hat{A}_{+} \hat{A}_{-} \pa \hat{B}_{3}
\nonu \\
&- & \frac{3}{(5+k)^2} \hat{A}_{+} \hat{A}_{-} \pa T^{(1)}
-\frac{24}{(5+k)^2} \hat{A}_{+} \hat{B}_3 V_{+}^{(2)}
-\frac{4}{(5+k)^2} i \hat{A}_{+} \hat{G}_{21}  V^{(\frac{3}{2})} 
\nonu \\
&+& \frac{4}{(5+k)^2} i \hat{A}_{+} \hat{G}_{21} \hat{G}_{22}
+\frac{8}{(5+k)^3} i \hat{A}_{+} \pa \hat{A}_{-} \hat{A}_3
-\frac{64}{(5+k)^3} i \hat{A}_{+} \pa \hat{A}_{-} \hat{B}_3
\nonu \\
& + & \frac{2k}{(5+k)^2} i \hat{A}_{+} \pa  V_{+}^{(2)}
-\frac{2(321+2600k+1493k^2+1598k^3)}{3(5+k)^3(3+7k)(19+23k)} 
\hat{A}_{+} \pa^2 \hat{A}_{-}
\nonu \\
&+ & \frac{8}{(5+k)} i \hat{B}_3 W^{(3)}
+\frac{16}{(5+k)^2} i \hat{B}_3 U^{(\frac{3}{2})} V^{(\frac{3}{2})}
+\frac{16}{(5+k)^2} i \hat{B}_3  T_{+}^{(\frac{3}{2})}  T_{-}^{(\frac{3}{2})} 
\nonu \\
&-& \frac{8(37+2k)}{3(5+k)^2} \hat{B}_3 \hat{B}_3 W^{(2)}
-\frac{24}{(5+k)^2} \hat{B}_3 \hat{B}_3 T^{(2)}
+\frac{16(-19+k)}{9(5+k)^2} i \hat{B}_3 \hat{B}_3 \hat{B}_3 T^{(1)}
\nonu \\
& + & \frac{4}{(5+k)} \hat{B}_3 \hat{B}_3 T^{(1)} T^{(1)} 
-\frac{12}{(5+k)^2} \hat{B}_3 \hat{B}_3 \pa T^{(1)}
-\frac{8}{(5+k)} i \hat{B}_3 T^{(1)} W^{(2)}
\nonu \\
&- & \frac{8}{(5+k)} i \hat{B}_3  T^{(1)} T^{(2)} 
+\frac{4}{(5+k)^2} i \hat{B}_3 \pa   W^{(2)}
-\frac{4(-57+31k+20k^2)}{(5+k)^2(19+23k)} i \hat{B}_3 \pa T^{(2)}
\nonu \\
& - & \frac{2(2811+7033k+5297k^2+1587k^3)}{3(5+k)^2(3+7k)(19+23k)} i
\hat{B}_3 \pa^2 T^{(1)}
-\frac{12}{(5+k)^2} i \hat{B}_{-} T_{-}^{(\frac{3}{2})}  V^{(\frac{3}{2})} 
\nonu \\
& - & \frac{8(-25+k)}{3(5+k)^2} \hat{B}_{-} \hat{B}_3 V_{-}^{(2)}
+\frac{4}{(5+k)^2} i \hat{B}_{-} \hat{G}_{12}  V^{(\frac{3}{2})} 
-\frac{12}{(5+k)^2} i \hat{B}_{-} \hat{G}_{22}  T_{-}^{(\frac{3}{2})} 
\nonu \\
& + & \frac{4}{(5+k)} i \hat{B}_{-} T^{(1)}  V_{-}^{(2)}
+
\frac{(-18+k)}{(5+k)^2} i \hat{B}_{-} \pa  V_{-}^{(2)}
+\frac{8}{(5+k)^2} i \hat{B}_{+}  T_{+}^{(\frac{3}{2})}  U^{(\frac{3}{2})}  
\nonu \\
&+ & \frac{8(-25+k)}{3(5+k)^2} \hat{B}_{+} \hat{B}_3  U_{+}^{(2)} 
+ \frac{8(26+k)}{3(5+k)^2} \hat{B}_{+} \hat{B}_{-}  W^{(2)} 
+ \frac{28}{(5+k)^2} \hat{B}_{+} \hat{B}_{-}  T^{(2)} 
\nonu \\
&- & \frac{8(-49+8k+k^2)}{(5+k)^3} \hat{B}_{+} \hat{B}_{-} \hat{B}_3 \hat{B}_3 
-\frac{8(-34+k)}{3(5+k)^2} i \hat{B}_{+} \hat{B}_{-} \hat{B}_3 T^{(1)}
\nonu \\
&- & \frac{2}{(5+k)}  \hat{B}_{+} \hat{B}_{-} T^{(1)} T^{(1)}
+ \frac{2(-266+35k+4k^2)}{(5+k)^3} i  \hat{B}_{+} \hat{B}_{-} \pa \hat{B}_3
\nonu \\
& - &  \frac{3}{(5+k)^2}  \hat{B}_{+} \hat{B}_{-} \pa T^{(1)} 
+\frac{(-133+16k+2k^2)}{(5+k)^3} \hat{B}_{+} \hat{B}_{+} \hat{B}_{-} \hat{B}_{-}
\nonu \\
&- & \frac{8}{(5+k)^2} i \hat{B}_{+} \hat{G}_{11}  T_{+}^{(\frac{3}{2})}
-\frac{4}{(5+k)^2} i \hat{B}_{+} \hat{G}_{21}  U^{(\frac{3}{2})}
-\frac{4}{(5+k)} i \hat{B}_{+} T^{(1)}  U_{+}^{(2)} 
\nonu \\
& - & \frac{4(-150+21k+2k^2)}{(5+k)^3} i \hat{B}_{+} \pa \hat{B}_{-} \hat{B}_3
+ \frac{2(-83+2k)}{3(5+k)^2}  \hat{B}_{+} \pa \hat{B}_{-} T^{(1)}
\nonu \\
&+& \frac{(-5+k)}{(5+k)^2} i \hat{B}_{+} \pa  U_{+}^{(2)} 
\nonu \\
& - & \frac{(-50730-166427k-142985k^2-14521k^3+287k^4)}{3(5+k)^3(3+7k)(19+23k)} 
\hat{B}_{+} \pa^2 \hat{B}_{-}
\nonu \\
&- & \frac{4}{(5+k)^2} \hat{G}_{11} \pa   V^{(\frac{3}{2})}
+\frac{4(3+k)}{(5+k)^2} \hat{G}_{12} \pa  T_{+}^{(\frac{3}{2})}
-\frac{4}{(5+k)^2} \hat{G}_{22} \pa  U^{(\frac{3}{2})}
\nonu \\
&- & \frac{8(-3+k)}{(19+23k)} T^{(1)} \pa T^{(2)}
+\frac{8(3+4k)}{(5+k)(3+7k)} \hat{T} W^{(2)}
+\frac{8(3+4k)}{(5+k)(3+7k)} \hat{T} T^{(2)}
\nonu \\
&- & \frac{8(3+4k)^2}{(5+k)(3+7k)^2} \hat{T} \hat{T} 
-\frac{16(3+4k)}{(5+k)^2(3+7k)} \hat{T} \hat{A}_3 \hat{A}_3
\nonu \\
& - & \frac{384}{(5+k)^2(19+23k)} \hat{T} \hat{A}_3 \hat{B}_3
- \frac{16(-12-55k+k^2)}{(5+k)^2(3+7k)(19+23k)} 
\hat{T} \hat{A}_{+} \hat{A}_{-}
\nonu \\
& + &  \frac{16(399+1291k+1056k^2+56k^3)}{(5+k)^2(3+7k)(19+23k)} 
\hat{T} \hat{B}_{3} \hat{B}_{3}
\nonu \\
&+ & \frac{16(63+159k+92k^2)}{(5+k)(3+7k)(19+23k)} i 
\hat{T} \hat{B}_3 T^{(1)}
\nonu \\
& - &  \frac{8(684+2115k+1689k^2+70k^3)}{(5+k)^2(3+7k)(19+23k)} 
\hat{T} \hat{B}_{+} \hat{B}_{-}
\nonu \\
&- &   \frac{16(-21-31k+6k^2)}{(5+k)(3+7k)(19+23k)} i 
\hat{T} \pa \hat{A}_3 
- \frac{48(114+356k+277k^2+7k^3)}{(5+k)^2(3+7k)(19+23k)} i 
\hat{T} \pa \hat{B}_3 
\nonu \\
&-& \frac{8}{(19+23k)} \hat{T} \pa T^{(1)} 
+\frac{4(3+k)}{(5+k)^2} \pa  U^{(\frac{3}{2})}  V^{(\frac{3}{2})}
-\frac{8(3+k)}{(5+k)^2} i \pa \hat{A}_3 T^{(2)} 
\nonu \\
&+ & \frac{8}{(5+k)^3} i \pa \hat{A}_3 \hat{A}_3 \hat{A}_3 
-\frac{48}{(5+k)^3} i \pa \hat{A}_3 \hat{A}_3 \hat{B}_3 
+\frac{8(23+4k)}{(5+k)^3} i \pa \hat{A}_3 \hat{B}_3 \hat{B}_3
\nonu \\
& -&  \frac{16}{(5+k)^2}  \pa \hat{A}_3 \hat{B}_3 T^{(1)}
+ \frac{2(-13+2k)}{(5+k)^3} i  \pa \hat{A}_3 \hat{B}_{+} \hat{B}_{-}
-\frac{52(-1+k)}{3(5+k)^3} \pa \hat{A}_3 \pa \hat{A}_3
\nonu \\
&- &  \frac{4(32+7k+3k^2)}{3(5+k)^3} \pa \hat{A}_3 \pa \hat{B}_3
- \frac{2(6+k)}{(5+k)^2} i \pa \hat{A}_3 \pa T^{(1)}
-\frac{2(9+k)}{(5+k)^2} i \pa \hat{A}_{-}  U_{-}^{(2)}
\nonu \\
& + & \frac{6(3+k)}{(5+k)^2} i \pa \hat{A}_{+} V_{+}^{(2)}
+\frac{34}{(5+k)^3} i \pa \hat{A}_{+} \hat{A}_{-} \hat{A}_3
-\frac{2(-8+k)}{(5+k)^3} i \pa \hat{A}_{+} \hat{A}_{-} \hat{B}_3
\nonu \\
&+& \frac{3}{(5+k)^2} \pa \hat{A}_{+} \hat{A}_{-} T^{(1)}
-\frac{2(1+8k)}{3(5+k)^3} \pa \hat{A}_{+} \pa \hat{A}_{-}
+\frac{8(29+k)}{3(5+k)^2} i \pa \hat{B}_3 W^{(2)}
\nonu \\
& + & \frac{24}{(5+k)^2} i \pa \hat{B}_3
T^{(2)}
-\frac{8(-49+7k+k^2)}{(5+k)^3} i \pa \hat{B}_3 \hat{B}_3 \hat{B}_3 
\nonu \\
& - & \frac{2}{(5+k)} i \pa \hat{B}_3 T^{(1)} T^{(1)}
-\frac{2(-281+44k)}{3(5+k)^3} \pa \hat{B}_3 \pa \hat{B}_3
+\frac{2(-2+k)}{(5+k)^2} i \pa \hat{B}_3 \pa T^{(1)}
\nonu \\
&+ & \frac{(163+5k)}{3(5+k)^2} i \pa \hat{B}_{-}  V_{-}^{(2)}
-\frac{(-103+7k)}{3(5+k)^2} i \pa \hat{B}_{+} U_{+}^{(2)}
+ \frac{8(-31+k)}{3(5+k)^2} \pa \hat{B}_3
 \hat{B}_3 T^{(1)}
\nonu \\
& + & 
\frac{2(-300+43k+4k^2)}{(5+k)^3} i \pa \hat{B}_{+} \hat{B}_{-} 
\hat{B}_3 -\frac{(-163+4k)}{3(5+k)^2} \pa \hat{B}_{+} \hat{B}_{-} T^{(1)}
\nonu \\
&- & \frac{2(-644+101k+6k^2)}{3(5+k)^3} \pa \hat{B}_{+} \pa \hat{B}_{-}
+\frac{4}{(5+k)^2} \pa \hat{G}_{11}  V^{(\frac{3}{2})}
\nonu \\
&+& \frac{4}{(5+k)^2} \pa \hat{G}_{22}  U^{(\frac{3}{2})}
+\frac{48(3+k)(16+41k+5k^2)}{(5+k)^2(3+7k)(19+23k)} i \pa \hat{T} \hat{A}_3
\nonu \\
& - &
\frac{8k(-267-125k+80k^2+14k^3)}{(5+k)^2(3+7k)(19+23k)} i \pa \hat{T} 
\hat{B}_3 -\frac{4(3+k)}{(5+k)^2} \pa \hat{G}_{21}  T_{-}^{(\frac{3}{2})}
\nonu \\
&+& \frac{8(3+k)(-12+9k+k^2)}{(5+k)(3+7k)(19+23k)}  \pa \hat{T} 
T^{(1)}
+\frac{2(411+665k+82k^2)}{3(5+k)^3(3+7k)} \pa^2 \hat{A}_3 \hat{A}_3
\nonu \\
&+ & \frac{2(832-581k-47k^2+138k^3)}{3(5+k)^3(19+23k)} 
\pa^2 \hat{A}_3 \hat{B}_3
+\frac{(3+2k)}{(5+k)^2} i \pa^2 \hat{A}_3 T^{(1)}
\nonu \\
&-& \frac{(-1581-3362k-571k^2+1264k^3)}{3(5+k)^3(3+7k)(19+23k)}
\pa^2 \hat{A}_{+} \hat{A}_{-}
\nonu \\
& - & \frac{2(40242+137827k+138114k^2+35211k^3+2954k^4)}{
3(5+k)^3(3+7k)(19+23k)} \pa^2 \hat{B}_{3} \hat{B}_3
\nonu \\
& - & \frac{(13731+30290k+20099k^2+7268k^3)}{
9(5+k)^2(3+7k)(19+23k)} i \pa^2 \hat{B}_{3} T^{(1)}
\nonu \\
& +& \frac{(5244+12926k+42689k^2+39488k^3+2289k^4)}{
3(5+k)^3(3+7k)(19+23k)} \pa^2 \hat{B}_{+} \hat{B}_{-}
\nonu \\
&+ & \frac{8(1+k)(3+k)}{(5+k)(3+7k)} \pa W^{(3)}
+\frac{(8+3k)}{(5+k)^2} \pa^2 W^{(2)}
+ \frac{5(1+2k)}{3(5+k)^2} \pa^2 T^{(2)}
\nonu \\
&+ &  \frac{2(21+52k+2k^2)}{3(5+k)^2(3+7k)} \pa^2 \hat{T}
-\frac{2(5085+9783k+1043k^2+901k^3+12k^4)}{3(5+k)^3(3+7k)(19+23k)} i
\pa^3 \hat{A}_3
\nonu \\
&+&  
\frac{(23313+70147k+64674k^2+12815k^3+775k^4+56k^5)}{
3(5+k)^3(3+7k)(19+23k)} i
\pa^3 \hat{B}_3
\nonu \\
&-& \left. \frac{(-321-446k-169k^2+4k^3)}{
3(5+k)(3+7k)(19+23k)} \pa^3 T^{(1)}
\right](w)  +\cdots.
\label{opeopeope}
\eea
The first order pole in the last OPE of (\ref{opeopeope})
contains composite field with spin-$4$ with vanishing $U(1)$ charge. 
Of course, the analysis done in Appendix $A$ can be done here 
similarly.
For given any OPE, one can determine the nonzero descendant fields with correct 
coefficients. Then any singular terms consist of these descendant fields and 
a couple of (quasi) primary fields.
The singular terms subtracted by the descendant fields should behave as 
a quasi primary or primary field. So one should calculate the OPE between 
the stress energy tensor $\hat{T}(z)$ and the above reduced 
singular terms. Then the third order singular terms of this OPE vanish.
If the fourth or higher singular terms of this OPE are not vanishing, then 
it will provide a quasi primary field. If they are vanishing, then 
one has a primary field as usual.  

\section{The nontrivial 
OPEs between  higher spin-$\frac{3}{2}$ current, $U^{(\frac{3}{2})}(z)$,  
and 
other $12$ higher spin currents}

Now we move to other ${\cal N}=2$ multiplet
and list the OPEs as follows
\bea
U^{(\frac{3}{2})}(z) \, U^{(\frac{5}{2})}(w) & = & \frac{1}{(z-w)^2} \, 
\left[ -\frac{4(-3+k)}{3(5+k)^2} \hat{A}_{+} \hat{B}_{-} \right](w)
\nonu \\
& + & \frac{1}{(z-w)} \, \left[
\frac{2}{(5+k)} i \hat{B}_{-} U_{-}^{(2)}  -\frac{2(7+k)}{(5+k)^2}
\hat{B}_{-} \pa \hat{A}_{+}  \right. \nonu \\
& + & \left. \frac{2}{(5+k)} i \hat{A}_{+}
U_{+}^{(2)}  -\frac{(12+k)}{3(5+k)} \hat{G}_{11} \hat{G}_{11}
\right](w) +\cdots,
\nonu \\
U^{(\frac{3}{2})}(z) \, V^{(\frac{3}{2})}(w) & = & 
-\frac{1}{(z-w)^3} \, \left[ \frac{6k}{(5+k)} \right]
\nonu \\
& + & \frac{1}{(z-w)^2} \, \left[ -\frac{6}{(5+k)} i \hat{A}_3  
+\frac{2k}{(5+k)} i \hat{B}_3 -T^{(1)} \right](w) 
\nonu \\
& + & \frac{1}{(z-w)} \, \left[ \frac{1}{2} \pa \{ U^{(\frac{3}{2})} \, 
V^{(\frac{3}{2})} \}_{-2}  - W^{(2)} \right](w) +\cdots,
\nonu \\
U^{(\frac{3}{2})}(z) \, V_{+}^{(2)}(w) & = & 
\frac{1}{(z-w)^2} \,  \left[ \frac{(5+2k)}{(5+k)} 
T_{+}^{(\frac{3}{2})} \right](w) 
\nonu \\
& + & \frac{1}{(z-w)} \, \left[ \frac{1}{3} \pa \{U^{(\frac{3}{2})} \, 
V_{+}^{(2)} \}_{-2}  +\frac{1}{2} 
{\bf P_{+}^{(\frac{5}{2})} } + W_{+}^{(\frac{5}{2})} \right](w) +\cdots,
\nonu \\
U^{(\frac{3}{2})}(z) \, V_{-}^{(2)}(w) & = & 
\frac{1}{(z-w)^2} \, \left[ -\frac{(8+k)}{(5+k)} \hat{G}_{12} +
\frac{(8+k)}{(5+k)} T_{-}^{(\frac{3}{2})}  \right](w) 
\nonu \\
& + & \frac{1}{(z-w)} \, \left[ \frac{1}{3} \pa \{ U^{(\frac{3}{2})} 
\, V_{-}^{(2)}\}_{-2} 
+\frac{1}{2} {\bf P_{-}^{(\frac{5}{2})} } \right](w) +\cdots, 
\nonu \\
U^{(\frac{3}{2})}(z) \, V^{(\frac{5}{2})}(w) & = & 
\frac{1}{(z-w)^3} \, \left[ -\frac{8(-4+k)}{(5+k)^2} i 
\hat{A}_3  +\frac{8k(8+k)}{3(5+k)^2} i \hat{B}_3  -\frac{4(-4+k)}{3(5+k)}
T^{(1)} \right](w)
\nonu \\
& + & \frac{1}{(z-w)^2} \, \left[ \frac{1}{2} {\bf P^{(2)}} +
\frac{16(30+61k+31k^2+4k^3)}{
3(5+k)^2(3+7k)} \hat{T} -\frac{8(2+k)}{3(5+k)} T^{(2)} 
\right. \nonu \\
&- &  \frac{8(2+k)}{3(5+k)} W^{(2)}  
+\frac{16(2+k)}{3(5+k)^2} \hat{A}_{+} \hat{A}_{-}
+\frac{16(2+k)}{3(5+k)^2} \hat{A}_3 \hat{A}_3
-\frac{32(2+k)}{3(5+k)^2} \hat{A}_3 \hat{B}_3
\nonu \\
&+ & \left. \frac{16(2+k)}{3(5+k)^2} \hat{B}_{+} \hat{B}_{-} 
+
\frac{16(2+k)}{3(5+k)^2} \hat{B}_3 \hat{B}_3
+\frac{16(2+k)}{3(5+k)^2} i \pa \hat{A}_3
 + \frac{16(2+k)}{3(5+k)^2} i  \pa \hat{B}_3 \right](w)
\nonu \\
& + & \frac{1}{(z-w)} \, \left[ \frac{1}{2} {\bf S^{(3)}} +\frac{1}{8} 
{\bf  \pa 
P^{(2)}}  -W^{(3)} -\frac{4(-3+k)}{(19+23k)} T^{(1)} \hat{T} 
-\frac{2}{(5+k)} i \hat{A}_{+} V_{+}^{(2)}
\right. \nonu \\
& + & \frac{(-7+4k)}{3(5+k)^2} \hat{A}_{+} \pa \hat{A}_{-}
+\frac{(23+4k)}{3(5+k)^2} \hat{A}_{-} \pa \hat{A}_{+}
+\frac{2(-11+2k)}{3(5+k)^2} \hat{A}_3 \pa \hat{B}_3 
\nonu \\
&- & \frac{4}{(5+k)} i \hat{A}_3 T^{(2)}
-\frac{4}{(5+k)} i \hat{A}_3 W^{(2)}
+\frac{8(-1+k)}{3(5+k)^2} \hat{A}_3 \pa \hat{A}_3
\nonu \\
&+ & \frac{2(8+7k)}{3(5+k)^2} \hat{B}_{+} \pa \hat{B}_{-}
-\frac{2k}{(5+k)^2} \hat{B}_{-} \pa \hat{B}_{+}
+\frac{8k(9+7k)}{(3+7k)(19+23k)} i \hat{B}_3 \hat{T}
\nonu \\
&+ & \frac{4}{(5+k)} i \hat{B}_3 T^{(2)}
-\frac{2(17+10k)}{3(5+k)^2} \hat{B}_3 \pa \hat{A}_3
-\frac{1}{(5+k)} i \hat{B}_3 \pa T^{(1)}
\nonu \\
& + & \frac{(24+17k+16k^2)}{3(5+k)(3+7k)} \pa \hat{T}
-\frac{(17+4k)}{6(5+k)} \pa T^{(2)}
-\frac{(11+4k)}{6(5+k)} \pa W^{(2)}
\nonu \\
&- & \frac{4}{(5+k)^2} i \pa^2 \hat{A}_3
+\frac{8(1+2k)}{3(5+k)^2} i \pa^2 \hat{B}_3
-\frac{2}{3(5+k)} \pa^2 T^{(1)}
-\frac{2}{(5+k)} U^{(\frac{3}{2})} V^{(\frac{3}{2})}
\nonu \\
&+ & \frac{8}{(5+k)^2} i \hat{A}_{+} \hat{A}_{-} \hat{A}_3 
+\frac{8}{(5+k)^2} i \hat{A}_3 \hat{A}_3 \hat{A}_3
-\frac{16}{(5+k)^2} i \hat{A}_3 \hat{A}_3 \hat{B}_3
\nonu \\
&+ & \frac{8}{(5+k)^2} i \hat{A}_3 \hat{B}_{+} \hat{B}_{-}
+\frac{8}{(5+k)^2} i \hat{A}_3 \hat{B}_3 \hat{B}_3
-\frac{4(2+k)}{3(5+k)^2} i \hat{B}_{-} \hat{B}_{+} \hat{B}_3
\nonu \\
&+ &
\frac{4(2+k)}{3(5+k)^2} i \hat{B}_3 \hat{B}_{+} \hat{B}_{-} 
-\frac{1}{2(5+k)} T^{(1)} \hat{B}_{+} \hat{B}_{-}
+\frac{1}{2(5+k)} \hat{B}_{-} T^{(1)} \hat{B}_{+}
\nonu \\
& - & \left. \frac{2}{(5+k)} T_{+}^{(\frac{3}{2})} 
T_{-}^{(\frac{3}{2})}
+\frac{8(375+965k+533k^2+71k^3)}
{(5+k)^2(3+7k)(19+23k)} i \hat{A}_3 \hat{T} 
\right](w) +\cdots, 
\nonu \\
U^{(\frac{3}{2})}(z) \, W^{(2)}(w) & = & 
\frac{1}{(z-w)^2} \, \left[ \frac{3(4+k)}{(5+k)} U^{(\frac{3}{2})} \right](w)
\nonu \\
& + & \frac{1}{(z-w)} \, \frac{1}{3} \pa \{ U^{(\frac{3}{2})} \, W^{(2)} \}_{-2}(w)
+\cdots,
\nonu \\
U^{(\frac{3}{2})}(z) \, W_{+}^{(\frac{5}{2})}(w) & = & 
-\frac{1}{(z-w)^3} 
\, \left[ \frac{8k(8+k)}{3(5+k)^2} i \hat{B}_{-} \right] (w) 
\nonu \\
& - & \frac{1}{(z-w)^2} \, \left[ \frac{8(2+k)}{3(5+k)} U_{+}^{(2)} \right](w) 
\nonu \\
&+ & \frac{1}{(z-w)} \, \left[ -\frac{1}{2} {\bf Q_{+}^{(3)} } +
\frac{1}{2(5+k)} i T^{(1)}  \pa \hat{B}_{-} 
-\frac{4}{(5+k)} i \hat{A}_3 U_{+}^{(2)}
\right. \nonu \\
&- & \frac{(11+2k)}{(5+k)^2} \hat{A}_3 \pa \hat{B}_{-}
-\frac{4 k(75+117k+14k^2)}{(5+k)(3+7k)(19+23k)} 
i \hat{B}_{-} \hat{T}
\nonu \\
& + & \frac{(15+2k)}{(5+k)^2}  \hat{B}_{-} \pa \hat{A}_3
-\frac{k}{(5+k)^2} \hat{B}_{-} \pa \hat{B}_3
-\frac{1}{2(5+k)} i \hat{B}_{-} \pa T^{(1)}
\nonu \\
&+ & \frac{k}{(5+k)^2} \hat{B}_3 \pa \hat{B}_{-}
-\frac{(5+4k)}{6(5+k)} \pa U_{+}^{(2)}
-\frac{4k}{3(5+k)^2} i \pa^2 \hat{B}_{-}
\nonu \\
&+ & \left. \frac{2}{(5+k)} \hat{G}_{11} T_{+}^{(\frac{3}{2})}
-\frac{2}{(5+k)} T_{+}^{(\frac{3}{2})} U^{(\frac{3}{2})}
-\frac{2}{(5+k)} i \hat{B}_{-} T^{(2)}
\right](w) +\cdots,
\nonu \\
U^{(\frac{3}{2})}(z) \, W_{-}^{(\frac{5}{2})}(w) & = & 
\frac{1}{(z-w)^3} 
\, \left[ \frac{8(5+2k)}{(5+k)^2} i \hat{A}_{+} \right] (w) 
\nonu \\
& - & \frac{1}{(z-w)^2} \, \left[ 
\frac{4(7+k)}{3(5+k)} U_{-}^{(2)} \right] (w) 
\nonu \\
&+ & \frac{1}{(z-w)} \, \left[ \frac{1}{2} {\bf Q_{-}^{(3)}} +
\frac{4(138+343k+113k^2)}{(5+k)(3+7k)(19+23k)} i \hat{A}_{+} \hat{T} 
-\frac{2}{(5+k)} i \hat{A}_{+} T^{(2)}
\right. \nonu \\
&- & \frac{7}{(5+k)^2} \hat{A}_{+} \pa \hat{A}_3
+\frac{(8+3k)}{(5+k)^2} 
i \hat{A}_{+} \pa \hat{B}_{3}
+\frac{1}{2(5+k)} i \hat{A}_{+} \pa T^{(1)}
\nonu \\
& + & \frac{3}{(5+k)^2}  \hat{A}_3 \pa \hat{A}_{+}
+\frac{4}{(5+k)} i  \hat{B}_3 U_{-}^{(2)}
-\frac{(8+3k)}{(5+k)^2}  \hat{B}_3 \pa \hat{A}_{+}
\nonu \\
&- & \frac{(11+2k)}{6(5+k)} \pa U_{-}^{(2)}
+\frac{4}{(5+k)^2} i \pa^2 \hat{A}_{+}
-\frac{2}{(5+k)} \hat{G}_{11} \hat{G}_{12}
\nonu \\
&+ &  \frac{2}{(5+k)} \hat{G}_{11} T_{-}^{(\frac{3}{2})}
+\frac{2}{(5+k)} \hat{G}_{12} U^{(\frac{3}{2})}
-\frac{2}{(5+k)} T_{-}^{(\frac{3}{2})} U^{(\frac{3}{2})}
\nonu \\
& + & \frac{4}{(5+k)^2} i \hat{A}_{+} \hat{A}_{+} \hat{A}_{-} 
+\frac{4}{(5+k)^2} i \hat{A}_{+} \hat{A}_3 \hat{A}_3
-\frac{8}{(5+k)^2} i \hat{A}_{+} \hat{A}_3 \hat{B}_3
\nonu \\
& + & \frac{4}{(5+k)^2} i \hat{A}_{+} \hat{B}_{+} \hat{B}_{-}
+ \frac{4}{(5+k)^2} i \hat{A}_{+} \hat{B}_3 \hat{B}_3 
-\frac{1}{2(5+k)} T^{(1)} \hat{A}_{+} \hat{A}_3
\nonu \\
& + & \left. \frac{1}{2(5+k)} \hat{A}_3 T^{(1)} \hat{A}_{+} 
\right](w) +\cdots,
\nonu \\
U^{(\frac{3}{2})}(z) \, W^{(3)}(w) & = & 
\frac{1}{(z-w)^3} 
\,  \left[  -\frac{4(-3+k)}{(5+k)^2} \hat{G}_{11}
+\frac{4(-3+k)(345+296k+55k^2)}{
3(5+k)^2(19+23k)} U^{(\frac{3}{2})} \right](w) 
\nonu \\
& + & \frac{1}{(z-w)^2} \, 
\left[ \frac{(19+5k)}{2(5+k)} {\bf Q^{(\frac{5}{2})} } 
+ \frac{(21+5k)}{(5+k)}  U^{(\frac{5}{2})} 
+\frac{12}{(5+k)^2} i \hat{A}_{+} \hat{G}_{21}
\right. \nonu \\
\nonu \\
&+ & \frac{4(9+2k)}{(5+k)^2} i \hat{A}_{+} T_{+}^{(\frac{3}{2})} 
+\frac{2(404+443k+107k^2)}{
(5+k)^2(19+23k)} i \hat{A}_3 U^{(\frac{3}{2})}
\nonu \\
&+ & \frac{(-5+3k)}{(5+k)^2} i \hat{B}_{-} \hat{G}_{12}
+\frac{1}{(5+k)} i \hat{B}_{-} T_{-}^{(\frac{3}{2})}
-\frac{2(133+537k+152k^2)}{(5+k)^2(19+23k)} i
\hat{B}_3 U^{(\frac{3}{2})}
\nonu \\
&- & \frac{(-199+k+12k^2)}{(95+134k+23k^2)} T^{(1)} U^{(\frac{3}{2})} 
\nonu \\
&+& \left. \frac{4(-3+k)}{3(5+k)^2} \pa \hat{G}_{11}
+\frac{(-2121-2020k-359k^2+12k^3)}{
3(5+k)^2(19+23k)} \pa U^{(\frac{3}{2})}
\right](w)
\nonu  \\
& + & \frac{1}{(z-w)} \, \left[ 
\frac{(11+k)}{2(5+k)} {\bf \pa Q^{(\frac{5}{2})}}
+\frac{1}{(5+k)} i {\bf P_{-}^{(\frac{5}{2})}} \hat{B}_{-}
+\frac{2}{(5+k)} i {\bf Q^{(\frac{5}{2})} }  \hat{A}_3
\right. \nonu \\
&- &
\frac{2}{(5+k)} i {\bf Q^{(\frac{5}{2})}} \hat{B}_3
+\frac{3(4+k)}{(5+k)^2} \hat{A}_{-} \hat{A}_{+} U^{(\frac{3}{2})}
+ \frac{9}{(5+k)^2} \hat{B}_{+} \hat{B}_{-} U^{(\frac{3}{2})}
\nonu \\
&- & \frac{9}{(5+k)^2} \hat{B}_{-} \hat{B}_{+} U^{(\frac{3}{2})} 
-\frac{(-5+k)}{(5+k)^2} \hat{B}_{-} \hat{B}_3 \hat{G}_{12}
+\frac{(29+k)}{3(5+k)^2} i \hat{B}_{-} \pa \hat{G}_{12} 
\nonu \\
&- & \frac{(29+k)}{3(5+k)^2} i \hat{B}_{-} \pa T_{-}^{(\frac{3}{2})} 
-\frac{(20+3k)}{(5+k)^2} \hat{A}_{+} \hat{A}_{-} U^{(\frac{3}{2})}
+\frac{6}{(5+k)^2} \hat{A}_{+} \hat{A}_3 \hat{G}_{21}
\nonu \\
&+ & \frac{2(11+2k)}{(5+k)^2} \hat{A}_{+} \hat{A}_3 T_{+}^{(\frac{3}{2})} 
+  \frac{8}{(5+k)^2} \hat{A}_{+} \hat{B}_3 T_{+}^{(\frac{3}{2})} 
- \frac{6}{(5+k)^2}  \hat{A}_3 \hat{A}_{+} \hat{G}_{21}
\nonu \\
&- & \frac{2(15+2k)}{(5+k)^2} \hat{A}_3 \hat{A}_{+} T_{+}^{(\frac{3}{2})}
+\frac{4}{(5+k)} i \hat{A}_3 U^{(\frac{5}{2})} 
-\frac{20}{(5+k)^2} \hat{A}_3 \hat{A}_3 U^{(
\frac{3}{2})}  
\nonu \\
&+ & \frac{24}{(5+k)^2} \hat{A}_3 \hat{B}_3 U^{(\frac{3}{2})} 
-\frac{4}{(5+k)} i \hat{B}_3 U^{(\frac{5}{2})}
+\frac{1}{(5+k)} \hat{B}_3 \hat{B}_{-} \hat{G}_{12}
\nonu \\
&- & \frac{4}{(5+k)^2} \hat{B}_3 \hat{B}_3 U^{(\frac{3}{2})} 
-\frac{38(1+21k+8k^2)}{3(5+k)^2(19+23k)} i
\hat{B}_3 \pa U^{(\frac{3}{2})}
\nonu \\
& + & 
\frac{(79+15k-4k^2)}{(5+k)(19+23k)} T^{(1)} \pa U^{(\frac{3}{2})}
+ \frac{(107+21k)}{3(5+k)^2} i \pa \hat{B}_{-} U_{-}^{(\frac{3}{2})}
\nonu \\
& + & \frac{(55+6k)}{3(5+k)^2} \pa^2 \hat{G}_{11}
+\frac{(9+k)}{(5+k)} \pa U^{(\frac{5}{2})}
-\frac{2(128+315k+31k^2)}{3(5+k)^2(19+23k)} 
i \hat{A}_3 \pa U^{(\frac{3}{2})}
\nonu \\
&- & \frac{2}{(5+k)} \hat{G}_{21} U_{-}^{(2)}  
+\frac{2}{(5+k)} T_{+}^{(\frac{3}{2})} U_{-}^{(2)}
+\frac{2}{(5+k)} \hat{G}_{12} U_{+}^{(2)}
\nonu \\
&- & \frac{2(46+9k)}{3(5+k)^2} i T_{-}^{(\frac{3}{2})} 
\pa \hat{B}_{-}
+\frac{8(-69-134k+2k^2+7k^3)}{(5+k)(3+7k)(19+23k)}
U^{(\frac{3}{2})} \hat{T}
\nonu \\
& + & \left. \frac{4}{(5+k)} U^{(\frac{3}{2})} T^{(2)} 
-\frac{1}{(5+k)} \pa T^{(1)} U^{(\frac{3}{2})}
\right](w) +\cdots.
\label{expexpope}
\eea
The first OPE of (\ref{expexpope}) has $\hat{G}_{11} \hat{G}_{11}(w)$
which can be written in terms of a derivative of $\hat{A}_{+} \hat{B}_{-}(w)$.
Note that there exists the $(k-3)$ factor in the third order pole of the 
last OPE of (\ref{expexpope}).

\section{
The nontrivial 
OPEs between  higher spin-$2$ current, $U_{+}^{(2)}(z)$,  
and 
other $11$ higher spin currents}

The corresponding OPEs are as follows
\bea
U_{+}^{(2)}(z) \, U_{-}^{(2)}(w) & = & \frac{1}{(z-w)^2} \,
\left[ \frac{2(3+k)}{(5+k)^2}  \hat{A}_{+} \hat{B}_{-} \right](w)
\nonu \\
& + & \frac{1}{(z-w)} \, \left[ \frac{2}{(5+k)} i \hat{B}_{-}
U_{-}^{(2)}  -\frac{2(7+k)}{(5+k)^2}  \hat{B}_{-} \pa \hat{A}_{+}
+ \frac{2}{(5+k)} i \hat{A}_{+} U_{+}^{(2)} \right. \nonu \\
& - & \left. \hat{G}_{11} \hat{G}_{11} 
\right](w)
+\cdots,
\nonu \\
U_{+}^{(2)}(z) \, U^{(\frac{5}{2})}(w) & = & \frac{1}{(z-w)^2} \,
\left[  \frac{2(6+k)}{3(5+k)^2} i  \hat{B}_{-} \hat{G}_{11} 
-\frac{16(3+k)}{3(5+k)^2} i \hat{B}_{-} U^{(\frac{3}{2})} \right](w)
\nonu \\
&+ & \frac{1}{(z-w)} \, \left[\
 \frac{8}{(5+k)^2} \hat{A}_3 \hat{B}_{-} \hat{G}_{11}
-\frac{4(9+4k)}{3(5+k)^2} \hat{B}_3 \hat{B}_{-} U^{(\frac{3}{2})}
\right. \nonu \\
&- & \frac{4}{(5+k)^2} i \hat{B}_{-} \pa U^{(\frac{3}{2})}
+\frac{2(12+k)}{3(5+k)^2} i \pa \hat{B}_{-} \hat{G}_{11}
+\frac{4(9+4k)}{3(5+k)^2} \hat{B}_{-} \hat{B}_3 U^{(\frac{3}{2})}
\nonu \\
&- & \left. 
\frac{1}{(5+k)} i \hat{B}_{-} {\bf Q^{(\frac{5}{2})} } 
-\frac{2}{(5+k)} \hat{G}_{11} U_{+}^{(2)}
-\frac{8}{(5+k)^2} \hat{G}_{11} \hat{A}_3 \hat{B}_{-} 
\right](w) + \cdots,
\nonu  \\
U_{+}^{(2)}(z) \, V^{(\frac{3}{2})}(w) 
& = & -\frac{1}{(z-w)^2} \, \left[ \frac{(8+k)}{(5+k)} \hat{G}_{21}
+ \frac{(8+k)}{(5+k)} T_{+}^{(\frac{3}{2})} \right](w)
\nonu \\
& + & \frac{1}{(z-w)} \, \left[ 
\frac{1}{2} {\bf P_{+}^{(\frac{5}{2})}}  +\frac{2}{3} \{ U_{+}^{(2)} \, 
V^{(\frac{3}{2})}\}_{-2}
\right](w) +\cdots, 
\nonu \\
U_{+}^{(2)}(z) \, V_{+}^{(2)}(w) 
& = & -\frac{1}{(z-w)^2} \, \left[ \frac{2(7+k)}{(5+k)^2} 
\hat{A}_{-} \hat{B}_{-} \right] (w)
\nonu \\
& + & \frac{1}{(z-w)} \, 
\left[ -\frac{2}{(5+k)} i \hat{A}_{-} U_{+}^{(2)} 
-\frac{2(3+k)}{(5+k)^2}  \hat{A}_{-} \pa \hat{B}_{-}
-\frac{2}{(5+k)} i \hat{B}_{-} V_{+}^{(2)}  \right. \nonu \\
&- &  \left. \frac{2}{(5+k)} \hat{G}_{21} \hat{G}_{21}
\right](w) +\cdots,
\nonu \\
U_{+}^{(2)}(z) \, V_{-}^{(2)}(w) 
& = & \frac{1}{(z-w)^3} \, \left[ \frac{4k(8+k)}{(5+k)^2} i \hat{B}_3 
\right](w) 
\nonu \\
& + & 
\frac{1}{(z-w)^2} \, \left[ \frac{1}{2}  {\bf P^{(2)} }  
+\frac{4(-60-77k+7k^2+4k^3)}{3(5+k)^2(3+7k)} \hat{T} 
-\frac{2(-4+k)}{3(5+k)} T^{(2)}
\right. \nonu \\
& - & \frac{2(4+k)}{(5+k)} W^{(2)}
+ \frac{4(-4+k)}{3(5+k)^2} \hat{A}_{+} \hat{A}_{-}
+\frac{4(-4+k)}{3(5+k)^2} \hat{A}_3 \hat{A}_3 
-\frac{8(-4+k)}{3(5+k)^2} \hat{A}_3 \hat{B}_3
\nonu \\
&+ & \frac{4(-4+k)}{3(5+k)^2} \hat{B}_{+} \hat{B}_{-} 
+
\frac{4(-4+k)}{3(5+k)^2} \hat{B}_3 \hat{B}_3
\nonu \\
&+ & \left. \frac{4(-4+k)}{3(5+k)^2} i \pa \hat{A}_3
+\frac{2(-8+26k+3k^2)}{3(5+k)^2} i \pa \hat{B}_3
\right](w)
\nonu \\
& + & 
\frac{1}{(z-w)} \, 
\left[ \frac{1}{2} {\bf S^{(3)} } 
+\frac{1}{2} {\bf P^{(3)} }
+\frac{1}{4} {\bf \pa P^{(2)} }
-W^{(3)}
-\frac{4}{(19+23k)} T^{(1)} \hat{T}
\right. 
\nonu \\
&- & \frac{2}{(5+k)} i \hat{A}_{+} V_{+}^{(2)}  
+\frac{(-23+2k)}{3(5+k)^2} \hat{A}_{+} \pa \hat{A}_{-}
+\frac{(7+2k)}{3(5+k)^2} \hat{A}_{-} \pa \hat{A}_{+}
\nonu \\
&+ & \frac{8(195+554k+563k^2+92k^3}{(5+k)^2(3+7k)(19+23k)} i 
\hat{A}_3 \hat{T}
-\frac{4}{(5+k)} i \hat{A}_3 T^{(2)} 
-\frac{4}{(5+k)} i \hat{A}_3 W^{(2)}
\nonu \\
& + &  \frac{4(-10+k)}{3(5+k)^2} \hat{A}_3 \pa \hat{A}_3
+  \frac{2(5+4k)}{3(5+k)^2} \hat{A}_3 \pa \hat{B}_3
-\frac{2(4+10k+k^2)}{3(5+k)^2} \hat{B}_{+} \pa \hat{B}_{-}
\nonu \\
&+ & 
\frac{2(-4+12k+k^2)}{3(5+k)^2} \hat{B}_{-} \pa \hat{B}_{+}
+\frac{8k(69+103k+14k^2)}{(5+k)(3+7k)(19+23k)} i \hat{B}_3 \hat{T}
\nonu \\
& + & \frac{4}{(5+k)} i \hat{B}_3 T^{(2)} -\frac{2(1+8k)}{3(5+k)^2}
\hat{B}_3 \pa \hat{A}_3 +\frac{4(-4+15k+k^2)}{3(5+k)^2} \hat{B}_3 \pa 
\hat{B}_3
\nonu \\
& 
- & \frac{i}{(5+k)} \hat{B}_3 \pa T^{(1)}
+\frac{(-24-53k+8k^2)}{3(5+k)(3+7k)} \pa \hat{T}
-\frac{(1+2k)}{6(5+k)} \pa T^{(2)}
\nonu \\
& - & \frac{(9+2k)}{2(5+k)} \pa W^{(2)}
-\frac{4}{(5+k)^2} i \pa^2 \hat{A}_3-\frac{2}{3(5+k)} \pa^2 T^{(1)}
-\frac{2}{(5+k)} U^{(\frac{3}{2})} V^{(\frac{3}{2})}
\nonu \\
&+ & \frac{8}{(5+k)^2} i \hat{A}_{+} \hat{A}_{-} \hat{A}_3 
+\frac{8}{(5+k)^2} i \hat{A}_3 \hat{A}_3 \hat{A}_3
-\frac{16}{(5+k)^2} i \hat{A}_3 \hat{A}_3 \hat{B}_3 
\nonu \\
& + & \frac{8}{(5+k)^2} i \hat{A}_3 \hat{B}_{+} \hat{B}_{-}
+\frac{8}{(5+k)^2} i \hat{A}_3 \hat{B}_3 \hat{B}_3
+\frac{2k(14+k)}{3(5+k)^2} i \hat{B}_{-} \hat{B}_{+} \hat{B}_3
\nonu \\
&- & \frac{2k(14+k)}{3(5+k)^2} i \hat{B}_3 \hat{B}_{+} \hat{B}_{-}
-\frac{1}{2(5+k)} T^{(1)} \hat{B}_{+} \hat{B}_{-}
+\frac{1}{2(5+k)} \hat{B}_{-} T^{(1)} \hat{B}_{+}
\nonu \\
&-& \left. \frac{2}{(5+k)} T_{+}^{(\frac{3}{2})} T_{-}^{(\frac{3}{2})}
\right](w) +\cdots,
\nonu \\
U_{+}^{(2)}(z) \, V^{(\frac{5}{2})}(w)  & = & 
\frac{1}{(z-w)^3} \, \left[ \frac{8(2+k)(8+k)}{3(5+k)^2} \hat{G}_{21}+
\frac{8(4+k)}{3(5+k)^2} T_{+}^{(\frac{3}{2})} \right](w) \nonu \\
& + & \frac{1}{(z-w)^2} \, 
\left[ -\frac{(34+7k)}{3(5+k)} {\bf P_{+}^{(\frac{5}{2})}} 
-\frac{(30+7k)}{3(5+k)} W_{+}^{(\frac{5}{2})}
-\frac{2(30+7k)}{3(5+k)^2} i \hat{A}_{-} U^{(\frac{3}{2})}
\right. \nonu \\
&- &  \frac{12}{(5+k)^2} i 
\hat{A}_3 \hat{G}_{21} -\frac{12}{(5+k)^2} i \hat{A}_3 T_{+}^{(\frac{3}{2})} 
+\frac{2(-2+k)}{(5+k)^2} i \hat{B}_{-} \hat{G}_{22}
\nonu \\
& - & \frac{2(14+3k)}{(5+k)^2} i \hat{B}_{-} V^{(\frac{3}{2})}
+  \frac{4(8+k)}{(5+k)^2} i \hat{B}_3 \hat{G}_{21} + 
\frac{4(36+7k)}{3(5+k)^2} i \hat{B}_3 T_{+}^{(\frac{3}{2})}
\nonu \\
&- & 
 \frac{2}{(5+k)} T^{(1)} \hat{G}_{21} -\frac{2}{(5+k)} T^{(1)}
T_{+}^{(\frac{3}{2})} +\frac{4(56+23k+2k^2)}{9(5+k)^2} \pa \hat{G}_{21}
\nonu \\
& + & \left. \frac{8(25+6k)}{9(5+k)^2} \pa T_{+}^{(\frac{3}{2})}
\right](w)
\nonu \\
& + &
\frac{1}{(z-w)} \, \left[
-\frac{14(7+k)}{15(5+k)} \pa {\bf P_{+}^{(\frac{5}{2})}} 
+ \frac{1}{2} {\bf S_{+}^{(\frac{7}{2})}}  - \frac{1}{(5+k)}
i {\bf Q^{(\frac{5}{2})}} \hat{A}_{-} \right. \nonu \\
 & -& \frac{(12740 k^4+301736 k^3+1526637 k^2+2216314 k+937537)}
{45(5+k)^2(19+23k)(47+35k)} 
\hat{A}_{-} \hat{A}_{+} \hat{G}_{21} \nonu \\
&+ &  \frac{4}{(5+k)^2}  \hat{A}_{-} \hat{A}_{+} T_{+}^{(\frac{3}{2})}
\frac{8}{(5+k)^2} \hat{A}_{-} \hat{A}_3 U^{(\frac{3}{2})} -
\frac{4}{(5+k)^2}   \hat{A}_{-}  
\hat{B}_{-} \hat{T}_{-}^{(\frac{3}{2})} \nonu \\
& - & \frac{8}{(5+k)^2} 
\hat{A}_{-} \hat{B}_3 \hat{G}_{11} 
-\frac{8}{(5+k)^2}
\hat{A}_{-} \hat{B}_3 U^{(\frac{3}{2})}
-\frac{16(3+k)}{15(5+k)^2} i \hat{A}_{-} \pa \hat{G}_{11}
\nonu \\
& - & \frac{4(48+11k)}{15(5+k)^2} i \hat{A}_{-} \pa U^{(\frac{3}{2})}
+\frac{(-494-437k+43k^2+6k^3)}{
(5+k)^2(95+115k)} \hat{B}_{+} \hat{B}_{-} \hat{G}_{21}
\nonu \\
&- & \frac{(54+11k)}{3(5+k)^2} \hat{B}_{+} \hat{B}_{-} T_{+}^{(\frac{3}{2})} 
-\frac{2}{(5+k)} i \hat{B}_{-} V^{(\frac{5}{2})}
-\frac{8}{(5+k)^2} \hat{B}_{-} \hat{A}_3 \hat{G}_{22}
\nonu \\
&+& \frac{(1254+1357k-43k^2-6k^3)}{(5+k)^2(95+115k)}
\hat{B}_{-} \hat{B}_{+} \hat{G}_{21}
+\frac{(78+11k)}{3(5+k)^2} \hat{B}_{-} \hat{B}_{+} T_{+}^{(\frac{3}{2})}
\nonu \\
&- & \frac{4(6+5k)}{5(5+k)^2} \hat{B}_{-} \hat{B}_3 \hat{G}_{22}
+\frac{2(88+13k)}{5(5+k)^2} \hat{B}_{-} \hat{B}_3 V^{(\frac{3}{2})}
-\frac{4(12+5k)}{15(5+k)^2} i \hat{B}_{-} \pa \hat{G}_{22}
\nonu \\
&- & \frac{8(1+k)}{15(5+k)^2} i \hat{B}_{-} \pa V^{(\frac{3}{2})} 
\nonu \\
& + &  \frac{(12740 k^4+301736 k^3+1671537 k^2+2530594 k+1098277)}{
45(5+k)^2(19+23k)(47+35k)}
\hat{A}_{+} \hat{A}_{-} \hat{G}_{21}
 \nonu \\
& - & \frac{4}{(5+k)} i \hat{A}_3 W_{+}^{(\frac{5}{2})} 
+\frac{4}{(5+k)^2} \hat{A}_3 \hat{A}_3 \hat{G}_{21}
+\frac{4}{(5+k)^2} \hat{A}_3 \hat{A}_3 T_{+}^{(\frac{3}{2})}
\nonu \\
& -& \frac{8}{(5+k)^2} \hat{A}_3 \hat{B}_3 \hat{G}_{21}
-\frac{8}{(5+k)^2} \hat{A}_3 \hat{B}_3 T_{+}^{(\frac{3}{2})}
-\frac{4(53+233k+56k^2)}{(5+k)^2(285+345k)} i \hat{A}_3 \pa \hat{G}_{21}
\nonu \\
&+ & \frac{4(15+4k)}{15(5+k)^2} 
i \hat{A}_3 \pa T_{+}^{(\frac{3}{2})} 
+\frac{4(6+5k)}{5(5+k)^2} \hat{B}_3 \hat{B}_{-} \hat{G}_{22} 
-\frac{2(88+13k)}{5(5+k)^2} \hat{B}_3 \hat{B}_{-} V^{(\frac{3}{2})}
\nonu \\
&+ & \frac{4}{(5+k)^2} \hat{B}_3 \hat{B}_3 \hat{G}_{21}
+\frac{4}{(5+k)^2} \hat{B}_3 \hat{B}_3 T_{+}^{(\frac{3}{2})} 
\nonu \\
& + & \frac{4(152+301k+91k^2+2k^3)}{
(5+k)^2(95+115k)} i \hat{B}_3 \pa \hat{G}_{21}
+\frac{4k}{3(5+k)^2} i \hat{B}_3 \pa T_{+}^{(\frac{3}{2})}
\nonu \\
&+ & \frac{2(-49-19k+2k^2)}{(5+k)(95+115k)}
 T^{(1)} \pa \hat{G}_{21}
-\frac{2}{(5+k)} T^{(1)} \pa T_{+}^{(\frac{3}{2})}
-\frac{2(36+7k)}{15(5+k)} \pa W_{+}^{(\frac{5}{2})}
\nonu \\
&- & \frac{2(-9+2k)}{15(5+k)^2} i \pa \hat{A}_{-} U^{(\frac{3}{2})} 
-\frac{2(25+4k)}{5(5+k)^2} i \pa \hat{A}_3 T_{+}^{(\frac{3}{2})}
\nonu \\
& - & \frac{(43+173k+6k^2)}{(5+k)(95+115k)} \pa T^{(1)} \hat{G}_{21}
+  \frac{1}{(5+k)} \pa T^{(1)}  T_{+}^{(\frac{3}{2})}
\nonu \\
& + & 
\frac{2(195 k^4+53859 k^3+376352 k^2+597627 k+251275)}{
45(5+k)(805 k^3+5771 k^2+9623 k+4465)}
\pa^2   T_{+}^{(\frac{3}{2})}
\nonu \\
& + & \frac{4(266 k^4+5371 k^3+29551 k^2+52181 k+27255 )}{
3(5+k)^2(19+23k)(47+35k)} \hat{G}_{21} \hat{T}
\nonu \\
&+ & \frac{4(6370 k^4+159688 k^3+775635 k^2+1087214 k+439793 )}{45
(5+k)^2(19+23k)(47+35k)} i \hat{G}_{21} \pa \hat{A}_3
\nonu \\
&- & \frac{20(-663-916k-331k^2+2k^3)}{
3(5+k)(893+1746k+805k^2)} T_{+}^{(\frac{3}{2})} \hat{T}
-\frac{2}{(5+k)}  T_{+}^{(\frac{3}{2})} W^{(2)}
\nonu \\
&+ & \frac{2}{(5+k)} \hat{G}_{11} V_{+}^{(2)}
+
\frac{2(7+4k)}{5(5+k)^2} i \hat{G}_{11} \pa \hat{A}_{-} 
+\frac{2}{(5+k)} U^{(\frac{3}{2})}  V_{+}^{(2)}
\nonu \\
&-& \left. \frac{2}{(5+k)} V^{(\frac{3}{2})}  U_{+}^{(2)} 
-\frac{2}{(5+k)} \hat{G}_{21} W^{(2)} 
 \right](w) +\cdots, 
\nonu \\
U_{+}^{(2)}(z) \, W^{(2)}(w) & = &
-\frac{1}{(z-w)^3} \, \left[ \frac{2k(8+k)}{(5+k)^2} i \hat{B}_{-} \right] (w)
\nonu \\
& + & \frac{1}{(z-w)^2} \, \left[ \frac{2(4+k)}{(5+k)} U_{+}^{(2)} +
\frac{1}{2} \pa \{ U_{+}^{(2)} \, W^{(2)} \}_{-3} \right](w)
\nonu \\
& + & \frac{1}{(z-w)} \, \left[-\frac{1}{2} {\bf Q_{+}^{(3)}} +
\frac{1}{2(5+k)} i T^{(1)} \pa \hat{B}_{-}
-\frac{4}{(5+k)} i \hat{A}_3 U_{+}^{(2)} \right. \nonu \\
& - & \frac{(11+2k)}{(5+k)^2}
\hat{A}_3  \pa \hat{B}_{-} 
-\frac{4k(75+117k+14k^2)}{(5+k)(3+7k)(19+23k)} i \hat{B}_{-} \hat{T}
-\frac{2}{(5+k)} i \hat{B}_{-} T^{(2)}
\nonu \\
&- & \frac{k}{(5+k)^2} \hat{B}_{-} \pa \hat{B}_3 
-\frac{1}{2(5+k)} i \hat{B}_{-} \pa T^{(1)}
+\frac{k}{(5+k)^2} \hat{B}_3 \pa \hat{B}_{-}
+\frac{(9+2k)}{2(5+k)} \pa U_{+}^{(2)}
\nonu \\
&- & \frac{k(12+k)}{3(5+k)^2} i \pa^2 \hat{B}_{-} 
-\frac{2}{(5+k)} T_{+}^{(\frac{3}{2})} U^{(\frac{3}{2})}
+\frac{(15+2k)}{2(5+k)^2} i \hat{A}_{+} \hat{A}_{-} \hat{B}_{-}
\nonu \\
& - & \left.
\frac{(15+2k)}{2(5+k)^2} i \hat{A}_{-} \hat{A}_{+} \hat{B}_{-}
+\frac{2}{(5+k)} \hat{G}_{11} T_{+}^{(\frac{3}{2})}
\right](w) +\cdots,
\nonu \\
U_{+}^{(2)}(z) \, W_{+}^{(\frac{5}{2})}(w) & = &
-\frac{1}{(z-w)^2} \, \left[ \frac{2(16+k)}{3(5+k)^2} 
i \hat{B}_{-} \hat{T}_{+}^{(\frac{3}{2})} \right] (w) 
\nonu \\
& + & \frac{1}{(z-w)} \, \left[ 
-\frac{8(8+k)}{3(5+k)^2} \hat{B}_3 \hat{B}_{-} \hat{G}_{21}
-\frac{2}{(5+k)} i \hat{B}_{-} W_{+}^{(\frac{5}{2})}
-\frac{4(11+k)}{3(5+k)^2} i \hat{B}_{-} \pa 
T_{+}^{(\frac{3}{2})}
\right. 
\nonu \\
& + & \frac{14(8+k)}{3(5+k)^2} i \pa \hat{B}_{-} \hat{G}_{21}
+\frac{4(7+k)}{3(5+k)^2} i \pa \hat{B}_{-} T_{+}^{(\frac{3}{2})}
-\frac{2}{(5+k)} \hat{G}_{21} U_{+}^{(2)}
\nonu \\
&+& \left. \frac{8(8+k)}{3(5+k)^2} \hat{G}_{21} \hat{B}_{-} \hat{B}_3
\right](w) + \cdots, 
\nonu \\
U_{+}^{(2)}(z) \, W_{-}^{(\frac{5}{2})}(w) & = &
\frac{1}{(z-w)^3} \, \left[ -\frac{4(7+k)(3+2k)}{3(5+k)^2} \hat{G}_{11}
+ \frac{4(11+3k)}{3(5+k)^2} U^{(\frac{3}{2})} \right](w) 
\nonu \\
& + & \frac{1}{(z-w)^2} \, \left[ 
\frac{(30+7k)}{3(5+k)} {\bf Q^{(\frac{5}{2})}} +\frac{(38+7k)}{3(5+k)}
U^{(\frac{5}{2})}
+ \frac{2(6+k)}{(5+k)^2} i \hat{A}_{+} \hat{G}_{21}
\right. \nonu \\
& + & \frac{4(28+5k)}{3(5+k)^2} i \hat{A}_{+} T_{+}^{(\frac{3}{2})}
+\frac{4(47+7k)}{3(5+k)^2} i \hat{A}_3 U^{(\frac{3}{2})}
+\frac{2(-10+k)}{3(5+k)^2} i \hat{B}_{-} \hat{G}_{12}
\nonu \\
&+ & \frac{4}{3(5+k)} i \hat{B}_{-} T_{-}^{(\frac{3}{2})} 
-\frac{8(7+2k)}{3(5+k)^2} i \hat{B}_3 U^{(\frac{3}{2})}
+\frac{2}{(5+k)} T^{(1)}  U^{(\frac{3}{2})}
\nonu \\
&-& \left. \frac{4(30+17k+2k^2)}{9(5+k)^2} \pa \hat{G}_{11}
-\frac{8(25+4k)}{9(5+k)^2} \pa U^{(\frac{3}{2})}
\right](w) \nonu \\
& + & \frac{1}{(z-w)} \left[ 
\frac{2(51+7k)}{15(5+k)} \pa {\bf Q^{(\frac{5}{2})}}
+\frac{1}{(5+k)} i {\bf P_{-}^{(\frac{5}{2})}} \hat{B}_{-}+
\frac{1}{2} {\bf Q^{(\frac{7}{2})} }
+\frac{2}{(5+k)} i {\bf Q^{(\frac{5}{2})}} \hat{A}_3
\right. \nonu \\
&- & \frac{2}{(5+k)} i {\bf Q^{(\frac{5}{2})} } \hat{B}_3
+\frac{(281+52k)}{15(5+k)^2} \hat{A}_{-} \hat{A}_{+} U^{(\frac{3}{2})}
 -\frac{2(12+k)}{5(5+k)^2} \hat{B}_{+} \hat{B}_{-} \hat{G}_{11}
\nonu \\
& + & 
\frac{4}{(5+k)^2} \hat{B}_{+} \hat{B}_{-} U^{(\frac{3}{2})}
-\frac{4}{(5+k)^2} \hat{B}_{-} \hat{A}_{+} \hat{G}_{22}
 +\frac{4}{(5+k)^2} \hat{B}_{-} \hat{A}_{+}  V^{(\frac{3}{2})}
\nonu \\
&- & \frac{8}{(5+k)^2} \hat{B}_{-} \hat{A}_3 \hat{G}_{12}
 +\frac{8}{(5+k)^2} \hat{B}_{-} \hat{A}_3  T_{-}^{(\frac{3}{2})}
+ \frac{2(12+k)}{5(5+k)^2} \hat{B}_{-} \hat{B}_{+} \hat{G}_{11}
\nonu \\
& + & \frac{4(9+k)}{5(5+k)^2} i \hat{B}_{-} \pa \hat{G}_{12}
- \frac{4(9+k)}{5(5+k)^2} i \hat{B}_{-} \pa  T_{-}^{(\frac{3}{2})}
-\frac{(281+52k)}{15(5+k)^2} \hat{A}_{+} \hat{A}_{-} U^{(\frac{3}{2})}
\nonu \\
&+ & \frac{2(-177+188k+37k^2)}{
(5+k)^2(95+115k)} \hat{A}_{+} \hat{A}_3 \hat{G}_{21}
+\frac{2(121+20k)}{
15(5+k)^2} \hat{A}_{+} \hat{A}_3 T_{+}^{(\frac{3}{2})}
\nonu \\
& +& \frac{8}{
(5+k)^2} \hat{A}_{+} \hat{B}_3 T_{+}^{(\frac{3}{2})}
+ \frac{4(249+199k+26k^2)}{(5+k)^2(95+115k)} i \hat{A}_{+} \pa \hat{G}_{21}
\nonu \\
&+ & \frac{(354-376k-74k^2)}{
(5+k)^2(95+115k)} \hat{A}_3 \hat{A}_{+} \hat{G}_{21}
-\frac{2(181+20k)}{
15(5+k)^2} \hat{A}_3 \hat{A}_{+} T_{+}^{(\frac{3}{2})}
\nonu \\
&+ & \frac{4}{(5+k)} i \hat{A}_3 U^{(\frac{5}{2})} 
-\frac{16}{(5+k)^2} \hat{A}_3 \hat{A}_3 U^{(\frac{3}{2})}
+\frac{16}{(5+k)^2} \hat{A}_3 \hat{B}_3 U^{(\frac{3}{2})}
\nonu \\
&+& \frac{24}{5(5+k)^2} i \hat{A}_3 \pa \hat{G}_{11}
- \frac{8(12+k)}{15(5+k)^2} i \hat{B}_3 \pa \hat{G}_{11}
- \frac{4(10+k)}{15(5+k)^2} i \hat{B}_3 \pa  U^{(\frac{3}{2})}
\nonu \\
& + &  \frac{4}{5(5+k)} T^{(1)} \pa \hat{G}_{11}
+ \frac{2}{(5+k)} T^{(1)} \pa  U^{(\frac{3}{2})}
- \frac{36}{5(5+k)^2} i \pa \hat{A}_3  \hat{G}_{11}
\nonu \\
& - & \frac{8(14+k)}{15(5+k)^2} i \pa \hat{B}_{-}  \hat{G}_{12}
+ \frac{2(56+19k)}{15(5+k)^2} i \pa \hat{B}_{-}   T_{-}^{(\frac{3}{2})}
- \frac{6}{5(5+k)} \pa T^{(1)} \hat{G}_{11}
\nonu \\
& - & \frac{(9633+70346k+195461k^2+112360k^3+5980k^4)}{
45(5+k)^2(893+1746k+805k^2)}  \pa^2 \hat{G}_{11}
\nonu \\
& - & \frac{(80599+71021k+10330k^2)}{
90(5+k)(235+222k+35k^2)} \pa^2  U^{(\frac{3}{2})}
+\frac{2(59+7k)}{15(5+k)} \pa U^{(\frac{5}{2})}
\nonu \\
& + & \frac{4(53+6k)}{15(5+k)^2} i \hat{A}_3 \pa  U^{(\frac{3}{2})}
+\frac{2}{(5+k)}  T_{+}^{(\frac{3}{2})}  U_{-}^{(2)}
+\frac{2}{(5+k)}  T_{-}^{(\frac{3}{2})}  U_{+}^{(2)}
\nonu \\
& - & \frac{4(573+1542k+853k^2+92k^3)}{
(5+k)(19+23k)(47+35k)} \hat{G}_{11} \hat{T}
+\frac{4(30+181k+35k^2)}{(5+k)(3+7k)(47+35k)}
U^{(\frac{3}{2})} \hat{T}
\nonu \\
&+ &  \frac{2}{(5+k)}  U^{(\frac{3}{2})}  T^{(2)}
-\frac{2(140+37k)}{15(5+k)^2} i U^{(\frac{3}{2})} \pa \hat{B}_3 -
\frac{1}{(5+k)}  U^{(\frac{3}{2})} \pa T^{(1)} 
\nonu \\
&+& \left.  \frac{8(34+5k)}{15(5+k)^2} i \hat{A}_{+} \pa  T_{+}^{(\frac{3}{2})}
\right](w) + \cdots, 
\nonu \\
U_{+}^{(2)}(z) \, W^{(3)}(w) & = &
\frac{1}{(z-w)^4} \, \left[ \frac{2k(1509+1724k+835k^2+84k^3)}{
(5+k)^3(19+23k)} i \hat{B}_{-} \right] (w) 
\nonu \\
& + & \frac{1}{(z-w)^3} \, \left[ 
\frac{2(114-853k+13k^2)}{
(5+k)(285+402k+69k^2)} U_{+}^{(2)}
-\frac{4(1+23k+3k^2)}{(5+k)^3} \hat{A}_3 \hat{B}_{-}
\right. 
\nonu \\
&+ & \left. \frac{12k}{(5+k)^3} \hat{B}_{-} \hat{B}_3
+\frac{6k}{(5+k)^3} i \pa \hat{B}_{-} - \frac{10}{(5+k)^2} 
i T^{(1)} \hat{B}_{-}
\right](w)
\nonu \\
& + & \frac{1}{(z-w)^2} \, \left[ 
\frac{(25+6k)}{2(5+k)} {\bf Q_{+}^{(3)}} +\frac{1}{(5+k)} i {\bf P^{(2)}} 
\hat{B}_{-}
\right. 
\nonu \\
&+ & \frac{25}{2(5+k)^2} i T^{(1)} \pa \hat{B}_{-} 
+\frac{2(602+623k+89k^2)}{(5+k)^2(19+23k)} i \hat{A}_3 U^{(\frac{3}{2})}
\nonu \\
& + & \frac{(93+30k+4k^2)}{(5+k)^3} \hat{A}_3 \pa \hat{B}_{-}
+\frac{(259-9k-16k^2)}{
(95+134k+23k^2)} T^{(1)} U^{(\frac{3}{2})}
\nonu \\
& + & \frac{2(285+12662k+23609k^2+6896k^3+336k^4)}{
3(5+k)^2(3+7k)(19+23k)} i \hat{B}_{-} \hat{T}
+\frac{(121+17k)}{3(5+k)^2} i \hat{B}_{-} T^{(2)}
\nonu \\
&- & \frac{(59+13k)}{(5+k)^2} i \hat{B}_{-} W^{(2)}
-\frac{27}{2(5+k)^2} i \hat{B}_{-} \pa T^{(1)}
-\frac{2(209+691k+126k^2)}{(5+k)^2(19+23k)} i \hat{B}_3 U_{+}^{(2)}
\nonu \\
&+ & \frac{k(35-2k)}{(5+k)^3} \hat{B}_3 \pa \hat{B}_{-} 
-\frac{(6175+11357k+2274k^2)}{
6(5+k)^2(19+23k)} \pa U_{+}^{(2)}
\nonu \\
& - & \frac{(10+193k+42k^2)}{6(5+k)^3} i \pa^2 \hat{B}_{-}
-\frac{2(6+k)}{(5+k)^2} \hat{G}_{11} \hat{G}_{21}
-\frac{2(23+6k)}{(5+k)^2} \hat{G}_{11} T_{+}^{(\frac{3}{2})}
\nonu \\
&+ & \frac{4(7+k)}{(5+k)^2} \hat{G}_{21} U^{(\frac{3}{2})}
+ \frac{2(25+6k)}{(5+k)^2} T_{+}^{(\frac{3}{2})}  U^{(\frac{3}{2})}
-\frac{(479+250k+30k^2)}{6(5+k)^3} i \hat{A}_{+} \hat{A}_{-} \hat{B}_{-}
\nonu \\
&+ & \frac{(547+350k+30k^2)}{6(5+k)^3} i \hat{A}_{-} \hat{A}_{+} \hat{B}_{-} 
+ \frac{2(-67+19k)}{3(5+k)^3} i \hat{A}_3 \hat{A}_3 \hat{B}_{-}
\nonu \\
& - & \frac{4(-31+7k)}{3(5+k)^3} 
i \hat{A}_3 \hat{B}_{-} \hat{B}_3
+ \frac{4(-10-77k+6k^2)}{6(5+k)^3} 
i \hat{B}_{+} \hat{B}_{-} \hat{B}_{-}
\nonu \\
&+ & \frac{(-10-35k+2k^2)}{2(5+k)^3} 
i \hat{B}_{-} \hat{B}_{+} \hat{B}_{-}
- \frac{10(-1+k)}{3(5+k)^3} 
i \hat{B}_{-} \hat{B}_3 \hat{B}_3
\nonu \\
& -& \left.  \frac{8}{(5+k)^2} T^{(1)} \hat{A}_3 \hat{B}_{-}
+ \frac{8}{(5+k)^2} T^{(1)} \hat{B}_{-} \hat{B}_3 \right](w)
\nonu \\
& + & \frac{1}{(z-w)} \, 
\left[ 
\frac{2}{(5+k)} i \hat{A}_3 {\bf Q_{+}^{(3)}}
-\frac{5(21+4k)}{6(5+k)} i \hat{B}_3 {\bf Q_{+}^{(3)}}
-\frac{(81+20k)}{12(5+k)} i \hat{B}_{-} {\bf S^{(3)}}
\right. 
\nonu \\
&- & \frac{(81+20k)}{12(5+k)} i \hat{B}_{-} {\bf P^{(3)}}
+\frac{(93+20k)}{12(5+k)} \hat{B}_{-} \hat{B}_3 {\bf P^{(2)}}
+ \frac{(93+20k)}{24(5+k)} i \pa \hat{B}_{-} {\bf P^{(2)}}
\nonu \\
&+ & \frac{1}{2(5+k)} i \hat{B}_{-} \pa  {\bf P^{(2)}}
+\frac{(5+2k)}{2(5+k)} \pa {\bf Q_{+}^{(3)}}
\nonu \\
&+ &
\frac{(93+20k)}{6(5+k)} U_{+}^{(2)} W^{(2)}  
+\frac{(117+20k)}{6(5+k)} T^{(2)} U_{+}^{(2)}
+\frac{2(5+2k)}{(5+k)^2} \pa  T_{+}^{(\frac{3}{2})} U^{(\frac{3}{2})}
 \nonu \\
&+ & \frac{2(5+2k)}{(5+k)^2}  T_{+}^{(\frac{3}{2})} \pa U^{(\frac{3}{2})}
+\frac{5(33+10k)}{3(5+k)^2} i \pa \hat{A}_3  U_{+}^{(2)}
-\frac{2(-7+13k)}{(5+k)(19+23k)} i \hat{A}_3 \pa  U_{+}^{(2)}
\nonu \\
&+ & \frac{8}{(5+k)^2} i
\hat{A}_3  T_{+}^{(\frac{3}{2})}  U^{(\frac{3}{2})}
-\frac{(141+20k)}{3(5+k)^2} \hat{A}_3 \hat{A}_3  U_{+}^{(2)}
-\frac{2(117+20k)}{3(5+k)^3} \hat{A}_3 \hat{A}_3 \hat{A}_3 \hat{B}_{-}
\nonu \\
&-& \frac{4(125+21k)}{3(5+k)^3} i \pa \hat{A}_3 \hat{A}_3 \hat{B}_{-} 
+\frac{(-3094-95k+100k^2)}{
12(5+k)^3} i \hat{A}_3 \hat{A}_3 \pa \hat{B}_{-}
\nonu \\
& + & \frac{(-2542-263k+100k^2)}{6(5+k)^3} 
\hat{A}_3 \hat{A}_3 \hat{B}_{-} \hat{B}_{3}
+ \frac{(93+20k)}{3(5+k)^2} i \hat{A}_3 \hat{A}_3 \hat{B}_{-} T^{(1)}
\nonu \\
&+& \frac{16}{(5+k)^2} \hat{A}_3 \hat{B}_3   U_{+}^{(2)}
+\frac{(117+20k)}{3(5+k)^2} \hat{A}_3 \hat{B}_{-} W^{(2)}
+\frac{(93+20k)}{3(5+k)^2} \hat{A}_3 \hat{B}_{-} T^{(2)}
\nonu \\
&-& \frac{(-35601-46772k+55011k^2+48566k^3+5952k^4)}{
6(5+k)^3(3+7k)(19+23k)} \pa^2 \hat{A}_3 \hat{B}_{-}
\nonu \\
&- & \frac{(4394+2001k+236k^2)}{
12(5+k)^3} \pa \hat{A}_3 \pa \hat{B}_{-}  
\nonu \\
& + & \frac{(51615+174084k+158471k^2+40886k^3+3812k^4)}{
2(5+k)^3(3+7k)(19+23k)} \hat{A}_3 \pa^2 \hat{B}_{-}
\nonu \\
&+ & \frac{(2562+1561k+260k^2)}{
6(5+k)^3} i \pa \hat{A}_3 \hat{B}_{-} \hat{B}_3  
+ \frac{(-613-239k+20k^2)}{3(5+k)^3} i \hat{A}_3 \pa \hat{B}_{-} 
\hat{B}_3
\nonu \\
&- & \frac{(3418+1839k+380k^2)}{12(5+k)^3} 
i \hat{A}_3 \hat{B}_{-} \pa \hat{B}_3 
+\frac{(-226-75k+20k^2)}{
6(5+k)^3} \hat{A}_3 \hat{B}_{-} \hat{B}_3 \hat{B}_3
\nonu \\
&+ & \frac{4(93+20k)}{3(5+k)^2} 
i \hat{A}_3 \hat{B}_{-} \hat{B}_3 T^{(1)}   
-\frac{(93+20k)}{3(5+k)^2} \pa \hat{A}_3 \hat{B}_{-} T^{(1)}
-\frac{(99+20k)}{(5+k)^2} \hat{A}_3 \pa \hat{B}_{-} T^{(1)}
\nonu \\
& - & \frac{2}{(5+k)^2} \hat{A}_3 \hat{B}_{-} \pa T^{(1)} 
-\frac{(1854+919k+220k^2)}{24(5+k)^3}
\hat{A}_3 \hat{B}_{+} \hat{B}_{-} \hat{B}_{-}
\nonu \\
& -& \frac{8}{(5+k)^2} i \hat{A}_3 \hat{G}_{11}  T_{+}^{(\frac{3}{2})}  
+\frac{8}{(5+k)^2} i \hat{A}_3 \hat{G}_{21}  U^{(\frac{3}{2})}
-\frac{5(21+4k)}{3(5+k)^2} \hat{A}_{+} \hat{A}_{-}  U_{+}^{(2)}
\nonu \\
&- & \frac{2(117+20k)}{3(5+k)^3} \hat{A}_{+} \hat{A}_{-} \hat{A}_3 
\hat{B}_{-} +\frac{5(41+8k)}{6(5+k)^3} i \pa \hat{A}_{+} \hat{A}_{-} 
\hat{B}_{-}
-\frac{(377+80k)}{6(5+k)^3} i 
\hat{A}_{+} \pa \hat{A}_{-} \hat{B}_{-}
\nonu \\
& + & \frac{(398+577k+100k^2)}{
12(5+k)^3} i \hat{A}_{+} \hat{A}_{-} \pa \hat{B}_{-}
+\frac{(-130+297k+100k^2)}{6(5+k)^3} 
\hat{A}_{+} \hat{A}_{-} \hat{B}_{-} \hat{B}_{3}
\nonu \\
& + & \frac{(93+20k)}{3(5+k)^2} i
\hat{A}_{+} \hat{A}_{-} \hat{B}_{-} T^{(1)} 
-\frac{(81+20k)}{3(5+k)^2} \hat{A}_{+} \hat{B}_{-}  V_{+}^{(2)}
+\frac{4}{(5+k)^2} i \hat{A}_{+} \hat{G}_{21}  T_{+}^{(\frac{3}{2})} 
\nonu \\
& - & \frac{(5454+1847k+220k^2)}{
72(5+k)^2} i \pa \hat{B}_3  U_{+}^{(2)}
+  \frac{(9291+10747k+1820k^2)}{
6(5+k)^2(19+23k)} i  \hat{B}_3  \pa U_{+}^{(2)}
\nonu \\
&-& \frac{10(21+4k)}{3(5+k)^2} i \hat{B}_3  T_{+}^{(\frac{3}{2})}   U^{(\frac{3}{2})}
+ \frac{11(-306-11k+20k^2)}{36(5+k)^2} 
\hat{B}_3 \hat{B}_3 U_{+}^{(2)}
\nonu \\
&+ & \frac{10(21+4k)}{3(5+k)^2} i \hat{B}_3 \hat{G}_{11}  T_{+}^{(\frac{3}{2})}
-\frac{8}{(5+k)^2} i \hat{B}_3 \hat{G}_{21}  U^{(\frac{3}{2})}
+\frac{(93+20k)}{3(5+k)} i \hat{B}_3 T^{(1)}  U_{+}^{(2)}
\nonu \\
& + & \frac{(81+20k)}{6(5+k)} i \hat{B}_{-} W^{(3)}
-\frac{(9450+3899k+460k^2)}{72(5+k)^2} i \pa \hat{B}_{-} W^{(2)}
\nonu \\
& + & \frac{(29+8k)}{4(5+k)^2} i \hat{B}_{-} \pa W^{(2)}
+\frac{(81+20k)}{3(5+k)^2} i \hat{B}_{-}  U^{(\frac{3}{2})}  V^{(\frac{3}{2})}
\nonu \\
& - & \frac{(-1426+113k+100k^2)}{24(5+k)^2} i \pa \hat{B}_{-} T^{(2)}
+ \frac{(449+128k)}{12(5+k)^2}  i \hat{B}_{-} \pa T^{(2)}
\nonu \\
& + & \frac{(81+20k)}{3(5+k)^2} i \hat{B}_{-} T_{+}^{(\frac{3}{2})} 
T_{-}^{(\frac{3}{2})}
\nonu \\
& - & \frac{(177498+3239199k+6063075k^2+2778861k^3+
537931k^4+39836k^5)}{
108(5+k)^3(3+7k)(19+23k)} i \pa^3 \hat{B}_{-}
\nonu \\
&-& \frac{(5562+2915k+460k^2)}{36(5+k)^2} 
\hat{B}_{-} \hat{B}_3 W^{(2)}
-\frac{(-1618-23k+100k^2)}{12(5+k)^2} \hat{B}_{-} \hat{B}_3 T^{(2)}
\nonu \\
&-& \frac{(-155610+1274721k+3069108k^2+1669509k^3+
407260k^4+28160k^5)}{36(5+k)^3(3+7k)(19+23k)} 
\nonu \\
&\times & \pa^2 \hat{B}_{-} \hat{B}_3
\nonu \\
&-& \frac{(-105906+1510145k+3428928k^2+1874637k^3+
449824k^4+41040k^5)}{36(5+k)^3(3+7k)(19+23k)} 
\nonu \\
&\times &  
\hat{B}_{-} \pa^2 \hat{B}_3
\nonu \\
& - & \frac{(822+937k+312k^2+80k^3)}{
12(5+k)^3} i \pa \hat{B}_{-} \hat{B}_3 \hat{B}_3
\nonu \\
& - & \frac{(1514+803k+192k^2+80k^3)}{
18(5+k)^3} \hat{B}_{-} \hat{B}_3 \hat{B}_3 \hat{B}_3
\nonu \\
& - & \frac{(-1134+359k+220k^2)}{
36(5+k)^2} i \hat{B}_{-} \hat{B}_3 \hat{B}_3 T^{(1)}
+ \frac{(756+719k+220k^2)}{36(5+k)^2} \pa \hat{B}_{-} 
\hat{B}_3 T^{(1)} 
\nonu \\
&- & \frac{(5706+1799k+220k^2)}{
72(5+k)^2} \hat{B}_{-} \pa \hat{B}_3 T^{(1)}
-\frac{(81+20k)}{3(5+k)^2} \hat{B}_{-} \hat{B}_3 \pa T^{(1)}
\nonu \\
&+& \frac{(93+20k)}{6(5+k)} \hat{B}_{-} \hat{B}_3 T^{(1)} T^{(1)}
+  \frac{(-18+599k+220k^2)}{144(5+k)^2} \hat{B}_{-} \hat{B}_{-}  V_{-}^{(2)}
\nonu \\
&- & \frac{4}{(5+k)^2} i \hat{B}_{-} \hat{G}_{12}  T_{+}^{(\frac{3}{2})} 
-\frac{4}{(5+k)^2} i \hat{B}_{-} \hat{G}_{12} \hat{G}_{21}
-\frac{4}{(5+k)^2} i \hat{B}_{-} \hat{G}_{21}  T_{-}^{(\frac{3}{2})} 
\nonu \\
& - & \frac{(93+20k)}{6(5+k)} i \hat{B}_{-} T^{(1)} W^{(2)} 
- \frac{(93+20k)}{6(5+k)} i \hat{B}_{-} T^{(1)} T^{(2)} 
\nonu \\
&-& \frac{(72063+224082k+201327k^2+51532k^3+3680k^4)}{
12(5+k)^2(3+7k)(19+23k)} i \pa^2 \hat{B}_{-} T^{(1)}
\nonu \\
&-& \frac{5(15+4k)}{6(5+k)^2} i \pa \hat{B}_{-} \pa T^{(1)} 
\nonu \\
& - & \frac{(62181+162582k+133925k^2+
46708k^3+5520k^4)}{
18(5+k)^2(3+7k)(19+23k)} i \hat{B}_{-} \pa^2 T^{(1)}
\nonu \\
& + & \frac{(93+20k)}{12(5+k)} i \pa \hat{B}_{-} T^{(1)} T^{(1)} 
-\frac{(2214+1079k+220k^2)}{72(5+k)^2} 
\hat{B}_{+} \hat{B}_{-}  U_{+}^{(2)}
\nonu \\
&+ & \frac{(-386+21k+20k^2)}{24(5+k)^3} i 
\pa \hat{B}_{+} \hat{B}_{-} \hat{B}_{-}
- \frac{(-216-11k+20k^2)}{6(5+k)^3} i 
 \hat{B}_{+} \pa \hat{B}_{-} \hat{B}_{-}
\nonu \\
&-& \frac{(-302+47k+60k^2)}{
12(5+k)^3}  \hat{B}_{+} \hat{B}_{-} \hat{B}_{-} \hat{B}_3
+  \frac{(4446+1559k+220k^2)}{
144(5+k)^2}  \hat{B}_{+} \hat{B}_{-} \hat{B}_{-} T^{(1)}
\nonu \\
&- & \frac{2(7+2k)}{(5+k)^2} \pa \hat{G}_{11}  T_{+}^{(\frac{3}{2})}   
-\frac{2(3+2k)}{(5+k)^2} \hat{G}_{11} \pa  T_{+}^{(\frac{3}{2})}
-\frac{4(6+k)}{3(5+k)^2} \pa \hat{G}_{11} \hat{G}_{21}
\nonu \\
& + & \frac{2}{(5+k)} \hat{G}_{21}  U^{(\frac{5}{2})}
+\frac{8}{3(5+k)} \hat{G}_{21} \pa  U^{(\frac{3}{2})}
-\frac{1}{(5+k)} \pa T^{(1)}  U_{+}^{(2)}
\nonu \\
&-& \frac{(-139-7k+8k^2)}{(5+k)(19+23k)} T^{(1)} \pa  U_{+}^{(2)}
-\frac{(5301+13977k+11480k^2+1840k^3)}{
3(5+k)(3+7k)(19+23k)} \hat{T} U_{+}^{(2)} 
\nonu \\
&-& \frac{2(9585+27429k+18152k^2+2176k^3)}{
3(5+k)^2(3+7k)(19+23k)} \hat{T} \hat{A}_3 \hat{B}_{-}
\nonu \\
&+ & \frac{(-4902+9241k+27015k^2+9596k^3+672k^4)}{
6(5+k)^2(3+7k)(19+23k)} i \pa \hat{T} \hat{B}_{-}
\nonu \\
& + & \frac{(17214+214775k+367181k^2+129408k^3+13680k^4)}{
12(5+k)^2(3+7k)(19+23k)}  i \hat{T} \pa \hat{B}_{-}
\nonu \\
&+& \frac{(-7410+105791k+224869k^2+97856k^3+13680k^4)}{
6(5+k)^2(3+7k)(19+23k)} \hat{T} \hat{B}_{-} \hat{B}_3
\nonu \\
&+& \left. \frac{(5787+15879k+11736k^2+1840k^3)}{
3(5+k)(3+7k)(19+23k)} i \hat{T} \hat{B}_{-} T^{(1)}
\right](w) + \cdots. 
\label{opeagain}
\eea
The first OPE of (\ref{opeagain}) has $\hat{G}_{11} \hat{G}_{11}(w)$
which can be written in terms of a derivative of $\hat{A}_{+} \hat{B}_{-}(w)$.
Similarly,  
The fourth OPE of (\ref{opeagain}) has $\hat{G}_{21} \hat{G}_{21}(w)$
which can be written in terms of a derivative of $\hat{A}_{-} \hat{B}_{-}(w)$.
The first order pole in the last OPE of (\ref{opeagain})
contains composite field with spin-$4$ with $U(1)$ charge of 
$\frac{2k}{(5+k)}$. The second order pole contains a composite field of 
spin-$3$ and each term can be seen in the Table $5$ of \cite{Ahn1311}.
There is no $T^{(1)} T^{(1)} \hat{B}_{-}(w)$ term in the second order pole. 

\section{
The nontrivial 
OPEs between  higher spin-$2$ current, $U_{-}^{(2)}(z)$,  
and 
other $10$ higher spin currents
}

We present the corresponding OPEs as follows:
\bea
U_{-}^{(2)}(z) \, U^{(\frac{5}{2})}(w) & = & 
\frac{1}{(z-w)^2} \, \left[ \frac{2(3+2k)}{3(5+k)^2} i \hat{A}_{+} (
\hat{G}_{11} +
 U^{(\frac{3}{2})}) \right](w) 
\nonu \\
& + & \frac{1}{(z-w)} \, \left[ -\frac{2}{(5+k)} i \hat{A}_{+}  
U^{(\frac{5}{2})} -\frac{4(3+2k)}{3(5+k)^2} \hat{A}_{+} \hat{A}_3  
U^{(\frac{3}{2})}  
-\frac{4}{(5+k)^2} \hat{A}_{+} \hat{B}_{-} \hat{G}_{12}
\right. \nonu \\
&+ & \frac{8(1+k)}{3(5+k)^2} i \hat{A}_{+} \pa U^{(\frac{3}{2})}
+ \frac{4(3+2k)}{3(5+k)^2} 
\hat{A}_3 \hat{A}_{+}  U^{(\frac{3}{2})}
+\frac{2(9+2k)}{3(5+k)^2} i \pa \hat{A}_{+} \hat{G}_{11}
\nonu \\
& + & \left. \frac{4}{(5+k)^2} \hat{G}_{12} \hat{A}_{+} \hat{B}_{-}
+\frac{2}{(5+k)} \hat{G}_{11} U_{-}^{(2)} +\frac{4}{(5+k)}
 U^{(\frac{3}{2})}  U_{-}^{(2)}  \right](w)
+ \cdots,
\nonu \\
U_{-}^{(2)}(z) \, V^{(\frac{3}{2})}(w) & = & 
-\frac{1}{(z-w)^2} \, \left[ \frac{(5+2k)}{(5+k)}  T_{-}^{(\frac{3}{2})} 
\right](w)
\nonu \\
& + & \frac{1}{(z-w)} \, \left[ \frac{1}{2} {\bf P_{-}^{(\frac{5}{2})} }
- W_{-}^{(\frac{5}{2})} -\frac{2(5+2k)}{3(5+k)} \pa   T_{-}^{(\frac{3}{2})}
\right](w) +\cdots,
\nonu \\
U_{-}^{(2)}(z) \, V_{+}^{(2)}(w) & = &  -\frac{1}{(z-w)^4} \, 
\left[ \frac{6k(5+2k)}{(5+k)^2} \right]
\nonu \\
& - & \frac{1}{(z-w)^3} \, \left[ \frac{12(5+2k)}{(5+k)^2} i \hat{A}_3 
\right](w)
\nonu \\
&+ & \frac{1}{(z-w)^2} \, \left[ -\frac{1}{2} {\bf P^{(2)}}
-\frac{4(75+310k+184k^2+25k^3)}{
3(5+k)^2(3+7k)} \hat{T} +\frac{2(-4+k)}{3(5+k)} T^{(2)}  
\right. \nonu \\
&+ & \frac{2(4+k)}{(5+k)} W^{(2)} 
-\frac{16(-1+k)}{3(5+k)^2} \hat{A}_1 \hat{A}_1
 -\frac{16(-1+k)}{3(5+k)^2} \hat{A}_2 \hat{A}_2
-\frac{4(-13+4k)}{3(5+k)^2} \hat{A}_3 \hat{A}_3
\nonu \\
&+ & \frac{8(-4+k)}{3(5+k)^2} \hat{A}_3 \hat{B}_3
-\frac{4(5+k)}{3(5+k)^2} \hat{B}_1 \hat{B}_1
-\frac{4(5+k)}{3(5+k)^2} \hat{B}_2 \hat{B}_2
+\frac{4(-5+2k)}{3(5+k)^2} \hat{B}_3 \hat{B}_3
\nonu \\
&- & \left. \frac{6(5+2k)}{(5+k)^2} i \pa \hat{A}_3 
-\frac{4}{(5+k)} i T^{(1)} \hat{A}_3
+ \frac{4}{(5+k)} i T^{(1)} \hat{B}_3
\right](w)
\nonu \\
& + & \frac{1}{(z-w)} \, \left[ 
-\frac{1}{2} {\bf S^{(3)}} +\frac{1}{2} {\bf P^{(3)}}
-\frac{1}{4} \pa {\bf P^{(2)}}
-\frac{4}{(19+23k)} T^{(1)} \hat{T} \right. \nonu \\
& - & 
\frac{3}{(5+k)} i T^{(1)} \pa \hat{A}_3
+\frac{(11-8k)}{3(5+k)^2} \hat{A}_{+} \pa \hat{A}_{-} 
-\frac{(1+8k)}{3(5+k)^2} \hat{A}_{-} \pa \hat{A}_{+}
\nonu \\
&- & \frac{8(600+1625k+866k^2+113k^3)}{
(5+k)^2(3+7k)(19+23k)} 
i \hat{A}_3  \hat{T} + \frac{4}{(5+k)} 
i \hat{A}_3 T^{(2)} 
\nonu \\
& - & \frac{2(-35+8k)}{3(5+k)^2} \hat{A}_3 \pa \hat{A}_3 
-  \frac{1}{(5+k)} i \hat{A}_3 \pa T^{(1)}
-\frac{2}{(5+k)} i \hat{B}_{+} U_{+}^{(2)} 
\nonu \\
& - & \frac{(26+k)}{3(5+k)^2} \hat{B}_{+} \pa \hat{B}_{-} 
-\frac{k}{(5+k)^2} \hat{B}_{-} \pa \hat{B}_{+}
- \frac{8k(51+55k)}{
(5+k)(3+7k)(19+23k)} 
i \hat{B}_3  \hat{T}
\nonu \\
& - &  \frac{4}{(5+k)} 
i \hat{B}_3 T^{(2)} 
+\frac{4}{(5+k)} i \hat{B}_3 W^{(2)}
+  \frac{10(-1+k)}{3(5+k)^2} \hat{B}_3 \pa \hat{A}_3 
\nonu \\
& + &   \frac{4}{(5+k)} i \hat{B}_3 \pa T^{(1)}
-\frac{(39+166k+50k^2)}{3(5+k)(3+7k)} \pa \hat{T}
+\frac{(-17+2k)}{6(5+k)} \pa T^{(2)}
\nonu \\
&+ & \frac{(5+2k)}{2(5+k)} \pa W^{(2)} 
-\frac{2(7+2k)}{(5+k)^2} i \pa^2 \hat{A}_3
-\frac{13}{3(5+k)^2} i \pa^2 \hat{B}_3
-\frac{2}{3(5+k)} \pa^2 T^{(1)}
\nonu \\
&+ & \frac{2}{(5+k)} \hat{G}_{12} \hat{G}_{21}
+\frac{4}{(5+k)} \hat{G}_{21} T_{-}^{(\frac{3}{2})}
-\frac{2}{(5+k)} U^{(\frac{3}{2})} V^{(\frac{3}{2})}
-\frac{8}{(5+k)^2} i \hat{A}_{+} \hat{A}_{-} \hat{A}_3
\nonu \\
&- & \frac{8}{(5+k)^2} i \hat{A}_3 \hat{A}_3 \hat{A}_3+\frac{16}{(5+k)^2} i
 \hat{A}_3 \hat{A}_3 \hat{B}_3 -
\frac{(17+k)}{3(5+k)^2} i 
\hat{A}_3 \hat{B}_{+} \hat{B}_{-} \nonu \\
&- & \frac{8}{(5+k)^2} i \hat{A}_3 \hat{B}_3 \hat{B}_3
+\frac{(-7+k)}{3(5+k)^2} i \hat{B}_{-} \hat{A}_3 \hat{B}_{+} 
- \frac{(-13+4k)}{3(5+k)^2} i \hat{B}_{-} \hat{B}_{+} \hat{B}_3
\nonu \\
&+& \left. \frac{(-13+4k)}{3(5+k)^2} i \hat{B}_3 \hat{B}_{+} \hat{B}_{-}
+\frac{2}{(5+k)} T_{+}^{(\frac{3}{2})} T_{-}^{(\frac{3}{2})}
\right](w) +\cdots,
\nonu \\
U_{-}^{(2)}(z) \, V_{-}^{(2)}(w) & = & 
-\frac{1}{(z-w)^2} \, 
\left[ \frac{2(7+k)}{(5+k)^2} \hat{A}_{+} \hat{B}_{+} \right] (w)
\nonu \\
& 
+ & \frac{1}{(z-w)} \, \left[ \frac{2}{(5+k)} i \hat{A}_{+} V_{-}^{(2)}
+\frac{2(3+k)}{(5+k)^2} \hat{A}_{+} \pa \hat{B}_{+} 
+\frac{2}{(5+k)} i \hat{B}_{+} U_{-}^{(2)}
\right. \nonu \\
& + & \left. \hat{G}_{12} \hat{G}_{12} 
\right](w) +  \cdots, 
\nonu \\
U_{-}^{(2)}(z) \, V^{(\frac{5}{2})}(w) & = & 
\frac{1}{(z-w)^3} \, \left[ -\frac{8(12+9k+k^2)}{3(5+k)^2} \hat{G}_{12}
+ \frac{4(11+3k)}{3(5+k)^2} 
T_{-}^{(\frac{3}{2})}  \right](w)
\nonu \\
& + & \frac{1}{(z-w)^2} \, \left[ \frac{2(13+4k)}{3(5+k)} {\bf 
P_{-}^{(\frac{5}{2})}} -\frac{(25+8k)}{3(5+k)}  W_{-}^{(\frac{5}{2})}  
+\frac{2}{(5+k)^2} i \hat{A}_{+} \hat{G}_{22}
\right. \nonu \\
&- & \frac{2(19+4k)}{(5+k)^2} i \hat{A}_{+} V^{(\frac{3}{2})}
-\frac{8(9+k)}{3(5+k)^2} i \hat{A}_3 \hat{G}_{12}
+\frac{8(9+k)}{3(5+k)^2} i \hat{A}_3 T_{-}^{(\frac{3}{2})}
\nonu \\
&- & 
\frac{2(5+2k)}{(5+k)^2} i \hat{B}_{+} \hat{G}_{11}
-  \frac{28(3+k)}{3(5+k)^2} i \hat{B}_{+} U^{(\frac{3}{2})}
+\frac{4(5+2k)}{(5+k)^2} i \hat{B}_3 T_{-}^{(\frac{3}{2})}
\nonu \\
&-& \left. \frac{4k(13+2k)}{9(5+k)^2} \pa \hat{G}_{12}
-\frac{4(55+14k)}{9(5+k)^2} \pa T_{-}^{(\frac{3}{2})}
\right](w)
\nonu \\
& + & \frac{1}{(z-w)} \, \left[ 
\frac{4(13+4k)}{15(5+k)} \pa {\bf P_{-}^{(\frac{5}{2})}}
+\frac{1}{2} {\bf S_{-}^{(\frac{7}{2})}} 
+\frac{2}{(5+k)} i {\bf P_{-}^{(\frac{5}{2})}} \hat{A}_3
-\frac{2}{(5+k)} i {\bf P_{-}^{(\frac{5}{2})}} \hat{B}_3
\right. \nonu \\
&- & \frac{2(2457+3253k+560k^2)}{15(5+k)^2(19+23k)} 
\hat{A}_{-} \hat{A}_{+} \hat{G}_{12}
+\frac{4}{(5+k)^2} \hat{B}_{+} \hat{A}_{+} \hat{G}_{21} 
\nonu \\
& + & 
\frac{4}{(5+k)^2} \hat{B}_{+} \hat{A}_{+} T_{+}^{(\frac{3}{2})}
+\frac{8}{(5+k)^2} \hat{B}_{+} \hat{A}_3 \hat{G}_{11}
+\frac{8}{(5+k)^2} \hat{B}_{+} \hat{A}_3 U^{(\frac{3}{2})}
\nonu \\
&+ & \frac{2(8260 k^4-2464 k^3-1061761 k^2-2346650 k-1276557)}{
45(5+k)^2(19+23k)(47+35k)} 
\hat{B}_{+} \hat{B}_{-} \hat{G}_{12}
\nonu \\
&- &  \frac{2(2280 k^4-49129 k^3-965999 k^2-2008307 k-1125397)}{
45(5+k)^2(19+23k)(47+35k)} 
\hat{B}_{+} \hat{B}_{-}  T_{-}^{(\frac{3}{2})}
\nonu \\
& + & \frac{2(4+5k)}{5(5+k)^2} 
\hat{B}_{+} \hat{B}_3 \hat{G}_{11}
-\frac{2(54+k)}{15(5+k)^2} \hat{B}_{+} \hat{B}_3  U^{(\frac{3}{2})}
-\frac{4(3+5k)}{5(5+k)^2} i \hat{B}_{+} \pa \hat{G}_{11}
\nonu \\
&- & \frac{4(42+23k)}{15(5+k)^2} i \hat{B}_{+} \pa  U^{(\frac{3}{2})}  
\nonu \\
& - & \frac{2(8260 k^4-2464 k^3-1061761 k^2-2346650 k-1276557)}{
45(5+k)^2(19+23k)(47+35k)} 
\hat{B}_{-} \hat{B}_{+} \hat{G}_{12}
\nonu \\
&+ & \frac{2(2280 k^4-49129 k^3-1038449 k^2-2165447 k-1205767)}{
45(5+k)^2(19+23k)(47+35k)}  
\hat{B}_{-} \hat{B}_{+} T_{-}^{(\frac{3}{2})}
\nonu \\
 &+ & \frac{2}{(5+k)} i \hat{A}_{+}  V^{(\frac{5}{2})}   
+\frac{2(2457+3253k+560k^2)}{15(5+k)^2(19+23k)} \hat{A}_{+} 
\hat{A}_{-}  \hat{G}_{12} \nonu \\
& - &  \frac{4}{(5+k)^2} \hat{A}_{+} \hat{A}_{-} 
T_{-}^{(\frac{3}{2})}
+\frac{4(15+2k)}{5(5+k)^2} \hat{A}_{+} \hat{A}_3 \hat{G}_{22}
-\frac{2(89+16k)}{5(5+k)^2}  \hat{A}_{+} \hat{A}_3  V^{(\frac{3}{2})} 
\nonu \\
& + & \frac{8}{(5+k)^2} \hat{A}_{+} \hat{B}_3 \hat{G}_{22}
-\frac{8}{(5+k)^2} \hat{A}_{+} \hat{B}_3  V^{(\frac{3}{2})} 
-\frac{4(15+4k)}{15(5+k)^2} i \hat{A}_{+} \pa \hat{G}_{22}
\nonu \\
&- & \frac{4(1+4k)}{15(5+k)^2} i \hat{A}_{+} \pa  V^{(\frac{3}{2})}  
-\frac{4(15+2k)}{5(5+k)^2} \hat{A}_3 \hat{A}_{+} \hat{G}_{22}
+\frac{2(109+16k)}{5(5+k)^2}  \hat{A}_3 \hat{A}_{+}  V^{(\frac{3}{2})}
\nonu \\
&- & \frac{4}{(5+k)} i \hat{A}_3 W_{-}^{(\frac{5}{2})}
-\frac{4}{(5+k)^2} \hat{A}_3 \hat{A}_3   T_{-}^{(\frac{3}{2})}
+ \frac{8}{(5+k)^2} \hat{A}_3 \hat{B}_3   T_{-}^{(\frac{3}{2})}
\nonu \\
&+ & \frac{8(3+2k)(83+55k)}{
15(5+k)^2(19+23k)} i \hat{A}_3 \pa \hat{G}_{12} 
-\frac{4(47+16k)}{15(5+k)^2} i \hat{A}_3 \pa  T_{-}^{(\frac{3}{2})}
\nonu \\
& + & \frac{2(-6+k)}{15(5+k)^2}  \hat{B}_3 \hat{B}_{+}  U^{(\frac{3}{2})}
-  \frac{4}{(5+k)^2}  \hat{B}_3 \hat{B}_3 T_{-}^{(\frac{3}{2})}
-\frac{2(4+5k)}{5(5+k)^2} \hat{B}_3 \hat{B}_{+} \hat{G}_{11}
\nonu \\
& + &  \frac{8(3 k^3-13 k^2-454 k-418)}{
15(5+k)^2(19+23k)} i \hat{B}_3 \pa \hat{G}_{12}
\nonu \\ 
& + & \frac{4(80+19k)}{15(5+k)^2} i \hat{B}_3 \pa  T_{-}^{(\frac{3}{2})}
+\frac{4(-53-44k+k^2)}{
5(5+k)(19+23k)} T^{(1)} \pa \hat{G}_{12}
\nonu \\
& - & \frac{8(1272 + 2258 k + 1227 k^2 + 133 k^3)}{
3(5+k)(19+23k)(47+35k)} \hat{G}_{12} \hat{T}
+\frac{2}{(5+k)} T^{(1)} \pa T_{-}^{(\frac{3}{2})}
\nonu \\
& + &  \frac{4(7315 k^4+362 k^3-985702 k^2-2180078 k-1180113)}{
45(5+k)^2(19+23k)(47+35k)} i \hat{G}_{12} \pa \hat{B}_3 
\nonu \\
& - & \frac{6(-53-44k+k^2)}{5(5+k)(19+23k)} \hat{G}_{12} \pa T^{(1)}
\nonu \\
& - & \frac{4(70 k^4+3418 k^3+24403 k^2+24696 k+7497)}{
3(5+k)(3+7k)(19+23k)(47+35k)}  T_{-}^{(\frac{3}{2})} \hat{T}
\nonu \\
&+ &  \frac{2(231+68k)}{15(5+k)^2} i  T_{-}^{(\frac{3}{2})}  \pa \hat{A}_3 
+\frac{2}{(5+k)} T_{-}^{(\frac{3}{2})} T^{(2)}
\nonu \\
&- & \frac{2(4560 k^4-105503 k^3-2020162 k^2-4181791 k-2331164)}{
45(5+k)^2(19+23k)(47+35k)} i  T_{-}^{(\frac{3}{2})}  \pa \hat{B}_3 
\nonu \\
&- & \frac{3}{(5+k)} T_{-}^{(\frac{3}{2})}  \pa  T^{(1)} 
-\frac{2}{(5+k)} \hat{G}_{11} V_{-}^{(2)}
-\frac{2}{(5+k)}  U^{(\frac{3}{2})} V_{-}^{(2)} 
\nonu \\
& -&  \left. \frac{2}{(5+k)}  V^{(\frac{3}{2})} U_{-}^{(2)} 
-\frac{2(19+8k)}{15(5+k)} \pa W_{-}^{(\frac{5}{2})}
\right](w) +\cdots, 
\nonu \\
U_{-}^{(2)}(z) \, W^{(2)}(w) & = & 
\frac{1}{(z-w)^3} \, \left[
\frac{6(5+2k)}{(5+k)^2} i \hat{A}_{+} \right](w)
\nonu \\
& + & \frac{1}{(z-w)^2} \, \left[ 
\frac{2(4+k)}{(5+k)} U_{-}^{(2)} +
\frac{3(5+2k)}{(5+k)^2} i \pa \hat{A}_{+}
\right](w) 
\nonu \\
& + & \frac{1}{(z-w)} \, \left[ 
\frac{1}{2} {\bf Q_{-}^{(3)}}
+\frac{4(138+343k+113k^2)}{(5+k)(3+7k)(19+23k)} 
i \hat{A}_{+} \hat{T}
-\frac{2}{(5+k)} i \hat{A}_{+} T^{(2)}
\right. \nonu \\
&-& \frac{7}{(5+k)^2} \hat{A}_{+} \pa \hat{A}_3 
+ \frac{(8+3k)}{(5+k)^2} \hat{A}_{+} \pa \hat{B}_3
+\frac{1}{2(5+k)} i \hat{A}_{+} \pa T^{(1)}
\nonu \\
 &+ & \frac{4}{(5+k)} i \hat{B}_3 U_{-}^{(2)} 
-\frac{(8+3k)}{(5+k)^2} \hat{B}_3 \pa \hat{A}_{+}
+\frac{(9+2k)}{2(5+k)} \pa U_{-}^{(2)}
\nonu \\
& - & \frac{2}{(5+k)} \hat{G}_{11} \hat{G}_{12}
+\frac{2}{(5+k)} \hat{G}_{11}  T_{-}^{(\frac{3}{2})}
+\frac{2}{(5+k)} \hat{G}_{12} U^{(\frac{3}{2})}
\nonu \\
&- & \frac{2}{(5+k)}   T_{-}^{(\frac{3}{2})}  U^{(\frac{3}{2})}
+\frac{4}{(5+k)^2} i \hat{A}_{+} \hat{A}_{+} \hat{A}_{-}
-\frac{2(7+2k)}{(5+k)^2} i \hat{A}_{+} \hat{A}_3 \hat{A}_3
\nonu \\
&- & \frac{8}{(5+k)^2} i \hat{A}_{+} \hat{A}_3 \hat{B}_3 
+\frac{4}{(5+k)^2} i \hat{A}_{+} \hat{B}_{+} \hat{B}_{-}
+\frac{4}{(5+k)^2} i \hat{A}_{+} \hat{B}_3 \hat{B}_3
\nonu \\
&+ & \frac{2(9+2k)}{(5+k)^2} i \hat{A}_3 \hat{A}_{+} \hat{A}_3
-\frac{(15+4k)}{(5+k)^2} \hat{A}_3 \pa \hat{A}_{+} 
\nonu \\
&-& \left. \frac{1}{2(5+k)} T^{(1)} \hat{A}_{+} \hat{A}_3
+\frac{1}{2(5+k)} \hat{A}_3 T^{(1)} \hat{A}_{+}
\right](w) +\cdots,
\nonu \\ 
U_{-}^{(2)}(z) \, W_{+}^{(\frac{5}{2})}(w) & = & 
\frac{1}{(z-w)^3} \, \left[ -\frac{8(2+k)(6+k)}{3(5+k)^2} \hat{G}_{11}
+ \frac{4(11+3k)}{3(5+k)^2} U^{(\frac{3}{2})} \right](w)
\nonu \\
& + & \frac{1}{(z-w)^2} \, \left[ 
-\frac{(27+8k)}{3(5+k)} {\bf Q^{(\frac{5}{2})}}
-\frac{(35+8k)}{3(5+k)} U^{(\frac{5}{2})}
-\frac{6}{(5+k)^2} i \hat{A}_{+} \hat{G}_{21}
\right. \nonu \\
&- & \frac{2(13+4k)}{(5+k)^2} i \hat{A}_{+} T_{+}^{(\frac{3}{2})}  
-\frac{4(5+2k)}{(5+k)^2} i \hat{A}_3  U^{(\frac{3}{2})} 
- \frac{2(3+2k)}{(5+k)^2} i \hat{B}_{-} T_{-}^{(\frac{3}{2})}  
\nonu \\
&-&  \frac{4k}{(5+k)^2} i \hat{B}_3  U^{(\frac{3}{2})} 
+\frac{2}{(5+k)} T^{(1)}  U^{(\frac{3}{2})} 
-\frac{4(8+k)(3+2k)}{9(5+k)^2} \pa \hat{G}_{11}
\nonu \\
& + & \left.  \frac{8(13+3k)}{9(5+k)^2} \pa  U^{(\frac{3}{2})}
\right](w) 
\nonu \\ 
& + & 
\frac{1}{(z-w)} \, \left[ 
-\frac{(1052+331k+18k^2)}{15(5+k)(14+k)} \pa {\bf Q^{\frac{5}{2}} }
+\frac{1}{(5+k)} i {\bf P_{-}^{(\frac{5}{2})}} \hat{B}_{-}
\right. \nonu \\
& + & \frac{1}{2}  {\bf Q^{(\frac{7}{2})}} 
-\frac{(20+k)}{10(14+k)} {\bf P^{(2)}}   U^{(\frac{3}{2})} 
+ \frac{(20+k)}{10(14+k)}  U^{(\frac{3}{2})}   {\bf P^{(2)}}
\nonu \\
&- & \frac{18}{5(5+k)^2} \hat{A}_{-} \hat{A}_{+} \hat{G}_{11} 
+ \frac{(-746-5k+4k^2)}{15(5+k)^2(14+k)} 
\hat{A}_{-} \hat{A}_{+}  U^{(\frac{3}{2})}
\nonu \\
&-&  \frac{2(12+k)}{5(5+k)^2} \hat{B}_{+} \hat{B}_{-} \hat{G}_{11} 
+ \frac{5(2+k)}{3(5+k)^2} 
\hat{B}_{+} \hat{B}_{-}  U^{(\frac{3}{2})}
-\frac{4}{(5+k)^2} \hat{B}_{-} \hat{A}_{+} \hat{G}_{22} 
\nonu \\
& + &  \frac{4}{(5+k)^2} 
\hat{B}_{-} \hat{A}_{+}  V^{(\frac{3}{2})}
-\frac{2}{(5+k)} i  \hat{B}_{-}  W_{-}^{(\frac{5}{2})}
-\frac{8}{(5+k)^2} \hat{B}_{-} \hat{A}_3 \hat{G}_{12} 
\nonu \\
& + & \frac{8}{(5+k)^2} \hat{B}_{-} \hat{A}_3  T_{-}^{(\frac{3}{2})}  
+ \frac{2(12+k)}{5(5+k)^2} \hat{B}_{-} \hat{B}_{+} \hat{G}_{11} 
 -   \frac{(-2+5k)}{3(5+k)^2} 
\hat{B}_{-} \hat{B}_{+}  U^{(\frac{3}{2})}
\nonu \\
& + & \frac{4(268+79k+4k^2)}{15(5+k)^2(14+k)} 
i \hat{B}_{-} \pa \hat{G}_{12}
-\frac{8(309+122k+7k^2)}{15(5+k)^2(14+k)} 
i \hat{B}_{-} \pa T_{-}^{(\frac{3}{2})}
\nonu \\
& - & \frac{2}{(5+k)} i  \hat{A}_{+}  W_{+}^{(\frac{5}{2})}
+ \frac{18}{5(5+k)^2} \hat{A}_{+} \hat{A}_{-} \hat{G}_{11} 
 -   \frac{(-2426-125k+4k^2)}{15(5+k)^2(14+k)} 
\hat{A}_{+} \hat{A}_{-}  U^{(\frac{3}{2})}
\nonu \\
&- &  \frac{2(1726+607k+34k^2)}{
15(5+k)^2(14+k)} \hat{A}_{+} \hat{A}_3  T_{+}^{(\frac{3}{2})} 
- \frac{4(13+418k+37k^2)}{15(5+k)^2(19+23k)} 
i \hat{A}_{+} \pa \hat{G}_{21}
\nonu \\
& - & \frac{4(548+211k+12k^2)}{15(5+k)^2(14+k)} 
i \hat{A}_{+} \pa  T_{+}^{(\frac{3}{2})} 
+ \frac{2(1726+607k+34k^2)}{
15(5+k)^2(14+k)} \hat{A}_3 \hat{A}_{+}  T_{+}^{(\frac{3}{2})} 
\nonu \\
&+&  \frac{4}{(5+k)^2} 
\hat{A}_3 \hat{A}_3  U^{(\frac{3}{2})}
- \frac{8}{(5+k)^2} 
\hat{A}_3 \hat{B}_3  U^{(\frac{3}{2})}
+\frac{24}{5(5+k)^2} 
i \hat{A}_3 \pa \hat{G}_{11}
\nonu \\
& + &  \frac{4}{(5+k)} i  \hat{B}_3  U^{(\frac{5}{2})}
+ \frac{4}{(5+k)^2} 
\hat{B}_3 \hat{B}_3  U^{(\frac{3}{2})}
-\frac{8(12+k)}{15(5+k)^2} 
i \hat{B}_3 \pa \hat{G}_{11}
\nonu \\
& + & \frac{4}{5(5+k)} 
T^{(1)} \pa \hat{G}_{11}
+ \frac{4(25+2k)}{5(5+k)(14+k)} 
T^{(1)} \pa  U^{(\frac{3}{2})} 
\nonu \\
&+& \frac{2(-272+73k+37k^2)}{5(5+k)^2(19+23k)} 
i \pa \hat{A}_{+}  \hat{G}_{21}
- \frac{2(-536+52k+7k^2)}{15(5+k)^2(14+k)} 
i \pa \hat{B}_{-}  \hat{G}_{12}
\nonu \\
&+&  \frac{2(-116+82k+7k^2)}{15(5+k)^2(14+k)} 
 i \pa \hat{B}_{-}  T_{-}^{(\frac{3}{2})} 
-\frac{6}{5(5+k)} \pa T^{(1)} \hat{G}_{11}
\nonu \\
& - & \frac{(110+7k)}{5(5+k)(14+k)} \pa T^{(1)}   U^{(\frac{3}{2})}  
\nonu \\
& - & \frac{(5980 k^4+84185 k^3+218876 k^2+222421 k+103398)}{
45(5+k)^2(19+23k)(47+35k)} \pa^2 \hat{G}_{11}
\nonu \\
&+&  \frac{(140 k^4+3843 k^3+21886 k^2-28299 k-93018)}{
45(5+k)^2(14+k)(47+35k)} \pa^2 U^{(\frac{3}{2})}  
\nonu \\
&- & \frac{2(744+197k+10k^2)}{15(5+k)(14+k)} \pa U^{(\frac{5}{2})} 
-\frac{8(659+245k+14k^2)}{
15(5+k)^2(14+k)} i \hat{A}_3 \pa  U^{(\frac{3}{2})}
\nonu \\
& + & \frac{2}{(5+k)} \hat{G}_{21} U_{-}^{(2)}
+ \frac{2}{(5+k)}  T_{+}^{(\frac{3}{2})}   U_{-}^{(2)}
-\frac{2}{(5+k)} \hat{G}_{12} U_{+}^{(2)}
\nonu \\
&+& \frac{2}{(5+k)}  T_{-}^{(\frac{3}{2})}   U_{+}^{(2)}
-\frac{4(573+1542k+853k^2+92k^3)}{(5+k)(19+23k)(47+35k)} 
\hat{G}_{11} \hat{T} 
\nonu \\
&+& \left. \frac{4(171+333k+70k^2)}{
(5+k)(3+7k)(47+35k)}  U^{(\frac{3}{2})} \hat{T}
-\frac{2}{(5+k)}  U^{(\frac{3}{2})} T^{(2)}
\right](w) + \cdots,
\label{u-2w5half+} \\
U_{-}^{(2)}(z) \, W_{-}^{(\frac{5}{2})}(w) & = & 
\frac{1}{(z-w)^2} \, \left[ -\frac{2(13+2k)}{3(5+k)^2} i \hat{A}_{+} 
\hat{G}_{12}  +\frac{2(13+2k)}{3(5+k)^2} i \hat{A}_{+} T_{-}^{(\frac{3}{2})} 
\right](w)
\nonu \\
& + & 
\frac{1}{(z-w)} \, \left[ 
\frac{(27+10k)}{3(5+k)^2}
\hat{B}_{+} \hat{A}_{+} \hat{G}_{11}
+\frac{8(4+k)}{3(5+k)^2} \hat{B}_{+} \hat{A}_{+}  U^{(\frac{3}{2})}
\right. \nonu \\
&- & \frac{8(2+k)}{3(5+k)^2} \hat{A}_{+} \hat{A}_3
T_{-}^{(\frac{3}{2})} 
+\frac{2(7+2k)}{3(5+k)^2} \hat{A}_3 \hat{A}_{+} \hat{G}_{12}
+\frac{8(2+k)}{3(5+k)^2} \hat{A}_3 \hat{A}_{+} T_{-}^{(\frac{3}{2})} 
\nonu \\
& -& \frac{(27+10k)}{3(5+k)^2} \hat{G}_{11} \hat{A}_{+} \hat{B}_{+}
-\frac{8(4+k)}{3(5+k)^2}  U^{(\frac{3}{2})} \hat{A}_{+} \hat{B}_{+}
+\frac{2}{(5+k)} i \hat{A}_{+} W_{-}^{(\frac{5}{2})}
\nonu \\
&-& \left. \frac{2}{(5+k)} \hat{G}_{12} U_{-}^{(2)} 
-\frac{2(7+2k)}{3(5+k)^2} \hat{G}_{12} \hat{A}_{+} \hat{A}_3
\right](w) +\cdots,
\nonu \\
U_{-}^{(2)}(z) \, W^{(3)}(w) & = & 
\frac{1}{(z-w)^4} \, 
\left[ \frac{12(289+1049k+636k^2+96k^3)}{(5+k)^3(19+23k)} 
i \hat{A}_{+} \right] (w)
\nonu \\
& + & \frac{1}{(z-w)^3} \, \left[ \frac{2(-127+85k+72k^2)}{
(5+k)^2(19+23k)} U_{-}^{(2)} -\frac{12(5+2k)}{(5+k)^3} 
\hat{A}_{+} \hat{A}_3  
\right. \nonu \\
& + & \left. \frac{44k}{(5+k)^3} \hat{A}_{+} \hat{B}_3
+\frac{6(5+2k)}{(5+k)^3} i \pa \hat{A}_{+}
+\frac{6}{(5+k)^2} i T^{(1)} \hat{A}_{+}
\right](w) 
\nonu \\
& + & \frac{1}{(z-w)^2} \left[ 
\frac{(23+6k)}{2(5+k)} {\bf Q_{-}^{(3)}}
-\frac{(-297-37k+16k^2)}{(5+k)(19+23k)} T^{(1)} U_{-}^{(2)}
\right. \nonu \\
& + & \frac{(37+6k)}{2(5+k)^2} i T^{(1)} \pa \hat{A}_{+}
+\frac{4(2406+7168k+4613k^2+831k^3)}{(5+k)^2(3+7k)(19+23k)} 
i \hat{A}_{+} \hat{T}
\nonu \\
& - & \frac{2(12+5k)}{(5+k)^2} i \hat{A}_{+} T^{(2)}
-\frac{2(19+5k)}{(5+k)^2} i \hat{A}_{+} W^{(2)}
-\frac{(127+54k)}{(5+k)^3} \hat{A}_{+} \pa \hat{A}_3
\nonu \\
& + & \frac{(200+79k+6k^2)}{(5+k)^3} \hat{A}_{+} \pa \hat{B}_3
-\frac{(23+6k)}{2(5+k)^2} i \hat{A}_{+} \pa T^{(1)}
\nonu \\
& - & \frac{2(291+629k+118k^2)}{(5+k)^2(19+23k)} i
\hat{A}_3 U_{-}^{(2)}
-\frac{(7+6k)}{(5+k)^3} \hat{A}_3 \pa \hat{A}_{+}
\nonu \\
& + &
 \frac{2(684+523k+35k^2)}{(5+k)^2(19+23k)} i
\hat{B}_3 U_{-}^{(2)}
\nonu \\
&- &  \frac{(204+65k+6k^2)}{(5+k)^3} \hat{B}_3 \pa \hat{A}_{+}
+\frac{(1399+2589k+650k^2)}{2(5+k)^2(19+23k)}
\pa  U_{-}^{(2)}
\nonu \\
&- & \frac{4(9+2k)}{(5+k)^2} \hat{G}_{11} \hat{G}_{12} 
+\frac{6(7+2k)}{(5+k)^2} \hat{G}_{11}  T_{-}^{(\frac{3}{2})}
+\frac{16(4+k)}{(5+k)^3} i \pa^2 \hat{A}_{+}
\nonu \\
& + & 
\frac{2(15+2k)}{(5+k)^2} \hat{G}_{12}  U^{(\frac{3}{2})}
-\frac{2(23+6k)}{(5+k)^2}  T_{-}^{(\frac{3}{2})}  U^{(\frac{3}{2})}
\nonu \\
&+& \frac{4(13+6k)}{(5+k)^3} i \hat{A}_{+} \hat{A}_{+} \hat{A}_{-}
-\frac{8(13+4k)}{(5+k)^3} i \hat{A}_{+} \hat{A}_3 \hat{B}_3
+ \frac{24(4+k)}{(5+k)^3} i \hat{A}_{+} \hat{B}_{+} \hat{B}_{-}
\nonu \\
& + & \frac{4(19+2k)}{(5+k)^3}  i \hat{A}_{+} \hat{B}_3 \hat{B}_3
+\frac{4(7+6k)}{(5+k)^3}  i \hat{A}_3 \hat{A}_{+} \hat{A}_3
\nonu \\
&- & \left. \frac{8}{(5+k)^2} T^{(1)}   \hat{A}_{+} \hat{A}_3 
+\frac{8}{(5+k)^2} T^{(1)}  \hat{A}_{+} \hat{B}_3
\right](w)
\nonu \\
& + & \frac{1}{(z-w)} \, \left[ 
\frac{2}{(5+k)} i \hat{A}_3 {\bf Q_{-}^{(3)}}
-\frac{2}{(5+k)} i \hat{B}_3  {\bf Q_{-}^{(3)}}
+\frac{(5+2k)}{2(5+k)} \pa  {\bf Q_{-}^{(3)}}
\right.
\nonu \\
& + & 
\frac{2}{(5+k)}   U_{-}^{(2)} W^{(2)} +\frac{5(5+2k)}{2(5+k)^2} \pa^2
 U_{-}^{(2)} + \frac{2}{(5+k)} T^{(2)}  U_{-}^{(2)}
 \nonu \\
& + & \frac{2}{(5+k)} \pa  T_{-}^{(\frac{3}{2})}  U^{(\frac{3}{2})}
-\frac{2(5+2k)}{(5+k)^2}  T_{-}^{(\frac{3}{2})}  \pa U^{(\frac{3}{2})}
-\frac{2(5+2k)}{(5+k)^2} i \pa \hat{A}_3   U_{-}^{(2)}
\nonu \\
&-& \frac{24(1+3k)}{(5+k)(19+23k)} i \hat{A}_3  \pa  U_{-}^{(2)}
-\frac{8}{(5+k)^2} i \hat{A}_3   T_{-}^{(\frac{3}{2})}  U^{(\frac{3}{2})}
-\frac{8}{(5+k)^2} \hat{A}_3 \hat{A}_3   U_{-}^{(2)} 
\nonu \\
&+ & \frac{8}{(5+k)^2} i \hat{A}_3 \hat{G}_{11}  T_{-}^{(\frac{3}{2})} 
-\frac{8}{(5+k)^2} i \hat{A}_3 \hat{G}_{11} \hat{G}_{12}
+\frac{8}{(5+k)^2} i \hat{A}_3 \hat{G}_{12}  U^{(\frac{3}{2})}
\nonu \\
&-& \frac{2(10+3k)}{(5+k)^2} i \pa \hat{A}_{+} W^{(2)}
-  \frac{2(3+k)}{(5+k)^2} i  \hat{A}_{+} \pa W^{(2)}
-\frac{2(7+3k)}{(5+k)^2}  i \pa \hat{A}_{+} T^{(2)}
\nonu \\
&-& \frac{2k}{(5+k)^2}  i  \hat{A}_{+} \pa T^{(2)}
-\frac{(27801+84372k+67849k^2+23426k^3+
3108k^4)}{6(5+k)^3(3+7k)(19+23k)} i \pa^3 \hat{A}_{+}
\nonu \\
&+& \frac{8}{(5+k)^2} \hat{A}_{+} \hat{A}_3  T^{(2)}
+\frac{(5349+14788k+8801k^2+2514k^3)}{(5+k)^3(3+7k)(19+23k)} 
\pa^2 \hat{A}_{+} \hat{A}_3
\nonu \\
&-& \frac{2(7+10k)}{(5+k)^3} \pa \hat{A}_{+} \pa \hat{A}_3
-\frac{(-3525-6956k+1199k^2+1350k^3)}{
(5+k)^3(3+7k)(19+23k)} \hat{A}_{+} \pa^2 \hat{A}_3 
\nonu \\
&+& \frac{8(-2+k)}{(5+k)^3} i \pa \hat{A}_{+} \hat{A}_3 \hat{A}_3
+\frac{4(3+4k)}{(5+k)^3} i \hat{A}_{+} \hat{A}_3 \pa \hat{A}_3
-\frac{16}{(5+k)^3} \hat{A}_{+} \hat{A}_3 \hat{A}_3 \hat{A}_3
\nonu \\
&+& \frac{48}{(5+k)^3} \hat{A}_{+} \hat{A}_3 \hat{A}_3 \hat{B}_3
-\frac{12(3+k)}{(5+k)^3} i \pa \hat{A}_{+} \hat{A}_3 \hat{B}_3
-\frac{4(3+4k)}{(5+k)^3} i \hat{A}_{+} \pa \hat{A}_{3} \hat{B}_3 
\nonu \\
&- & \frac{4(2+k)}{(5+k)^3} i \hat{A}_{+} \hat{A}_3 \pa \hat{B}_3  
-\frac{12(3+k)}{(5+k)^3} i \hat{A}_{+} \hat{A}_3 \hat{B}_3 \hat{B}_3
-\frac{16}{(5+k)^3} \hat{A}_{+} \hat{A}_3 \hat{B}_{+} \hat{B}_{-}
\nonu \\
&-& \frac{6}{(5+k)^2} \pa \hat{A}_{+} \hat{A}_3 T^{(1)}
-\frac{2}{(5+k)^2} \hat{A}_{+} \hat{A}_3 \pa T^{(1)} 
-\frac{8}{(5+k)^2} \hat{A}_{+} \hat{A}_{-}  U_{-}^{(2)}
\nonu \\
& + & \frac{16(2+k)}{(5+k)^3} i \pa \hat{A}_{+} \hat{A}_{+} \hat{A}_{-} 
+\frac{4(1+2k)}{(5+k)^3} i \hat{A}_{+} \hat{A}_{+} \pa \hat{A}_{-}
-\frac{16}{(5+k)^3}  \hat{A}_{+} \hat{A}_{+} \hat{A}_{-} \hat{A}_3
\nonu \\
&+& \frac{16}{(5+k)^3}  \hat{A}_{+} \hat{A}_{+} \hat{A}_{-} \hat{B}_3
-\frac{8}{(5+k)^2}  \hat{A}_{+} \hat{B}_{3} T^{(2)}
\nonu \\
&-& \frac{(27162+86205k+66826k^2+15677k^3+966k^4)}{
3(5+k)^3(3+7k)(19+23k)} \pa^2 \hat{A}_{+} \hat{B}_3
\nonu \\
&+& \frac{2(84+17k)}{3(5+k)^3} \pa \hat{A}_{+} \pa \hat{B}_3
\nonu \\
& + & \frac{(-12456-26223k+3272k^2+8965k^3+966k^4)}{
3(5+k)^3(3+7k)(19+23k)} \hat{A}_{+} \pa^2 \hat{B}_3
\nonu \\
&+& \frac{4(13+k)}{(5+k)^3} i \pa \hat{A}_{+} \hat{B}_3 \hat{B}_3
+\frac{4(2+k)}{(5+k)^3} i \hat{A}_{+} \hat{B}_3 \pa \hat{B}_3
+\frac{16}{(5+k)^3} \hat{A}_{+} \hat{B}_3 \hat{B}_3 \hat{B}_3
\nonu \\
&+& \frac{6}{(5+k)^2} \pa \hat{A}_{+} \hat{B}_3 T^{(1)}
+  \frac{2}{(5+k)^2}  \hat{A}_{+} \hat{B}_3 \pa T^{(1)}
+\frac{8}{(5+k)^2} i \pa \hat{A}_{+} \hat{B}_{+} \hat{B}_{-}
\nonu \\
&+ & \frac{4(-1+2k)}{(5+k)^3} i    \hat{A}_{+} \pa \hat{B}_{+} \hat{B}_{-}
+ \frac{4(11+2k)}{(5+k)^3} i \hat{A}_{+} \hat{B}_{+} \pa \hat{B}_{-}
+\frac{16}{(5+k)^3}  \hat{A}_{+} \hat{B}_{+} \hat{B}_{-} \hat{B}_3
\nonu \\
&- & \frac{4}{(5+k)^2} i \hat{A}_{+} \hat{G}_{12} \hat{G}_{21} 
-\frac{4}{(5+k)^2} i \hat{A}_{+} \hat{G}_{21}  T_{-}^{(\frac{3}{2})}
+\frac{(11+2k)}{2(5+k)^2} i \pa^2 \hat{A}_{+} T^{(1)}
\nonu \\
&- & \frac{5}{(5+k)^2} i \pa \hat{A}_{+} \pa T^{(1)}
- \frac{(-1+2k)}{2(5+k)^2} i  \hat{A}_{+} \pa^2 T^{(1)}
+\frac{2(2+3k)}{(5+k)^2} i \pa \hat{B}_3  U_{-}^{(2)}
\nonu \\
&- & \frac{2(-133+20k+17k^2)}{(5+k)^2(19+23k)} i  \hat{B}_3  \pa U_{-}^{(2)}
+\frac{8}{(5+k)^2} i \hat{B}_3  T_{-}^{(\frac{3}{2})}  U^{(\frac{3}{2})}
\nonu \\
& + & \frac{8}{(5+k)^2} \hat{B}_3 \hat{B}_3   U_{-}^{(2)}
-\frac{8}{(5+k)^2} i \hat{B}_3 \hat{G}_{11}  T_{-}^{(\frac{3}{2})} 
+\frac{8}{(5+k)^2} i \hat{B}_3 \hat{G}_{11} \hat{G}_{12}
\nonu \\
&-& \frac{8}{(5+k)^2} i \hat{B}_3 \hat{G}_{12}  U^{(\frac{3}{2})}
-
\frac{4}{(5+k)^2} \hat{B}_{+} \hat{B}_{-}  U_{-}^{(2)}
+\frac{2(5+2k)}{(5+k)^2} \pa \hat{G}_{11}  T_{-}^{(\frac{3}{2})}
\nonu \\
&+& \frac{2(3+2k)}{(5+k)^2}  \hat{G}_{11}  \pa T_{-}^{(\frac{3}{2})}
-\frac{2(9+2k)}{3(5+k)^2} \pa \hat{G}_{11} \hat{G}_{12}
-\frac{2(3+2k)}{(5+k)^2} \hat{G}_{11} \pa \hat{G}_{12}
\nonu \\
&+& \frac{2}{(5+k)} \hat{G}_{12}  U^{(\frac{5}{2})}
+ \frac{2(5+2k)}{(5+k)^2} \pa \hat{G}_{12}  U^{(\frac{3}{2})}
-\frac{2(1+2k)}{3(5+k)^2}  \hat{G}_{12}  \pa U^{(\frac{3}{2})}
\nonu \\
&+ & \frac{1}{(5+k)} \pa T^{(1)}   U_{-}^{(2)}
-\frac{(-139-7k+8k^2)}{(5+k)(19+23k)} T^{(1)} \pa  U_{-}^{(2)}
\nonu \\
& - & 
\frac{4(96+251k+127k^2)}{(5+k)(3+7k)(19+23k)} \hat{T}  U_{-}^{(2)}
+\frac{4(690+2276k+1653k^2+295k^3)}{(5+k)^2(3+7k)(19+23k)} i
\pa \hat{T} \hat{A}_{+}
\nonu \\
&+ & \frac{4(474+1301k+924k^2+241k^3)}{(5+k)^2(3+7k)(19+23k)} 
i \hat{T} \pa \hat{A}_{+}
\nonu \\
& - & \frac{16(138+343k+113k^2)}{
(5+k)^2(3+7k)(19+23k)} \hat{T} \hat{A}_{+} \hat{A}_3
\nonu \\
&+& \left. \frac{16(138+343k+113k^2)}{
(5+k)^2(3+7k)(19+23k)} \hat{T} \hat{A}_{+} \hat{B}_3
\right](w) + \cdots.
\label{spope}
\eea
In (\ref{u-2w5half+}), 
the fourth and fifth terms in the first order pole
have relative minus sign.
It is easy to calculate the following OPE
\bea
 U^{(\frac{3}{2})}(z) \, {\bf P^{(2)}}(w) & = &
\frac{1}{(z-w)^2} \,
\left[\frac{16(2+k)(8+k)}{3(5+k)^2} \, U^{(\frac{3}{2})} \right](w)
\nonu \\
&+& \frac{1}{(z-w)} \,
\left[ 
\frac{4(19+2k)}{3(5+k)}  U^{(\frac{5}{2})}
+\frac{8(19+2k)}{3(5+k)^2} i \hat{A}_{+}  T_{+}^{(\frac{3}{2})}
\right. \nonu \\
&+& \frac{8(47+4k)}{3(5+k)^2} i \hat{A}_3  U^{(\frac{3}{2})}
-\frac{8}{3(5+k)} i \hat{B}_{-} \hat{G}_{12} +\frac{8}{3(5+k)} i
\hat{B}_{-}  T_{-}^{(\frac{3}{2})}
\nonu \\
&-& 
\frac{8(14+k)}{3(5+k)^2} i \hat{B}_3  U^{(\frac{3}{2})}
+\frac{4}{(5+k)} T^{(1)}  U^{(\frac{3}{2})}
+\frac{8(-20+15k+2k^2)}{9(5+k)^2} \pa  U^{(\frac{3}{2})}
\nonu \\
&+& \left. \frac{2(19+2k)}{3(5+k)} {\bf Q^{(\frac{5}{2})}}
\right](w) +
\cdots.
\label{u3halfp2}
\eea
From this OPE (\ref{u3halfp2}),
one obtains the following identity (See also the reference $[63]$ 
of \cite{Ahn1311})
\bea
[ U^{(\frac{3}{2})}, {\bf P^{(2)}}](w) = -\frac{1}{2} \pa^2 
\{ U^{(\frac{3}{2})} \, 
 {\bf P^{(2)}} \}_{-2}(w) + \pa \{ U^{(\frac{3}{2})} \, 
 {\bf P^{(2)}} \}_{-1}(w).
\nonu
\eea
Then 
the fourth and fifth terms in the first order pole of (\ref{u-2w5half+})
can be written in terms of derivatives of the composite fields appearing in the 
singular terms in (\ref{u3halfp2}).

The fourth OPE of (\ref{spope}) has $\hat{G}_{12} \hat{G}_{12}(w)$
which can be written in terms of a derivative of $\hat{A}_{+} \hat{B}_{+}(w)$.
The first order pole in the last OPE of (\ref{spope})
contains composite field with spin-$4$ with $U(1)$ charge of 
$-\frac{6}{(5+k)}$.  
The second order pole contains a composite field of spin-$3$ which 
can be seen from the Table $5$ of \cite{Ahn1311}.

\section{
The nontrivial 
OPEs between  higher spin-$\frac{5}{2}$ current, $U^{(\frac{5}{2})}(z)$,  
and 
other $9$ higher spin currents}

We continue to present the corresponding OPEs as follows:
\bea
U^{(\frac{5}{2})}(z) \, U^{(\frac{5}{2})}(w) & = &
\frac{1}{(z-w)^3} \, 
\left[ \frac{16(-3+13k+4k^2)}{9(5+k)^3} \hat{A}_{+} 
\hat{B}_{-} \right](w) \nonu \\
& - & \frac{1}{(z-w)^2} \,
\left[ \frac{4(-3+13k+4k^2)}{9(5+k)^2} \hat{G}_{11} \hat{G}_{11} \right] (w)
\nonu \\
& + & 
\frac{1}{(z-w)} \left[  
\frac{2}{(5+k)} i \hat{B}_{-} {\bf Q_{-}^{(3)}}
-\frac{4}{(5+k)} U^{(\frac{3}{2})} U^{(\frac{5}{2})}
+ \frac{4}{(5+k)} U_{+}^{(2)} U_{-}^{(2)} 
\right.
\nonu \\
& - & \frac{4(7+4k)}{3(5+k)^2} i \pa \hat{A}_{+}  U_{+}^{(2)}
+ \frac{8(1+k)}{3(5+k)^2} i  \hat{A}_{+}  \pa U_{+}^{(2)}
-\frac{16}{(5+k)^3} \hat{A}_{+} \hat{A}_3 \hat{A}_3 \hat{B}_{-}
\nonu \\
&+& \frac{12}{(5+k)^3} i \pa \hat{A}_{+} \hat{A}_3 \hat{B}_{-}
-\frac{28}{(5+k)^3} i  \hat{A}_{+} \pa \hat{A}_3 \hat{B}_{-}
+ \frac{32}{(5+k)^3} \hat{A}_{+}  \hat{A}_{3}  \hat{B}_{-} \hat{B}_{3}
\nonu \\
&- & \frac{16}{(5+k)^3} \hat{A}_{+}  \hat{A}_{+}  \hat{A}_{-} \hat{B}_{-}
+  \frac{8}{(5+k)^2} \hat{A}_{+}  \hat{B}_{-}  W^{(2)}
 + 
\frac{16}{(5+k)^2} \hat{A}_3 \hat{B}_{-}  U_{-}^{(2)} 
\nonu \\
&+& \frac{2(3121+2929k+848k^2+92k^3)}{3(5+k)^3(19+23k)} 
\pa^2 \hat{A}_{+} \hat{B}_{-}
-\frac{4(3+18k+2k^2)}{3(5+k)^3} \pa \hat{A}_{+} \pa \hat{B}_{-}
\nonu \\
&+& \frac{4(1798+1771k+447k^2+46k^3)}{
3(5+k)^3(19+23k)} \hat{A}_{+} \pa^2 \hat{B}_{-}
+\frac{4(-8+k)}{(5+k)^3} i \pa \hat{A}_{+} \hat{B}_{-} \hat{B}_3
\nonu \\
&- & \frac{4k}{(5+k)^3} i \hat{A}_{+} \hat{B}_{-} \pa \hat{B}_3
-\frac{16}{(5+k)^3} \hat{A}_{+} \hat{B}_{-} \hat{B}_3 \hat{B}_3 
-\frac{6}{(5+k)^2} \pa \hat{A}_{+} \hat{B}_{-} T^{(1)}
\nonu \\
&+ & \frac{6}{(5+k)^2} \hat{A}_{+} \hat{B}_{-} \pa T^{(1)}  
-\frac{16}{(5+k)^3} \hat{A}_{+} \hat{B}_{+} \hat{B}_{-} \hat{B}_{-}  
-\frac{52}{3(5+k)^2} i \pa \hat{B}_{-}   U_{-}^{(2)} 
\nonu \\
&+& \frac{38}{3(5+k)^2} i  \hat{B}_{-}  \pa  U_{-}^{(2)} 
-\frac{8}{(5+k)^2} i 
\hat{B}_{-}  T_{-}^{(\frac{3}{2})}  U^{(\frac{3}{2})}
-\frac{16}{(5+k)^2} \hat{B}_{-} \hat{B}_3  U_{-}^{(2)} 
\nonu \\
&+ & \frac{8}{(5+k)^2} i \hat{B}_{-} \hat{G}_{11}  T_{-}^{(\frac{3}{2})} 
-\frac{8}{(5+k)^2} i \hat{B}_{-} \hat{G}_{11} \hat{G}_{12}
+\frac{8}{(5+k)^2} i \hat{B}_{-} \hat{G}_{12}  U^{(\frac{3}{2})}
\nonu \\
&+ & \frac{8(3+2k)}{3(5+k)^2} \pa \hat{G}_{11}   U^{(\frac{3}{2})} 
- \frac{8(3+2k)}{3(5+k)^2}  \hat{G}_{11}  \pa  U^{(\frac{3}{2})} 
-\frac{32(23+13k)}{(5+k)^2(19+23k)} \hat{T} \hat{A}_{+} \hat{B}_{-}
\nonu \\
&+& \left. \frac{8}{(5+k)^2} \hat{G}_{11} \pa \hat{G}_{11} 
-\frac{16(2+k)}{3(5+k)^2}  U^{(\frac{3}{2})}   \pa  U^{(\frac{3}{2})}  
\right](w) + \cdots,
\nonu \\
U^{(\frac{5}{2})}(z) \, V^{(\frac{3}{2})}(w) & = & 
\frac{1}{(z-w)^3} \, \left[ \frac{8(5+2k)}{(5+k)^2} i \hat{A}_3
+\frac{8k(-2+k)}{3(5+k)^2} i \hat{B}_3 -\frac{4(-2+k)}{3(5+k)}
T^{(1)}  \right](w)
\nonu \\
& + & \frac{1}{(z-w)^2} \, \left[ \frac{1}{2} {\bf P^{(2)}} 
+ \frac{4(15+23k+16k^2)}{
3(5+k)(3+7k)} \hat{T} -\frac{8(2+k)}{3(5+k)} T^{(2)}
-\frac{4(5+2k)}{3(5+k)} W^{(2)}  
\right. \nonu \\
&+ & \frac{16(-1+k)}{3(5+k)^2} \hat{A}_1 \hat{A}_1
+ \frac{16(-1+k)}{3(5+k)^2} \hat{A}_2 \hat{A}_2
+ \frac{4(-13+4k)}{3(5+k)^2} \hat{A}_3 \hat{A}_3
\nonu \\
&- & \frac{8(-4+k)}{3(5+k)^2} \hat{A}_3 \hat{B}_3
+\frac{4(5+k)}{3(5+k)^2} \hat{B}_1 \hat{B}_1
+ \frac{4(5+k)}{3(5+k)^2} \hat{B}_2 \hat{B}_2
-\frac{4(-5+2k)}{3(5+k)^2} \hat{B}_3 \hat{B}_3
\nonu \\
& + & \frac{8(5+2k)}{(5+k)^2} i \pa \hat{A}_3
+  \frac{8(-2+k)}{3(5+k)^2} i \pa \hat{B}_3
\nonu \\
&-& \left.  \frac{4(-2+k)}{3(5+k)} i \pa T^{(1)}
+  \frac{4}{(5+k)} i T^{(1)} \hat{A}_3
- \frac{4}{(5+k)} i T^{(1)} \hat{B}_3
\right](w)
\nonu \\
& + & \frac{1}{(z-w)} \, \left[ 
\frac{3}{8} \pa {\bf P^{(2)}}
- W^{(3)}
-\frac{4(-3+k)}{(19+23k)} T^{(1)} \hat{T}
+\frac{4}{(5+k)} i T^{(1)} \pa \hat{A}_3
\right. \nonu \\
& + & \frac{(-5+4k)}{(5+k)^2} \hat{A}_{+} \pa \hat{A}_{-} 
+ \frac{(-1+4k)}{(5+k)^2} \hat{A}_{-} \pa \hat{A}_{+}
+\frac{8(75+178k+71k^2)}{(5+k)(3+7k)(19+23k)} 
i  \hat{A}_3 \hat{T}
\nonu \\
& - & \frac{4}{(5+k)} i \hat{A}_3 T^{(2)}
+ \frac{8(-4+k)}{(5+k)^2} \hat{A}_3 \pa \hat{A}_3
+\frac{2}{(5+k)} i \hat{A}_3 \pa T^{(1)} 
\nonu \\
&+ & \frac{2}{(5+k)} i \hat{B}_{+} U_{+}^{(2)}
-\frac{2(-9-5k+2k^2)}{3(5+k)^2} \hat{B}_{+} \pa \hat{B}_{-}
+ \frac{2(9-2k+2k^2)}{3(5+k)^2} \hat{B}_{-} \pa \hat{B}_{+}
\nonu \\
&+& \frac{8k(9+7k)}{(3+7k)(19+23k)} 
i  \hat{B}_3 \hat{T}
+ \frac{4}{(5+k)} i \hat{B}_3 T^{(2)}
- \frac{4}{(5+k)} i \hat{B}_3 W^{(2)}
\nonu \\
& - &  \frac{2(-3+2k)}{(5+k)^2} \hat{B}_3 \pa \hat{A}_3
+ \frac{4(9-5k+2k^2)}{3(5+k)^2} \hat{B}_3 \pa \hat{B}_3
-\frac{5}{(5+k)} i \hat{B}_3 \pa T^{(1)} 
\nonu \\
& + & \frac{(18+39k+16k^2)}{(5+k)(3+7k)} \pa \hat{T}
-\frac{(5+4k)}{2(5+k)} \pa  T^{(2)}
-\frac{(7+4k)}{2(5+k)} \pa  W^{(2)}
\nonu \\
& + & \frac{8(3+k)}{(5+k)^2} i \pa^2 \hat{A}_3
-\frac{2(-3+k)}{3(5+k)} \pa^2 T^{(1)}
-\frac{2}{(5+k)} \hat{G}_{12} \hat{G}_{21}
-\frac{4}{(5+k)} \hat{G}_{21}   T_{-}^{(\frac{3}{2})} 
\nonu \\
& + & \frac{2}{(5+k)}  U^{(\frac{3}{2})}  V^{(\frac{3}{2})}
+\frac{8}{(5+k)^2} i \hat{A}_{+} \hat{A}_{-} \hat{A}_3
+\frac{8}{(5+k)^2} i \hat{A}_3 \hat{A}_3 \hat{A}_3
\nonu \\
& - & \frac{16}{(5+k)^2} i \hat{A}_3 \hat{A}_3 \hat{B}_3
+ \frac{7}{(5+k)^2} i \hat{A}_3 \hat{B}_{+} \hat{B}_{-}
+\frac{8}{(5+k)^2} i \hat{A}_3 \hat{B}_3 \hat{B}_3
\nonu \\
&+ & \frac{1}{(5+k)^2} i \hat{B}_{-} \hat{A}_3 \hat{B}_{+}
+ \frac{4k(-1+k)}{3(5+k)^2} i \hat{B}_{-} \hat{B}_{+} \hat{B}_3
- \frac{4(-1+k)k}{3(5+k)^2} i \hat{B}_3 \hat{B}_{+} \hat{B}_{-}
\nonu \\
& + & \left. \frac{1}{2(5+k)}  T^{(1)} \hat{B}_{+} \hat{B}_{-}
- \frac{1}{2(5+k)}  \hat{B}_{-} T^{(1)}  \hat{B}_{+}
- \frac{2}{(5+k)}    T_{+}^{(\frac{3}{2})}  T_{-}^{(\frac{3}{2})} 
\right](w) + \cdots,
\nonu \\
U^{(\frac{5}{2})}(z) \, V_{+}^{(2)}(w) & = & 
\frac{1}{(z-w)^3} \, \left[ -\frac{8(12+9k+k^2)}{3(5+k)^2}
\hat{G}_{21} + \frac{4(11+k)}{3(5+k)^2} 
T_{+}^{(\frac{3}{2})}  \right](w) 
\nonu \\
& + & \frac{1}{(z-w)^2} \, \left[ 
\frac{(31+8k)}{3(5+k)} {\bf P_{+}^{(\frac{5}{2})}}
+\frac{(35+8k)}{3(5+k)} W_{+}^{(\frac{5}{2})}
+\frac{2}{(5+k)^2} i \hat{A}_{-} \hat{G}_{11}
\right. \nonu \\
& + & \frac{2(11+4k)}{(5+k)^2} i \hat{A}_{-} U^{(\frac{3}{2})}
-\frac{8(9+k)}{3(5+k)^2} i \hat{A}_3 \hat{G}_{21}
-\frac{4(33+8k)}{3(5+k)^2} i \hat{A}_3 T_{+}^{(\frac{3}{2})} 
\nonu \\
&+ & 
\frac{2(27+8k)}{3(5+k)^2} i \hat{B}_{-} V^{(\frac{3}{2})}
+\frac{4k}{(5+k)^2} i \hat{B}_3  T_{+}^{(\frac{3}{2})} 
-\frac{2}{(5+k)} T^{(1)}  T_{+}^{(\frac{3}{2})} 
\nonu \\
&-& \left. \frac{4(8+k)(9+4k)}{9(5+k)^2} \pa \hat{G}_{21} 
-\frac{4(17+9k)}{9(5+k)^2} \pa  T_{+}^{(\frac{3}{2})} 
\right](w)
\nonu \\
& + & \frac{1}{(z-w)} \, \left[
\frac{1}{(5+k)} i {\bf R^{(\frac{5}{2})}} \hat{B}_{-} 
+\frac{(21+8k)}{5(5+k)} \pa {\bf P_{+}^{(\frac{5}{2})}}
-\frac{1}{2} {\bf S_{+}^{(\frac{7}{2})}}
\right. \nonu \\
& + & \frac{(14980 k^4+215092 k^3+753549 k^2+915638 k+351209)}{
15(5+k)^2(19+23k)(47+35k)} \hat{A}_{-} \hat{A}_{+} \hat{G}_{21}
\nonu \\
&- & \frac{(45+16k)}{5(5+k)^2}   \hat{A}_{-} \hat{A}_{+}  T_{+}^{(\frac{3}{2})} 
-\frac{2}{(5+k)} i \hat{A}_{-}  U^{(\frac{5}{2})}
-\frac{8(-2+k)}{5(5+k)^2} \hat{A}_{-} \hat{A}_3  U^{(\frac{3}{2})}
\nonu \\
&+ & \frac{4}{(5+k)^2} \hat{A}_{-} \hat{B}_{-}  T_{-}^{(\frac{3}{2})} 
+\frac{8}{(5+k)^2}  \hat{A}_{-} \hat{B}_3 \hat{G}_{11}
+\frac{2(39+8k)}{15(5+k)^2} i  
\hat{A}_{-} \pa \hat{G}_{11}
\nonu \\
& + & \frac{2(131+52k)}{15(5+k)^2} i \hat{A}_{-} \pa U^{(\frac{3}{2})}
-\frac{(-684-667k+43k^2+6k^3)}{5(5+k)^2(19+23k)}  
\hat{B}_{+} \hat{B}_{-} \hat{G}_{21}
\nonu \\
&+ & \frac{(12+k)}{(5+k)^2} \hat{B}_{+} \hat{B}_{-}  T_{+}^{(\frac{3}{2})} 
+\frac{8}{(5+k)^2}  \hat{B}_{-} \hat{A}_3 \hat{G}_{22} 
-\frac{8}{(5+k)^2} \hat{B}_{-} \hat{A}_3  V^{(\frac{3}{2})}
\nonu \\
&+ & \frac{(-684-667k+43k^2+6k^3)}{
5(5+k)^2(19+23k)}   \hat{B}_{-} \hat{B}_{+} \hat{G}_{21}
-\frac{(12+k)}{(5+k)^2}  \hat{B}_{-} \hat{B}_{+}  T_{+}^{(\frac{3}{2})} 
\nonu \\
&+ & \frac{2(12+5k)}{5(5+k)^2} \hat{B}_{-} \hat{B}_3 \hat{G}_{22}
-\frac{2(43+13k)}{5(5+k)^2} \hat{B}_{-} \hat{B}_3  V^{(\frac{3}{2})}
+\frac{4(27+5k)}{15(5+k)^2} i \hat{B}_{-} \pa \hat{G}_{22}
\nonu \\
&+ & \frac{2(9+14k)}{15(5+k)^2} i \hat{B}_{-} \pa   V^{(\frac{3}{2})}
\nonu \\
& - & \frac{(14980 k^4+215092 k^3+753549 k^2+915638 k+351209)}{
15(5+k)^2(19+23k)(47+35k)}  
\hat{A}_{+} \hat{A}_{-} \hat{G}_{21}
\nonu \\
& + & \frac{(65+16k)}{5(5+k)^2} 
\hat{A}_{+} \hat{A}_{-}  T_{+}^{(\frac{3}{2})} 
+\frac{8(-2+k)}{5(5+k)^2} \hat{A}_3  \hat{A}_{-} U^{(\frac{3}{2})}
\nonu \\
& - & \frac{4(489+599k+58k^2)}{
5(5+k)^2(19+23k)} i \hat{A}_3 \pa  \hat{G}_{21}
-\frac{32}{5(5+k)} i \hat{A}_3  \pa T_{+}^{(\frac{3}{2})} 
\nonu \\
& + & \frac{4}{(5+k)} i \hat{B}_3  W_{+}^{(\frac{5}{2})} 
-\frac{2(12+5k)}{5(5+k)^2} \hat{B}_3 \hat{B}_{-} \hat{G}_{22}
+\frac{2(63+13k)}{5(5+k)^2} 
\hat{B}_3 \hat{B}_{-} V^{(\frac{3}{2})} 
\nonu \\
&- & \frac{4(-1064-1127k+43k^2+6k^3)}{
15(5+k)^2(19+23k)} i \hat{B}_3 \pa \hat{G}_{21} 
+\frac{8(9+2k)}{3(5+k)^2} i \hat{B}_3 \pa  T_{+}^{(\frac{3}{2})} 
\nonu \\
&- & \frac{2(-49-19k+2k^2)}{5(5+k)(19+23k)} 
T^{(1)} \pa \hat{G}_{21}
+\frac{(19+8k)}{5(5+k)} \pa  W_{+}^{(\frac{5}{2})} 
\nonu \\
& + & \frac{3(-49-19k+2k^2)}{5(5+k)(19+23k)} \pa T^{(1)} \hat{G}_{21}
-\frac{3}{(5+k)} \pa T^{(1)}  T_{+}^{(\frac{3}{2})}
\nonu \\
& - & \frac{(130 k^4+23831 k^3+57003 k^2+21273 k-18525)
}{15(5+k)^2(19+23k)(47+35k)} \pa^2  T_{+}^{(\frac{3}{2})}
\nonu \\
& - & \frac{4(1862 k^4+19425 k^3+49348 k^2+39765 k+8316)}{
3(5+k)(3+7k)(19+23k)(47+35k)} \hat{G}_{21} \hat{T}
\nonu \\
& - & 
\frac{2}{(5+k)} \hat{G}_{21} T^{(2)}
\nonu \\
& - & \frac{8(3745 k^4+55243 k^3+202515 k^2+251609 k+96368)}{
15(5+k)^2(19+23k)(47+35k)} i \hat{G}_{21} \pa \hat{A}_3 
\nonu \\
&+ & \frac{8(-318+329k+380k^2+5k^3)}{
3(5+k)(19+23k)(47+35k)}  T_{+}^{(\frac{3}{2})}  \hat{T}
-\frac{2}{(5+k)}  T_{+}^{(\frac{3}{2})} W^{(2)}
\nonu \\
&- & \frac{2(17+4k)}{5(5+k)^2} i \hat{G}_{11} \pa \hat{A}_{-} 
+\frac{2}{(5+k)} U^{(\frac{3}{2})} V_{+}^{(2)} 
\nonu \\
& + & \left. \frac{2}{(5+k)} \hat{G}_{22} U_{+}^{(2)} 
-\frac{2}{(5+k)}  V^{(\frac{3}{2})} U_{+}^{(2)}
\right](w) + \cdots,
\nonu \\
U^{(\frac{5}{2})}(z) \, V_{-}^{(2)}(w) & = & 
\frac{1}{(z-w)^3} \,
\left[ \frac{4(1+k)(13+2k)}{3(5+k)^2} \hat{G}_{12}
+\frac{4(11+3k)}{3(5+k)^2} T_{-}^{(\frac{3}{2})}  
\right](w)
\nonu \\
& + & \frac{1}{(z-w)^2} \,
\left[ 
-\frac{(29+7k)}{3(5+k)} {\bf P_{-}^{(\frac{5}{2})}}
+\frac{(30+7k)}{3(5+k)} W_{-}^{(\frac{5}{2})}
-\frac{2(8+k)}{(5+k)^2} i \hat{A}_{+} \hat{G}_{22}
\right. \nonu \\
& + & \frac{4(27+5k)}{3(5+k)^2} i \hat{A}_{+} V^{(\frac{3}{2})}
+\frac{4(8+k)}{(5+k)^2} i \hat{A}_3 \hat{G}_{12}
-\frac{4(8+k)}{(5+k)^2} i \hat{A}_3 T_{-}^{(\frac{3}{2})} 
\nonu \\
&+ & 
\frac{2(-2+k)}{(5+k)^2} i \hat{B}_{+} \hat{G}_{11}
+ \frac{2(16+5k)}{(5+k)^2} i \hat{B}_{+} U^{(\frac{3}{2})}
-\frac{16(3+k)}{3(5+k)^2} i \hat{B}_3  T_{-}^{(\frac{3}{2})}  
\nonu \\
&+& \left. \frac{4(-1+k)(25+4k)}{9(5+k)^2} \pa \hat{G}_{12} 
+\frac{4(91+22k)}{9(5+k)^2} \pa  T_{-}^{(\frac{3}{2})} 
\right](w)
\nonu \\
& + & \frac{1}{(z-w)} \,
\left[  -\frac{(29+7k)}{5(5+k)} \pa {\bf P_{-}^{(\frac{5}{2})}} 
-\frac{1}{2} {\bf S_{-}^{(\frac{7}{2})}}
-\frac{2}{(5+k)} i {\bf P_{-}^{(\frac{5}{2})}} \hat{A}_3
+\frac{2}{(5+k)} i {\bf P_{-}^{(\frac{5}{2})}} \hat{B}_3
\right. 
\nonu \\
&- & \frac{(1520 k^4-70051 k^3-883214 k^2-1723627 k-925888)}{
15(5+k)^2(19+23k)(47+35k)} 
\hat{A}_{-} \hat{A}_{+}  T_{-}^{(\frac{3}{2})}
\nonu \\
& - & \frac{4}{(5+k)^2} \hat{B}_{+} \hat{A}_{+} \hat{G}_{21}
-\frac{4}{(5+k)^2}  \hat{B}_{+} \hat{A}_{+} T_{+}^{(\frac{3}{2})}
+\frac{2}{(5+k)} i \hat{B}_{+} U^{(\frac{5}{2})}
\nonu \\
&- & \frac{8}{(5+k)^2} \hat{B}_{+} \hat{A}_3 \hat{G}_{11}-
\frac{8}{(5+k)^2} \hat{B}_{+} \hat{A}_3  U^{(\frac{3}{2})}  +
\frac{(10+k)}{5(5+k)^2} \hat{B}_{+} \hat{B}_{-}  T_{-}^{(\frac{3}{2})}
\nonu \\
&+ & \frac{4(14+k)}{5(5+k)^2} \hat{B}_{+} \hat{B}_3   U^{(\frac{3}{2})}
+\frac{2(-42+25k)}{15(5+k)^2} i \hat{B}_{+} \pa \hat{G}_{11}
+\frac{2(154+71k)}{15(5+k)^2} i \hat{B}_{+} \pa  U^{(\frac{3}{2})}
\nonu \\
&- & \frac{(-10+k)}{5(5+k)^2} \hat{B}_{-} \hat{B}_{+} T_{-}^{(\frac{3}{2})} 
\nonu \\
& + & \frac{(1520 k^4-70051 k^3-834914 k^2-1618867 k-872308)}{
15(5+k)^2(19+23k)(47+35k)} \hat{A}_{+} \hat{A}_{-} T_{-}^{(\frac{3}{2})} 
\nonu \\
&+ & \frac{2(89+16k)}{5(5+k)^2} 
\hat{A}_{+} \hat{A}_3   V^{(\frac{3}{2})}
-\frac{8}{(5+k)^2} \hat{A}_{+} \hat{B}_3 \hat{G}_{22}
+\frac{8}{(5+k)^2}  \hat{A}_{+} \hat{B}_3  V^{(\frac{3}{2})}
\nonu \\
&+ & \frac{2(-10+k)}{5(5+k)^2} i \hat{A}_{+} \pa \hat{G}_{22} +
\frac{12(4+k)}{5(5+k)^2} i \hat{A}_{+} \pa  V^{(\frac{3}{2})} 
-\frac{2(109+16k)}{5(5+k)^2} \hat{A}_3 \hat{A}_{+}   V^{(\frac{3}{2})} 
\nonu \\
&+ & \frac{4}{(5+k)} i \hat{A}_3  W_{-}^{(\frac{5}{2})} 
+\frac{4}{(5+k)^2} \hat{A}_3 \hat{A}_3  T_{-}^{(\frac{3}{2})}
-\frac{8}{(5+k)^2} \hat{A}_3 \hat{B}_3    T_{-}^{(\frac{3}{2})}
\nonu \\
& -& \frac{4(-8+5k)(3+7k)}{5(5+k)^2(19+23k)} 
i \hat{A}_3 \pa \hat{G}_{12} +\frac{4(17+11k)}{15(5+k)^2} 
i \hat{A}_3 \pa  
T_{-}^{(\frac{3}{2})} 
\nonu \\
&+ & \frac{4}{(5+k)^2} \hat{B}_3 \hat{B}_3  T_{-}^{(\frac{3}{2})}  
-\frac{8(-418-454k-13k^2+3k^3)}{
15(5+k)^2(19+23k)} i \hat{B}_3 \pa \hat{G}_{12}
\nonu \\
&- & \frac{4(80+19k)}{15(5+k)^2} i \hat{B}_3 \pa   T_{-}^{(\frac{3}{2})}
-\frac{4(-53-44k+k^2)}{5(5+k)(19+23k)} T^{(1)} \pa \hat{G}_{12}
-\frac{2}{(5+k)} T^{(1)} \pa  T_{-}^{(\frac{3}{2})}
\nonu \\
& & + \frac{8(1272+2258k+1227k^2+133k^3)}{
3(5+k)(19+23k)(47+35k)} \hat{G}_{12} \hat{T}
-\frac{4(4+k)}{5(5+k)^2} \hat{B}_3 \hat{B}_{+}   U^{(\frac{3}{2})}
\nonu \\
& - & \frac{8(4060 k^4+22851 k^3-153461 k^2-409299 k-231959)}{
15(5+k)^2(19+23k)(47+35k)} i 
\hat{G}_{12} \pa \hat{A}_3
\nonu \\
&+ & \frac{6(-53-44k+k^2)}{
5(5+k)(19+23k)} \hat{G}_{12} \pa T^{(1)}  
\nonu \\
& + & \frac{4(70 k^4+3418 k^3+24403 k^2+24696 k+7497)}{
3(5+k)(3+7k)(19+23k)(47+35k)}  T_{-}^{(\frac{3}{2})} \hat{T}
\nonu \\
& - & \frac{2}{(5+k)}  T_{-}^{(\frac{3}{2})} T^{(2)}
+\frac{3}{(5+k)} T_{-}^{(\frac{3}{2})} \pa T^{(1)} 
\nonu \\
& + & \frac{2(1520 k^4-132841 k^3-1253657 k^2-2301367 k-1185751)}{
15(5+k)^2(19+23k)(47+35k)}  i T_{-}^{(\frac{3}{2})} \pa \hat{A}_3
\nonu \\
&+ & \frac{3}{(5+k)}  T_{-}^{(\frac{3}{2})} \pa T^{(1)} 
-\frac{2(14+5k)}{5(5+k)^2} i \hat{G}_{11} \pa \hat{B}_{+}
+\frac{2}{(5+k)}  U^{(\frac{3}{2})} V_{-}^{(2)}
\nonu \\
&- & \frac{2}{(5+k)} \hat{G}_{22}  U_{-}^{(2)}
+\frac{2}{(5+k)} V^{(\frac{3}{2})}  U_{-}^{(2)}
+\frac{(26+7k)}{5(5+k)} \pa   W_{-}^{(\frac{5}{2})}
\nonu \\
&-&  \frac{2(55+9k)}{5(5+k)^2} i \pa \hat{A}_{+} \hat{G}_{22}
\nonu \\
& + & \left.
\frac{4(8120 k^4+69327 k^3-133867 k^2-522867 k-321649)}{
15(5+k)^2(19+23k)(47+35k)} i \pa \hat{A}_3 \hat{G}_{12}
\right](w)+ \cdots,
\nonu  \\
U^{(\frac{5}{2})}(z) \, V^{(\frac{5}{2})}(w) & = & 
-\frac{1}{(z-w)^5} \, \left[ \frac{8k(37+33k+4k^2)}{(5+k)^3} \right]
\nonu \\
& + & \frac{1}{(z-w)^4} \, \left[-\frac{8(37+33k+4k^2)}{
(5+k)^3} i \hat{A}_3  + 
\frac{8k(37+33k+4k^2)}{3(5+k)^3} i \hat{B}_3 
\right. \nonu \\
& + & \left.  \frac{4(11+3k)}{3(5+k)^2} T^{(1)} 
\right](w) 
\nonu \\
& + & 
\frac{1}{(z-w)^3} \, \left[ 
\frac{2(-3+k)}{3(5+k)} {\bf P^{(2)}}
-\frac{8(315+1038k+587k^2+48k^3)}{9(5+k)^2(3+7k)} \hat{T}
\right. \nonu \\
& - & \frac{4(-3+k)(-5+2k)}{9(5+k)^2} T^{(2)}
-\frac{8(-37+3k+3k^2)}{9(5+k)^2} W^{(2)}
\nonu \\
& - & \frac{8(33+77k+6k^2)}{9(5+k)^3} \hat{A}_1 \hat{A}_1
-\frac{8(33+77k+6k^2)}{9(5+k)^3} \hat{A}_2 \hat{A}_2
\nonu \\
& - & \frac{8(-39+86k+6k^2)}{9(5+k)^3} \hat{A}_3 \hat{A}_3
+\frac{16(60+77k+2k^2)}{9(5+k)^3} \hat{A}_3 \hat{B}_3
\nonu \\
&- & \frac{8(3+k)(35+6k)}{9(5+k)^3} \hat{B}_1 \hat{B}_1  
-\frac{8(3+k)(35+6k)}{9(5+k)^3} \hat{B}_2 \hat{B}_2
-  \frac{56(15+8k)}{9(5+k)^3} \hat{B}_3 \hat{B}_3
\nonu \\
&+ & \frac{4}{(5+k)} T^{(1)} T^{(1)}
-\frac{4(37+33k+4k^2)}{(5+k)^3} i \pa \hat{A}_3
+\frac{4k(37+33k+4k^2)}{3(5+k)^3} i \pa \hat{B}_3
\nonu \\
&+& \left. \frac{2(11+3k)}{3(5+k)^2} \pa T^{(1)}
+ \frac{8}{3(5+k)^2} i T^{(1)} \hat{A}_3 
-\frac{8(-5+2k)}{3(5+k)^2} i T^{(1)} \hat{B}_3
\right](w)
\nonu \\
& + & \frac{1}{(z-w)^2 } \, 
\left[ 
\frac{(-3+k)}{3(5+k)} {\bf S^{(3)}}
+\frac{(11+3k)}{(5+k)} {\bf P^{(3)}}
+\frac{(-3+k)}{3(5+k)} \pa {\bf P^{(2)}}
\right. \nonu \\
& - & \frac{10(3+k)}{3(5+k)} W^{(3)}
-\frac{4(41+29k)}{(5+k)(19+23k)} T^{(1)} \hat{T}
+\frac{4}{(5+k)} T^{(1)} W^{(2)}
\nonu \\
&- & \frac{76}{3(5+k)^2} i T^{(1)} \pa \hat{A}_3 
-\frac{2(-15+11k)}{3(5+k)^2} i T^{(1)} \pa \hat{B}_3 
+\frac{4}{(5+k)} T^{(1)} \pa T^{(1)}
\nonu \\
& - & \frac{4(31+7k)}{3(5+k)^2} i \hat{A}_{+} V_{+}^{(2)}
-\frac{2(-399+73k+12k^2)}{
9(5+k)^3} \hat{A}_{+} \pa \hat{A}_{-}
\nonu \\
& - & \frac{2(447+175k+12k^2)}{
9(5+k)^3} \hat{A}_{-} \pa \hat{A}_{+}
-\frac{8(4347+12102k+5819k^2+352k^3)}{
3(5+k)^2(3+7k)(19+23k)} i \hat{A}_3 \hat{T} 
\nonu \\
&+ & \frac{8(9+k)}{3(5+k)^2} i \hat{A}_3 T^{(2)}  
-\frac{8(-2+k)}{3(5+k)^2} i \hat{A}_3 W^{(2)} 
-\frac{8(-168+65k+6k^2)}{9(5+k)^3} \hat{A}_3 \pa \hat{A}_3 
\nonu \\
& + & \frac{20}{(5+k)^2} i \hat{A}_3 \pa T^{(1)}
-\frac{16(7+2k)}{3(5+k)^2} i 
\hat{B}_{+} U_{+}^{(2)} -\frac{8(42+43k+3k^2)}{
9(5+k)^3} \hat{B}_{+} \pa \hat{B}_{-}
\nonu \\
&- & \frac{8(42-5k+3k^2)}{
9(5+k)^3} \hat{B}_{-} \pa \hat{B}_{+}
+\frac{8k(1431+3354k+1631k^2+140k^3)}{
3(5+k)^2(3+7k)(19+23k)} i \hat{B}_3 \hat{T}
\nonu \\
& + & \frac{16(3+k)}{3(5+k)^2} i \hat{B}_3 T^{(2)}
-\frac{8(-4+k)}{3(5+k)^2} i \hat{B}_3 W^{(2)} 
-\frac{8(84+23k)}{9(5+k)^3} \hat{B}_3 \pa \hat{B}_3 
\nonu \\
&+ & \frac{2(17+3k)}{3(5+k)^2} i \hat{B}_3 \pa T^{(1)} 
-\frac{2(504+2691k+1207k^2+96k^3)}{9(5+k)^2(3+7k)} \pa \hat{T}
\nonu \\
& - & \frac{(363+59k+4k^2)}{9(5+k)^2} \pa T^{(2)}
-\frac{(-55+21k+12k^2)}{9(5+k)^2} \pa W^{(2)}
\nonu \\
& - & \frac{8(7+k)(1+2k)}{3(5+k)^3} i \pa^2 \hat{A}_3
+\frac{16k(5+11k+k^2)}{9(5+k)^3} i \pa^2 \hat{B}_3
-\frac{4(37+9k)}{9(5+k)^2} \pa^2 T^{(1)}
\nonu \\
&- & \frac{4(3+k)}{(5+k)^2} \hat{G}_{11} \hat{G}_{22}
+\frac{4(27+5k)}{3(5+k)^2} \hat{G}_{11} V^{(\frac{3}{2})} 
+\frac{8(8+k)}{3(5+k)^2} \hat{G}_{12} \hat{G}_{21}
\nonu \\
& + & \frac{4(29+5k)}{3(5+k)^2} \hat{G}_{21}  T_{-}^{(\frac{3}{2})}
+\frac{28(3+k)}{3(5+k)^2} \hat{G}_{22} U^{(\frac{3}{2})} 
-
\frac{4(11+3k)}{3(5+k)^2}  U^{(\frac{3}{2})}   V^{(\frac{3}{2})} 
\nonu \\
& - & 
\frac{16(18+k)}{3(5+k)^3} i \hat{A}_{+} \hat{A}_{-} 
\hat{A}_3 -\frac{2(-465-193k+2k^2)}{9(5+k)^3} i 
\hat{A}_{+} \hat{A}_{-} \hat{B}_3
\nonu \\
&+ & \frac{2(-177-121k+2k^2)}{9(5+k)^3} 
\hat{A}_{-} \hat{A}_{+} \hat{B}_3 
-\frac{8(27+2k)}{3(5+k)^3} i \hat{A}_3 \hat{A}_3 \hat{A}_3
\nonu \\
& + & \frac{8(54+7k)}{3(5+k)^3} i \hat{A}_3 \hat{A}_3 \hat{B}_3
+  \frac{2(-201+115k+10k^2)}{9(5+k)^3} i \hat{A}_3 \hat{B}_{+} \hat{B}_{-}
\nonu \\
& - & \frac{8(27+8k)}{3(5+k)^3} i \hat{A}_3 \hat{B}_3 \hat{B}_3
-\frac{2(303+211k+10k^2)}{
9(5+k)^3} i \hat{B}_{-} \hat{A}_3 \hat{B}_{+}
\nonu \\
&+ & \frac{16k}{(5+k)^3} i   \hat{B}_{-} \hat{B}_{+} \hat{B}_3 
+\frac{8k}{(5+k)^3} i \hat{B}_3 \hat{B}_3 \hat{B}_3
-\frac{8}{(5+k)^2} T^{(1)} \hat{A}_{+} \hat{A}_{-}
\nonu \\
& - & \frac{4}{(5+k)^2} T^{(1)}  \hat{A}_3 \hat{A}_3 
+  \frac{8}{(5+k)^2} T^{(1)}  \hat{A}_3 \hat{B}_3
- \frac{4}{(5+k)^2} T^{(1)}  \hat{B}_3 \hat{B}_3
\nonu \\
&-& \left. \frac{8}{(5+k)^2} \hat{B}_{-} T^{(1)} \hat{B}_{+} 
-\frac{4(-3+k)}{3(5+k)^2}  T_{+}^{(\frac{3}{2})}  T_{-}^{(\frac{3}{2})}
\right](w) 
\nonu \\
& + & \frac{1}{(z-w)} \left[
\frac{1}{2}   \{ \hat{G}_{21} \, 
S_{-}^{(\frac{7}{2})} \}_{-1} +\frac{2}{(5+k)} i \hat{A}_3  
{\bf P^{(3)}} +\frac{1}{2(5+k)} i \hat{A}_3 \pa {\bf P^{(2)}}
\right. 
\nonu \\
& - & \frac{2}{(5+k)} i \hat{B}_3 {\bf P^{(3)}}
-\frac{1}{2(5+k)} i \hat{B}_3 \pa  {\bf P^{(2)}}
-\frac{(7+k)}{3(5+k)} \pa {\bf S^{(3)}}
+\frac{(34+7k)}{5(5+k)} \pa   {\bf P^{(3)}}
\nonu \\
& + & \frac{(-1+2k)}{20(5+k)} \pa^2  {\bf P^{(2)}}
\nonu \\
&+ & \frac{2}{(5+k)} W^{(2)} W^{(2)}
+\frac{2}{(5+k)}  U^{(\frac{3}{2})}  V^{(\frac{5}{2})} 
+\frac{4}{(5+k)^2}  U^{(\frac{3}{2})}  \pa V^{(\frac{3}{2})} 
\nonu \\
& - & \frac{2}{(5+k)}   U_{-}^{(2)}  V_{+}^{(2)}
- \frac{2}{(5+k)}   U_{+}^{(2)}  V_{-}^{(2)}
+\frac{8(11+2k)}{3(5+k)^2}  T_{+}^{(\frac{3}{2})}  \pa T_{-}^{(\frac{3}{2})}
\nonu \\
& -& \frac{4}{(5+k)^2} \hat{A}_3 \hat{A}_3 W^{(2)} 
+\frac{32}{(5+k)^3} \hat{A}_3 \hat{A}_3 \hat{B}_{+} \hat{B}_{-}
-\frac{2}{(5+k)}  U^{(\frac{5}{2})}  V^{(\frac{3}{2})} 
\nonu \\
&-& \frac{4(-7736-10241k-923k^2+18k^3)}{15(5+k)^3(19+23k)}  
i \hat{A}_3 \hat{A}_3 \pa \hat{B}_{3}
-\frac{4}{(5+k)} i \hat{A}_3 W^{(3)} 
\nonu \\
&- &  \frac{2(157+311k+6k^2)}{5(5+k)^2(19+23k)} \hat{A}_3 \hat{A}_3
\pa T^{(1)}  
+\frac{8}{(5+k)^2} \hat{A}_3 \hat{B}_3 W^{(2)}
\nonu \\
& + & \frac{8(31+98k+3k^2)}{5(5+k)^2(19+23k)} \hat{A}_3 \hat{B}_3
\pa T^{(1)} -\frac{8}{(5+k)^2} \hat{A}_3 \hat{B}_{-}   V_{-}^{(2)}
+\frac{8}{(5+k)^2} \hat{A}_3 \hat{B}_{+}   U_{+}^{(2)}
\nonu \\
&-& \frac{4(4153+5764k+595k^2)}{
15(5+k)^3(19+23k)} i \hat{A}_3 \hat{B}_{+} \pa \hat{B}_{-}
+\frac{8}{(5+k)^2} i \hat{A}_3 \hat{G}_{11}  V^{(\frac{3}{2})} 
\nonu \\
&- & \frac{8}{(5+k)^2} i \hat{A}_3 \hat{G}_{11} \hat{G}_{22} 
+\frac{8}{(5+k)^2} i \hat{A}_3 \hat{G}_{22}  U^{(\frac{3}{2})} 
\nonu \\
& + & \frac{8(85-298k-437k^2+2k^3)}{
5(5+k)^3(19+23k)} i \hat{A}_3 \pa \hat{B}_3 \hat{B}_3 
\nonu \\
&+&  \frac{4(2687+3656k+785k^2)}{
15(5+k)^3(19+23k)} i \hat{A}_3 \pa \hat{B}_{+} \hat{B}_{-} 
-\frac{4(1+k)}{(5+k)^2} i \hat{A}_3 \pa W^{(2)}
\nonu \\
&- & \frac{4(7+11k)}{15(5+k)^2} i \hat{A}_3 \pa T^{(2)}
-\frac{4(-307-291k+4k^2)}{
5(5+k)^2(19+23k)} \hat{A}_3 \pa \hat{B}_3 T^{(1)}
\nonu \\
& + & \frac{4(161738+350863k+303145k^2+101693k^3+3801k^4)}{
15(5+k)^3(19+23k)(47+35k)} \hat{A}_3 \pa^2 \hat{B}_3 
\nonu \\ 
&- & \frac{2(-558-791k+275k^2)}{
15(5+k)^2(19+23k)} i \hat{A}_3 \pa^2 T^{(1)} 
-\frac{4}{(5+k)^2} i \hat{A}_{-}  T_{-}^{(\frac{3}{2})} U^{(\frac{3}{2})} 
\nonu \\
&+ & \frac{8}{(5+k)^2} \hat{A}_{-} \hat{B}_3  U_{-}^{(2)}
+\frac{8}{(5+k)^2} \hat{A}_{+} \hat{A}_3  V_{+}^{(2)}
-\frac{8}{(5+k)^2} \hat{A}_{+} \hat{A}_{-} W^{(2)}
\nonu \\
&-& \frac{4}{(5+k)^2} \hat{A}_{+} \hat{A}_{-} T^{(2)}
+\frac{8}{(5+k)^3}  \hat{A}_{+} \hat{A}_{-} \hat{A}_3 \hat{A}_3
-\frac{16}{(5+k)^3}  \hat{A}_{+} \hat{A}_{-} \hat{A}_3 \hat{B}_3
\nonu \\
&+& \frac{40}{(5+k)^3}  \hat{A}_{+} \hat{A}_{-} \hat{B}_3 \hat{B}_3
+ \frac{24}{(5+k)^3}  \hat{A}_{+} \hat{A}_{-} \hat{B}_{+} \hat{B}_{-}
\nonu \\
& - & \frac{4(3869+5621k+1580k^2)}{
15(5+k)^3(19+23k)} i  \hat{A}_{+} \hat{A}_{-} \pa \hat{A}_3
-\frac{6(-11+27k+2k^2)}{
5(5+k)^2(19+23k)} \hat{A}_{+} \hat{A}_{-} \pa T^{(1)}
\nonu \\
&+& \frac{8}{(5+k)^3} \hat{A}_{+} \hat{A}_{+} \hat{A}_{-} \hat{A}_{-}
-\frac{8}{(5+k)^2} \hat{A}_{+} \hat{B}_3  V_{+}^{(2)}
-\frac{8}{(5+k)^2} i \hat{A}_{+} \hat{G}_{21} \hat{G}_{22}
\nonu \\
&+& \frac{4}{(5+k)^2} i \hat{A}_{+} \hat{G}_{22}  T_{+}^{(\frac{3}{2})}
+\frac{32(-113-122k+55k^2)}{
15(5+k)^3(19+23k)} i \hat{A}_{+} \pa \hat{A}_{-} \hat{A}_{3}
\nonu \\
&+& \frac{8(589+760k-95k^2+6k^3)}{
15(5+k)^3(19+23k)}  i \hat{A}_{+} \pa \hat{A}_{-} \hat{B}_{3}
\nonu \\
& - & \frac{4(17+3k)}{5(5+k)^2} i \hat{A}_{+} \pa  V_{+}^{(2)}
+\frac{4(-201-203k+2k^2)}{
5(5+k)^2(19+23k)} \hat{A}_{+} \pa \hat{A}_{-} T^{(1)}
\nonu \\
& - & \frac{2(42532 k^5+68824 k^4-934489 k^3-2457151 k^2-1872375 k-405765)}{
15(5+k)^3(3+7k)(19+23k)(47+35k)} \nonu \\
& \times & \hat{A}_{+} \pa^2 \hat{A}_{-}
- \frac{4}{(5+k)^2} \hat{B}_3 \hat{B}_3 W^{(2)}
-\frac{6(-11+27k+2k^2)}{5(5+k)^2(19+23k)}  \hat{B}_3 \hat{B}_3 \pa T^{(1)}
\nonu \\
&+ & \frac{8}{(5+k)^2} i \hat{B}_3 \hat{G}_{11} \hat{G}_{22}
-\frac{8}{(5+k)^2} i \hat{B}_3 \hat{G}_{22}  U^{(\frac{3}{2})} 
+\frac{4(8+k)}{3(5+k)^2} i \hat{B}_3 \pa W^{(2)}
\nonu \\
&+& \frac{8(20+7k)}{15(5+k)^2} i \hat{B}_3 \pa T^{(2)} 
+\frac{2(448+877k+169k^2)}{5(5+k)^2(19+23k)} i \hat{B}_3 \pa^2 T^{(1)}
\nonu \\
& - & \frac{4}{(5+k)^2} i \hat{B}_{-}  T_{-}^{(\frac{3}{2})}  V^{(\frac{3}{2})} 
-\frac{4}{(5+k)^2} i \hat{B}_{-} \hat{G}_{22}  T_{-}^{(\frac{3}{2})}
-\frac{8}{(5+k)^2} \hat{B}_{+} \hat{B}_3  U_{+}^{(2)}
\nonu \\
& - & \frac{8}{(5+k)^2} \hat{B}_{+} \hat{B}_{-} W^{(2)}
+\frac{4}{(5+k)^2} \hat{B}_{+} \hat{B}_{-} T^{(2)}
-\frac{8}{(5+k)^2} i \hat{B}_3 \hat{G}_{11}   V^{(\frac{3}{2})} 
\nonu \\
& - & \frac{4(-456-923k-601k^2+6k^3)}{5(5+k)^3(19+23k)} 
i  \hat{B}_{+} \hat{B}_{-} \pa \hat{B}_3
\nonu \\
& + & \frac{4}{(5+k)^2} i \hat{B}_{+} \hat{G}_{11}   T_{+}^{(\frac{3}{2})}
+\frac{8}{(5+k)^2} i \hat{B}_{+} \hat{G}_{11} \hat{G}_{21} 
\nonu \\
&+ & \frac{8(-836-623k+319k^2+6k^3)}{15(5+k)^3(19+23k)} i   
\hat{B}_{+} \pa \hat{B}_{-} \hat{B}_3
\nonu \\
& - & \frac{28(11+4k)}{15(5+k)^2} i \hat{B}_{+} \pa U_{+}^{(2)}
-\frac{2(157+311k+6k^2)}{5(5+k)^2(19+23k)} \hat{B}_{+} 
\hat{B}_{-} \pa T^{(1)}
\nonu \\
& + & 
\frac{2(20923 k^5-1054648 k^4-3673323 k^3-4337567 k^2-2083444 k-328845)}{
15(5+k)^3(3+7k)(19+23k)(47+35k)} \nonu \\
& \times & \hat{B}_{+} \pa^2 \hat{B}_{-} +\frac{4(17+3k)}{5(5+k)^2}
\hat{G}_{11} \pa  V^{(\frac{3}{2})} 
+\frac{12(-1+k)}{5(5+k)^2} \hat{G}_{11} \pa \hat{G}_{22}
\nonu \\
&+ & \frac{12(-53-44k+k^2)}{5(5+k)(19+23k)} 
\hat{G}_{12} \pa  T_{+}^{(\frac{3}{2})}
+ \frac{8(-53-44k+k^2)}{5(5+k)^2(19+23k)} \hat{B}_{+} \pa \hat{B}_{-} 
T^{(1)}
\nonu \\
& + & \frac{4(980 k^4+16883 k^3+85035 k^2+120589 k+48673)}{
15(5+k)^2(19+23k)(47+35k)} \hat{G}_{12} \pa \hat{G}_{21}
\nonu \\
& - & \frac{2}{(5+k)} \hat{G}_{21}  W_{-}^{(\frac{5}{2})}
+\frac{4(124+19k)}{15(5+k)^2} \hat{G}_{21} \pa  T_{-}^{(\frac{3}{2})}
+\frac{4(42+23k)}{15(5+k)^2} \hat{G}_{22} \pa  U^{(\frac{3}{2})} 
\nonu \\
& + & \frac{2}{(5+k)} T^{(1)} \pa W^{(2)}
-\frac{4}{(5+k)} \left( \hat{T} W^{(2)} -\frac{3}{10} \pa^2 W^{(2)} \right)
\nonu \\
& -& \frac{8(-90-79k-448k^2+5k^3)}{3(5+k)(19+23k)(47+35k)}
\left( \hat{T} T^{(2)} -\frac{3}{10} \pa^2 T^{(2)} \right)
\nonu \\
&- & \frac{32(1908+7704k+9625k^2+3822k^3+473k^4)}{3(5+k)(3+7k)(19+23k)(47+35k)} 
\left( \hat{T} \hat{T} -\frac{3}{10} \pa^2 \hat{T} \right)
\nonu \\
&-& \frac{16(-2958-2584k-33k^2+133k^3)}{
3(5+k)^2(19+23k)(47+35k)} \left( \hat{T} \hat{A}_3 \hat{A}_3 -\frac{3}{10} 
\pa^2 (\hat{A}_3 \hat{A}_3) \right)
\nonu \\
&+ &  \frac{32(-420+1703k+1611k^2+28k^3)}{
3(5+k)^2(19+23k)(47+35k)} \left( \hat{T} \hat{A}_3 \hat{B}_3 -\frac{3}{10} 
\pa^2 (\hat{A}_3 \hat{B}_3) \right)
\nonu \\
& - & \frac{16(-2+k)}{
(5+k)(19+23k)} i \left( \hat{T} T^{(1)} \hat{A}_3 -\frac{3}{10} 
\pa^2 ( T^{(1)} \hat{A}_3) \right)
\nonu \\
&-& \frac{16(-1272-2k+735k^2+77k^3)}{
3(5+k)^2(19+23k)(47+35k)} \left( \hat{T} \hat{B}_3 \hat{B}_3 -\frac{3}{10} 
\pa^2 (\hat{B}_3 \hat{B}_3) \right)
\nonu \\
& - & \frac{16(-4+k)}{
(5+k)(19+23k)} i \left( \hat{T} T^{(1)} \hat{B}_3 -\frac{3}{10} 
\pa^2 ( T^{(1)} \hat{B}_3) \right)
\nonu \\
&+& \frac{8(-405+4926k+10777k^2+8316k^3+1862k^4)}{
3(5+k)^2(3+7k)(19+23k)(47+35k)} \hat{T} \hat{A}_{+} \hat{A}_{-}
\nonu \\
&-& \frac{4(-405+4926k+10777k^2+8316k^3+1862k^4)}{
5(5+k)^2(3+7k)(19+23k)(47+35k)} \pa^2 (\hat{A}_{+} \hat{A}_{-})
\nonu \\
&-& \frac{8(7632+23319k+23260k^2+10731k^3+1862k^4)}{
3(5+k)^2(3+7k)(19+23k)(47+35k)} \hat{T} \hat{B}_{+} \hat{B}_{-}
\nonu \\
&+& \frac{4(7632+23319k+23260k^2+10731k^3+1862k^4)}{
5(5+k)^2(3+7k)(19+23k)(47+35k)} \pa^2 (\hat{B}_{+} \hat{B}_{-})
\nonu \\
& - & \frac{4(174510 k^4+1716425 k^3+4970751 k^2+5259883 k+1567623)}{
15(5+k)^2(3+7k)(19+23k)(47+35k)} i \hat{T} \pa \hat{A}_3
\nonu \\
& + & \frac{4(30240 k^5+419499 k^4+1376297 k^3+1649817 k^2+760275 k+116568)}{
15(5+k)^2(3+7k)(19+23k)(47+35k)}  i 
\nonu \\
&\times & \hat{T} \pa \hat{B}_3
\nonu \\
&- & \frac{2(13059+25162k+7951k^2+680k^3)}{
15(5+k)(19+23k)(47+35k)} \hat{T} \pa T^{(1)}
-\frac{4(14+3k)}{3(5+k)^2} \pa U^{(\frac{3}{2})} V^{(\frac{3}{2})}  
\nonu \\
&- & \frac{4(23+5k)}{3(5+k)^2} \pa  T_{+}^{(\frac{3}{2})}  T_{-}^{(\frac{3}{2})}
-\frac{12}{(5+k)^2} i \pa \hat{A}_3 W^{(2)}
+\frac{16(37+11k)}{15(5+k)^2} i \pa \hat{A}_3 T^{(2)}
\nonu \\
&-& \frac{4(9097+11713k+700k^2)}{15(5+k)^3(19+23k)} 
i \pa \hat{A}_3 \hat{A}_3 \hat{A}_3 
\nonu \\
& + & \frac{8(7091+9896k+1433k^2+12k^3)}{
15(5+k)^3(19+23k)} i \pa \hat{A}_3 \hat{A}_3 \hat{B}_3 
\nonu \\
& + & \frac{4(-307-291k+4k^2)}{5(5+k)^2(19+23k)} 
\pa \hat{A}_3 \hat{A}_3 T^{(1)} 
\nonu \\
& - & \frac{4(745+1353k+492k^2+8k^3)}{5(5+k)^3(19+23k)} 
i \pa \hat{A}_3 \hat{B}_3 \hat{B}_3 
\nonu \\
& - & \frac{16(-53-44k+k^2)}{5(5+k)^2(19+23k)} 
\pa \hat{A}_3 \hat{B}_3 T^{(1)} 
-\frac{4(7403+11039k+2960k^2)}{
15(5+k)^3(19+23k)} i \pa \hat{A}_3 \hat{B}_{+} \hat{B}_{-}
\nonu \\
& - & \frac{2(12152 k^4+71316 k^3-108291 k^2-609594 k-483467)}{
15(5+k)^3(19+23k)(47+35k)} \pa \hat{A}_3 \pa \hat{A}_3
\nonu \\
&-& \frac{4(6678 k^4-41340 k^3-623981 k^2-1018710 k-371275)}{
15(5+k)^3(19+23k)(47+35k)} \pa \hat{A}_3 \pa \hat{B}_3
\nonu \\
& + &  \frac{2(117+211k+218k^2)}{
5(5+k)^2(19+23k)} i \pa \hat{A}_3 \pa T^{(1)}
+\frac{10}{(5+k)^2} i \pa \hat{A}_{-}  U_{-}^{(2)}
\nonu \\
&+& \frac{8(-1307-1523k+220k^2)}{15(5+k)^3(19+23k)} i
\pa \hat{A}_{+} \hat{A}_{-} \hat{A}_3 -
\frac{2(99+16k)}{5(5+k)^2} i \pa \hat{A}_{+}  V_{+}^{(2)}
\nonu \\
& + & \frac{8(2869+4090k+595k^2+6k^3)}{
15(5+k)^3(19+23k)} i \pa \hat{A}_{+} \hat{A}_{-} \hat{B}_3
\nonu \\
& + & \frac{4(-201-203k+2k^2)}{
5(5+k)^2(19+23k)} \pa \hat{A}_{+} \hat{A}_{-} T^{(1)}
\nonu \\
& - & \frac{2(85064 k^5+1246868 k^4+3718801 k^3+4143319 k^2+
1773075 k+188865)}{15(5+k)^3(3+7k)(19+23k)(47+35k)} 
\nonu \\
& \times & \pa \hat{A}_{+} \pa \hat{A}_{-}
-\frac{4(1+2k)}{3(5+k)^2} i \pa \hat{B}_3 W^{(2)}
-\frac{8(5+13k)}{15(5+k)^2} i \pa \hat{B}_3 T^{(2)}
\nonu \\
&+& \frac{4(-1976-1433k+1009k^2+6k^3)}{
15(5+k)^3(19+23k)} i \pa \hat{B}_3 \hat{B}_3 \hat{B}_3 
\nonu \\
& + & 
\frac{16(-53-44k+k^2)}{5(5+k)^2(19+23k)} \pa \hat{B}_3 \hat{B}_3 T^{(1)}
\nonu \\
& -& \frac{2(1260 k^5+9329 k^4+156008 k^3+462595 k^2+633694 k+376398)}{
15(5+k)^3(19+23k)(47+35k)} \pa \hat{B}_3  \pa \hat{B}_3 
\nonu \\
& + & \frac{2(209-677k-156k^2+6k^3)}{
5(5+k)^2(19+23k)} i \pa \hat{B}_3  \pa T^{(1)}
-\frac{2k}{(5+k)^2} i \pa \hat{B}_{-}  V_{-}^{(2)}
\nonu \\
& -& \frac{2(64+11k)}{15(5+k)^2} i \pa \hat{B}_{+}  U_{+}^{(2)}
+ \frac{16(-418-454k-13k^2+3k^3)}{15(5+k)^3(19+23k)} 
i \pa \hat{B}_{+} \hat{B}_{-} \hat{B}_3
\nonu \\
&+ & \frac{8(-53-44k+k^2)}{5(5+k)^2(19+23k)} 
\pa \hat{B}_{+} \hat{B}_{-} T^{(1)} 
\nonu \\
& - & 
\frac{2(237209 k^5+1163897 k^4+3158515 k^3+4546537 k^2+3042192 k+743418
)}{15(5+k)^3(3+7k)(19+23k)(47+35k)} \nonu \\
& \times & \pa \hat{B}_{+} \pa \hat{B}_{-}
+\frac{4(111+19k)}{15(5+k)^2} 
\pa \hat{G}_{11}  V^{(\frac{3}{2})} 
-\frac{4(13+7k)}{5(5+k)^2} \pa \hat{G}_{11} \hat{G}_{22}
\nonu \\
& - & \frac{4(-625-661k-78k^2+2k^3)}{
5(5+k)^2(19+23k)} \pa \hat{G}_{12} T_{+}^{(\frac{3}{2})}
\nonu \\
& - & \frac{4(-638-729k-45k^2+2k^3)}{
5(5+k)^2(19+23k)} \pa \hat{G}_{12} \hat{G}_{21}
\nonu \\
& - & \frac{4(25 k^4-9360 k^3-66754 k^2-96016 k-46007)}{
15(5+k)^2(19+23k)(47+35k)} \pa \hat{G}_{21}  T_{-}^{(\frac{3}{2})}
\nonu \\
&+& \frac{4(14+k)}{5(5+k)^2} \pa \hat{G}_{22}  U^{(\frac{3}{2})} 
+\frac{2}{(5+k)} \pa T^{(1)} W^{(2)}
-\frac{3}{2(5+k)} \pa T^{(1)} \pa T^{(1)}
\nonu \\
& + & \frac{4(-19839-53284k-15823k^2+470k^3)}{15(5+k)^2(3+7k)(19+23k)} i 
\pa \hat{T} \hat{A}_{3}
\nonu \\
&+ & \frac{16(-1254-2350k+799k^2+1324k^3+133k^4)}{
15(5+k)^2(3+7k)(19+23k)} i \pa \hat{T} \hat{B}_3 
\nonu \\
&+ & \frac{8(-258-154k+3k^2)}{
15(5+k)(19+23k)} \pa \hat{T} T^{(1)}
\nonu \\
& - & \frac{8(3038 k^4+44254 k^3+37331 k^2-37396 k-18583)}{
15(5+k)^3(19+23k)(47+35k)} 
\pa^2 \hat{A}_3 \hat{A}_3
\nonu \\
&+ & \frac{4(10171 k^4+148502 k^3+433686 k^2+498022 k+216587)}{
15(5+k)^3(19+23k)(47+35k)} \pa^2 \hat{A}_3 \hat{B}_3 
\nonu \\
& - & \frac{2(1742+2363k+161k^2)}{
15(5+k)^2(19+23k)} i \pa^2 \hat{A}_3 T^{(1)}
\nonu \\
&-& \frac{2(42532 k^5+864094 k^4+4950425 k^3+9134315 k^2+6380415 k+1438515)}{
15(5+k)^3(3+7k)(19+23k)(47+35k)}
\nonu \\
& \times & \pa^2 \hat{A}_{+} \hat{A}_{-}
\nonu \\
& + & \frac{4(420 k^5+1033 k^4-96484 k^3-324570 k^2-291992 k-45319)}{
15(5+k)^3(19+23k)(47+35k)} 
\pa^2 \hat{B}_{3} \hat{B}_3
\nonu \\
&- & \frac{2(1762+1657k+655k^2+12k^3)}{
15(5+k)^2(19+23k)} i \pa^2 \hat{B}_3 T^{(1)}
\nonu \\
& - & \frac{2(4557 k^5-235466 k^4+446077 k^3+2165225 k^2+1842818 k+414573)}{
15(5+k)^3(3+7k)(19+23k)(47+35k)} 
\nonu \\
& \times & \pa^2 \hat{B}_{+} \hat{B}_{-}
\nonu \\
& + & \frac{3}{(5+k)} \pa^2 T^{(1)} T^{(1)}
-\frac{4(17+4k)}{15(5+k)} \pa W^{(3)}
-\frac{2(56+25k+3k^2)}{15(5+k)^2} \pa^2 W^{(2)}
\nonu \\
&- & \frac{2(640 k^4+33669 k^3+236599 k^2+401583 k+196477)}{
15(5+k)^2(19+23k)(47+35k)} 
 \pa^2 T^{(2)}
\nonu \\
&- & 
\frac{2(46372 k^5+781978 k^4+3550583 k^3+5476445 k^2+3013401 k+371709)}{
15(5+k)^2(3+7k)(19+23k)(47+35k)} 
\nonu \\
&\times & \pa^2 \hat{T}
-  \frac{2
}{45(5+k)^3(3+7k)(19+23k)(47+35k)}
(17675 k^5-332695 k^4
\nonu \\
& - & 10621152 k^3-42158960 k^2-
50572019 k-15635265) i \pa^3 \hat{A}_3
\nonu \\
&- & \frac{2}{45(5+k)^3(3+7k)(19+23k)(47+35k)}(
39375 k^6+759694 k^5 \nonu \\
& + & 7947192 k^4 
 +  23250813 k^3+25433166 k^2
+10270601 k+1178367) i \pa^3 \hat{B}_3  
\nonu \\
&+& \left. \frac{(86771+108025k+10036k^2-7231k^3+195k^4)}{
15(5+k)^2(19+23k)(47+35k)} \pa^3 T^{(1)} 
\right](w) 
+ \cdots,
\nonu \\
U^{(\frac{5}{2})}(z) \, W^{(2)}(w) & = & 
\frac{1}{(z-w)^3} \, \left[  \frac{4(-3+k)}{3(5+k)^2} 
\hat{G}_{11} + \frac{4(-3+k)}{3(5+k)^2} U^{(\frac{3}{2})} \right](w)
\nonu \\
& + & \frac{1}{(z-w)^2} \left[ 
\frac{(19+5k)}{2(5+k)} {\bf Q^{(\frac{5}{2})}}
+\frac{(22+5k)}{(5+k)} U^{(\frac{5}{2})}
+\frac{4(6+k)}{3(5+k)^2} i \hat{A}_{+} \hat{G}_{21} 
\right. \nonu \\
&+ & \frac{28(3+k)}{3(5+k)^2} i \hat{A}_{+}  T_{+}^{(\frac{3}{2})}
+\frac{4(5+2k)}{(5+k)^2} i \hat{A}_3  U^{(\frac{3}{2})}
+ \frac{4(3+2k)}{3(5+k)^2} i \hat{B}_{-}  T_{-}^{(\frac{3}{2})}
\nonu \\
& + &  \frac{4k}{(5+k)^2} i \hat{B}_3  U^{(\frac{3}{2})}
-\frac{2}{(5+k)} T^{(1)}  U^{(\frac{3}{2})}
+\frac{4(-3+k)}{3(5+k)^2} \pa \hat{G}_{11}
 \nonu \\
&-&\left. \frac{4(8+k)}{3(5+k)^2} \pa U^{(\frac{3}{2})}
\right](w)
\nonu \\ 
& + & \frac{1}{(z-w)} \left[
\frac{3(3+k)}{2(5+k)} \pa {\bf Q^{(\frac{5}{2})}}
-\frac{1}{(5+k)} i {\bf P_{-}^{(\frac{5}{2})}} \hat{B}_{-}
-\frac{4}{(5+k)^2} \hat{A}_{-} \hat{A}_{+}   U^{(\frac{3}{2})}
\right. \nonu \\
& + & \frac{2}{(5+k)} i \hat{B}_{-}   W_{-}^{(\frac{5}{2})}
+\frac{4}{(5+k)^2} \hat{B}_{-} \hat{B}_3 \hat{G}_{12}
-\frac{4}{(5+k)^2} i \hat{B}_{-} \pa \hat{G}_{12}
\nonu \\
& + & \frac{8(4+k)}{3(5+k)^2} i \hat{B}_{-} \pa  T_{-}^{(\frac{3}{2})}
+  \frac{2}{(5+k)} i \hat{A}_{+} W_{+}^{(\frac{5}{2})}
+\frac{4}{(5+k)^2} \hat{A}_{+} \hat{A}_{-}  U^{(\frac{3}{2})}
\nonu \\
& - & \frac{4}{(5+k)^2} \hat{A}_{+} \hat{A}_3 \hat{G}_{21}
+\frac{2(1+2k)}{(5+k)^2} \hat{A}_{+} \hat{A}_3  T_{+}^{(\frac{3}{2})} 
+ \frac{4(8+k)}{3(5+k)^2} i \hat{A}_{+} \pa  \hat{G}_{21}
\nonu \\
& + &  \frac{20(3+k)}{3(5+k)^2} i \hat{A}_{+} \pa  T_{+}^{(\frac{3}{2})} 
+ \frac{4}{(5+k)^2} \hat{A}_3 \hat{A}_{+} \hat{G}_{21}
-\frac{2(1+2k)}{(5+k)^2} \hat{A}_3 \hat{A}_{+}  T_{+}^{(\frac{3}{2})} 
\nonu \\
& - &  \frac{4}{(5+k)^2} \hat{B}_3 \hat{B}_{-} \hat{G}_{12}
+  \frac{4(2+k)}{(5+k)^2} i \hat{B}_3 \pa  U^{(\frac{3}{2})}
- \frac{2}{(5+k)} T^{(1)} \pa  U^{(\frac{3}{2})}
\nonu \\
&+ & \frac{8}{(5+k)^2} i \pa \hat{B}_3   U^{(\frac{3}{2})} 
+\frac{2(-3+k)}{3(5+k)^2} \pa^2 \hat{G}_{11}
+ \frac{9}{(5+k)^2} \pa^2   U^{(\frac{3}{2})}
\nonu \\
&+& \left. \frac{(10+3k)}{(5+k)} \pa   U^{(\frac{5}{2})} 
+\frac{4(3+2k)}{(5+k)^2} i \hat{A}_3 \pa  U^{(\frac{3}{2})}   
-\frac{2}{(5+k)} T_{+}^{(\frac{3}{2})}  U_{-}^{(2)} \right](w) 
+\cdots,
\nonu \\
U^{(\frac{5}{2})}(z) \, W_{+}^{(\frac{5}{2})}(w) & = & 
-\frac{1}{(z-w)^4} \, 
\left[ \frac{32k(10+8k+k^2)}{3(5+k)^3} i \hat{B}_{-} \right](w)
\nonu \\
& + & \frac{1}{(z-w)^3} \, \left[ 
\frac{8(7+4k)}{9(5+k)^2} U_{+}^{(2)}
-\frac{16(57+28k+4k^2)}{9(5+k)^3} \hat{A}_3 \hat{B}_{-}
\right. 
\nonu \\
& - &  \frac{16k(-4+k)}{3(5+k)^3} \hat{B}_{-} \hat{B}_3
 \nonu \\
&-& \left. \frac{8k(16+17k+2k^2)}{3(5+k)^3} i \pa \hat{B}_{-} 
-\frac{8(-4+k)}{3(5+k)^2} i T^{(1)} \hat{B}_{-}
\right](w)
\nonu \\
&+ & \frac{1}{(z-w)^2} \left[ -\frac{(31+10k)}{3(5+k)}
{\bf Q_{+}^{(3)}}  
-\frac{1}{(5+k)} i {\bf P^{(2)}} \hat{B}_{-}
+\frac{4}{(5+k)} T^{(1)} U_{+}^{(2)}
\right. \nonu \\
&- & \frac{8(7+4k)}{3(5+k)^2} i \hat{A}_3   U_{+}^{(2)}
-\frac{2(483+292k+52k^2)}{9(5+k)^3} \hat{A}_3 \pa \hat{B}_{-}
\nonu \\
& - & \frac{8(285+3563k+5819k^2+2041k^3+140k^4)}{
3(5+k)^2(3+7k)(19+23k)} i \hat{B}_{-} \hat{T} 
-\frac{4(31+7k)}{3(5+k)^2} i \hat{B}_{-} T^{(2)}
\nonu \\
&+ & \frac{4(32+7k)}{3(5+k)^2} i \hat{B}_{-} W^{(2)} 
+\frac{2(315+236k+20k^2)}{
9(5+k)^3} \hat{B}_{-} \pa \hat{A}_3
\nonu \\
& + & \frac{2(60+85k+2k^2)}{3(5+k)^3} \hat{B}_{-} \pa \hat{B}_3 
+\frac{(73+2k)}{3(5+k)^2} i \hat{B}_{-} \pa T^{(1)}
-\frac{8k}{(5+k)^2} i \hat{B}_3   U_{+}^{(2)}
\nonu \\
& - & \frac{2k(53+10k)}{3(5+k)^3} \hat{B}_3 \pa \hat{B}_{-}
+\frac{(253+106k)}{9(5+k)^2} \pa  U_{+}^{(2)}
+\frac{4(25+9k)}{3(5+k)^2} \hat{G}_{11}  T_{+}^{(\frac{3}{2})}
\nonu \\
& - & 
\frac{8(-15+16k+21k^2+2k^3)}{9(5+k)^3} i \pa^2 \hat{B}_{-}
+\frac{4(6+k)}{3(5+k)^2} \hat{G}_{11} \hat{G}_{21}
\nonu \\
& - & \frac{4k}{(5+k)^2} \hat{G}_{21}  U^{(\frac{3}{2})}
+\frac{4(11+k)}{3(5+k)^2}  T_{+}^{(\frac{3}{2})}  U^{(\frac{3}{2})}
-\frac{8(13+6k)}{3(5+k)^3} i \hat{A}_{+} \hat{A}_{-} \hat{B}_{-}
\nonu \\
&- & \frac{8(-13+4k)}{3(5+k)^3} i 
\hat{A}_3 \hat{A}_3 \hat{B}_{-} 
+\frac{16(-4+k)}{3(5+k)^3} i \hat{A}_3 \hat{B}_{-} \hat{B}_3
-\frac{8}{3(5+k)^2} i \hat{B}_{+} \hat{B}_{-} \hat{B}_{-}
\nonu \\
& + & \frac{8(-5+2k)}{3(5+k)^3} i \hat{B}_{-} \hat{B}_3 \hat{B}_3
+\frac{8}{(5+k)^2} T^{(1)} \hat{A}_3 \hat{B}_{-} 
\nonu \\
&+ & \left. \frac{(29+10k)}{3(5+k)^2} T^{(1)}
\hat{B}_{-} \hat{B}_3
-\frac{(53+10k)}{3(5+k)^2} \hat{B}_3 T^{(1)} \hat{B}_{-}
\right](w) \nonu \\
& + &  
   \frac{1}{(z-w)} \left[ 
\frac{(22+9k)}{3(5+k)} i \hat{B}_3 {\bf Q_{+}^{(3)}}
+\frac{(22+9k)}{6(5+k)} i \hat{}B_{-} {\bf S^{(3)}}
+\frac{(28+9k)}{6(5+k)} i \hat{B}_{-} {\bf P^{(3)}}
\right. \nonu \\
&- & \frac{(22+9k)}{6(5+k)} \hat{B}_{-} \hat{B}_3   {\bf P^{(2)}}
 -\frac{(34+9k)}{12(5+k)} i \pa \hat{B}_{-}  {\bf P^{(2)}}
-\frac{1}{4(5+k)} i \hat{B}_{-} \pa  {\bf P^{(2)}}
\nonu \\
& - & \frac{(12+5k)}{3(5+k)} \pa  {\bf Q_{+}^{(3)}}
\nonu \\
&- & \frac{(16+9k)}{3(5+k)} U_{+}^{(2)} W^{(2)}
-\frac{(28+9k)}{3(5+k)} T^{(2)}  U_{+}^{(2)}
-\frac{2}{(5+k)}  T_{+}^{(\frac{3}{2})}  U^{(\frac{5}{2})}
\nonu \\ 
&+& \frac{4}{(5+k)^2} \pa  T_{+}^{(\frac{3}{2})}  U^{(\frac{3}{2})}
+\frac{4(14+k)}{3(5+k)^2}   T_{+}^{(\frac{3}{2})} \pa  U^{(\frac{3}{2})}
-\frac{2(22+9k)}{3(5+k)^2} i \pa \hat{A}_3  U_{+}^{(2)}
\nonu \\
& -& \frac{4(3+2k)}{(5+k)^2} i   \hat{A}_3  \pa U_{+}^{(2)}
+\frac{2(22+9k)}{3(5+k)^2} \hat{A}_3 \hat{A}_3  U_{+}^{(2)}
+\frac{4(22+9k)}{3(5+k)^3} \hat{A}_3 \hat{A}_3 \hat{A}_3 \hat{B}_{-} 
\nonu \\
& + & \frac{4(57+14k)}{3(5+k)^3} i \pa \hat{A}_3 \hat{A}_3 \hat{B}_{-}
+\frac{4(252+130k+11k^2)}{3(5+k)^3} i \hat{A}_3 \hat{A}_3 \pa \hat{B}_{-}
\nonu \\
& + & \frac{8(226+138k+11k^2)}{3(5+k)^3}  
\hat{A}_3 \hat{A}_3 \hat{B}_{-} \hat{B}_{3}
-\frac{2(22+9k)}{3(5+k)^2} i \hat{A}_3 \hat{A}_3 \hat{B}_{-} T^{(1)} 
\nonu \\
& - & \frac{2(22+9k)}{3(5+k)^2} \hat{A}_3 \hat{B}_{-} W^{(2)} 
-\frac{2(22+9k)}{3(5+k)^2} \hat{A}_3 \hat{B}_{-} T^{(2)} 
\nonu \\
& + & \frac{(-13206-25117k+7310k^2+11619k^3+1886k^4)}{
3(5+k)^3(3+7k)(19+23k)} \pa^2 \hat{A}_3 \hat{B}_{-}  
\nonu \\
& - &  \frac{(44898+159545k+164002k^2+62969k^3+774k^4)}{
3(5+k)^3(3+7k)(19+23k)}  \hat{A}_3 \pa^2 \hat{B}_{-} 
\nonu \\
&+ & \frac{4(140+101k+4k^2)}{3(5+k)^3} i \pa \hat{A}_3 \hat{B}_{-} \hat{B}_3
+\frac{2(1314+1011k+124k^2)}{
3(5+k)^3} i \hat{A}_3 \pa \hat{B}_{-} \hat{B}_3
\nonu \\
&-& \frac{2(242+177k+19k^2)}{
(5+k)^3} i \hat{A}_3  \hat{B}_{-} \pa \hat{B}_3
+\frac{4(298+235k+31k^2)}{3(5+k)^3}  
\hat{A}_3 \hat{B}_{-} \hat{B}_3 \hat{B}_3
\nonu \\
&- & \frac{8(22+9k)}{3(5+k)^2} i \hat{A}_3 \hat{B}_{-} \hat{B}_3 T^{(1)} 
+\frac{2(31+9k)}{3(5+k)^2} \pa \hat{A}_3 \hat{B}_{-}  T^{(1)} 
+\frac{2(26+9k)}{(5+k)^2} \hat{A}_3 \pa \hat{B}_{-}  T^{(1)}  
\nonu \\
& - & \frac{2}{(5+k)^2} \hat{A}_3 \hat{B}_{-} \pa T^{(1)}  
-\frac{(784+616k+75k^2)}{
3(5+k)^3} \hat{A}_3 \hat{B}_{+} \hat{B}_{-} \hat{B}_{-}
\nonu \\
&+& \frac{4}{(5+k)^2} \hat{A}_{-} \hat{B}_{-}  U_{-}^{(2)}
+\frac{2(22+9k)}{3(5+k)^2} \hat{A}_{+} \hat{A}_{-}  U_{+}^{(2)}
+ \frac{4(22+9k)}{3(5+k)^3} 
\hat{A}_{+} \hat{A}_{-} \hat{A}_{3} \hat{B}_{-}
\nonu \\
& - & \frac{(78+25k)}{3(5+k)^3} i \pa \hat{A}_{+} \hat{A}_{-} \hat{B}_{-}  
+ \frac{(54+29k)}{3(5+k)^3} i  \hat{A}_{+} \pa \hat{A}_{-} \hat{B}_{-} 
\nonu \\
& + & 
\frac{2(238+175k+22k^2)}{3(5+k)^3} i  \hat{A}_{+}  \hat{A}_{-} \pa \hat{B}_{-} 
+\frac{4(298+213k+22k^2)}{3(5+k)^3} \hat{A}_{+} \hat{A}_{-} \hat{B}_{-} 
\hat{B}_3
\nonu \\
&- & \frac{2(22+9k)}{3(5+k)^2} i  \hat{A}_{+} \hat{A}_{-} \hat{B}_{-} T^{(1)}  
+\frac{2(28+9k)}{3(5+k)^2}  \hat{A}_{+} \hat{B}_{-}  V_{+}^{(2)}
\nonu \\
&- & \frac{(608+580k+75k^2)}{9(5+k)^2} i \pa \hat{B}_3  U_{+}^{(2)} 
- \frac{(110+57k)}{3(5+k)^2} i  \hat{B}_3  \pa U_{+}^{(2)} 
\nonu \\
&+& \frac{4(22+9k)}{3(5+k)^2} i \hat{B}_3  T_{+}^{(\frac{3}{2})}  U^{(\frac{3}{2})}
+\frac{2(1070+733k+75k^2)}{9(5+k)^2} 
\hat{B}_3 \hat{B}_3  U_{+}^{(2)} 
\nonu \\
&- & \frac{4(22+9k)}{3(5+k)^2} i 
\hat{B}_3 \hat{G}_{11}  T_{+}^{(\frac{3}{2})}  
-\frac{2(22+9k)}{3(5+k)} i \hat{B}_3 T^{(1)}   U_{+}^{(2)}
-\frac{(28+9k)}{3(5+k)} i \hat{B}_{-} W^{(3)}
\nonu \\
&- & \frac{(170+331k+48k^2)}{9(5+k)^2} i \pa \hat{B}_{-} W^{(2)}
- \frac{(-58+7k)}{6(5+k)^2} i  \hat{B}_{-} \pa W^{(2)}
\nonu \\
& - & 
\frac{2(22+9k)}{3(5+k)^2} i \hat{B}_{-}  U^{(\frac{3}{2})}  V^{(\frac{3}{2})}
-\frac{(366+237k+22k^2)}{3(5+k)^2} i \pa \hat{B}_{-} T^{(2)}
\nonu \\
&- & \frac{(254+81k)}{6(5+k)^2}  i \hat{B}_{-} \pa T^{(2)}
-\frac{2(22+9k)}{3(5+k)^2} i \hat{B}_{-}  T_{+}^{(\frac{3}{2})}  T_{-}^{(\frac{3}{2})}
\nonu \\
& - & \frac{(1012662+2910798k+3006261k^2+
1518313k^3+323453k^4+18513k^5)}{
54(5+k)^3(3+7k)(19+23k)} i 
\nonu \\
& \times & \pa^3 \hat{B}_{-}
- \frac{2(542+451k+48k^2)}{
9(5+k)^2} \hat{B}_{-} \hat{B}_3 W^{(2)}
-\frac{2(386+249k+22k^2)}{3(5+k)^2} \hat{B}_{-} \hat{B}_3  T^{(2)}
\nonu \\
& - & \frac{2(356706+1237359k+1475724k^2+
707068k^3 + 117398k^4 + 6009k^5)}{
9(5+k)^3(3+7k)(19+23k)} \nonu \\
& \times & \pa^2 \hat{B}_{-} \hat{B}_3
-  \frac{(3814+3065k+434k^2+9k^3)}{3(5+k)^3} \pa \hat{B}_{-} 
\pa \hat{B}_3
\nonu \\
& - & \frac{(346446+872313k+802528k^2+409492k^3+
109558k^4+7671k^5)}{
9(5+k)^3(3+7k)(19+23k)}  \hat{B}_{-} \pa^2 \hat{B}_3
\nonu \\
&- & \frac{2(660+458k+53k^2)}{3(5+k)^3} i \pa \hat{B}_{-} \hat{B}_3 
\hat{B}_3
\nonu \\
& - & \frac{2(938+679k+75k^2)}{9(5+k)^2} i  \hat{B}_{-}  \hat{B}_3 
\hat{B}_3 T^{(1)} +\frac{(1606+1277k+150k^2)}{9(5+k)^2} 
\pa \hat{B}_{-} \hat{B}_3 T^{(1)}
\nonu \\
&- & \frac{(560+517k+75k^2)}{9(5+k)^2}  \hat{B}_{-} \pa \hat{B}_3 T^{(1)}
+
\frac{2(19+9k)}{3(5+k)^2}  \hat{B}_{-}  \hat{B}_3 \pa T^{(1)}
\nonu \\
&- & \frac{(22+9k)}{3(5+k)} \hat{B}_{-} \hat{B}_3 T^{(1)} T^{(1)}
+\frac{(872+652k+75k^2)}{18(5+k)^2} \hat{B}_{-} \hat{B}_{-}  V_{-}^{(2)}
\nonu \\
&+& \frac{(22+9k)}{3(5+k)} i \hat{B}_{-} T^{(1)} W^{(2)} 
+\frac{(22+9k)}{3(5+k)} i \hat{B}_{-} T^{(1)} T^{(2)} 
\nonu \\
& + & \frac{(13710+44947k+48300k^2+18095k^3+1656k^4)}{
6(5+k)^2(3+7k)(19+23k)} i \pa^2 \hat{B}_{-} T^{(1)}
\nonu \\
&+ & \frac{(22+9k)}{3(5+k)^2} i \pa \hat{B}_{-} \pa T^{(1)} 
 +\frac{2(2716+2137k+301k^2+9k^3)}{3(5+k)^3} 
i \hat{B}_{-} \pa \hat{B}_3 
\hat{B}_3 
\nonu \\
& + & \frac{(18570+56503k+53180k^2+16275k^3+2484k^4)}{
9(5+k)^2(3+7k)(19+23k)} i  \hat{B}_{-} \pa^2 T^{(1)} 
\nonu \\
& - & \frac{(22+9k)}{6(5+k)} i \pa \hat{B}_{-} T^{(1)} T^{(1)}
-\frac{(740+598k+75k^2)}{9(5+k)^2} \hat{B}_{+} \hat{B}_{-} U_{+}^{(2)}
\nonu \\
& -& \frac{(3210+2575k+372k^2+9k^3)}{6(5+k)^3}  
i \pa \hat{B}_{+} \hat{B}_{-} \hat{B}_{-}
\nonu \\
& + & \frac{(2588+2109k+310k^2+9k^3)}{
3(5+k)^3} i  \hat{B}_{+} \pa \hat{B}_{-} \hat{B}_{-}
\nonu \\
& + & 
\frac{(2086+1667k+248k^2+9k^3)}{3(5+k)^3}
 \hat{B}_{+} \hat{B}_{-} \hat{B}_{-} \hat{B}_3
+\frac{4(3+4k)}{3(5+k)^2} \pa \hat{G}_{11}  T_{+}^{(\frac{3}{2})}
\nonu \\
&+ & \frac{(608+544k+75k^2)}{
18(5+k)^2}  \hat{B}_{+} \hat{B}_{-} \hat{B}_{-} T^{(1)}
+
\frac{2}{(5+k)} \hat{G}_{11}  W_{+}^{(\frac{5}{2})}
\nonu \\
& + & \frac{20(3+k)}{3(5+k)^2} \hat{G}_{11} \pa  T_{+}^{(\frac{3}{2})}
-
\frac{4}{(5+k)^2} \pa \hat{G}_{11} \hat{G}_{21}
+\frac{4(8+k)}{3(5+k)^2} \hat{G}_{11} \pa \hat{G}_{21} 
\nonu \\
&- & \frac{4k}{(5+k)^2} \hat{G}_{21} \pa  U^{(\frac{3}{2})}
+\frac{2}{(5+k)} \pa T^{(1)}  U_{+}^{(2)}
+\frac{2}{(5+k)} T^{(1)} \pa  U_{+}^{(2)}
\nonu \\
&+ & \frac{2(66+97k+36k^2)}{3(5+k)(3+7k)} \hat{T}  U_{+}^{(2)} 
+\frac{4(2586+7351k+4589k^2+828k^3)}{
3(5+k)^2(3+7k)(19+23k)} \hat{T} \hat{A}_3 \hat{B}_{-} 
\nonu \\
&-  & \frac{(-1938+8768k+19977k^2+7243k^3+560k^4)}{
3(5+k)^2(3+7k)(19+23k)} i  \pa \hat{T}  \hat{B}_{-} 
\nonu \\
&+ & \frac{2(13338+34161k+28787k^2+11064k^3+1240k^4)}{
3(5+k)^2(3+7k)(19+23k)} i \hat{T} \pa \hat{B}_{-} 
\nonu \\
& + & \frac{4(16986+49087k+47415k^2+17510k^3+1520k^4)}{
3(5+k)^2(3+7k)(19+23k)} \hat{T} \hat{B}_{-} \hat{B}_3
\nonu \\
&-& \left. \frac{2(1314+3933k+3539k^2+828k^3)}{
3(5+k)(3+7k)(19+23k)} i \hat{T} \hat{B}_{-} T^{(1)} 
\right](w) + \cdots,
\nonu \\ 
U^{(\frac{5}{2})}(z) \, W_{-}^{(\frac{5}{2})}(w) & = & 
-\frac{1}{(z-w)^4} \, \left[ \frac{16(28+19k+2k^2)}{
(5+k)^3} i \hat{A}_{+} \right](w) 
\nonu \\
&+& \frac{1}{(z-w)^3} \left[ -\frac{8(-10+9k)}{9(5+k)^2} U_{-}^{(2)} 
+\frac{16(7+k)}{(5+k)^3} \hat{A}_{+} \hat{A}_3
\right. \nonu \\
& + &  
\frac{32(-6-17k+2k^2)}{9(5+k)^3} \hat{A}_{+} \hat{B}_3
-\frac{8(35+20k+2k^2)}{(5+k)^3} i \pa \hat{A}_{+}
\nonu \\
&+ & \left.
\frac{16(-2+k)}{3(5+k)^2} i T^{(1)} \hat{A}_{+}
\right](w)
\nonu \\
& + & \frac{1}{(z-w)^2} \, \left[ 
-\frac{2(19+4k)}{3(5+k)} {\bf Q_{-}^{(3)}}
-\frac{4}{(5+k)} T^{(1)} U_{-}^{(2)} -\frac{74}{3(5+k)^2} i T^{(1)}
\pa \hat{A}_{+}
\right. \nonu \\
& - & \frac{8(5358+15169k+8385k^2+1226k^3)}{
3(5+k)^2(3+7k)(19+23k)} i \hat{A}_{+} \hat{T}
+\frac{4(29+8k)}{3(5+k)^2} i \hat{A}_{+} W^{(2)}
\nonu \\
&+ & \frac{4(137+44k)}{3(5+k)^3} \hat{A}_{+} \pa \hat{A}_3 
+\frac{2(15+8k)}{3(5+k)^2} i \hat{A}_{+} \pa T^{(1)}
+\frac{8(5+2k)}{3(5+k)^2} i \hat{A}_3  U_{-}^{(2)}
\nonu \\
&+ & \frac{4(263+108k+8k^2)}{
3(5+k)^3} \hat{A}_3 \pa \hat{A}_{+} 
-\frac{8(31+4k)}{3(5+k)^2} i \hat{B}_3   U_{-}^{(2)}
-\frac{2(151+66k)}{9(5+k)^2} \pa U_{-}^{(2)} 
\nonu \\
&+ & \frac{4(37+6k)}{3(5+k)^2} \hat{G}_{11} \hat{G}_{12}  
-\frac{32}{3(5+k)} \hat{G}_{11}   T_{-}^{(\frac{3}{2})}
-\frac{8(13+k)}{3(5+k)^2} \hat{G}_{12}  U^{(\frac{3}{2})}
\nonu \\
&+ & \frac{8(19+4k)}{3(5+k)^2}   T_{-}^{(\frac{3}{2})}  U^{(\frac{3}{2})}
-\frac{8(31+10k)}{3(5+k)^3} i \hat{A}_{+} \hat{A}_{+} \hat{A}_{-}
\nonu \\
& + & \frac{16(48+25k+2k^2)}{3(5+k)^3} i  \hat{A}_{+} \hat{A}_3 \hat{A}_3
\nonu \\
&- &  \frac{4(66+13k+20k^2)}{9(5+k)^3}   i  
\hat{A}_{+} \hat{A}_3 \hat{B}_3
-\frac{2(1020+317k+4k^2)}{9(5+k)^3} i \hat{A}_{+} \hat{B}_{+} \hat{B}_{-}
\nonu \\
&- & \frac{32(10+k)}{3(5+k)^3} i   \hat{A}_{+} \hat{B}_3 \hat{B}_3
-\frac{16(59+30k+2k^2)}{3(5+k)^3} i   \hat{A}_3 \hat{A}_{+} \hat{A}_3
\nonu \\
&+ & \frac{4(438+97k+20k^2)}{9(5+k)^3}  i   \hat{A}_3 \hat{A}_{+} \hat{B}_3
+\frac{2(468+209k+4k^2)}{
9(5+k)^3}  i   \hat{B}_{-} \hat{A}_{+} \hat{B}_{+}
\nonu \\
&+& \left. \frac{8}{(5+k)^2}   T^{(1)} \hat{A}_{+} \hat{A}_3
-\frac{8}{(5+k)^2}  T^{(1)} \hat{A}_{+} \hat{B}_3 
\right](w) 
\nonu \\
& + &  \frac{1}{(z-w)} \, \left[ 
-\frac{2}{(5+k)} i \hat{A}_3  {\bf Q_{-}^{(3)}}
+\frac{2}{(5+k)} i \hat{B}_3  {\bf Q_{-}^{(3)}}
-\frac{(15+4k)}{3(5+k)} \pa  {\bf Q_{-}^{(3)}} 
\right. \nonu \\
&- & \frac{2}{(5+k)}   U_{-}^{(2)} W^{(2)}
-\frac{(45+22k)}{3(5+k)^2} \pa^2  U_{-}^{(2)}
-\frac{2}{(5+k)} T^{(2)}  U_{-}^{(2)}
\nonu \\
&+ & \frac{4(15+4k)}{3(5+k)^2} \pa  U_{-}^{(2)}   U^{(\frac{3}{2})}
+  \frac{4(15+4k)}{3(5+k)^2}   U_{-}^{(2)}   \pa U^{(\frac{3}{2})}
+\frac{16(4+k)}{3(5+k)^2} i \pa \hat{A}_3  U_{-}^{(2)} 
\nonu \\
& + & \frac{2(-7+2k)}{3(5+k)^2} i  \hat{A}_3  \pa U_{-}^{(2)} 
+\frac{8}{(5+k)^2} i \hat{A}_3  T_{-}^{(\frac{3}{2})}  U^{(\frac{3}{2})}
+\frac{4}{(5+k)^2} \hat{A}_3 \hat{A}_3  U_{-}^{(2)}
\nonu \\
& +& \frac{8}{(5+k)^2} \hat{A}_3 \hat{B}_3  U_{-}^{(2)}
-\frac{8}{(5+k)^2} i \hat{A}_3 \hat{G}_{11} T_{-}^{(\frac{3}{2})}
+\frac{8}{(5+k)^2} i \hat{A}_3 \hat{G}_{11} \hat{G}_{12}
\nonu \\
& - & \frac{8}{(5+k)^2} i \hat{A}_3 \hat{G}_{12}  U^{(\frac{3}{2})}
+\frac{2}{(5+k)} i \hat{A}_{+} W^{(3)}
+\frac{2(37+10k)}{3(5+k)^2} i \pa \hat{A}_{+} W^{(2)}
\nonu \\
& +& \frac{2(22+7k)}{3(5+k)^2} i \hat{A}_{+}  \pa W^{(2)}
+\frac{2(7+2k)}{(5+k)^2} i \pa \hat{A}_{+} T^{(2)}
+\frac{6(3+k)}{(5+k)^2} i \hat{A}_{+} \pa  T^{(2)}
\nonu \\
&+& \frac{(42237+121467k+72569k^2+18101k^3+1946k^4)}{
3(5+k)^3(3+7k)(19+23k)} i \pa^3 \hat{A}_{+}
\nonu \\
& - & \frac{8}{(5+k)^2} \hat{A}_{+} \hat{A}_3  T^{(2)}
-\frac{2(1839+4018k+895k^2+844k^3)}{
(5+k)^3(3+7k)(19+23k)} \pa^2 \hat{A}_{+} \hat{A}_3
\nonu \\
&+& \frac{8(14+5k)}{3(5+k)^3} \pa \hat{A}_{+} \pa \hat{A}_3
+\frac{2(411+12146k+25371k^2+6484k^3)}{3(5+k)^3(3+7k)(19+23k)}  
\hat{A}_{+} \pa^2 \hat{A}_3
\nonu \\
&- & \frac{40(1+k)}{3(5+k)^3} i \pa \hat{A}_{+} \hat{A}_3 \hat{A}_3
-\frac{8(19+10k)}{3(5+k)^3} i \hat{A}_{+} \hat{A}_3 \pa \hat{A}_3
+\frac{16}{(5+k)^3} \hat{A}_{+} \hat{A}_3 \hat{A}_3 \hat{A}_3 
\nonu \\
&- & \frac{48}{(5+k)^3} \hat{A}_{+} \hat{A}_3 \hat{A}_3 \hat{B}_3 
+\frac{4(71+17k)}{3(5+k)^3} i \pa  \hat{A}_{+} \hat{A}_3 \hat{B}_3
+ \frac{4(47+17k)}{3(5+k)^3} i   \hat{A}_{+} \pa \hat{A}_3 \hat{B}_3
\nonu \\
&+&  \frac{16(11+2k)}{3(5+k)^3} i   \hat{A}_{+}  \hat{A}_3 \pa \hat{B}_3
+\frac{48}{(5+k)^3} \hat{A}_{+} \hat{A}_3 \hat{B}_3 \hat{B}_3 
+\frac{16}{(5+k)^3} \hat{A}_{+} \hat{A}_3 \hat{B}_{+} \hat{B}_{-} 
\nonu \\
&+ & \frac{6}{(5+k)^2} \pa  \hat{A}_{+} \hat{A}_3 T^{(1)}
+\frac{2}{(5+k)^2} \hat{A}_{+} \pa \hat{A}_3 T^{(1)} 
+ \frac{4}{(5+k)^2} \hat{A}_{+}  \hat{A}_3 \pa T^{(1)} 
\nonu \\
&+& \frac{4}{(5+k)^2} \hat{A}_{+} \hat{A}_{-}  U_{-}^{(2)}
-\frac{20(13+4k)}{3(5+k)^3} i \pa \hat{A}_{+} \hat{A}_{+} \hat{A}_{-}
-\frac{8(11+5k)}{3(5+k)^3} i  \hat{A}_{+} \hat{A}_{+} \pa \hat{A}_{-}
\nonu \\
& + & \frac{16}{(5+k)^3} \hat{A}_{+} \hat{A}_{+}  \hat{A}_{-} \hat{A}_3
- \frac{16}{(5+k)^3} \hat{A}_{+} \hat{A}_{+}  \hat{A}_{-} \hat{B}_3
+\frac{8}{(5+k)^2} \hat{A}_{+} \hat{B}_3 T^{(2)}
\nonu \\
&+& \frac{2(16203+54736k+47583k^2+11218k^3+1288k^4)}{
3(5+k)^3(3+7k)(19+23k)} \pa^2 \hat{A}_{+} \hat{B}_3
\nonu \\
&- & \frac{4(39+22k+2k^2)}{3(5+k)^3}  
\pa \hat{A}_{+} \pa \hat{B}_3
-\frac{2(101+17k)}{3(5+k)^3} i  \hat{A}_{+}  \hat{B}_{+} \pa \hat{B}_{-}
\nonu \\
& - & \frac{2(-2409-266k+11631k^2+6176k^3)}{
3(5+k)^3(3+7k)(19+23k)} \hat{A}_{+} \pa^2 \hat{B}_3
\nonu \\
&- & \frac{4(61+7k)}{3(5+k)^3} i \pa \hat{A}_{+} \hat{B}_3 \hat{B}_3 
-\frac{4(53+5k)}{3(5+k)^3} i \hat{A}_{+} \hat{B}_3 \pa \hat{B}_3
-\frac{16}{(5+k)^3}  \hat{A}_{+} \hat{B}_3 \hat{B}_3  \hat{B}_3
\nonu \\
& - & \frac{6}{(5+k)^2} \pa \hat{A}_{+} \hat{B}_3 T^{(1)}
-\frac{2}{(5+k)^2}  \hat{A}_{+} \pa \hat{B}_3 T^{(1)}
-\frac{4}{(5+k)^2}  \hat{A}_{+}  \hat{B}_3 \pa T^{(1)}
\nonu \\
&- & \frac{8(26+5k)}{3(5+k)^3} i \pa \hat{A}_{+} \hat{B}_{+} \hat{B}_{-}
-\frac{2(53+17k)}{3(5+k)^3} i  \hat{A}_{+} \pa \hat{B}_{+} \hat{B}_{-}
\nonu \\
&- & \frac{16}{(5+k)^3} \hat{A}_{+} \hat{B}_{+} \hat{B}_{-} \hat{B}_3
+\frac{4}{(5+k)^2} i \hat{A}_{+} \hat{G}_{12} \hat{G}_{21}
+\frac{4}{(5+k)^2} i \hat{A}_{+} \hat{G}_{21} T_{-}^{(\frac{3}{2})}
\nonu \\
&- & \frac{2(35+123k+4k^2)}{(5+k)^2(19+23k)} i \pa^2 \hat{A}_{+} T^{(1)}
-\frac{2}{(5+k)^2} i  \pa \hat{A}_{+} \pa T^{(1)}
\nonu \\
& + & \frac{(531+311k+160k^2)}{3(5+k)^2(19+23k)} i \hat{A}_{+} \pa^2 T^{(1)}
-\frac{4(25+4k)}{3(5+k)^2} i \pa \hat{B}_3  U_{-}^{(2)}
\nonu \\
& - & \frac{2(47+8k)}{3(5+k)^2} i  \hat{B}_3  \pa U_{-}^{(2)}
\nonu \\
& -& \frac{8}{(5+k)^2} i \hat{B}_3  T_{-}^{(\frac{3}{2})}  U^{(\frac{3}{2})}
- \frac{12}{(5+k)^2}  \hat{B}_3  \hat{B}_3  U_{-}^{(2)}
+\frac{8}{(5+k)^2} i \hat{B}_3 \hat{G}_{11}  T_{-}^{(\frac{3}{2})}  
\nonu \\
& - & \frac{8}{(5+k)^2} i \hat{B}_3 \hat{G}_{11} \hat{G}_{12}
+\frac{8}{(5+k)^2} i \hat{B}_3 \hat{G}_{12}  U^{(\frac{3}{2})}
-\frac{2}{(5+k)} \hat{G}_{11} W_{-}^{(\frac{5}{2})}
\nonu \\
&- & \frac{4(11+2k)}{3(5+k)^2} \pa \hat{G}_{11} T_{-}^{(\frac{3}{2})}
- \frac{8(10+3k)}{3(5+k)^2}  \hat{G}_{11} \pa T_{-}^{(\frac{3}{2})}
+\frac{8(7+k)}{3(5+k)^2} \pa \hat{G}_{11} \hat{G}_{12}
\nonu \\
&+& \frac{4(15+4k)}{3(5+k)^2}  \hat{G}_{11} \pa \hat{G}_{12}
-\frac{4(15+4k)}{3(5+k)^2} \pa \hat{G}_{12}  U^{(\frac{3}{2})}
+\frac{4(-3+2k)}{3(5+k)^2} \hat{G}_{12} \pa  U^{(\frac{3}{2})}
\nonu \\
&- & \frac{2}{(5+k)} \pa T^{(1)}   U_{-}^{(2)}
- \frac{2}{(5+k)}  T^{(1)} \pa  U_{-}^{(2)}
+\frac{4(3+4k)}{(5+k)(3+7k)} \hat{T}  U_{-}^{(2)}
\nonu \\
&- & \frac{2(9249+26792k+16113k^2+2314k^3)}{
3(5+k)^2(3+7k)(19+23k)} i \pa \hat{T} \hat{A}_{+}
\nonu \\
&- & \frac{4(5511+15811k+8652k^2+1364k^3)}{3(5+k)^2(3+7k)(19+23k)} 
i \hat{T} \pa \hat{A}_{+}
\nonu \\
& + & \frac{16(129+295k+50k^2)}{(5+k)^2(3+7k)(19+23k)} 
\hat{T} \hat{A}_{+} \hat{A}_3
-\frac{16(138+373k+183k^2)}{(5+k)^2(3+7k)(19+23k)}
\hat{T} \hat{A}_{+} \hat{B}_3
\nonu \\
&+& \left. \frac{16(-3+k)}{(5+k)(19+23k)} i \hat{T} \hat{A}_{+} T^{(1)}
\right](w) 
+\cdots,
\nonu \\
U^{(\frac{5}{2})}(z) \, W^{(3)}(w) & = & 
\frac{1}{(z-w)^4} \, \left[ -\frac{4(218 k^4+2688 k^3+10195 k^2+13278 k+3585)}{
3(5+k)^3(19+23k)} \hat{G}_{11} \right. \nonu \\
&- & \left.  \frac{4(7 k^3-639 k^2-1937 k-1091)}{
(5+k)^3(19+23k)} U^{(\frac{3}{2})} 
    \right](w)
\nonu \\
& + & 
\frac{1}{(z-w)^3} \, 
\left[ \frac{(-6229-2564k+289k^2+152k^3)}{
3(5+k)^2(19+23k)} {\bf Q^{(\frac{5}{2})}} 
\right. \nonu \\
& + & \frac{2(-3735-2212k+279k^2+76k^3)}{
3(5+k)^2(19+23k)} U^{(\frac{5}{2})}
+\frac{8(42+2k+k^2)}{3(5+k)^3} i \hat{A}_{+} \hat{G}_{21}
\nonu \\
& -& \frac{8k(4+k)}{3(5+k)^3} i \hat{A}_{+}  T_{+}^{(\frac{3}{2})}
-\frac{8(8+k)(3+2k)}{3(5+k)^3} i \hat{A}_3 \hat{G}_{11}
\nonu \\
&+& 
\frac{4(7+13k+2k^2)}{3(5+k)^3} i \hat{B}_{-} \hat{G}_{12}
- \frac{8(-1+8k+2k^2)}{3(5+k)^3} i \hat{B}_{-}  T_{-}^{(\frac{3}{2})}
\nonu \\
&+ & \frac{8(12+23k+2k^2)}{3(5+k)^3} i \hat{B}_3 \hat{G}_{11}
-\frac{8(-17-8k+2k^2)}{3(5+k)^3} i \hat{B}_3   U^{(\frac{3}{2})}
\nonu \\
& - & \frac{4(3+2k)}{3(5+k)^2} T^{(1)} \hat{G}_{11}
-\frac{4(17+4k)}{3(5+k)^2} T^{(1)}  U^{(\frac{3}{2})}
+\frac{8(-10+k)(7+2k)}{3(5+k)^3} i \hat{A}_3   U^{(\frac{3}{2})}
\nonu \\
&- & \frac{4(218 k^4+2734 k^3+9727 k^2+14102 k+4611)}{
9(5+k)^3(19+23k)} \pa \hat{G}_{11}
\nonu \\
& - & \left. \frac{4(53 k^3-785 k^2-3193 k-2003)}{
3(5+k)^3(19+23k)} \pa  U^{(\frac{3}{2})}
\right](w)
\nonu \\
& + & 
\frac{1}{(z-w)^2} \, \left[ \frac{(-21233-11116k+377k^2+364k^3)}{
15(5+k)^2(19+23k)} \pa {\bf Q^{(\frac{5}{2})}} \right. \nonu \\
&- &
\frac{(43+8k)}{3(5+k)^2} i {\bf P_{-}^{(\frac{5}{2})}} \hat{B}_{-} 
+\frac{(29+7k)}{2(5+k)} {\bf Q^{(\frac{7}{2})}}
+\frac{(13+5k)}{3(5+k)^2} i {\bf Q^{(\frac{5}{2})}} \hat{A}_3
\nonu \\
&- & \frac{(13+5k)}{3(5+k)^2}  i {\bf Q^{(\frac{5}{2})}} \hat{B}_3
-\frac{4(264+112k+5k^2)}{
15(5+k)^3} \hat{A}_{-} \hat{A}_{+}
 \hat{G}_{11}
\nonu \\
&+& \frac{(129+465k+64k^2)}{
15(5+k)^3}  \hat{A}_{-} \hat{A}_{+} U^{(\frac{3}{2})}
-\frac{2(1254+309k+31k^2)}{
15(5+k)^3} \hat{B}_{+} \hat{B}_{-}
 \hat{G}_{11}
\nonu \\
&-& \frac{(-420-263k+41k^2)}{
15(5+k)^3}  \hat{B}_{+} \hat{B}_{-} U^{(\frac{3}{2})}
-\frac{2(57+13k)}{
(5+k)^3} \hat{B}_{-} \hat{A}_{+}
 \hat{G}_{22}
\nonu \\
& - &  \frac{4(21+k)}{
3(5+k)^3}  \hat{B}_{-} \hat{A}_{+} V^{(\frac{3}{2})}
+
\frac{(51+5k)}{3(5+k)^2} i \hat{B}_{-}  W_{-}^{(\frac{5}{2})}
-\frac{4(153+29k)}{3(5+k)^3}  \hat{B}_{-} \hat{A}_3
 \hat{G}_{12}
\nonu \\
& + &  \frac{4(213+53k)}{
3(5+k)^3}  \hat{B}_{-} \hat{A}_3 T_{-}^{(\frac{3}{2})}
+\frac{2(1074+249k+31k^2)}{15(5+k)^3}  \hat{B}_{-} \hat{B}_{+}
 \hat{G}_{11}
\nonu \\
& + & \frac{(690-113k+41k^2)}{
15(5+k)^3}  \hat{B}_{-} \hat{B}_{+} U^{(\frac{3}{2})}
- \frac{16k}{(5+k)^3}  \hat{B}_{-} \hat{B}_3
 \hat{G}_{12} 
\nonu \\
& - &
 \frac{2(244+303k+83k^2)}{
15(5+k)^3}  \hat{B}_{-} \hat{B}_3 T_{-}^{(\frac{3}{2})}
-\frac{8}{(5+k)^2} i   \hat{B}_{-} T^{(1)}
 \hat{G}_{12} 
\nonu \\
&+ &   \frac{4(-541+273k+73k^2)}{45(5+k)^3} i \hat{B}_{-} \pa
 \hat{G}_{12} 
-\frac{4(-1201+143k+123k^2)}{45(5+k)^3} i \hat{B}_{-} \pa T_{-}^{(\frac{3}{2})}
\nonu \\
& - & \frac{6(9+2k)}{(5+k)^2} i \hat{A}_{+}  W_{+}^{(\frac{5}{2})}
+\frac{4(264+72k+5k^2)}{15(5+k)^3}  \hat{A}_{+} \hat{A}_{-}
 \hat{G}_{11} 
\nonu \\
& - &  \frac{(-2791-135k+64k^2)}{
15(5+k)^3}  \hat{A}_{+} \hat{A}_{-} U^{(\frac{3}{2})}
-\frac{16(16+5k)}{3(5+k)^3}  
\hat{A}_{+} \hat{A}_3 T_{+}^{(\frac{3}{2})}
\nonu \\
& + & \frac{2(-10839+15689k+3987k^2+547k^3)}{
15(5+k)^3(19+23k)} \hat{A}_{+} \hat{A}_3
 \hat{G}_{21} 
\nonu \\
& + & \frac{8(8+k)}{(5+k)^3}  \hat{A}_{+} \hat{B}_3
 \hat{G}_{21} + \frac{16(16+5k)}{3(5+k)^3}
\hat{A}_{+} \hat{B}_3 T_{+}^{(\frac{3}{2})}
+\frac{4}{(5+k)^2} i  \hat{A}_{+} T^{(1)}
 \hat{G}_{21} 
\nonu \\
&- &   \frac{4(-4563+4903k+2199k^2+29k^3)}{15(5+k)^3(19+23k)} i \hat{A}_{+} \pa
 \hat{G}_{21} 
\nonu \\
&
- & \frac{8(129+42k+5k^2)}{15(5+k)^3} i \hat{A}_{+} \pa T_{+}^{(\frac{3}{2})}
- \frac{8(-243+156k+10k^2)}{45(5+k)^3} i \hat{A}_3 \pa
 \hat{G}_{11} 
\nonu \\
& - &  \frac{2(-2859+26109k+4907k^2+547k^3)}{
15(5+k)^3(19+23k)}  \hat{A}_3 \hat{A}_{+}
 \hat{G}_{21} 
\nonu \\
& - & \frac{10(166+399k+77k^2)}{
3(5+k)^2(19+23k)} i \hat{A}_3  U^{(\frac{5}{2})}
- \frac{8(-15+4k)}{3(5+k)^3}  \hat{A}_3 \hat{A}_3
 \hat{G}_{11} 
\nonu \\
& - &  \frac{4(-35+3k)}{3(5+k)^3} \hat{A}_3 \hat{A}_3 U^{(\frac{3}{2})}
+  \frac{16(-6+k)}{3(5+k)^3} \hat{A}_3 \hat{B}_3
 \hat{G}_{11} 
- \frac{8(45+7k)}{3(5+k)^3} \hat{A}_3 \hat{B}_3 U^{(\frac{3}{2})}
\nonu \\
&- & \frac{8}{(5+k)^2}  i \hat{A}_3 T^{(1)}
 \hat{G}_{11} 
-\frac{8}{(5+k)^2} i \hat{A}_3 T^{(1)} U^{(\frac{3}{2})}
\nonu \\
& + & \frac{2(2375+1869k+298k^2)}{
3(5+k)^2(19+23k)} i \hat{B}_3 U^{(\frac{5}{2})}
+\frac{2(244+303k+83k^2)}{
15(5+k)^3} \hat{B}_3 \hat{B}_{-} T_{-}^{(\frac{3}{2})}
\nonu \\
& + &  \frac{8(-3+2k)}{3(5+k)^3} \hat{B}_3 \hat{B}_3
 \hat{G}_{11} +
\frac{4(55+17k)}{3(5+k)^3} \hat{B}_3 \hat{B}_3 U^{(\frac{3}{2})} 
+\frac{8}{(5+k)^2} i \hat{B}_3 T^{(1)}
 \hat{G}_{11} 
\nonu \\
&+& \frac{8}{(5+k)^2} i \hat{B}_3 T^{(1)} U^{(\frac{3}{2})} 
- \frac{8(864-6k+11k^2)}{45(5+k)^3} i \hat{B}_3 \pa
 \hat{G}_{11} 
\nonu \\
& - & \frac{4(-13+k)(250+81k)}{
45(5+k)^3} i \hat{B}_3 \pa  U^{(\frac{3}{2})}
-\frac{(-319+17k+20k^2)}{(5+k)(19+23k)} T^{(1)}  U^{(\frac{5}{2})} 
\nonu \\
& + & \frac{4(12+11k)}{15(5+k)^2}  T^{(1)} \pa
 \hat{G}_{11}  +  \frac{2(53+13k)}{3(5+k)^2}  T^{(1)} \pa
  U^{(\frac{3}{2})}
\nonu \\
& + &  \frac{2(324+207k+10k^2)}{
15(5+k)^3}  i \pa \hat{A}_{+} T_{+}^{(\frac{3}{2})}
- \frac{2(364+333k+53k^2)}{
15(5+k)^3}  i \pa \hat{B}_{-} \hat{G}_{12}
\nonu \\
&- &   \frac{2(81+53k)}{15(5+k)^2}  \pa T^{(1)} 
 \hat{G}_{11}
- \frac{5(43+11k)}{3(5+k)^2}  \pa T^{(1)}
  U^{(\frac{3}{2})}
\nonu \\
& - & \frac{(25760 k^5+389695 k^4+932092 k^3+593696 k^2+364440 k+815157)}
{45(5+k)^3(19+23k)(47+35k)} \pa^2 \hat{G}_{11}
\nonu \\
& + & \frac{(2135 k^3+141152 k^2+695500 k+646807)}{45(5+k)^3(47+35k)} 
\pa^2   U^{(\frac{3}{2})}
\nonu \\
& + & \frac{(364 k^3+1023 k^2-17956 k-25887)}{
15(5+k)^2(19+23k)} \pa  U^{(\frac{5}{2})}
+\frac{2(74+19k)}{3(5+k)^2}   T_{+}^{(\frac{3}{2})} U_{-}^{(2)}
\nonu \\
& - & \frac{4(3919+1325k+24k^2)}{
45(5+k)^3} i \hat{A}_3 \pa    U^{(\frac{3}{2})}
+\frac{20(7+k)}{3(5+k)^2} \hat{G}_{21} U_{-}^{(2)}
\nonu \\
& - & \frac{10(11+k)}{3(5+k)^2}  \hat{G}_{12} U_{+}^{(2)}
+\frac{2(29+7k)}{(5+k)^2}  T_{-}^{(\frac{3}{2})} U_{+}^{(2)}
\nonu \\
&- & \frac{4(11564 k^5+193683 k^4+901082 k^3+1406170 k^2+725478 k+109791)}{
3(5+k)^2(3+7k)(19+23k)(47+35k)} 
\nonu \\
&\times & \hat{G}_{11} \hat{T} 
+  \frac{4(-9+2k)}{3(5+k)^2} \hat{G}_{11} T^{(2)}
+\frac{4(9+2k)}{(5+k)^2}  \hat{G}_{11} W^{(2)}
\nonu \\
& + & \frac{4(37625 k^4+347542 k^3+970624 k^2+859106 k+227391)}{
3(5+k)^2(3+7k)(19+23k)(47+35k)}  U^{(\frac{3}{2})} \hat{T}
\nonu \\
&- & \left. \frac{8(8+k)}{3(5+k)^2}  U^{(\frac{3}{2})} T^{(2)}
-\frac{2(13+3k)}{(5+k)^2}  U^{(\frac{3}{2})} W^{(2)}
\right](w)
\nonu \\
& + & 
 \frac{1}{(z-w)} \, 
 \{ U^{(\frac{5}{2})} \, 
 W^{(3)} \}_{-1}(w)
+\cdots.
\label{u5halfw3}
\eea
The first order pole in the first OPE of (\ref{u5halfw3})
contains composite field with spin-$4$ with $U(1)$ charge of 
$\frac{2(-3+k)}{(5+k)}$.
The $\hat{G}_{11} \hat{G}_{11}(w)$ can be written in terms of 
a derivative of $\hat{A}_{+} \hat{B}_{-}(w)$.
The first order pole in the fifth OPE has a composite field with spin-$4$
with vanishing $U(1)$ charge.
Also the fifth OPE has a term of the first order pole in 
(\ref{g21s7half-}) where 
the higher spin-$4$ current was constructed.
The $(k-3)$ factor appears in $P^{(2)}(w)$, $T^{(2)}(w)$ and $S^{(3)}(w)$ terms.
The composite fields of spin-$2$ except the first term
can be seen from the Table $3$ of \cite{Ahn1311}.
The $(k-3)$ factor appears in the third order pole in the sixth OPE.
The first order pole in the seventh OPE has a composite field with spin-$4$
with $U(1)$ charge of $\frac{2k}{(5+k)}$.
 The first order pole in the eighth OPE has a composite field with spin-$4$
with $U(1)$ charge of $-\frac{6}{(5+k)}$.
The first order pole in the last OPE has a composite field with 
spin-$\frac{9}{2}$
with $U(1)$ charge of $\frac{(-3+k)}{(5+k)}$ and this will appear in 
Appendix $L$ separately.
The $13$ composite fields of spin-$\frac{5}{2}$ in the third order pole
can be found in Table $4$ of \cite{Ahn1311}. 

\section{
The nontrivial 
OPEs between  higher spin-$\frac{3}{2}$ current, $V^{(\frac{3}{2})}(z)$,  
and 
other $8$ higher spin currents
}

Now the OPEs containing third ${\cal N}=2$ multiplet are given by  
\bea
V^{(\frac{3}{2})}(z) \, V^{(\frac{5}{2})}(w) & = & 
\frac{1}{(z-w)^2} \, \left[ \frac{4(-3+k)}{3(5+k)^2} 
\hat{A}_{-} \hat{B}_{+} \right] (w) 
\nonu \\
& + & \frac{1}{(z-w)} \, \left[ -\frac{2}{(5+k)} i \hat{B}_{+}
V_{+}^{(2)} +\frac{2(7+k)}{(5+k)^2} 
\hat{B}_{+} \pa \hat{A}_{-} 
-\frac{2}{(5+k)} i \hat{A}_{-} V_{-}^{(2)}
\right. \nonu \\
&+ & \left. \frac{(12+k)}{3(5+k)} \hat{G}_{22} \hat{G}_{22} 
\right](w)
+\cdots, 
\nonu \\
V^{(\frac{3}{2})}(z) \, W^{(2)}(w) & = & 
\frac{1}{(z-w)^2} \, \left[ \frac{3(4+k)}{(5+k)} V^{(\frac{3}{2})} \right](w)
\nonu \\
& + & \frac{1}{(z-w)} \, \left[ \frac{(4+k)}{(5+k)} \pa V^{(\frac{3}{2})} 
\right](w)
+ \cdots,
\nonu \\
V^{(\frac{3}{2})}(z) \, W_{+}^{(\frac{5}{2})}(w) & = & 
\frac{1}{(z-w)^3} \, \left[ \frac{8(5+2k)}{(5+k)^2} i \hat{A}_{-} \right] (w)
\nonu \\
& - & \frac{1}{(z-w)^2} \, \left[ \frac{4(7+k)}{3(5+k)} V_{+}^{(2)} \right] (w)
\nonu \\
& + & 
\frac{1}{(z-w)} \, \left[ 
\frac{1}{2} {\bf R_{+}^{(3)}} 
+\frac{4(138+343k+113k^2)}{
(5+k)(3+7k)(19+23k)} i \hat{A}_{-} \hat{T}
-\frac{2}{(5+k)} i \hat{A}_{-} T^{(2)} 
\right. \nonu \\
&- & \frac{9}{(5+k)^2}  \hat{A}_{-} \pa \hat{A}_3
-\frac{(8+3k)}{(5+k)^2} \hat{A}_{-} \pa \hat{B}_3 
-\frac{1}{2(5+k)} i \hat{A}_{-} \pa T^{(1)} 
\nonu \\
&- & \frac{3}{(5+k)^2} \hat{A}_3 \pa \hat{A}_{-} 
-\frac{4}{(5+k)} i \hat{B}_3   V_{+}^{(2)}
+\frac{(8+3k)}{(5+k)^2} \hat{B}_3 \pa   \hat{A}_{-} 
-\frac{(11+2k)}{6(5+k)} \pa  V_{+}^{(2)}
\nonu \\
&+& \frac{2}{(5+k)} \hat{G}_{22} \hat{G}_{21}
+\frac{2}{(5+k)} \hat{G}_{22}   T_{+}^{(\frac{3}{2})}
+\frac{2}{(5+k)} \hat{G}_{21}   V^{(\frac{3}{2})}
+ \frac{2}{(5+k)}  T_{+}^{(\frac{3}{2})}   V^{(\frac{3}{2})}
\nonu \\
&+ & \frac{4}{(5+k)^2} i \hat{A}_{-} \hat{A}_3 \hat{A}_3 
-\frac{8}{(5+k)^2} i \hat{A}_{-} \hat{A}_3 \hat{B}_3
+\frac{4}{(5+k)^2} i \hat{A}_{-} \hat{B}_{-} \hat{B}_{+}  
\nonu \\
&+ & \frac{4}{(5+k)^2} i \hat{A}_{-} \hat{B}_3 \hat{B}_3 
+ \frac{4}{(5+k)^2} i \hat{A}_{+} \hat{A}_{-} \hat{A}_{-}
-\frac{1}{2(5+k)} T^{(1)} \hat{A}_{-} \hat{A}_3
\nonu \\
&+ & \left.
\frac{1}{2(5+k)} \hat{A}_3 T^{(1)} \hat{A}_{-} 
\right](w) +\cdots, 
\nonu \\
V^{(\frac{3}{2})}(z) \, W_{-}^{(\frac{5}{2})}(w) & = & 
-\frac{1}{(z-w)^3} \, 
\left[ \frac{8k(8+k)}{3(5+k)^2} i \hat{B}_{+} \right](w)
\nonu \\
& - & \frac{1}{(z-w)^2} \,  \left[ \frac{8(2+k)}{3(5+k)} V_{-}^{(2)}
\right] (w) 
\nonu \\
& + & \frac{1}{(z-w)} \, \left[ 
-\frac{1}{2} {\bf R_{-}^{(3)}}  
-\frac{1}{2(5+k)} i T^{(1)} \pa \hat{B}_{+}
+\frac{4}{(5+k)} i \hat{A}_3 V_{-}^{(2)}
\right. \nonu \\
&+ & \frac{(11+2k)}{(5+k)^2}   \hat{A}_3 \pa \hat{B}_{+}
-\frac{4k(75+117k+14k^2)}{
(5+k)(3+7k)(19+23k)} i \hat{B}_{+} \hat{T}
\nonu \\
&- & \frac{2}{(5+k)}  i \hat{B}_{+} T^{(2)}
+\frac{1}{2(5+k)}   i \hat{B}_{+} \pa T^{(1)}
-\frac{k}{(5+k)^2}  \hat{B}_3 \pa \hat{B}_{+}
\nonu \\
&- & \frac{(5+4k)}{6(5+k)} \pa V_{-}^{(2)}  
-\frac{5k}{6(5+k)^2} i \pa^2 \hat{B}_{+}
+\frac{2}{(5+k)} \hat{G}_{22}  T_{-}^{(\frac{3}{2})}
+\frac{2}{(5+k)}  T_{-}^{(\frac{3}{2})}  V^{(\frac{3}{2})}
\nonu \\
&+ & \frac{(15+2k)}{2(5+k)^2} i   \hat{A}_{-}  \hat{A}_{+}
  \hat{B}_{+}  
-  \frac{(15+2k)}{2(5+k)^2} i   \hat{A}_{+}  \hat{A}_{-}
  \hat{B}_{+}  
- \frac{k}{2(5+k)^2} i   \hat{B}_{-}  \hat{B}_{+}
  \hat{B}_{+}  
\nonu \\
&+ & \left. \frac{k}{2(5+k)^2}  i \hat{B}_{+}  \hat{B}_{-}
  \hat{B}_{+}   
\right](w) + \cdots, 
\nonu \\
V^{(\frac{3}{2})}(z) \, W^{(3)}(w) & = & 
-\frac{1}{(z-w)^3} \,
\left[\frac{4(-3+k)}{(5+k)^2} \hat{G}_{22} +\frac{4(-3+k)(345+296k+55k^2)}{
3(5+k)^2(19+23k)}  V^{(\frac{3}{2})}  
\right](w) 
\nonu \\
& + & \frac{1}{(z-w)^2} \, \left[ 
-\frac{(19+5k)}{2(5+k)} {\bf R^{(\frac{5}{2})}} 
+\frac{(21+5k)}{(5+k)} V^{(\frac{5}{2})} 
-\frac{12}{(5+k)^2} i \hat{A}_{-} \hat{G}_{12}
\right. \nonu \\
& + & \frac{2(1+k)}{(5+k)^2}  i \hat{A}_{-}  T_{-}^{(\frac{3}{2})}
-\frac{2(394+675k+77k^2)}{
(5+k)^2(19+23k)} i \hat{A}_3  V^{(\frac{3}{2})}
+\frac{(33+5k)}{(5+k)^2} i \hat{B}_{+} \hat{G}_{21}
\nonu \\
&+ & \frac{3(11+3k)}{(5+k)^2} i \hat{B}_{+}  T_{+}^{(\frac{3}{2})}
+\frac{2(551+481k+78k^2)}{(5+k)^2(19+23k)} 
i \hat{B}_3  V^{(\frac{3}{2})}
\nonu \\
&+& \frac{(161-47k-12k^2)}{(5+k)(19+23k)} T^{(1)} V^{(\frac{3}{2})} 
+\frac{4(-3+k)}{3(5+k)^2} \pa \hat{G}_{22}
\nonu \\
& -& \left. \frac{(243+692k+141k^2+4k^3)}{
(5+k)^2(19+23k)} \pa  V^{(\frac{3}{2})}
\right](w) 
\nonu \\
& + & \frac{1}{(z-w)} \, \left[ 
-\frac{(11+k)}{2(5+k)}
\pa {\bf R^{(\frac{5}{2})}}
+\frac{2}{(5+k)} i  {\bf R^{(\frac{5}{2})}} \hat{A}_3
- \frac{2}{(5+k)} i  {\bf R^{(\frac{5}{2})}} \hat{B}_3
\right. \nonu \\
&+& \frac{1}{(5+k)} i {\bf P_{-}^{(\frac{5}{2})} } \hat{A}_{-} 
-\frac{(-2+k)}{(5+k)^2} \hat{A}_{-} \hat{A}_{+}  V^{(\frac{3}{2})}
-\frac{2}{(5+k)} i \hat{A}_{-}   W_{-}^{(\frac{5}{2})}
\nonu \\
&- &  \frac{2(1+k)}{(5+k)^2} \hat{A}_{-} \hat{A}_3  T_{-}^{(\frac{3}{2})}
+\frac{4}{(5+k)^2} i  \hat{A}_{-} \pa \hat{G}_{12}
-\frac{2(25+k)}{3(5+k)^2} i  \hat{A}_{-} \pa  T_{-}^{(\frac{3}{2})}
\nonu \\
&+ & \frac{1}{8(5+k)^2}  \hat{B}_{+} \hat{A}_3 \hat{G}_{21}
+\frac{1}{8(5+k)^2}  \hat{B}_{+} \hat{A}_3  T_{+}^{(\frac{3}{2})}
-\frac{(15+2k)}{(5+k)^2}  \hat{B}_{+} \hat{B}_{-}  V^{(\frac{3}{2})}
\nonu \\
&+ & \frac{(21+k)}{(5+k)^2}  \hat{B}_{+} \hat{B}_3 \hat{G}_{21}
+\frac{(25+3k)}{(5+k)^2}  \hat{B}_{+} \hat{B}_3  T_{+}^{(\frac{3}{2})}
+ \frac{(1+k)}{(5+k)^2}  i \hat{B}_{+} \pa \hat{G}_{21}
\nonu \\
&+ & \frac{1}{(5+k)} i   \hat{B}_{+} \pa  T_{+}^{(\frac{3}{2})}
+ \frac{(15+2k)}{(5+k)^2}  \hat{B}_{-} \hat{B}_{+}  V^{(\frac{3}{2})}
+ \frac{(6+k)}{(5+k)^2}  \hat{A}_{+} \hat{A}_{-}  V^{(\frac{3}{2})}
\nonu \\
&- & \frac{4}{(5+k)} i 
\hat{A}_3   V^{(\frac{5}{2})}
+\frac{2(1+k)}{(5+k)^2}  \hat{A}_3 \hat{A}_{-}  T_{-}^{(\frac{3}{2})}
+\frac{4}{(5+k)^2} \hat{A}_3 \hat{A}_3  V^{(\frac{3}{2})}
\nonu \\
& +& \frac{8}{(5+k)^2} \hat{A}_3 \hat{B}_3  V^{(\frac{3}{2})}
-\frac{2(622+913k+31k^2)}{
3(5+k)^2(19+23k)} i \hat{A}_3 \pa  V^{(\frac{3}{2})}
+\frac{4}{(5+k)} i \hat{B}_3  V^{(\frac{5}{2})}
\nonu \\
& - &  \frac{(29+k)}{(5+k)^2}  \hat{B}_3 \hat{B}_{+} \hat{G}_{21}
- \frac{3(11+k)}{(5+k)^2}  \hat{B}_3 \hat{B}_{+}  T_{+}^{(\frac{3}{2})}
- \frac{12}{(5+k)^2}  \hat{B}_3 \hat{B}_3  V^{(\frac{3}{2})}
\nonu \\
& - & \frac{2(-133+101k+14k^2)}{
3(5+k)^2(19+23k)} i \hat{B}_3 \pa   V^{(\frac{3}{2})}
-\frac{(-41+31k+4k^2)}{
(5+k)(19+23k)}  T^{(1)} \pa   V^{(\frac{3}{2})}
\nonu \\
&+& \frac{(7+k)}{(5+k)} \pa  V^{(\frac{5}{2})}
-\frac{2}{(5+k)^2} i \pa  \hat{A}_{-} \hat{G}_{12}
+\frac{1}{(5+k)} \pa T^{(1)}  V^{(\frac{3}{2})}
\nonu \\
&- & \frac{(-7+5k)}{3(5+k)^2} \pa^2 \hat{G}_{22}
+\frac{(-35+2k)}{3(5+k)^2} \pa^2  V^{(\frac{3}{2})}
-\frac{4}{(5+k)} \hat{G}_{21} V_{-}^{(2)}
\nonu \\
&- & \frac{2}{(5+k)}  T_{+}^{(\frac{3}{2})} V_{-}^{(2)}
+\frac{2}{(5+k)} \hat{G}_{12}  V_{+}^{(2)} 
\nonu \\
&-& \left. \frac{4}{(5+k)}  V^{(\frac{3}{2})} T^{(2)}
-\frac{8(-69-134k+2k^2+7k^3)}{(5+k)(3+7k)(19+23k)}  V^{(\frac{3}{2})} 
\hat{T}
\right](w) +\cdots. 
\label{resultope}
\eea
In the first OPE of (\ref{resultope}), the $(k-3)$ factor appears in the 
second order pole.
Also the third order pole of last OPE contains the $(k-3)$ factor.

\section{
The nontrivial 
OPEs between  higher spin-$2$ current, $V_{+}^{(2)}(z)$,  
and 
other $7$ higher spin currents
}

We present the following OPEs
\bea
V_{+}^{(2)}(z) \, V_{-}^{(2)}(w) & = & 
\frac{1}{(z-w)^2} \,
\left[ \frac{2(3+k)}{(5+k)^2} \hat{A}_{-} \hat{B}_{+} \right] (w)
\nonu \\
&+ & \frac{1}{(z-w)} \, \left[
-\frac{2}{(5+k)} i
\hat{B}_{+} V_{+}^{(2)} +\frac{2(7+k)}{(5+k)^2} \hat{B}_{+} \pa \hat{A}_{-}
-\frac{2}{(5+k)} i
\hat{A}_{-}  V_{-}^{(2)}
\right. \nonu \\
& + & \left.
\frac{2}{(5+k)} \hat{G}_{22} \hat{G}_{22} 
\right](w)
+\cdots,
\nonu \\
V_{+}^{(2)}(z) \, V^{(\frac{5}{2})}(w) & = & 
\frac{1}{(z-w)^2} \, \left[ 
\frac{2(3+2k)}{3(5+k)^2} i \hat{A}_{-} \hat{G}_{22}
+\frac{8(9+k)}{3(5+k)^2} i \hat{A}_{-}  V^{(\frac{3}{2})}
\right](w)
\nonu \\
& + & \frac{1}{(z-w)} \, \left[- \frac{1}{(5+k)} 
\hat{A}_{-} {\bf R^{(\frac{5}{2})}} 
-\frac{4(15+2k)}{3(5+k)^2} \hat{A}_{-} \hat{A}_3  V^{(\frac{3}{2})}
+\frac{2}{(5+k)^2}  \hat{A}_{-} \hat{B}_{+}
\hat{G}_{21}
\right. \nonu \\
&- & \frac{2}{(5+k)^2} i \hat{A}_{-} T^{(1)} \hat{G}_{22}  
+ \frac{4(15+2k)}{3(5+k)^2} \hat{A}_3  \hat{A}_{-}   V^{(\frac{3}{2})}
+ \frac{2(9+2k)}{3(5+k)^2} i \pa \hat{A}_{-} \hat{G}_{22} 
\nonu \\
& - & \left.
\frac{2}{(5+k)^2} \hat{G}_{21} \hat{A}_{-} \hat{B}_{+}
+ \frac{2}{(5+k)^2} i \hat{G}_{22} T^{(1)} \hat{A}_{-}
+\frac{2}{(5+k)} \hat{G}_{22} V_{+}^{(2)} 
\right](w) + \cdots,
\nonu \\
V_{+}^{(2)}(z) \, W^{(2)}(w) & = & 
\frac{1}{(z-w)^3} \, \left[ \frac{6(5+2k)}{(5+k)^2} i \hat{A}_{-} 
\right](w)
\nonu \\
& + & \frac{1}{(z-w)^2} \, \left[  \frac{3(5+2k)}{(5+k)^2} i \pa \hat{A}_{-} 
+ \frac{2(4+k)}{(5+k)}  V_{+}^{(2)}  \right](w)  
\nonu \\
& + & \frac{1}{(z-w)} \, \left[  
\frac{1}{2} 
{\bf R_{+}^{(3)}}
+\frac{1}{2(5+k)} i T^{(1)} \pa \hat{A}_{-}
+\frac{4(138+343k+113k^2)}{
(5+k)(3+7k)(19+23k)} i \hat{A}_{-} \hat{T}
\right. \nonu \\
&- & \frac{2}{(5+k)} i  \hat{A}_{-} T^{(2)}  
- \frac{9}{(5+k)^2}   \hat{A}_{-} \pa \hat{A}_3
- \frac{(8+3k)}{(5+k)^2}   \hat{A}_{-} \pa \hat{B}_3
- \frac{1}{2(5+k)}   i \hat{A}_{-} \pa  T^{(1)}
\nonu \\
&- & \frac{3}{(5+k)^2} \hat{A}_3 \pa  \hat{A}_{-}
+\frac{(8+3k)}{(5+k)^2} \hat{B}_3 \pa  \hat{A}_{-}
+ \frac{(9+2k)}{2(5+k)} \pa  V_{+}^{(2)}
+ \frac{(5+2k)}{(5+k)^2} i \pa^2 \hat{A}_{-} 
\nonu \\
&- & \frac{2}{(5+k)} \hat{G}_{21} \hat{G}_{22}
+ \frac{2}{(5+k)} \hat{G}_{21}  V^{(\frac{3}{2})}
+\frac{2}{(5+k)} \hat{G}_{22}  T_{+}^{(\frac{3}{2})}
+ \frac{2}{(5+k)}   T_{+}^{(\frac{3}{2})}  V^{(\frac{3}{2})}
\nonu \\
&+ & \frac{4}{(5+k)^2}    i \hat{A}_{+}    \hat{A}_{-}    \hat{A}_{-}
+  \frac{4}{(5+k)^2}    i \hat{A}_{-}    \hat{A}_3    \hat{A}_3
+ \frac{4}{(5+k)^2}    i \hat{A}_{-}    \hat{B}_{+}    \hat{B}_{-}
\nonu \\
&+ &   \left. 
\frac{4}{(5+k)^2}    i \hat{A}_{-}    \hat{B}_3    \hat{B}_3
-\frac{8}{(5+k)^2}    i \hat{A}_3    \hat{A}_{-}    \hat{B}_3
-\frac{4}{(5+k)} i \hat{B}_3 
 V_{+}^{(2)}
\right](w) +\cdots,
\nonu \\
V_{+}^{(2)}(z) \, W_{+}^{(\frac{5}{2})}(w) & = & 
\frac{1}{(z-w)^2} \, \left[ -\frac{2(13+2k)}{3(5+k)^2} i \hat{A}_{-} 
\hat{G}_{21} -\frac{2(13+2k)}{3(5+k)^2} 
i \hat{A}_{-}  T_{+}^{(\frac{3}{2})}  \right](w)
\nonu \\
& + & \frac{1}{(z-w)} \, \left[ 
\frac{2}{(5+k)} i \hat{A}_{-} W_{+}^{(\frac{5}{2})}
-\frac{4}{(5+k)^2} i \hat{A}_{-} \pa \hat{G}_{21}
-\frac{8(4+k)}{3(5+k)^2}  i \hat{A}_{-} \pa  T_{+}^{(\frac{3}{2})}
\right. \nonu \\
&- & \left. \frac{2(7+2k)}{3(5+k)^2} i \pa \hat{A}_{-} \hat{G}_{21}
+  \frac{8(2+k)}{3(5+k)^2} i \pa \hat{A}_{-} T_{+}^{(\frac{3}{2})}
-\frac{2}{(5+k)} \hat{G}_{21} V_{+}^{(2)}
\right](w) +\cdots,
\nonu \\
V_{+}^{(2)}(z) \, W_{-}^{(\frac{5}{2})}(w) & = & 
\frac{1}{(z-w)^3} \, \left[ \frac{8(2+k)(6+k)}{3(5+k)^2} \hat{G}_{22}
+\frac{4(11+3k)}{3(5+k)^2}   V^{(\frac{3}{2})}  \right](w) 
\nonu \\
&+& \frac{1}{(z-w)^2} \, \left[
-\frac{(27+8k)}{3(5+k)} {\bf R^{(\frac{5}{2})}}
+\frac{(35+8k)}{3(5+k)}  V^{(\frac{5}{2})}
-\frac{6}{(5+k)^2} i \hat{A}_{-} \hat{G}_{12} 
\right. \nonu \\
& + & \frac{8(1+k)}{3(5+k)^2} i \hat{A}_{-} T_{-}^{(\frac{3}{2})}
-\frac{8(10+k)}{3(5+k)^2} i \hat{A}_3  V^{(\frac{3}{2})} 
+ \frac{2(35+8k)}{3(5+k)^2} i \hat{B}_{+} \hat{G}_{21}
\nonu \\
&+&  \frac{4(22+7k)}{3(5+k)^2} i \hat{B}_{+} T_{+}^{(\frac{3}{2})}
+ \frac{4(35+11k)}{3(5+k)^2} i \hat{B}_3  V^{(\frac{3}{2})} 
-\frac{2}{(5+k)} T^{(1)}  V^{(\frac{3}{2})} 
\nonu \\
&+& \left.  \frac{4(24+19k+2k^2)}{9(5+k)^2} 
\pa \hat{G}_{22} - \frac{8(22+5k)}{9(5+k)^2} \pa  V^{(\frac{3}{2})}  \right](w) 
\nonu \\
& + &  \frac{1}{(z-w)} \, \left[ -\frac{1}{2} {\bf R^{(\frac{7}{2})}}  
+ \frac{(-25+4k)}{90} {\bf P^{(2)}}  V^{(\frac{3}{2})}  
+  \frac{(25-4k)}{90} V^{(\frac{3}{2})}  {\bf P^{(2)}}
\right. \nonu \\
& + & \frac{2}{(5+k)} i {\bf R^{(\frac{5}{2})}} \hat{A}_3
- \frac{2}{(5+k)} i {\bf R^{(\frac{5}{2})}} \hat{B}_3
-\frac{(-1339-118k+8k^2)}{135(5+k)} {\bf R^{(\frac{5}{2})}} T^{(1)}
\nonu \\
&+ & \frac{(-1339-118k+8k^2)}{135(5+k)} T^{(1)} {\bf R^{(\frac{5}{2})}}
+\frac{1}{(5+k)} i {\bf P_{-}^{(\frac{5}{2})}} \hat{A}_{-}
\nonu \\
& + & \frac{8}{(5+k)^2} \hat{A}_{+} \hat{A}_{-}  V^{(\frac{3}{2})}
-\frac{(-1198-127k+8k^2)}{
45(5+k)^2}  \hat{B}_{-} \hat{B}_{+}  V^{(\frac{3}{2})}
\nonu \\
&+& \frac{4}{(5+k)^2}  \hat{B}_{+} \hat{A}_{-} \hat{G}_{11}
+  \frac{4}{(5+k)^2}  \hat{B}_{+} \hat{A}_{-} U^{(\frac{3}{2})}
+ \frac{8}{(5+k)^2}  \hat{B}_{+} \hat{A}_3 \hat{G}_{21}
\nonu \\
&+ &   \frac{8}{(5+k)^2}  \hat{B}_{+} \hat{A}_3 
T_{+}^{(\frac{3}{2})}
+ \frac{(-1018-127k+8k^2)}{
45(5+k)^2}  \hat{B}_{+} \hat{B}_{-}  V^{(\frac{3}{2})}
\nonu \\
&+ & \frac{2(25760 k^5-384988 k^4-3753855 k^3+12475630 k^2+37920721 k+
20787216)}{405(5+k)^2(19+23k)(47+35k)}   
\nonu \\
& \times & \hat{B}_{+} \hat{B}_3 \hat{G}_{21}
-\frac{2(-772-43k+8k^2)}{
45(5+k)^2}  \hat{B}_{+} \hat{B}_3 T_{+}^{(\frac{3}{2})}
\nonu \\
& - & \frac{8(-2983-3723k-189k^2+55k^3)}{
45(5+k)^2(19+23k)} i \hat{B}_{+} \pa \hat{G}_{21}
\nonu \\
& - & \frac{8(-193-22k+2k^2)}{45(5+k)^2} 
i \hat{B}_{+} \pa  T_{+}^{(\frac{3}{2})}
\nonu \\
&- &  \frac{2}{(5+k)} i \hat{A}_{-}   W_{-}^{(\frac{5}{2})}
-\frac{2(53+20k)}{15(5+k)^2} \hat{A}_{-} \hat{A}_3   T_{-}^{(\frac{3}{2})}
+ \frac{8}{(5+k)^2} \hat{A}_{-} \hat{B}_3   T_{-}^{(\frac{3}{2})}
\nonu \\
& + &  \frac{4}{(5+k)^2} i \hat{A}_{-} \pa \hat{G}_{12} 
-\frac{68}{5(5+k)^2} i \hat{A}_{-} \pa   T_{-}^{(\frac{3}{2})}
+ \frac{2(53+20k)}{15(5+k)^2} \hat{A}_3 \hat{A}_{-}   T_{-}^{(\frac{3}{2})}
\nonu \\
&+& \frac{4}{(5+k)^2} \hat{A}_3 \hat{A}_3  V^{(\frac{3}{2})}
+ \frac{8}{(5+k)^2} \hat{A}_3 \hat{B}_3  V^{(\frac{3}{2})}
+  \frac{8(9+2k)}{15(5+k)^2} i \hat{A}_3 \pa \hat{G}_{22} 
\nonu \\
& + &  \frac{4}{(5+k)} i \hat{B}_3   V^{(\frac{5}{2})}
\nonu \\
& - & 
 \frac{2(25760 k^5-384988 k^4-3753855 k^3+13779730 k^2+40749241 k+22233876)}{
405(5+k)^2(19+23k)(47+35k)}   
 \nonu \\
& \times & \hat{B}_3 \hat{B}_{+} \hat{G}_{21} 
+  \frac{2(-952-43k+8k^2)}{
45(5+k)^2} \hat{B}_3 \hat{B}_{+}  T_{+}^{(\frac{3}{2})}
-\frac{12}{(5+k)^2} \hat{B}_3 \hat{B}_3  V^{(\frac{3}{2})} 
\nonu \\
&- & \frac{8k}{5(5+k)^2} i \hat{B}_3 \pa  \hat{G}_{22} 
 -\frac{16(-56-14k+k^2)}{45(5+k)^2} i \hat{B}_3   \pa V^{(\frac{3}{2})}
-\frac{8}{45} T^{(1)} \pa  V^{(\frac{3}{2})} 
\nonu \\
&- & \frac{2(3+8k)}{3(5+k)^2} i \pa \hat{A}_3  V^{(\frac{3}{2})} 
- \frac{(-95+8k)}{45(5+k)} i \pa T^{(1)}  V^{(\frac{3}{2})} 
\nonu \\
&- & \frac{(1680 k^4-17204 k^3-204397 k^2+444476 k+929245)}{
405(5+k)^2(47+35k)} \pa^2  V^{(\frac{3}{2})}
\nonu \\
&- & \frac{4(-422-23k+4k^2)}{135(5+k)} \pa  V^{(\frac{5}{2})}
+\frac{4}{5(5+k)} T^{(1)} \pa \hat{G}_{22}
+\frac{2}{(5+k)} \hat{G}_{12} V_{+}^{(2)}
\nonu \\
&- & \frac{2}{(5+k)^2} i \hat{G}_{12} \pa \hat{A}_{-} 
-\frac{2}{(5+k)}   T_{-}^{(\frac{3}{2})} V_{+}^{(2)}
-\frac{2}{(5+k)} \hat{G}_{21}  V_{-}^{(2)} 
\nonu \\
& - & \frac{4(12880 k^5-172019 k^4-1918890 k^3+3721091 k^2+
13826720 k+7869990)}{405(5+k)^2(19+23k)(47+35k)} \nonu \\
& \times & i 
\hat{G}_{21} \pa \hat{B}_{+}
-\frac{2}{(5+k)}   T_{+}^{(\frac{3}{2})}  V_{-}^{(2)} 
\nonu \\
& + & \frac{8(1368+2493k+1178k^2+133k^3)}{
3(5+k)(19+23k)(47+35k)} \hat{G}_{22} \hat{T}
-\frac{4(9+2k)}{5(5+k)^2} i \hat{G}_{22} \pa \hat{A}_3 
\nonu \\
& + & \frac{12 k}{5(5+k)^2} i \hat{G}_{22} \pa \hat{B}_3 
-\frac{6}{5(5+k)} \hat{G}_{22} \pa T^{(1)}
+ \frac{12(51+99k+28k^2)}{(
5+k)(3+7k)(47+35k)}  V^{(\frac{3}{2})} \hat{T}
\nonu \\
&-& \left. \frac{2}{(5+k)}  V^{(\frac{3}{2})} T^{(2)}  
\right](w) +\cdots,
\label{v2+w5half-}  \\
V_{+}^{(2)}(z) \, W^{(3)}(w) & = & 
-\frac{1}{(z-w)^4} \, \left[ \frac{12(270+893k+437k^2+50k^3)}{
(5+k)^3(19+23k)} i \hat{A}_{-} \right] (w)
\nonu \\
& + & 
\frac{1}{(z-w)^3} \, \left[ 
-\frac{2(-127+85k+72k^2)}{(5+k)^2(19+23k)} V_{+}^{(2)}
-\frac{12(-3+2k)}{(5+k)^3} \hat{A}_{-} \hat{A}_3
\right. \nonu \\
& + & \left. \frac{4(40+19k)}{(5+k)^3} \hat{A}_{-} \hat{B}_3
-\frac{6(-3+2k)}{(5+k)^3} i \pa \hat{A}_{-}
-\frac{10}{(5+k)^2} i T^{(1)} \hat{A}_{-}
\right](w) 
\nonu \\
& + & 
\frac{1}{(z-w)^2} \, \left[ -\frac{(25+6k)}{2(5+k)} {\bf R_{+}^{(3)}} 
+\frac{1}{(5+k)} i {\bf P^{(2)}} \hat{A}_{-} 
 \right. 
\nonu  \\
& + & \frac{(221-55k-16k^2)}{(5+k)(19+23k)} T^{(1)}  V_{+}^{(2)}
+\frac{3(-3+2k)}{2(5+k)^2}  i T^{(1)} \pa \hat{A}_{-}
\nonu \\
& - & \frac{4(7476+20801k+11060k^2 + 1343k^3)}{
3(5+k)^2(3+7k)(19+23k)} i \hat{A}_{-} \hat{T}
+\frac{2(47+13k)}{3(5+k)^2} i  \hat{A}_{-} T^{(2)}
\nonu \\
&+ & \frac{6(4+k)}{(5+k)^2}   i  \hat{A}_{-} W^{(2)}
+ \frac{(251+58k)}{(5+k)^3}     \hat{A}_{-} \pa \hat{A}_3
+  \frac{(448+251k+18k^2)}{3(5+k)^3}     \hat{A}_{-} \pa \hat{B}_3
\nonu \\
& - &  \frac{(25+6k)}{2(5+k)^2}  i   \hat{A}_{-} 
\pa  T^{(1)} 
-\frac{2(329+675k+118k^2)}{(
5+k)^2(19+23k)} i \hat{A}_3  V_{+}^{(2)}
+\frac{3(17-10k)}{(5+k)^3} \hat{A}_3 \pa  \hat{A}_{-} 
\nonu \\
&- &  \frac{(724+185k+18k^2)}{3(5+k)^3} \hat{B}_3 \pa  \hat{A}_{-}
-\frac{(2107+4933k+1582k^2)}{6(5+k)^2(19+23k)} \pa  V_{+}^{(2)}
\nonu \\
& + & \frac{10(-1+4k)}{3(5+k)^3} i \pa^2 \hat{A}_{-} 
+\frac{8}{(5+k)} \hat{G}_{21} \hat{G}_{22} 
-\frac{2(17+2k)}{(5+k)^2} \hat{G}_{21}   V^{(\frac{3}{2})}
\nonu \\
&- & \frac{2(23+6k)}{(5+k)^2} \hat{G}_{22}   T_{+}^{(\frac{3}{2})} 
-\frac{2(25+6k)}{(5+k)^2}  T_{+}^{(\frac{3}{2})}   V^{(\frac{3}{2})}
-\frac{4(53+10k)}{3(5+k)^3} i \hat{A}_{+}  \hat{A}_{-}  \hat{A}_{-} 
\nonu \\
& - & 
\frac{4(53+10k)}{3(5+k)^3}   i \hat{A}_{-}  \hat{A}_3  \hat{A}_3
+\frac{8(53+10k)}{3(5+k)^3}   i \hat{A}_{-}  \hat{A}_3  \hat{B}_3
-\frac{16(17+4k)}{3(5+k)^3}   i \hat{A}_{-}  \hat{B}_{+}  \hat{B}_{-}
\nonu \\
&- & \left. \frac{4(53+10k)}{3(5+k)^3}   i \hat{A}_{-}  \hat{B}_3  \hat{B}_3
+\frac{2(608+507k+127k^2)}{
(5+k)^2(19+23k)} i \hat{B}_3  V_{+}^{(2)}
\right](w) 
\nonu \\
& + & 
\frac{1}{(z-w)} \, \left[ 
\frac{2}{(5+k)} i \hat{A}_3  {\bf R_{+}^{(3)}} 
-\frac{1}{(5+k)} i \hat{A}_{-} {\bf S^{(3)}}
+\frac{1}{(5+k)} i \hat{A}_{-} {\bf P^{(3)}} 
\right. \nonu \\
&+& \frac{1}{2(5+k)} i \hat{A}_{-} \pa {\bf P^{(2)}}
-\frac{2}{(5+k)} i \hat{B}_3  {\bf R_{+}^{(3)}} 
-\frac{(5+2k)}{2(5+k)} \pa  {\bf R_{+}^{(3)}} 
\nonu \\
&- & \frac{(21+10k)}{2(5+k)^2} \pa^2   V_{+}^{(2)} 
-\frac{4}{(5+k)}  T^{(2)}   V_{+}^{(2)} 
-\frac{2(5+2k)}{(5+k)^2} \pa  T_{+}^{(\frac{3}{2})}   V^{(\frac{3}{2})}
\nonu \\
&- & \frac{2(5+2k)}{(5+k)^2}  T_{+}^{(\frac{3}{2})} \pa  V^{(\frac{3}{2})}
-\frac{2(-5+2k)}{(5+k)^2} i \pa \hat{A}_3  V_{+}^{(2)}  
-\frac{4(11+73k+18k^2)}{(5+k)^2(19+23k)} i \hat{A}_3  \pa V_{+}^{(2)} 
\nonu \\
& + & \frac{8}{(5+k)^2} i \hat{A}_3   T_{+}^{(\frac{3}{2})}   V^{(\frac{3}{2})}
+\frac{8}{(5+k)^2} \hat{A}_3 \hat{A}_3  V_{+}^{(2)}  
-\frac{8}{(5+k)^2} i \hat{A}_3 \hat{G}_{21} \hat{G}_{22} 
\nonu \\
&+& \frac{8}{(5+k)^2} i \hat{A}_3 \hat{G}_{22}  T_{+}^{(\frac{3}{2})}
+\frac{6(3+k)}{(5+k)^2} i \pa \hat{A}_{-} W^{(2)}
-\frac{7}{(5+k)^2} i  \hat{A}_{-} \pa W^{(2)}
\nonu \\
&-& \frac{4}{(5+k)^2} i \hat{A}_{-}  U^{(\frac{3}{2})}  V^{(\frac{3}{2})}
+\frac{2(8+3k)}{(5+k)^2} i \pa \hat{A}_{-}  T^{(2)} 
+\frac{(5+4k)}{3(5+k)^2} i  \hat{A}_{-} \pa  T^{(2)} 
\nonu \\
&+ & \frac{4}{(5+k)^2} i \hat{A}_{-}  T_{+}^{(\frac{3}{2})}  T_{-}^{(\frac{3}{2})}
\nonu \\
& + & \frac{(85455+253320k+205195k^2+73714k^3+7024k^4)}{
18(5+k)^3(3+7k)(19+23k)} i \pa^3 \hat{A}_{-}
\nonu \\
&-& \frac{(45+563k+322k^2)}{
(5+k)^3(19+23k)} \pa^2 \hat{A}_{-} \hat{A}_3
+\frac{2(79+38k)}{3(5+k)^3} \pa \hat{A}_{-} \pa \hat{A}_3
\nonu \\
& + & \frac{(2981+3755k+1058k^2)}{
3(5+k)^3(19+23k)} \hat{A}_{-} \pa^2 \hat{A}_3
-\frac{8(-2+k)}{(5+k)^3} i \pa \hat{A}_{-} \hat{A}_3 \hat{A}_3
\nonu \\
&- & \frac{8(19+2k)}{3(5+k)^3} i \hat{A}_{-} \pa \hat{A}_3  \hat{A}_3
+\frac{16}{(5+k)^3} \hat{A}_{-} \hat{A}_3 \hat{A}_3 \hat{B}_3
+\frac{4(17+3k)}{(5+k)^3} i \pa \hat{A}_{-} \hat{A}_3 \hat{B}_3
\nonu \\
&+ & \frac{4(56+13k)}{3(5+k)^3} i  \hat{A}_{-} \pa \hat{A}_3 \hat{B}_3
- \frac{4(-17+2k)}{3(5+k)^3} i  \hat{A}_{-}  \hat{A}_3 \pa \hat{B}_3
\nonu \\
&-& \frac{32}{(5+k)^3} \hat{A}_{-} \hat{A}_3 \hat{B}_3 \hat{B}_3
+\frac{6}{(5+k)^2} \pa \hat{A}_{-} \hat{A}_3 T^{(1)}
-\frac{2}{(5+k)^2}  \hat{A}_{-} \pa \hat{A}_3 T^{(1)}
\nonu \\
&-& \frac{4}{(5+k)^2}  \hat{A}_{-}  \hat{A}_3 \pa T^{(1)}
-\frac{8}{(5+k)^2} \hat{A}_{-} \hat{B}_3 W^{(2)}
\nonu \\
& - & \frac{(2714+2823k+759k^2+46k^3)}{(5+k)^3(19+23k)} \pa^2 \hat{A}_{-} 
\hat{B}_3 -\frac{10(4+k)}{(5+k)^3} \pa \hat{A}_{-} \pa \hat{B}_3
\nonu \\
&+& \frac{(-4418-921k+1631k^2+138k^3)}{
3(5+k)^3(19+23k)} \hat{A}_{-} \pa^2 \hat{B}_3
-\frac{4(21+k)}{(5+k)^3} i \pa \hat{A}_{-} \hat{B}_3 \hat{B}_3
\nonu \\
&- & \frac{4(-13+7k)}{3(5+k)^3} i \hat{A}_{-} \pa \hat{B}_3 \hat{B}_3
+\frac{16}{(5+k)^3} \hat{A}_{-}  \hat{B}_3 \hat{B}_3  \hat{B}_3 
-\frac{6}{(5+k)^2}  \pa \hat{A}_{-}  \hat{B}_3 T^{(1)}
\nonu \\
&+& \frac{8}{(5+k)^2}   \hat{A}_{-}  \pa \hat{B}_3 T^{(1)}
-\frac{2}{(5+k)^2}  \hat{A}_{-}   \hat{B}_3 \pa T^{(1)}
+\frac{4}{(5+k)^2} \hat{A}_{-} \hat{B}_{+}  U_{+}^{(2)} 
\nonu \\
& - & \frac{8(4+k)}{(5+k)^3} i \pa \hat{A}_{-} \hat{B}_{+} \hat{B}_{-}
-\frac{2(47+7k)}{3(5+k)^3} i \hat{A}_{-} \pa \hat{B}_{+} \hat{B}_{-}
-\frac{2(-37+13k)}{3(5+k)^3} i \hat{A}_{-}  \hat{B}_{+} \pa \hat{B}_{-}
\nonu \\
& +& \frac{16}{(5+k)^3}  \hat{A}_{-}  \hat{B}_{+}  \hat{B}_{-} \hat{B}_3 
-\frac{4}{(5+k)^2} i \hat{A}_{-} \hat{G}_{12} T_{+}^{(\frac{3}{2})}
+\frac{4}{(5+k)^2} i \hat{A}_{-} \hat{G}_{21} T_{-}^{(\frac{3}{2})}
\nonu \\
&+& \frac{(-131-161k+46k^2)}{2(5+k)^2(19+23k)} \pa^2 \hat{A}_{-} T^{(1)}
+\frac{3}{(5+k)^2} i \pa \hat{A}_{-} \pa T^{(1)}
\nonu \\
& - & \frac{(-253-71k+138k^2)}{6(5+k)^2(19+23k)}  i \hat{A}_{-} \pa^2 T^{(1)}
+\frac{8}{(5+k)^2} \hat{A}_{+} \hat{A}_{-}  V_{+}^{(2)} 
\nonu \\
&-& \frac{2(11+4k)}{3(5+k)^3} i \pa \hat{A}_{+}  \hat{A}_{-} \hat{A}_{-} 
- \frac{2(41+16k)}{3(5+k)^3} i  \hat{A}_{+} \pa \hat{A}_{-} \hat{A}_{-}
+\frac{16}{(5+k)^3}  \hat{A}_{+}  \hat{A}_{-} \hat{A}_{-} \hat{B}_3 
\nonu \\
&+& \frac{2(14+3k)}{(5+k)^2} i \pa \hat{B}_3  V_{+}^{(2)} 
+\frac{2(-19-166k+29k^2)}{(5+k)^2(19+23k)}  
i  \hat{B}_3  \pa V_{+}^{(2)} 
\nonu \\
&- & \frac{8}{(5+k)^2} i \hat{B}_3   T_{+}^{(\frac{3}{2})}  
V^{(\frac{3}{2})} -\frac{8}{(5+k)^2} \hat{B}_3 \hat{B}_3    V_{+}^{(2)} 
+\frac{8}{(5+k)^2} i \hat{B}_3 \hat{G}_{21} \hat{G}_{22} 
\nonu \\
&- & \frac{8}{(5+k)^2} i \hat{B}_3 \hat{G}_{22} T_{+}^{(\frac{3}{2})}  
+\frac{4}{(5+k)^2} \hat{B}_{+} \hat{B}_{-}  V_{+}^{(2)} 
+\frac{4}{(5+k)^2} i \hat{B}_{+} \hat{G}_{21}  T_{+}^{(\frac{3}{2})} 
\nonu \\
&+& \frac{2}{(5+k)} \hat{G}_{21}  V^{(\frac{5}{2})} 
-\frac{2(5+2k)}{(5+k)^2} \pa \hat{G}_{21}  V^{(\frac{3}{2})} 
+\frac{2(-7+2k)}{3(5+k)^2}   \hat{G}_{21} \pa  V^{(\frac{3}{2})} 
\nonu \\
&+& \frac{2(3+2k)}{(5+k)^2} \pa \hat{G}_{21} \hat{G}_{22}
+ \frac{2(15+2k)}{3(5+k)^2} \hat{G}_{21} \pa \hat{G}_{22}
-\frac{2(7+2k)}{(5+k)^2} \pa \hat{G}_{22} T_{+}^{(\frac{3}{2})} 
\nonu \\
&-& \frac{2(3+2k)}{(5+k)^2}  \hat{G}_{22} \pa T_{+}^{(\frac{3}{2})} 
+\frac{1}{(5+k)} \pa T^{(1)}  V_{+}^{(2)} 
-\frac{(-101+39k+8k^2)}{(5+k)(19+23k)}  T^{(1)}  \pa V_{+}^{(2)} 
\nonu \\
&+& \frac{8(48+97k+29k^2)}{(5+k)(3+7k)(19+23k)} \hat{T}  V_{+}^{(2)}
\nonu \\
& - & \frac{2(3057+9077k+5012k^2+620k^3)}{3(5+k)^2(3+7k)(19+23k)} i
\pa \hat{T} \hat{A}_{-}
\nonu \\
&-& \frac{4(474+1244k+855k^2+241k^3)}{(5+k)^2(3+7k)(19+23k)} i
 \hat{T} \pa \hat{A}_{-}
-\frac{96}{(5+k)^2(19+23k)} \hat{T} \hat{A}_{-} \hat{A}_3
\nonu \\
&+& \left. \frac{32(23+12k)}{(5+k)^2(19+23k)} 
\hat{T} \hat{A}_{-} \hat{B}_3 -\frac{8}{(5+k)(19+23k)} i
\hat{T} \hat{A}_{-} T^{(1)}
 \right](w) +\cdots.
\label{opetoday}
\eea
As explained before, the 
the combination of the second and third terms in the first order pole of
(\ref{v2+w5half-}) can be written in terms of 
derivatives of known composite fields.
One has the following OPE
\bea
 V^{(\frac{3}{2})}(z) \, {\bf P^{(2)}}(w) & = &
\frac{1}{(z-w)^2} \,
\left[\frac{16(2+k)(8+k)}{3(5+k)^2} \, V^{(\frac{3}{2})} \right](w)
\nonu \\
&+& \frac{1}{(z-w)} \,
\left[ 
-\frac{4(19+2k)}{3(5+k)}  V^{(\frac{5}{2})}
-\frac{8(8+k)}{(5+k)^2} i \hat{B}_{+}  T_{+}^{(\frac{3}{2})}
\right. \nonu \\
&-& \frac{24}{(5+k)^2} i \hat{A}_3  V^{(\frac{3}{2})}
-\frac{8(8+k)}{(5+k)^2} i \hat{B}_{+} \hat{G}_{21} 
\nonu \\
&-& 
\frac{8(8+k)}{(5+k)^2} i \hat{B}_3  V^{(\frac{3}{2})}
-\frac{4}{(5+k)} T^{(1)}  V^{(\frac{3}{2})}
+\frac{8(8+k)(7+2k)}{9(5+k)^2} \pa  V^{(\frac{3}{2})}
\nonu \\
&+& \left. \frac{2(19+2k)}{3(5+k)} {\bf R^{(\frac{5}{2})}}
\right](w) +
\cdots.
\label{v3halfp2}
\eea
From (\ref{v3halfp2}), the following commutator can be expressed as 
follows:
\bea
[ V^{(\frac{3}{2})}, {\bf P^{(2)}}](w) = -\frac{1}{2} \pa^2 
\{ V^{(\frac{3}{2})} \, 
 {\bf P^{(2)}} \}_{-2}(w) + \pa \{ V^{(\frac{3}{2})} \, 
 {\bf P^{(2)}} \}_{-1}(w).
\nonu
\eea
Therefore, one does not see any mixture between different 
${\cal N}=4$ multiplets in (\ref{v2+w5half-}).
Similarly, 
one has
the following OPE
\bea
T^{(1)}(z) \, {\bf R^{(\frac{5}{2})}}(w) & = &
\frac{1}{(z-w)^2} \, \left[ \frac{8(-3+k)}{3(5+k)} (\hat{G}_{22} - 
V^{(\frac{3}{2})}) \right](w)  
+\frac{1}{(z-w)} \,  {\bf R^{(\frac{5}{2})}}(w)
+\cdots.
\nonu
\eea 
Then one can reexpress the commutator
as follows:
\bea
[ T^{(1)}, {\bf R^{(\frac{5}{2})}}](w) = -\frac{1}{2} \pa^2 
\{ T^{(1)} \, 
 {\bf R^{(\frac{5}{2})}} \}_{-2}(w) + \pa \{ T^{(1)} \, 
 {\bf R^{(\frac{5}{2})}} \}_{-1}(w).
\label{t1r5half}
\eea
The combination of the sixth and seventh terms in the first order pole of
(\ref{v2+w5half-}) can be written in terms of 
derivatives of known composite fields.

In the first OPE of (\ref{opetoday}), the expression $\hat{G}_{22} 
\hat{G}_{22}(w)$ can be expressed as a derivative of $\hat{A}_{-} 
\hat{B}_{+}(w)$. 
The first order pole in the last OPE of (\ref{opetoday})
contains composite field with spin-$4$ with $U(1)$ charge of 
$\frac{6}{(5+k)}$. 
The second order pole contains a composite field of spin-$3$ which 
can be seen from the Table $5$ of \cite{Ahn1311}.

\section{
The nontrivial 
OPEs between  higher spin-$2$ current, $V_{-}^{(2)}(z)$,  
and 
other $7$ higher spin currents
}

We present the following OPEs
\bea
V_{-}^{(2)}(z) \, V^{(\frac{5}{2})}(w) & = &
\frac{1}{(z-w)^2} \, \left[  \frac{2(6+k)}{3(5+k)^2} i \hat{B}_{+} 
\hat{G}_{22}  -\frac{2(6+k)}{3(5+k)^2} 
i \hat{B}_{+}  V^{(\frac{3}{2})} \right](w) 
\nonu \\
& + &  
\frac{1}{(z-w)} \, \left[ 
\frac{4(6+k)}{3(5+k)^2} i \pa \hat{B}_{+}  V^{(\frac{3}{2})}
-\frac{4}{3(5+k)} i  \hat{B}_{+}  \pa V^{(\frac{3}{2})}
- \frac{2(12+k)}{3(5+k)^2} i  \pa \hat{B}_{+}   \hat{G}_{22}
\right. \nonu \\
& - & \frac{4}{(5+k)^2} i   \hat{B}_{+}  \pa  \hat{G}_{22}
+ \frac{2}{(5+k)} i   \hat{B}_{+}  V^{(\frac{5}{2})} 
-\frac{2}{(5+k)} \hat{G}_{22} V_{-}^{(2)}
\nonu \\
&+ & \left. \frac{4}{(5+k)}  V^{(\frac{3}{2})}  V_{-}^{(2)}
\right](w) +\cdots,
\nonu \\
V_{-}^{(2)}(z) \, W^{(2)}(w) & = &
-\frac{1}{(z-w)^3} \, \left[
\frac{2k(8+k)}{(5+k)^2} i \hat{B}_{+} \right] (w)
\nonu \\
& + & \frac{1}{(z-w)^2} \, \left[ \frac{2(4+k)}{(5+k)} V_{-}^{(2)} 
-\frac{k(8+k)}{(5+k)^2} i \pa \hat{B}_{+}
\right](w)
\nonu \\
& + &  \frac{1}{(z-w)} \, \left[
-\frac{1}{2} {\bf R_{-}^{(3)}}
+\frac{4}{(5+k)} i \hat{A}_3 V_{-}^{(2)}
+\frac{(11+2k)}{(5+k)^2} \hat{A}_3 \pa  \hat{B}_{+}
\right. \nonu \\
&- & \frac{4k(75+117k+14k^2)}{
(5+k)(3+7k)(19+23k)} i \hat{B}_{+} \hat{T}
-\frac{2}{(5+k)} i \hat{B}_{+} T^{(2)}
-\frac{(15+2k)}{(5+k)^2} \hat{B}_{+} \pa \hat{A}_3
\nonu \\
&+& \frac{k}{(5+k)^2} \hat{B}_{+} \pa \hat{B}_3
+\frac{1}{2(5+k)}  i \hat{B}_{+} \pa T^{(1)}
-\frac{k}{(5+k)^2}  \hat{B}_3 \pa \hat{B}_{+}
\nonu \\
& + & \frac{(9+2k)}{2(5+k)} \pa  V_{-}^{(2)}
-\frac{k(12+k)}{3(5+k)^2} i \pa^2 \hat{B}_{+}
+\frac{2}{(5+k)} \hat{G}_{22}  T_{-}^{(\frac{3}{2})} 
+\frac{2}{(5+k)} T_{-}^{(\frac{3}{2})} V^{(\frac{3}{2})} 
\nonu \\
&-& \left. \frac{1}{2(5+k)} 
T^{(1)} \hat{B}_{+} \hat{B}_3 + \frac{1}{2(5+k)} \hat{B}_3
T^{(1)} \hat{B}_{+}
\right](w) + \cdots,
\nonu \\
V_{-}^{(2)}(z) \, W_{+}^{(\frac{5}{2})}(w) & = &
\frac{1}{(z-w)^3} \, \left[ \frac{4(7+k)(3+2k)}{3(5+k)^2} \hat{G}_{22}+
\frac{4(11+3k)}{3(5+k)^2}  
V^{(\frac{3}{2})} 
\right](w)
\nonu \\
&+& \frac{1}{(z-w)^2} \, \left[
\frac{(30+7k)}{3(5+k)} {\bf R^{(\frac{5}{2})}} 
-\frac{(38+7k)}{3(5+k)} V^{(\frac{5}{2})}
+\frac{2(6+k)}{(5+k)^2} i \hat{A}_{-} \hat{G}_{12}
\right. \nonu \\
& - & \frac{2(6+k)}{(5+k)^2} i \hat{A}_{-}   T_{-}^{(\frac{3}{2})} 
-\frac{12}{(5+k)^2} i \hat{A}_3 V^{(\frac{3}{2})}  
-\frac{4(8+k)}{(5+k)^2} i \hat{B}_{+} \hat{G}_{21}
 \nonu \\
& - & \frac{2(16+3k)}{(5+k)^2} i \hat{B}_{+}   T_{+}^{(\frac{3}{2})} 
-\frac{4(8+k)}{(5+k)^2} i \hat{B}_3 V^{(\frac{3}{2})} 
-\frac{2}{(5+k)} T^{(1)}  V^{(\frac{3}{2})} 
\nonu \\
&+& \left.  \frac{4(30+17k+2k^2)}{
9(5+k)^2} \pa \hat{G}_{22}  + \frac{8(13+3k)}{9(5+k)^2} \pa 
 V^{(\frac{3}{2})}  \right](w) 
\nonu \\
&+& \frac{1}{(z-w)} \, \left[ 
\frac{2(27+7k)}{15(5+k)} \pa {\bf R^{(\frac{5}{2})}}
-\frac{1}{2} {\bf  R^{(\frac{7}{2})} }
+\frac{1}{(5+k)} i {\bf P_{-}^{(\frac{5}{2})}} \hat{A}_{-}
\right. \nonu \\
&- &  \frac{22}{5(5+k)^2} \hat{A}_{-} \hat{A}_3  T_{-}^{(\frac{3}{2})} 
+  \frac{4}{(5+k)^2} \hat{A}_{-} \hat{B}_{+}  \hat{G}_{11} 
+  \frac{4}{(5+k)^2} \hat{A}_{-} \hat{B}_{+}    U^{(\frac{3}{2})} 
\nonu \\
& + & \frac{8}{(5+k)^2} \hat{A}_{-} \hat{B}_3    T_{-}^{(\frac{3}{2})}  
+  \frac{4(11+k)}{3(5+k)^2} i \hat{A}_{-} \pa \hat{G}_{12} 
-\frac{4(46+5k)}{15(5+k)^2} i \hat{A}_{-} \pa    T_{-}^{(\frac{3}{2})} 
\nonu \\
&+& \frac{2}{(5+k)} i \hat{B}_{+} W_{+}^{(\frac{5}{2})}
+\frac{(8+k)}{5(5+k)^2} \hat{B}_{+} \hat{B}_{-}    V^{(\frac{3}{2})} 
\nonu \\
& + & \frac{4(-551-679k-45k^2+3k^3)}{5(5+k)^2(19+23k)}  
\hat{B}_{+} \hat{B}_3  \hat{G}_{21} 
\nonu \\
& - & \frac{8(304+451k+70k^2+3k^3)}{
15(5+k)^2(19+23k)} i \hat{B}_{+} \pa  \hat{G}_{21} 
-\frac{4(33+8k)}{15(5+k)^2}  i \hat{B}_{+} \pa    T_{+}^{(\frac{3}{2})}
\nonu \\
&- & \frac{(-12+k)}{5(5+k)^2}  
\hat{B}_{-} \hat{B}_{+}    V^{(\frac{3}{2})}   
+\frac{4}{(5+k)} i \hat{A}_3  V^{(\frac{5}{2})} 
+\frac{22}{5(5+k)^2} \hat{A}_3 \hat{A}_{-}    T_{-}^{(\frac{3}{2})}
\nonu \\
& + & \frac{8(9+2k)}{15(5+k)^2} i \hat{A}_3 \pa \hat{G}_{22} 
+\frac{4(25+4k)}{15(5+k)^2} i \hat{A}_3 \pa  V^{(\frac{3}{2})} 
\nonu \\
&-& \frac{4(-551-679k-45k^2+3k^3)}{
5(5+k)^2(19+23k)}  \hat{B}_3 \hat{B}_{+}  \hat{G}_{21} 
-\frac{8k}{5(5+k)^2} i \hat{B}_3 \pa \hat{G}_{22} 
\nonu \\
& - & \frac{4(44+3k)}{5(5+k)^2} i \hat{B}_3 \pa  V^{(\frac{3}{2})} 
+\frac{4}{5(5+k)} T^{(1)} \pa \hat{G}_{22} 
-\frac{2}{(5+k)}  T^{(1)} \pa  V^{(\frac{3}{2})}
\nonu \\
&- & \frac{2(44+7k)}{15(5+k)} \pa  V^{(\frac{5}{2})}
-\frac{4(9+2k)}{5(5+k)^2} i \pa \hat{A}_3 \hat{G}_{22}
- \frac{2(15+4k)}{5(5+k)^2} i \pa \hat{A}_3  V^{(\frac{3}{2})}
\nonu \\
& -& \frac{2(32+7k)}{5(5+k)^2} i \pa \hat{B}_{+}  T_{+}^{(\frac{3}{2})}
+ \frac{12k}{5(5+k)^2} i \pa \hat{B}_3  \hat{G}_{22}
-\frac{6}{5(5+k)} \pa  T^{(1)} \hat{G}_{22}
\nonu \\
& + & \frac{1}{(5+k)} \pa  T^{(1)}   V^{(\frac{3}{2})}
\nonu \\
& + & \frac{(6370 k^4+46693 k^3+162836 k^2+193913 k+68856
)}{45(5+k)^2(19+23k)(47+35k)} \pa^2 \hat{G}_{22}
\nonu \\
&+ & \frac{2(8047+8257k+1350k^2)}{
45(5+k)^2(47+35k)}  \pa^2   V^{(\frac{3}{2})}
-\frac{2}{(5+k)}  T_{+}^{(\frac{3}{2})}  V_{-}^{(2)}
-\frac{2}{(5+k)}  T_{-}^{(\frac{3}{2})}  V_{+}^{(2)}
\nonu \\
& + & \frac{8(1368+2493k+1178k^2+133k^3)}{
3(5+k)(19+23k)(47+35k)}  \hat{G}_{22} \hat{T}
\nonu \\
&+& \left. \frac{4(12+145k+49k^2)}{(
5+k)(3+7k)(47+35k)}  V^{(\frac{3}{2})} \hat{T} 
+\frac{2}{(5+k)}  V^{(\frac{3}{2})} T^{(2)}
\right](w) + \cdots,
\nonu \\
V_{-}^{(2)}(z) \, W_{-}^{(\frac{5}{2})}(w) & = &
\frac{1}{(z-w)^2} \, \left[ \frac{2(16+k)}{3(5+k)^2} 
i \hat{B}_{+}  T_{-}^{(\frac{3}{2})}\right] (w)
\nonu \\
& + & \frac{1}{(z-w)} \, \left[ 
\frac{8(8+k)}{3(5+k)^2} \hat{A}_3 \hat{B}_{+} \hat{G}_{12}
-\frac{2(8+k)}{(5+k)^2} \hat{B}_3 \hat{B}_{+} \hat{G}_{12}
-\frac{2}{(5+k)} i \hat{B}_{+}  W_{-}^{(\frac{5}{2})}  
\right. \nonu \\
& + & 
\frac{2(8+k)}{(5+k)^2}  \hat{B}_{+} \hat{B}_3 \hat{G}_{12}
-\frac{2}{(5+k)} \hat{G}_{12} V_{-}^{(2)} 
-\frac{8(8+k)}{3(5+k)^2} \hat{G}_{12}  \hat{A}_3 \hat{B}_{+} 
\nonu \\
&+ & \left. \frac{4(7+k)}{3(5+k)^2} 
\hat{B}_3 \hat{B}_{+}   T_{-}^{(\frac{3}{2})}
- \frac{4(7+k)}{3(5+k)^2} 
\hat{B}_{+} \hat{B}_3   T_{-}^{(\frac{3}{2})}
+\frac{4(11+k)}{3(5+k)^2} i \hat{B}_{+} \pa T_{-}^{(\frac{3}{2})}
\right](w) \nonu \\
& + & \cdots,
\nonu \\
V_{-}^{(2)}(z) \, W^{(3)}(w) & = &
-\frac{1}{(z-w)^4} \, \left[ \frac{2k(2687+3264k+973k^2+84k^3)}{
(5+k)^3(19+23k)} i \hat{B}_{+} \right] (w)
\nonu \\
& + & \frac{1}{(z-w)^3} \, \left[-\frac{2(114-853k+13k^2)}{
3(5+k)^2(19+23k)} V_{-}^{(2)}
-\frac{4(-9+11k+k^2)}{(5+k)^3} \hat{A}_3 \hat{B}_{+}  
\right. \nonu \\
& +&   \left. \frac{4k(5+2k)}{(5+k)^3} \hat{B}_{+} \hat{B}_3  
-\frac{2k(5+2k)}{(5+k)^3} i \pa \hat{B}_{+}
+\frac{2(-3+2k)}{(5+k)^2} i T^{(1)} \hat{B}_{+}
\right](w) 
\nonu \\
& + & \frac{1}{(z-w)^2} \, \left[ -\frac{(23+6k)}{2(5+k)}  {\bf R_{-}^{(3)} }
-\frac{(-183+101k+16k^2)}{(5+k)(19+23k)}  T^{(1)} V_{-}^{(2)}
\right. \nonu \\
&+ & \frac{(31+8k)}{2(5+k)^2} i  T^{(1)} \pa \hat{B}_{+} 
+ \frac{6(112+113k+45k^2)}{(
5+k)^2(19+23k)} i \hat{A}_3  V_{-}^{(2)}
\nonu \\
& + & \frac{3(25+22k+4k^2)}{(5+k)^3} \hat{A}_3 \pa \hat{B}_{+}
\nonu \\
& - & \frac{2(285+4328k+7515k^2+2120k^3+112k^4)}{
(5+k)^2(3+7k)(19+23k)} i \hat{B}_{+} \hat{T}
-\frac{(33+7k)}{(5+k)^2} i \hat{B}_{+} T^{(2)}
\nonu \\
&+ & \frac{(41+9k)}{(5+k)^2}  i \hat{B}_{+} W^{(2)}
-\frac{(203+114k+14k^2)}{(5+k)^3} \hat{B}_{+} \pa \hat{A}_3
\nonu \\
& + &  \frac{(10+39k+6k^2)}{(5+k)^3} \hat{B}_{+} \pa \hat{B}_3
-\frac{(29+4k)}{2(5+k)^2}  i \hat{B}_{+} \pa T^{(1)}
\nonu \\
& - &  \frac{6(57+215k+42k^2)}{(
5+k)^2(19+23k)} i \hat{B}_3  V_{-}^{(2)}
- \frac{k(19+2k)}{(5+k)^3} \hat{B}_3 \pa \hat{B}_{+}
\nonu \\
& + & \frac{(4617+8863k+1538k^2)}{
6(5+k)^2(19+23k)} \pa  V_{-}^{(2)}
-\frac{2k(51+10k)}{3(5+k)^3} i \pa^2 \hat{B}_{+}
\nonu \\
& + & \frac{2(6+k)}{(5+k)^2} \hat{G}_{12} \hat{G}_{22}
-\frac{4(7+k)}{(5+k)^2} \hat{G}_{12}  V^{(\frac{3}{2})}
+\frac{6(7+2k)}{(5+k)^2} \hat{G}_{22}  T_{-}^{(\frac{3}{2})}
\nonu \\
&+ &  \frac{2(23+6k)}{(5+k)^2}  T_{-}^{(\frac{3}{2})}  V^{(\frac{3}{2})}
 -  \frac{2(9+7k)}{(5+k)^3} i \hat{A}_{+} \hat{A}_{-} \hat{B}_{+}
-\frac{2(-19+5k)}{(5+k)^3}  i \hat{A}_3 \hat{A}_3 \hat{B}_{+}
\nonu \\
& + & \frac{4(-7+k)}{(5+k)^3} i \hat{A}_3 \hat{B}_{+} \hat{B}_3
- \frac{2}{(5+k)^2} i \hat{B}_{+} \hat{B}_{+} \hat{B}_{-}
+ \frac{2(-5+3k)}{(5+k)^3} i \hat{B}_{+} \hat{B}_3 \hat{B}_3
\nonu \\
&+ & \left. \frac{8}{(5+k)^2} T^{(1)} \hat{A}_3 \hat{B}_{+}
-\frac{8}{(5+k)^2}   T^{(1)} \hat{B}_{+} \hat{B}_3
\right](w) 
\nonu \\
& + & \frac{1}{(z-w)} \, \left[ 
\frac{2}{(5+k)} i \hat{A}_3 {\bf R_{-}^{(3)}}
-\frac{2}{(5+k)} i \hat{B}_3 {\bf R_{-}^{(3)} }
-\frac{(5+2k)}{2(5+k)} \pa  {\bf R_{-}^{(3)}}
\right.
\nonu \\
&+& \frac{2}{(5+k)}  V_{-}^{(2)} W^{(2)}
+\frac{5(21+4k)}{6(5+k)^2} \pa^2  V_{-}^{(2)}
-\frac{2}{(5+k)} T^{(2)}  V_{-}^{(2)}
 \nonu \\
&+& \frac{2(5+2k)}{(5+k)^2} \pa  T_{-}^{(\frac{3}{2})}  V^{(\frac{3}{2})}
+ \frac{2(5+2k)}{(5+k)^2}   T_{-}^{(\frac{3}{2})}  \pa V^{(\frac{3}{2})}
+\frac{2(8+3k)}{(5+k)^2} i \pa \hat{A}_3  V_{-}^{(2)}
\nonu \\
& +& \frac{2(1+3k)(-41+11k)}{(5+k)^2(19+23k)} i  \hat{A}_3  \pa V_{-}^{(2)}
-\frac{8}{(5+k)^2} i \hat{A}_3   T_{-}^{(\frac{3}{2})}  V^{(\frac{3}{2})}
\nonu \\
&+& \frac{4(6+k)}{(5+k)^3} i \pa \hat{A}_3 \hat{A}_3 \hat{B}_{+}
-\frac{2(-17+7k)}{(5+k)^3} i \hat{A}_3 \hat{A}_3 \pa \hat{B}_{+} 
-\frac{8}{(5+k)^2} \hat{A}_3 \hat{B}_3  V_{-}^{(2)} 
\nonu \\
&-& \frac{8}{(5+k)^2} \hat{A}_3 \hat{B}_{+}  T^{(2)} 
-\frac{(1539+4506k+6567k^2+5108k^3+532k^4)}{(5+k)^3(3+7k)(19+23k)} 
\pa^2 \hat{A}_3 \hat{B}_{+}
\nonu \\
&- & \frac{2(64+8k+k^2)}{(5+k)^3} 
\pa \hat{A}_3 \pa \hat{B}_{+} +
\frac{12}{(5+k)^2} \hat{A}_3 \hat{A}_3  V_{-}^{(2)} 
\nonu \\
& + & \frac{(3591+24804k+37689k^2+16010k^3+2590k^4)}{
3(5+k)^3(3+7k)(19+23k)} \hat{A}_3 \pa^2 \hat{B}_{+}
\nonu \\
&-& \frac{4(6+k)}{(5+k)^3} i \pa \hat{A}_3 \hat{B}_{+} \hat{B}_3
+ \frac{8(-8+k)}{(5+k)^3} i  \hat{A}_3 \pa \hat{B}_{+} \hat{B}_3
+ \frac{36}{(5+k)^3} i  \hat{A}_3  \hat{B}_{+} \pa \hat{B}_3
\nonu \\
& + & \frac{6}{(5+k)^2} \hat{A}_3  \pa \hat{B}_{+} T^{(1)} 
+ \frac{2}{(5+k)^2} \hat{A}_3   \hat{B}_{+} \pa T^{(1)} 
-\frac{8}{(5+k)^2} i \hat{A}_3 \hat{G}_{22}  T_{-}^{(\frac{3}{2})}
\nonu \\
&-& \frac{2(9+k)}{(5+k)^3} i \pa \hat{A}_{+} \hat{A}_{-} \hat{B}_{+} 
- \frac{2(9+k)}{(5+k)^3} i  \hat{A}_{+} \pa \hat{A}_{-} \hat{B}_{+} 
-\frac{2(-11+5k)}{(5+k)^3} i  \hat{A}_{+}  \hat{A}_{-} \pa \hat{B}_{+} 
\nonu \\
&-& \frac{2(7+2k)}{(5+k)^2} i \pa \hat{B}_3  V_{-}^{(2)} 
- \frac{80k}{(5+k)(19+23k)} i  \hat{B}_3  \pa V_{-}^{(2)} 
+\frac{8}{(5+k)^2} i \hat{B}_3  T_{-}^{(\frac{3}{2})}  V^{(\frac{3}{2})}
\nonu \\
&-& \frac{4}{(5+k)^2} \hat{B}_3 \hat{B}_3  V_{-}^{(2)} 
+\frac{8}{(5+k)^2} i \hat{B}_3 \hat{G}_{22}  T_{-}^{(\frac{3}{2})} 
+\frac{7(3+k)}{(5+k)^2} i \pa \hat{B}_{+} W^{(2)}
\nonu \\
&+& \frac{(9+k)}{(5+k)^2} i \hat{B}_{+} \pa W^{(2)}
-\frac{(25+k)}{(5+k)^2} i \pa \hat{B}_{+} T^{(2)}
-\frac{3(-1+k)}{(5+k)^2} i  \hat{B}_{+} \pa T^{(2)}
\nonu \\
&+& \frac{(-41040-17541k+135996k^2+78991k^3+8990k^4+336k^5)}{
18(5+k)^3(3+7k)(19+23k)} i \pa^3 \hat{B}_{+}
\nonu \\
&+& \frac{8}{(5+k)^2} \hat{B}_{+} \hat{B}_3  T^{(2)}
\nonu \\
& - & \frac{(-5130-5247k+15402k^2+16001k^3+658k^4)}{
3(5+k)^3(3+7k)(19+23k)} \pa^2 \hat{B}_{+} \hat{B}_3
\nonu \\
&+& \frac{(4788+8937k+1460k^2-539k^3+630k^4)}{
3(5+k)^3(3+7k)(19+23k)}  \hat{B}_{+} \pa^2 \hat{B}_3
\nonu \\
&- & \frac{2(-45+13k)}{(5+k)^3} 
i \pa \hat{B}_{+} \hat{B}_3 \hat{B}_3
- \frac{8(-3+4k)}{(5+k)^3} 
i  \hat{B}_{+} \pa \hat{B}_3 \hat{B}_3
\nonu \\
& + & \frac{8(-15+8k)}{3(5+k)^3} 
\hat{B}_{+} \hat{B}_3 \hat{B}_3 \hat{B}_3
-\frac{2(-15+8k)}{(5+k)^3} 
\hat{B}_{+} \hat{B}_{+} \hat{B}_{-} \hat{B}_3
\nonu \\
& - & \frac{6}{(5+k)^2} \pa \hat{B}_{+} \hat{B}_3 T^{(1)}
-\frac{2}{(5+k)^2} \hat{B}_{+} \hat{B}_3 \pa T^{(1)}
-\frac{4}{(5+k)^2} \hat{B}_{+} \hat{B}_{-}  V_{-}^{(2)} 
\nonu \\
&+& \frac{2(-9+8k)}{(5+k)^3} 
i \pa \hat{B}_{+} \hat{B}_{+} \hat{B}_{-}
-\frac{(3+10k)}{(5+k)^3} i \hat{B}_{+} \hat{B}_{+} \pa \hat{B}_{-}
\nonu \\
&-& \frac{4}{(5+k)^2} i \hat{B}_{+} \hat{G}_{21}   T_{-}^{(\frac{3}{2})}
+\frac{(9+4k)}{2(5+k)^2} i \pa^2 \hat{B}_{+} T^{(1)}
-\frac{(-1+2k)}{(5+k)^2} i \pa \hat{B}_{+} \pa T^{(1)}
\nonu \\
&- & \frac{5}{2(5+k)^2} i \hat{B}_{+} \pa^2 T^{(1)}  
+\frac{2}{(5+k)} \hat{G}_{12}  V^{(\frac{5}{2})}
-\frac{8(7+k)}{3(5+k)^2} \hat{G}_{12} \pa  V^{(\frac{3}{2})}
\nonu \\
&+ & \frac{4(6+k)}{3(5+k)^2} \hat{G}_{12} \pa \hat{G}_{22}  
+\frac{2(5+2k)}{(5+k)^2} \pa \hat{G}_{22}  T_{-}^{(\frac{3}{2})}
+\frac{2(3+2k)}{(5+k)^2} \hat{G}_{22} \pa  T_{-}^{(\frac{3}{2})}
\nonu \\
&- & \frac{1}{(5+k)} \pa T^{(1)}   V_{-}^{(2)} 
- \frac{(-101+39k+8k^2)}{(5+k)(19+23k)} T^{(1)} \pa  V_{-}^{(2)}
\nonu \\
&-& \frac{4(57+199k+90k^2)}{(5+k)(3+7k)(19+23k)} \hat{T} V_{-}^{(2)}
-\frac{16k(75+117k+14k^2)}{(5+k)^2(3+7k)(19+23k)} \hat{T} \hat{A}_3 \hat{B}_{+}
\nonu \\
&-& \frac{2(3+k)(171+571k+532k^2+56k^3)}{(5+k)^2(3+7k)(19+23k)} 
i \pa \hat{T} \hat{B}_{+}
\nonu \\
&-& \frac{2(-855-786k+1509k^2+608k^3)}{
(5+k)^2(3+7k)(19+23k)} i \hat{T} \pa \hat{B}_{+}
\nonu \\
&+& \left. \frac{16k(75+117k+14k^2)}{(5+k)^2(3+7k)(19+23k)} 
\hat{T} \hat{B}_{+} \hat{B}_3
\right](w) +\cdots.
\label{toope}
\eea
The first order pole in the last OPE of (\ref{toope})
contains composite field with spin-$4$ with $U(1)$ charge of 
$-\frac{2k}{(5+k)}$. 
The second order pole has a composite field of spin-$3$ starting from the 
second term. The field contents can be seen from the Table $5$ in 
\cite{Ahn1311}. Note that there is no $ T^{(1)} T^{(1)} \hat{B}_{+}(w)$ term 
in the second order pole. 
One can also analyze each singular term in terms of descendant fields plus
(quasi) primary fields as done before.  

\section{
The nontrivial 
OPEs between  higher spin-$\frac{5}{2}$ current, $V^{(\frac{5}{2})}(z)$,  
and 
other $7$ higher spin currents
}

The corresponding OPEs are given by
\bea
V^{(\frac{5}{2})}(z) \, V^{(\frac{5}{2})}(w) & = &
-\frac{1}{(z-w)^3} \, \left[ \frac{16(-111+7k+2k^2)}{
9(5+k)^3} \hat{A}_{-} \hat{B}_{+} \right] (w) 
\nonu \\
&+& 
\frac{1}{(z-w)^2} \, \frac{1}{2} \pa \{ V^{(\frac{5}{2})} \, V^{(\frac{5}{2})} 
\}_{-3}(w) 
\nonu  \\
& + & 
\frac{1}{(z-w)} \left[  
\frac{2}{(5+k)} i \hat{A}_{-} {\bf R_{-}^{(3)}}
-\frac{4}{(5+k)} V^{(\frac{3}{2})} V^{(\frac{5}{2})}
+ \frac{4}{(5+k)} V_{+}^{(2)} V_{-}^{(2)} 
\right.
\nonu \\
& + & \frac{52}{3(5+k)^2} i \pa \hat{A}_{-}  V_{-}^{(2)}
- \frac{38}{3(5+k)^2} i  \hat{A}_{-}  \pa V_{-}^{(2)}
-\frac{16}{(5+k)^3} \hat{A}_{-} \hat{A}_3 \hat{A}_3 \hat{B}_{+}
\nonu \\
&-& \frac{4(-1+2k)}{(5+k)^3} i  \hat{A}_{-} \hat{A}_3 \pa \hat{B}_{+}
+\frac{4(-9+2k)}{(5+k)^3} i  \hat{A}_{-} \pa \hat{A}_3 \hat{B}_{+}
+ \frac{32}{(5+k)^3} \hat{A}_{-}  \hat{A}_{3}  \hat{B}_{+} \hat{B}_{3}
\nonu \\
&- & \frac{16}{(5+k)^3} \hat{A}_{+}  \hat{A}_{-}  \hat{A}_{-} \hat{B}_{+}
+  \frac{8}{(5+k)^2} \hat{A}_{-}  \hat{B}_{+}  W^{(2)}
 + 
\frac{16}{(5+k)^2} \hat{A}_{-} \hat{A}_3  V_{-}^{(2)} 
\nonu \\
&+& \frac{4(1786+2513k+303k^2+12k^3)}{3(5+k)^3(19+23k)} 
\pa^2 \hat{A}_{-} \hat{B}_{+}
\nonu \\
& - & \frac{4(3211+4874k+989k^2+70k^3)}{3(5+k)^3(19+23k)} 
\pa \hat{A}_{-} \pa \hat{B}_{+}
\nonu \\
&-&
\frac{4(3154+4463k+537k^2+24k^3)}{
3(5+k)^3(19+23k)} i \pa \hat{A}_{-} \hat{B}_{+} \hat{B}_3
\nonu \\
&- & \frac{4(4+k)}{(5+k)^3} i \hat{A}_{-} \hat{B}_{+} \pa \hat{B}_3
-\frac{16}{(5+k)^3} \hat{A}_{-} \hat{B}_{+} \hat{B}_3 \hat{B}_3 
-\frac{6}{(5+k)^2} \pa \hat{A}_{-} \hat{B}_{+} T^{(1)}
\nonu \\
&- & \frac{6}{(5+k)^2} \hat{A}_{-} \hat{B}_{+} \pa T^{(1)}  
-\frac{16}{(5+k)^3} \hat{A}_{-} \hat{B}_{+} \hat{B}_{+} \hat{B}_{-}  
+\frac{4(13+2k)}{3(5+k)^2} i \pa \hat{B}_{+}   V_{+}^{(2)} 
\nonu \\
&-& \frac{4}{3(5+k)} i  \hat{B}_{+}  \pa  V_{+}^{(2)} 
-\frac{8}{(5+k)^2} i 
\hat{A}_{-}  T_{-}^{(\frac{3}{2})}  V^{(\frac{3}{2})}
-\frac{16}{(5+k)^2} \hat{A}_{-} \hat{B}_3  V_{-}^{(2)} 
\nonu \\
&- & \frac{8}{(5+k)^2} i \hat{A}_{-} \hat{G}_{22}  T_{-}^{(\frac{3}{2})} 
-\frac{8(1577+2203k+234k^2+12k^3)}{
3(5+k)^3(19+23k)} i \hat{A}_{-} \pa \hat{B}_{+} \hat{B}_3
\nonu \\
&- & \frac{8(6+k)}{3(5+k)^2} \pa \hat{G}_{22}   V^{(\frac{3}{2})} 
+ \frac{8(6+k)}{3(5+k)^2}  \hat{G}_{22}  \pa  V^{(\frac{3}{2})} 
-\frac{16(19+29k+2k^2)}{(5+k)^2(19+23k)} \hat{T} \hat{A}_{-} \hat{B}_{+}
\nonu \\
&+&  \frac{8}{(5+k)^2} \hat{G}_{22} \pa \hat{G}_{22} 
-\frac{8(7+k)}{3(5+k)^2}  V^{(\frac{3}{2})}   \pa  V^{(\frac{3}{2})}  
\nonu \\
&-& \left. \frac{2(3154+4463k+537k^2+24k^3)}{
3(5+k)^2(19+23k)} i \hat{B}_3 \hat{G}_{22} \hat{G}_{22} 
-\frac{3}{(5+k)} T^{(1)} \hat{G}_{22} \hat{G}_{22}
\right](w) + \cdots,
\nonu \\
 V^{(\frac{5}{2})}(z) \, W^{(2)}(w) & = &
\frac{1}{(z-w)^3} \, \left[ \frac{4(-3+k)}{3(5+k)^2} \hat{G}_{22} -
 \frac{4(-3+k)}{3(5+k)^2}  V^{(\frac{3}{2})}
\right](w) \nonu \\
&+ &  \frac{1}{(z-w)^2} \, \left[
-\frac{(19+5k)}{2(5+k)} {\bf R^{(\frac{5}{2})}} 
+\frac{(22+5k)}{(5+k)}  V^{(\frac{5}{2})}
-\frac{4(6+k)}{3(5+k)^2} i \hat{A}_{-} \hat{G}_{12}
\right. \nonu \\
&+ & \frac{4(6+k)}{3(5+k)^2}   i \hat{A}_{-}  T_{-}^{(\frac{3}{2})}
+\frac{12}{(5+k)^2} i \hat{A}_3  V^{(\frac{3}{2})}
+\frac{4(8+k)}{(5+k)^2} i \hat{B}_{+} \hat{G}_{21}
\nonu \\
& + &  \frac{4(27+5k)}{3(5+k)^2}  i \hat{B}_{+} 
T_{+}^{(\frac{3}{2})}
+\frac{2}{(5+k)} T^{(1)}  V^{(\frac{3}{2})}
+\frac{4(-3+k)}{3(5+k)^2} \pa \hat{G}_{22} 
\nonu \\
& - & \left. \frac{4(5+2k)}{3(5+k)^2} \pa   V^{(\frac{3}{2})} 
\right](w) 
\nonu \\
& + &  \frac{1}{(z-w)} \, \left[ -\frac{3(3+k)}{2(5+k)} 
\pa {\bf R^{(\frac{5}{2})}} 
-\frac{1}{(5+k)} i {\bf P_{-}^{(\frac{5}{2})}} \hat{A}_{-}
-\frac{4(11+k)}{3(5+k)^2} 
i \hat{A}_{-} \pa \hat{G}_{12}
\right. \nonu \\
&+& \frac{4(11+k)}{3(5+k)^2} 
i \hat{A}_{-} \pa T_{-}^{(\frac{3}{2})}
-\frac{2}{(5+k)} i \hat{B}_{+} W_{+}^{(\frac{5}{2})}
+\frac{8(8+k)}{3(5+k)^2} i \hat{B}_{+} \pa \hat{G}_{21}
\nonu \\
& + & \frac{8(9+2k)}{3(5+k)^2}  i \hat{B}_{+} \pa T_{+}^{(\frac{3}{2})}
+\frac{12}{(5+k)^2} i \hat{A}_3 \pa  V^{(\frac{3}{2})} 
+\frac{4(8+k)}{(5+k)^2} i \hat{B}_3 \pa  V^{(\frac{3}{2})} 
\nonu \\
& + & \frac{2}{(5+k)} T^{(1)} \pa V^{(\frac{3}{2})}  
+\frac{3(4+k)}{(5+k)} \pa  V^{(\frac{5}{2})} 
+\frac{2(8+k)}{(5+k)^2} i \pa  \hat{B}_{+}  \hat{G}_{21}
\nonu \\
&+ & \frac{2(4+k)}{(5+k)^2}  i \pa  \hat{B}_{+}  T_{+}^{(\frac{3}{2})}
+\frac{4(-1+k)}{3(5+k)^2} \pa^2 \hat{G}_{22}
-\frac{(20+7k)}{3(5+k)^2} \pa^2  V^{(\frac{3}{2})} 
\nonu \\
&+ & \left. \frac{2}{(5+k)} \hat{G}_{21}  V_{-}^{(2)}
+ \frac{2}{(5+k)} T_{+}^{(\frac{3}{2})}   V_{-}^{(2)}
\right](w)
+\cdots,
\nonu \\
V^{(\frac{5}{2})}(z) \, W_{+}^{(\frac{5}{2})}(w) & = &
\frac{1}{(z-w)^4} \, 
\left[ \frac{16(17+17k+2k^2)}{(5+k)^3} i \hat{A}_{-} \right] (w)
\nonu \\
& + & \frac{1}{(z-w)^3} \, \left[ 
\frac{8(13+2k)}{9(5+k)^2} V_{+}^{(2)} 
-\frac{16(-2+k)}{(5+k)^3} \hat{A}_{-} \hat{A}_3
+\frac{16(108+20k+k^2)}{9(5+k)^3} \hat{A}_{-} \hat{B}_3
\right. \nonu \\
&+& \left. \frac{8(19+16k+2k^2)}{
(5+k)^3} i \pa  \hat{A}_{-} 
+\frac{8(-2+k)}{3(5+k)^2} i T^{(1)} \hat{A}_{-}
\right](w) 
\nonu \\
&+ &  \frac{1}{(z-w)^2} \, \left[ 
\frac{(37+8k)}{3(5+k)} {\bf R_{+}^{(3)}} 
-\frac{1}{(5+k)} i {\bf P^{(2)}} \hat{A}_{-}
-\frac{4}{(5+k)} T^{(1)} V_{+}^{(2)} 
\right. \nonu \\
&- & \frac{(35+8k)}{3(5+k)^2} i T^{(1)} \pa \hat{A}_{-} 
+\frac{8(4935+13246k+6359k^2+720k^3)}{
3(5+k)^2(3+7k)(19+23k)} i \hat{A}_{-} \hat{T}
\nonu \\
&- & \frac{8(17+3k)}{3(5+k)^2}  i \hat{A}_{-} T^{(2)} 
-\frac{4(5+2k)}{3(5+k)^2}  i \hat{A}_{-} W^{(2)} 
-\frac{2(191+40k)}{(5+k)^3} \hat{A}_{-} \pa \hat{A}_3 
\nonu \\
&- &  \frac{2(-204+13k+8k^2)}{9(5+k)^3} \hat{A}_{-} \pa \hat{B}_3
+ \frac{(55+16k)}{3(5+k)^2} i \hat{A}_{-} \pa T^{(1)}
-\frac{24}{(5+k)^2} i \hat{A}_3 V_{+}^{(2)}
\nonu \\
&+ & \frac{2(35+8k)}{(5+k)^3} \hat{A}_3 \pa  \hat{A}_{-} 
-\frac{8(13+2k)}{3(5+k)^2} i \hat{B}_3 V_{+}^{(2)}
+\frac{2(1008+233k+16k^2)}{
9(5+k)^3} \hat{B}_3  \pa  \hat{A}_{-} 
\nonu \\
&+ & \frac{(25+14k)}{9(5+k)^2} \pa   V_{+}^{(2)} 
+\frac{4(-27+15k+4k^2)}{3(5+k)^3} i \pa^2 \hat{A}_{-}
+\frac{4(31+4k)}{3(5+k)^2} \hat{G}_{22} \hat{G}_{21}
\nonu \\
&+& \frac{8(17+3k)}{3(5+k)^2} 
\hat{G}_{22}  T_{+}^{(\frac{3}{2})}
-\frac{4(17+2k)}{3(5+k)^2} \hat{G}_{21}  V^{(\frac{3}{2})} 
-\frac{8(4+k)}{3(5+k)^2}  T_{+}^{(\frac{3}{2})}  V^{(\frac{3}{2})} 
\nonu \\
&+ & \frac{16(17+3k)}{3(5+k)^3} 
i \hat{A}_{-} \hat{A}_3 \hat{A}_3 
-\frac{32(17+3k)}{3(5+k)^3} 
i \hat{A}_{-} \hat{A}_3 \hat{B}_3 
+\frac{40(9+2k)}{3(5+k)^3} 
i \hat{A}_{-} \hat{B}_{-} \hat{B}_{+} 
\nonu \\
& + & \left.  \frac{16(17+3k)}{3(5+k)^3} 
i \hat{A}_{-} \hat{B}_3 \hat{B}_3 
+  \frac{16(17+3k)}{3(5+k)^3} 
i \hat{A}_{+} \hat{A}_{-} \hat{A}_{-} 
\right](w)
\nonu \\
&+ &  \frac{1}{(z-w)} \, \left[ 
\frac{1}{(5+k)} i \hat{A}_{-} {\bf P^{(3)}}
-\frac{1}{(5+k)} i \pa \hat{A}_{-} {\bf P^{(2)}}
-\frac{1}{4(5+k)} i  \hat{A}_{-} \pa {\bf P^{(2)}}
\right. \nonu \\
& + & \frac{(15+4k)}{3(5+k)} \pa {\bf R_{+}^{(3)}}
\nonu \\
&+ & \frac{2}{(5+k)}   V_{+}^{(2)} W^{(2)}
+\frac{(31+6k)}{3(5+k)^2} \pa^2  V_{+}^{(2)}
+ \frac{2}{(5+k)}   T^{(2)} V_{+}^{(2)} 
\nonu \\
&+& \frac{2}{(5+k)}  T_{+}^{(\frac{3}{2})}  V^{(\frac{5}{2})} 
-\frac{4}{(5+k)^2}  \pa T_{+}^{(\frac{3}{2})}  V^{(\frac{3}{2})} 
-\frac{4(11+2k)}{3(5+k)^2}   T_{+}^{(\frac{3}{2})} \pa V^{(\frac{3}{2})} 
\nonu \\
&- & \frac{24}{(5+k)^2} i \pa \hat{A}_3   V_{+}^{(2)}
- \frac{12}{(5+k)^2} i  \hat{A}_3 \pa  V_{+}^{(2)}
-\frac{4}{(5+k)^2} \hat{A}_3 \hat{A}_3  V_{+}^{(2)} 
\nonu \\
& + & \frac{8}{(5+k)^2} \hat{A}_3 \hat{B}_3  V_{+}^{(2)} 
+\frac{2(11+2k)}{3(5+k)^2} i \pa \hat{A}_{-} W^{(2)}
-\frac{2(13+4k)}{3(5+k)^2} i  \hat{A}_{-} \pa W^{(2)}
\nonu \\
&- & \frac{2(21+2k)}{3(5+k)^2} i \pa \hat{A}_{-} T^{(2)}
- \frac{2(39+8k)}{3(5+k)^2} i  \hat{A}_{-} \pa T^{(2)}
\nonu \\
&- & \frac{4(10065+29260k+18704k^2+3616k^3+107k^4)}{3(5+k)^3(3+7k)(19+23k)} 
i \pa^3 \hat{A}_{-} 
\nonu \\
& + & \frac{2(349+801k+104k^2)}{(5+k)^3(19+23k)} \pa^2 \hat{A}_{-} 
\hat{A}_3
-\frac{4(31+6k)}{(5+k)^3} \pa \hat{A}_{-} \pa \hat{A}_3
\nonu \\
& - & \frac{2(1703+2139k+448k^2)}{(5+k)^3(19+23k)}  
\hat{A}_{-} \pa^2 \hat{A}_3
+\frac{4(15+2k)}{3(5+k)^3} i \pa \hat{A}_{-} \hat{A}_3 \hat{A}_3 
\nonu \\
&+& \frac{8(39+8k)}{3(5+k)^3} i  \hat{A}_{-} \pa \hat{A}_3 \hat{A}_3
-\frac{8(15+2k)}{3(5+k)^3} i \pa \hat{A}_{-} \hat{A}_3 \hat{B}_3 
-\frac{64(6+k)}{3(5+k)^3} i  \hat{A}_{-} \pa \hat{A}_3 \hat{B}_3
\nonu \\
& - & \frac{16(15+4k)}{3(5+k)^3} i  \hat{A}_{-} \hat{A}_3 \pa \hat{B}_3 
+\frac{2(4674+7449k+1819k^2+104k^3)}{3(5+k)^3(19+23k)} 
\pa^2 \hat{A}_{-} \hat{B}_3
\nonu \\
&-& \frac{4(-31+7k+2k^2)}{3(5+k)^3} \pa \hat{A}_{-} \pa \hat{B}_3
+ \frac{2(-418-273k-143k^2+12k^3)}{
3(5+k)^3(19+23k)}  \hat{A}_{-} \pa^2 \hat{B}_3
\nonu \\
&+ & \frac{4(15+2k)}{3(5+k)^3}  i \pa \hat{A}_{-} \hat{B}_3 \hat{B}_3  
+\frac{8(39+8k)}{3(5+k)^3}  i  \hat{A}_{-} \pa \hat{B}_3 \hat{B}_3  
+\frac{4}{(5+k)^2} \hat{A}_{-} \pa \hat{B}_3 T^{(1)}
\nonu \\
& - & \frac{4}{(5+k)^2} \hat{A}_{-} \hat{B}_3 \pa T^{(1)}
+\frac{4}{(5+k)^2} \hat{A}_{-} \hat{B}_{-}  V_{-}^{(2)}
+\frac{4}{(5+k)^2} \hat{A}_{-} \hat{B}_{+}  U_{+}^{(2)} 
\nonu \\
&+& \frac{4(25+6k)}{3(5+k)^3} i \pa \hat{A}_{-} \hat{B}_{+} \hat{B}_{-} 
+ \frac{4(55+12k)}{3(5+k)^3} i \hat{A}_{-} \pa \hat{B}_{+} \hat{B}_{-} 
\nonu \\
& + & \frac{4(43+12k)}{3(5+k)^3} i \hat{A}_{-}  \hat{B}_{+} \pa \hat{B}_{-} 
\nonu \\
&- & \frac{(235+663k+104k^2)}{3(5+k)^2(19+23k)} i \pa^2 \hat{A}_{-} T^{(1)}
+\frac{(411+347k+172k^2)}{3(5+k)^2(19+23k)} i \hat{A}_{-} \pa^2 T^{(1)}
\nonu \\
&- & \frac{4}{(5+k)^2} \hat{A}_{+} \hat{A}_{-}  V_{+}^{(2)} 
+\frac{4(39+8k)}{3(5+k)^3} i \pa \hat{A}_{+} \hat{A}_{-} \hat{A}_{-}
+\frac{8(27+5k)}{3(5+k)^3} i  \hat{A}_{+} \pa \hat{A}_{-} \hat{A}_{-}
\nonu \\
&+ & \frac{4}{(5+k)^2} i \pa \hat{B}_3  V_{+}^{(2)} 
-\frac{4(6+k)}{(5+k)^2} i  \hat{B}_3  \pa V_{+}^{(2)} 
-\frac{4}{(5+k)^2} \hat{B}_3 \hat{B}_3  V_{+}^{(2)} 
\nonu \\
&-& \frac{4}{(5+k)^2} \hat{B}_{+} \hat{B}_{-}  V_{+}^{(2)} 
+\frac{2}{(5+k)} \hat{G}_{21}  V^{(\frac{5}{2})} 
-\frac{4}{(5+k)^2} \pa \hat{G}_{21}  V^{(\frac{3}{2})} 
\nonu \\
& - & \frac{8(10+k)}{3(5+k)^2}  \hat{G}_{21}  \pa V^{(\frac{3}{2})} 
-\frac{4(13+2k)}{3(5+k)^2} \pa \hat{G}_{21} \hat{G}_{22}
-\frac{8(6+k)}{3(5+k)^2} \hat{G}_{21} \pa \hat{G}_{22}
\nonu \\
&-& \frac{2}{(5+k)} \hat{G}_{22} W_{+}^{(\frac{5}{2})}
+\frac{4(9+2k)}{3(5+k)^2} \pa \hat{G}_{22}  T_{+}^{(\frac{3}{2})}
+\frac{8(9+2k)}{3(5+k)^2} \hat{G}_{22}  \pa T_{+}^{(\frac{3}{2})}
\nonu \\
&- & \frac{2}{(5+k)} \pa T^{(1)}  V_{+}^{(2)} 
-\frac{2}{(5+k)}  T^{(1)} \pa  V_{+}^{(2)} 
-\frac{4(3+4k)}{(5+k)(3+7k)} \hat{T}  V_{+}^{(2)} 
\nonu \\
&+& \frac{4(4653+12699k+6962k^2+904k^3)}{3(5+k)^2(3+7k)(19+23k)} 
i \pa \hat{T} \hat{A}_{-}
\nonu \\
&+& \frac{4(3285+8535k+3470k^2+352k^3)}{3(5+k)^2(3+7k)(19+23k)} 
i \hat{T} \pa \hat{A}_{-}
-\frac{48(-4+k)}{(5+k)^2(19+23k)} \hat{T} \hat{A}_{-} \hat{A}_3
\nonu \\
&-& \left. \frac{16k(8+k)}{(5+k)^2(19+23k)} \hat{T} \hat{A}_{-}
\hat{B}_3
+ \frac{8(-4+k)}{(5+k)(19+23k)} i \hat{T} \hat{A}_{-} T^{(1)} 
\right](w) + \cdots,
\nonu \\
V^{(\frac{5}{2})}(z) \, W_{-}^{(\frac{5}{2})}(w) & = &
\frac{1}{(z-w)^4} \,
\left[ \frac{8k(59+37k+4k^2)}{3(5+k)^3} i \hat{B}_{+} \right](w) 
\nonu \\
&+& 
\frac{1}{(z-w)^3} \, \left[ 
\frac{4(-61+9k)}{9(5+k)^2} V_{-}^{(2)} 
+\frac{16(-45+35k+2k^2)}{
9(5+k)^3} \hat{A}_3 \hat{B}_{+}
\right. \nonu \\
&-& \left. \frac{32k(2+k)}{
3(5+k)^3}  \hat{B}_{+} \hat{B}_3
+\frac{4k(67+41k+4k^2)}{
3(5+k)^3} i \pa \hat{B}_{+}
-\frac{16(-4+k)}{3(5+k)^2} i T^{(1)} \hat{B}_{+}
\right](w)
\nonu \\
& + & \frac{1}{(z-w)^2} \, \left[
\frac{2(16+5k)}{3(5+k)} {\bf R_{-}^{(3)}}
 -\frac{(-15+65k+27k^2+2k^3)}{
3k(5+k)(8+k)} i {\bf P^{(2)}} \hat{B}_{+} 
\right. \nonu \\
& + &  \frac{(-15+65k+27k^2+2k^3)}{
3k(5+k)(8+k)} i \hat{B}_{+} {\bf P^{(2)}} 
\nonu \\
&+ & \frac{4}{(5+k)} T^{(1)} V_{-}^{(2)} 
-\frac{2(10+9k)}{3(5+k)^2} i T^{(1)} \pa \hat{B}_{+}
-\frac{8(19+8k)}{3(5+k)^2} i \hat{A}_3  V_{-}^{(2)} 
\nonu \\
&- & \frac{4(402+95k+26k^2)}{9(5+k)^3} 
\hat{A}_3 \pa \hat{B}_{+}
+\frac{4(372+337k+34k^2)}{9(5+k)^3}  \hat{B}_{+} \pa \hat{A}_3
\nonu \\
& + & \frac{8(285+3638k+5936k^2+2055k^3+140k^4)}{3(5+k)^2(3+7k)(19+23k)} 
\hat{B}_{+} \hat{T}
\nonu \\
&+& \frac{4(32+7k)}{3(5+k)^2} i \hat{B}_{+} T^{(2)}
-\frac{4(32+7k)}{3(5+k)^2} i \hat{B}_{+} W^{(2)}
\nonu \\
&- & \frac{4(-30+22k+7k^2)}{3(5+k)^3} \hat{B}_{+} \pa \hat{B}_3
+  \frac{2(36+k)}{3(5+k)^2} i \hat{B}_{+} \pa T^{(1)}
+\frac{8(8+k)}{3(5+k)^2} i \hat{B}_3 V_{-}^{(2)}
\nonu \\
&- &  \frac{4k(-6+k)}{3(5+k)^3} \hat{B}_3 \pa \hat{B}_{+}
+\frac{2(-240-1080k+438k^2+215k^3+16k^4)}{
9k(5+k)^2(8+k)} \pa  V_{-}^{(2)}
\nonu \\
&+ & \frac{4(6+k)}{3(5+k)^2} \hat{G}_{22} \hat{G}_{12} 
-\frac{8(17+5k)}{3(5+k)^2} \hat{G}_{22}  T_{-}^{(\frac{3}{2})}
+\frac{4(7+k)}{(5+k)^2} \hat{G}_{12}  V^{(\frac{3}{2})} 
\nonu \\
& - & \frac{8(16+5k)}{3(5+k)^2} T_{-}^{(\frac{3}{2})}  V^{(\frac{3}{2})}
+\frac{40(1+k)}{3(5+k)^3} i \hat{A}_{-} \hat{A}_{+} \hat{B}_{+}
+ \frac{8(-13+4k)}{3(5+k)^3} i \hat{A}_3 \hat{A}_3 \hat{B}_{+}
\nonu \\
& + &  \frac{8}{3(5+k)^2} i \hat{B}_{-} \hat{B}_{+} \hat{B}_{+}
-  \frac{8(-5+2k)}{3(5+k)^3} i \hat{B}_{+} \hat{B}_3 \hat{B}_3
-  \frac{16(-4+k)}{3(5+k)^3} i \hat{B}_3 \hat{A}_3 \hat{B}_{+}
\nonu \\
&- &  \left.  \frac{8}{(5+k)^2} T^{(1)} \hat{A}_3 \hat{B}_{+}
+  \frac{8}{(5+k)^2} T^{(1)} \hat{B}_{+} \hat{B}_3
\right](w)
\nonu \\
& + & \frac{1}{(z-w)} \, \left[ 
-\frac{2}{(5+k)} i \hat{A}_3 {\bf R_{-}^{(3)}}
+\frac{2}{(5+k)} i \hat{B}_3  {\bf R_{-}^{(3)}}
+\frac{(12+5k)}{3(5+k)} \pa  {\bf R_{-}^{(3)}}
\right. \nonu \\
&- & \frac{2}{(5+k)}  V_{-}^{(2)} W^{(2)} -\frac{(78+11k)}{3(5+k)^2}
\pa^2  V_{-}^{(2)} +\frac{2}{(5+k)}  T^{(2)}  V_{-}^{(2)}   
\nonu \\
&-& \frac{4(12+5k)}{3(5+k)^2} \pa  T_{-}^{(\frac{3}{2})}  V^{(\frac{3}{2})}
-\frac{4(12+5k)}{3(5+k)^2}   T_{-}^{(\frac{3}{2})} \pa  V^{(\frac{3}{2})}
-\frac{8(11+4k)}{3(5+k)^2} i \pa \hat{A}_3   V_{-}^{(2)}
\nonu \\
&- & \frac{2(23+16k)}{3(5+k)^2} i \hat{A}_3 \pa  V_{-}^{(2)} 
+\frac{8}{(5+k)^2} i \hat{A}_3  T_{-}^{(\frac{3}{2})}   V^{(\frac{3}{2})}
-\frac{16}{(5+k)^2} \hat{A}_3 \hat{A}_3  V_{-}^{(2)}  
\nonu \\
&- &  \frac{8(29+k)}{3(5+k)^3} i \pa \hat{A}_3 \hat{A}_3 \hat{B}_{+}
+ \frac{16}{(5+k)^2}  \hat{A}_3 \hat{B}_{3}  V_{-}^{(2)} 
+ \frac{8}{(5+k)^2}  \hat{A}_3 \hat{B}_{+}  T^{(2)} 
\nonu \\
& + & \frac{2(4590+15651k+20358k^2+13027k^3+1442k^4)}{
3(5+k)^3(3+7k)(19+23k)} \pa^2 \hat{A}_3 \hat{B}_{+}
\nonu \\
&+ & \frac{4(49+5k)}{3(5+k)^3} i \pa \hat{A}_3 \hat{B}_{+} \hat{B}_3 
- \frac{4(17+13k)}{3(5+k)^3} i  \hat{A}_3 \pa \hat{B}_{+} \hat{B}_3 
+\frac{8(2+k)}{3(5+k)^3} i  \hat{A}_3  \hat{B}_{+} \pa \hat{B}_3 
\nonu \\
&- & \frac{2}{(5+k)^2} \pa  \hat{A}_3  \hat{B}_{+} T^{(1)}
-\frac{6}{(5+k)^2} \hat{A}_3  \pa \hat{B}_{+} T^{(1)}
-\frac{4}{(5+k)^2} \hat{A}_3   \hat{B}_{+} \pa T^{(1)}
\nonu \\
& + & \frac{8}{(5+k)^2} i \hat{A}_3 \hat{G}_{22}   T_{-}^{(\frac{3}{2})}
-\frac{4}{(5+k)^2} \hat{A}_{+} \hat{A}_{-}  V_{-}^{(2)}
+\frac{8(2+k)}{3(5+k)^3} i \pa \hat{A}_{+} \hat{A}_{-} \hat{B}_{+}
\nonu \\
& + & \frac{8(-4+k)}{3(5+k)^3} i  \hat{A}_{+} \pa \hat{A}_{-} \hat{B}_{+}
+\frac{4(16+5k)}{3(5+k)^2} i \pa \hat{B}_3  V_{-}^{(2)}
-\frac{2(-5+2k)}{3(5+k)^2} i  \hat{B}_3  \pa V_{-}^{(2)}
\nonu \\
&- & \frac{8}{(5+k)^2} i \hat{B}_3  T_{-}^{(\frac{3}{2})}  V^{(\frac{3}{2})}
-\frac{8}{(5+k)^2} i \hat{B}_3  \hat{G}_{22} T_{-}^{(\frac{3}{2})}
-\frac{2}{(5+k)} i \hat{B}_{+} W^{(3)}
\nonu \\
&- & \frac{2(40+11k)}{3(5+k)^2} i \pa \hat{B}_{+} W^{(2)}
-\frac{2(28+5k)}{3(5+k)^2} i \hat{B}_{+} \pa  W^{(2)}
+ \frac{2(8+k)}{(5+k)^2} i \pa \hat{B}_{+} T^{(2)}
\nonu \\
&+ & \frac{6(3+k)}{(5+k)^2} i \hat{B}_{+} \pa T^{(2)} 
-\frac{8}{(5+k)^2} \hat{B}_{+} \hat{B}_3  T^{(2)}
+ \frac{20(2+k)}{3(5+k)^3}  \hat{B}_{+} \hat{B}_{+}  \hat{B}_{-} \hat{B}_3
\nonu \\
&+ & \frac{2(1140+7424k+8947k^2+3134k^3+7k^4)}{
3(5+k)^3(3+7k)(19+23k)} \pa^2 \hat{B}_{+} \hat{B}_3
\nonu \\
&- & \frac{2(-285-3695k+9502k^2+15325k^3+2877k^4)}{
9(5+k)^3(3+7k)(19+23k)} \hat{B}_{+} \pa^2 \hat{B}_3
\nonu \\
&+ & \frac{4(10+3k)}{(5+k)^3} i \pa \hat{B}_{+} \hat{B}_3 \hat{B}_3
+ \frac{20}{3(5+k)^2} i \hat{B}_{+} \pa \hat{B}_3 \hat{B}_3
-\frac{80(2+k)}{9(5+k)^3} \hat{B}_{+} \hat{B}_3 \hat{B}_3 \hat{B}_3 
\nonu \\
&+ & \frac{6}{(5+k)^2} \pa \hat{B}_{+} \hat{B}_3 T^{(1)} 
+\frac{2}{(5+k)^2}  \hat{B}_{+} \pa \hat{B}_3 T^{(1)} 
+\frac{4}{(5+k)^2}  \hat{B}_{+}  \hat{B}_3 \pa T^{(1)} 
\nonu \\
&- & \frac{10(-1+k)}{3(5+k)^3} i \pa \hat{B}_{+} \hat{B}_{+} \hat{B}_{-}
+\frac{2(5+2k)}{(5+k)^3} i \hat{B}_{+} \hat{B}_{+} \pa \hat{B}_{-}
\nonu \\
& + & \frac{4}{(5+k)^2} i \hat{B}_{+} \hat{G}_{21} T_{-}^{(\frac{3}{2})}
-\frac{(177+110k+61k^2)}{(5+k)^2(19+23k)} i \pa^2 \hat{B}_{+} T^{(1)}
\nonu \\
& - & \frac{2}{(5+k)^2} i \pa \hat{B}_{+} \pa T^{(1)}
+\frac{(210+757k+47k^2)}{3(5+k)^2(19+23k)} i \hat{B}_{+} \pa^2 T^{(1)}
\nonu \\
&- & \frac{4(8+k)}{3(5+k)^2} \pa \hat{G}_{12} \hat{G}_{22} 
+\frac{4}{(5+k)^2}  \hat{G}_{12} \pa \hat{G}_{22} 
+\frac{2}{(5+k)} \hat{G}_{22} W_{-}^{(\frac{5}{2})}
\nonu \\
& - & \frac{4(5+4k)}{3(5+k)^2} \pa \hat{G}_{22} T_{-}^{(\frac{3}{2})}
- \frac{8(10+3k)}{3(5+k)^2}  \hat{G}_{22} \pa T_{-}^{(\frac{3}{2})}
+\frac{2}{(5+k)} \pa T^{(1)}  V_{-}^{(2)}
\nonu \\
&+& \frac{2}{(5+k)}  T^{(1)} \pa  V_{-}^{(2)}
+\frac{12k}{(5+k)(3+7k)} \hat{T}  V_{-}^{(2)}
+\frac{4(7+k)}{(5+k)^2} \hat{G}_{12} \pa  V^{(\frac{3}{2})}
\nonu \\
& + & \frac{16(9+123k+180k^2+14k^3)}{(5+k)^2(3+7k)(19+23k)} 
\hat{T} \hat{A}_3 \hat{B}_{+} 
\nonu \\
& + & \frac{2(285+5180k+9029k^2+3518k^3+280k^4)}{
3(5+k)^2(3+7k)(19+23k)} i \pa \hat{T} \hat{B}_{+} 
\nonu \\
& + & \frac{4(570+4492k+6961k^2+2479k^3+140k^4)}{
3(5+k)^2(3+7k)(19+23k)} i  \hat{T} \pa \hat{B}_{+} 
\nonu \\
&-& \left. \frac{16k(45+47k+14k^2)}{(5+k)^2(3+7k)(19+23k)} 
\hat{T} \hat{B}_{+} \hat{B}_3 
-\frac{16(-3+k)}{(5+k)(19+23k)} i \hat{T} \hat{B}_{+} T^{(1)}
\right](w) +\cdots,
\nonu \\
V^{(\frac{5}{2})}(z) \, W^{(3)}(w) & = &
\frac{1}{(z-w)^4} \, \left[ \frac{4(2559+12378k+10609k^2+2688k^3+218k^4)}{
3(5+k)^3(19+23k)} \hat{G}_{22} \right. \nonu \\
&+ & \left. 
\frac{8(-1959+1029k+1328k^2+208k^3)}{3(5+k)^3(19+23k)}   V^{(\frac{3}{2})}
\right](w) 
\nonu \\
& + & \frac{1}{(z-w)^3} \, \left[ 
\frac{(-4367+108k+795k^2+152k^3)}{
3(5+k)^2(19+23k)} {\bf R^{(\frac{5}{2})}} 
\right. \nonu \\
&- & \frac{2(-2139+24k+647k^2+76k^3)}{
3(5+k)^2(19+23k)}  V^{(\frac{5}{2})}
-\frac{8(-33+5k+k^2)}{3(5+k)^3} i \hat{A}_{-} \hat{G}_{12}
\nonu \\
&- &  \frac{8(29+4k)}{3(5+k)^3} i \hat{A}_{-}  T_{-}^{(\frac{3}{2})}
- \frac{8(6+25k+2k^2)}{3(5+k)^3} i \hat{A}_3 \hat{G}_{22}
+ \frac{8(-1+8k)}{3(5+k)^3} i \hat{A}_3    V^{(\frac{3}{2})}
\nonu \\
&- &
\frac{8(75+k)}{3(5+k)^3} i \hat{B}_{+} \hat{G}_{21}
+   \frac{4(-117+7k+6k^2)}{3(5+k)^3} i \hat{B}_{+}  T_{+}^{(\frac{3}{2})}
\nonu \\
&- &   \frac{8(34+16k+k^2)}{3(5+k)^3} i \hat{B}_3    V^{(\frac{3}{2})}
-\frac{4(-3+4k)}{3(5+k)^2} T^{(1)} \hat{G}_{22}
-\frac{4(-3+4k)}{3(5+k)^2} T^{(1)}  V^{(\frac{3}{2})}
\nonu \\
&+&  \frac{4(3813+13402k+10049k^2+2734k^3+218k^4)}{
9(5+k)^3(19+23k)} \pa \hat{G}_{22}
\nonu \\
& + & \left. \frac{8(-135+3370k+1413k^2+116k^3)}{
9(5+k)^3(19+23k)} \pa  V^{(\frac{3}{2})}
+  \frac{8(12+17k+4k^2)}{3(5+k)^3} i \hat{B}_3 \hat{G}_{22}
\right](w) 
\nonu \\
& + & \frac{1}{(z-w)^2} \, \left[ 
\frac{(-11011+3804k+3459k^2+364k^3)}{
15(5+k)^2(19+23k)} \pa {\bf R^{(\frac{5}{2})}}
-\frac{(29+7k)}{2(5+k)} {\bf R^{(\frac{7}{2})}}
\right. \nonu \\
& + & \frac{(25+k)}{3(5+k)^2} i {\bf R^{(\frac{5}{2})}} \hat{A}_3 
- \frac{(25+k)}{3(5+k)^2} i {\bf R^{(\frac{5}{2})}} \hat{B}_3 
- \frac{(37+10k)}{3(5+k)^2} i {\bf P_{-}^{(\frac{5}{2})}} \hat{A}_{-} 
\nonu \\
& + & 
\frac{8(6+5k)}{3(5+k)^3} \hat{A}_{-} \hat{A}_{+} \hat{G}_{22}
-\frac{(-425+243k+84k^2)}{15(5+k)^3} \hat{A}_{-} \hat{A}_{+} V^{(\frac{3}{2})}
\nonu \\
& + & \frac{2(19+5k)}{3(5+k)^2} i \hat{A}_{-}  W_{-}^{(\frac{5}{2})}
+  \frac{4(49+17k+k^2)}{3(5+k)^3} \hat{A}_{-} \hat{A}_3 \hat{G}_{12}
\nonu \\
& - & \frac{2(1117+261k)}{15(5+k)^3} \hat{A}_{-} \hat{A}_3  T_{-}^{(\frac{3}{2})}
+ \frac{24}{(5+k)^2} \hat{A}_{-} \hat{B}_{+} \hat{G}_{11}
-  \frac{2(27+7k)}{3(5+k)^3} \hat{A}_{-} \hat{B}_{+}  U^{(\frac{3}{2})}
\nonu \\
& - &  \frac{16(8+k)}{(5+k)^3} \hat{A}_{-} \hat{B}_3 \hat{G}_{12}
+ \frac{16(54+13k)}{3(5+k)^3} \hat{A}_{-} \hat{B}_3   T_{-}^{(\frac{3}{2})}
- \frac{8}{(5+k)^2} i \hat{A}_{-} T^{(1)} \hat{G}_{12}
\nonu \\
& - &  \frac{4(111+70k+2k^2)}{9(5+k)^3} 
i \hat{A}_{-} \pa \hat{G}_{12}
+  \frac{4(-77+134k)}{45(5+k)^3} i \hat{A}_{-} \pa   T_{-}^{(\frac{3}{2})}
\nonu \\
& + & \frac{(57+11k)}{(5+k)^2} i \hat{B}_{+}  W_{+}^{(\frac{5}{2})}
+  \frac{8(23+2k)}{3(5+k)^3} \hat{B}_{+} \hat{A}_3 \hat{G}_{21}
+ \frac{16(19+4k)}{3(5+k)^3} \hat{B}_{+} \hat{A}_3   T_{+}^{(\frac{3}{2})}
\nonu \\
& - &  \frac{2(-360+271k+63k^2)}{
15(5+k)^3} \hat{B}_{+} \hat{B}_{-} \hat{G}_{22}
+  \frac{(246+165k+11k^2)}{
15(5+k)^3} \hat{B}_{+} \hat{B}_{-}  V^{(\frac{3}{2})}
\nonu \\
& + &  \frac{2(-8018-8236k+509k^2+349k^3+42k^4)}{
5(5+k)^3(19+23k)} \hat{B}_{+} \hat{B}_3 \hat{G}_{21}
\nonu \\
& + &  
\frac{2(-4+k)(106+33k)}{15(5+k)^3} \hat{B}_{+} \hat{B}_3   T_{+}^{(\frac{3}{2})}
+  \frac{4}{(5+k)^2} i \hat{B}_{+} T^{(1)} \hat{G}_{21} 
\nonu \\
& - &  \frac{4(-1273-546k-66k^2+349k^3+42k^4)}{15(5+k)^3(19+23k)} 
i \hat{B}_{+} \pa \hat{G}_{21}
\nonu \\
& + &  \frac{4(-622-93k+9k^2)}{15(5+k)^3} i \hat{B}_{+} \pa   T_{+}^{(\frac{3}{2})}
+ \frac{2(-8+21k)}{5(5+k)^2} 
\hat{B}_{-} \hat{B}_{+} \hat{G}_{22}
\nonu \\
&- &  \frac{(-1914-395k+11k^2)}{
15(5+k)^3} \hat{B}_{-} \hat{B}_{+}  V^{(\frac{3}{2})}
+  \frac{(15+323k+84k^2)}{
15(5+k)^3} \hat{A}_{+} \hat{A}_{-}  V^{(\frac{3}{2})}
\nonu \\
&+ & \frac{2(1640+1679k+443k^2)}{
3(5+k)^2(19+23k)} i \hat{A}_3  V^{(\frac{5}{2})}
-  \frac{4(13+17k+k^2)}{
3(5+k)^3} \hat{A}_3 \hat{A}_{-} \hat{G}_{12} 
\nonu \\
&+ &  \frac{2(1117+261k)}{
15(5+k)^3} \hat{A}_3 \hat{A}_{-}  T_{-}^{(\frac{3}{2})}
+\frac{32(-3+k)}{
3(5+k)^3} \hat{A}_3 \hat{A}_3 \hat{G}_{22} 
+  \frac{4(47+5k)}{
3(5+k)^3} \hat{A}_3 \hat{A}_3  V^{(\frac{3}{2})}
\nonu \\
& - & \frac{16(-3+k)}{
3(5+k)^3} \hat{A}_3 \hat{B}_3 \hat{G}_{22} 
-  \frac{8(13+3k)}{
3(5+k)^3} \hat{A}_3 \hat{B}_3  V^{(\frac{3}{2})}
+  \frac{8}{
(5+k)^2} i \hat{A}_3 T^{(1)} \hat{G}_{22} 
\nonu \\
&- &   \frac{8}{
(5+k)^2} i \hat{A}_3 T^{(1)}  V^{(\frac{3}{2})}
+ \frac{8(423+48k+32k^2)}{45(5+k)^3} i \hat{A}_3 \pa \hat{G}_{22}
\nonu \\
&+ & \frac{4(835+883k+84k^2)}{45(5+k)^3} i 
\hat{A}_3 \pa  V^{(\frac{3}{2})}  
- \frac{2(437+2105k+392k^2)}{
3(5+k)^2(19+23k)} i \hat{B}_3  V^{(\frac{5}{2})}
\nonu \\
& - & \frac{2(-11894-9608k+1987k^2+1047k^3+
126k^4)}{15(5+k)^3(19+23k)}  \hat{B}_3 \hat{B}_{+} \hat{G}_{21}
\nonu \\
& - & \frac{2(336+134k+33k^2)}{
15(5+k)^3}  \hat{B}_3 \hat{B}_{+}  T_{+}^{(\frac{3}{2})}
-\frac{16(-3+k)}{3(5+k)^3}  \hat{B}_3 \hat{B}_3 \hat{G}_{22}
\nonu \\
&+& \frac{4(-21+k)}{
3(5+k)^3}  \hat{B}_3 \hat{B}_3  V^{(\frac{3}{2})}
-\frac{8}{(5+k)^2}  i \hat{B}_3 T^{(1)} \hat{G}_{22}
+ \frac{8}{
(5+k)^2}  i \hat{B}_3 T^{(1)}  V^{(\frac{3}{2})}
\nonu \\
&-& \frac{8(-720+36k+23k^2)}{45(5+k)^3} i \hat{B}_3 \pa \hat{G}_{22}
-\frac{4(4114+1075k+9k^2)}{45(5+k)^3} i 
\hat{B}_3 \pa  V^{(\frac{3}{2})} 
\nonu \\
&-&  \frac{(-281+63k+20k^2)}{
(5+k)(19+23k)}  T^{(1)}  V^{(\frac{5}{2})}
+  \frac{4(42+k)}{15(5+k)^2} T^{(1)} \pa \hat{G}_{22}
\nonu \\
& - & \frac{2(59+11k)}{3(5+k)^2} T^{(1)} \pa  V^{(\frac{3}{2})} 
-\frac{(-17527-6126k+3093k^2+364k^3)}{
15(5+k)^2(19+23k)} \pa  V^{(\frac{5}{2})}
\nonu \\
& - & \frac{8(324+229k+26k^2)}{15(5+k)^3} i \pa \hat{A}_3 \hat{G}_{22}
-  \frac{2(111+43k)}{15(5+k)^2} \pa T^{(1)}  \hat{G}_{22}
\nonu \\
&  + &   \frac{(221+53k)}{3(5+k)^2} \pa T^{(1)}  V^{(\frac{3}{2})} 
\nonu \\
& + & \frac{2(14245 k^5+240343 k^4+801672 k^3+798115 k^2-229783 k-430388)
}{45(5+k)^3(19+23k)(47+35k)} \nonu \\
& \times & \pa^2 \hat{G}_{22} 
\nonu \\
& - & \frac{(-123626-168274k-42863k^2+525k^3)
}{45(5+k)^3(47+35k)} \pa^2  V^{(\frac{3}{2})} 
+ \frac{2(-40+3k)}{3(5+k)^2} \hat{G}_{21} V_{-}^{(2)} 
\nonu \\
& - & \frac{2(80+17k)}{3(5+k)^2}    T_{+}^{(\frac{3}{2})}
 V_{-}^{(2)} + \frac{16(7+k)}{3(5+k)^2}
\hat{G}_{12}  V_{+}^{(2)}  
-\frac{2(29+7k)}{(5+k)^2}  T_{-}^{(\frac{3}{2})}  V_{+}^{(2)} 
\nonu \\
& +& \frac{8(5537 k^5+90748 k^4+438605 k^3+771974 k^2+510696 k+107172)}{
3(5+k)^2(3+7k)(19+23k)(47+35k)} \hat{G}_{22} \hat{T}
\nonu \\
& - & \frac{4(9+2k)}{(5+k)^2} \hat{G}_{22} W^{(2)}
-\frac{4(-3+2k)}{3(5+k)^2} \hat{G}_{22} T^{(2)}
\nonu \\
& + & \frac{4(50827 k^4+352522 k^3+640696 k^2+331582 k+44229)}{
3(5+k)^2(3+7k)(19+23k)(47+35k)}  V^{(\frac{3}{2})}  \hat{T}
\nonu \\
&-& \left. \frac{2(13+3k)}{(5+k)^2} 
  V^{(\frac{3}{2})} W^{(2)}
+  \frac{8(5+2k)}{3(5+k)^2} 
  V^{(\frac{3}{2})} T^{(2)} 
\right](w)
\nonu \\
& + &  \frac{1}{(z-w)} \, 
 \{ V^{(\frac{5}{2})} \, 
 W^{(3)} \}_{-1}(w)
+ \cdots.
\label{v5halfw3}  
\eea
The first order pole in the first  OPE of (\ref{v5halfw3})
contains composite field with spin-$4$ with $U(1)$ charge of 
$-\frac{2(-3+k)}{(5+k)}$. 
Although $\hat{G}_{22} \hat{G}_{22}(w)$ can be written in terms of 
derivative of $ \hat{A}_{-} \hat{B}_{+}(w)$, 
the expression $\hat{G}_{22} \pa \hat{G}_{22}(w)$ is an independent quantity
we should consider.
Note the $(k-3)$ factor in the third order pole in the second OPE.
The third OPE has a composite field with spin-$4$ with 
 $U(1)$ charge of 
$\frac{6}{(5+k)}$. 
For the fourth OPE, the corresponding $U(1)$ charge is given by
$-\frac{2k}{(5+k)}$.
For the last OPE, the corresponding $U(1)$ charge of spin-$\frac{9}{2}$ 
is given by
$-\frac{(-3+k)}{(5+k)}$ and the complete expression will appear in Appendix
$L$.
The $13$ composite fields of spin-$\frac{5}{2}$ in the third order pole
can be seen from the Table $4$ of \cite{Ahn1311}.

\section{
The nontrivial 
OPEs between  the fourth ${\cal N}=2$ multiplet in (\ref{lowesthigher})  
}

Now the last OPEs corresponding to the last ${\cal N}=2$ multiplet 
in (\ref{lowesthigher})
are as follows:
\bea 
W^{(2)}(z) \, W^{(2)}(w) & = &
\frac{1}{(z-w)^4} \, \left[ \frac{18k(4+k)}{(5+k)^2} \right] 
\nonu \\
& + & \frac{1}{(z-w)^2} \, \left[ \frac{4(4+k)}{(5+k)} W^{(2)}\right] (w) 
\nonu \\
& + &  \frac{1}{(z-w)} \,  \left[ \frac{2(4+k)}{(5+k)} \pa W^{(2)} \right] (w)
 + \cdots,
\nonu \\
 W^{(2)}(z) \, W_{+}^{(\frac{5}{2})}(w) & = &
\frac{1}{(z-w)^3} \, \left[-\frac{8(2+k)(8+k)}{3(5+k)^2} \hat{G}_{21}
- \frac{4(-3+k)}{3(5+k)^2}  T_{+}^{(\frac{3}{2})} \right](w)  
\nonu \\
& + & \frac{1}{(z-w)^2} \, \left[-\frac{(-3+k)}{6(5+k)} 
{\bf P_{+}^{(\frac{5}{2})}} + 
\frac{(31+7k)}{3(5+k)} W_{+}^{(\frac{5}{2})} 
-\frac{8(16+10k+k^2)}{9(5+k)^2} \pa \hat{G}_{21} 
\right. \nonu \\
&-& \left. \frac{4(-3+k)}{9(5+k)^2} \pa   T_{+}^{(\frac{3}{2})}   
\right](w) 
\nonu \\
&+& \frac{1}{(z-w)} \, \left[ 
-\frac{(-3+k)}{15(5+k)} \pa {\bf P_{+}^{(\frac{5}{2})}}
-\frac{1}{2} {\bf S_{+}^{(\frac{7}{2})}}
\right. \nonu \\
& +& \frac{(12740 k^4+171326 k^3+321255 k^2+70732 k-85841)}{
45(5+k)^2(19+23k)(47+35k)} 
\hat{A}_{-} \hat{A}_{+} \hat{G}_{21}
\nonu \\
&+& \frac{4}{(5+k)^2} \hat{A}_{-} \hat{A}_{+}  T_{+}^{(\frac{3}{2})}
-\frac{2}{(5+k)} i \hat{A}_{-}  V^{(\frac{5}{2})}
+\frac{2(17+4k)}{5(5+k)^2} \hat{A}_{-} \hat{A}_3 \hat{G}_{11}
 \nonu \\
&+ & \frac{4}{(5+k)^2}   \hat{A}_{-} \hat{B}_{-}  T_{-}^{(\frac{3}{2})}
+\frac{8}{(5+k)^2} \hat{A}_{-} \hat{B}_3 \hat{G}_{11}
+ \frac{16(3+k)}{15(5+k)^2} i \hat{A}_{-} \pa \hat{G}_{11}
\nonu \\
&- &   \frac{8(1+2k)}{
15(5+k)^2} i \hat{A}_{-} \pa   U^{(\frac{3}{2})}
-\frac{(-684-667k+43k^2+6k^3)}{
5(5+k)^2(19+23k)} \hat{B}_{+} \hat{B}_{-} \hat{G}_{21}
\nonu \\
& + &   
\frac{(12+k)}{(5+k)^2}   \hat{B}_{+} \hat{B}_{-}  T_{+}^{(\frac{3}{2})}
+ \frac{2}{(5+k)} i \hat{B}_{-}  V^{(\frac{5}{2})} 
+\frac{8}{(5+k)^2}  \hat{B}_{-} \hat{A}_3 \hat{G}_{22}
\nonu \\
&+ & \frac{(-1064-1127k+43k^2+6k^3)}{
5(5+k)^2(19+23k)} \hat{B}_{-} \hat{B}_{+} \hat{G}_{21}
-\frac{(16+k)}{(5+k)^2}   \hat{B}_{-} \hat{B}_{+}  T_{+}^{(\frac{3}{2})}
\nonu \\
&+ &  \frac{2(22+5k)}{5(5+k)^2}  \hat{B}_{-} \hat{B}_3 \hat{G}_{22}
+  \frac{4(12+5k)}{15(5+k)^2} i \hat{B}_{-} \pa \hat{G}_{22} 
+  \frac{8(1+k)}{
15(5+k)^2} i \hat{B}_{-} \pa   V^{(\frac{3}{2})} 
\nonu \\
& - &  \frac{(12740 k^4+171326 k^3+321255 k^2+70732 k-85841)}{
45(5+k)^2(19+23k)(47+35k)}  
\hat{A}_{+} \hat{A}_{-} \hat{G}_{21}
\nonu \\
& + & \frac{4}{(5+k)} i \hat{A}_3  W_{+}^{(\frac{5}{2})}
-\frac{2(17+4k)}{5(5+k)^2} \hat{A}_3 \hat{A}_{-} \hat{G}_{11} 
+ \frac{4(53+233k+56k^2)}{
15(5+k)^2(19+23k)} i \hat{A}_3 \pa \hat{G}_{21} 
\nonu \\
&- &   \frac{4(15+4k)}{
15(5+k)^2} i \hat{A}_3 \pa   T_{+}^{(\frac{3}{2})}
+ \frac{4}{(5+k)} i \hat{B}_3  W_{+}^{(\frac{5}{2})}
-\frac{2(22+5k)}{5(5+k)^2} \hat{B}_3 \hat{B}_{-} \hat{G}_{22}
\nonu \\
& - &  \frac{4(-1064-1127k+43k^2+6k^3)}{
15(5+k)^2(19+23k)} i \hat{B}_3 \pa \hat{G}_{21} 
+  \frac{4(18+k)}{
3(5+k)^2} i \hat{B}_3 \pa   T_{+}^{(\frac{3}{2})}
\nonu \\
& - & \frac{2(-49-19k+2k^2)}{
5(5+k)(19+23k)}  T^{(1)} \pa \hat{G}_{21} 
+ \frac{2}{
(5+k)} T^{(1)} \pa   T_{+}^{(\frac{3}{2})}
+\frac{2(31+7k)}{15(5+k)} \pa  W_{+}^{(\frac{5}{2})}
\nonu \\
&+ &   \frac{4(1+2k)}{
5(5+k)^2} i \pa \hat{A}_{-}    U^{(\frac{3}{2})} 
+  \frac{2(-15+4k)}{
5(5+k)^2} i \pa \hat{A}_3    T_{+}^{(\frac{3}{2})} 
- \frac{4(-4+k)}{
5(5+k)^2} i \pa \hat{B}_{-}    V^{(\frac{3}{2})} 
\nonu \\
&+ &  \frac{3(-49-19k+2k^2)}{
5(5+k)(19+23k)}  \pa T^{(1)}  \hat{G}_{21} 
- \frac{3}{
(5+k)} \pa T^{(1)}    T_{+}^{(\frac{3}{2})}
\nonu \\
& - & \frac{(390 k^4+27218 k^3+26679 k^2-90056 k-109155)}{
45(5+k)^2(19+23k)(47+35k)}  \pa^2  T_{+}^{(\frac{3}{2})}
\nonu \\
& - & \frac{4(1862 k^4+19425 k^3+49348 k^2+39765 k+8316)}{
3(5+k)(3+7k)(19+23k)(47+35k)} \hat{G}_{21} \hat{T}
\nonu \\
&-  & \frac{2}{(5+k)} \hat{G}_{21} T^{(2)}  
\nonu \\
& - &  \frac{4(6370 k^4+94483 k^3+245394 k^2+171563 k+8474)}{
45(5+k)^2(19+23k)(47+35k)} i \hat{G}_{21} \pa \hat{A}_3
\nonu \\
& + & \frac{4(70 k^4-1895 k^3-8828 k^2-10515 k-1908)}{
3(5+k)(3+7k)(19+23k)(47+35k)}  T_{+}^{(\frac{3}{2})} \hat{T}
\nonu \\
&- & \left. \frac{2}{(5+k)}  T_{+}^{(\frac{3}{2})} T^{(2)}  
-\frac{2}{(5+k)} \hat{G}_{11} V_{+}^{(2)}
+\frac{2}{(5+k)} \hat{G}_{22} U_{+}^{(2)}
\right](w) + \cdots,
\nonu \\ 
 W^{(2)}(z) \, W_{-}^{(\frac{5}{2})}(w) & = &
\frac{1}{(z-w)^3} \, \left[-\frac{8(2+k)(8+k)}{3(5+k)^2} \hat{G}_{12}
+ \frac{4(-3+k)}{3(5+k)^2}  T_{-}^{(\frac{3}{2})} \right](w)  
\nonu \\
& + & \frac{1}{(z-w)^2} \, \left[\frac{(-3+k)}{6(5+k)} 
{\bf P_{-}^{(\frac{5}{2})}} + 
\frac{(31+7k)}{3(5+k)} W_{-}^{(\frac{5}{2})} 
-\frac{8(16+10k+k^2)}{9(5+k)^2} \pa \hat{G}_{12} 
\right. \nonu \\
&+& \left. \frac{4(-3+k)}{9(5+k)^2} \pa   T_{-}^{(\frac{3}{2})}   
\right](w) 
\nonu \\
&+& \frac{1}{(z-w)} \, \left[ 
\frac{(-3+k)}{15(5+k)} \pa {\bf P_{-}^{(\frac{5}{2})} }
+\frac{1}{2} {\bf S_{-}^{(\frac{7}{2})}}
-\frac{4}{(5+k)^2} \hat{A}_{-} \hat{A}_{+}  T_{-}^{(\frac{3}{2})} 
\right. \nonu \\
& + & \frac{4}{(5+k)^2} \hat{B}_{+} \hat{A}_{+}  T_{+}^{(\frac{3}{2})}
-\frac{2}{(5+k)} i \hat{B}_{+}  U^{(\frac{5}{2})} 
+ \frac{8}{(5+k)^2} \hat{B}_{+} \hat{A}_3  \hat{G}_{11} 
\nonu \\
& + & \frac{8}{(5+k)^2} \hat{B}_{+} \hat{A}_3  U^{(\frac{3}{2})}  
\nonu \\
& + &   \frac{2(5980 k^4+24965 k^3-51087 k^2-147485 k-27973)}{
45(5+k)^2(19+23k)(47+35k)} 
\hat{B}_{+} \hat{B}_{-}  \hat{G}_{12} 
\nonu \\
& - &  
\frac{4}{(5+k)^2} \hat{B}_{+} \hat{B}_{-}  T_{-}^{(\frac{3}{2})} 
+\frac{2(14+5k)}{5(5+k)^2} \hat{B}_{+} \hat{B}_3  \hat{G}_{11} 
\nonu \\
& - &  \frac{4(1140 k^4+37018 k^3+128337 k^2+100412 k-32703)}{
45(5+k)^2(19+23k)(47+35k)} 
\hat{B}_{+} \hat{B}_3  U^{(\frac{3}{2})}  
\nonu \\
&- & \frac{4(-6+5k)}{
15(5+k)^2} i \hat{B}_{+} \pa \hat{G}_{11}
+  \frac{4(43+2k)}{
15(5+k)^2} i \hat{B}_{+} \pa  U^{(\frac{3}{2})} 
\nonu \\
&- &  \frac{2(5980 k^4+24965 k^3-51087 k^2-147485 k-27973)}{
45(5+k)^2(19+23k)(47+35k)} \hat{B}_{-} \hat{B}_{+}  \hat{G}_{12} 
\nonu \\
& + & \frac{2}{(5+k)} i \hat{A}_{+}  V^{(\frac{5}{2})} 
+ \frac{4(3+2k)}{5(5+k)^2} \hat{A}_{+} \hat{A}_3  V^{(\frac{3}{2})} 
+  \frac{8}{(5+k)^2} \hat{A}_{+} \hat{B}_3 \hat{G}_{22} 
\nonu \\
&- & \frac{8}{(5+k)^2}  \hat{A}_{+} \hat{B}_3  V^{(\frac{3}{2})}  
- \frac{4(15+4k)}{
15(5+k)^2} i \hat{A}_{+} \pa \hat{G}_{22}
-  \frac{4(1+4k)}{
15(5+k)^2} i \hat{A}_{+} \pa  V^{(\frac{3}{2})} 
\nonu \\
&- &  \frac{4(-7+2k)}{5(5+k)^2} \hat{A}_3 \hat{A}_{+}  V^{(\frac{3}{2})} 
-\frac{4}{(5+k)} i \hat{A}_3  W_{-}^{(\frac{5}{2})} 
-\frac{4}{(5+k)^2} \hat{A}_3 \hat{A}_3  T_{-}^{(\frac{3}{2})} 
\nonu \\
& + & \frac{8}{(5+k)^2} \hat{A}_3 \hat{B}_3  T_{-}^{(\frac{3}{2})} 
+ \frac{16(172+223k+55k^2)}{
15(5+k)^2(19+23k)} i \hat{A}_3 \pa \hat{G}_{12}
\nonu \\
& - & \frac{4(57+16k)}{
15(5+k)^2} i \hat{A}_3 \pa  T_{-}^{(\frac{3}{2})} 
-\frac{4}{(5+k)} i \hat{B}_3  W_{-}^{(\frac{5}{2})} 
- \frac{2(14+5k)}{5(5+k)^2} \hat{B}_3 \hat{B}_{+}  \hat{G}_{11} 
\nonu \\
& + &  
\frac{4(1140 k^4+37018 k^3+55887 k^2-56728 k-113073)}{
45(5+k)^2(19+23k)(47+35k)} \hat{B}_3 \hat{B}_{+}  U^{(\frac{3}{2})} 
\nonu \\
& - &  \frac{4}{(5+k)^2} \hat{B}_3 \hat{B}_3  T_{-}^{(\frac{3}{2})} 
+ \frac{8(-418-454k-13k^2+3k^3)}{
15(5+k)^2(19+23k)} i \hat{B}_3 \pa \hat{G}_{12}
\nonu \\
&- &  \frac{4(-30+k)}{
15(5+k)^2} i \hat{B}_3 \pa  T_{-}^{(\frac{3}{2})} 
+ 
\frac{4(-53-44k+k^2)}{
5(5+k)(19+23k)}  T^{(1)} \pa \hat{G}_{12}
\nonu \\
&- & \frac{8(1272+2258k+1227k^2+133k^3)}{
3(5+k)(19+23k)(47+35k)} \hat{G}_{12} \hat{T}
+  \frac{2}{
(5+k)} T^{(1)} \pa  T_{-}^{(\frac{3}{2})} 
\nonu \\
& + & \frac{4(5035 k^4+27791 k^3+24972 k^2+19087 k+68471)}{
45(5+k)^2(19+23k)(47+35k)}  i \hat{G}_{12} \pa \hat{B}_3
\nonu \\
& - & \frac{6(-53-44k+k^2)}{
5(5+k)(19+23k)}  \hat{G}_{12} \pa T^{(1)}
- \frac{4(3+2k)(83+55k)}{
5(5+k)^2(19+23k)} i \hat{G}_{12} \pa \hat{A}_3
\nonu \\
& - & \frac{4(70 k^4+3418 k^3+24403 k^2+24696 k+7497)}{
3(5+k)(3+7k)(19+23k)(47+35k)}  T_{-}^{(\frac{3}{2})} \hat{T}
\nonu \\
&+& \frac{2}{(5+k)}  T_{-}^{(\frac{3}{2})}  T^{(2)}
+
\frac{2(37+16k)}{5(5+k)^2}  i T_{-}^{(\frac{3}{2})} \pa 
\hat{A}_3
+\frac{2(-10+k)}{5(5+k)^2}   i T_{-}^{(\frac{3}{2})} \pa 
\hat{B}_3
\nonu \\
&- & \frac{3}{(5+k)}  T_{-}^{(\frac{3}{2})} \pa T^{(1)}
- \frac{2}{(5+k)} \hat{G}_{11} V_{-}^{(2)} 
\nonu \\
& + & \frac{4(1140 k^4+29773 k^3+11193 k^2-127621 k-145221)}{
45(5+k)^2(19+23k)(47+35k)}  i 
U^{(\frac{3}{2})} \pa \hat{B}_{+} 
\nonu \\
&+& \left. \frac{2}{(5+k)}  \hat{G}_{22} U_{-}^{(2)} 
+\frac{2(31+7k)}{15(5+k)} \pa  W_{-}^{(\frac{5}{2})} 
+\frac{2(25+4k)}{5(5+k)^2} i \pa \hat{A}_{+} \hat{G}_{22} 
\right](w) + \cdots,
\nonu \\
 W^{(2)}(z) \, W^{(3)}(w) & = &
\frac{1}{(z-w)^4} \, \left[-\frac{24(169+553k+341k^2+29k^3)}{
(5+k)^3(19+23k)} i \hat{A}_3 
\right. \nonu \\
& - & \left. \frac{8k(268+775k+280k^2+21k^3)}{
(5+k)^3(19+23k)} i \hat{B}_3
+\frac{12(-3+k)(4-7k+k^2)}{5+k)^2(19+23k)} T^{(1)} 
\right](w) 
\nonu \\
& + & \frac{1}{(z-w)^3} \, \left[ -\frac{4(21+43k+10k^2)}{
(5+k)^2(3+7k)} \hat{T} 
-\frac{2(-3+k)}{(5+k)^2} T^{(2)} 
+\frac{2}{(5+k)} W^{(2)}
\right. \nonu \\
& - & \frac{4(19+7k)}{(5+k)^3} \hat{A}_{+} \hat{A}_{-}
-\frac{28(1+k)}{(5+k)^3} \hat{A}_3 \hat{A}_3
+ \frac{8(16+k+k^2)}{(5+k)^3} \hat{A}_3 \hat{B}_3
\nonu \\
& - & \frac{4(1+k)(7+k)}{(5+k)^3} \hat{B}_{+} \hat{B}_{-}
-\frac{28(1+k)}{(5+k)^3} \hat{B}_3 \hat{B}_3
-\frac{2}{(5+k)} T^{(1)} T^{(1)} 
\nonu \\
& - & \frac{4(19+7k)}{(5+k)^3} i \pa \hat{A}_3
-\frac{4(1+k)(7+k)}{(5+k)^3} i \pa \hat{B}_3
+ \frac{4(-3+k)}{(5+k)^2} i T^{(1)} \hat{A}_3 
\nonu \\
&+ & \left. \frac{4(-3+k)}{(5+k)^2} i T^{(1)} \hat{B}_3
\right](w)
\nonu \\
& + & \frac{1}{(z-w)^2} \, \left[-\frac{(13+3k)}{(5+k)} {\bf S^{(3)}}
-\frac{2}{(5+k)} i {\bf P^{(2)}} \hat{B}_3  
+\frac{2}{(5+k)} i {\bf P^{(2)}} \hat{A}_3  
\right. \nonu \\
\nonu \\
& + & W^{(3)} +\frac{8(-84-199k+63k^2+14k^3)}{
(5+k)(3+7k)(19+23k)} T^{(1)} \hat{T}
-\frac{4}{(5+k)} T^{(1)} T^{(2)}
\nonu \\
& - & \frac{16(-3+k)}{(19+23k)} T^{(1)} W^{(2)} 
-\frac{2(2+k)}{(5+k)^2} i T^{(1)} \pa \hat{A}_3
-\frac{2(23+4k)}{(5+k)^2} i T^{(1)} \pa \hat{B}_3
\nonu \\
& - & \frac{1}{(5+k)} T^{(1)} \pa T^{(1)} 
+\frac{10}{(5+k)} i \hat{A}_{+} V_{+}^{(2)}
-\frac{4(19+4k)}{(5+k)^3} \hat{A}_{+} \pa \hat{A}_{-}  
\nonu \\
&-& \frac{2(7+k)}{(5+k)^2} i \hat{A}_{-} U_{-}^{(2)}
-\frac{16(1965+5170k+2971k^2+334k^3)}{
3(5+k)^2(3+7k)(19+23k)} i \hat{A}_3 \hat{T}
\nonu \\
&+ & \frac{40}{3(5+k)} i \hat{A}_3 T^{(2)}
+ \frac{8(13+5k)}{(5+k)(19+23k)} i \hat{A}_3 W^{(2)}
+ \frac{2(3+2k)}{(5+k)^2} i \hat{A}_3 \pa T^{(1)}
\nonu \\
&- & \frac{(47+11k)}{(5+k)^2} i \hat{B}_{+} U_{+}^{(2)}
 -\frac{(116+181k+27k^2)}{
3(5+k)^3} \hat{B}_{+} \pa \hat{B}_{-}
+ \frac{5(1+k)}{(5+k)^2} i \hat{B}_{-} V_{-}^{(2)}
\nonu \\
& + & \frac{(-40+127k+21k^2)}{
3(5+k)^3} \hat{B}_{-} \pa \hat{B}_{+}
\nonu \\
& - & \frac{16(285+2735k+3143k^2+1031k^3+42k^4)}{
3(5+k)^2(3+7k)(19+23k)} i \hat{B}_3 \hat{T}
-\frac{16(17+k)}{3(5+k)^2} i \hat{B}_3 T^{(2)}
\nonu \\
& + &  \frac{8(247+256k+49k^2)}{(5+k)^2(19+23k)} i \hat{B}_3 W^{(2)}
+ \frac{4(113+109k+15k^2)}{3(5+k)^3} \hat{B}_3 \pa \hat{A}_3
\nonu \\
&+ & \frac{10(4+k)}{(5+k)^2} i \hat{B}_3 \pa T^{(1)}
-\frac{2(13+k)}{(5+k)^2} 
\pa \hat{T} -\frac{(17+3k)}{(5+k)^2} \pa W^{(2)}
\nonu \\
&- & \frac{2(121+15k)}{(5+k)^3} i \pa^2 \hat{A}_3 
- \frac{2(19+182k+22k^2)}{3(5+k)^3} i \pa^2 \hat{B}_3 
+\frac{2(-27+k)}{3(5+k)^2} \pa^2 T^{(1)} 
\nonu \\
&+& \frac{4(10+k)}{(5+k)^2} \hat{G}_{12} \hat{G}_{21}
-\frac{2(-3+k)}{(5+k)^2} \hat{G}_{12}   T_{+}^{(\frac{3}{2})}
+\frac{2(29+5k)}{(5+k)^2} \hat{G}_{21}  T_{-}^{(\frac{3}{2})}
\nonu \\
&- & \frac{2(62+25k)}{3(5+k)^3} i \hat{A}_{+} \hat{A}_{-} \hat{A}_3
-\frac{16(5+7k)}{3(5+k)^3} i \hat{A}_{+} \hat{A}_{-} \hat{B}_3
-\frac{2(61+8k)}{3(5+k)^3} i \hat{A}_{-} \hat{A}_{+} \hat{A}_3
\nonu \\
&- & \frac{14(11+k)}{3(5+k)^3} i \hat{A}_3 \hat{A}_{+} \hat{A}_{-}
-\frac{8(59+10k)}{3(5+k)^3} i \hat{A}_3 \hat{A}_3 \hat{A}_3
+ \frac{24(16+k)}{(5+k)^3} i \hat{A}_3 \hat{A}_3 \hat{B}_3
\nonu \\
& - & \frac{2(124+107k+12k^2)}{3(5+k)^3} i \hat{A}_3 \hat{B}_{+} \hat{B}_{-}
- \frac{200}{(5+k)^3} i \hat{A}_3 \hat{B}_3 \hat{B}_3
\nonu \\
& + &  \frac{2(-148+43k+12k^2)}{
3(5+k)^3} i \hat{B}_{-} \hat{A}_3 \hat{B}_{+}
+  \frac{2(19+8k)}{
3(5+k)^3} i \hat{B}_{-} \hat{B}_{+} \hat{B}_3
\nonu \\
& - & \frac{2(59+16k)}{
3(5+k)^3} i \hat{B}_3 \hat{B}_{+} \hat{B}_{-}
+  \frac{8(-10+k)}{
3(5+k)^3} i \hat{B}_3 \hat{B}_3 \hat{B}_3
+ \frac{8}{
(5+k)^2}  T^{(1)} \hat{A}_{+} \hat{A}_{-}
\nonu \\
&+&  \frac{4}{
(5+k)^2}  T^{(1)} \hat{A}_3 \hat{A}_3
+ \frac{8}{
(5+k)^2}  T^{(1)} \hat{A}_3 \hat{B}_3
-\frac{12}{
(5+k)^2}  T^{(1)} \hat{B}_3 \hat{B}_3
\nonu \\
&+ & \left.  \frac{4(13+3k)}{
(5+k)^2}   T_{+}^{(\frac{3}{2})}  T_{-}^{(\frac{3}{2})}
\right](w)
\nonu \\
& + &  \frac{1}{(z-w)} \, 
\left[ 
\frac{1}{(5+k)} i \hat{A}_3 \pa {\bf P^{(2)}}
+\frac{1}{(5+k)} i \hat{A}_{-} {\bf Q_{-}^{(3)}}
+\frac{2}{(5+k)} i \hat{B}_3  {\bf S^{(3)}}
\right. \nonu \\
& + &  \frac{2}{(5+k)} i \hat{B}_3 {\bf P^{(3)}}
-\frac{1}{(5+k)} \hat{B}_3 \hat{B}_3 {\bf P^{(2)}}
-\frac{1}{(5+k)} i \hat{B}_3 \pa  {\bf P^{(2)}}
-\frac{1}{(5+k)} i \hat{B}_{+}  {\bf Q_{+}^{(3)}}
 \nonu \\
&+& \frac{1}{2(5+k)} \hat{B}_{+} \hat{B}_{-} {\bf P^{(2)}}
+\frac{1}{2(5+k)} i \pa \hat{B}_3  {\bf P^{(2)}}
-\frac{(9+2k)}{2(5+k)} \pa  {\bf S^{(3)}}
-\frac{3}{2(5+k)} \pa  {\bf P^{(3)}}
\nonu \\
&+ &  \frac{1}{(5+k)} W^{(2)} W^{(2)}
+\frac{6}{(5+k)^2}  U^{(\frac{3}{2})}  \pa V^{(\frac{3}{2})} 
+\frac{2}{(5+k)}   U_{-}^{(2)}  V_{+}^{(2)}
\nonu \\
&+& \frac{4}{(5+k)}   U_{+}^{(2)}  V_{-}^{(2)}
+\frac{2}{(5+k)} T^{(2)} W^{(2)}
+\frac{1}{(5+k)} T^{(2)} T^{(2)}
\nonu \\
& + & \frac{2(9+2k)}{(5+k)^2}  T_{+}^{(\frac{3}{2})}  \pa T_{-}^{(\frac{3}{2})}
-\frac{2(-16+5k)}{(5+k)^3}  i  \hat{A}_{+} \hat{A}_{-} \pa \hat{B}_3
\nonu \\
& - &  \frac{4}{(5+k)^2} \hat{A}_3 \hat{A}_3 W^{(2)} 
-\frac{4}{(5+k)^2} \hat{A}_3 \hat{A}_3 T^{(2)} 
+\frac{4}{(5+k)^3} \hat{A}_3 \hat{A}_3 \hat{A}_3 \hat{A}_3  
\nonu \\
& + & \frac{64}{(5+k)^3} \hat{A}_3 \hat{A}_3 \hat{B}_3 \hat{B}_3 
-\frac{8}{(5+k)^2} i  \hat{A}_3 \hat{A}_3 \hat{B}_3 T^{(1)}
-\frac{20}{(5+k)^3} \hat{A}_3 \hat{A}_3 \hat{B}_{+} \hat{B}_{-}
\nonu \\
&+& \frac{4(109+2k)}{3(5+k)^3}  
i \hat{A}_3 \hat{A}_3 \pa \hat{B}_{3} 
+ \frac{4}{(5+k)^2} \hat{A}_3 \hat{A}_3
\pa T^{(1)}
+\frac{16(-4+k)}{3(5+k)^3}  \hat{A}_3 \hat{B}_3 \hat{B}_3 \hat{B}_3 
\nonu \\
& - & \frac{8}{(5+k)^2} i  \hat{A}_3 \hat{B}_3 \hat{B}_3 T^{(1)}
+\frac{12}{(5+k)^2} \hat{A}_3 \hat{B}_{-}   V_{-}^{(2)}
-\frac{4}{(5+k)^2} \hat{A}_3 \hat{B}_{+}   U_{+}^{(2)}
\nonu \\
& - & \frac{8(-2+k)}{(5+k)^3}  \hat{A}_3 \hat{B}_{+} \hat{B}_{-} \hat{B}_3 
+\frac{12}{(5+k)^2} i  \hat{A}_3 \hat{B}_{+} \hat{B}_{-} T^{(1)} 
- \frac{2(55+32k)}{
3(5+k)^3} i \hat{A}_3 \hat{B}_{+} \pa \hat{B}_{-}
\nonu \\
& + &  \frac{8(-37+k)}{
3(5+k)^3} i \hat{A}_3 \pa \hat{B}_3 \hat{B}_3 
-\frac{4}{
(5+k)^2} \hat{A}_3 \pa \hat{B}_3 T^{(1)}
- \frac{28(11+k)}{
3(5+k)^3} i \hat{A}_3 \pa \hat{B}_{+} \hat{B}_{-} 
\nonu \\
& - & \frac{8(-4+25k+9k^2)}{(5+k)^2(19+23k)} i \hat{A}_3 \pa W^{(2)}
+  \frac{8(11+k)}{3(5+k)^2} i \hat{A}_3 \pa T^{(2)}
\nonu \\
&  - & \frac{2(-11005-10597k+5630k^2+690k^3)}{
15(5+k)^3(19+23k)} \hat{A}_3 \pa^2 \hat{B}_3 
+ \frac{2k}{
(5+k)^2} i \hat{A}_3 \pa^2 T^{(1)} 
\nonu \\
&-& \frac{4}{(5+k)^2} i \hat{A}_{-}  T_{-}^{(\frac{3}{2})} U^{(\frac{3}{2})} 
- \frac{8}{(5+k)^2} \hat{A}_{-} \hat{B}_3  U_{-}^{(2)}
+\frac{4}{(5+k)^2} i \hat{A}_{-} \hat{G}_{11}  T_{-}^{(\frac{3}{2})}
\nonu \\
&- & \frac{4}{(5+k)^2} i \hat{A}_{-} \hat{G}_{11} \hat{G}_{12}
+\frac{4}{(5+k)^2} i \hat{A}_{-} \hat{G}_{12}  U^{(\frac{3}{2})} 
+\frac{3}{(5+k)^2} i \hat{A}_{-} \pa  U_{-}^{(2)}
\nonu \\
&- & \frac{4}{(5+k)^2} \hat{A}_{+} \hat{A}_{-} W^{(2)} 
+\frac{16}{(5+k)^3}  \hat{A}_{+} \hat{A}_{-} \hat{A}_3 \hat{B}_3
-\frac{8}{(5+k)^2} i  \hat{A}_{+} \hat{A}_{-} \hat{B}_3 T^{(1)}
\nonu \\
&+& 
 \frac{8}{(5+k)^3}  \hat{A}_{+} \hat{A}_{-} \hat{B}_{+} \hat{B}_{-}
 -  \frac{2(27+8k)}{
(5+k)^3} i  \hat{A}_{+} \hat{A}_{-} \pa \hat{A}_3
\nonu \\
& + & \frac{7}{
(5+k)^2} \hat{A}_{+} \hat{A}_{-} \pa T^{(1)}
- \frac{4}{(5+k)^3} \hat{A}_{+} \hat{A}_{+} \hat{A}_{-} \hat{A}_{-}
+\frac{8}{(5+k)^2} \hat{A}_{+} \hat{B}_3  V_{+}^{(2)}
\nonu \\
&-& 
\frac{8(19+2k)}{
3(5+k)^3} i \hat{A}_{+} \pa \hat{A}_{-} \hat{A}_{3}
- \frac{4(1+8k)}{
3(5+k)^3}  i \hat{A}_{+} \pa \hat{A}_{-} \hat{B}_{3}
+ \frac{2(9+k)}{(5+k)^2} i \hat{A}_{+} \pa  V_{+}^{(2)}
\nonu \\
&+&  \frac{(3715+1963 k+532 k^2)}{
15(5+k)^3(19+23k)} \hat{A}_{+} \pa^2 \hat{A}_{-}
-\frac{4}{(5+k)} i \hat{B}_3 W^{(3)}
\nonu \\
& - & \frac{8}{(5+k)^2} i \hat{B}_3  U^{(\frac{3}{2})}  V^{(\frac{3}{2})}
-\frac{2(53+4k)}{3(5+k)^2} i \hat{B}_3 \pa T^{(2)} 
\nonu \\
& - & \frac{8}{(5+k)^2} i  \hat{B}_3 
T_{+}^{(\frac{3}{2})}  \pa T_{-}^{(\frac{3}{2})}
+\frac{4(13+2k)}{3(5+k)^2} \hat{B}_3 \hat{B}_3 W^{(2)}
-\frac{20}{(5+k)^3} \hat{B}_3 \hat{B}_3 T^{(2)}
\nonu \\
&-& \frac{8}{9(5+k)} i \hat{B}_3 \hat{B}_3 \hat{B}_3 T^{(1)}  
-\frac{2}{(5+k)} \hat{B}_3 \hat{B}_3 T^{(1)} T^{(1)}
+\frac{4}{(5+k)} i \hat{B}_3 T^{(1)} W^{(2)} 
\nonu \\
&+& \frac{4}{(5+k)} i \hat{B}_3 T^{(1)} T^{(2)}
+\frac{2(247+175k+52k^2)}{(5+k)^2(19+23k)} i \hat{B}_3 \pa W^{(2)}
\nonu \\
& + & \frac{(11265+37142k+37441k^2+11868k^3)}{15(5+k)^2(3+7k)(19+23k)} 
i \hat{B}_3 \pa^2 T^{(1)}
+\frac{4(-1+k)}{3(5+k)^2} \hat{B}_{-} \hat{B}_3  V_{-}^{(2)}
\nonu \\
& - & \frac{2}{(5+k)} i \hat{B}_{-} T^{(1)} V_{-}^{(2)}
+\frac{(3+k)}{(5+k)^2} i \hat{B}_{-} \pa  V_{-}^{(2)}
-\frac{4}{(5+k)^2} i \hat{B}_{+}  T_{+}^{(\frac{3}{2})}  U^{(\frac{3}{2})} 
\nonu \\
& - & \frac{4(-1+k)}{3(5+k)^2} \hat{B}_{+} \hat{B}_3  U_{+}^{(2)}
-  \frac{4(14+k)}{3(5+k)^2} \hat{B}_{+} \hat{B}_{-} W^{(2)}
-\frac{4}{(5+k)^2} \hat{B}_{+} \hat{B}_{-} T^{(2)}
\nonu \\
&+ & \frac{4(1+k)(7+k)}{(5+k)^3} \hat{B}_{+} \hat{B}_{-} \hat{B}_3 
\hat{B}_3
+ \frac{4(-10+k)}{3(5+k)^2} i \hat{B}_{+} \hat{B}_{-} \hat{B}_3 T^{(1)}
\nonu \\
& + & \frac{1}{(5+k)} \hat{B}_{+} \hat{B}_{-} T^{(1)} T^{(1)}
-\frac{2(-21+17k+2k^2)}{(5+k)^3} 
i  \hat{B}_{+} \hat{B}_{-} \pa \hat{B}_3
\nonu \\
&+& \frac{1}{(5+k)^2} \hat{B}_{+} 
\hat{B}_{-} \pa T^{(1)}
- \frac{(-21+16k+2k^2)}{2(5+k)^3}
 \hat{B}_{+} \hat{B}_{+} \hat{B}_{-} \hat{B}_{-} 
\nonu \\
&+& \frac{4}{(5+k)^2} i \hat{B}_{+} \hat{G}_{11}   T_{+}^{(\frac{3}{2})}
+\frac{2}{(5+k)} i \hat{B}_{+} T^{(1)} U^{(\frac{3}{2})} 
 -  \frac{3k}{(5+k)^2} i \hat{B}_{+} \pa U_{+}^{(2)}
\nonu \\
&+& 
 \frac{2(10+59k+6k^2)}{3(5+k)^3} i   
\hat{B}_{+} \pa \hat{B}_{-} \hat{B}_3
- \frac{(-35+2k)}{3(5+k)^2} \hat{B}_{+} \pa \hat{B}_{-} 
T^{(1)}
\nonu \\
& - & 
\frac{(5971 k^4+54614 k^3+119505 k^2+90516 k+22230)}{
15(5+k)^3(3+7k)(19+23k)} \hat{B}_{+} \pa^2 \hat{B}_{-}
\nonu \\
&+& \frac{2}{(5+k)} \hat{G}_{12} W_{+}^{(\frac{5}{2})}
- \frac{4(-3+k)}{3(5+k)^2} 
\hat{G}_{12} \pa  T_{+}^{(\frac{3}{2})}
 +  \frac{4(8+k)}{
3(5+k)^2} \hat{G}_{12} \pa \hat{G}_{21}
\nonu \\
&- & \frac{2}{(5+k)} \hat{G}_{21}  W_{-}^{(\frac{5}{2})}
+\frac{8(6+k)}{3(5+k)^2} \hat{G}_{21} \pa  T_{-}^{(\frac{3}{2})}
- \frac{8(-3+k)}{(19+23k)} T^{(1)} \pa W^{(2)}
\nonu \\
&-&\frac{4(3+4k)}{(5+k)(3+7k)} 
\left( \hat{T} W^{(2)} -\frac{3}{10} \pa^2 W^{(2)} \right)
\nonu \\
& -& \frac{4(3+4k)}{(5+k)(3+7k)}
\left( \hat{T} T^{(2)} -\frac{3}{10} \pa^2 T^{(2)} \right)
\nonu \\
&+ & \frac{4(3+4k)^2}{(5+k)(3+7k)^2} 
\left( \hat{T} \hat{T} -\frac{3}{10} \pa^2 \hat{T} \right)
\nonu \\
&+& \frac{8(3+4k)}{
(5+k)^2(3+7k)} \left( \hat{T} \hat{A}_3 \hat{A}_3 -\frac{3}{10} 
\pa^2 (\hat{A}_3 \hat{A}_3) \right)
\nonu \\
&+ &  \frac{192}{
(5+k)^2(19+23k)} \left( \hat{T} \hat{A}_3 \hat{B}_3 -\frac{3}{10} 
\pa^2 (\hat{A}_3 \hat{B}_3) \right)
\nonu \\
&-& \frac{8(-57+131k+320k^2+56k^3)}{
(5+k)^2(3+7k)(19+23k)} \left( \hat{T} \hat{B}_3 \hat{B}_3 -\frac{3}{10} 
\pa^2 (\hat{B}_3 \hat{B}_3) \right)
\nonu \\
& - & \frac{8(63+159k+92k^2)}{
(5+k)(3+7k)(19+23k)} i \left( \hat{T} T^{(1)} \hat{B}_3 -\frac{3}{10} 
\pa^2 ( T^{(1)} \hat{B}_3) \right)
\nonu \\
& - & 
\frac{24(9+k)}{
(5+k)^2(19+23k)} \hat{T} \hat{A}_{+} \hat{A}_{-}
\nonu \\
&+& \frac{36(9+k)}{
5(5+k)^2(19+23k)} \pa^2 (\hat{A}_{+} \hat{A}_{-})
\nonu \\
&+& \frac{8(114+365k+301k^2+14k^3)}{
(5+k)^2(3+7k)(19+23k)} \hat{T} \hat{B}_{+} \hat{B}_{-}
\nonu \\
&-& \frac{12(114+365k+301k^2+14k^3)}{
5(5+k)^2(3+7k)(19+23k)} \pa^2 (\hat{B}_{+} \hat{B}_{-})
\nonu \\
&-& \frac{64(42+134 k+119 k^2+23 k^3)}{
(5+k)^2(3+7k)(19+23k)} i \hat{T} \pa \hat{A}_3
\nonu \\
& - & \frac{8(2+k)(-57+131 k+168 k^2)}{
(5+k)^2(3+7k)(19+23k)}  i \hat{T} \pa \hat{B}_3
\nonu \\
&- & \frac{4(21+19k+58k^2)}{
(5+k)(3+7k)(19+23k)} \hat{T} \pa T^{(1)}
\nonu \\
&+& \frac{6}{(5+k)^2} \pa U^{(\frac{3}{2})} V^{(\frac{3}{2})}  
+ \frac{2(9+2k)}{(5+k)^2} \pa  T_{+}^{(\frac{3}{2})}  T_{-}^{(\frac{3}{2})}
+\frac{8}{(5+k)} i \pa \hat{A}_3 W^{(2)}
\nonu \\
& + & \frac{8(2+k)}{(5+k)^2} i \pa \hat{A}_3 T^{(2)}
- \frac{40(13+2k)}{3(5+k)^3} 
i \pa \hat{A}_3 \hat{A}_3 \hat{A}_3 
+  \frac{32(25+2k)}{
3(5+k)^3} i \pa \hat{A}_3 \hat{A}_3 \hat{B}_3 
\nonu \\
& - & \frac{16(19+2k)}{3(5+k)^3} 
i \pa \hat{A}_3 \hat{B}_3 \hat{B}_3 
 +  \frac{8}{(5+k)^2} 
\pa \hat{A}_3 \hat{B}_3 T^{(1)} 
-\frac{6(7+2k)}{
(5+k)^3} i \pa \hat{A}_3 \hat{B}_{+} \hat{B}_{-}
\nonu \\
&+ &
\frac{8(330+827 k+421k^2)}{
15(5+k)^3(3+7k)} \pa \hat{A}_3 \pa \hat{A}_3
\nonu \\
&-& \frac{4(-1495-862 k-1210 k^2+345 k^3)}{
15(5+k)^3(19+23k)} \pa \hat{A}_3 \pa \hat{B}_3
\nonu \\
& + &  \frac{2(-6+k)}{
(5+k)^2} i \pa \hat{A}_3 \pa T^{(1)}
-\frac{2}{(5+k)} i \pa \hat{A}_{-}  U_{-}^{(2)}
+\frac{2(16+3k)}{(5+k)^2} i \pa \hat{A}_{+}  V_{+}^{(2)}
\nonu \\
&-& \frac{2(91+8k)}{3(5+k)^3} i
\pa \hat{A}_{+} \hat{A}_{-} \hat{A}_3
- \frac{2(62+25k)}{
3(5+k)^3} i \pa \hat{A}_{+} \hat{A}_{-} \hat{B}_3
 +  \frac{1}{
(5+k)^2} \pa \hat{A}_{+} \hat{A}_{-} T^{(1)}
\nonu \\
& + & \frac{2(-6475-3277k+122k^2)}{15(5+k)^3(19+23k)} 
 \pa \hat{A}_{+} \pa \hat{A}_{-}
+\frac{20(2+k)}{3(5+k)^2} i \pa \hat{B}_3 W^{(2)}
\nonu \\
& - & \frac{44}{(5+k)^2} i \pa \hat{B}_3 T^{(2)}
+ \frac{4(1+26k+3k^2)}{
3(5+k)^3} i \pa \hat{B}_3 \hat{B}_3 \hat{B}_3 
\nonu \\
& - &  
\frac{4(-1+k)}{3(5+k)^2} \pa \hat{B}_3 \hat{B}_3 T^{(1)}
+ \frac{1}{(5+k)} i \pa \hat{B}_3 T^{(1)} T^{(1)}
\nonu \\
& + & \frac{(798 k^4+33700 k^3-87977 k^2-85646 k+1425)}{
15(5+k)^3(3+7k)(19+23k)} \pa \hat{B}_3  \pa \hat{B}_3 
\nonu \\
&- & 
 \frac{2(2355+5531k+4413k^2+1909k^3)}{
5(5+k)^2(3+7k)(19+23k)} i \pa \hat{B}_3  \pa T^{(1)}
+\frac{(7+11k)}{3(5+k)^2} i \pa \hat{B}_{-}  V_{-}^{(2)}
\nonu \\
& -& \frac{(101+13k)}{3(5+k)^2} i \pa \hat{B}_{+}  U_{+}^{(2)}
- \frac{4(25+32k+3k^2)}{3(5+k)^3} 
i \pa \hat{B}_{+} \hat{B}_{-} \hat{B}_3
\nonu \\
&+ & \frac{2(-19+k)}{3(5+k)^2} 
\pa \hat{B}_{+} \hat{B}_{-} T^{(1)} 
\nonu \\
& + & 
\frac{2(2919 k^4+54221 k^3+116450 k^2+84209 k+20805)}{15(5+k)^3(3+7k)(19+23k)} 
 \pa \hat{B}_{+} \pa \hat{B}_{-}
\nonu \\
&+& 
 \frac{4(8+k)}{
3(5+k)^2} \pa \hat{G}_{12} \hat{G}_{21}
+ \frac{4(3+k)}{
(5+k)^2} \pa \hat{G}_{21}  T_{-}^{(\frac{3}{2})}
-\frac{4}{(5+k)} \pa T^{(1)} T^{(2)}
\nonu \\
&+& \frac{1}{(5+k)} \pa T^{(1)} \pa T^{(1)} - 
 \frac{16(654+1526k+689k^2+29k^3)}{3(5+k)^2(3+7k)(19+23k)} i 
\pa \hat{T} \hat{A}_{3}
\nonu \\
&- & \frac{4(570+4561k+4999k^2+1768k^3+84k^4)}{
3(5+k)^2(3+7k)(19+23k)} i \pa \hat{T} \hat{B}_3 
\nonu \\
&+ & \frac{4(-39+8k+2k^2)}{
(5+k)(19+23k)} \pa \hat{T} T^{(1)}
 -  \frac{2(105-13 k+166 k^2)}{
15(5+k)^3(3+7k)} 
\pa^2 \hat{A}_3 \hat{A}_3
\nonu \\
&+ & \frac{2(1380 k^3+12065 k^2+23342 k+10625)}{
15(5+k)^3(19+23k)} \pa^2 \hat{A}_3 \hat{B}_3 
+ \frac{3}{
(5+k)^2} i \pa^2 \hat{A}_3 T^{(1)}
\nonu \\
&-& \frac{4(-395-721k+6k^2)}{
5(5+k)^3(19+23k)}
 \pa^2 \hat{A}_{+} \hat{A}_{-}
\nonu \\
&+&  \frac{4(1477 k^4-2255 k^3+22417 k^2+21651 k+1710)}{
15(5+k)^3(3+7k)(19+23k)} 
\pa^2 \hat{B}_{3} \hat{B}_3
\nonu \\
&- & \frac{(6256 k^3+144667 k^2+159934 k+38595)}{
45(5+k)^2(3+7k)(19+23k)} i \pa^2 \hat{B}_3 T^{(1)}
\nonu \\
& + & \frac{(2079 k^4-9899 k^3-54710 k^2-55031 k-15675)}{
15(5+k)^3(3+7k)(19+23k)} 
\pa^2 \hat{B}_{+} \hat{B}_{-}
\nonu \\
& - & \frac{1}{(5+k)} \pa^2 T^{(1)} T^{(1)}
+\frac{(13+2k)}{(5+k)} \pa W^{(3)}
-\frac{(255+481k+118k^2)}{10(5+k)^2(3+7k)} \pa^2 W^{(2)}
\nonu \\
&- & \frac{3(-15-83k+16k^2)}{
10(5+k)^2(3+7k)} \pa^2 T^{(2)}
\nonu \\
& + & \frac{(-90+237k+2117k^2+778k^3)}{
15(5+k)^2(3+7k)^2} \pa^2 \hat{T}
\nonu \\
& + & \frac{(4648 k^4+43872 k^3+117393 k^2+134186 k+47841)}{
9(5+k)^3(3+7k)(19+23k)} i  \pa^3 \hat{A}_3
\nonu \\
&+ & \frac{(336 k^5+6512 k^4+35290 k^3+147379 k^2+110672 k-16473)}{
18(5+k)^3(3+7k)(19+23k)} i \pa^3 \hat{B}_3  
\nonu \\
&-& \left. \frac{(28 k^4+335 k^3+673 k^2-2923 k-921)}{
3(5+k)^2(3+7k)(19+23k)} \pa^3 T^{(1)} 
\right]  + \cdots, 
\nonu \\
 W_{+}^{(\frac{5}{2})}(z) \, W_{+}^{(\frac{5}{2})}(w) & = &
-\frac{1}{(z-w)^3} \, 
\left[ \frac{16(131+51k+4k^2)}{9(5+k)^3} \hat{A}_{-} \hat{B}_{-}\right] (w) 
\nonu \\
& - & \frac{1}{(z-w)^2}  \, 
\left[ \frac{4(131+51k+4k^2)}{9(5+k)^2} \hat{G}_{21} \hat{G}_{21} \right] (w) 
\nonu \\
& + &  \frac{1}{(z-w)} \, \left[
 \frac{4}{(5+k)}  U_{+}^{(2)}
 V_{+}^{(2)} +
\frac{4(7+4k)}{3(5+k)^2} 
i \pa \hat{A}_{-}  U_{+}^{(2)} -
\frac{8(1+k)}{3(5+k)^2} 
i  \hat{A}_{-}  \pa U_{+}^{(2)}  \right. \nonu \\
& - & \frac{4(51+24k+2k^2)}{
3(5+k)^3} \pa^2 \hat{A}_{-} \hat{B}_{-}
+ \frac{4(-27+8k+2k^2)}{
3(5+k)^3} \pa \hat{A}_{-} \pa \hat{B}_{-}
\nonu \\
&- & \frac{4(48+19k+2k^2)}{
3(5+k)^3}  \hat{A}_{-} \pa^2 \hat{B}_{-}
-\frac{4(7+2k)}{3(5+k)^2} i \pa \hat{B}_{-}  V_{+}^{(2)}
+\frac{4(11+k)}{3(5+k)^2} i  \hat{B}_{-}  \pa V_{+}^{(2)}
\nonu \\
& + & \frac{4}{(5+k)} \hat{G}_{21}  W_{+}^{(\frac{5}{2})}
- \frac{8(9+k)}{3(5+k)^2} \pa \hat{G}_{21}  T_{+}^{(\frac{3}{2})}
+\frac{8(9+k)}{3(5+k)^2} \hat{G}_{21}  \pa T_{+}^{(\frac{3}{2})}
\nonu \\
&+ & \left. \frac{8(8+k)}{3(5+k)^2} \hat{G}_{21} \pa \hat{G}_{21}  
+\frac{8}{(5+k)} T_{+}^{(\frac{3}{2})} \pa T_{+}^{(\frac{3}{2})}
 \right](w) +\cdots, 
\nonu \\
 W_{+}^{(\frac{5}{2})}(z)  \, W_{-}^{(\frac{5}{2})}(w) & = &
-\frac{1}{(z-w)^5} \,  \left[ \frac{8k(67+39k+4k^2)}{(5+k)^3} \right] 
\nonu \\
& + & \frac{1}{(z-w)^4} \, \left[ \frac{8(67+39k+4k^2)}{(5+k)^3} 
i \hat{A}_3 + 
\frac{8k(67+39k+4k^2)}{3(5+k)^3} i \hat{B}_3 \right. \nonu \\
& + & \left.  
\frac{4(-3+k)}{3(5+k)^2} T^{(1)}  \right](w)
\nonu \\
& + & \frac{1}{(z-w)^3} \, \left[ 
\frac{2(-3+k)}{3(5+k)} {\bf P^{(2)}}
-\frac{8(321+1193k+604k^2+48k^3)}{
9(5+k)^2(3+7k)} \hat{T} 
\right. \nonu \\
& - & \frac{8(-9+k)(-3+k)}{9(5+k)^2} T^{(2)}
-\frac{4(-1+7k+2k^2)}{
3(5+k)^2} W^{(2)}
\nonu \\
& - & \frac{8(-19+49k+6k^2)}{
9(5+k)^3} \hat{A}_{+} \hat{A}_{-}
- \frac{8(71+58k+6k^2)}{
9(5+k)^3} \hat{A}_3 \hat{A}_3
\nonu \\
& + & \frac{16(8+97k)}{
9(5+k)^3} \hat{A}_3 \hat{B}_3
- \frac{8(107+25k)}{
9(5+k)^3} \hat{B}_{+} \hat{B}_{-}
-\frac{8(107+28k+12k^2)}{
9(5+k)^3} \hat{B}_3 \hat{B}_3
\nonu \\
&+ & \frac{4}{(5+k)} T^{(1)} T^{(1)} 
+\frac{4(641+253k+24k^2)}{
9(5+k)^3} i \pa \hat{A}_3
\nonu \\
& + & \frac{4(-214+151k+117k^2+12k^3)}{
9(5+k)^3} i \pa \hat{B}_3
+\frac{2(-3+k)}{3(5+k)^2} \pa T^{(1)}
\nonu \\
&-& \left. \frac{8(-7+2k)}{3(5+k)^2} i  T^{(1)} \hat{A}_3
-\frac{8(-11+4k)}{3(5+k)^2} i  T^{(1)} \hat{B}_3
\right](w) 
\nonu \\
& + & \frac{1}{(z-w)^2} \, \left[
\frac{(13+3k)}{(5+k)} {\bf S^{(3)}}
+\frac{(-3+k)}{3(5+k)} {\bf P^{(3)}}
+ \frac{(-3+k)}{3(5+k)} \pa {\bf P^{(2)}}
\right. \nonu \\
& + & \frac{2}{(5+k)} i {\bf P^{(2)}} \hat{B}_3
- \frac{2}{(5+k)} i {\bf P^{(2)}} \hat{A}_3
-\frac{2(26+5k)}{3(5+k)} W^{(3)}
\nonu \\
&- & \frac{4(-135-348k+307k^2)}{3(5+k)(3+7k)(19+23k)} 
T^{(1)} \hat{T} + \frac{4}{(5+k)} T^{(1)} T^{(2)}
+ \frac{10}{(5+k)^2} i T^{(1)} \pa \hat{A}_3
\nonu \\
&+ &   \frac{10(18+k)}{3(5+k)^2} i T^{(1)} \pa \hat{B}_3
+ \frac{4}{(5+k)} T^{(1)} \pa T^{(1)} 
-\frac{4(37+7k)}{3(5+k)^2} i \hat{A}_{+} V_{+}^{(2)} 
\nonu \\
&- & \frac{4(-131+92k+12k^2)}{9(5+k)^3}  \hat{A}_{+} \pa \hat{A}_{-} 
+  \frac{4(11+2k)}{3(5+k)^2}  i \hat{A}_{-}  U_{-}^{(2)} 
\nonu \\
& + &  \frac{8(5217+14234k+8201k^2+1088k^3)}{
3(5+k)^2(3+7k)(19+23k)} 
i \hat{A}_3 \hat{T}
-\frac{40}{3(5+k)} i \hat{A}_3 T^{(2)} 
\nonu \\
& - & \frac{8}{(5+k)} i \hat{A}_3 W^{(2)}
+  \frac{4(421+524k+42k^2)}{
9(5+k)^3}  \hat{A}_3 \pa \hat{B}_3
-\frac{2(-1+8k)}{
3(5+k)^2}  i \hat{A}_3 \pa T^{(1)}
\nonu \\
& + &  \frac{8(17+4k)}{3(5+k)^2}  i \hat{B}_{+}  U_{+}^{(2)} 
+  \frac{2(-302+101k+12k^2)}{9(5+k)^3}   \hat{B}_{+} \pa \hat{B}_{-}
\nonu \\
& - &  \frac{4(5+4k)}{3(5+k)^2}  i \hat{B}_{-}  V_{-}^{(2)} 
- \frac{2(-20+63k+4k^2)}{3(5+k)^3}   \hat{B}_{-} \pa \hat{B}_{+}
\nonu \\
& + &  \frac{8(570+6223k+8340k^2+2779k^3+140k^4)}{
3(5+k)^2(3+7k)(19+23k)} 
i \hat{B}_3 \hat{T}
+ \frac{16(17+k)}{3(5+k)^2} i \hat{B}_3 T^{(2)} 
\nonu \\
& - & \frac{8(13+3k)}{(5+k)^2} i \hat{B}_3 W^{(2)}
-\frac{2(46+21k)}{
3(5+k)^2}  i \hat{B}_3 \pa T^{(1)}
\nonu \\
&- & \frac{2(363+1636k+1193k^2+96k^3)}{9(5+k)^2(3+7k)} \pa \hat{T}
-\frac{(-3+k)(-27+4k)}{9(5+k)^2} \pa   T^{(2)} 
\nonu \\
& - & \frac{(-57+5k+4k^2)}{3(5+k)^2} \pa W^{(2)}
+ \frac{2(1451+241k+12k^2)}{9(5+k)^3} i \pa^2 \hat{A}_3
\nonu \\
& + &  \frac{2(-181+594k+120k^2+8k^3)}{9(5+k)^3} i \pa^2 \hat{B}_3
-\frac{4(-39+k)}{9(5+k)^2} \pa^2 T^{(1)} 
\nonu \\
&- & \frac{4(10+k)}{(5+k)^2} \hat{G}_{12} \hat{G}_{21} 
+ \frac{4(-3+k)}{3(5+k)^2} \hat{G}_{12}  T_{+}^{(\frac{3}{2})}
-\frac{4(45+7k)}{3(5+k)^2} \hat{G}_{21}  T_{-}^{(\frac{3}{2})}
\nonu \\
& + & \frac{2(20+19k)}{3(5+k)^3} i \hat{A}_{+} \hat{A}_{-} \hat{A}_3
-\frac{2(107-32k+42k^2)}{9(5+k)^3} i \hat{A}_{+} \hat{A}_{-} \hat{B}_3
\nonu \\
& + &  \frac{2(349+170k+12k^2)}{
9(5+k)^3} i \hat{A}_{-} \hat{A}_{+} \hat{A}_3
+  \frac{2(227+136k+42k^2)}{
9(5+k)^3} i \hat{A}_{-} \hat{A}_{+} \hat{B}_3
\nonu \\
& - &  \frac{2(-191+107k+12k^2)}{
9(5+k)^3} i \hat{A}_3 \hat{A}_{+} \hat{A}_{-}
+  \frac{8(59+10k)}{
3(5+k)^3} i \hat{A}_3 \hat{A}_3 \hat{A}_3
\nonu \\
& - &   \frac{24(16+k)}{
(5+k)^3} i \hat{A}_3 \hat{A}_3 \hat{B}_3
+  \frac{200}{
(5+k)^3} i \hat{A}_3 \hat{B}_3 \hat{B}_3
+  \frac{32(17+4k)}{
3(5+k)^3} i \hat{B}_{-} \hat{A}_3 \hat{B}_{+}
\nonu \\
& + &  \frac{2(181+62k+24k^2)}{
9(5+k)^3} i \hat{B}_{-} \hat{B}_{+} \hat{B}_3
-  \frac{2(61+38k+24k^2)}{
9(5+k)^3} i \hat{B}_3 \hat{B}_{+} \hat{B}_{-}
\nonu \\
& - & \frac{8(-10+k)}{
3(5+k)^3} i \hat{B}_3 \hat{B}_3 \hat{B}_3
- \frac{8}{
(5+k)^2} T^{(1)} \hat{A}_{+} \hat{A}_{-}
-\frac{4}{
(5+k)^2} T^{(1)} \hat{A}_3 \hat{A}_3
\nonu \\
&- & \left. \frac{8}{
(5+k)^2} T^{(1)} \hat{A}_3 \hat{B}_3
+  \frac{12}{
(5+k)^2} T^{(1)} \hat{B}_3 \hat{B}_3 
-  \frac{4(13+3k)}{(5+k)^2} 
T_{+}^{(\frac{3}{2})}  T_{-}^{(\frac{3}{2})}
\right](w) 
\nonu \\
& + &  \frac{1}{(z-w)} \, 
\left[ 
\frac{1}{2}   \{ \hat{G}_{21} \, 
S_{-}^{(\frac{7}{2})} \}_{-1}
-\frac{1}{(5+k)} i \hat{A}_3 \pa {\bf P^{(2)}}
-\frac{1}{(5+k)} i \hat{A}_{-} {\bf Q_{-}^{(3)} }
\right. \nonu \\
& -& \frac{16}{3(5+k)} i \hat{B}_3  {\bf S^{(3)}}
- \frac{16}{3(5+k)} i \hat{B}_3 {\bf P^{(3)}}
+\frac{8}{3(5+k)} \hat{B}_3 \hat{B}_3 {\bf P^{(2)}}
\nonu \\
&+& \frac{5}{3(5+k)} i \hat{B}_{-} {\bf R_{-}^{(3)} }
+\frac{8}{3(5+k)} i \hat{B}_{+}  {\bf Q_{+}^{(3)} }
-\frac{4}{3(5+k)} \hat{B}_{+} \hat{B}_{-} {\bf P^{(2)}}
\nonu \\
&
- & 
\frac{4}{3(5+k)} i \pa \hat{B}_3 {\bf P^{(2)}}
+\frac{(16+3k)}{3(5+k)} \pa {\bf S^{(3)}}
+\frac{(32+k)}{15(5+k)} \pa  {\bf P^{(3)}}
\nonu \\
&+ & \frac{1}{(5+k)} i \hat{B}_3 \pa  {\bf P^{(2)}}
+\frac{(-3+k)}{10(5+k)} \pa^2  {\bf P^{(2)}}
\nonu \\
&- &  \frac{8}{3(5+k)} W^{(2)} W^{(2)} 
-\frac{16}{3(5+k)^2}  U^{(\frac{3}{2})}  \pa V^{(\frac{3}{2})} 
-\frac{2}{(5+k)}   U_{-}^{(2)}  V_{+}^{(2)}
\nonu \\
&-&  \frac{22}{3(5+k)}   U_{+}^{(2)}  V_{-}^{(2)} 
-\frac{16}{3(5+k)} T^{(2)} W^{(2)}
-\frac{2}{3(5+k)} T^{(2)} T^{(2)}
\nonu \\
&-& \frac{4}{3(5+k)^2}  
T_{+}^{(\frac{3}{2})}  \pa T_{-}^{(\frac{3}{2})} 
-\frac{4}{(5+k)} i \hat{A}_3 W^{(3)} +
\frac{32}{3(5+k)^2} \hat{A}_3 \hat{A}_3 W^{(2)} 
\nonu \\
&+& \frac{20}{3(5+k)^2} \hat{A}_3 \hat{A}_3 T^{(2)} 
-\frac{32}{3(5+k)^3} \hat{A}_3 \hat{A}_3 \hat{A}_3 \hat{A}_3 
+\frac{64}{3(5+k)^2} i  \hat{A}_3 \hat{A}_3 \hat{B}_3 T^{(1)}
\nonu \\
& -& \frac{4(18 k^3+917 k^2+17959 k+14304)}{15(5+k)^3(19+23k)}  
i \hat{A}_3 \hat{A}_3 \pa \hat{B}_{3}
\nonu \\
& - & \frac{2(157+311k+6k^2)}{5(5+k)^2(19+23k)} \hat{A}_3 \hat{A}_3
\pa T^{(1)} +\frac{8}{(5+k)^2} \hat{A}_3 \hat{B}_3 T^{(2)}
\nonu \\
& - & \frac{128(-4+k)}{9(5+k)^3} 
\hat{A}_3 \hat{B}_3 \hat{B}_3 \hat{B}_3 
+\frac{64}{3(5+k)^2} i \hat{A}_3 \hat{B}_3 \hat{B}_3  T^{(1)}
\nonu \\
& + & \frac{4(-223-149k+6k^2)}{5(5+k)^2(19+23k)} \hat{A}_3 \hat{B}_3
\pa T^{(1)}
- \frac{80}{3(5+k)^2} \hat{A}_3 \hat{B}_{-}   V_{-}^{(2)}
\nonu \\
&+& \frac{56}{3(5+k)^2} \hat{A}_3 \hat{B}_{+}   U_{+}^{(2)}
+\frac{64(-18+k)}{3(5+k)^3} \hat{A}_3 \hat{B}_{+} \hat{B}_{-} \hat{B}_3
-\frac{32}{(5+k)^2} i  \hat{A}_3 \hat{B}_{+} \hat{B}_{-} T^{(1)}
\nonu \\
& + &  \frac{4(-14413-14194k+4235k^2)}{
15(5+k)^3(19+23k)} i \hat{A}_3 \hat{B}_{+} \pa \hat{B}_{-}
+ \frac{8}{(5+k)^2} i \hat{A}_3 \hat{G}_{11}  V^{(\frac{3}{2})} 
\nonu \\
& - &  \frac{8}{(5+k)^2} i \hat{A}_3 \hat{G}_{11} \hat{G}_{22} 
+\frac{8}{(5+k)^2} i \hat{A}_3 \hat{G}_{22}  U^{(\frac{3}{2})} 
\nonu \\
& + & \frac{8(-6395-8659k-966k^2+6k^3)}{
15(5+k)^3(19+23k)} i \hat{A}_3 \pa \hat{B}_3 \hat{B}_3 
\nonu \\
& - & \frac{4(-877-981k+4k^2)}{
5(5+k)^2(19+23k)} \hat{A}_3 \pa \hat{B}_3 T^{(1)}
\nonu \\
& + &  \frac{2(67409+82597k+1800k^2)}{
15(5+k)^3(19+23k)} i \hat{A}_3 \pa \hat{B}_{+} \hat{B}_{-} 
\nonu \\
& - & \frac{4(23+3k)}{3(5+k)^2} i \hat{A}_3 \pa W^{(2)}
-  \frac{4(177+31k)}{15(5+k)^2} i \hat{A}_3 \pa T^{(2)}
\nonu \\
&+& \frac{2}{
45(5+k)^3(3+7k)(19+23k)(47+35k)} 
(800562 k^5+10880618 k^4
\nonu \\
& - & 6649745 k^3 
 -  58864247 
k^2-52472701 k-12431631) \hat{A}_3 \pa^2 \hat{B}_3 
\nonu \\
& -& \frac{2(1442+1405k+711k^2)}{
15(5+k)^2(19+23k)} i \hat{A}_3 \pa^2 T^{(1)} 
+ \frac{16}{(5+k)^2} \hat{A}_{-} \hat{B}_3  U_{-}^{(2)}
\nonu \\
& - &\frac{4}{(5+k)^2} i \hat{A}_{-} \hat{G}_{11}  T_{-}^{(\frac{3}{2})} 
+\frac{4}{(5+k)^2} i \hat{A}_{-} \hat{G}_{11} \hat{G}_{12}
-\frac{4}{(5+k)^2} i \hat{A}_{-} \hat{G}_{12}  U^{(\frac{3}{2})} 
\nonu \\
&+ & \frac{(-5+2k)}{3(5+k)^2} i \hat{A}_{-} \pa  U_{-}^{(2)}
+\frac{8}{(5+k)^2} \hat{A}_{+} \hat{A}_3  V_{+}^{(2)}
+\frac{20}{3(5+k)^2} \hat{A}_{+} \hat{A}_{-} W^{(2)}
\nonu \\
&-& \frac{4}{3(5+k)^2} \hat{A}_{+} \hat{A}_{-} T^{(2)}
-\frac{16}{3(5+k)^3}  \hat{A}_{+} \hat{A}_{-} \hat{A}_3 \hat{A}_3
-\frac{32}{(5+k)^3}  \hat{A}_{+} \hat{A}_{-} \hat{A}_3 \hat{B}_3
\nonu \\
&+& \frac{592}{3(5+k)^3}  \hat{A}_{+} \hat{A}_{-} \hat{B}_3 \hat{B}_3
+\frac{64}{3(5+k)^2} i \hat{A}_{+} \hat{A}_{-} \hat{B}_3 T^{(1)}
- \frac{104}{(5+k)^3}  \hat{A}_{+} \hat{A}_{-} \hat{B}_{+} \hat{B}_{-}
\nonu \\
&+&  \frac{2(10217+14293k+1440k^2)}{
15(5+k)^3(19+23k)} i  \hat{A}_{+} \hat{A}_{-} \pa \hat{A}_3
-\frac{3(73+169k+4k^2)}{
5(5+k)^2(19+23k)} \hat{A}_{+} \hat{A}_{-} \pa T^{(1)}
\nonu \\
& - & \frac{6(1596+1805k-255k^2+4k^3)}{5(5+k)^3(19+23k)}  
i  \hat{A}_{+} \hat{A}_{-} \pa \hat{B}_3
\nonu \\
&+&  \frac{16}{3(5+k)^3} \hat{A}_{+} \hat{A}_{+} \hat{A}_{-} \hat{A}_{-}
-\frac{88}{3(5+k)^2} \hat{A}_{+} \hat{B}_3  V_{+}^{(2)}
-\frac{4}{(5+k)^2} i \hat{A}_{+} \hat{G}_{21} \hat{G}_{22}
\nonu \\
&+& \frac{4}{(5+k)^2} i \hat{A}_{+} \hat{G}_{22}  T_{+}^{(\frac{3}{2})}
+\frac{8(2873+3917k+680k^2)}{
15(5+k)^3(19+23k)} i \hat{A}_{+} \pa \hat{A}_{-} \hat{A}_{3}
\nonu \\
&+&
 \frac{8(-836-395k+595k^2+6k^3)}{
15(5+k)^3(19+23k)}  i \hat{A}_{+} \pa \hat{A}_{-} \hat{B}_{3}
\nonu \\
& + & \frac{4(-201-203k+2k^2)}{
5(5+k)^2(19+23k)} \hat{A}_{+} \pa \hat{A}_{-} T^{(1)}
 -  \frac{56(9+k)}{15(5+k)^2} i \hat{A}_{+} \pa  V_{+}^{(2)}
\nonu \\
& - & \frac{4(21266 k^5+217532 k^4+293807 k^3+17629 k^2-41773 k-1605)}{
15(5+k)^3(3+7k)(19+23k)(47+35k)} \hat{A}_{+} \pa^2 \hat{A}_{-}
\nonu \\
&+ & \frac{20}{3(5+k)} i \hat{B}_3 W^{(3)}
+\frac{64}{3(5+k)^2} i \hat{B}_3 
 U^{(\frac{3}{2})}   V^{(\frac{3}{2})} 
+\frac{64}{3(5+k)^2} i \hat{B}_3  T_{+}^{(\frac{3}{2})}  T_{-}^{(\frac{3}{2})}
\nonu \\
&-&  \frac{32(37+2k)}{9(5+k)^2} \hat{B}_3 \hat{B}_3 W^{(2)}
-\frac{36}{(5+k)^2}  \hat{B}_3 \hat{B}_3 T^{(2)}
\nonu \\
& - & \frac{32(-133+16k+2k^2)}{9(5+k)^3}
\hat{B}_3 \hat{B}_3 \hat{B}_3 \hat{B}_3
+\frac{64(-19+k)}{27(5+k)^2} i \hat{B}_3 \hat{B}_3 \hat{B}_3 T^{(1)}
\nonu \\
&+& \frac{16}{3(5+k)}  \hat{B}_3 \hat{B}_3 T^{(1)} T^{(1)}
-\frac{2(-289+13k+18k^2)}{15(5+k)^2(19+23k)}  
\hat{B}_3 \hat{B}_3 \pa T^{(1)}
\nonu \\
& - & \frac{8}{(5+k)^2} i \hat{B}_3 \hat{G}_{11}   V^{(\frac{3}{2})} 
+  \frac{8}{(5+k)^2} i \hat{B}_3 \hat{G}_{11} \hat{G}_{22}
-\frac{8}{(5+k)^2} i \hat{B}_3 \hat{G}_{22}  U^{(\frac{3}{2})} 
\nonu \\
&-& \frac{32}{3(5+k)} i \hat{B}_3 T^{(1)} W^{(2)}
-\frac{32}{3(5+k)} i \hat{B}_3 T^{(1)} T^{(2)}
-\frac{4(29+9k)}{3(5+k)^2} i \hat{B}_3 \pa W^{(2)}
\nonu \\
&+& \frac{16(60+k)}{15(5+k)^2} i \hat{B}_3 \pa T^{(2)} 
-\frac{2(26661 k^3+54367 k^2+70367 k+27429)}{
45(5+k)^2(3+7k)(19+23k)} i \hat{B}_3 \pa^2 T^{(1)}
\nonu \\
&-& \frac{32}{3(5+k)^2} i  \hat{B}_{-}  T_{-}^{(\frac{3}{2})}  V^{(\frac{3}{2})} 
-\frac{32(-25+k)}{9(5+k)^2} \hat{B}_{-} \hat{B}_3  V_{-}^{(2)}
-\frac{32}{3(5+k)^2} i \hat{B}_{-} \hat{G}_{22}  T_{-}^{(\frac{3}{2})}
\nonu \\
& + &  \frac{16}{3(5+k)} i \hat{B}_{-} T^{(1)}  V_{-}^{(2)}
-\frac{(49+4k)}{3(5+k)^2} i \hat{B}_{-} \pa V_{-}^{(2)}
+\frac{32}{3(5+k)^2} i \hat{B}_{+}  T_{+}^{(\frac{3}{2})}   U^{(\frac{3}{2})}
\nonu \\
& + & \frac{8(-109+4k)}{9(5+k)^2} \hat{B}_{+} \hat{B}_3  U_{+}^{(2)}
+  \frac{4(199+8k)}{9(5+k)^2} \hat{B}_{+} \hat{B}_{-} W^{(2)}
+\frac{140}{3(5+k)^2} \hat{B}_{+} \hat{B}_{-} T^{(2)}
\nonu \\
& + & \frac{32(-6+k)(14+k)}{3(5+k)^3} \hat{B}_{+} \hat{B}_{-} \hat{B}_3
\hat{B}_3 -\frac{32(-34+k)}{9(5+k)^2} i \hat{B}_{+} \hat{B}_{-} \hat{B}_3 T^{(1)}
\nonu \\
&- & \frac{8}{3(5+k)}  \hat{B}_{+} \hat{B}_{-} T^{(1)} T^{(1)}
-\frac{2(-912-1181k-397k^2+12k^3)}{5(5+k)^3(19+23k)} 
i  \hat{B}_{+} \hat{B}_{-} \pa \hat{B}_3
\nonu \\
& - & \frac{3(-117-61k+4k^2)}{5(5+k)^2(19+23k)} \hat{B}_{+} 
\hat{B}_{-} \pa T^{(1)}
-  \frac{20}{3(5+k)^2} i \hat{B}_{+} \hat{G}_{11}   T_{+}^{(\frac{3}{2})}
\nonu \\
& + & \frac{4}{(5+k)^2} i \hat{B}_{+} \hat{G}_{11} \hat{G}_{21} 
-\frac{16}{3(5+k)} i \hat{B}_{+} T^{(1)} U_{+}^{(2)}  
\nonu \\
& + & \frac{4(-44327-48986k+6113k^2+932k^3)}{15(5+k)^3(19+23k)} i   
\hat{B}_{+} \pa \hat{B}_{-} \hat{B}_3
\nonu \\
& + & \frac{4(-15869-18467k+478k^2)}{45(5+k)^2(19+23k)} 
\hat{B}_{+} \pa \hat{B}_{-} 
T^{(1)}
+ \frac{4(13+22k)}{15(5+k)^2} i \hat{B}_{+} \pa U_{+}^{(2)}
\nonu \\
& + & 
\frac{2
}{
45(5+k)^3(3+7k)(19+23k)(47+35k)}
(590009 k^5+6059319 k^4 \nonu \\
& + & 32182447 k^3
+  60931353 k^2+44708052 k+11028780) 
 \hat{B}_{+} \pa^2 \hat{B}_{-} 
\nonu \\
& + & \frac{4(17+3k)}{5(5+k)^2}
\hat{G}_{11} \pa  V^{(\frac{3}{2})} 
+\frac{4(2+3k)}{5(5+k)^2} \hat{G}_{11} \pa \hat{G}_{22}
-\frac{2}{(5+k)} \hat{G}_{12} W_{+}^{(\frac{5}{2})}
\nonu \\
& + &
\frac{4(-2670-2707k-236k^2+9k^3)}{15(5+k)^2(19+23k)} 
\hat{G}_{12} \pa  T_{+}^{(\frac{3}{2})}
\nonu \\
& + & \frac{4(980 k^4+4808 k^3-37755 k^2-102326 k-58487)}{
15(5+k)^2(19+23k)(47+35k)} \hat{G}_{12} \pa \hat{G}_{21}
\nonu \\
& - & \frac{2}{(5+k)} \hat{G}_{21}  W_{-}^{(\frac{5}{2})}
+\frac{4(36+11k)}{15(5+k)^2} \hat{G}_{21} \pa  T_{-}^{(\frac{3}{2})}
-\frac{4(21+4k)}{5(5+k)^2} \hat{G}_{22} \pa  U^{(\frac{3}{2})} 
\nonu \\
&+& \frac{2}{(5+k)}  T^{(1)} \pa T^{(2)}
+\frac{32(3+4k)}{3(5+k)(3+7k)} 
\left( \hat{T} W^{(2)} -\frac{3}{10} \pa^2 W^{(2)} \right)
\nonu \\
& -& \frac{4(70 k^4-29587 k^3-66503 k^2-53821 k-13935)}{
3(5+k)(3+7k)(19+23k)(47+35k)}
\left( \hat{T} T^{(2)} -\frac{3}{10} \pa^2 T^{(2)} \right)
\nonu \\
&- & \frac{8(13244 k^5+193192 k^4+588979 k^3+664250 k^2+318567 k+55044)}{
3(5+k)(3+7k)^2(19+23k)(47+35k)} 
\nonu \\
& \times & \left( \hat{T} \hat{T} -\frac{3}{10} \pa^2 \hat{T} \right)
\nonu \\
&-& \frac{8(1862 k^4+24521 k^3+74188 k^2+79021 k+23988)}{
3(5+k)^2(3+7k)(19+23k)(47+35k)} 
\nonu \\
& \times &
\left( \hat{T} \hat{A}_3 \hat{A}_3 -\frac{3}{10} 
\pa^2 (\hat{A}_3 \hat{A}_3) \right)
\nonu \\
&+ &  \frac{16(1862 k^4+53151 k^3+85054 k^2+21615 k-3366)}{
3(5+k)^2(3+7k)(19+23k)(47+35k)} 
\nonu \\
& \times & 
\left( \hat{T} \hat{A}_3 \hat{B}_3 -\frac{3}{10} 
\pa^2 (\hat{A}_3 \hat{B}_3) \right)
\nonu \\
&+& \frac{8(13818 k^4+320915 k^3+755928 k^2+574663 k+142392)}{
3(5+k)^2(3+7k)(19+23k)(47+35k)} 
\nonu \\
& \times & \left( \hat{T} \hat{B}_3 \hat{B}_3 -\frac{3}{10} 
\pa^2 (\hat{B}_3 \hat{B}_3) \right)
\nonu \\
& + & \frac{32(153+372k+163k^2)}{
3(5+k)(3+7k)(19+23k)} i \left( \hat{T} T^{(1)} \hat{B}_3 -\frac{3}{10} 
\pa^2 ( T^{(1)} \hat{B}_3) \right)
\nonu \\
&-& \frac{8(1862 k^4+29456 k^3+49735 k^2+20590 k+1569)}{
3(5+k)^2(3+7k)(19+23k)(47+35k)} \hat{T} \hat{A}_{+} \hat{A}_{-}
\nonu \\
&+& \frac{4(1862 k^4+29456 k^3+49735 k^2+20590 k+1569)}{
5(5+k)^2(3+7k)(19+23k)(47+35k)} \pa^2 (\hat{A}_{+} \hat{A}_{-})
\nonu \\
&-& \frac{8(8232 k^4+234325 k^3+595738 k^2+500061 k+136224)}{
3(5+k)^2(3+7k)(19+23k)(47+35k)} \hat{T} \hat{B}_{+} \hat{B}_{-}
\nonu \\
&+& \frac{4(8232 k^4+234325 k^3+595738 k^2+500061 k+136224)}{
5(5+k)^2(3+7k)(19+23k)(47+35k)} \pa^2 (\hat{B}_{+} \hat{B}_{-})
\nonu \\
&+&
 \frac{4(206290 k^4+1708335 k^3+4279379 k^2+3866337 k+1062027)}{
15(5+k)^2(3+7k)(19+23k)(47+35k)} i \hat{T} \pa \hat{A}_3
\nonu \\
& + & \frac{4(30240 k^5+471999 k^4+153547 k^3-2181583 k^2-3010275 k-1169352)}{
15(5+k)^2(3+7k)(19+23k)(47+35k)}  
\nonu \\
& \times & i \hat{T} \pa \hat{B}_3
-  \frac{2(4760 k^4+9747 k^3-114653 k^2-255251 k-76443)}{
15(5+k)(3+7k)(19+23k)(47+35k)} \hat{T} \pa T^{(1)}
\nonu \\
& - & \frac{40}{3(5+k)^2} \pa U^{(\frac{3}{2})} V^{(\frac{3}{2})}  
-  \frac{4(46+9k)}{3(5+k)^2} \pa  T_{+}^{(\frac{3}{2})}  T_{-}^{(\frac{3}{2})}
\nonu \\
&- & 
\frac{8(6+k)}{(5+k)^2} i \pa \hat{A}_3 W^{(2)}
-\frac{4(29+2k)}{5(5+k)^2} i \pa \hat{A}_3 T^{(2)}
\nonu \\
&-& \frac{4(17313+23677k+3440k^2)}{15(5+k)^3(19+23k)} 
i \pa \hat{A}_3 \hat{A}_3 \hat{A}_3 
\nonu \\
& + & \frac{8(-14664-17959k-407k^2+12k^3)}{
15(5+k)^3(19+23k)} i \pa \hat{A}_3 \hat{A}_3 \hat{B}_3 
\nonu \\
& + & \frac{4(-307-291k+4k^2)}{5(5+k)^2(19+23k)} 
\pa \hat{A}_3 \hat{A}_3 T^{(1)} 
\nonu \\
& - & \frac{4(-26835-34361k-2434k^2+24k^3)}{15(5+k)^3(19+23k)} 
i \pa \hat{A}_3 \hat{B}_3 \hat{B}_3 
\nonu \\
&+& \frac{4(156-79k+97k^2)}{5(5+k)^2(19+23k)} 
\pa \hat{A}_3 \hat{B}_3 T^{(1)} 
+\frac{2(-5021-4723k+750k^2)}{
15(5+k)^3(19+23k)} i \pa \hat{A}_3 \hat{B}_{+} \hat{B}_{-}
\nonu \\
& - & \frac{2
}{
15(5+k)^3(3+7k)(19+23k)(47+35k)}
(85064 k^5+2516388 k^4 \nonu \\
& + & 10513005 k^3
+  17219163 k^2+12175111 k+
2902269) \pa \hat{A}_3 \pa \hat{A}_3
\nonu \\
&+& \frac{4
}{
15(5+k)^3(3+7k)(19+23k)(47+35k)}
(83594 k^5+1156309 k^4 \nonu \\
& + & 4855259 k^3+5691227 k^2+1291223 k-209460)
 \pa \hat{A}_3 \pa \hat{B}_3
\nonu \\
& + &  \frac{4(156-79k+97k^2)}{
5(5+k)^2(19+23k)} i \pa \hat{A}_3 \pa T^{(1)}
+\frac{2(29+4k)}{3(5+k)^2} i \pa \hat{A}_{-}  U_{-}^{(2)}
\nonu \\
& - & \frac{2(327+28k)}{15(5+k)^2} i \pa \hat{A}_{+}  V_{+}^{(2)}
+ \frac{2(16717+21993k+2720k^2)}{15(5+k)^3(19+23k)} i
\pa \hat{A}_{+} \hat{A}_{-} \hat{A}_3
\nonu \\
& + & \frac{2(10336+16975k+4795k^2+24k^3)}{
15(5+k)^3(19+23k)} i \pa \hat{A}_{+} \hat{A}_{-} \hat{B}_3
\nonu \\
& + & \frac{(-519-467k+8k^2)}{
5(5+k)^2(19+23k)} \pa \hat{A}_{+} \hat{A}_{-} T^{(1)}
\nonu \\
& - & \frac{2(85064 k^5+2064048 k^4+7312847 k^3+8842257 k^2+3808733 k+
494835)}{15(5+k)^3(3+7k)(19+23k)(47+35k)} 
\nonu \\
& \times & \pa \hat{A}_{+} \pa \hat{A}_{-}
-\frac{8(-59+5k)}{9(5+k)^2} i \pa \hat{B}_3 W^{(2)}
-\frac{4(-365+6k)}{15(5+k)^2} i \pa \hat{B}_3 T^{(2)}
\nonu \\
&+& \frac{4(-62966-70323k+7749k^2+926k^3)}{
15(5+k)^3(19+23k)} i \pa \hat{B}_3 \hat{B}_3 \hat{B}_3 
\nonu \\
& + & 
\frac{4(-22903-26239k+956k^2)}{45(5+k)^2(19+23k)} 
\pa \hat{B}_3 \hat{B}_3 T^{(1)}
-\frac{8}{3(5+k)} i \pa \hat{B}_3 T^{(1)} T^{(1)}
\nonu \\
& -& \frac{2}{15(5+k)^3(3+7k)(19+23k)(47+35k)}
(8820 k^6-83797 k^5-4098863 k^4 \nonu \\
& - & 21880453 k^3-37626583 k^2-
25119370 k-5996946)  \pa \hat{B}_3  \pa \hat{B}_3 
\nonu \\
& + & \frac{4(63 k^4+2553 k^3+23855 k^2+46063 k+18138)}{
15(5+k)^2(3+7k)(19+23k)} i \pa \hat{B}_3  \pa T^{(1)}
\nonu \\
&- &
\frac{2(-224+41k)}{9(5+k)^2} i \pa \hat{B}_{-}  V_{-}^{(2)}
\nonu \\
& +& \frac{2(1858+47k)}{45(5+k)^2} i \pa \hat{B}_{+}  U_{+}^{(2)}
\nonu \\
&- & \frac{2(1816 k^3+11399 k^2-98463 k-87666)}{15(5+k)^3(19+23k)} 
i \pa \hat{B}_{+} \hat{B}_{-} \hat{B}_3
\nonu \\
&- & \frac{(-60119-72707k+1768k^2)}{45(5+k)^2(19+23k)} 
\pa \hat{B}_{+} \hat{B}_{-} T^{(1)} 
\nonu \\
& - & 
\frac{2
}{15(5+k)^3(3+7k)(19+23k)(47+35k)}
(88249 k^5+4376295 k^4 \nonu \\
& + & 30184041 k^3
+  59761311 k^2+44337746 k+10977198
)  \pa \hat{B}_{+} \pa \hat{B}_{-}
\nonu \\
& - & \frac{8(4+k)}{5(5+k)^2} 
\pa \hat{G}_{11}  V^{(\frac{3}{2})} 
-\frac{4(3+2k)}{5(5+k)^2} \pa \hat{G}_{11} \hat{G}_{22}
\nonu \\
& - & \frac{4(-625-661k-78k^2+2k^3)}{
5(5+k)^2(19+23k)} \pa \hat{G}_{12} T_{+}^{(\frac{3}{2})}
\nonu \\
& - & \frac{4(-109+188k+95k^2+6k^3)}{
15(5+k)^2(19+23k)} \pa \hat{G}_{12} \hat{G}_{21}
\nonu \\
& - & \frac{4(25 k^4+14790 k^3+114426 k^2+210134 k+96873)}{
15(5+k)^2(19+23k)(47+35k)} \pa \hat{G}_{21}  T_{-}^{(\frac{3}{2})}
\nonu \\
&+& \frac{4(14+k)}{5(5+k)^2} \pa \hat{G}_{22}  U^{(\frac{3}{2})} 
+\frac{2}{(5+k)} \pa T^{(1)} W^{(2)}
-\frac{3}{2(5+k)} \pa T^{(1)} \pa T^{(1)}
\nonu \\
& + & \frac{4(5830 k^3+50327 k^2+71136 k+25551)}{15(5+k)^2(3+7k)(19+23k)} i 
\pa \hat{T} \hat{A}_{3}
\nonu \\
&+ & \frac{4(532 k^4+13716 k^3+44951 k^2+28130 k-741)}{
15(5+k)^2(3+7k)(19+23k)} i \pa \hat{T} \hat{B}_3 
\nonu \\
&+ & \frac{4(-318-961k-1453k^2+42k^3)}{
15(5+k)(3+7k)(19+23k)} \pa \hat{T} T^{(1)}
\nonu \\
& - & \frac{2(85064 k^5+1870988 k^4+5577225 k^3+5379343 k^2+1389691 k-84111)}{
15(5+k)^3(3+7k)(19+23k)(47+35k)} 
\nonu \\
& \times & \pa^2 \hat{A}_3 \hat{A}_3
\nonu \\
&- & \frac{2
}{
15(5+k)^3(3+7k)(19+23k)(47+35k)}
(320166 k^5+3643913 k^4
\nonu \\
& + & 12700087 k^3+16997887 k^2+8776847 k+
1479588)
 \pa^2 \hat{A}_3 \hat{B}_3 
\nonu \\
& - & \frac{(3969+4103k+274k^2)}{
15(5+k)^2(19+23k)} i \pa^2 \hat{A}_3 T^{(1)}
\nonu \\
&-& \frac{(85064 k^5+1333668 k^4-886069 k^3-7366947 k^2-6428527 k-1501725)}{
15(5+k)^3(3+7k)(19+23k)(47+35k)}
\nonu \\
& \times & \pa^2 \hat{A}_{+} \hat{A}_{-}
\nonu \\
& + & \frac{2}{45(5+k)^3(3+7k)(19+23k)(47+35k)}
(17640 k^6-480214 k^5-4153301 k^4\nonu \\
& - & 92493151 k^3-222665671 k^2-
171005815 k-40652472) 
\pa^2 \hat{B}_{3} \hat{B}_3
\nonu \\
&- & \frac{(1512 k^4+30106 k^3-863177 k^2-865880 k-179277)}{
135(5+k)^2(3+7k)(19+23k)} i \pa^2 \hat{B}_3 T^{(1)}
\nonu \\
& - & \frac{1}{45(5+k)^3(3+7k)(19+23k)(47+35k)}
(1001462 k^5+4935273 k^4 \nonu \\
& - & 33946625 k^3-107296233 k^2-
91917225 k-23954292)
\pa^2 \hat{B}_{+} \hat{B}_{-}
\nonu \\
& + & \frac{3}{(5+k)} \pa^2 T^{(1)} T^{(1)}
-\frac{2(79+8k)}{15(5+k)} \pa W^{(3)}
\nonu \\
& - & \frac{(-699-1019k-90k^2+42k^3)}{15(5+k)^2(3+7k)} \pa^2 W^{(2)}
\nonu \\
&- & \frac{(8960 k^5-228364 k^4+495121 k^3+2454919 k^2+2272731 k+549297)}{
15(5+k)^2(3+7k)(19+23k)(47+35k)} \pa^2 T^{(2)}
\nonu \\
& - & \frac{2}{15(5+k)^2(3+7k)^2(19+23k)(47+35k)}(
324604 k^6+6377712 k^5 \nonu \\
& + & 27846830 k^4+47488347 k^3+
36287997 k^2+11944809 k+1356237) \pa^2 \hat{T}
\nonu \\
& - & \frac{2
}{45(5+k)^3(3+7k)(19+23k)(47+35k)}
(628775 k^5+9007275 k^4 \nonu \\
& + & 35487668 k^3  +  63577800 k^2+
50760701 k+13719525) i  \pa^3 \hat{A}_3
\nonu \\
&- & \frac{2}{135(5+k)^3(3+7k)(19+23k)(47+35k)}(
118125 k^6+2745632 k^5 \nonu \\
& + & 14758386 k^4 
 +  3954874 k^3-62931417 k^2-91989642 k-38056374) i \pa^3 \hat{B}_3  
\nonu \\
&+&  \frac{(4095 k^5+269554 k^4+1185907 k^3+741839 k^2-
1089214 k-363981)}{
45(5+k)^2(3+7k)(19+23k)(47+35k)} \nonu \\
&\times & \left. \pa^3 T^{(1)} 
\right](w)+  \cdots, 
\nonu \\
W_{+}^{(\frac{5}{2})}(z) \, W^{(3)}(w) & = &
\frac{1}{(z-w)^4} \, \left[ -\frac{4(1833+15134k+13873k^2+3214k^3+218k^4)}{
3(5+k)^3(19+23k)} \hat{G}_{21} \right. \nonu \\
&- & \left. \frac{4(-3+k)(2249+2616k+487k^2)}{
3(5+k)^3(19+23k)}  T_{+}^{(\frac{3}{2})} \right](w) 
\nonu \\
& + & \frac{1}{(z-w)^3} \, \left[ 
\frac{(-3+k)(1861+1113k+152k^2)}{
3(5+k)^2(19+23k)} {\bf P_{+}^{(\frac{5}{2})}}
\right. \nonu \\
&+&  
\frac{2(-2361-1292k+193k^2+76k^3)}{
3(5+k)^2(19+23k)}  W_{+}^{(\frac{5}{2})}
-\frac{8(-9+6k+k^2)}{3(5+k)^3} i \hat{A}_{-} \hat{G}_{11}
\nonu \\
&-& \frac{8(13+11k+k^2)}{3(5+k)^3} i \hat{A}_{-}  U^{(\frac{3}{2})}
+ \frac{8(-22+3k)}{(5+k)^3}  i \hat{A}_3 \hat{G}_{21} 
\nonu \\
&- & \frac{4(-9+9k+2k^2)}{3(5+k)^3} i \hat{B}_{-} \hat{G}_{22}
+ \frac{4(41+17k+2k^2)}{3(5+k)^3} i \hat{B}_{-}  V^{(\frac{3}{2})}
\nonu \\
& - & 
 \frac{8(52-16k+k^2)}{3(5+k)^3}  i \hat{B}_3 \hat{G}_{21} 
+  \frac{8(-37+12k+3k^2)}{3(5+k)^3}  i \hat{B}_3  T_{+}^{(\frac{3}{2})}
\nonu \\
& + &  \frac{4(-36+k)}{3(5+k)^2}  T^{(1)} \hat{G}_{21} 
-  \frac{4(13+3k)}{(5+k)^2}  T^{(1)}  T_{+}^{(\frac{3}{2})}
+  \frac{8(-59-k+4k^2)}{3(5+k)^3}  i \hat{A}_3  T_{+}^{(\frac{3}{2})}
\nonu \\
& - & \frac{4(8787+23590k+13881k^2+3168k^3+218k^4)}{
9(5+k)^3(19+23k)} \pa \hat{G}_{21}
\nonu \\
&-& \left. \frac{4(-3+k)(501+158k+73k^2)}{
9(5+k)^3(19+23k)} \pa  T_{+}^{(\frac{3}{2})}
\right](w) 
\nonu \\
& + & \frac{1}{(z-w)^2} \, \left[ 
\frac{2(33+7k)}{3(5+k)^2} i {\bf R^{(\frac{5}{2})}} \hat{B}_{-} 
\right. \nonu \\
&-& \frac{(34+7k)}{2(5+k)} {\bf S_{+}^{(\frac{7}{2})}}
-\frac{(37+9k)}{(5+k)^2} i {\bf P_{+}^{(\frac{5}{2})} } \hat{A}_3
+ \frac{(37+9k)}{(5+k)^2} i {\bf P_{+}^{(\frac{5}{2})} } \hat{B}_3
\nonu \\
& + & \frac{4(15+4k)}{3(5+k)^2} i {\bf Q^{(\frac{5}{2})}} \hat{A}_{-} 
+ \frac{2(-3+k)(2687+1505k+182k^2)}{
15(5+k)^2(19+23k)} \pa {\bf P_{+}^{(\frac{5}{2})} } 
\nonu \\
& + & 
\frac{1
}{
45(5+k)^3(19+23k)(47+35k)}
(56980 k^5+1296952 k^4
\nonu \\
& + & 14948469 k^3+65606454 k^2+93061361 k+39507836) 
 \hat{A}_{-} \hat{A}_{+} \hat{G}_{21}
\nonu \\
& + & \frac{20(6+k)}{(5+k)^3}  \hat{A}_{-} \hat{A}_{+}  T_{+}^{(\frac{3}{2})}
-\frac{10(11+2k)}{3(5+k)^2} i \hat{A}_{-} U^{(\frac{5}{2})}
\nonu \\
&+& \frac{2(1094+645k+94k^2)}{
15(5+k)^3}  \hat{A}_{-} \hat{A}_3 \hat{G}_{11}
- \frac{8(11+2k)}{
3(5+k)^3}  \hat{A}_{-} \hat{A}_3  U^{(\frac{3}{2})}
\nonu \\
&+ &  \frac{2(49+9k)}{
3(5+k)^3}  \hat{A}_{-} \hat{B}_{-} \hat{G}_{12} 
+  \frac{2(71+13k)}{
(5+k)^3}  \hat{A}_{-} \hat{B}_{-}   T_{-}^{(\frac{3}{2})}
+  \frac{8(34+7k)}{
(5+k)^3}  \hat{A}_{-} \hat{B}_3 \hat{G}_{11} 
\nonu \\
& - &  \frac{8(4+7k)}{
3(5+k)^3}  \hat{A}_{-} \hat{B}_3  U^{(\frac{3}{2})}
-   \frac{4}{
(5+k)^2}  i \hat{A}_{-} T^{(1)}  U^{(\frac{3}{2})}
\nonu \\
& + &  \frac{8(1362+335k+37k^2)}{
45(5+k)^3} i \hat{A}_{-} \pa \hat{G}_{11}
+  \frac{4(2011+765k+96k^2)}{
45(5+k)^3} i \hat{A}_{-} \pa  U^{(\frac{3}{2})}
\nonu \\
& - & \frac{4(58+17k)}{
3(5+k)^3}  \hat{B}_{+} \hat{B}_{-} \hat{G}_{21} 
+  \frac{(1214+390k+27k^2)}{
3(5+k)^3}  \hat{B}_{+} \hat{B}_{-}  T_{+}^{(\frac{3}{2})}
\nonu \\
&+ &  \frac{5(19+5k)}{3(5+k)^2} i \hat{B}_{-} V^{(\frac{5}{2})}
+ \frac{8(34+7k)}{
(5+k)^3}  \hat{B}_{-} \hat{A}_3 \hat{G}_{22} 
+   \frac{8(19+2k)}{
3(5+k)^3}  \hat{B}_{-} \hat{A}_3  V^{(\frac{3}{2})}
\nonu \\
& - &  \frac{(1310+426k+27k^2)}{
3(5+k)^3}  \hat{B}_{-} \hat{B}_{+}  T_{+}^{(\frac{3}{2})}
+ \frac{2(1424+912k+115k^2)}{
15(5+k)^3}  \hat{B}_{-} \hat{B}_3 \hat{G}_{22} 
\nonu \\
& + &   \frac{8(14+k)}{
3(5+k)^3}  \hat{B}_{-} \hat{B}_3  V^{(\frac{3}{2})}
- \frac{4}{
(5+k)^2}  i \hat{B}_{-} T^{(1)}  V^{(\frac{3}{2})}
\nonu \\
& + &   \frac{4(1794+872k+95k^2)}{
45(5+k)^3} i \hat{B}_{-} \pa \hat{G}_{22}
-   \frac{4(2416+729k+33k^2)}{
45(5+k)^3} i \hat{B}_{-} \pa  V^{(\frac{3}{2})}
\nonu \\
& - & 
\frac{1
}{
45(5+k)^3(19+23k)(47+35k)}
(56980 k^5+1296952 k^4
\nonu \\
& + & 15624669 k^3+65817294 k^2+91087721 k+38114756) 
 \hat{A}_{+} \hat{A}_{-} \hat{G}_{21}
\nonu \\
& + &  \frac{2(439+328k+25k^2)}{
(5+k)^2(19+23k)} i \hat{A}_3 W_{+}^{(\frac{5}{2})}
\nonu \\
& - & \frac{2(1094+645k+94k^2)}{
15(5+k)^3}  \hat{A}_3 \hat{A}_{-} \hat{G}_{11} 
-  \frac{8(-13+2k)}{
(5+k)^3}  \hat{A}_3 \hat{A}_3 \hat{G}_{21} 
\nonu \\
& + &  \frac{8(15+2k)}{
3(5+k)^3}  \hat{A}_3 \hat{A}_3  T_{+}^{(\frac{3}{2})}
-  \frac{112}{
(5+k)^3}  \hat{A}_3 \hat{B}_3 \hat{G}_{21} 
-  \frac{16(15+2k)}{
3(5+k)^3}  \hat{A}_3 \hat{B}_3  T_{+}^{(\frac{3}{2})}
\nonu \\
&- &   \frac{16}{
(5+k)^2}  i \hat{A}_3 T^{(1)} \hat{G}_{21} 
+  \frac{4(-13398-11437k+1925k^2+392k^3)}{
15(5+k)^3(19+23k)} i \hat{A}_3 \pa \hat{G}_{21}
\nonu \\
& - &   \frac{4(-1140-251k+12k^2)}{
15(5+k)^3} i \hat{A}_3 \pa  T_{+}^{(\frac{3}{2})}
+ \frac{10(247+256k+49k^2)}{
(5+k)^2(19+23k)} i \hat{B}_3 W_{+}^{(\frac{5}{2})}
\nonu \\
& - &  \frac{2(1424+912k+115k^2)}{
15(5+k)^3}  \hat{B}_3 \hat{B}_{-} \hat{G}_{22} 
+ \frac{8(1+2k)}{
(5+k)^3}  \hat{B}_3 \hat{B}_3 \hat{G}_{21} 
\nonu \\
& + & \frac{8(15+2k)}{
3(5+k)^3}  \hat{B}_3 \hat{B}_3  T_{+}^{(\frac{3}{2})}
+  \frac{16}{
(5+k)^2}  i \hat{B}_3 T^{(1)} \hat{G}_{21} 
\nonu \\
& - &  \frac{4(-9272-11502k-1619k^2+245k^3+14k^4)}{
5(5+k)^3(19+23k)} i \hat{B}_3 \pa \hat{G}_{21}
\nonu \\
&  + &   \frac{4(546+184k+13k^2)}{
3(5+k)^3} i \hat{B}_3 \pa  T_{+}^{(\frac{3}{2})}
- \frac{20(-3+k)}{
(19+23k)} T^{(1)} W_{+}^{(\frac{5}{2})}
-\frac{}{} \pa \hat{A}_{-} 
\nonu \\
&- &  \frac{2(-146+1041k+165k^2+14k^3)}{
5(5+k)^2(19+23k)} T^{(1)} \pa \hat{G}_{21}
+  \frac{2(8+k)}{
(5+k)^2} T^{(1)} \pa  T_{+}^{(\frac{3}{2})}
\nonu \\
& +&  \frac{(-33983-36586k-4783k^2+364k^3)}{
15(5+k)^2(19+23k)} \pa  W_{+}^{(\frac{5}{2})}
\nonu \\
& - &  \frac{2(1981+975k+116k^2)}{
15(5+k)^3} i \pa \hat{A}_{-} U^{(\frac{3}{2})}
+  \frac{2(-1100+43k+44k^2)}{
15(5+k)^3} i \pa \hat{A}_3   T_{+}^{(\frac{3}{2})}
\nonu \\
& + & \frac{2(4576+1179k+53k^2)}{
15(5+k)^3} i \pa \hat{B}_{-} V^{(\frac{3}{2})}
\nonu \\
&+ &  \frac{2(-83068-96028k-6441k^2+1745k^3+126k^4)}{
15(5+k)^3(19+23k)} i \pa \hat{B}_3 \hat{G}_{21}
\nonu \\
& + &  \frac{(-14424-6121k+1945k^2+126k^3)}{
15(5+k)^2(19+23k)}  \pa T^{(1)} \hat{G}_{21}
- \frac{3(24+5k)}{
(5+k)^2} \pa T^{(1)}   T_{+}^{(\frac{3}{2})}
\nonu \\
& - & \frac{(2730 k^5+38761 k^4+3153135 k^3+19146734 k^2+28875081 k+11652795)}{
45(5+k)^3(19+23k)(47+35k)} \nonu \\
& \times & \pa^2  T_{+}^{(\frac{3}{2})}
\nonu \\
& - & 
\frac{4(11074 k^5+239071 k^4+1229030 k^3+2162333 k^2+1330050 k+205758)}{
3(5+k)^2(3+7k)(19+23k)(47+35k)} \nonu \\
& \times & \hat{G}_{21} \hat{T} 
-\frac{2(156+17k)}{3(5+k)^2} \hat{G}_{21} T^{(2)}
+ \frac{4(17+4k)}{(5+k)^2} \hat{G}_{21} W^{(2)}
\nonu \\
& - & \frac{4
}{
45(5+k)^3(19+23k)(47+35k)}
(28490 k^5+710216 k^4 \nonu \\
& + & 8041455 k^3+34518762 k^2+48095245 k+
19934116) 
  i \hat{G}_{21} \pa \hat{A}_3
\nonu \\
& + &  \frac{4(490 k^5-10465 k^4-96612 k^3-191527 k^2-104832 k+34110)}{
3(5+k)^2(3+7k)(19+23k)(47+35k)} T_{+}^{(\frac{3}{2})} \hat{T}
\nonu \\
& - &  \frac{2(50+11k)}{(5+k)^2} T_{+}^{(\frac{3}{2})} T^{(2)}
-\frac{2(32+7k)}{(5+k)^2} \hat{G}_{11} V_{+}^{(2)}
-\frac{6}{(5+k)}  U^{(\frac{3}{2})}   V_{+}^{(2)}
\nonu \\
&+ & \left.
\frac{2(32+7k)}{(5+k)^2} \hat{G}_{22} U_{+}^{(2)}
-\frac{6}{(5+k)}  V^{(\frac{3}{2})}   U_{+}^{(2)}
\right](w) 
\nonu \\
& + &  \frac{1}{(z-w)} \, 
 \{ W_{+}^{(\frac{5}{2})} \, 
 W^{(3)} \}_{-1}(w) +\cdots, 
\label{w5half+w3} 
\\
 W_{-}^{(\frac{5}{2})}(z)  \, W_{-}^{(\frac{5}{2})}(w) & = &
-\frac{1}{(z-w)^3} \, \left[ \frac{16(131+51k+4k^2)}{9(5+k)^3} 
\hat{A}_{+} \hat{B}_{+}\right] (w) 
\nonu \\
& - & \frac{1}{(z-w)^2} \, \left[ \frac{4(131+51k+4k^2)}{
9(5+k)^2} \hat{G}_{12} \hat{G}_{12} \right] (w) 
\nonu \\
& + &  \frac{1}{(z-w)} \, \left[ \frac{4}{(5+k)}  U_{-}^{(2)}
 V_{-}^{(2)} +
\frac{4(1+4k)}{3(5+k)^2} 
i \pa \hat{A}_{+}  V_{-}^{(2)} -
\frac{8(4+k)}{3(5+k)^2} 
i  \hat{A}_{+}  \pa V_{-}^{(2)}  \right. \nonu \\
& - & \frac{4(42+21k+2k^2)}{
3(5+k)^3} \pa^2 \hat{A}_{+} \hat{B}_{+}
+ \frac{4(-27+8k+2k^2)}{
3(5+k)^3} \pa \hat{A}_{+} \pa \hat{B}_{+}
\nonu \\
&- & \frac{4(57+22k+2k^2)}{
3(5+k)^3}  \hat{A}_{+} \pa^2 \hat{B}_{+}
-\frac{4(13+2k)}{3(5+k)^2} i \pa \hat{B}_{+}  U_{-}^{(2)}
+\frac{4}{3(5+k)} i  \hat{B}_{+}  \pa U_{-}^{(2)}
\nonu \\
& + & \frac{4}{(5+k)} \hat{G}_{12}  W_{-}^{(\frac{5}{2})}
+ \frac{8(9+k)}{3(5+k)^2} \pa \hat{G}_{12}  T_{-}^{(\frac{3}{2})}
-\frac{8(9+k)}{3(5+k)^2} \hat{G}_{12}  \pa T_{-}^{(\frac{3}{2})}
\nonu \\
&+ & \left. \frac{8(8+k)}{3(5+k)^2} \hat{G}_{12} \pa \hat{G}_{12}  
+\frac{8}{(5+k)} T_{-}^{(\frac{3}{2})} \pa T_{-}^{(\frac{3}{2})}
\right](w) +\cdots, 
\nonu \\
W_{-}^{(\frac{5}{2})}(z) \, W^{(3)}(w) & = &
\frac{1}{(z-w)^4} \, \left[ \frac{4(-1587+10766k+13597k^2+3214k^3+218k^4)}{
3(5+k)^3(19+23k)} \hat{G}_{12} 
\right. \nonu \\
&-& \left. \frac{4(-3+k)(3275+4086k+763k^2)}{
3(5+k)^3(19+23k)}   T_{-}^{(\frac{3}{2})}
\right](w) 
\nonu \\
& + & \frac{1}{(z-w)^3} \, \left[ 
\frac{(-3+k)(1823+1067k+152k^2)}{
3(5+k)^2(19+23k)} {\bf P_{-}^{(\frac{5}{2})}}
\right. \nonu \\
& - &  \frac{2(-2551-1598k+101k^2+76k^3)}{
3(5+k)^2(19+23k)}  W_{-}^{(\frac{5}{2})}
+ \frac{8(41+14k+k^2)}{3(5+k)^3} i \hat{A}_{+} \hat{G}_{22}
\nonu \\
& + &  \frac{8(11+8k+k^2)}{3(5+k)^3} i \hat{A}_{+}  V^{(\frac{3}{2})}
+\frac{40(-7+k)}{3(5+k)^3} i \hat{A}_3 \hat{G}_{12}
+\frac{8(29+k)}{3(5+k)^3} i \hat{A}_3  T_{-}^{(\frac{3}{2})}
\nonu \\
&+ & \frac{4(79+29k+2k^2)}{
3(5+k)^3} i \hat{B}_{+} \hat{G}_{11} 
- \frac{4(19+17k+2k^2)}{
3(5+k)^3} i \hat{B}_{+}  U^{(\frac{3}{2})}
\nonu \\
&- & \frac{8(-12-3k+k^2)}{3(5+k)^3} i \hat{B}_3 \hat{G}_{12}
-\frac{8(17+2k+k^2)}{3(5+k)^3} i \hat{B}_3  T_{-}^{(\frac{3}{2})}
\nonu \\
&+ &  \frac{4(-15+k)}{3(5+k)^2} T^{(1)} \hat{G}_{12}
+\frac{4}{(5+k)} T^{(1)}  T_{-}^{(\frac{3}{2})}
\nonu \\
& + & \frac{4(3695+18604k+15459k^2+3352k^3+218k^4)}{
9(5+k)^3(19+23k)} \pa \hat{G}_{12}
\nonu \\
& - & \left. \frac{4(-3+k)(2743+3328k+625k^2)}{
9(5+k)^3(19+23k)} \pa  T_{-}^{(\frac{3}{2})}
\right](w) 
\nonu \\
& + & \frac{1}{(z-w)^2} \, \left[ 
\frac{2(-3+k)(2611+1413k+182k^2)}{
15(5+k)^2(19+23k)} \pa {\bf P_{-}^{(\frac{5}{2})}}
-\frac{(32+7k)}{2(5+k)}  {\bf S_{-}^{(\frac{7}{2})}}
\right. \nonu \\
& + & \frac{(27+7k)}{(5+k)^2} i {\bf P_{-}^{(\frac{5}{2})}} \hat{A}_3
- \frac{(27+7k)}{(5+k)^2} i {\bf P_{-}^{(\frac{5}{2})}} \hat{B}_3
+\frac{16(-4+k)}{(5+k)^3} \hat{A}_{-} \hat{A}_{+} \hat{G}_{12}
\nonu \\
&+& \frac{4(151+32k)}{
3(5+k)^3} \hat{A}_{-} \hat{A}_{+}  T_{-}^{(\frac{3}{2})}
- \frac{2(121+21k)}{3(5+k)^3} \hat{B}_{+} \hat{A}_{+} \hat{G}_{21}
\nonu \\
&-& \frac{4(32+7k)}{
(5+k)^3} \hat{B}_{+} \hat{A}_{+}  T_{+}^{(\frac{3}{2})}
\nonu \\
&+& \frac{(51+13k)}{(5+k)^2} i \hat{B}_{+}  U^{(\frac{5}{2})}
-\frac{8(32+7k)}{(5+k)^3} \hat{B}_{+} \hat{A}_3 \hat{G}_{11}
-\frac{28(19+5k)}{3(5+k)^3}  
\hat{B}_{+} \hat{A}_3  U^{(\frac{3}{2})}
\nonu \\
& + & \frac{2(15960 k^5+783607 k^4+5999964 k^3+13936928 k^2+9776832 k-949059)}{
45(5+k)^3(19+23k)(47+35k)} 
\nonu \\
& \times & \hat{B}_{+} \hat{B}_{-}  T_{-}^{(\frac{3}{2})}
-\frac{2(1062+385k+53k^2)}{
15(5+k)^3} \hat{B}_{+} \hat{B}_3  U^{(\frac{3}{2})}
\nonu \\
& + &  \frac{4(-152+208k+45k^2)}{
15(5+k)^3} i \hat{B}_{+} \pa \hat{G}_{11}
+\frac{4(-466-30k+11k^2)}{
15(5+k)^3}  i \hat{B}_{+} \pa  U^{(\frac{3}{2})}
\nonu \\
&+ & \frac{14}{3(5+k)^2} \hat{B}_{-} \hat{B}_{+} \hat{G}_{12}  
\nonu \\
& - &  \frac{2(15960 k^5+783607 k^4+5480739 k^3+10335383 k^2
+3831897 k-3695034)}{45(5+k)^3(19+23k)(47+35k)} 
\nonu \\
& \times & \hat{B}_{-} \hat{B}_{+}  T_{-}^{(\frac{3}{2})}
- \frac{6(9+2k)}{(5+k)^2} i \hat{A}_{+}  V^{(\frac{5}{2})}
+\frac{2(1479+905k+116k^2)}{
15(5+k)^3} \hat{A}_{+} \hat{A}_3 V^{(\frac{3}{2})}
\nonu \\
& - & 
\frac{8(32+7k)}{(5+k)^3} \hat{A}_{+} \hat{B}_3 \hat{G}_{22}
+ \frac{56(11+2k)}{3(5+k)^3} \hat{A}_{+} \hat{B}_3 V^{(\frac{3}{2})}
\nonu \\
& + &  \frac{4(910+273k+38k^2)}{
15(5+k)^3} i \hat{A}_{+} \pa \hat{G}_{22}
- \frac{4(833+270k+32k^2)}{
15(5+k)^3} i \hat{A}_{+} \pa  V^{(\frac{3}{2})}
\nonu \\
& - & \frac{2(3019+1185k+116k^2)}{
15(5+k)^3} \hat{A}_3 \hat{A}_{+} V^{(\frac{3}{2})}
+ \frac{2(477+412k+71k^2)}{
(5+k)^2(19+23k)} i \hat{A}_3  W_{-}^{(\frac{5}{2})} 
\nonu \\
& + & \frac{16(-7+k)}{(5+k)^3} 
\hat{A}_3 \hat{A}_3 \hat{G}_{12} 
+  \frac{4(132+25k)}{3(5+k)^3} 
\hat{A}_3 \hat{A}_3   T_{-}^{(\frac{3}{2})}
+\frac{128}{(5+k)^3} \hat{A}_3 \hat{B}_3 \hat{G}_{12}
\nonu \\
&- & \frac{8(132+25k)}{3(5+k)^3}  \hat{A}_3 \hat{B}_3  T_{-}^{(\frac{3}{2})}
+\frac{16}{(5+k)^2} i \hat{A}_3 T^{(1)} \hat{G}_{12}
\nonu \\
& - &  \frac{8(17183+25390k+8137k^2+770k^3)}{
15(5+k)^3(19+23k)} i \hat{A}_3 \pa \hat{G}_{12}
\nonu \\
& - & \frac{4(2414+981k+112k^2)}{
15(5+k)^3} i \hat{A}_3 \pa  T_{-}^{(\frac{3}{2})}
+
\frac{2(1197+1196k+199k^2)}{
(5+k)^2(19+23k)} i \hat{B}_3  W_{-}^{(\frac{5}{2})} 
\nonu \\
&+&  \frac{2(2392+735k+53k^2)}{15(5+k)^3} 
\hat{B}_3 \hat{B}_{+} U^{(\frac{3}{2})}
-\frac{16(1+k)}{(5+k)^3} \hat{B}_3 \hat{B}_3 \hat{G}_{12}
\nonu \\
&+& \frac{4(132+25k)}{
3(5+k)^3}   \hat{B}_3 \hat{B}_3  T_{-}^{(\frac{3}{2})}
-\frac{16}{(5+k)^2} i \hat{B}_3 T^{(1)} \hat{G}_{12}
\nonu \\
& - & \frac{8(-16796-19979k-1544k^2+120k^3+21k^4)}{
15(5+k)^3(19+23k)} i \hat{B}_3 \pa \hat{G}_{12}
\nonu \\
& - & \frac{4(1430+258k+3k^2)}{
15(5+k)^3} i \hat{B}_3 \pa   T_{-}^{(\frac{3}{2})}
-
\frac{20(-3+k)}{
(19+23k)} T^{(1)}  W_{-}^{(\frac{5}{2})} 
\nonu \\
& - &  \frac{4(-1601-1569k-161k^2+7k^3)}{
5(5+k)^2(19+23k)} T^{(1)} \pa \hat{G}_{12}
-\frac{2(22+5k)}{(5+k)^2} T^{(1)} \pa  T_{-}^{(\frac{3}{2})}
\nonu \\
& + & 
\frac{8(5537 k^5+112602 k^4+525937 k^3+808206 k^2+461892 k+75582)}{
3(5+k)^2(3+7k)(19+23k)(47+35k)} \hat{G}_{12} \hat{T}
\nonu \\
& - & \frac{8(-15+k)}{3(5+k)^2} \hat{G}_{12} T^{(2)} 
-\frac{16(4+k)}{(5+k)^2} \hat{G}_{12} W^{(2)}
\nonu \\
& - & \frac{4(41720 k^5+836027 k^4+5457975 k^3+8086225 k^2-493871 k-4726188)}{
45(5+k)^3(19+23k)(47+35k)} \nonu \\
& \times & i \hat{G}_{12} \pa \hat{A}_3
+\frac{4(-29773-31087k-107k^2+130k^3+63k^4)}{
15(5+k)^3(19+23k)} i \hat{G}_{12} \pa \hat{B}_3
\nonu \\
&+& \frac{2(-17259-17381k-1219k^2+63k^3)}{
15(5+k)^2(19+23k)} \hat{G}_{12} \pa T^{(1)}
\nonu \\
&+&  \frac{4(490 k^5+26586 k^4+314641 k^3+1160964 k^2+1145571 k+354960)}{
3(5+k)^2(3+7k)(19+23k)(47+35k)}  T_{-}^{(\frac{3}{2})} \hat{T}
\nonu \\
&- & \frac{2(52+11k)}{(5+k)^2}  T_{-}^{(\frac{3}{2})} T^{(2)}
-\frac{2(4782+2543k+336k^2)}{
15(5+k)^3}  i T_{-}^{(\frac{3}{2})} \pa \hat{A}_3
\nonu \\
& + & \frac{2
}{45(5+k)^3(19+23k)(47+35k)}
(31920 k^5+1540649 k^4 \nonu \\
& + & 11372370 k^3+25883719 k^2+17350020 k
-2701818)   i T_{-}^{(\frac{3}{2})} \pa \hat{B}_3 
\nonu \\
&
+ &\frac{(86+19k)}{(5+k)^2} T_{-}^{(\frac{3}{2})} \pa T^{(1)}
+\frac{50(4+k)}{3(5+k)^2} \hat{G}_{11} V_{-}^{(2)} 
\nonu \\
&- & \frac{2(924+804k+115k^2)}{15(5+k)^3} i \hat{G}_{11}
\pa \hat{B}_{+} + \frac{2(81+17k)}{3(5+k)^2}  U^{(\frac{3}{2})}  V_{-}^{(2)} 
\nonu \\
& - & \frac{2(106+23k)}{3(5+k)^2} \hat{G}_{22}   U_{-}^{(2)}  
+ \frac{2(75+19k)}{3(5+k)^2}  V^{(\frac{3}{2})}    U_{-}^{(2)} 
\nonu \\
&- & \frac{(-20949-18110k-1517k^2+364k^3)}{
15(5+k)^2(19+23k)} \pa  W_{-}^{(\frac{5}{2})} 
\nonu \\
& - & \frac{2(2160+879k+94k^2)}{
15(5+k)^3} \pa \hat{A}_{+} \hat{G}_{22}
\nonu \\
&+ &  \frac{4
}{45(5+k)^3(19+23k)(47+35k)}
(41720 k^5+1078577 k^4 \nonu \\
& + & \left. 8044965 k^3+16746076 k^2
+10714429 k+184701)  i 
\pa \hat{A}_3 \hat{G}_{12} 
\right](w) 
\nonu \\
& + &  \frac{1}{(z-w)} \, 
 \{ W_{-}^{(\frac{5}{2})} \, 
 W^{(3)} \}_{-1}(w) +\cdots, 
\label{w5half-w3} \\
W^{(3)}(z) \, W^{(3)}(w) & = &
\frac{1}{(z-w)^6} \, 
\left[ \frac{8k(436 k^4+5665 k^3+25397 k^2+30515 k+6651)}{
(5+k)^4(19+23k)} \right]  
\nonu \\
& + & \frac{1}{(z-w)^4} \, \left[ 
-\frac{2(-3+k)(-61+7k)}{(5+k)(19+23k)} {\bf P^{(2)}}
\right. \nonu \\
& + & 
\frac{4}{3(5+k)^3(3+7k)(19+23k)^2}
\left(134764 k^6+1883036 k^5+9239691 k^4 \right. \nonu \\
& + & \left. 16022216 k^3 +  9216682 k^2-412572 k-986841 \right)
 \hat{T}
 \nonu \\
& + & \frac{8(-3+k)(-1438-2187k-414k^2+7k^3)}{
3(5+k)^3(19+23k)} T^{(2)}
\nonu \\
&+ &   \frac{8(3950+3052k+1000k^2+133k^3+7k^4)}{
(5+k)^3(19+23k)} W^{(2)}
\nonu \\
& + &  \frac{64(-1617-212k+2034k^2+627k^3+50k^4)}{
3(5+k)^4(19+23k)} \hat{A}_{+} \hat{A}_{-}
\nonu \\
& + &  \frac{8(9200 k^5+142885 k^4+605694 k^3+881935 k^2+661760 k+206826)}{
3(5+k)^4(19+23k)^2} \hat{A}_3 \hat{A}_3
\nonu \\
&- &   \frac{16(4922 k^5+102226 k^4+573765 k^3+878806 k^2+262283 k-74670)}{
3(5+k)^4(19+23k)^2} \hat{A}_3 \hat{B}_3
\nonu \\
& + &  \frac{4(428 k^4+6675 k^3+34401 k^2+28441 k-17313)}{
3(5+k)^4(19+23k)} \hat{B}_{+} \hat{B}_{-}
\nonu \\
& + &  \frac{4(25093 k^5+277883 k^4+1359738 k^3+1631252 k^2+304573 k-328947)}{
3(5+k)^4(19+23k)^2} \nonu \\
& \times & \hat{B}_3 \hat{B}_3
-   \frac{(608 k^4+7295 k^3+38991 k^2+97249 k+24913)}{
(5+k)^2(19+23k)^2} T^{(1)} T^{(1)}
\nonu \\
& + &  \frac{64(50 k^4+627 k^3+2034 k^2-212 k-1617)}{
3(5+k)^4(19+23k)} i \pa \hat{A}_3
\nonu \\
& + &  \frac{4(428 k^4+6675 k^3+34401 k^2+28441 k-17313)}{
3(5+k)^4(19+23k)} i \pa \hat{B}_3
\nonu \\
&+ &   \frac{8(58 k^4+3302 k^3-9909 k^2-67948 k-38075)}{
(5+k)^3(19+23k)^2} i T^{(1)} \hat{A}_3
\nonu \\
&+ & \left.   \frac{4(2043 k^4+7042 k^3+22178 k^2-27142 k+703)}{
(5+k)^3(19+23k)^2} i T^{(1)} \hat{B}_3 \right](w) 
\nonu \\
& + & \frac{1}{(z-w)^3} \, \left[-\frac{(-3+k)(-61+7k)}{(5+k)(19+23k)} 
\pa {\bf P^{(2)}} 
\right. \nonu \\
& - & \frac{(608 k^4+7295 k^3+38991 k^2+97249 k+24913)}{
(5+k)^2(19+23k)^2} T^{(1)} \pa T^{(1)}
\nonu \\
& + & \frac{32(50 k^4+627 k^3+2034 k^2-212 k+1617)}{
3(5+k)^4(19+23k)} \hat{A}_{+} \pa \hat{A}_{-}
\nonu \\
& + & \frac{32(50 k^4+627 k^3+2034 k^2-212 k-1617)}{
3(5+k)^4(19+23k)} \hat{A}_{-} \pa \hat{A}_{+}
\nonu \\
&- & \frac{8(4922 k^5+102226 k^4+573765 k^3+878806 k^2+262283 k-74670)}{
3(5+k)^4(19+23k)^2} \hat{A}_3 \pa \hat{B}_3 
\nonu \\
&+ & \frac{2(34937 k^5+439540 k^4+2277786 k^3+2939014 k^2+446753 k-657894)}{
3(5+k)^4(19+23k)^2}  \nonu \\
&\times & \hat{B}_{+} \pa \hat{B}_{-}
\nonu \\
&- & \frac{2k(5083 k^4+38742 k^3+147230 k^2+107830 k+54131)}{
(5+k)^4(19+23k)^2}  \hat{B}_{-} \pa \hat{B}_{+}
\nonu \\
&+ & \frac{2}{3(5+k)^3(3+7k)(19+23k)^2} \left(134764 k^6+
1883036 k^5+9239691 k^4 \right. \nonu \\
& + & \left. 16022216 k^3
+9216682 k^2-412572 k-986841 \right) \pa \hat{T} 
\nonu \\
&+& \frac{4(-3+k)(7 k^3-414 k^2-2187 k-1438)}{
3(5+k)^3(19+23k)} \pa T^{(2)}
\nonu \\
&+& \frac{4(7 k^4+133 k^3+1000 k^2+3052 k+3950)}{
(5+k)^3(19+23k)} \pa W^{(2)} 
\nonu \\
&+& \frac{2(25093 k^5+277883 k^4+1359738 k^3+1631252 k^2+304573 k-328947)}{
3(5+k)^4(19+23k)^2} 
\nonu \\
&\times & i \pa^2 \hat{B}_3
+ \frac{4(58 k^4+3302 k^3-9909 k^2-67948 k-38075)}{
(5+k)^3(19+23k)^2} i \pa T^{(1)} \hat{A}_3
\nonu \\
&+& \frac{2(2043 k^4+7042 k^3+22178 k^2-27142 k+703)}{
(5+k)^3(19+23k)^2} i \pa T^{(1)} \hat{B}_3
\nonu \\
&+& \frac{4(9200 k^5+142885 k^4+605694 k^3+881935 k^2+661760 k+206826)}{
3(5+k)^4(19+23k)^2} i 
\nonu \\
& \times & \hat{A}_{+} \hat{A}_{-} \hat{A}_3
\nonu \\
& - & \frac{4(4922 k^5+102226 k^4+573765 k^3+878806 k^2+262283 k-74670)}{
3(5+k)^4(19+23k)^2}   i 
\nonu \\
& \times & \hat{A}_{+} \hat{A}_{-} \hat{B}_3
\nonu \\
&- & \frac{4(9200 k^5+142885 k^4+605694 k^3+881935 k^2+661760 k+206826)}{
3(5+k)^4(19+23k)^2}  i 
\nonu \\
& \times & \hat{A}_{-} \hat{A}_{+} \hat{A}_3
\nonu \\
& + & \frac{4(4922 k^5+102226 k^4+573765 k^3+878806 k^2+262283 k-74670)}{
3(5+k)^4(19+23k)^2}  i 
\nonu \\
& \times & \hat{A}_{-} \hat{A}_{+} \hat{B}_3
\nonu \\
&- & \frac{2(25093 k^5+277883 k^4+1359738 k^3+1631252 k^2+304573 k-328947)}{
3(5+k)^4(19+23k)^2}  \nonu \\
& \times & i \hat{B}_{-} \hat{B}_{+} \hat{B}_3 
\nonu \\
&+& \frac{2(25093 k^5+277883 k^4+1359738 k^3+1631252 k^2+304573 k-328947)}{
3(5+k)^4(19+23k)^2}  \nonu \\
& \times & i \hat{B}_3 \hat{B}_{+} \hat{B}_{-}
\nonu \\
& - & \frac{(2043 k^4+7042 k^3+22178 k^2-27142 k+703)}{
(5+k)^3(19+23k)^2}  T^{(1)} \hat{B}_{+} \hat{B}_{-}
\nonu \\
&- & \frac{2(58 k^4+3302 k^3-9909 k^2-67948 k-38075)}{
(5+k)^3(19+23k)^2} \hat{A}_{+} T^{(1)} \hat{A}_{-}
\nonu \\
&+& \frac{2(58 k^4+3302 k^3-9909 k^2-67948 k-38075)}{
(5+k)^3(19+23k)^2}  \hat{A}_{-} T^{(1)} \hat{A}_{+}
\nonu \\
&+& \left. \frac{(2043 k^4+7042 k^3+22178 k^2-27142 k+703)}{
(5+k)^3(19+23k)^2}   \hat{B}_{-} T^{(1)} \hat{B}_{+}
\right](w) 
\nonu \\
& + & \frac{1}{(z-w)^2} \, 
\left[ 
-4   \{ \hat{G}_{21} \, 
S_{-}^{(\frac{7}{2})} \}_{-1}  +\frac{8(3+k)}{(5+k)^2} i \hat{A}_3  
{\bf P^{(3)}}
 -\frac{2(4+k)}{(5+k)^2} i \hat{A}_3 \pa {\bf P^{(2)}}
\right. \nonu \\ 
&- & \frac{2(7+2k)}{(5+k)^2} i \hat{A}_{-} {\bf Q_{-}^{(3)}}
-\frac{8(3+k)}{(5+k)^2} i \hat{B}_3 {\bf P^{(3)}}
+\frac{2(4+k)}{(5+k)^2} i \hat{B}_3 \pa {\bf P^{(2)}}
\nonu \\
&-& \frac{2(7+2k)}{(5+k)^2} i \hat{B}_{-}  {\bf R_{-}^{(3)}}
+\frac{16(-3+k)}{(19+23k)} \left( \hat{T} {\bf P^{(2)}} -
\frac{3}{10} \pa^2 {\bf P^{(2)}} \right)
\nonu \\
& + & \frac{4(6+k)}{(5+k)^2} i \pa \hat{B}_3  {\bf P^{(2)}}
+\frac{2(41+18k+2k^2)}{(5+k)^2} \pa  {\bf S^{(3)}}
-\frac{4(6+k)}{(5+k)^2} i \pa \hat{A}_3  {\bf P^{(2)}}
\nonu \\
& - & \frac{3(-3+k)(-61+7k)}{10(5+k)(19+23k)} \pa^2  {\bf P^{(2)}} 
+\frac{2(-125-16k+2k^2)}{5(5+k)^2} \pa {\bf P^{(3)}}
\nonu \\
&- &  \frac{8}{(5+k)^2} W^{(2)} W^{(2)} 
-\frac{8(131+56k+6k^2)}{(5+k)^2}  U^{(\frac{3}{2})}  \pa V^{(\frac{3}{2})} 
+ \frac{4(19+4k)}{(5+k)^2}   U_{-}^{(2)}  V_{+}^{(2)}
\nonu \\
& + & \frac{4(19+4k)}{(5+k)^2}   U_{+}^{(2)}  V_{-}^{(2)}
-\frac{8(14+3k)}{(5+k)^2} T^{(2)} T^{(2)}
-\frac{8(141+58k+6k^2)}{(5+k)^3}  T_{+}^{(\frac{3}{2})}  \pa T_{-}^{(\frac{3}{2})}
\nonu \\
& - & \frac{16(-69-13k+4k^2)}{(5+k)^2(19+23k)} i \hat{A}_3 W^{(3)} 
-\frac{32(13+3k)}{(5+k)^3} i \hat{A}_3   U^{(\frac{3}{2})}   V^{(\frac{3}{2})}
\nonu \\
& - & \frac{16(7+2k)}{(5+k)^3} \hat{A}_3 \hat{A}_3 W^{(2)} 
+\frac{16(7+2k)}{(5+k)^3} \hat{A}_3 \hat{A}_3 T^{(2)} 
+\frac{32(-7+k)}{(5+k)^4} \hat{A}_3 \hat{A}_3 \hat{A}_3 \hat{A}_3 
\nonu \\
& - & \frac{64(-11+k)}{(5+k)^4} \hat{A}_3 \hat{A}_3 \hat{A}_3 \hat{B}_3 
+ \frac{32}{(5+k)^3} i \hat{A}_3 \hat{A}_3 \hat{A}_3 T^{(1)}
+\frac{80(-59+k)}{7(5+k)^4} i \hat{A}_3 \hat{A}_3 \hat{B}_3 \hat{B}_3
\nonu \\
&-&  \frac{96}{(5+k)^3} i \hat{A}_3 \hat{A}_3 \hat{B}_3 T^{(1)}
-\frac{16(661+97k)}{7(5+k)^4} \hat{A}_3 \hat{A}_3 \hat{B}_{+} \hat{B}_{-}
\nonu \\
&+&  \frac{8(504 k^4-784 k^3-43263 k^2-1059648 k-836705)}{
105(5+k)^4(19+23k)}  
i \hat{A}_3 \hat{A}_3 \pa \hat{B}_{3}
\nonu \\
&+&  \frac{8(-615+589k+452k^2+12k^3)}{5(5+k)^3(19+23k)} \hat{A}_3 \hat{A}_3
\pa T^{(1)}  
+\frac{32(7+2k)}{(5+k)^3} \hat{A}_3 \hat{B}_3 W^{(2)}
\nonu \\
& -& \frac{32(17+2k)}{(5+k)^3} \hat{A}_3 \hat{B}_3 T^{(2)}
+\frac{64}{(5+k)^3} \hat{A}_3 \hat{B}_3 \hat{B}_3 \hat{B}_3 
+\frac{96}{(5+k)^3} i \hat{A}_3 \hat{B}_3 \hat{B}_3 T^{(1)} 
\nonu \\
&-& 
\frac{32(-925-548k+111k^2+6k^3)}{5(5+k)^3(19+23k)} \hat{A}_3 \hat{B}_3
\pa T^{(1)} +\frac{8(41+9k)}{(5+k)^3} \hat{A}_3 \hat{B}_{-}   V_{-}^{(2)}
\nonu \\
& - & \frac{8(31+7k)}{(5+k)^3} \hat{A}_3 \hat{B}_{+}   U_{+}^{(2)}
-\frac{16(5+17k)}{7(5+k)^4} \hat{A}_3 \hat{B}_{+} \hat{B}_{-} \hat{B}_3 
+\frac{32}{(5+k)^3} i \hat{A}_3 \hat{B}_{+} \hat{B}_{-} T^{(1)}
\nonu \\
& -& 
 \frac{8(14735 k^3+93928 k^2+125341 k+71420)}{
105(5+k)^4(19+23k)} i \hat{A}_3 \hat{B}_{+} \pa \hat{B}_{-}
\nonu \\
& - & \frac{32(9+2k)}{(5+k)^3} i \hat{A}_3 \hat{G}_{11}  V^{(\frac{3}{2})} 
+  \frac{64}{(5+k)^2} i \hat{A}_3 \hat{G}_{11} \hat{G}_{22} 
-\frac{8(17+5k)}{(5+k)^3} i \hat{A}_3 \hat{G}_{12}  T_{+}^{(\frac{3}{2})}
\nonu \\
&-& \frac{96}{(5+k)^3} i \hat{A}_3 \hat{G}_{12} \hat{G}_{21}
+\frac{8(1+k)}{(5+k)^3} i \hat{A}_3 \hat{G}_{21}  T_{-}^{(\frac{3}{2})}
-\frac{32(9+2k)}{(5+k)^3} i \hat{A}_3 \hat{G}_{22}  U^{(\frac{3}{2})} 
\nonu \\
& - & \frac{32(56135+89175k+31629k^2-2639k^3+84k^4)}{
105(5+k)^4(19+23k)} i \hat{A}_3 \pa \hat{B}_3 \hat{B}_3 
\nonu \\
& - & \frac{16(-2975-3124k-197k^2+8k^3)}{
5(5+k)^3(19+23k)} \hat{A}_3 \pa \hat{B}_3 T^{(1)}
\nonu \\
&-&  \frac{4(39130 k^3+450446 k^2+1018187 k+603115)}{
105(5+k)^4(19+23k)} i \hat{A}_3 \pa \hat{B}_{+} \hat{B}_{-} 
\nonu \\
& - & \frac{8(-5+6k+2k^2)}{(5+k)^3} i \hat{A}_3 \pa W^{(2)}
+  \frac{8(175+138k+44k^2)}{15(5+k)^3} i \hat{A}_3 \pa T^{(2)}
\nonu \\
& + & \frac{4}{105(5+k)^4(3+7k)(19+23k)^2(47+35k)}
(24345454 k^7+1167474 k^6 \nonu \\
& - & 360049599 k^5+5273994481 k^4+
21719859128 k^3+28463437412 k^2 \nonu \\
& + & 14803288521 k+2557666665) 
\hat{A}_3 \pa^2 \hat{B}_3 
\nonu \\ 
&+ & \frac{2(39930 k^4+507568 k^3+1283113 k^2+841234 k+342255)}{
15(5+k)^3(19+23k)^2} i \hat{A}_3 \pa^2 T^{(1)} 
\nonu \\
&+& \frac{24(9+2k)}{(5+k)^3} i \hat{A}_{-}  T_{-}^{(\frac{3}{2})} U^{(\frac{3}{2})} 
+\frac{48}{(5+k)^3} \hat{A}_{-} \hat{A}_3   U_{-}^{(2)}
- \frac{32(11+2k)}{(5+k)^3} \hat{A}_{-} \hat{B}_3  U_{-}^{(2)}
\nonu \\
&-& \frac{8(5+2k)}{(5+k)^3} i \hat{A}_{-} \hat{G}_{11}  T_{-}^{(\frac{3}{2})}
+\frac{8(8+k)}{(5+k)^3} i \hat{A}_{-} \hat{G}_{11} \hat{G}_{12}
+\frac{8(16+5k)}{(5+k)^3} i \hat{A}_{-} \hat{G}_{12}  U^{(\frac{3}{2})} 
\nonu \\
& - & \frac{8}{(5+k)^2} i \hat{A}_{-} T^{(1)}  U_{-}^{(2)}
-\frac{14(15+4k)}{(5+k)^3} i \hat{A}_{-} \pa U_{-}^{(2)}
-\frac{16(17+3k)}{(5+k)^3} \hat{A}_{+} \hat{A}_3  V_{+}^{(2)}
\nonu \\
& - & \frac{8(-3+k)}{(5+k)^3} \hat{A}_{+} \hat{A}_{-} W^{(2)}
+ \frac{8(48+7k)}{(5+k)^3} \hat{A}_{+} \hat{A}_{-} T^{(2)}
\nonu \\
& + & \frac{16(-33+2k)}{(5+k)^4}  \hat{A}_{+} \hat{A}_{-} \hat{A}_3 \hat{A}_3
- \frac{16(666+121k)}{7(5+k)^4}  
\hat{A}_{+} \hat{A}_{-} \hat{B}_3 \hat{B}_3
\nonu \\
& + & \frac{864}{(5+k)^4}  \hat{A}_{+} \hat{A}_{-} \hat{A}_3 \hat{B}_3 
+\frac{32}{(5+k)^3} i \hat{A}_{+} \hat{A}_{-} \hat{A}_3 T^{(1)}
\nonu \\
&-& \frac{32}{(5+k)^3} i \hat{A}_{+} \hat{A}_{-} \hat{B}_3 T^{(1)}
-\frac{48(106+9k)}{7(5+k)^4}  
\hat{A}_{+} \hat{A}_{-} \hat{B}_{+} \hat{B}_{-}
\nonu \\
& +& \frac{4(7120 k^3+48178 k^2+49807 k+22425)}{
15(5+k)^4(19+23k)} i  \hat{A}_{+} \hat{A}_{-} \pa \hat{A}_3
\nonu \\
&+ & \frac{4(1008 k^4-9310 k^3+56225 k^2-927343 k-885210)}{
105(5+k)^4(19+23k)}  i  \hat{A}_{+} \hat{A}_{-} \pa \hat{B}_3
\nonu \\
& + & \frac{2(-7115-4989k-262k^2+48k^3)}{
5(5+k)^3(19+23k)} \hat{A}_{+} \hat{A}_{-} \pa T^{(1)}
\nonu \\
&-& \frac{304}{(5+k)^4} \hat{A}_{+} \hat{A}_{+} \hat{A}_{-} \hat{A}_{-}
+\frac{16(17+3k)}{(5+k)^3} 
\hat{A}_{+} \hat{B}_3  V_{+}^{(2)}
+\frac{8(13+3k)}{(5+k)^3} i \hat{A}_{+} \hat{G}_{21}  V^{(\frac{3}{2})} 
\nonu \\
&+ & 
\frac{8(27+5k)}{(5+k)^3} i \hat{A}_{+} \hat{G}_{21} \hat{G}_{22}
- \frac{32}{(5+k)^2} i \hat{A}_{+} \hat{G}_{22}  T_{+}^{(\frac{3}{2})}
\nonu \\
& - & \frac{16(880 k^3+4403 k^2+23972 k+18985)}{
15(5+k)^4(19+23k)} i \hat{A}_{+} \pa \hat{A}_{-} \hat{A}_{3}
\nonu \\
&-& \frac{64(6 k^4-295 k^3-710 k^2-3866 k-3325)}{
15(5+k)^4(19+23k)}  i \hat{A}_{+} \pa \hat{A}_{-} \hat{B}_{3}
\nonu \\
& - & \frac{8(-2215-2299k-312k^2+8k^3)}{
5(5+k)^3(19+23k)} \hat{A}_{+} \pa \hat{A}_{-} T^{(1)}
\nonu \\ 
&-&  \frac{16(65+31k+4k^2)}{5(5+k)^3} i \hat{A}_{+} \pa  V_{+}^{(2)}
-\frac{80(-3+k)}{3(5+k)^4} \hat{B}_3 \hat{B}_3 \hat{B}_3 \hat{B}_3 
\nonu \\
& + & \frac{8}{15(5+k)^4(3+7k)(19+23k)(47+35k)}
(90944 k^6+1133349 k^5 \nonu \\
& + & 3586684 k^4 +  1421825 k^3-6479343 k^2
-6630554 k-1180545)  \hat{A}_{+} \pa^2 \hat{A}_{-}
\nonu \\
&+& \frac{16(171+95k+16k^2)}{(5+k)^2(19+23k)}
\hat{B}_3 W^{(3)}
+\frac{32(13+3k)}{(5+k)^3} i
\hat{B}_3  U^{(\frac{3}{2})}   V^{(\frac{3}{2})}
\nonu \\
& - & 
\frac{8(139+33k)}{7(5+k)^3} \hat{B}_3 \hat{B}_3 W^{(2)}
+\frac{8(197+23k)}{7(5+k)^3} \hat{B}_3 \hat{B}_3 T^{(2)}
\nonu \\
& -& \frac{16(83+5k)}{21(5+k)^3} i \hat{B}_3  \hat{B}_3 \hat{B}_3  T^{(1)}
+\frac{8(-3085-2781k-8k^2+12k^3)}{
5(5+k)^3(19+23k)}  \hat{B}_3 \hat{B}_3 \pa T^{(1)}
\nonu \\
& + & \frac{32(9+2k)}{(5+k)^3} i \hat{B}_3 \hat{G}_{11}   V^{(\frac{3}{2})} 
-  \frac{64}{(5+k)^2} i \hat{B}_3 \hat{G}_{11} \hat{G}_{22}
+\frac{8(17+5k)}{(5+k)^3} i \hat{B}_3 \hat{G}_{12}  T_{+}^{(\frac{3}{2})}
\nonu \\
&+& \frac{128}{(5+k)^3} i \hat{B}_3 \hat{G}_{12} \hat{G}_{21} 
-\frac{8(1+k)}{(5+k)^3} i \hat{B}_3 \hat{G}_{21}  T_{-}^{(\frac{3}{2})}
+\frac{32(9+2k)}{(5+k)^3} 
i \hat{B}_3 \hat{G}_{22}  U^{(\frac{3}{2})} 
\nonu \\
& - & \frac{8(2+7k+2k^2)}{(5+k)^3} i \hat{B}_3 \pa W^{(2)}
+ \frac{8(395+315k+19k^2)}{15(5+k)^3} i \hat{B}_3 \pa T^{(2)} 
\nonu \\
& - & 
\frac{(68901 k^4+1303622 k^3+4900034 k^2+6146006 k+1484565)}{
15(5+k)^3(19+23k)^2} i \hat{B}_3 \pa^2 T^{(1)}
\nonu \\
&- & 
\frac{24(9+2k)}{(5+k)^3} i \hat{B}_{-}  T_{-}^{(\frac{3}{2})}  V^{(\frac{3}{2})} 
-\frac{8(-41+9k)}{7(5+k)^3} \hat{B}_{-} \hat{B}_3 V_{-}^{(2)}
\nonu \\
& - & \frac{8(29+5k)}{(5+k)^3} i \hat{B}_{-} \hat{G}_{12}  V^{(\frac{3}{2})} 
 -  \frac{8(-31+k)}{7(5+k)^3} \hat{B}_{+} \hat{B}_{-} W^{(2)}
\nonu \\
&+& \frac{4(-3+k)}{(5+k)^3} i \hat{B}_{-} \hat{G}_{12} \hat{G}_{22}
+\frac{8(25+6k)}{(5+k)^3} i \hat{B}_{-} \hat{G}_{22}  T_{-}^{(\frac{3}{2})}
-\frac{8}{(5+k)^2} i \hat{B}_{-} T^{(1)}  V_{-}^{(2)}
\nonu \\
&- & \frac{2(-29-4k+4k^2)}{(5+k)^3} i \hat{B}_{-} \pa V_{-}^{(2)}
+ \frac{352(4+k)}{7(5+k)^3} \hat{B}_{+} \hat{B}_3  U_{+}^{(2)}
\nonu \\
& + & \frac{8(150+13k)}{7(5+k)^3} 
\hat{B}_{+} \hat{B}_{-} T^{(2)}
+\frac{8(13+5k)}{7(5+k)^3} i \hat{B}_{+} \hat{B}_{-} \hat{B}_3 T^{(1)}
\nonu \\
&+&  \frac{4(144 k^4-2434 k^3-16537 k^2-54829 k-46170)}{
15(5+k)^4(19+23k)} 
i  \hat{B}_{+} \hat{B}_{-} \pa \hat{B}_3
\nonu \\
&+& \frac{2(-7875-5719k-32k^2+48k^3)}{5(5+k)^3(19+23k)} \hat{B}_{+} 
\hat{B}_{-} \pa T^{(1)}
-  \frac{32}{(5+k)^2} i \hat{B}_{+} \hat{G}_{11}   T_{+}^{(\frac{3}{2})}
\nonu \\
& - & \frac{12(19+3k)}{(5+k)^3} 
i \hat{B}_{+} \hat{G}_{11} \hat{G}_{21} 
-\frac{8(13+3k)}{(5+k)^3} i \hat{B}_{+} \hat{G}_{21}  U^{(\frac{3}{2})}
\nonu \\
& -& \frac{16(168 k^4+917 k^3-4019 k^2-106928 k-97660)}{105(5+k)^4(19+23k)} i   
\hat{B}_{+} \pa \hat{B}_{-} \hat{B}_3
\nonu \\
&-& \frac{8(-5245-443k+1726k^2+56k^3)}{
35(5+k)^3(19+23k)} \hat{B}_{+} \pa \hat{B}_{-} 
T^{(1)}
\nonu \\
&-&  \frac{8(-50-16k+k^2)}{5(5+k)^3} i \hat{B}_{+} \pa U_{+}^{(2)}
\nonu \\
& - & 
\frac{1}{105(5+k)^4(3+7k)(19+23k)(47+35k)}(6781796 k^6+20616813 k^5
\nonu \\
& - & 259343187 k^4-865472370 k^3-870132206 k^2
-313763163 k-45693315) \nonu \\
& \times & \hat{B}_{+} \pa^2 \hat{B}_{-}  
- \frac{8}{(5+k)^2} \hat{G}_{11}  V^{(\frac{5}{2})}
-\frac{8(4+k)(175+36k)}{15(5+k)^3}
\hat{G}_{11} \pa  V^{(\frac{3}{2})} 
\nonu \\
& - & \frac{4(135+443k+72k^2)}{15(5+k)^3} \hat{G}_{11} \pa \hat{G}_{22}
+\frac{4(24+5k)}{(5+k)^2} \hat{G}_{12} W_{+}^{(\frac{5}{2})}
\nonu \\
&-&  \frac{4(72 k^4-1648 k^3-31821 k^2-129290 k-98085)}{15(5+k)^3(19+23k)} 
\hat{G}_{12} \pa  T_{+}^{(\frac{3}{2})}
\nonu \\
& - & \frac{4(7840 k^5+142064 k^4+838985 k^3+1087237 k^2-215121 k-575805)}{
15(5+k)^3(19+23k)(47+35k)} \nonu \\
& \times & \hat{G}_{12} \pa \hat{G}_{21}
-\frac{48(-3+k)}{(19+23k)} T^{(1)} W^{(3)}
\nonu \\
& + & \frac{4(22+5k)}{(5+k)^2} \hat{G}_{21}  W_{-}^{(\frac{5}{2})}
-\frac{4(856 k^3+19549 k^2+83698 k+48285)}{15(5+k)^3(19+23k)} 
\hat{G}_{21} \pa  T_{-}^{(\frac{3}{2})}
\nonu \\
&+& \frac{8}{(5+k)^2} \hat{G}_{22}  U^{(\frac{5}{2})}
+\frac{8(955+422k+48k^2)}{15(5+k)^3} \hat{G}_{22} \pa  U^{(\frac{3}{2})} 
\nonu \\
&-& \frac{4}{(5+k)} T^{(1)} \pa T^{(2)}
-\frac{16(-131+13k+32k^2+4k^3)}{(5+k)^2(19+23k)} 
\left( \hat{T} W^{(2)} -\frac{3}{10} \pa^2 W^{(2)} \right)
\nonu \\
& -& \frac{16(840 k^5+9460 k^4-10235 k^3-223163 k^2-288861 k-112617)}{
3(5+k)^2(3+7k)(19+23k)(47+35k)}
\nonu \\
& \times & \left( \hat{T} T^{(2)} -\frac{3}{10} \pa^2 T^{(2)} \right)
\nonu \\
&+ & \frac{32}{(5+k)^2(3+7k)^2(19+23k)^2
(47+35k)}(253708 k^7+3623943 k^6 \nonu \\
& + & 18746450 k^5+42987689 k^4+
46681491 k^3+24246670 k^2+5396307 k \nonu \\
& + & 276678) 
\left( \hat{T} \hat{T} -\frac{3}{10} \pa^2 \hat{T} \right)
\nonu \\
&+& \frac{32}{(5+k)^3(3+7k)(19+23k)^2(47+35k)}
(36064 k^6+422800 k^5\nonu \\
& + & 1683592 k^4+3851921 k^3+4848911 k^2+
2749739 k+484629) \nonu \\
& \times &  \left( \hat{T} \hat{A}_3 \hat{A}_3 -\frac{3}{10} 
\pa^2 (\hat{A}_3 \hat{A}_3) \right)
\nonu \\
&- &  \frac{64}{(5+k)^3(3+7k)(19+23k)^2(47+35k)}(36064 k^6+920150 k^5
\nonu \\ 
& + & 5585952 k^4+11939749 k^3+10938615 k^2+
4180289 k+491397)\nonu \\ 
& \times & \left( \hat{T} \hat{A}_3 \hat{B}_3 -\frac{3}{10} 
\pa^2 (\hat{A}_3 \hat{B}_3) \right)
\nonu \\
& + & \frac{64(-173-81k+56k^2)}{
(5+k)(19+23k)^2} i \left( \hat{T} T^{(1)} \hat{A}_3 -\frac{3}{10} 
\pa^2 ( T^{(1)} \hat{A}_3) \right)
\nonu \\
&+& \frac{16}{7(5+k)^3(3+7k)(19+23k)^2(47+35)}
(504896 k^6+7181300 k^5 \nonu \\
& + & 32866093 k^4+59240826 k^3+50421488 k^2+
20801666 k+2665491) \nonu \\
& \times &
\left( \hat{T} \hat{B}_3 \hat{B}_3 -\frac{3}{10} 
\pa^2 (\hat{B}_3 \hat{B}_3) \right)
\nonu \\
& - & \frac{64(-133-3k+6k^2)}{
(5+k)(19+23k)^2} i \left( \hat{T} T^{(1)} \hat{B}_3 -\frac{3}{10} 
\pa^2 ( T^{(1)} \hat{B}_3) \right)
\nonu \\
 &+& \frac{16(3136 k^5+51177 k^4+216244 k^3+322771 k^2+173238 k+13374)}{
(5+k)^3(3+7k)(19+23k)(47+35k)} 
\nonu \\
& \times & \hat{T} \hat{A}_{+} \hat{A}_{-}
\nonu \\
&-& \frac{24(3136 k^5+51177 k^4+216244 k^3+322771 k^2+173238 k+13374)}{
5(5+k)^3(3+7k)(19+23k)(47+35k)} \nonu \\
& \times & \pa^2 (\hat{A}_{+} \hat{A}_{-})
\nonu \\
&+& \frac{16(21952 k^5+383264 k^4+1713738 k^3+2327307 k^2+978656 k+59919)}{
7(5+k)^3(3+7k)(19+23k)(47+35k)} 
\nonu \\
&\times & \hat{T} \hat{B}_{+} \hat{B}_{-}
\nonu \\
&-& \frac{24(21952 k^5+383264 k^4+1713738 k^3+2327307 k^2+978656 k+59919)}{
35(5+k)^2(3+7k)(19+23k)(47+35k)} 
\nonu \\
&\times & \pa^2 (\hat{B}_{+} \hat{B}_{-})
- \frac{128(-3+k)^2}{(19+23k)^2} 
\left( \hat{T} T^{(1)} T^{(1)} -\frac{3}{10} \pa^2 (T^{(1)} T^{(1)}) \right)
\nonu \\
& + & \frac{32(75460 k^5+742190 k^4+3447221 k^3+7379728 k^2+7096893 k+
2043720)}{
15(5+k)^3(3+7k)(19+23k)(47+35k)} i \nonu \\
& \times & \hat{T} \pa \hat{A}_3
\nonu \\
& - & \frac{8}{105(5+k)^3(3+7k)(19+23k)(47+35k)}
(160720 k^6+2333072 k^5 \nonu \\
& - & 3965934 k^4-40462789 k^3-33400835 k^2+
29540049 k+20987325)   i \hat{T} \pa \hat{B}_3
\nonu \\
&+ & \frac{8(9520 k^5-34056 k^4-757491 k^3-2391097 k^2-2915481 k-908955)}{
15(5+k)^2(3+7k)(19+23k)(47+35k)} \nonu \\
& \times & \hat{T} \pa T^{(1)}
 +  \frac{8(169+64k+6k^2)}{(5+k)^3} 
\pa U^{(\frac{3}{2})} V^{(\frac{3}{2})}  
+  \frac{8(159+62k+6k^2)}{(5+k)^3} 
\pa  T_{+}^{(\frac{3}{2})}  T_{-}^{(\frac{3}{2})}
\nonu \\
& + & \frac{8(103+23k)}{(5+k)^3} i \pa \hat{A}_3 W^{(2)}
-\frac{8(825+607k+116k^2)}{15(5+k)^3} i \pa \hat{A}_3 T^{(2)}
\nonu \\
&+& \frac{16(20 k^3+8931 k^2+37779 k+24280)}{15(5+k)^4(19+23k)} 
i \pa \hat{A}_3 \hat{A}_3 \hat{A}_3 
\nonu \\
& - & \frac{32(24 k^4+1376 k^3+19107 k^2+39852 k+19705)}{
15(5+k)^4(19+23k)} i \pa \hat{A}_3 \hat{A}_3 \hat{B}_3 
\nonu \\
& - & \frac{16(-2595-2759k-312k^2+8k^3)}{5(5+k)^3(19+23k)} 
\pa \hat{A}_3 \hat{A}_3 T^{(1)} 
\nonu \\
& + & \frac{16(336 k^4+14294 k^3+106116 k^2-55725 k-117815)}{
105(5+k)^4(19+23k)} 
i \pa \hat{A}_3 \hat{B}_3 \hat{B}_3 
\nonu \\
& + & \frac{16(-1930-1859k-197k^2+8k^3)}{5(5+k)^3(19+23k)} 
\pa \hat{A}_3 \hat{B}_3 T^{(1)} 
\nonu \\
& + & \frac{4(69160 k^3+507214 k^2+529693 k+201435)}{
105(5+k)^4(19+23k)} i \pa \hat{A}_3 \hat{B}_{+} \hat{B}_{-}
\nonu \\
& + & \frac{4}{15(5+k)^4(3+7k)(19+23k)^2(47+35k)}
(8366848 k^7+148724107 k^6\nonu \\
& + & 1146036918 k^5 
 +  3566435038 k^4+
5695057128 k^3+4992746971 k^2
\nonu \\
& + & 2201162282 k+351698820)
 \pa \hat{A}_3 \pa \hat{A}_3
\nonu \\
&-& \frac{8}{
105(5+k)^4(3+7k)(19+23k)^2(47+35k)} (30404206 k^7+907869158 k^6
\nonu \\
& + & 8638455231 k^5+30587128586 k^4+49942627344 k^3+39751696406 k^2
\nonu \\ & + & 14337012683 k+1733997570) 
\pa \hat{A}_3 \pa \hat{B}_3
\nonu \\
& - &  \frac{8(7143 k^4+67298 k^3+144833 k^2+166976 k+102090)}{
5(5+k)^3(19+23k)^2} i \pa \hat{A}_3 \pa T^{(1)}
\nonu \\
& - & \frac{4(49+10k)}{(5+k)^3} i \pa \hat{A}_{-}  U_{-}^{(2)}
+\frac{4(725+146k+4k^2)}{5(5+k)^3} 
i \pa \hat{A}_{+}  V_{+}^{(2)}
\nonu \\
&-& \frac{4(3520 k^3+22442 k^2+9143 k+985)}{15(5+k)^4(19+23k)} i
\pa \hat{A}_{+} \hat{A}_{-} \hat{A}_3
\nonu \\
& - & \frac{4(96 k^4+2870 k^3+40105 k^2+143839 k+85880)}{
15(5+k)^4(19+23k)} i \pa \hat{A}_{+} \hat{A}_{-} \hat{B}_3
\nonu \\
& - & \frac{2(-6675-5981k-558k^2+32k^3)}{
5(5+k)^3(19+23k)} \pa \hat{A}_{+} \hat{A}_{-} T^{(1)}
\nonu \\
& + & \frac{4}{15(5+k)^4(3+7k)(19+23k)(47+35k)}
(363776 k^6+9155076 k^5 \nonu \\
& + & 67102712 k^4+171907699 k^3+
187798133 k^2+84736109 k+10434615) 
\nonu \\
& \times &  \pa \hat{A}_{+} \pa \hat{A}_{-}
\nonu \\
& - &  \frac{12(341+101k)}{7(5+k)^3} i \pa \hat{B}_3 W^{(2)}
+\frac{4(15735+4835k+1036k^2)}{105(5+k)^3} 
i \pa \hat{B}_3 T^{(2)}
\nonu \\
&-& \frac{8(24 k^4+16 k^3-8637 k^2-38664 k-27835)}{
15(5+k)^4(19+23k)} i \pa \hat{B}_3 \hat{B}_3 \hat{B}_3 
\nonu \\
& - & 
\frac{8(112 k^3-573 k^2-25946 k-28445)}{
35(5+k)^3(19+23k)} \pa \hat{B}_3 \hat{B}_3 T^{(1)}
\nonu \\
& -& \frac{2}{
105(5+k)^4(3+7k)(19+23k)^2(47+35k)}
(11360160 k^8+221786789 k^7 \nonu \\
& + & 2877632877 k^6 
 +  14682323833 k^5+40793175743 k^4+62689641815 k^3
\nonu \\ 
& + & 52229962795 k^2
+  21884867323 k+3327709305) \pa \hat{B}_3  \pa \hat{B}_3 
\nonu \\
& - & \frac{2(1104 k^5-10497 k^4-185978 k^3-270492 k^2+164898 k+432725)}{
5(5+k)^3(19+23k)^2} i 
\nonu \\
& \times & \pa \hat{B}_3  \pa T^{(1)}
\nonu \\
 & + & \frac{4(-71+124k+56k^2)}{7(5+k)^3} 
i \pa \hat{B}_{-}  V_{-}^{(2)}
+ \frac{4(-1265-206k+56k^2)}{35(5+k)^3} i \pa \hat{B}_{+}  U_{+}^{(2)}
\nonu \\
& - & \frac{4(672 k^4-1162 k^3-108041 k^2-417587 k-322240)}{
105(5+k)^4(19+23k)} 
i \pa \hat{B}_{+} \hat{B}_{-} \hat{B}_3
\nonu \\
&- & \frac{2(224 k^3-5056 k^2-34537 k-39885)}{35(5+k)^3(19+23k)} 
\pa \hat{B}_{+} \hat{B}_{-} T^{(1)} 
\nonu \\
& + & 
\frac{2}{105(5+k)^4(3+7k)(19+23k)(47+35k)}
(15979684 k^6+207360573 k^5 \nonu \\
&+ & 
937708201 k^4+1999113788 k^3+1951577980 k^2+740231791 k+69536415) \nonu \\
& \times & \pa \hat{B}_{+} \pa \hat{B}_{-}
\nonu \\
& + & \frac{8(-20+37k+8k^2)}{5(5+k)^3} 
\pa \hat{G}_{11}  V^{(\frac{3}{2})} 
+\frac{4(285+337k+48k^2)}{15(5+k)^3} \pa \hat{G}_{11} \hat{G}_{22}
\nonu \\
& + & \frac{4(48 k^4-592 k^3-22129 k^2-106450 k-94545)}{
15(5+k)^3(19+23k)} \pa \hat{G}_{12} T_{+}^{(\frac{3}{2})}
\nonu \\
& - & \frac{4(48 k^4+80 k^3-12591 k^2-91572 k-73805)}{
15(5+k)^3(19+23k)} \pa \hat{G}_{12} \hat{G}_{21}
\nonu \\
& + & \frac{4(200 k^5-13680 k^4-67217 k^3+188537 k^2+1238029 k+775035)}{
15(5+k)^3(19+23k)(47+35k)} \pa \hat{G}_{21}  T_{-}^{(\frac{3}{2})}
\nonu \\
&-& \frac{8(115+36k+4k^2)}{5(5+k)^3} 
\pa \hat{G}_{22}  U^{(\frac{3}{2})} 
+\frac{8(4+k)}{(5+k)^2} \pa T^{(1)} T^{(2)}
\nonu \\
& - & \frac{3(256 k^4-12201 k^3-99921 k^2-142375 k+8865)}{
10(5+k)^2(19+23k)^2} \pa T^{(1)} \pa T^{(1)}
\nonu \\
& - & \frac{16(3250 k^4+26924 k^3+48197 k^2-53093 k-28140)}{
15(5+k)^3(3+7k)(19+23k)} i 
\pa \hat{T} \hat{A}_{3}
\nonu \\
&+ & \frac{16(336 k^5+6628 k^4+45503 k^3+147215 k^2+151582 k+38190)}{
15(5+k)^3(3+7k)(19+23k)} i \pa \hat{T} \hat{B}_3 
\nonu \\
&- & \frac{8(56 k^4-604 k^3-7283 k^2-12564 k-4365)}{
5(5+k)^2(3+7k)(19+23k)} \pa \hat{T} T^{(1)}
\nonu \\
& + & \frac{4}{15(5+k)^4(3+7k)(19+23k)^2(47+35k)}
(8366848 k^7+193232907 k^6 \nonu \\
& + & 1019795928 k^5
+  1850940023 k^4+1030164808 k^3-443719079 k^2
\nonu \\
& - & 594260288 k-128783355) 
\pa^2 \hat{A}_3 \hat{A}_3
\nonu \\
&- & \frac{8}{
15(5+k)^4(3+7k)(19+23k)^2(47+35k)} (7547519 k^7+108461955 k^6
\nonu \\
& + & 530755252 k^5+1041212262 k^4+765694639 k^3+13514031 k^2
\nonu \\
& -& 167340802 k-35939640) \pa^2 \hat{A}_3 \hat{B}_3 
\nonu \\
& + & \frac{4(3159 k^4+77191 k^3+292596 k^2+228519 k+73715)}{
5(5+k)^3(19+23k)^2} i \pa^2 \hat{A}_3 T^{(1)}
\nonu \\
&+& \frac{2}{15(5+k)^4(3+7k)(19+23k)(47+35k)}
(363776 k^6+6162156 k^5 \nonu \\
& + & 29832318 k^4+50898513 k^3+32334403 k^2+
7566879 k+2248155)
\nonu \\
& \times & \pa^2 \hat{A}_{+} \hat{A}_{-}
\nonu \\
& - & \frac{2}{
105(5+k)^4(3+7k)(19+23k)^2(47+35k)} (7573440 k^8-104761069 k^7
\nonu \\
& - & 1803734457 k^6-15086499303 k^5-39975895483 k^4-43518293555 k^3
\nonu \\ 
& - & 17597053655 k^2+845252007 k+1168885035) 
\pa^2 \hat{B}_{3} \hat{B}_3
\nonu \\
&+& \frac{(30912 k^5+617421 k^4+1228270 k^3+4971332 k^2+6835346 k+9331375)}{
105(5+k)^3(19+23k)^2} i 
\nonu \\
& \times & \pa^2 \hat{B}_3 T^{(1)}
\nonu \\
& - & \frac{1}{105(5+k)^4(3+7k)(19+23k)(47+35k)}
(3928036 k^6+11205481 k^5 \nonu \\
& - & 268259605 k^4-934591616 k^3-
1021849884 k^2-410272289 k-64446315)
\nonu \\
& \times & \pa^2 \hat{B}_{+} \hat{B}_{-}
\nonu \\
& - & \frac{(2528 k^4+60877 k^3+394797 k^2+770995 k+106835)}{
10(5+k)^2(19+23k)^2} \pa^2 T^{(1)} T^{(1)}
\nonu \\
& - & \frac{4(10+33k+6k^2)}{5(5+k)^2} \pa W^{(3)}
\nonu \\
& + & 
\frac{2(21 k^4+261 k^3+2426 k^2+8201 k+11375)}{5(5+k)^3(19+23k)} \pa^2 W^{(2)}
\nonu \\
&- & \frac{2}{15(5+k)^3(3+7k)(19+23k)(47+35k)}
(4095 k^6+444673 k^5-1692586 k^4 \nonu \\
& - & 27257131 k^3-61412456 k^2-49615602 k-13640265) \pa^2 T^{(2)}
\nonu \\
& + & \frac{1}{
15(5+k)^3(3+7k)^2(19+23k)^2(47+35k)} (118969172 k^8+2478091972 k^7
\nonu \\
& + & 19051798253 k^6+65162953166 k^5+110920904947 k^4+98781173400 k^3
\nonu \\
& + & 44993496739 k^2+9138891030 k+479838825
) \pa^2 \hat{T}
\nonu \\
& + & \frac{4}{45(5+k)^4(3+7k)(19+23k)(47+35k)} (387800 k^6+11482590 k^5
\nonu \\
& + & 74103127 k^4+108578095 k^3-61846821 k^2-196076785 k-70068270) i  
\pa^3 \hat{A}_3
\nonu \\
&+ & \frac{1}{315(5+k)^4(3+7k)(19+23k)(47+35k)}(
2489200 k^7+70731668 k^6 \nonu \\
& + & 576838129 k^5
+  1846274021 k^4+1985160702 k^3\nonu \\
& + & 754936442 k^2+
804073729 k+523778925) i \pa^3 \hat{B}_3  
\nonu \\
&-&  \frac{2}{
15(5+k)^3(3+7k)(19+23k)(47+35k)} (5460 k^6+441072 k^5
\nonu \\
& + & \left.
4165651 k^4+14082572 k^3+18356678 k^2+6588412 k+259275) \pa^3 T^{(1)} 
\right](w)  
\nonu \\
& + &  \frac{1}{(z-w)} \, 
 \{ W^{(3)} \, 
 W^{(3)} \}_{-1}(w) +\cdots. 
\label{finalope}
\eea
In the second OPE of (\ref{finalope}), the $(k-3)$ factor appears 
in $T_{+}^{(\frac{3}{2})}(w)$ and $P^{(\frac{5}{2})}(w)$ (and their descendant 
fields). Similar behavior appears in the third OPE.
The first order pole in the fourth[sixth]  OPE of (\ref{finalope})
contains a composite field with spin-$4$ with vanishing $U(1)$ charge.
One sees also $(k-3)$ factor in the several places of singular terms.  
The first order pole in the fifth  OPE of (\ref{finalope})
contains a composite field with spin-$4$ with vanishing $U(1)$ charge of
$\frac{2(3+k)}{(5+k)}$.
In the fifth OPE, the expression $\hat{G}_{21} \hat{G}_{21}(w)$ can be 
written in terms of the derivative of $\hat{A}_{-} \hat{B}_{-}(w)$
according to the OPE $\hat{G}_{21}(z) \, \hat{G}_{21}(w)$.
Also the sixth OPE has a term  of first order pole in 
(\ref{g21s7half-}) where 
the higher spin-$4$ current was constructed.
The $(k-3)$ factor appears in $T^{(1)}(w)$, $T^{(2)}(w)$, $P^{(2)}(w)$ and 
$P^{(3)}(w)$ terms.

The first order pole in the seventh  OPE of (\ref{finalope})
contains a composite field with spin-$\frac{9}{2}$ with 
$U(1)$ charge of
$\frac{(3+k)}{(5+k)}$ and Appendix $L$ contains this composite field 
explicitly.
The $(k-3)$ factor appears in $T_{+}^{(\frac{3}{2})}(w)$ and 
$P_{+}^{(\frac{5}{2})}(w)$.
The $13$ composite fields of spin-$\frac{5}{2}$ in the third order pole 
are seen from the Table $4$ of \cite{Ahn1311}.
The first order pole in the eighth  OPE 
contains a composite field with spin-$4$ with $U(1)$ charge of
$-\frac{2(3+k)}{(5+k)}$.
The expression $\hat{G}_{12} \hat{G}_{12}(w)$ can be 
written in terms of the derivative of $\hat{A}_{+} \hat{B}_{+}(w)$
as before.
The first order pole in the ninth  OPE 
contains a composite field with spin-$\frac{9}{2}$ with 
$U(1)$ charge of
$-\frac{(3+k)}{(5+k)}$ which will appear in 
Appendix $L$.
Also the $13$ composite fields of spin-$\frac{5}{2}$ in the third order pole 
are seen from the Table $4$ of \cite{Ahn1311}.
The $(k-3)$ factor appears in $T_{-}^{(\frac{3}{2})}(w)$ and 
$P_{-}^{(\frac{5}{2})}(w)$.
The second order pole in the last  OPE of (\ref{finalope})
contains a composite field with spin-$4$ with vanishing $U(1)$ charge.
Also the last OPE has a term of the first order pole in 
(\ref{g21s7half-}) where 
the higher spin-$4$ current was constructed.
The $(k-3)$ factor appears in $P^{(2)}(w)$ and $T^{(2)}(w)$ terms.

\section{The composite fields with spin-$\frac{9}{2}$}
The first order singular term in the last OPE of 
(\ref{u5halfw3}) can be summarized by
the following complicated expression
\bea
& & \{ U^{(\frac{5}{2})} \, 
 W^{(3)} \}_{-1}(w)  = 
\left[ 
\frac{2 i}{(k+5)} \, \, \hat{A}_3 {\bf Q^{(\frac{7}{2})}} \right.
\nonu \\
&& +\frac{8}{(k+5)^2 
d(k)} 
(140 k^4+35816 k^3+133397 k^2+128682 k+39105)
\hat{A}_3 \, \, \hat{B}_3 {\bf Q^{(\frac{5}{2})}}
\nonu \\
&& + \frac{4}{(k+5)^2 d(k)} 
(140 k^4+35816 k^3+133397 k^2+128682 k+39105)
\, \, \hat{A}_{+} \hat{B}_3 {\bf P_{+}^{(\frac{5}{2})}}
\nonu \\
&&
-\frac{2}{(k+5)^2 d(k)} 
(140 k^4+35816 k^3+133397 k^2+128682 k+39105)
\, \, \hat{A}_{+} \hat{B}_{-} {\bf R^{(\frac{5}{2})}}
\nonu \\
&& -\frac{4 i }{(k+5) d(k)} (140 k^4+16496 k^3+77003 k^2+75822 k+23031)
\, \, \hat{B}_3 {\bf Q^{(\frac{7}{2})}}
\nonu \\
&& +
\frac{4}{(k+5)^2 d(k)} 
(140 k^5+36376 k^4+218701 k^3+493088 k^2+395253 k+108198)
\, \, \hat{B}_3 \hat{B}_3 {\bf Q^{(\frac{5}{2})}}
\nonu \\
&& +
\frac{i}{(k+5) d(k)} (140 k^4+35816 k^3+133397 k^2+128682 k+39105)
\, \, \hat{B}_{3} \hat{G}_{11}  {\bf   P^{(2)} }
\nonu \\
&& + \frac{12 i}{(k+5) d(k)} (4 k+3) (23 k+19) (35 k+47)
\, \, \hat{B}_{-} {\bf S_{-}^{(\frac{7}{2})}}
\nonu \\
&& + \frac{16}{(k+5)^2 d(k)} (70 k^4+3418 k^3+24403 k^2+24696 k+7497)
\, \,\hat{B}_{-} \hat{B}_3 {\bf P_{-}^{(\frac{5}{2})}}
\nonu \\
&&  -\frac{i}{2 (k+5) d(k)}
(140 k^4+35816 k^3+133397 k^2+128682 k+39105)
\,\, \hat{B}_{-} \hat{G}_{12} {\bf P^{(2)}}
 \nonu \\
&& -\frac{2}{(k+5)^2 d(k)} 
(140 k^5+36516 k^4+254517 k^3+626485 k^2+523935 k+147303)
\,\, \hat{B}_{+} \hat{B}_{-} {\bf Q^{(\frac{5}{2})}}
\nonu \\
&&+
\frac{1}{(k+5) d(k)} (140k^4+35816 k^3+133397 k^2+128682 k+39105 )
\,\, \hat{G}_{11}  {\bf   S^{(3)} }
\nonu \\
&& +
\frac{1}{(k+5) d(k)} (140k^4+35816 k^3+133397 k^2+128682 k+39105) 
\,\, \hat{G}_{11}  {\bf   P^{(3)} }
\nonu \\
&&
-\frac{1}{(k+5) d(k)} (140 k^4+35816 k^3+133397 k^2+128682 k+39105 )
\,\, \hat{G}_{12} {\bf Q_{+}^{(3)}}
\nonu \\
&& +
\frac{8 (k-3)}{(23 k+19)} \hat{T}  \,\, {\bf Q^{(\frac{5}{2})}} 
-\frac{i}{k+5} \,\, \pa \hat{A}_3  {\bf Q^{(\frac{5}{2})}} 
+ \frac{i (7 k+11)}{5 (k+5)^2} \,\, \hat{A}_3  \pa {\bf Q^{(\frac{5}{2})}} 
\nonu \\
&& +
\frac{2 i
}{(k+5)^2 d(k)}
 (140 k^4+35816 k^3+133397 k^2+128682 k+39105)
\,\, \hat{A}_{+} \pa {\bf P_{+}^{(\frac{5}{2})}}
\nonu \\
&& 
-\frac{i 
}{(k+5)^2 d(k)}
 (140 k^5+74876 k^4+430913 k^3+860169 k^2+668739 k+181611)
\,\, \pa \hat{B}_3   {\bf Q^{(\frac{5}{2})}} 
\nonu \\
&&
-\frac{i 
}{5 (k+5)^2 d(k)}
(2660 k^5+419404 k^4+2560459 k^3+6018209 k^2+4805133 k+1301967)
\nonu \\
& \times &  \hat{B}_3   \pa {\bf Q^{(\frac{5}{2})}} 
\nonu \\
&& 
-\frac{i
}{(k+5)^2 d(k)}
(280 k^5-5228 k^4-160454 k^3-456189 k^2-445800 k-139869)
\,\, \pa \hat{B}_{-} {\bf P_{-}^{(\frac{5}{2})}}
\nonu \\
&& +
\frac{3 i
}{5 (k+5)^2 d(k)}
(50820 k^4+525144 k^3+1090603 k^2+914774 k+265263)
\,\, \hat{B}_{-} \pa {\bf P_{-}^{(\frac{5}{2})}}
\nonu \\
&& +
\frac{1
}{15 (k+5)^2 (23 k+19) d(k)}
(
10640 k^7-157364 k^6-1561752 k^5+16723915 k^4+112161275 k^3  \nonu \\
&&  +123877935 k^2+55032633 k+8400294)
\,\, \pa^2 {\bf Q^{(\frac{5}{2})}} 
+\frac{3 (k+3)}{2 (k+5)} \,\, \pa {\bf Q^{(\frac{7}{2})}} 
\nonu \\
&& 
+\frac{2}{(k+5) d(k)} (140 k^4+35816 k^3+133397 k^2+128682 k+39105 )
\,\, U^{(\frac{5}{2})} W^{(2)} 
\nonu \\
&& + \frac{2}{(k+5)} \,\, U_{-}^{(2)} W_{+}^{(\frac{5}{2})}
-\frac{2}{(k+5) d(k)} (140 k^4+35816 k^3+133397 k^2+128682 k+39105)
\nonu \\
& & \times  U_{+}^{(2)} W_{-}^{(\frac{5}{2})}
+
\frac{6}{(k+5) d(k)}(140 k^4+10056 k^3+58205 k^2+58202 k+17673)
\,\, T^{(2)}  U^{(\frac{5}{2})}
\nonu \\
&& +\frac{96 i}{(k+5)^2 d(k)} (4 k+3) (23 k+19) (35 k+47)
\,\, \hat{A}_3 U^{(\frac{3}{2})} W^{(2)}
\nonu \\
&& + \frac{8 i}{(k+5)^2 d(k)} 
(140 k^4+35816 k^3+133397 k^2+128682 k+39105)
\,\, \hat{A}_3 T^{(2)}  U^{(\frac{3}{2})}
\nonu \\ 
&& +\frac{8 i}{(k+5)^2} \,\, \hat{A}_3 T_{-}^{(\frac{3}{2})}  U_{+}^{(2)}
\nonu \\
&& -\frac{8}{(k+5)^2 d(k)} (140 k^4+16496 k^3+77003 k^2+75822 k+23031)
\,\, \hat{A}_3 \hat{A}_3   U^{(\frac{5}{2})}
\nonu \\
&& -\frac{192 i}{(k+5)^3 d(k)} (4 k+3) (23 k+19) (35 k+47)
\,\, \hat{A}_3 \hat{A}_3 \hat{A}_3  U^{(\frac{3}{2})}
\nonu \\
&& -\frac{8 i}{(k+5)^3 d(k)} 
(140 k^4+35816 k^3+133397 k^2+128682 k+39105)
\,\, \hat{A}_3 \hat{A}_3 \hat{A}_3  \hat{G}_{11}
\nonu \\
&& -\frac{16 i}{(k+5)^3 d(k)} 
(140 k^5+36936 k^4+246045 k^3+688312 k^2+592821 k+168174)
\nonu \\
& & \times  \hat{A}_3 \hat{A}_3 \hat{B}_3  U^{(\frac{3}{2})}
 -\frac{8 i}{(k+5)^3 (4 k+3) (23 k+19) (35 k+47) d(k)} \nonu \\
&& \times (497840 k^8+85440936 k^7 
 +  
1872671072 k^6+10161354790 k^5+23829823389 k^4  \nonu \\
&& +   28711136752 k^3+18791474214 k^2+6395440986 k+891697221)
 \,\, \hat{A}_3 \hat{A}_3 \hat{B}_3  \hat{G}_{11}
\nonu \\
&& +
\frac{96 i}{(k+5)^3 d(k)} (70 k^4-3022 k^3+5605 k^2+7076 k+2139)
\,\, \hat{A}_3 \hat{A}_3 \hat{B}_{-}  T_{-}^{(\frac{3}{2})}
\nonu \\
&&
+\frac{4 i}{(k+5)^3 (4 k+3) (23 k+19) (35 k+47) d(k)} 
\nonu \\
&& \times 
(47040 k^8-37513644 k^7+840886228 k^6+5423499711 k^5+12596529387 k^4
 \nonu \\
&&  +14498360426 k^3+8974548138 k^2+2872046259 k+372394503)
\,\, \hat{A}_3 \hat{A}_3 \hat{B}_{-}  \hat{G}_{12}
\nonu \\
&& -\frac{4}{(k+5)^2 d(k)} 
(140 k^4+35816 k^3+133397 k^2+128682 k+39105)
\,\, \hat{A}_3 \hat{A}_3 \hat{G}_{11} T^{(1)}
\nonu \\
&& -\frac{8}{(k+5)^2} \,\, \hat{A}_3 \hat{B}_3 U^{(\frac{5}{2})}
 -\frac{16 i}{3 (k+5)^3 (4 k+3) (23 k+19) (35 k+47) d(k)} 
\nonu \\
& & \times  
(591920 k^8+5004048 k^7+1097146168 k^6+5625463204 k^5+10416145305 k^4
 \nonu \\
&&  +7776871100 k^3+1383741870 k^2-946321992 k-352719495)
\,\, \hat{A}_3 \hat{B}_3 \hat{B}_{3}  U^{(\frac{3}{2})}
\nonu \\
&& -\frac{8 i}{(k+5)^3 (4 k+3) (23 k+19) (35 k+47) d(k)} \nonu \\
&& \times 
(948640 k^8+198928716 k^7+1823573736 k^6+6773578985 k^5+12547923000 k^4
 \nonu \\
&&  +12323691666 k^3+6470480196 k^2+1675424385 k+158361156) 
\,\, \hat{A}_3 \hat{B}_3 \hat{B}_{3}  \hat{G}_{11}
\nonu  \\
&& -\frac{16}{(k+5)^2 d(k)} 
(140 k^4+35816 k^3+133397 k^2+128682 k+39105)
\,\, \hat{A}_3 \hat{B}_{3}  \hat{G}_{11} T^{(1)}
\nonu \\
&&-\frac{16}{(k+5)^2 d(k)} 
(140 k^4+35816 k^3+133397 k^2+128682 k+39105)
\,\, \hat{A}_3 \hat{B}_{3}  T^{(1)}  U^{(\frac{3}{2})}
\nonu \\
&& -\frac{4}{(k+5)^2 d(k)} 
(140 k^4+35816 k^3+133397 k^2+128682 k+39105)
\,\, \hat{A}_3 \hat{B}_{-}   W_{-}^{(\frac{5}{2})}
\nonu \\
&& \frac{8 i}{(k+5)^3 (4 k+3) (23 k+19) (35 k+47) d(k)} 
\nonu \\
&& \times (544880 k^8+45222492 k^7+893909820 k^6+4443224077 k^5
+8853561090 k^4  \nonu \\
&&  +7820804946 k^3+2792979348 k^2+33932637 k-139602474)
\,\, \hat{A}_3 \hat{B}_{-} \hat{B}_{3}  \hat{G}_{12}
\nonu \\
&& + \frac{12}{(k+5)^2 d(k)} 
(140 k^4+35816 k^3+133397 k^2+128682 k+39105)
\,\, \hat{A}_3 \hat{B}_{-}   \hat{G}_{12} T^{(1)}
\nonu \\
&&
+\frac{16 i}{3 (k+5)^3 (4 k+3) (23 k+19) (35 k+47) d(k)} \nonu \\
& &\times 
(1648360 k^8+358517964 k^7+3305690486 k^6+12587729885 k^5+24144352551 k^4
 \nonu \\
&&  +24903851566 k^3+14029424436 k^2+4048026777 k+462810807)
\,\, \hat{A}_3 \hat{B}_{+}  \hat{B}_{-}  U^{(\frac{3}{2})}
\nonu \\
&& + \frac{4 i}{(k+5)^3 (4 k+3) (23 k+19) (35 k+47) d(k)} \nonu \\
&& \times 
(1944320 k^8+359442188 k^7+3383801340 k^6+12351361993 k^5+21537719977 k^4
 \nonu \\
&&  +18572082294 k^3+7448315646 k^2+879252453 k-123279219)
\,\, \hat{A}_3 \hat{B}_{+} \hat{B}_{-}  \hat{G}_{11}
\nonu \\
&& +\frac{4 i}{(k+5)^2 d(k)} 
(140 k^4+35816 k^3+133397 k^2+128682 k+39105)
\,\,\hat{A}_{3}  \hat{G}_{11} W^{(2)}
\nonu \\
&& + \frac{4 i}{(k+5)^2 d(k)} 
(140 k^4+35816 k^3+133397 k^2+128682 k+39105)
\,\,\hat{A}_{3}  \hat{G}_{11} T^{(2)}
\nonu \\
&& -\frac{4 i}{(k+5)^2 d(k)} 
(140 k^4-41464 k^3-92179 k^2-82758 k-25191)
\,\, \hat{A}_{3}  \hat{G}_{12} U_{+}^{(2)}
\nonu \\
&& +\frac{8 i}{(k+5)^2} 
\,\,
\hat{A}_{3}  \hat{G}_{21} U_{-}^{(2)}
+\frac{4 i}{(k+5)^2}
\,\, \hat{A}_{-}  U^{(\frac{3}{2})} U_{-}^{(2)}
\nonu \\
&& -\frac{4 i}{(k+5)^2 d(k)} 
(140 k^4+35816 k^3+133397 k^2+128682 k+39105)
\,\, \hat{A}_{+}  U_{+}^{(2)}  V^{(\frac{3}{2})}
\nonu \\
&& +\frac{4 i}{(k+5)^2 d(k)} 
(140 k^4+35816 k^3+133397 k^2+128682 k+39105)
\,\,\hat{A}_{+}  T_{+}^{(\frac{3}{2})} W^{(2)}
\nonu \\
&& + \frac{4 i}{(k+5)^2 d(k)} 
(140 k^4+35816 k^3+133397 k^2+128682 k+39105)
\,\,\hat{A}_{+}  T_{+}^{(\frac{3}{2})} T^{(2)}
\nonu \\
&& -\frac{8 i}{(k+5)^3 d(k)} 
(140 k^4+35816 k^3+133397 k^2+128682 k+39105)
\,\,\hat{A}_{+} \hat{A}_3 \hat{A}_{3} T_{+}^{(\frac{3}{2})}
\nonu \\
&& -\frac{8 i}{(k+5)^3} \,\,
\hat{A}_{+} \hat{A}_3 \hat{A}_{3} \hat{G}_{21}
+\frac{16 i}{(k+5)^3}
\,\,\hat{A}_{+} \hat{A}_3 \hat{B}_{3} \hat{G}_{21}
\nonu \\
&&
+\frac{8 i}{(k+5)^3 d(k)} (140 k^4-41464 k^3-92179 k^2-82758 k-25191)
\,\,\hat{A}_{+} \hat{A}_3 \hat{B}_{-} V^{(\frac{3}{2})}
\nonu \\
&& -\frac{16 i}{(k+5)^3} \,\,
\hat{A}_{+} \hat{A}_3 \hat{B}_{-} \hat{G}_{22}
\nonu
\\
&& -\frac{8}{(k+5)^2 d(k)} 
(140 k^4+16496 k^3+77003 k^2+75822 k+23031)
\,\, \hat{A}_{+} \hat{A}_{-} U^{(\frac{5}{2})}
\nonu \\
&& +
\frac{16 i}{(k+5)^3 d(k)} 
(140 k^4-41464 k^3-92179 k^2-82758 k-25191)
\,\,\hat{A}_{+} \hat{A}_{-} \hat{A}_3 U^{(\frac{3}{2})}
\nonu \\
&& -\frac{8 i}{(k+5)^3 d(k)} 
(140 k^4+35816 k^3+133397 k^2+128682 k+39105)
\,\,\hat{A}_{+} \hat{A}_{-} \hat{A}_{3}  \hat{G}_{11}
\nonu \\
&& -\frac{16 i}{(k+5)^3 d(k)} 
(140 k^5+36796 k^4+248869 k^3+667703 k^2+569859 k+161217)
\nonu \\
& & \times \hat{A}_{+} \hat{A}_{-} \hat{B}_3 U^{(\frac{3}{2})}
-\frac{16 i}{(k+5)^3 (4 k+3) (23 k+19) (35 k+47) d(k)} 
\nonu \\
&& \times 
(248920 k^8+41142668 k^7+528083706 k^6+2394755261 k^5+4970672074 k^4
 \nonu \\
&&  +5232542394 k^3+2890613034 k^2+785331045 k+79180578)
\,\, \hat{A}_{+} \hat{A}_{-} \hat{B}_{3}  \hat{G}_{11}
\nonu \\
&& + \frac{64 i (7 k+3)}{(k+5)^3 d(k)} (5 k^3-448 k^2-79 k-90)
\,\, \hat{A}_{+} \hat{A}_{-} \hat{B}_{-} T_{-}^{(\frac{3}{2})}
\nonu \\
&& +\frac{4 i}{(k+5)^3 (4 k+3) (23 k+19) (35 k+47) d(k)} 
\nonu \\
&& \times (47040 k^8-37062844 k^7-286678392 k^6-1072641165 k^5
-2828714270 k^4  \nonu \\
&&  -4838637522 k^3-4523958024 k^2-2103177213 k-384088122)
\,\, \hat{A}_{+} \hat{A}_{-} \hat{B}_{-}  \hat{G}_{12}
\nonu \\
&& -\frac{4}{(k+5)^2 d(k)} 
(140 k^4+35816 k^3+133397 k^2+128682 k+39105)
\,\, \hat{A}_{+} \hat{A}_{-}  \hat{G}_{11} T^{(1)}
\nonu \\
&& -\frac{8 i}{(k+5)^3 d(k)} 
(140 k^4+35816 k^3+133397 k^2+128682 k+39105)
\,\,\hat{A}_{+} \hat{A}_{+} \hat{A}_{-}    T_{+}^{(\frac{3}{2})}
\nonu \\
&& -\frac{8 i}{(k+5)^3}
\,\, \hat{A}_{+} \hat{A}_{+} \hat{A}_{-}  \hat{G}_{21}
+
\frac{8 i}{3 (k+5)^3 (4 k+3) (23 k+19) (35 k+47) d(k)}
\nonu \\
&& \times 
(760480 k^8+343573692 k^7+1415036264 k^6+1293119453 k^5-714423684 k^4
 \nonu \\
&&  -146177462 k^3+2018900748 k^2+1741861413 k+425992824)
\,\, \hat{A}_{+} \hat{B}_{3} \hat{B}_{3}    T_{+}^{(\frac{3}{2})}
\nonu \\
&&-\frac{8 i}{(k+5)^3} \,\,
\hat{A}_{+} \hat{B}_{3} \hat{B}_{3}  \hat{G}_{21}
\nonu \\
&& -\frac{8}{(k+5)^2 d(k)} (140 k^4+35816 k^3+133397 k^2+128682 k+39105)
\,\, \hat{A}_{+} \hat{B}_{3} T^{(1)}    T_{+}^{(\frac{3}{2})}
\nonu \\
&& -\frac{8 i}{3 (k+5)^3 (4 k+3) (23 k+19) (35 k+47) d(k)} \nonu \\
&& \times  
(591920 k^8+13118448 k^7+770521408 k^6+5274859996 k^5+12181219005 k^4
 \nonu \\
&&  +11905315676 k^3+4892295246 k^2+414520992 k-146425779)
\,\, \hat{A}_{+} \hat{B}_{-} \hat{B}_{3}    V^{(\frac{3}{2})}
\nonu \\
&& +\frac{32 i}{(k+5)^3 d(k)} 
(140 k^4+35816 k^3+133397 k^2+128682 k+39105)
\,\, \hat{A}_{+} \hat{B}_{-} \hat{B}_{3}  \hat{G}_{22}
\nonu \\
&& -\frac{4}{(k+5)^2 d(k)} 
(140 k^4+35816 k^3+133397 k^2+128682 k+39105)
\,\,\hat{A}_{+} \hat{B}_{-} T^{(1)}    V^{(\frac{3}{2})}
\nonu \\
&& -\frac{4 i}{3 (k+5)^3 (4 k+3) (23 k+19) (35 k+47) d(k)} \nonu \\
&& \times  
(760480 k^8+353040492 k^7+3118022444 k^6+13050523937 k^5+30505426767 k^4
 \nonu \\
&&  +41425832350 k^3+31835377362 k^2+12817511541 k+2109254067)
\,\,\hat{A}_{+} \hat{B}_{+} \hat{B}_{-}    T_{+}^{(\frac{3}{2})}
\nonu \\
&& -\frac{8 i}{(k+5)^3 d(k)} 
(140 k^4-41464 k^3-92179 k^2-82758 k-25191)
\,\,\hat{A}_{+} \hat{B}_{+} \hat{B}_{-}  \hat{G}_{21}
\nonu \\
&& -\frac{4 i}{(k+5)^2 d(k)} 
(140 k^4+35816 k^3+133397 k^2+128682 k+39105)
\,\,\hat{A}_{+} \hat{G}_{11}    V_{+}^{(2)}
\nonu \\
&& +\frac{4 i}{(k+5)^2}
\,\, \hat{A}_{+} \hat{G}_{21}    W^{(2)}
\nonu \\
&& + \frac{8 i}{(k+5)^2 d(k)} 
(140 k^5+36656 k^4+251693 k^3+647094 k^2+546897 k+154260)
\,\,\hat{B}_3  U^{(\frac{3}{2})}  W^{(2)}
\nonu \\
&& +\frac{8 i}{(k+5)^2 d(k)} 
(140 k^5+36376 k^4+218701 k^3+493088 k^2+395253 k+108198)
\,\,\hat{B}_3  T^{(2)}  U^{(\frac{3}{2})} 
\nonu \\
&& -\frac{48 i}{(k+5)^2 d(k)} (4 k+3) (23 k+19) (35 k+47)
\,\,\hat{B}_3  T_{-}^{(\frac{3}{2})}  V_{+}^{(2)}
\nonu \\
&& -\frac{8 i}{(k+5)^2 d(k)} 
(140 k^4+35816 k^3+133397 k^2+128682 k+39105)
\,\,\hat{B}_3  T_{+}^{(\frac{3}{2})}  U_{-}^{(2)}
\nonu \\
&& +
\frac{4}{3 (k+5)^2 (4 k+3) (23 k+19) (35 k+47) d(k)} \nonu \\
&& \times  
(760480 k^8+353040492 k^7+2744760044 k^6+10871459777 k^5+25282633131 k^4
 \nonu \\
&&  
+34842759310 k^3+27228243450 k^2+11118168261 k+1850880591)
\,\,\hat{B}_3  \hat{B}_3  U^{(\frac{5}{2})} 
\nonu \\
&& -\frac{4}{3 (k+5)^2 (4 k+3) (23 k+19) (35 k+47) d(k)} \nonu \\
&& \times 
(1944320 k^8+364400988 k^7+4169195320 k^6+17887412677 k^5+
36398748222 k^4  \nonu \\
&&  +38467018946 k^3+21750146088 k^2+6195256821 k+684608058)
\,\,\hat{B}_{3} \hat{B}_{3} \hat{G}_{11} T^{(1)}
\nonu \\
&& -\frac{8}{(k+5)^2 d(k)} 
(140 k^5+36376 k^4+218701 k^3+493088 k^2+395253 k+108198)
\nonu \\
&& \times \hat{B}_3  \hat{B}_3  T^{(1)} U^{(\frac{3}{2})} 
+ \frac{16 i}{3 (k+5)^2 (4 k+3) (23 k+19) (35 k+47) d(k)} 
\nonu \\
&& \times (147980 k^8+1589112 k^7-58151123 k^6-469526636 k^5-
1783555167 k^4  
\nonu \\
&&  -3506805256 k^3-3443135025 k^2-1631401092 k-300278313)
\,\,\hat{B}_{3} \hat{G}_{11}    W^{(2)}
\nonu \\
&& +
\frac{8 i (7 k+3)}{(k+5)^2 (4 k+3) (23 k+19) (35 k+47) d(k)} \nonu \\
&& \times (35560 k^7+5926684 k^6+89563862 k^5+413352857 k^4
\nonu \\
&&  +816383138 k^3+769995976 k^2+348461712 k+61314291)
\,\,\hat{B}_{3} \hat{G}_{11}    T^{(2)}
\nonu \\
&& +
\frac{2 i}{(k+5) d(k)} (140 k^4+35816 k^3+133397 k^2+128682 k+39105)
\,\,\hat{B}_3 \hat{G}_{11} T^{(1)} T^{(1)}
\nonu \\
&& +
\frac{4 i}{3 (k+5)^2 (4 k+3) (23 k+19) (35 k+47) d(k)} \nonu \\
&& \times  (591920 k^8+10413648 k^7+1190448328 k^6+7207614532 k^5+15945189135 k^4  \nonu \\
&&  +16015061684 k^3+7562055714 k^2+1377026064 k+120879)
\,\,\hat{B}_3 \hat{G}_{12}  U_{+}^{(2)}
\nonu \\
&& -\frac{8 i}{(k+5)^2} \,\,\hat{B}_3 \hat{G}_{21}  U_{-}^{(2)}
\nonu \\
&& +\frac{4 i}{(k+5) d(k)} 
(140 k^4+35816 k^3+133397 k^2+128682 k+39105)
\,\,\hat{B}_3  T^{(1)}  U^{(\frac{5}{2})} 
\nonu \\
&& -\frac{4 i}{(k+5)^2 d(k)} 
(140 k^5+36376 k^4+257341 k^3+605876 k^2+500973 k+140346)
\,\,\hat{B}_{-}  U^{(\frac{3}{2})}   V_{-}^{(2)} 
\nonu \\
&& -\frac{4 i}{(k+5)^2 d(k)} 
(140 k^4+35816 k^3+133397 k^2+128682 k+39105)
\,\,\hat{B}_{-}   U_{-}^{(2)}  V^{(\frac{3}{2})} 
\nonu \\
&&  -\frac{16 i}{(k+5)^2 d(k)} 
(70 k^4+3418 k^3+24403 k^2+24696 k+7497)
\,\,\hat{B}_{-}  T_{-}^{(\frac{3}{2})}  W^{(2)} 
\nonu \\
&& -\frac{32 i (7 k+3)}{(k+5)^2 d(k)} (5 k^3-448 k^2-79 k-90)
\,\,\hat{B}_{-}  T_{-}^{(\frac{3}{2})}  T^{(2)} 
\nonu \\
&& -\frac{4}{3 (k+5)^2 (4 k+3) (23 k+19) (35 k+47) d(k)} \nonu \\
&& \times 
(591920 k^8+10413648 k^7+70661128 k^6+670422052 k^5+276808227 k^4
 \nonu \\
&&  -3734157436 k^3-6259346022 k^2-3721003776 k-774999549)
\,\,\hat{B}_{-} \hat{B}_3    W_{-}^{(\frac{5}{2})} 
\nonu \\
&& -\frac{8 i}{3 (k+5)^3 (4 k+3) (23 k+19) (35 k+47) d(k)} 
(591920 k^9+13710368 k^8+2114401456 k^7 \nonu \\
&& +18851061284 k^6+70638605743 k^5
+138598606415 k^4  \nonu \\
&&  +153928618502 k^3+97797887394 k^2+33294095451 k+4735617723)
\,\,\hat{B}_{-} \hat{B}_3 \hat{B}_3  T_{-}^{(\frac{3}{2})}
\nonu \\
&& 
-\frac{4 i}{3 (k+5)^3 (4 k+3) (23 k+19) (35 k+47) d(k)}  
(760480 k^9+346560732 k^8+1870881080 k^7 \nonu \\
&& +2311228805 k^6-10290721848 k^5-44486339834 k^4  \nonu \\
&&  -75146337612 k^3-64408152987 k^2-27655262028 k-4749384132)
\,\, \hat{B}_{-} \hat{B}_{3} \hat{B}_3 \hat{G}_{12}
\nonu \\
&& +\frac{8}{3 (k+5)^2 (4 k+3) (23 k+19) (35 k+47) d(k)} \nonu \\
&& \times 
(295960 k^8+1825824 k^7+93661214 k^6+27323996 k^5-1685125269 k^4
 \nonu \\
&& -4958737508 k^3-5551389816 k^2-2781549648 k-527283297)
\,\, \hat{B}_{-} \hat{B}_3 \hat{G}_{12} T^{(1)}
\nonu \\
&& + \frac{32}{(k+5)^2 d(k)} 
(70 k^4+3418 k^3+24403 k^2+24696 k+7497)
\,\, \hat{B}_{-} \hat{B}_3 T^{(1)}  T_{-}^{(\frac{3}{2})}
\nonu \\
&& -\frac{2 i}{3 (k+5)^2 (4 k+3) (23 k+19) (35 k+47) d(k)} \nonu \\
&& \times (1944320 k^8+360343788 k^7+3865929700 k^6+15338884081 k^5+28987719327 k^4  \nonu \\
&&  
+28173955358 k^3+14236952010 k^2+3390656229 k+258494355)
\,\, \hat{B}_{-} \hat{G}_{11}  V_{-}^{(2)} 
\nonu \\
&& +
\frac{2 i}{3 (k+5)^2 (4 k+3) (23 k+19) (35 k+47) d(k)} \nonu \\
&& \times 
(760480 k^8+351688092 k^7+4261141904 k^6+19464561605 k^5+45444395922 k^4
 \nonu \\
&&  +59938387954 k^3+44688092376 k^2+17547122277 k+2828461086)
\,\, \hat{B}_{-} \hat{G}_{12}  W^{(2)} 
\nonu \\
&&  -\frac{2 i}{(k+5)^2 (4 k+3) (23 k+19) (35 k+47) d(k)} \nonu \\ 
&& \times 
(47040 k^8-37964444 k^7-22281952 k^6+297965067 k^5+166873652 k^4
 \nonu \\
&&  -1274364506 k^3-2098326564 k^2-1215894429 k-249114744)
\,\, \hat{B}_{-} \hat{G}_{12}  T^{(2)} 
\nonu \\
&& -\frac{i}{(k+5) d(k)} 
(140 k^4+35816 k^3+133397 k^2+128682 k+39105)
\,\,\hat{B}_{-} \hat{G}_{12} T^{(1)} T^{(1)}
\nonu \\
&& +
\frac{48 i}{(k+5)^2 d(k)} (4 k+3) (23 k+19) (35 k+47)
\,\, \hat{B}_{-} \hat{G}_{22}  U_{-}^{(2)} 
\nonu \\
&& +
\frac{2 i}{(k+5) d(k)} 
(140 k^4+35816 k^3+133397 k^2+128682 k+39105)
\,\,\hat{B}_{-} T^{(1)}    W_{-}^{(\frac{5}{2})} 
\nonu \\
&& \frac{4 i}{(k+5)^2 d(k)} 
(140 k^5+36516 k^4+254517 k^3+626485 k^2+523935 k+147303)
\,\,\hat{B}_{+}  U^{(\frac{3}{2})}   U_{+}^{(2)} 
\nonu \\
&& 
-\frac{2}{3 (k+5)^2 (4 k+3) (23 k+19) (35 k+47) d(k)} \nonu \\
&& \times (760480 k^8+355745292 k^7+3817882724 k^6+
17654961881 k^5+42409837545 k^4  \nonu \\
&&  +57065305462 k^3+42987018630 k^2+16953036309 k+2737827837)
\,\,\hat{B}_{+} \hat{B}_{-}  U^{(\frac{5}{2})} 
\nonu \\
&& -\frac{8 i}{3 (k+5)^3 (4 k+3) (23 k+19) (35 k+47) d(k)} \nonu \\
&& \times  
(8874880 k^8+2495898132 k^7+23836583156 k^6+97375525535 k^5+
207832291737 k^4  \nonu \\
&&  +249406985986 k^3+169686171966 k^2+61329181419 k+9195139413)
\,\,\hat{B}_{+} \hat{B}_{-}  \hat{B}_3 U^{(\frac{3}{2})} 
\nonu \\
&& -\frac{4 i}{3 (k+5)^3 (4 k+3) (23 k+19) (35 k+47) d(k)} 
(1352400 k^9+372105580 k^8+7067446908 k^7
\nonu \\
&& +49250719265 k^6+
167659300796 k^5 +   301955782524 k^4  
+296224110280 k^3 \nonu \\
&& +158226710421 k^2+42600264840 k+4373207226)
\,\, \hat{B}_{+} \hat{B}_{-}  \hat{B}_3 \hat{G}_{11}
\nonu \\
&& \frac{2 i}{3 (k+5)^3 (4 k+3) (23 k+19) (35 k+47) d(k)} 
(591920 k^9+13710368 k^8+2133335056 k^7
\nonu \\
&& +17964516044 k^6+69094176871 k^5
+140976180503 k^4+161367298166 k^3 \nonu \\
&& +104448800634 k^2+35902947987 k+
5130845235)
\,\, \hat{B}_{+} \hat{B}_{-}  \hat{B}_{-}  T_{-}^{(\frac{3}{2})}
\nonu \\
&& +
\frac{i}{3 (k+5)^3 (4 k+3) (23 k+19) (35 k+47) d(k)} 
(2112880 k^9+709199512 k^8+6413082608 k^7
\nonu \\
&& +23688837826 k^6
+32539226153 k^5  
-20913548630 k^4-117613685066 k^3 \nonu \\
&& -136151922606 k^2-67363490007 k-12550399470
)
\,\,  \hat{B}_{+} \hat{B}_{-}  \hat{B}_{-} \hat{G}_{12}
\nonu \\
&& +\frac{2}{3 (k+5)^2 (4 k+3) (23 k+19) (35 k+47) d(k)} \nonu \\
&& \times 
 (1944320 k^8+357638988 k^7+2419544620 k^6+6376317817 k^5
+6637721277 k^4  \nonu \\
&&  -631663834 k^3-6128957082 k^2-4143555099 k-886826367)
\,\, \hat{B}_{+}  \hat{B}_{-} \hat{G}_{11} T^{(1)}
\nonu \\
&& +
\frac{4}{(k+5)^2 d(k)} 
(140 k^5+36516 k^4+254517 k^3+626485 k^2+523935 k+147303)
\nonu \\
&& \times  \hat{B}_{+}  \hat{B}_{-}  T^{(1)}  U^{(\frac{3}{2})} 
+
\frac{2 i}{3 (k+5)^2 (4 k+3) (23 k+19) (35 k+47) d(k)} \nonu \\ 
&& \times 
(1944320 k^8+360343788 k^7+3119404900 k^6+10980755761 k^5+18542132055 k^4
 \nonu \\ 
&&  +15007809278 k^3+5022684186 k^2-8030331 k-258252597)
\,\,\hat{B}_{+} \hat{G}_{11}  U_{+}^{(2)} 
\nonu \\
&& -\frac{2}{(k+5) d(k)} 
(140 k^4+35816 k^3+133397 k^2+128682 k+39105)
\,\,\hat{G}_{11} W^{(3)}
\nonu \\
&& -\frac{4}{(k+5)^2 d(k)} 
(140 k^4+35816 k^3+133397 k^2+128682 k+39105)
\,\,\hat{G}_{11}  U^{(\frac{3}{2})}   V^{(\frac{3}{2})} 
\nonu  \\
&& -\frac{4}{(k+5)^2 d(k)} 
(140 k^4+35816 k^3+133397 k^2+128682 k+39105)
\,\,\hat{G}_{11}  T_{+}^{(\frac{3}{2})}   T_{-}^{(\frac{3}{2})} 
\nonu \\
&& -\frac{4}{(k+5)^2 d(k)} 
(140 k^4+35816 k^3+133397 k^2+128682 k+39105)
\,\,\hat{G}_{11} \hat{G}_{12}  T_{+}^{(\frac{3}{2})}
\nonu \\
&& +\frac{4}{(k+5)^2} \,\,\hat{G}_{11} \hat{G}_{12} \hat{G}_{21}
+ \frac{4}{(k+5)^2}
 \,\,\hat{G}_{11} \hat{G}_{21}   T_{-}^{(\frac{3}{2})}  
\nonu \\
&& +
\frac{2}{(k+5) d(k)} (140 k^4+35816 k^3+133397 k^2+128682 k+39105)
\,\,\hat{G}_{11} T^{(1)} W^{(2)}
\nonu \\
&& +
\frac{2}{(k+5) d(k)} (140 k^4+35816 k^3+133397 k^2+128682 k+39105)
\,\,\hat{G}_{11} T^{(1)} T^{(2)}
\nonu \\
&& 
-\frac{4}{(k+5)^2 d(k)} (140 k^4+35816 k^3+133397 k^2+128682 k+39105)
\,\,\hat{G}_{12}  T_{+}^{(\frac{3}{2})}   U^{(\frac{3}{2})} 
\nonu \\
&& +
\frac{2}{(k+5) d(k)} (140 k^4+35816 k^3+133397 k^2+128682 k+39105)
\,\,\hat{G}_{12} T^{(1)} U_{+}^{(2)}
\nonu \\
&&+
\frac{4}{(k+5) (7 k+3) (23 k+19) d(k)} 
(1960 k^7-51856 k^6-3075662 k^5-16332408 k^4
 \nonu \\
&&   -38167075 k^3-38830209 k^2-18038331 k-3189051)
\,\, \hat{T}  U^{(\frac{5}{2})} 
\nonu \\
&& +
\frac{16 i}{(k+5)^2 (7 k+3) (35 k+47) d(k)} 
(4900 k^6-5483100 k^5-26897469 k^4  \nonu \\
&&  -48846745 k^3-46361901 k^2-22377807 k-4324158)
\,\, \hat{T} \hat{A}_3 U^{(\frac{3}{2})}
\nonu \\
&&-\frac{8 i}{(k+5)^2 (7 k+3) (23 k+19) (35 k+47) d(k)} \nonu \\
&& \times 
(180320 k^8-1437352 k^7+112055156 k^6+1087499002 k^5+3360953593 k^4
 \nonu \\
&&  +4786671064 k^3+3522752130 k^2+1309761486 k+194846121)
\,\,\hat{T} \hat{A}_3 \hat{G}_{11}
\nonu \\
&&
-\frac{8 i (4 k+3)}{(k+5)^2 (7 k+3) d(k)} 
(140 k^4+35816 k^3+133397 k^2+128682 k+39105)
\,\, \hat{T} \hat{A}_{+} T_{+}^{(\frac{3}{2})}
\nonu \\
&& -\frac{8 i}{(k+5)^2} \,\, \hat{T} \hat{A}_{+} \hat{G}_{21}
-\frac{16 i}{(k+5)^2 (7 k+3) (35 k+47) d(k)} \nonu \\
&& \times (19600 k^7+5182660 k^6+47202368 k^5+180877867 k^4
 \nonu \\
&&  +335825221 k^3+306757389 k^2+135197199 k+23132520)  
\,\, \hat{T} \hat{B}_3 U^{(\frac{3}{2})}
\nonu \\
&&
-\frac{4 i}{(k+5)^2 (7 k+3) (23 k+19) (35 k+47) d(k)} \nonu \\
&& \times  
(90160 k^8+121683884 k^7+1359836772 k^6+5547998009 k^5+11729943100 k^4
 \nonu \\
&&  +14367980562 k^3+10135550376 k^2+3797129745 k+584932752)
\,\,\hat{T} \hat{B}_{-} \hat{G}_{12}
\nonu \\
&& -\frac{4}{(k+5) (7 k+3) (23 k+19) d(k)} 
(92 k^2+159 k+63) \nonu \\
&& \times (140 k^4+35816 k^3+133397 k^2+128682 k+39105)
\,\,\hat{T}  \hat{G}_{11} T^{(1)}
\nonu \\
&& +
\frac{2 
}{3 (k+5)^2 d(k)}
(140 k^5-79404 k^4-837327 k^3-1890041 k^2-1634049 k-479583)
\nonu \\
&& \times \pa  U^{(\frac{3}{2})} W^{(2)} 
-\frac{2 (3 k+11)}{(k+5)^2}  \,\, U^{(\frac{3}{2})} \pa W^{(2)} 
+
\frac{1}{45 (k+5)^3 (23 k+19) (35 k+47) d(k)} \nonu \\
&& \times (3379600 k^8-616758140 k^7-11364369892 k^6-70453481393 k^5-
203439434260 k^4  \nonu \\
&&  -334269296210 k^3-307584719000 k^2-145515073161 k-27521284104)
\,\,\pa^3  U^{(\frac{3}{2})}
\nonu \\
&&  \frac{1}{15 (k+5)^2 (23 k+19) d(k)} 
(10640 k^7+11196 k^6-6567088 k^5+63421585 k^4  \nonu \\
&&  +368596575 k^3+506779975 k^2+284850093 k+58674456)
\,\,\pa^2  U^{(\frac{5}{2})}
\nonu \\
&&  + \frac{4}{(k+5)^2} \,\, \pa T^{(2)}  U^{(\frac{3}{2})}
\nonu \\
&& -\frac{4
}{(k+5)^2 d(k)}
 (k+8) (140 k^4+10056 k^3+58205 k^2+58202 k+17673)
\,\,T^{(2)} \pa U^{(\frac{3}{2})}
\nonu \\
&& +
\frac{2
}{3 (k+5)^2 d(k)}
 (980 k^5-94948 k^4-587353 k^3-1095027 k^2-939807 k-281421)
\,\,\pa  T_{-}^{(\frac{3}{2})} U_{+}^{(2)}
\nonu \\
&&  +
\frac{2 
}{(k+5)^2 d(k)}
(420 k^5-7492 k^4-35221 k^3-12427 k^2-29835 k-15597)
 \,\,T_{-}^{(\frac{3}{2})} \pa U_{+}^{(2)}
\nonu \\
&& +
 \frac{2 
}{3 (k+5)^2 d(k)}
(980 k^5-16408 k^4+308327 k^3+1332078 k^2+
1234107 k+359856)
\,\,\pa  T_{+}^{(\frac{3}{2})}  U_{-}^{(2)}
\nonu \\
&& +
\frac{2 
}{(k+5)^2 d(k)}
(420 k^5-6512 k^4+99571 k^3+582988 k^2+553779 k+161694)
 \,\,T_{+}^{(\frac{3}{2})}  \pa U_{-}^{(2)}
\nonu  \\
&&  +
\frac{2 i 
}{15 (k+5)^3 (7 k+3) (35 k+47) d(k)} \nonu \\
&& \times (1234800 k^8-13719020 k^7+497521024 k^6+6778408251 k^5
+24682779185 k^4  \nonu \\
&&  +39759062330 k^3+32982030330 k^2+13803473367 k+2313433413)
\,\,\pa^2 \hat{A}_3  U^{(\frac{3}{2})}
\nonu \\
&& -\frac{2 i 
}{15 (k+5)^3 d(k)} (1120 k^6-40372 k^5-322500 k^4  \nonu \\
&&  -12486055 k^3-42822545 k^2-38908425 k-11354031)
\,\,\pa \hat{A}_3  \pa U^{(\frac{3}{2})}
\nonu \\
&& 
-\frac{4 i
}{15 (k+5)^3 (35 k+47) d(k)}
(9800 k^7+469980 k^6-3561326 k^5-19194959 k^4
 \nonu \\
&&  -112141437 k^3-219973537 k^2-200387745 k-60113520)
 \,\,\hat{A}_3  \pa^2 U^{(\frac{3}{2})}
\nonu \\
&& 
-\frac{6 i 
}{(k+5)^2 d(k)}
(140 k^5-2264 k^4-16447 k^3+30206 k^2+28325 k+6396)
\,\, \pa  \hat{A}_3   U^{(\frac{5}{2})}
\nonu \\
&& -\frac{2 i (247 k^2+1145 k+222)}{5 (k+5)^2 (23 k+19)}
\,\, \hat{A}_3 \pa  U^{(\frac{5}{2})}
\nonu \\
&& +
\frac{8 
}{5 (k+5)^3 d(k)}
(1540 k^5-27144 k^4-238773 k^3-298246 k^2-337737 k-126684)
\,\,\pa \hat{A}_3 \hat{A}_3   U^{(\frac{3}{2})}
\nonu \\
&& 
-\frac{4 
}{15 (k+5)^3 d(k)}
(6860 k^5-522396 k^4-3257247 k^3-8604749 k^2-8047833 k-2453571)
\nonu \\
&& \hat{A}_3 \hat{A}_3   \pa U^{(\frac{3}{2})}
+
 \frac{48 
}{5 (k+5)^3 d(k)}
(420 k^4+55928 k^3+249807 k^2+245086 k+74451)
\,\,\pa \hat{A}_3 \hat{A}_3  \hat{G}_{11} 
\nonu \\
&& 
-\frac{4 
}{15 (k+5)^3 d(k)}
(4900 k^5-317660 k^4-2223833 k^3-11955477 k^2-12048291 k-3684231)
\nonu \\
&&  \hat{A}_3 \hat{A}_3  \pa \hat{G}_{11} 
\nonu \\
&& +
\frac{8 
}{5 (k+5)^3 d(k)}
(420 k^5+535148 k^4+3210083 k^3+6677633 k^2+5347461 k+1482399)
\nonu \\
&& \times \pa \hat{A}_3 \hat{B}_3  U^{(\frac{3}{2})}
-\frac{16 
}{15 (k+5)^3 (4 k+3) (23 k+19) (35 k+47) d(k)} \nonu \\
&& \times 
(19061000 k^8+1619455740 k^7+14480900326 k^6+53662510369 k^5
+98738236143 k^4  \nonu \\
&&  +95262338966 k^3+48227837892 k^2+11626557597 k+935387055)
\,\, \hat{A}_3 \pa \hat{B}_3  U^{(\frac{3}{2})}
\nonu \\
&& +
\frac{8 
}{15 (k+5)^3 d(k)}
(9940 k^5+1857356 k^4+14694527 k^3+38641273 k^2+32968977 k+9358047)
\nonu \\
&& \hat{A}_3 \hat{B}_3  \pa U^{(\frac{3}{2})}
 +
\frac{4 
}{5 (k+5)^3 (4 k+3) (23 k+19) (35 k+47) d(k)} \nonu \\
&& \times
(470400 k^8-364768040 k^7-603895180 k^6+3607998882 k^5
+7951334591 k^4  \nonu \\
&&  -1334704064 k^3-11863423002 k^2-8672781906 k-1962224433)
\,\, \pa \hat{A}_3 \hat{B}_3  \hat{G}_{11} 
\nonu \\
&& 
-\frac{4 
}{5 (k+5)^3 (4 k+3) (23 k+19) (35 k+47) d(k)} \nonu \\
&& \times
(7016800 k^8+596511860 k^7+7597279816 k^6+31705992311 k^5
+55455767116 k^4  \nonu \\
&&  +39765011582 k^3+5544323844 k^2-5933170449 k-2038728096)
\,\,\hat{A}_3 \pa \hat{B}_3  \hat{G}_{11} 
\nonu \\
&& +
\frac{8 
}{5 (k+5)^3 (4 k+3) (23 k+19) (35 k+47) d(k)} \nonu \\
&& \times 
(8584800 k^8+1200164000 k^7+19975496852 k^6+99303364528 k^5
+222631364345 k^4  \nonu \\
&&  +260103602372 k^3+165977671722 k^2+55184677524 k+7517786769)
\,\,\hat{A}_3  \hat{B}_3  \pa \hat{G}_{11} 
\nonu \\
&& +
\frac{4 
}{5 (k+5)^3 d(k)}
(4900 k^5-707980 k^4-3276133 k^3-5413241 k^2-4275783 k-1219491)
\nonu \\
&& \times
\pa \hat{A}_3 \hat{B}_{-}  T_{-}^{(\frac{3}{2})}
\nonu \\
&& +
\frac{4 
}{5 (k+5)^3 d(k)}
(4060 k^5-67756 k^4+119237 k^3+2183467 k^2+1992315 k+541917)
\nonu \\
&& \times 
\hat{A}_3 \pa \hat{B}_{-}  T_{-}^{(\frac{3}{2})}
\nonu \\
&& +
\frac{4 
}{15 (k+5)^3 d(k)}
(14980 k^5-203468 k^4-551573 k^3-1622457 k^2-2655639 k-1057347)
\nonu \\
&& \hat{A}_3  \hat{B}_{-}  \pa T_{-}^{(\frac{3}{2})}
-\frac{2 
}{5 (k+5)^3 (4 k+3) (23 k+19) (35 k+47) d(k)} \nonu \\
&& \times 
(9486400 k^8-4134830840 k^7-38004588740 k^6-141047020930 k^5
-284220635911 k^4  \nonu \\
&& -339247630336 k^3-237701134374 k^2-89935717806 k-14162631543)
\,\,\pa \hat{A}_3  \hat{B}_{-} \hat{G}_{12} 
\nonu \\
&& 
-\frac{2 
}{5 (k+5)^3 (4 k+3) (23 k+19) (35 k+47) d(k)} \nonu \\
&& \times 
(18541600 k^8+1174304320 k^7+13964129652 k^6
+67589343656 k^5+138180069773 k^4  \nonu \\
&&  +125788164036 k^3+46937475042 k^2+1575432540 k-2066526459)
\,\, \hat{A}_3  \pa \hat{B}_{-} \hat{G}_{12} 
\nonu \\
&& 
-\frac{4 
}{15 (k+5)^3 d(k)}
(11480 k^5-26748 k^4+7107726 k^3+26433415 k^2+24876324 k+7468623)
\nonu \\
&& \hat{A}_3  \pa \hat{B}_{-} \pa \hat{G}_{12} 
 + 
\frac{2 i 
}{5 (k+5)^3 (7 k+3) (23 k+19) (35 k+47) d(k)} (1803200 k^9-47958120 k^8
\nonu \\
&& +182464900 k^7+3948116342 k^6
+14669967223 k^5+39472158705 k^4  \nonu \\
&&  +66534195690 k^3+60320589960 k^2+26788403691 k+4592413449)
\,\,\pa^2  \hat{A}_3  \pa \hat{G}_{11} 
\nonu \\
&& 
-\frac{4 i
}{15 (k+5)^3 d(k)}
(1400 k^6-9060 k^5-362658 k^4+2260369 k^3+4356093 k^2
\nonu \\
&& +2652291 k+563517)
 \pa \hat{A}_3  \pa \hat{G}_{11} 
+
\frac{4 i 
}{15 (k+5)^3 (7 k+3) (23 k+19) (35 k+47) d(k)} \nonu \\
&& \times
(2028600 k^9-16549960 k^8+2110311830 k^7+30553694730 k^6
+176136678127 k^5  \nonu \\
&&  +438509397691 k^4+544894411356 k^3+360020540160 k^2+121745083671 k+16584979635) \nonu \\
&& \hat{A}_3  \pa^2 \hat{G}_{11} 
-\frac{4 i 
}{(k+5)^2 d(k)}
(140 k^4+35816 k^3+133397 k^2+128682 k+39105)
\,\,\pa \hat{A}_3 \hat{G}_{11} T^{(1)}
\nonu \\
&& 
-\frac{32 i 
}{5 (k+5)^2 d(k)}
(280 k^4+42652 k^3+182203 k^2+178074 k+54099)
\,\,\hat{A}_3 \pa \hat{G}_{11} T^{(1)}
\nonu \\
&& -\frac{24 i}{5 (k+5)^2}
\,\,\hat{A}_3 \hat{G}_{11} \pa T^{(1)}
-\frac{12 i}{(k+5)^2}
\,\,\hat{A}_3 \pa T^{(1)} U^{(\frac{3}{2})}
-\frac{8 i (k+4)}{(k+5)^2}
\,\,\pa \hat{A}_{+} W_{+}^{(\frac{5}{2})}
\nonu \\
&& -\frac{2 i (2 k+7)}{(k+5)^2}
\,\,\hat{A}_{+} \pa W_{+}^{(\frac{5}{2})}
 +
\frac{2 i
}{15 (k+5)^3 (7 k+3) d(k)}
(68740 k^6-7111044 k^5+15887955 k^4
 \nonu \\
&&  +175463669 k^3+258198261 k^2+144510003 k+28676376)
\,\, \pa^2 \hat{A}_{+} T_{+}^{(\frac{3}{2})}
\nonu \\
&& +
\frac{2 i
}{15 (k+5)^3 d(k)}
(1400 k^6-9900 k^5-902774 k^4
 \nonu \\
&&  +1922231 k^3+11589970 k^2+11111211 k+3290994)
\,\,\pa \hat{A}_{+} \pa T_{+}^{(\frac{3}{2})}
\nonu \\
&& -\frac{2 i
}{15 (k+5)^3 (7 k+3) d(k)}
 (9800 k^7-155820 k^6+2402574 k^5+8054479 k^4
 \nonu \\
&&  +6209808 k^3-9169794 k^2-12463038 k-3901689)
\,\, \hat{A}_{+} \pa^2 T_{+}^{(\frac{3}{2})}
\nonu \\
&& 
-\frac{16 
}{5 (k+5)^3 d(k)}
(700 k^5-11880 k^4+251061 k^3+1008494 k^2+930777 k+272052)
\,\,\pa \hat{A}_{+} \hat{A}_3 T_{+}^{(\frac{3}{2})}
\nonu \\
&& +
\frac{8 
}{(k+5)^3 d(k)}
(140 k^4+35816 k^3+133397 k^2+128682 k+39105)
\,\,\hat{A}_{+} \pa \hat{A}_3 T_{+}^{(\frac{3}{2})}
\nonu \\
&& -\frac{16 (5 k+21)}{5 (k+5)^3}
\,\,\hat{A}_{+}  \hat{A}_3 \pa T_{+}^{(\frac{3}{2})}
-\frac{8 \left(37 k^2+533 k+108\right)}{5 (k+5)^3 (23 k+19)}
\,\,\pa \hat{A}_{+} \hat{A}_3 \hat{G}_{21}
-\frac{4}{(k+5)^3} \,\,\hat{A}_{+} \pa \hat{A}_3 \hat{G}_{21}
\nonu \\
&& 
-\frac{8 
}{5 (k+5)^3 (23 k+19) d(k)}
(7280 k^6-107228 k^5-4114704 k^4  \nonu \\
&&  -10753121 k^3-7707223 k^2-2517903 k-250389)
\,\, \hat{A}_{+}  \hat{A}_3 \pa \hat{G}_{21}
\nonu \\
&& +
\frac{4 
}{(k+5)^3 d(k)}
(1120 k^5-18252 k^4+48 k^3+596999 k^2+556038 k+151371)
\,\, \pa \hat{A}_{+}  \hat{A}_{-}  U^{(\frac{3}{2})}
\nonu \\
&& +
\frac{8 (3 k+4)}{(k+5)^3}
\,\,\hat{A}_{+}  \pa \hat{A}_{-}  U^{(\frac{3}{2})}
\nonu \\
&& +
\frac{16
}{3 (k+5)^3 d(k)}
 (k+8) (140 k^4+16496 k^3+77003 k^2+75822 k+23031)
\,\,\hat{A}_{+}  \hat{A}_{-}  \pa U^{(\frac{3}{2})}
\nonu \\
&& -\frac{2
}{(k+5)^3 d(k)}
 (700 k^4+24520 k^3+215833 k^2+220530 k+66933)
\,\,\pa \hat{A}_{+}  \hat{A}_{-} \hat{G}_{11}
\nonu \\
&& 
-\frac{2
}{(k+5)^3 d(k)}
 (980 k^4-212968 k^3-419677 k^2-367866 k-112041)
\,\, \hat{A}_{+}  \pa \hat{A}_{-} \hat{G}_{11}
\nonu \\
&& 
-\frac{4 
}{3 (k+5)^3 d(k)}
(980 k^5-61348 k^4-25141 k^3-716139 k^2-782811 k-242541)
\,\, \hat{A}_{+}   \hat{A}_{-} \pa \hat{G}_{11}
\nonu \\
&& +
\frac{8
}{5 (k+5)^3 d(k)}
 (1400 k^5-22780 k^4+752834 k^3+2950767 k^2+2762328 k+817839)
\nonu \\
&& \times \pa \hat{A}_{+} \hat{B}_3 T_{+}^{(\frac{3}{2})}
+
\frac{4 
}{3 (k+5)^3 (4 k+3) (23 k+19) (35 k+47) d(k)} \nonu \\
&& \times 
(760480 k^8+358450092 k^7+3771218204 k^6+17901271505 k^5
+43868661051 k^4  \nonu \\
&&  +59538632494 k^3+44924392074 k^2+17689874517 k+2849654655)
 \,\,\hat{A}_{+} \pa \hat{B}_3 T_{+}^{(\frac{3}{2})}
\nonu \\
&& +
\frac{16
}{15 (k+5)^3 d(k)}
 (3500 k^5+329940 k^4+1956801 k^3+5460973 k^2+4453227 k+1200951)
\nonu \\ 
&& \times \hat{A}_{+}  \hat{B}_3 \pa T_{+}^{(\frac{3}{2})}
+
\frac{8 
}{5 (k+5)^3 (23 k+19) d(k)}
(10360 k^6+1386044 k^5+18558138 k^4  \nonu \\
&&  +76719497 k^3+80889019 k^2+30169623 k+2581551)
\,\, \pa \hat{A}_{+}   \hat{B}_{3} \pa \hat{G}_{21}
\nonu \\
&& +
\frac{4
}{(k+5)^3 d(k)}
(140 k^5-2544 k^4-139599 k^3-386972 k^2-369999 k-114678)
\,\, \hat{A}_{+}   \pa \hat{B}_{3}  \hat{G}_{21}
\nonu \\
&& +
\frac{32 
}{15 (k+5)^3 (4 k+3) (23 k+19) (35 k+47) d(k)} \nonu \\
&& \times 
(2268700 k^8+202982640 k^7+2346522269 k^6+11510578586 k^5
+26786881518 k^4  \nonu \\
&&  +32016549688 k^3+20563007781 k^2+6808596822 k+924394572)
\,\, \hat{A}_{+}    \hat{B}_{3}  \pa \hat{G}_{21}
\nonu \\
&& 
-\frac{4 
}{5 (k+5)^3 d(k)}
(700 k^5-241760 k^4-2551763 k^3-5427852 k^2-4649523 k-1366542)
\nonu \\
&& \times \pa \hat{A}_{+}    \hat{B}_{-}  V^{(\frac{3}{2})}
-\frac{4 
}{3 (k+5)^3 (4 k+3) (23 k+19) (35 k+47) d(k)} \nonu \\
&& \times 
(760480 k^8-26983908 k^7-64167616 k^6+674763761 k^5+4606914054 k^4
 \nonu \\
&&  +10886161738 k^3+11471198736 k^2+5619012705 k+1054566594)
\,\, \hat{A}_{+}   \pa \hat{B}_{-}  V^{(\frac{3}{2})}
\nonu \\
&& +
\frac{4 
}{15 (k+5)^3 d(k)}
(3500 k^5-137380 k^4-1197271 k^3-1813719 k^2-1656981 k-539109)
 \nonu \\
&& \times \hat{A}_{+}    \hat{B}_{-}  \pa V^{(\frac{3}{2})}
\nonu \\
&& 
-\frac{2 
}{5 (k+5)^3 d(k)}
(3500 k^5+258820 k^4+2940049 k^3+10047515 k^2+9046599 k+2616165)
\nonu \\
&& \times  \pa \hat{A}_{+}    \hat{B}_{-} \hat{G}_{22} 
\nonu \\
&& 
-\frac{2 
}{(k+5)^3 d(k)}
(700 k^5-10060 k^4+330269 k^3+1614775 k^2+1546443 k+458937)
\nonu \\
&& \times \hat{A}_{+}  \pa   \hat{B}_{-} \hat{G}_{22}  
-\frac{2 
}{15 (k+5)^3 (4 k+3) (23 k+19) (35 k+47) d(k)} \nonu \\
& & \times
(41414800 k^8-1475700380 k^7-9036556680 k^6-3227653537 k^5
+61045077905 k^4  \nonu \\
&&  +149451032706 k^3+147891609090 k^2+68185056123 k+12226082085)
\,\, \hat{A}_{+}     \hat{B}_{-} \pa \hat{G}_{22} 
\nonu \\
&& +
\frac{2 i (111 k^3+1201 k^2+5867 k+1473)}{5 (k+5)^3 (23 k+19)}
\,\, \pa^2 \hat{A}_{+} \hat{G}_{21} 
\nonu \\
&& 
-\frac{2 i 
}{15 (k+5)^3 (23 k+19) d(k)}
(16660 k^7-39396 k^6-1629265 k^5+22676117 k^4
 \nonu \\
&&  +352920705 k^3+588927247 k^2+351412080 k+74502324)
\,\, \pa \hat{A}_{+} \pa \hat{G}_{21} 
\nonu \\
&& +
\frac{i 
}{15 (k+5)^3 (23 k+19) d(k)}
(2240 k^7-413804 k^6-932724 k^5-173892089 k^4
 \nonu \\
&&  -425309143 k^3-224059911 k^2-6756609 k+18026784)
\,\,\hat{A}_{+} \pa^2 \hat{G}_{21} 
\nonu \\
&& -\frac{4 i 
}{(k+5)^2 d(k)}
(140 k^4+35816 k^3+133397 k^2+128682 k+39105)
\,\, \hat{A}_{+} \pa \hat{G}_{21} T^{(1)}
\nonu \\
&& +
\frac{6 i}{(k+5)^2}
\,\, \hat{A}_{+}  \hat{G}_{21} \pa T^{(1)}
+ 
\frac{2 i
}{15 (k+5)^3 (7 k+3) (35 k+47) d(k)} \nonu \\
&& \times  
(1484700 k^8+310713620 k^7+3579506201 k^6+18214097691 k^5
+49964011726 k^4  \nonu \\
&&  +77939715586 k^3+66066383901 k^2+28392989391 k+4876382304)
\,\, \pa^2 \hat{B}_{3}  U^{(\frac{3}{2})}
\nonu \\
&& 
-\frac{2 i 
}{45 (k+5)^3 (4 k+3) (23 k+19) (35 k+47) d(k)} \nonu \\
&& \times (15386000 k^9+11800202260 k^8+100865124784 k^7
+496274132403 k^6+1241539210451 k^5
 \nonu \\
&& 
+1448454287474 k^4+489587787382 k^3-412104385545 k^2-384278283225 k
-88554266496)
\nonu \\
&& \pa \hat{B}_{3}  \pa U^{(\frac{3}{2})}
+
\frac{8 i 
}{15 (k+5)^3 (35 k+47) d(k)}
 (4900 k^7+10661140 k^6+133075777 k^5+767600755 k^4
 \nonu \\
&& +2150393230 k^3+2917404010 k^2+1791605421 k+414489519)
\,\, \hat{B}_{3}  \pa^2 U^{(\frac{3}{2})}
\nonu \\
&& +
 \frac{2 i 
}{3 (k+5)^2 (4 k+3) (23 k+19) (35 k+47) d(k)} \nonu \\
&& \times 
(7353920 k^8-475452852 k^7-3428937752 k^6-11431169183 k^5
-24048680154 k^4  \nonu \\
&&  -32703232390 k^3-26246203176 k^2-11091269295 k-1899126702)
\,\, \pa \hat{B}_{3}   U^{(\frac{5}{2})}
\nonu \\
&& 
-\frac{2 i 
}{5 (k+5)^2 (23 k+19) d(k)} 
(48440 k^6+13486716 k^5+92805558 k^4
 \nonu \\
&&  +256405093 k^3+282099639 k^2+141258987 k+26843751) 
\,\, \hat{B}_{3}   \pa U^{(\frac{5}{2})}
\nonu \\
&& +
\frac{8 
}{5 (k+5)^3 d(k)}
(7280 k^5+1282692 k^4+10441236 k^3+28431529 k^2+24484542 k+6976485)
\nonu \\
&& \pa \hat{B}_3 \hat{B}_{3}   U^{(\frac{3}{2})}
 +
\frac{4 
}{45 (k+5)^3 (4 k+3) (23 k+19) (35 k+47) d(k)} \nonu \\
&& \times
(46491200 k^9+11041766680 k^8+156839442100 k^7+1007342316654 k^6
+3330833770487 k^5  \nonu \\
&&  
+5976579261377 k^4+5980068284026 k^3+
3338637094848 k^2+970434837099 k
\nonu \\
&& +113648078025)
\hat{B}_3 \hat{B}_{3}   \pa U^{(\frac{3}{2})}
+
 \frac{4 
}{15 (k+5)^3 (4 k+3) (23 k+19) (35 k+47) d(k)} \nonu \\
& & \times
(6762000 k^9+1890790300 k^8+40417466340 k^7+320788990901 k^6
+1203896324150 k^5  \nonu \\
&& +2335165490664 k^4+2458337434540 k^3+1424658770265 k^2+427358270610 k
\nonu \\
&& +51649200342)
\pa \hat{B}_3 \hat{B}_{3} \hat{G}_{11}
 +
\frac{4 
}{45 (k+5)^3 (4 k+3) (23 k+19) (35 k+47) d(k)}
\nonu \\
&& \times
(9721600 k^9+1897151340 k^8+33524902580 k^7
+186745477433 k^6+449259266475 k^5  \nonu \\
&&  
+521843454412 k^4+263423364570 k^3+1900396305 k^2-47225762505 k-
12698631954)
\nonu \\
&& \hat{B}_3 \hat{B}_{3} \pa \hat{G}_{11}
 +
\frac{2 i }
{15 (k+5)^3 (4 k+3) (23 k+19) (35 k+47) d(k)}
\nonu \\
&& \times
(5860400 k^9-632913260 k^8-10923563468 k^7-60695730897 k^6
-155310019072 k^5  \nonu \\
&& 
-196539933238 k^4-103437014840 k^3+5752668495 k^2+25811112204 k
\nonu \\
&& +
7074548316)
\,\, \pa^2 \hat{B}_{3}  \hat{G}_{11}
+
\frac{2 i}
{45 (k+5)^3 (4 k+3) (23 k+19) (35 k+47) d(k)} \nonu \\
&& \times
 (25440800 k^9-2200852500 k^8+14362370692 k^7+161018202313 k^6+
637884593013 k^5  \nonu \\
&& +1339766220122 k^4+1591944361470 k^3+1069620889005 k^2+380642976729 k
\nonu \\
&& +56254237236)
\pa \hat{B}_{3}  \pa \hat{G}_{11}
-\frac{i 
}{5 (k+5)^3 (4 k+3) (23 k+19) (35 k+47) d(k)} \nonu \\
&& \times
(2382800 k^9-526442100 k^8-7800137408 k^7-63449924583 k^6
-222775296289 k^5  \nonu \\
&& 
-391081127552 k^4-367447778946 k^3-187054353303 k^2-47640576285 k-4482802494
)
\nonu \\
&& 
\hat{B}_{3}  \pa^2 \hat{G}_{11}
 +
\frac{2 i 
}{3 (k+5)^2 (4 k+3) (23 k+19) (35 k+47) d(k)} \nonu \\
&& \times
(1944320 k^8+356286588 k^7+2069614480 k^6+4074098845 k^5
 \nonu \\
&& 
+685515888 k^4-8451400390 k^3-11704777716 k^2-6211317483 k-1201113252)
\,\, \pa \hat{B}_{3}   \hat{G}_{11} T^{(1)}
\nonu \\
&& 
-\frac{4 i
}{15 (k+5)^2 d(k)}
 (700 k^5+182440 k^4+2222089 k^3+5875122 k^2+5186853 k+1517184)
\nonu \\
&& \times \hat{B}_{3}   \pa \hat{G}_{11}  T^{(1)}
 +
\frac{18 i
}{5 (k+5)^2 d(k)}
 (140 k^4-15704 k^3-16987 k^2-12278 k-3759)
\,\,\hat{B}_{3}    \hat{G}_{11}  \pa T^{(1)}
\nonu \\
&& +
\frac{4 i
}{(k+5)^2 d(k)}
 (140 k^5+36376 k^4+218701 k^3+493088 k^2+395253 k+108198)
\,\,\pa \hat{B}_{3}     T^{(1)} U^{(\frac{3}{2})}
\nonu \\
&& +
 \frac{32 i
}{(k+5)^2 d(k)}
 (70 k^4+3418 k^3+24403 k^2+24696 k+7497)
\,\, \hat{B}_{3}    \pa T^{(1)} U^{(\frac{3}{2})}
\nonu \\
&& 
-\frac{8 i 
}{3 (k+5)^2 d(k)}
(k+11) (140 k^4+35816 k^3+133397 k^2+128682 k+39105)
\,\,\hat{B}_{3}   T^{(1)} \pa U^{(\frac{3}{2})}
\nonu \\
&& +
\frac{i 
}{3 (k+5)^2 (4 k+3) (23 k+19) (35 k+47) d(k)} \nonu \\
&& \times
(2873360 k^8-73242876 k^7-75177200 k^6+2448403711 k^5+12638726883 k^4
 \nonu \\
&&  +26451000746 k^3+26216876742 k^2+12399196131 k+2276873415)
\,\,\pa \hat{B}_{-}  W_{-}^{(\frac{5}{2})}
\nonu \\
&& 
-\frac{i 
}{5 (k+5)^2 d(k)}
(700 k^5+136940 k^4+2634077 k^3+5887441 k^2+5086767 k+1497771)
\nonu \\
&& \times \hat{B}_{-}  \pa W_{-}^{(\frac{5}{2})}
 +
\frac{2 i 
}{5 (k+5)^3 (4 k+3) (23 k+19) (35 k+47) d(k)} \nonu \\
&& \times 
(4057200 k^9-63705180 k^8+489093176 k^7+9681385839 k^6
+33307244179 k^5  \nonu \\
&&  
+58682973860 k^4+69954417718 k^3+54597316527 k^2+23680981935 k+4213781946)
\nonu \\
&& 
\pa^2 \hat{B}_{-}   T_{-}^{(\frac{3}{2})}
+
\frac{2 i 
}{45 (k+5)^3 (4 k+3) (23 k+19) (35 k+47) d(k)} \nonu \\
&& \times
(33300400 k^9-786710820 k^8-33588858352 k^7-290927970583 k^6
-993483522543 k^5  \nonu \\
&& -1611200366930 k^4-1255564971246 k^3-400981470939 k^2+7639614765 k+22181319288)
\nonu \\
&& \pa \hat{B}_{-}   \pa T_{-}^{(\frac{3}{2})}
-\frac{2 i 
}{15 (k+5)^3 d(k)}
(4620 k^6-303272 k^5+284069 k^4
 \nonu \\
&&  +8676218 k^3+15970639 k^2+14596332 k+4814190)
\,\,\hat{B}_{-}   \pa^2  T_{-}^{(\frac{3}{2})}
\nonu \\
&& +
\frac{8 
}{15 (k+5)^3 (4 k+3) (23 k+19) (35 k+47) d(k)} \nonu \\
&& \times (2959600 k^9+73706640 k^8+11545714280 k^7+98331882092 k^6
+362088180609 k^5  \nonu \\
&&  +714852825760 k^4+813787059582 k^3+535190823576 k^2+189365958513 k+28033088148)
\nonu \\
&& \pa \hat{B}_{-} \hat{B}_3  T_{-}^{(\frac{3}{2})}
-\frac{4 
}{15 (k+5)^3 (4 k+3) (23 k+19) (35 k+47) d(k)}
\nonu \\
&& \times
(2959600 k^9+74471040 k^8+11144074160 k^7
+96440879660 k^6+379188540111 k^5  \nonu \\
&&  +766459994185 k^4+839932780782 k^3+507421225350 k^2+159556508451 k+20487965445)
\nonu \\
&&  \hat{B}_{-} \pa \hat{B}_3  T_{-}^{(\frac{3}{2})} 
-\frac{8 
}{45 (k+5)^3 (4 k+3) (23 k+19) (35 k+47) d(k)} \nonu \\
&& \times (2959600 k^9+101695440 k^8+19045114580 k^7+
175807742744 k^6+636727641732 k^5  \nonu \\
&&  +1124931558655 k^4+1021710570792 k^3+469455346242 k^2+92458487880 k+3125306511)
\nonu \\
&& \hat{B}_{-} \hat{B}_3 \pa  T_{-}^{(\frac{3}{2})} 
+
\frac{2 
}{15 (k+5)^3 (4 k+3) (23 k+19) (35 k+47) d(k)} \nonu \\
&& \times
(7604800 k^9+3426113320 k^8+15945441380 k^7
+3054363022 k^6-174821814481 k^5  \nonu \\
&& 
-640438290020 k^4-1105790461418 k^3-988332991854 k^2-440401746177 k
-77901949692)
\nonu \\
&&  \pa \hat{B}_{-} \hat{B}_3 \hat{G}_{12} 
-\frac{4 
}{15 (k+5)^3 (4 k+3) (23 k+19) (35 k+47) d(k)}
\nonu \\
&& \times
(5282200 k^9+1703800980 k^8+15780335450 k^7
+69113521403 k^6+128782205148 k^5  \nonu \\
&&  +41652610828 k^4-191399345898 k^3-274418359281 k^2-144889920396 k-27794092626)
\nonu \\
&& \hat{B}_{-} \pa \hat{B}_3 \hat{G}_{12} 
+
\frac{4 
}{45 (k+5)^3 (4 k+3) (23 k+19) (35 k+47) d(k)} \nonu \\
&& \times
(2959600 k^9-144460960 k^8+9893488880 k^7+106213653384 k^6+
441031556737 k^5  \nonu \\
&&  +1004785859086 k^4+1365157612142 k^3+1068003114072 k^2+440137613385 k
\nonu \\
&& +74061506250)
 \hat{B}_{-}  \hat{B}_3 \pa \hat{G}_{12} 
-\frac{i 
}{15 (k+5)^3 (4 k+3) (7 k+3) (23 k+19) (35 k+47) d(k)} \nonu \\
&& \times 
(191139200 k^{10}-7749094640 k^9-104356934388 k^8-456281526544 k^7
-1568848331935 k^6  \nonu \\
&& -4321638913263 k^5-7748968573414 k^4-8266727299902 k^3-5064223665339 k^2
\nonu \\
&&  -1650190901283 k-222320046492)
\,\, \pa^2 \hat{B}_{-}  \hat{G}_{12} 
\nonu \\
&& +
\frac{i 
}{45 (k+5)^3 (4 k+3) (23 k+19) (35 k+47) d(k)} \nonu \\
&& \times 
(8624000 k^9+271288360 k^8-10676772836 k^7-1583308734 k^6+
252803831981 k^5  \nonu \\
&& +1070453891465 k^4+2053467640822 k^3+1959542792508 k^2+910299502305 k
\nonu \\
&& +165543166449)
\pa \hat{B}_{-}  \pa \hat{G}_{12} 
+
\frac{i 
}{30 (k+5)^3 (4 k+3) (7 k+3) (23 k+19) (35 k+47) d(k)} \nonu \\
&& \times
(190688400 k^{10}+867286700 k^9+198758036316 k^8+2131461414493 k^7
+9101858753056 k^6  \nonu \\
&&  +21156148690905 k^5+29777074153388 k^4+25934015893359 k^3+
13602199981392 k^2 \nonu \\
&&  +3925828133487 k+478031947704)
\,\,\hat{B}_{-}  \pa^2 \hat{G}_{12} 
\nonu \\
&& +
\frac{i 
}{3 (k+5)^2 (4 k+3) (23 k+19) (35 k+47) d(k)} 
(1183840 k^8-4868304 k^7
\nonu \\
&& -1281676804 k^6-11894418124 k^5
-39419175033 k^4-63880286876 k^3-53959409322 k^2
\nonu \\
&& -22938686928 k-3904221249)
\,\, \pa \hat{B}_{-}   \hat{G}_{12} T^{(1)} 
\nonu \\
&& -\frac{2 i 
}{15 (k+5)^2 d(k)}
(700 k^5+151920 k^4-278967 k^3-3121816 k^2-3109023 k-879138)
\nonu \\
&& \times \hat{B}_{-}   \pa \hat{G}_{12} T^{(1)} 
\nonu \\
&& -\frac{3 i
}{5 (k+5)^2 d(k)}
 (8820 k^4-259576 k^3-471093 k^2-522554 k-200769)
\,\,\hat{B}_{-}   \hat{G}_{12} \pa  T^{(1)} 
\nonu \\
&& +
\frac{12 i 
}{(k+5)^2 d(k)}
(140 k^4+16496 k^3+77003 k^2+75822 k+23031)
\,\,\pa \hat{B}_{-}    T^{(1)}  T_{-}^{(\frac{3}{2})} 
\nonu \\
&& +
\frac{4 i 
}{(k+5)^2 d(k)}
(140 k^4-22144 k^3-35785 k^2-29898 k-9117)
\,\,\hat{B}_{-}   \pa  T^{(1)}  T_{-}^{(\frac{3}{2})} 
\nonu \\
&&  +
\frac{4 i 
}{3 (k+5)^2 d(k)}
(140 k^5+36656 k^4+464213 k^3+1267428 k^2+
1128357 k+331074)
\nonu \\
&& \times \hat{B}_{-}     T^{(1)}  \pa T_{-}^{(\frac{3}{2})} 
-\frac{2 
}{15 (k+5)^3 (4 k+3) (23 k+19) (35 k+47) d(k)}
\nonu \\
&& \times (95510800 k^8+38408244420 k^7+361495888280 k^6+
1447326547991 k^5  \nonu \\
&& 
+3038417391789 k^4+3598554205282 k^3+2423190016938 k^2+868591207971 k
\nonu \\
&& +129301479009)
\pa \hat{B}_{+}  \hat{B}_{-}  U^{(\frac{3}{2})}
+
\frac{2 
}{3 (k+5)^3 (4 k+3) (23 k+19) (35 k+47) d(k)}
\nonu \\
&& \times 
(31442320 k^8+7036384908 k^7+70989011288 k^6+301296738077 k^5
 \nonu \\
&&  +658945042677 k^4+804343163854 k^3+554229345378 k^2+202320835929 k+30585453201)
\nonu \\
&&  \hat{B}_{+} \pa \hat{B}_{-}  U^{(\frac{3}{2})} 
-\frac{2 
}{45 (k+5)^3 (4 k+3) (23 k+19) (35 k+47) d(k)} \nonu \\
&& \times (46491200 k^9+11012013880 k^8+142506104620 k^7+
769772348910 k^6+2101544593793 k^5  \nonu \\
&& 
+3054220516967 k^4+2294334973126 k^3+773375682684 k^2+34548832413 k-26917074873)
\nonu \\
&&
 \hat{B}_{+}  \hat{B}_{-}  \pa U^{(\frac{3}{2})}
-\frac{2 
}{15 (k+5)^3 (4 k+3) (23 k+19) (35 k+47) d(k)} \nonu \\
&& \times 
(6762000 k^9+1908430300 k^8+48627633900 k^7+353432296805 k^6
+1197949691060 k^5  \nonu \\
&&  +2119728948768 k^4+2021447952040 k^3+1033567097601 k^2+258684609600 k
\nonu \\
&& +22958463798)
 \pa \hat{B}_{+}  \hat{B}_{-} \hat{G}_{11}
+
 \frac{2 
}{3 (k+5)^3 (4 k+3) (23 k+19) (35 k+47) d(k)} \nonu \\
&& \times (1352400 k^9+377797420 k^8+8252904804 k^7+59585630441 k^6
+203203717514 k^5  \nonu \\
&& 
+362421262734 k^4+349580133100 k^3+182023203441 k^2+47043609966 k+4471146756)
\nonu \\
&&  \hat{B}_{+}  \pa \hat{B}_{-} \hat{G}_{11}
-\frac{2 
}{45 (k+5)^3 (4 k+3) (23 k+19) (35 k+47) d(k)}
\nonu \\
&& \times
(9721600 k^9+1970357340 k^8+23136647360 k^7+172242034025 k^6
+639273998700 k^5  \nonu \\
&& +1183893904429 k^4+1129231469520 k^3+552428950563 k^2+122884486020 k+7574943339)
\nonu \\
&& \hat{B}_{+}  \hat{B}_{-} \pa \hat{G}_{11}
\nonu \\
&& +
\frac{2 
}{3 (k+5)^2 d(k)}
(1540 k^5-64244 k^4-112145 k^3+612249 k^2+
622737 k+174879)
\,\, \pa \hat{G}_{11} W^{(2)}
\nonu \\
&& +
\frac{1
}{(k+5)^2 d(k)}
(140 k^4-118744 k^3-317755 k^2-294198 k-89487 )
\,\,\hat{G}_{11} \pa W^{(2)}
\nonu \\
&& +
\frac{2
}{(k+5)^2 d(k)}
 (140 k^5-16404 k^4-41507 k^3-228111 k^2-224289 k-66933)
\,\,\pa \hat{G}_{11} T^{(2)}
\nonu \\
&& 
-\frac{3 
}{(k+5)^2 d(k)}
(420 k^4+55928 k^3+249807 k^2+245086 k+74451)
\,\,\hat{G}_{11} \pa T^{(2)}
\nonu \\
&& 
\frac{1
}{45 (k+5)^3 (7 k+3) (23 k+19) (35 k+47) d(k)}
\nonu \\
&& \times 
(6252400 k^{10}+66733660 k^9-3102835140 k^8-29936667307 k^7
+11944510994 k^6  \nonu \\
&& +388806577419 k^5+676758319416 k^4+204081928107 k^3-276067361862 k^2
\nonu \\
&&  -216379218375 k-44095955472 ) 
\,\, \pa^3 \hat{G}_{11}
\nonu \\
&& +
\frac{1
}{15 (k+5)^2 (7 k+3) (23 k+19) d(k)} \nonu \\
&& \times (
553840 k^7+44985756 k^6+915326484 k^5+3691085405 k^4
 \nonu \\
&&  +6135700380 k^3+5003354970 k^2+1999123848 k+313751637)
\,\, \pa^2 \hat{G}_{11} T^{(1)}
\nonu \\
&& +
\frac{2 (k+18)}{15 (k+5)^2} \,\, \pa \hat{G}_{11} \pa T^{(1)}
-\frac{2 
}{15 (k+5)^2 (7 k+3) (23 k+19) d(k)} \nonu \\
&& \times 
(415380 k^7-60187208 k^6-327315477 k^5-770052990 k^4
 \nonu \\
&&  -1196865970 k^3-1121328120 k^2-540611109 k-102874266)
\,\, \hat{G}_{11} \pa^2 T^{(1)}
\nonu \\
&& -\frac{8
}{3 (k+5)^2 d(k)}
 (280 k^5-14328 k^4+2913 k^3+191954 k^2+193041 k+56736)
\,\,\pa \hat{G}_{12} U_{+}^{(2)}
\nonu \\
&& +
\frac{1}
{(k+5)^2 d(k)}
(280 k^5-4668 k^4+369210 k^3+1205279 k^2+1126128 k+338031)
\,\,\hat{G}_{12}  \pa U_{+}^{(2)}
\nonu \\
&& +
\frac{4 (k+8)}{3 (k+5)^2} 
\,\, \pa \hat{G}_{21}  U_{-}^{(2)}
+\frac{4}{k+5} \,\,\hat{G}_{21}  \pa U_{-}^{(2)}
-\frac{3 (3 k+11)}{(k+5)^2} 
\,\, \pa^2 T^{(1)} U^{(\frac{3}{2})}
-\frac{k-19}{3 (k+5)^2}  \,\, \pa T^{(1)} \pa U^{(\frac{3}{2})}
\nonu \\
&& +
\frac{2 
}{3 (k+5)^2 d(k)}
(700 k^5-12300 k^4+182253 k^3+721091 k^2+650451 k+186885)
\,\, T^{(1)} \pa^2 U^{(\frac{3}{2})}
\nonu \\
&& -\frac{1}{k+5} \,\, \pa T^{(1)}  U^{(\frac{5}{2})}
-\frac{12 k^2+k-199}{(k+5) (23 k+19)}  \,\, T^{(1)}  \pa U^{(\frac{5}{2})}
\nonu \\
&& 
+
\frac{12 (175 k^3+1069 k^2+1687 k+513)}{(k+5)^2 (7 k+3) (35 k+47)}
\,\, \pa \hat{T}  U^{(\frac{3}{2})}
-\frac{4 
}{3 (k+5)^2 (7 k+3) (23 k+19) (35 k+47) d(k)} \nonu \\
&& \times
(592900 k^8-261590280 k^7-3358410661 k^6-14782673878 k^5
 \nonu \\
&& 
-34491090345 k^4-44904443924 k^3-32524028751 k^2-12287294046 k-1901958327)
\nonu \\ 
&& 
\hat{T}  \pa U^{(\frac{3}{2})} 
-\frac{2 
}{(k+5)^2 (7 k+3) (23 k+19) (35 k+47) d(k)} \nonu \\
&& \times
(540960 k^9-4041576 k^8-27372140 k^7+545792458 k^6+4762996989 k^5
 \nonu \\
&&  
+11888154956 k^4+12729357194 k^3+6483721926 k^2+1439928189 k+80290116)
\nonu \\
&& \pa \hat{T} \hat{G}_{11} 
-\frac{4 
}{3 (k+5)^2 (7 k+3) (23 k+19) (35 k+47) d(k)} \nonu \\
&& \times
(537040 k^9-72604 k^8-220189512 k^7-117877545 k^6+3323246033 k^5
\label{9halfone}
 \\
&& \left.  
+6277062454 k^4+1036583346 k^3-4078755189 k^2-2879912043 k-578339676)
 \hat{T} \pa \hat{G}_{11} \right](w), 
\nonu
\eea
where we introduce
\bea
d(k) \equiv \left(140 k^4-2824 k^3+20609 k^2+22962 k+6957 \right).
\label{dk}
\eea
Compared to the last three expressions below, the equation (\ref{9halfone}) 
is rather complicated. The common denominator appearing in the most of 
the terms denoted by $d(k)$ in (\ref{dk}) is not a simple polynomial of $k$.
As seen in the fusion rules in section $4$, this is the only case where
the second higher spin currents in (\ref{nexthigher}) 
with integer or half integer spins appear. 
When one looks at the first couple of terms  in 
(\ref{9halfone}), the spin-$\frac{3}{2}$ currents of 
the large ${\cal N}=4$ nonlinear superconformal algebra are multiplied by
the higher spin currents with integer spin.

The first order singular term in the last OPE of
(\ref{v5halfw3}) can be summarized by
the following expression
\bea
& & \{ V^{(\frac{5}{2})} \, 
 W^{(3)} \}_{-1}(w)  = 
\left[ 
\frac{2 i}{(k+5)} \,\,\hat{A}_3 {\bf R^{(\frac{7}{2})}}
+\frac{i}{(k+5)} \,\,\hat{A}_{-}  {\bf S_{-}^{(\frac{7}{2})}}
-\frac{2 i}{(k+5)} \,\,\hat{B}_3 {\bf R^{(\frac{7}{2})}}
\right. \nonu \\
&& -\frac{16}{3 (k+5)^2} \,\,\hat{B}_3 \hat{B}_3 {\bf R^{(\frac{5}{2})}}
 -\frac{16}{3 (k+5)^2} \,\,\hat{B}_{+} \hat{B}_3  {\bf P_{+}^{(\frac{5}{2})}}
+ \frac{8}{3 (k+5)^2}  \,\, \hat{B}_{+} \hat{B}_{-} {\bf R^{(\frac{5}{2})}}
+ \frac{8 (k-3)}{(23 k+19)} \,\, \hat{T}  {\bf R^{(\frac{5}{2})}}
\nonu \\
& & -\frac{i}{(k+5)} \,\,\pa \hat{A}_3 {\bf R^{(\frac{5}{2})}}
+ \frac{i (3 k+23)}{5 (k+5)^2} \,\,\hat{A}_3 \pa {\bf R^{(\frac{5}{2})}}
-\frac{i (2 k+11)}{(k+5)^2} \,\,\pa \hat{A}_{-}  {\bf P_{-}^{(\frac{5}{2})}}
\nonu \\
&& -\frac{3 i (2 k+9)}{5 (k+5)^2} \,\,\hat{A}_{-} \pa  {\bf P_{-}^{(\frac{5}{2})}}
+ \frac{i (3 k+23)}{3 (k+5)^2} \,\,\pa \hat{B}_3  {\bf R^{(\frac{5}{2})}}
-\frac{i (9 k+149)}{15 (k+5)^2} \,\,\hat{B}_3  \pa {\bf R^{(\frac{5}{2})}}
\nonu \\
&& +
\frac{8 i}{3 (k+5)^2} \,\,\pa \hat{B}_{+} {\bf P_{+}^{(\frac{5}{2})}}
-\frac{8 i}{3 (k+5)^2} \,\,\hat{B}_{+} \pa {\bf P_{+}^{(\frac{5}{2})}}
+ \frac{(152 k^3-171 k^2-4370 k-7407)}{30 (k+5)^2 (23 k+19)} \,\,\pa^2
 {\bf R^{(\frac{5}{2})}}
\nonu \\
&& -\frac{3 (k+3)}{2 (k+5)} \,\,\pa  {\bf R^{(\frac{7}{2})}}
\nonu \\
&&
+\frac{2}{(k+5)} \,\,
V_{-}^{(2)} W_{+}^{(\frac{5}{2})}
-\frac{4}{(k+5)} \,\,T^{(2)} 
 V^{(\frac{5}{2})}
+\frac{8 i}{(k+5)^2} \,\,\hat{A}_3  V^{(\frac{3}{2})} W^{(2)}
\nonu \\\
&&+ 
\frac{8 i}{(k+5)^2}  \,\,\hat{A}_3  T_{-}^{(\frac{3}{2})} V_{+}^{(2)}
+\frac{20}{(k+5)^2} \,\,\hat{A}_3 \hat{A}_3   V^{(\frac{5}{2})}
-\frac{16 i}{(k+5)^3} \,\,\hat{A}_3 \hat{A}_3 \hat{A}_3  V^{(\frac{3}{2})}
\nonu \\
&& + \frac{80 i}{3 (k+5)^3}
 \,\,\hat{A}_3 \hat{A}_3 \hat{B}_3  V^{(\frac{3}{2})}
+\frac{32 i}{3 (k+5)^3}\,\,  \hat{A}_3 \hat{A}_3 \hat{B}_{+}  T_{+}^{(\frac{3}{2})}
+\frac{56 i}{3 (k+5)^3}\,\,  \hat{A}_3 \hat{A}_3 \hat{B}_{+} \hat{G}_{21}
\nonu \\
&& -\frac{24}{(k+5)^2}\,\,  \hat{A}_3 \hat{B}_3  V^{(\frac{5}{2})}
-\frac{208 i}{3 (k+5)^3}\,\,  \hat{A}_3 \hat{B}_3 \hat{B}_{3}  
V^{(\frac{3}{2})}
+\frac{128 i}{3 (k+5)^3}\,\,  \hat{A}_3 \hat{B}_{+} \hat{B}_{3}  T_{+}^{(\frac{3}{2})}
\nonu \\
&& +
\frac{80 i}{3 (k+5)^3}\,\, \hat{A}_3 \hat{B}_{+} \hat{B}_{3} \hat{G}_{21}
-\frac{8 i}{(k+5)^2}\,\, \hat{A}_3 \hat{G}_{21}  V_{-}^{(2)}
-\frac{4 i}{(k+5)^2}\,\, \hat{A}_{-}  U_{-}^{(2)}  V^{(\frac{3}{2})}
\nonu \\
&& + \frac{4 i}{(k+5)^2}\,\,  \hat{A}_{-}  T_{-}^{(\frac{3}{2})} T^{(2)}
+ \frac{8}{(k+5)^2}
\,\,\hat{A}_{-} \hat{A}_3  W_{-}^{(\frac{5}{2})}
-\frac{8 i}{(k+5)^3}
 \,\,\hat{A}_{-}  \hat{A}_3 \hat{A}_3 T_{-}^{(\frac{3}{2})}
\nonu \\
&& -\frac{8 i}{(k+5)^3}\,\,  \hat{A}_{-}  \hat{A}_3 \hat{A}_3 \hat{G}_{12}
-\frac{16 i}{(k+5)^3}\,\,  \hat{A}_{-}  \hat{A}_3 \hat{B}_3 T_{-}^{(\frac{3}{2})}
+\frac{16 i}{(k+5)^3}\,\,  \hat{A}_{-}  \hat{A}_3 \hat{B}_3 \hat{G}_{12}
\nonu \\
&& -\frac{16 i}{(k+5)^3}\,\, \hat{A}_{-}  
\hat{A}_3 \hat{B}_{+} U^{(\frac{3}{2})}
+\frac{24 i}{(k+5)^3}\,\,  \hat{A}_{-}  \hat{B}_3 \hat{B}_3 T_{-}^{(\frac{3}{2})}
-\frac{8 i}{(k+5)^3}\,\,  \hat{A}_{-}  \hat{B}_3 \hat{B}_3 \hat{G}_{12}
\nonu \\
&& + \frac{16 i}{(k+5)^3} \,\,  \hat{A}_{-}  
\hat{B}_{+} \hat{B}_{3} U^{(\frac{3}{2})}
+\frac{16 i}{(k+5)^3}  \,\,\hat{A}_{-}  \hat{B}_{+} \hat{B}_3 \hat{G}_{11}
-\frac{8 i}{(k+5)^3}  \,\,\hat{A}_{-}  \hat{B}_{+} \hat{B}_{-} T_{-}^{(\frac{3}{2})}
\nonu \\
& & -\frac{16 i}{(k+5)^3} \,\, \hat{A}_{-}  \hat{B}_{+} \hat{B}_{-} \hat{G}_{12}
-\frac{4 i}{(k+5)^2} \,\,\hat{A}_{-}  
 \hat{G}_{11} V_{-}^{(2)}
+\frac{4 i}{(k+5)^2} \,\,\hat{A}_{-}  
 \hat{G}_{12} W^{(2)}
\nonu \\
& & + \frac{4 i}{(k+5)^2}\,\,  \hat{A}_{-}  
 \hat{G}_{12} T^{(2)}
+ \frac{4 i}{(k+5)^2} \,\, \hat{A}_{-}  
 \hat{G}_{22} U_{-}^{(2)}
+ \frac{4}{(k+5)^2} \,\, \hat{A}_{+} \hat{A}_{-}  V^{(\frac{5}{2})}
\nonu \\
&&  -\frac{8 i}{(k+5)^3}\,\, 
\hat{A}_{+}  \hat{A}_{-} \hat{A}_{-} T_{-}^{(\frac{3}{2})}
-\frac{8 i}{(k+5)^3} \,\,  \hat{A}_{+}  \hat{A}_{-} \hat{A}_{-} \hat{G}_{12}
-\frac{64 i}{3 (k+5)^3} 
 \,\,\hat{A}_{+}  
\hat{A}_{-} \hat{B}_{3} V^{(\frac{3}{2})}
\nonu \\
&& +  \frac{16 i}{(k+5)^3}
 \,\,\hat{A}_{+}  \hat{A}_{-} \hat{B}_{3} \hat{G}_{22}
+\frac{56 i}{3 (k+5)^3}
 \,\,\hat{A}_{+}  \hat{A}_{-} \hat{B}_{+} T_{+}^{(\frac{3}{2})}
+\frac{80 i}{3 (k+5)^3} 
 \,\,\hat{A}_{+}  \hat{A}_{-} \hat{B}_{+} \hat{G}_{21}
\nonu \\
&& +
 \frac{8 i}{3 (k+5)^2} \,\,\hat{B}_{3} V^{(\frac{3}{2})} W^{(2)}
+\frac{32 i}{3 (k+5)^2}  \,\,\hat{B}_{3}  T^{(2)}  V^{(\frac{3}{2})}
-\frac{8 i}{(k+5)^2}  \,\,\hat{B}_{3} T_{-}^{(\frac{3}{2})}  V_{+}^{(2)}
\nonu \\
&& -\frac{32 i}{3 (k+5)^2}  \,\,\hat{B}_{3} T_{+}^{(\frac{3}{2})}  V_{-}^{(2)}
+ \frac{44}{3 (k+5)^2} \,\,\hat{B}_{3} \hat{B}_{3} V^{(\frac{5}{2})}
-\frac{32}{3 (k+5)^2}\,\,
\hat{B}_{3} \hat{B}_{3} T^{(1)} V^{(\frac{3}{2})}
\nonu \\
&& -\frac{8 i}{3 (k+5)^2} 
 \,\,\hat{B}_{3}  
 \hat{G}_{21} V_{-}^{(2)}
-\frac{28 i}{3 (k+5)^2} \,\, \hat{B}_{-} V^{(\frac{3}{2})}  V_{-}^{(2)}
-\frac{16 i}{3 (k+5)^2}  \,\,\hat{B}_{+} T_{+}^{(\frac{3}{2})}  W^{(2)}
\nonu \\
&&  + \frac{16 i}{3 (k+5)^2}\,\,  \hat{B}_{+} U_{+}^{(2)}  V^{(\frac{3}{2})} 
-\frac{16 i}{3 (k+5)^2}\,\,  \hat{B}_{+} T_{+}^{(\frac{3}{2})}  T^{(2)}
+ \frac{8 i (4 k+33)}{3 (k+5)^3} 
 \,\,\hat{B}_{+}  \hat{B}_{3} \hat{B}_{3} T_{+}^{(\frac{3}{2})}
\nonu \\
&& + \frac{32 i (k+9)}{3 (k+5)^3} 
 \,\,\hat{B}_{+}  \hat{B}_{3} \hat{B}_{3} \hat{G}_{21}
+ \frac{32}{3 (k+5)^2} \,\,  \hat{B}_{+} \hat{B}_{3} \hat{G}_{21} T^{(1)} 
+ \frac{32}{3 (k+5)^2}  \,\,\hat{B}_{+}  \hat{B}_{3} T^{(1)} T_{+}^{(\frac{3}{2})}
\nonu \\
&& -\frac{4}{3 (k+5)^2} \,\, \hat{B}_{+} \hat{B}_{-} V^{(\frac{5}{2})}
+ \frac{8 i}{3 (k+5)^3} \,\, \hat{B}_{+}  \hat{B}_{-} \hat{B}_{3} V^{(\frac{3}{2})}
+ \frac{16}{3 (k+5)^2} \,\, \hat{B}_{+}  \hat{B}_{-} T^{(1)} V^{(\frac{3}{2})}
\nonu \\
&& -\frac{2 i (4 k+13)}{3 (k+5)^3} \,\, \hat{B}_{+}  \hat{B}_{+} \hat{B}_{-} 
T_{+}^{(\frac{3}{2})}
-\frac{2 i (4 k+1)}{3 (k+5)^3} 
 \,\,\hat{B}_{+}  \hat{B}_{+} \hat{B}_{-} \hat{G}_{21}
-\frac{28 i}{3 (k+5)^2} \,\,\hat{B}_{+} \hat{G}_{21} W^{(2)}
\nonu \\
&& -\frac{16 i}{3 (k+5)^2} \,\,\hat{B}_{+} \hat{G}_{21} T^{(2)}
-\frac{4}{(k+5)^2}  \,\,\hat{G}_{21} \hat{G}_{22} T_{-}^{(\frac{3}{2})}
-\frac{8 \left(7 k^3+2 k^2-134 k-69\right)}{(k+5) (7 k+3) (23 k+19)}
\,\,\hat{T}  V^{(\frac{5}{2})}
\nonu \\
&& -\frac{48 i (9 k+17)}{(k+5)^2 (35 k+47)}
\,\,\hat{T} \hat{A}_3  V^{(\frac{3}{2})}
-\frac{32 i \left(133 k^3+1178 k^2+2493 k+1368\right)}{
3 (k+5)^2 (23 k+19) (35 k+47)}
\,\,\hat{T} \hat{A}_3 \hat{G}_{22}
\nonu \\
& & -\frac{8 i \left(70 k^4+3418 k^3+24403 k^2+24696 k+7497\right)}{
3 (k+5)^2 (7 k+3) (23 k+19) (35 k+47)}
\,\, \hat{T} \hat{A}_3 T_{-}^{(\frac{3}{2})}
\nonu \\
&& -\frac{8 i \left(1862 k^4+27636 k^3+67171 k^2+57786 k+15669\right)}{
3 (k+5)^2 (7 k+3) (23 k+19) (35 k+47)}
\,\,\hat{T} \hat{A}_{-} \hat{G}_{12}
\nonu \\
&& +\frac{16 i (k+21) (7 k-5)}{3 (k+5)^2 (7 k+3) (35 k+47)} 
\,\,\hat{T} \hat{B}_3  V^{(\frac{3}{2})}
+ \frac{32 i \left(133 k^3+1178 k^2+2493 k+1368\right)}{
3 (k+5)^2 (23 k+19) (35 k+47)}
\,\,\hat{T} \hat{B}_3  \hat{G}_{22}
\nonu \\
&& + \frac{32 i (4 k+3)}{3 (k+5)^2 (7 k+3)} 
\,\,\hat{T} \hat{B}_{+} T_{+}^{(\frac{3}{2})}
+\frac{8 i (37 k+21)}{3 (k+5)^2 (7 k+3)} 
\,\,\hat{T} \hat{B}_{+}  \hat{G}_{21}
\nonu \\
&& +\frac{2 (k+7)}{(k+5)^2} \,\, \pa 
V^{(\frac{3}{2})} W^{(2)}
-\frac{2 (3 k+11)}{(k+5)^2}  \,\,V^{(\frac{3}{2})} \pa W^{(2)}
\nonu \\
&& +
\frac{(76810 k^4+629970 k^3+2551461 k^2+4681236 k+3176039)}{45 (k+5)^3 (23 k+19) (35 k+47)} \,\, \pa^3  V^{(\frac{3}{2})}
\nonu \\
&& 
-\frac{2 \left(38 k^3-171 k^2-3927 k-3986\right)}{15 (k+5)^2 (23 k+19)}
\,\,\pa^2   V^{(\frac{5}{2})} 
-\frac{4}{(k+5)^2} \,\,\pa T^{(2)}  V^{(\frac{3}{2})}
+\frac{8 (2 k+5)}{3 (k+5)^2} \,\,T^{(2)}  \pa V^{(\frac{3}{2})}
\nonu \\
&& 
-\frac{6 (k+3)}{(k+5)^2} \,\,\pa  T_{-}^{(\frac{3}{2})}
V_{+}^{(2)}
-\frac{6 (k+3)}{(k+5)^2} \,\,T_{-}^{(\frac{3}{2})}
\pa V_{+}^{(2)} -\frac{2 (5 k+24)}{3 (k+5)^2} \,\,\pa  
T_{+}^{(\frac{3}{2})} V_{-}^{(2)}
\nonu \\
&&  -\frac{6 (k+4)}{(k+5)^2} \,\,T_{+}^{(\frac{3}{2})} \pa V_{-}^{(2)}
-\frac{2 i \left(420 k^3+1039 k^2-4212 k-7591\right)}{5 (k+5)^3 (35 k+47)}
\,\,\pa^2 \hat{A}_3 V^{(\frac{3}{2})} 
\nonu \\
&& -\frac{2 i \left(12 k^2+349 k+1277\right)}{15 (k+5)^3} 
 \,\,\pa \hat{A}_3 \pa V^{(\frac{3}{2})} 
+ \frac{4 i \left(420 k^3+3044 k^2+3497 k+1437\right)}{15 (k+5)^3 (35 k+47)}
 \,\,\hat{A}_3 \pa^2 V^{(\frac{3}{2})} 
\nonu \\
&&  +
\frac{2 i (5 k+28)}{(k+5)^2} \,\,\pa \hat{A}_3   V^{(\frac{5}{2})} 
+ \frac{6 i \left(71 k^2-25 k+116\right)}{5 (k+5)^2 (23 k+19)} 
\,\, \hat{A}_3   \pa V^{(\frac{5}{2})} 
+ \frac{8 (11 k+60)}{5 (k+5)^3} \,\,\pa \hat{A}_3 \hat{A}_3  V^{(\frac{3}{2})} 
\nonu \\
&& -\frac{4 (19 k+85)}{15 (k+5)^3}
\,\,\hat{A}_3 \hat{A}_3  \pa V^{(\frac{3}{2})} 
-\frac{32 (k+2)}{5 (k+5)^3} \,\,\pa \hat{A}_3 \hat{A}_3  \hat{G}_{22}
+\frac{16 (14 k-27)}{15 (k+5)^3} \,\,\hat{A}_3 \hat{A}_3  \pa \hat{G}_{22}
\nonu \\
&& -\frac{8 (11 k+20)}{5 (k+5)^3} \,\,\pa 
\hat{A}_3 \hat{B}_3   V^{(\frac{3}{2})} 
-\frac{8 (16 k+53)}{5 (k+5)^3} \,\,\hat{A}_3 \pa \hat{B}_3  V^{(\frac{3}{2})} 
+ \frac{8 (13 k+67)}{5 (k+5)^3}\,\,  \hat{A}_3 \hat{B}_3  \pa V^{(\frac{3}{2})} 
\nonu \\
&& +
\frac{32 (k-3)}{5 (k+5)^3}\,\,  \pa \hat{A}_3 \hat{B}_3  \hat{G}_{22}
+ \frac{16 (3 k-5)}{5 (k+5)^3}\,\, \hat{A}_3 \pa \hat{B}_3  \hat{G}_{22}
-\frac{16 (5 k-9)}{5 (k+5)^3} \,\,\hat{A}_3  \hat{B}_3  \pa \hat{G}_{22}
\nonu \\
&& -\frac{80}{3 (k+5)^3} \,\,\pa \hat{A}_3 \hat{B}_{+} T_{+}^{(\frac{3}{2})}
+\frac{64 (k+6)}{5 (k+5)^3} \,\,\hat{A}_3 \pa \hat{B}_{+} T_{+}^{(\frac{3}{2})}
+\frac{64 (k+6)}{5 (k+5)^3} \,\,\hat{A}_3 \hat{B}_{+} \pa T_{+}^{(\frac{3}{2})}
\nonu \\
&& 
-\frac{140}{3 (k+5)^3}  \,\,\pa \hat{A}_3 \hat{B}_{+}  \hat{G}_{21}
+ \frac{8 \left(6 k^3+25 k^2+807 k+608\right)}{5 (k+5)^3 (23 k+19)}
\,\,\hat{A}_3  \pa \hat{B}_{+}  \hat{G}_{21}
\nonu \\
&& -\frac{16 \left(2 k^3-30 k^2-491 k-399\right)}{5 (k+5)^3 (23 k+19)}
 \,\,\hat{A}_3 \hat{B}_{+}  \pa \hat{G}_{21}
\nonu \\
&& -\frac{4 i \left(11830 k^4+129003 k^3+335611 k^2+296749 k+80199\right)}{
15 (k+5)^3 (23 k+19) (35 k+47)}
\,\,\pa^2 \hat{A}_3 \hat{G}_{22}
\nonu \\
&& -\frac{4 i \left(16 k^2+24 k+261\right)}{15 (k+5)^3}
\,\,\pa \hat{A}_3 \pa \hat{G}_{22}
\nonu \\
&& + \frac{4 i \left(3850 k^4+22312 k^3+115709 k^2+
199570 k+97603\right)}{5 (k+5)^3 (23 k+19) (35 k+47)}
 \,\,\hat{A}_3 \pa^2 \hat{G}_{22}
\nonu \\
&&  + \frac{24 i}{5 (k+5)^2} \,\,\hat{A}_3 \pa \hat{G}_{22} T^{(1)}
+ \frac{24 i}{5 (k+5)^2}  \,\,\hat{A}_3  \hat{G}_{22}  \pa T^{(1)}
-\frac{12 i}{(k+5)^2} \,\,\hat{A}_3 \pa T^{(1)}   V^{(\frac{3}{2})} 
\nonu \\
&& +
\frac{2 i (k+9)}{(k+5)^2} \,\,\pa \hat{A}_{-}  W_{-}^{(\frac{5}{2})} 
+ \frac{2 i (3 k-1)}{5 (k+5)^2}  \,\,\hat{A}_{-}  \pa W_{-}^{(\frac{5}{2})} 
\nonu \\
&& +
\frac{2 i \left(700 k^5+144745 k^4+2231388 k^3+5681302 k^2+4497168 k+1106217\right)}{15 (k+5)^3 (7 k+3) (23 k+19) (35 k+47)}
\,\,\pa^2 \hat{A}_{-}  T_{-}^{(\frac{3}{2})}
\nonu \\
&& + \frac{2 i (53 k+211)}{15 (k+5)^3}  \,\,\pa \hat{A}_{-}  \pa 
T_{-}^{(\frac{3}{2})}
\nonu \\
&& +
\frac{4 i \left(1365 k^5-7997 k^4+34446 k^3+256586 k^2+191373 k+50787\right)}{15 (k+5)^3 (7 k+3) (23 k+19) (35 k+47)} \,\,\hat{A}_{-}  
\pa^2 
T_{-}^{(\frac{3}{2})}
\nonu \\
&& + \frac{88}{5 (k+5)^3} 
\,\,\pa \hat{A}_{-} \hat{A}_3  T_{-}^{(\frac{3}{2})}
-\frac{4 (16 k-3)}{5 (k+5)^3} \,\,\hat{A}_{-} \pa \hat{A}_3  T_{-}^{(\frac{3}{2})}
+ \frac{8 (16 k+45)}{15 (k+5)^3} \,\,\hat{A}_{-} \hat{A}_3  \pa T_{-}^{(\frac{3}{2})}
\nonu \\
&& +\frac{24}{(k+5)^3} \,\,\pa \hat{A}_{-} \hat{A}_3 \hat{G}_{12}
+ \frac{16 \left(55 k^2+683 k+552\right)}{5 (k+5)^3 (23 k+19)}
 \,\,\hat{A}_{-} \pa \hat{A}_3 \hat{G}_{12}
\nonu \\
&& -\frac{32 \left(55 k^2+223 k+172\right)}{15 (k+5)^3 (23 k+19)}
 \,\,\hat{A}_{-} \hat{A}_3 \pa \hat{G}_{12}
+ \frac{8 (60 k+307)}{15 (k+5)^3}  \,\,
\pa \hat{A}_{-} \hat{B}_3  T_{-}^{(\frac{3}{2})}
\nonu \\
&& 
+ \frac{52 (3 k+10)}{5 (k+5)^3} \,\, \hat{A}_{-} \pa 
\hat{B}_3  T_{-}^{(\frac{3}{2})}
+ \frac{8 \left(3 k^3-128 k^2-1124 k-893\right)}{5 (k+5)^3 (23 k+19)}
 \,\,\hat{A}_{-} \pa \hat{B}_3 \hat{G}_{12}
\nonu \\
&& + \frac{8 (17 k+64)}{5 (k+5)^3}
 \,\,\hat{A}_{-} \hat{B}_3  \pa T_{-}^{(\frac{3}{2})}
-\frac{8 (k+8)}{(k+5)^3} \,\,\pa \hat{A}_{-} \hat{B}_3 \hat{G}_{12}
\nonu \\
&&
-\frac{16 (k+6) \left(k^2+28 k+19\right)}{5 (k+5)^3 (23 k+19)} 
 \,\,\hat{A}_{-} \hat{B}_3 \pa \hat{G}_{12}
-\frac{2 (k+9)}{(k+5)^3} \,\,\pa \hat{A}_{-} \hat{B}_{+}  U^{(\frac{3}{2})} 
\nonu \\
&& 
+ \frac{2 (57 k-107)}{15 (k+5)^3} \,\, \hat{A}_{-} \hat{B}_{+}  \pa U^{(\frac{3}{2})} 
+ \frac{8 (k+4)}{(k+5)^3}  \,\,\pa \hat{A}_{-} \hat{B}_{+}  \hat{G}_{11}
+ \frac{4 (5 k+66)}{5 (k+5)^3}  \,\,\hat{A}_{-} \pa \hat{B}_{+} \hat{G}_{11}
\nonu \\
&& + \frac{8 (30 k+19)}{15 (k+5)^3} \,\, \hat{A}_{-} \hat{B}_{+} \pa \hat{G}_{11}
-\frac{2 (41 k+109)}{5 (k+5)^3} \,\, \hat{A}_{-} \pa \hat{B}_{+}  U^{(\frac{3}{2})}
\nonu \\
&& + \frac{4 i \left(1862 k^5+53851 k^4+502837 k^3+1086773 k^2+829641 k+198900\right)}{3 (k+5)^3 (7 k+3) (23 k+19) (35 k+47)}
\,\,\pa^2 \hat{A}_{-} \hat{G}_{12}
\nonu \\
&& -\frac{4 i (k+11)}{3 (k+5)^2} \,\,\pa \hat{A}_{-} \pa \hat{G}_{12}
\nonu \\
&& + \frac{2 i \left(12495 k^5-129899 k^4+194276 k^3+1400108 k^2+1543957 k+450183\right)}{15 (k+5)^3 (7 k+3) (23 k+19) (35 k+47)} 
\,\,\hat{A}_{-} \pa^2 \hat{G}_{12}
\nonu \\
&& 
-\frac{4 i}{(k+5)^2} \pa \hat{A}_{-} \hat{G}_{12} T^{(1)}
+ \frac{8 i \left(k^2-44 k-53\right)}{5 (k+5)^2 (23 k+19)}
\hat{A}_{-} \pa \hat{G}_{12} T^{(1)}
\nonu \\
&& -\frac{6 i}{(k+5)^2} \,\,\hat{A}_{-} \pa T^{(1)}  T_{-}^{(\frac{3}{2})}
+ \frac{4 i}{(k+5)^2} \,\,\hat{A}_{-}  T^{(1)}  \pa T_{-}^{(\frac{3}{2})}
+ \frac{4 (16 k+69)}{5 (k+5)^3} \,\,\pa \hat{A}_{+} \hat{A}_{-}  V^{(\frac{3}{2})} 
\nonu \\
&& + \frac{8 (k+4)}{(k+5)^3} \,\,\hat{A}_{+} \pa \hat{A}_{-}  V^{(\frac{3}{2})} 
-\frac{16 (8 k+47)}{15 (k+5)^3} \,\,\hat{A}_{+} \hat{A}_{-}  \pa V^{(\frac{3}{2})} 
-\frac{4 (4 k-5)}{5 (k+5)^3} \,\,\pa \hat{A}_{+} \hat{A}_{-} \hat{G}_{22}
\nonu 
\\
&& -\frac{8}{(k+5)^3}  \,\,\hat{A}_{+} \pa \hat{A}_{-} \hat{G}_{22}
+ \frac{232 k}{15 (k+5)^3} \,\, \hat{A}_{+} \hat{A}_{-} \pa \hat{G}_{22}
-\frac{4 i \left(3 k^2+98 k+31\right)}{5 (k+5)^2 (23 k+19)}
\hat{A}_{-} \hat{G}_{12} \pa T^{(1)}
\nonu \\
&& 
-\frac{2 i \left(735 k^4-25683 k^3-162575 k^2-217529 k-70428\right)}{5 (k+5)^3 (7 k+3) (35 k+47)} \,\,\pa^2 \hat{B}_3  V^{(\frac{3}{2})} 
\nonu \\
&& +\frac{2 i \left(21 k^2+671 k+1594\right)}{45 (k+5)^3} \,\,\pa  
\hat{B}_3  \pa V^{(\frac{3}{2})} 
-\frac{4 i \left(210 k^3+10102 k^2+44277 k+45389\right)}{45 (k+5)^3 (35 k+47)}  \,\,\hat{B}_3  \pa^2 V^{(\frac{3}{2})} 
\nonu \\
&&  -\frac{2 i (12 k+53)}{3 (k+5)^2} \,\,\pa \hat{B}_3  V^{(\frac{5}{2})} 
-\frac{2 i \left(624 k^2+2225 k-1691\right)}{15 (k+5)^2 (23 k+19)}
 \,\,\hat{B}_3  \pa V^{(\frac{5}{2})} 
+ \frac{16 (24 k+37)}{15 (k+5)^3} \,\,\pa \hat{B}_3  \hat{B}_3 V^{(\frac{3}{2})} 
\nonu \\
&& -\frac{4 (257 k+1591)}{45 (k+5)^3} \,\,
\hat{B}_3  \hat{B}_3 \pa V^{(\frac{3}{2})} 
-\frac{16 (3 k-5)}{5 (k+5)^3} \,\,\pa \hat{B}_3  \hat{B}_3 \hat{G}_{22}
+ \frac{16 k}{15 (k+5)^3} \,\,\hat{B}_3  \hat{B}_3 \pa \hat{G}_{22}
\nonu \\
&& +
\frac{4 i \left(19075 k^4+87562 k^3-48383 k^2-348072 k-217170\right)}{15 (k+5)^3 (23 k+19) (35 k+47)} \,\,\pa^2 \hat{B}_3  \hat{G}_{22}
\nonu \\
&& + \frac{4 i \left(9 k^2+112 k+180\right)}{15 (k+5)^3} \pa \hat{B}_3 \pa 
\hat{G}_{22}
-\frac{4 i \left(8330 k^4+59147 k^3
+102724 k^2+79255 k+27588\right)}{15 (k+5)^3 (23 k+19) (35 k+47)}
\nonu \\
&& \times \hat{B}_3 \pa^2 \hat{G}_{22}
-\frac{24 i}{5 (k+5)^2} \,\,\hat{B}_3  \pa \hat{G}_{22} T^{(1)}
-\frac{24 i}{5 (k+5)^2} \,\,\hat{B}_3   \hat{G}_{22} \pa T^{(1)}
+ \frac{16 i}{3 (k+5)^2} \,\,\pa \hat{B}_3 T^{(1)}  V^{(\frac{3}{2})} 
\nonu \\
&& +\frac{12 i}{(k+5)^2} \,\,\hat{B}_3 \pa T^{(1)}  V^{(\frac{3}{2})} 
+ \frac{i (9 k+37)}{(k+5)^2} \,\,\pa \hat{B}_{+}   W_{+}^{(\frac{5}{2})} 
+ \frac{i (3 k+17)}{(k+5)^2}  \,\,\hat{B}_{+}  \pa  W_{+}^{(\frac{5}{2})} 
\nonu \\
&& +
\frac{2 i \left(189 k^3+1398 k^2+2891 k+762\right)}{15 (k+5)^3 (7 k+3)}
\,\,\pa^2 \hat{B}_{+}  T_{+}^{(\frac{3}{2})}
-\frac{2 i \left(63 k^2-406 k-3674\right)}{45 (k+5)^3}
\,\,\pa \hat{B}_{+}  \pa T_{+}^{(\frac{3}{2})}
\nonu \\
&&  +
\frac{2 i \left(252 k^3+491 k^2-9842 k-4641\right)}{45 (k+5)^3 (7 k+3)}
 \,\,\hat{B}_{+}  \pa^2 T_{+}^{(\frac{3}{2})}
-\frac{8 (4 k-21)}{15 (k+5)^3} \,\,\pa \hat{B}_{+} \hat{B}_3  T_{+}^{(\frac{3}{2})}
\nonu \\
&& 
+ \frac{4 (4 k+21)}{3 (k+5)^3} \,\,\hat{B}_{+} \pa \hat{B}_3  T_{+}^{(\frac{3}{2})}
\nonu \\
&& 
-\frac{64 (19 k+89)}{45 (k+5)^3} \,\,\hat{B}_{+} \hat{B}_3  \pa T_{+}^{(\frac{3}{2})}
-\frac{8 \left(18 k^3-385 k^2-1064 k-741\right)}{15 (k+5)^3 (23 k+19)}
\,\,\pa \hat{B}_{+} \hat{B}_3  \hat{G}_{21}
\nonu \\
&& + \frac{4 (7 k+15)}{3 (k+5)^3}\,\,   \hat{B}_{+} \pa \hat{B}_3  \hat{G}_{21}
+  \frac{16 \left(18 k^3-500 k^2-4379 k-3401\right)}{45 (k+5)^3 (23 k+19)}
\,\,\hat{B}_{+} \hat{B}_3  \pa \hat{G}_{21}
\nonu \\
&& +\frac{4 (21 k+83)}{3 (k+5)^3} \,\,\pa \hat{B}_{+} \hat{B}_{-}  V^{(\frac{3}{2})} 
+\frac{4 (18 k+61)}{3 (k+5)^3}  \,\,\hat{B}_{+} \pa \hat{B}_{-}  V^{(\frac{3}{2})} 
+\frac{8 (7 k-22)}{9 (k+5)^3}  \,\,\hat{B}_{+} \hat{B}_{-} \pa  V^{(\frac{3}{2})} 
\nonu \\
&&  +\frac{24}{(k+5)^3} \,\,\pa \hat{B}_{+}  \hat{B}_{-}  \hat{G}_{22}
+\frac{8}{(k+5)^3} \,\,\hat{B}_{+}  \pa \hat{B}_{-}  \hat{G}_{22}
 +\frac{8 k}{3 (k+5)^3} \,\,\hat{B}_{+}  \hat{B}_{-}  \pa \hat{G}_{22}
\nonu \\
&&  +
\frac{2 i \left(378 k^5+3123 k^4+3490 k^3-66660 k^2-98308 k-31863\right)}{
15 (k+5)^3 (7 k+3) (23 k+19)}
\,\,\pa^2 \hat{B}_{+} \hat{G}_{21}
\nonu \\
&& 
+ \frac{2 i \left(54 k^4+243 k^3+14123 k^2+105038 k+77444\right)}{45 (k+5)^3 (23 k+19)}  \,\,\pa \hat{B}_{+} \pa \hat{G}_{21}
\nonu \\
&& 
-\frac{i \left(1512 k^5+11484 k^4-7577 k^3+235909 k^2+307885 k+94107\right)}{45 (k+5)^3 (7 k+3) (23 k+19)}  \,\, \hat{B}_{+} \pa^2 \hat{G}_{21}
\nonu \\
&& 
-\frac{16 i}{3 (k+5)^2} \,\,\pa \hat{B}_{+} \hat{G}_{21} T^{(1)}
+\frac{6 i}{(k+5)^2} \,\,\hat{B}_{+} \hat{G}_{21} \pa T^{(1)}
-\frac{16 i}{3 (k+5)^2} \,\,\pa \hat{B}_{+} T^{(1)}  T_{+}^{(\frac{3}{2})}
\nonu \\
&&  +
\frac{4 (k+7)}{(k+5)^2} \,\,\hat{G}_{12} \pa V_{+}^{(2)}
-\frac{2 (k+14)}{3 (k+5)^2}  \,\,\pa \hat{G}_{21}  V_{-}^{(2)}
+ \frac{2 (k-8)}{(k+5)^2}  \,\, \hat{G}_{21} \pa V_{-}^{(2)}
\nonu \\
&& -\frac{4 (2 k+7)}{(k+5)^2} \,\,\pa \hat{G}_{22} W^{(2)} 
-\frac{4}{(k+5)^2} \,\,\hat{G}_{22} \pa W^{(2)}
-\frac{8 (k-3)}{3 (k+5)^2}  \,\,\pa \hat{G}_{22} T^{(2)} 
\nonu \\
&& -\frac{4}{(k+5)^2} \,\, \hat{G}_{22} \pa T^{(2)}
\nonu \\
&& -\frac{1
}{45 (k+5)^3 (7 k+3) (23 k+19) (35 k+47)}
(33320 k^6+1221157 k^5+11750912 k^4
\nonu \\
&& +40155014 k^3+54877868 k^2
+28673661 k+4637268)
 \pa^3 \hat{G}_{22}
\nonu \\
&& -\frac{4 (k-12)}{15 (k+5)^2} \,\,\pa^2 \hat{G}_{22} T^{(1)}
+ \frac{2 (11 k-12)}{15 (k+5)^2} \,\,\pa \hat{G}_{22} \pa T^{(1)}
-\frac{2 (9 k+22)}{5 (k+5)^2} \,\,\hat{G}_{22} \pa^2 T^{(1)}
\nonu \\
&& +\frac{3 (3 k+11)}{(k+5)^2} \,\,\pa^2 T^{(1)}  V^{(\frac{3}{2})}
-\frac{(k+13)}{3 (k+5)^2}  \,\,\pa T^{(1)}  \pa V^{(\frac{3}{2})}
-\frac{2 (4 k+13)}{3 (k+5)^2}  \,\,T^{(1)} \pa^2  V^{(\frac{3}{2})}
\nonu \\
&& +\frac{1}{(k+5)} \,\,\pa T^{(1)} V^{(\frac{5}{2})}
-\frac{(12 k^2+47 k-161)}{(k+5) (23 k+19)} \,\,T^{(1)} \pa  V^{(\frac{5}{2})}
\nonu  \\
&& + \frac{4 \left(567 k^3+3295 k^2+4987 k+1659\right)}{(k+5)^2 (7 k+3) (35 k+47)} \,\,\pa \hat{T}   V^{(\frac{3}{2})}
\nonu \\
&& -\frac{4 \left(1337 k^4+72278 k^3+321296 k^2+413762 k+130191\right)}{3 (k+5)^2 (7 k+3) (23 k+19) (35 k+47)}  \,\,\hat{T}   \pa V^{(\frac{3}{2})}
\nonu \\
&& +
\frac{8 \left(931 k^5+11438 k^4+50140 k^3+89409 k^2+64079 k+14991\right)}{(k+5)^2 (7 k+3) (23 k+19) (35 k+47)}
\,\,\pa \hat{T} \hat{G}_{22}
\nonu \\
&& + \left.
 \frac{8 \left(1813 k^5+37548 k^4+155200 k^3+211946 k^2+92283 k+9018\right)}{3 (k+5)^2 (7 k+3) (23 k+19) (35 k+47)}
\,\,\hat{T} \pa \hat{G}_{22} \right](w).
\label{9halftwo}
\eea
The $(k-3)$ factor appears in seventh term  in (\ref{9halftwo}).

The first order singular term in (\ref{w5half+w3}) can be summarized by
\bea
& & \{ W_{+}^{(\frac{5}{2})} \, 
 W^{(3)} \}_{-1}(w)  =  \left[
\frac{i}{(k+5)} \,\, \hat{A}_{-} {\bf Q^{(\frac{7}{2})}}
+  \frac{i}{(k+5)} \,\, \hat{B}_{-} {\bf R^{(\frac{7}{2})}}
+ \frac{8 (k-3)}{(23 k+19)} \,\, \hat{T}  {\bf P_{+}^{(\frac{5}{2})}}
 \right. \nonu \\
&& -\frac{i}{(k+5)} \,\,\pa \hat{A}_3  {\bf P_{+}^{(\frac{5}{2})}}
-\frac{i (5 k+21)}{(k+5)^2} \,\,\hat{A}_3  \pa {\bf P_{+}^{(\frac{5}{2})}}
+ \frac{2 i}{(k+5)^2} \,\,\pa \hat{A}_{-} {\bf Q^{(\frac{5}{2})}}
+\frac{2 i (8 k+29)}{5 (k+5)^2} \,\,\hat{A}_{-} \pa {\bf Q^{(\frac{5}{2})}}
\nonu \\
&& +\frac{i}{(k+5)} \,\, \pa \hat{B}_3 {\bf P_{+}^{(\frac{5}{2})}}
+ \frac{i (5 k+21)}{(k+5)^2} \,\,\hat{B}_3 \pa {\bf P_{+}^{(\frac{5}{2})}}
+ \frac{2 i}{(k+5)^2} \,\,\pa \hat{B}_{-}  {\bf R^{(\frac{5}{2})}}
+\frac{2 i (7 k+32)}{5 (k+5)^2} \,\,  \hat{B}_{-}  \pa {\bf R^{(\frac{5}{2})}}
\nonu \\
&& + \frac{(152 k^3+1301 k^2+2642 k-2619)}{30 (k+5)^2 (23 k+19)}
\,\,\pa^2 {\bf P_{+}^{(\frac{5}{2})}}
-\frac{(3 k+14)}{2 (k+5)} \,\,\pa  {\bf S_{+}^{(\frac{7}{2})}}
\nonu \\
&& -\frac{8 i}{(k+5)^2} \,\,\hat{A}_3  U^{(\frac{3}{2})} V_{+}^{(2)}
+\frac{8 i}{(k+5)^2} \,\,
\hat{A}_3  U_{+}^{(2)}  V^{(\frac{3}{2})}
+\frac{4}{(k+5)^2} \,\,\hat{A}_3 \hat{A}_3  W_{+}^{(\frac{5}{2})}
\nonu \\
&& 
-\frac{8}{(k+5)^2}  \,\,\hat{A}_3 \hat{B}_3  W_{+}^{(\frac{5}{2})}
+ \frac{8}{(k+5)^2} \,\,\hat{A}_3 \hat{B}_{-}  V^{(\frac{5}{2})}
-\frac{4 i}{(k+5)^2} \,\,\hat{A}_{-} T^{(2)} U^{(\frac{3}{2})} 
\nonu  \\
&& + \frac{4 i}{(k+5)^2} \,\,\hat{A}_{-} T_{-}^{(\frac{3}{2})}   U_{+}^{(2)}
+\frac{8 i}{(k+5)^3} \,\,\hat{A}_{-} \hat{A}_3 \hat{A}_3  U^{(\frac{3}{2})} 
-\frac{16 i}{(k+5)^3}\,\,
\hat{A}_{-} \hat{A}_3 \hat{B}_3  U^{(\frac{3}{2})} 
\nonu \\
&& + \frac{16 i}{(k+5)^3} \,\,\hat{A}_{-} \hat{A}_3 \hat{B}_{-}  
T_{-}^{(\frac{3}{2})} 
-\frac{16 i}{(k+5)^3} \,\,\hat{A}_{-} \hat{A}_3 \hat{B}_{-} \hat{G}_{12}
-\frac{8}{(k+5)^2} \,\,\hat{A}_{-} \hat{B}_3  U^{(\frac{5}{2})}
\nonu  \\
&&  + \frac{8 i}{(k+5)^3} \,\,  \hat{A}_{-} \hat{B}_3 \hat{B}_{3}  
U^{(\frac{3}{2})} 
+ \frac{4}{(k+5)^2} \,\, \hat{A}_{-} \hat{B}_{-}  W_{-}^{(\frac{5}{2})}
-\frac{16 i}{(k+5)^3} \,\,\hat{A}_{-} \hat{B}_{-} \hat{B}_3  T_{-}^{(\frac{3}{2})} 
\nonu \\
&& -\frac{8 i}{(k+5)^3}  \,\,\hat{A}_{-} \hat{B}_{+} \hat{B}_{-} \hat{G}_{11}
+ \frac{4 i}{(k+5)^2} \,\,\hat{A}_{-} \hat{G}_{21}  U_{-}^{(2)}
+ \frac{4}{(k+5)^2}  \,\,\hat{A}_{+} \hat{A}_{-}  W_{+}^{(\frac{5}{2})}
\nonu \\
&&  + \frac{16 i}{(k+5)^3}\,\,  \hat{A}_{+} \hat{A}_{-} \hat{A}_{-}  U^{(\frac{3}{2})} 
+ \frac{8 i}{(k+5)^3} \,\, \hat{A}_{+} \hat{A}_{-} \hat{B}_{-}  V^{(\frac{3}{2})} 
-\frac{8 i}{(k+5)^3}  \,\,\hat{A}_{+} \hat{A}_{-} \hat{B}_{-} \hat{G}_{22} 
\nonu \\
&& +
\frac{4 i}{(k+5)^2}  \,\,\hat{A}_{+} \hat{G}_{21}  V_{+}^{(2)}
+ \frac{8 i}{(k+5)^2} \,\,\hat{B}_3  U^{(\frac{3}{2})}  V_{+}^{(2)}
-\frac{8 i}{(k+5)^2} \,\,\hat{B}_3  U_{+}^{(2)}  V^{(\frac{3}{2})} 
\nonu \\
&& +
\frac{4}{(k+5)^2} \,\,\hat{B}_3 \hat{B}_3  W_{+}^{(\frac{5}{2})}
-\frac{4 i}{(k+5)^2}  \,\,\hat{B}_{-}  T^{(2)}  V^{(\frac{3}{2})} 
+\frac{4 i}{(k+5)^2} \,\, \hat{B}_{-}  T_{-}^{(\frac{3}{2})}   V_{+}^{(2)}
\nonu \\
&& -\frac{4 i}{(k+5)^2} \,\, \hat{B}_{-} \hat{G}_{12}  V_{+}^{(2)}
-\frac{4 i}{(k+5)^2} \,\, \hat{B}_{-} \hat{G}_{21}  V_{-}^{(2)}
+ \frac{4}{(k+5)^2} \,\,\hat{B}_{+} \hat{B}_{-}  W_{+}^{(\frac{5}{2})}
\nonu \\
&& 
-\frac{8 i}{(k+5)^3}  \,\,\hat{B}_{+} \hat{B}_{-} \hat{B}_{-}  V^{(\frac{3}{2})} 
-\frac{4 i}{(k+5)^2}  \,\,\hat{B}_{+} \hat{G}_{21}  U_{+}^{(2)}
+ \frac{2}{(k+5)} \,\,\hat{G}_{21} W^{(3)}
\nonu \\
&& + \frac{8 \left(k^2+4 k-7\right)}{(k+5) (23 k+19)} \hat{T}  
W_{+}^{(\frac{5}{2})}
+  \frac{48 i (3 k+1)}{(k+5)^2 (23 k+19)} 
\hat{T} \hat{A}_3 \hat{G}_{21}
\nonu \\
&& -\frac{8 i \left(92 k^3+853 k^2+1542 k+573\right)}{(k+5)^2 (23 k+19) (35 k+47)} \,\,\hat{T} \hat{A}_{-} \hat{G}_{11}
+ \frac{160 i k}{(k+5)^2 (23 k+19)} \,\,\hat{T} \hat{B}_3 \hat{G}_{21}
\nonu \\
&& 
-\frac{8 i \left(49 k^2+145 k+12\right)}{(k+5)^2 (7 k+3) (35 k+47)}
\,\,\hat{T} \hat{B}_{-}  V^{(\frac{3}{2})} 
-\frac{16 i \left(133 k^3+1178 k^2+2493 k+1368\right)}{3 (k+5)^2 (23 k+19) (35 k+47)}  \,\,\hat{T} \hat{B}_{-} \hat{G}_{22}
\nonu \\
&&  +\frac{16 (k-3)}{(k+5) (23 k+19)} \,\,\hat{T} \hat{G}_{21} T^{(1)}
+ \frac{8 i \left(70 k^2+333 k+171\right)}{(k+5)^2 (7 k+3) (35 k+47)}
\hat{T} \hat{A}_{-}  U^{(\frac{3}{2})} 
\nonu \\
&& +
\frac{4 \left(19 k^3-199 k^2-1407 k-1317\right)}{15 (k+5)^2 (23 k+19)}
\,\,\pa^2  W_{+}^{(\frac{5}{2})}
+ \frac{2}{(k+5)} \,\,\pa  U^{(\frac{3}{2})}   V_{+}^{(2)}
-\frac{6}{(k+5)}  \,\,U^{(\frac{3}{2})} \pa  V_{+}^{(2)}
\nonu \\
&& -\frac{6}{(k+5)}  \,\,\pa U_{+}^{(2)}  V^{(\frac{3}{2})}
+ \frac{2}{(k+5)} \,\,U_{+}^{(2)}  \pa V^{(\frac{3}{2})}
-\frac{2 (7 k+30)}{(k+5)^2}  \,\,\pa T_{+}^{(\frac{3}{2})} T^{(2)}  
\nonu \\
&& -\frac{2 (3 k+14)}{(k+5)^2}  \,\,T_{+}^{(\frac{3}{2})} \pa T^{(2)} 
\nonu \\
&& -\frac{(24570 k^6+534074 k^5-1450675 k^4-24521426 k^3
-56410142 k^2-43292856 k-7977465)}{45 (k+5)^3 (7 k+3) (23 k+19) (35 k+47)} 
\nonu \\
&& \times \pa^3  T_{+}^{(\frac{3}{2})} 
\nonu \\
&& +
\frac{2 i (5 k+23)}{(k+5)^2} \,\,\pa \hat{A}_3  W_{+}^{(\frac{5}{2})}
-\frac{2 i \left(31 k^2+16 k-119\right)}{(k+5)^2 (23 k+19)}
 \,\,\hat{A}_3  \pa W_{+}^{(\frac{5}{2})}
+ \frac{2 i \left(12 k^2+51 k-14\right)}{5 (k+5)^3} \,\, \pa^2 \hat{A}_3   
T_{+}^{(\frac{3}{2})}
\nonu \\
&& 
-\frac{2 i \left(28 k^2+79 k-196\right)}{15 (k+5)^3}  \,\,\pa \hat{A}_3   
\pa T_{+}^{(\frac{3}{2})}
+ \frac{8 i \left(4 k^2-13 k-60\right)}{15 (k+5)^3}
 \,\,\hat{A}_3   
\pa^2 T_{+}^{(\frac{3}{2})}
+ \frac{8 (2 k+15)}{3 (k+5)^3} \,\,\hat{A}_3 \hat{A}_3  \pa T_{+}^{(\frac{3}{2})}
\nonu \\
&& -\frac{4 (4 k-15)}{(k+5)^3}  \,\,\pa \hat{A}_3 \hat{A}_3 \hat{G}_{21}
-\frac{16 (k-13)}{3 (k+5)^3}  \,\,\hat{A}_3 \hat{A}_3 \pa \hat{G}_{21}
-\frac{16 (2 k+15)}{3 (k+5)^3} \,\,\hat{A}_3 \hat{B}_3  \pa T_{+}^{(\frac{3}{2})}
\nonu \\
&&  + \frac{12 (k-4)}{(k+5)^3}  \,\,\pa \hat{A}_3 \hat{B}_3 \hat{G}_{21}
-\frac{4 k}{(k+5)^3}  \,\,\hat{A}_3 \pa \hat{B}_3 \hat{G}_{21}
-\frac{16 (k+17)}{3 (k+5)^3} \,\, \hat{A}_3 \hat{B}_3 \pa \hat{G}_{21}
\nonu \\
&& -\frac{4 (4 k+45)}{5 (k+5)^3} \,\, 
\pa \hat{A}_{3} \hat{B}_{-}   V^{(\frac{3}{2})} 
+ \frac{8 (k+6)}{(k+5)^3} \,\,\hat{A}_{3} \pa \hat{B}_{-}   V^{(\frac{3}{2})} 
+ \frac{16 (2 k+5)}{15 (k+5)^3} \,\,\hat{A}_{3} \hat{B}_{-} \pa   V^{(\frac{3}{2})} 
\nonu \\
&&  +\frac{8 (13 k+71)}{5 (k+5)^3} \,\,\pa \hat{A}_3 \hat{B}_{-} \hat{G}_{22}
+\frac{8 (3 k+14)}{(k+5)^3}  \,\,\hat{A}_3 \pa \hat{B}_{-} \hat{G}_{22}
+\frac{8 (49 k+228)}{15 (k+5)^3}  \,\,\hat{A}_3 \hat{B}_{-} \pa \hat{G}_{22}
\nonu \\
&&
-\frac{2 i \left(168 k^3+1507 k^2+6637 k+2398\right)}{5 (k+5)^3 (23 k+19)}
\pa^2 \hat{A}_3 \hat{G}_{21}
\nonu \\
&& +
\frac{2 i \left(168 k^3+3142 k^2-9923 k-8957\right)}{15 (k+5)^3 (23 k+19)}
\pa \hat{A}_3 \pa \hat{G}_{21}
\nonu \\
&&  +
\frac{i \left(672 k^3+6197 k^2-5206 k-11707\right)}{15 (k+5)^3 (23 k+19)}
\hat{A}_3 \pa^2 \hat{G}_{21}
-\frac{2 i}{(k+5)^2} \pa \hat{A}_3 \hat{G}_{21} T^{(1)}
-\frac{8 i}{(k+5)^2} \hat{A}_3 \pa \hat{G}_{21} T^{(1)}
\nonu \\
&& -\frac{10 i}{(k+5)^2}  \,\,\hat{A}_3 \hat{G}_{21} \pa T^{(1)}
-\frac{2 i \left(4410 k^4+43547 k^3+141745 k^2+163081 k+49017\right)}{5 (k+5)^3 (7 k+3) (35 k+47)} \,\,\pa^2 \hat{A}_{-}  U^{(\frac{3}{2})} 
\nonu  \\
&& 
-\frac{2 i \left(8 k^2+73 k+107\right)}{15 (k+5)^3} 
\,\,\pa \hat{A}_{-}  \pa U^{(\frac{3}{2})} 
+ \frac{2 i \left(910 k^3+7327 k^2+24441 k+19884\right)}{
15 (k+5)^3 (35 k+47)}
\,\,\hat{A}_{-}  \pa^2 U^{(\frac{3}{2})} 
\nonu \\
&& -\frac{2 i (4 k+17)}{(k+5)^2} \,\,\pa \hat{A}_{-}  U^{(\frac{5}{2})} 
-\frac{2 i (2 k+9)}{5 (k+5)^2} \,\,\hat{A}_{-}  \pa U^{(\frac{5}{2})} 
+ \frac{8 (2 k+3)}{(k+5)^3} \,\,\pa \hat{A}_{-} \hat{A}_3  U^{(\frac{3}{2})} 
\nonu \\
&& 
-\frac{12 (8 k+39)}{5 (k+5)^3}  \,\,\hat{A}_{-} \pa \hat{A}_3  U^{(\frac{3}{2})} 
-\frac{16 (k+8)}{5 (k+5)^3}  \,\,\hat{A}_{-} \hat{A}_3  \pa U^{(\frac{3}{2})} 
+ \frac{72}{5 (k+5)^3} \,\,\hat{A}_{-} \pa \hat{A}_3 \hat{G}_{11}
\nonu \\
&& -\frac{48}{5 (k+5)^3} \,\,\hat{A}_{-} \hat{A}_3 \pa \hat{G}_{11}
-\frac{8 (2 k+3)}{(k+5)^3}  \,\,\pa \hat{A}_{-}  \hat{B}_3  U^{(\frac{3}{2})} 
-\frac{4 (k-30)}{5 (k+5)^3} \,\,\hat{A}_{-}  \pa \hat{B}_3  U^{(\frac{3}{2})} 
\nonu \\
&& -\frac{16 (7 k-10)}{15 (k+5)^3}\,\, \hat{A}_{-}  \hat{B}_3  \pa U^{(\frac{3}{2})} 
+\frac{8 (3 k+14)}{(k+5)^3}\,\,  \pa \hat{A}_{-} \hat{B}_3 \hat{G}_{11}
+\frac{16 (7 k+34)}{5 (k+5)^3}\,\, \hat{A}_{-} \pa \hat{B}_3 \hat{G}_{11}
\nonu \\
&& +\frac{8 (47 k+234)}{15 (k+5)^3}\,\, \hat{A}_{-} \hat{B}_3 \pa \hat{G}_{11}
+\frac{2 (25 k+137)}{5 (k+5)^3}\,\,  \pa \hat{A}_{-} \hat{B}_{-}  T_{-}^{(\frac{3}{2})}
+\frac{38 (k+9)}{5 (k+5)^3}\,\,  \hat{A}_{-} \pa \hat{B}_{-}  T_{-}^{(\frac{3}{2})}
\nonu \\
&& +\frac{2 (107 k+369)}{15 (k+5)^3}\,\,  \hat{A}_{-} \hat{B}_{-}  
\pa T_{-}^{(\frac{3}{2})} 
+\frac{2 (3 k+5)}{(k+5)^3}\,\, \pa \hat{A}_{-} \hat{B}_{-} \hat{G}_{12}
+\frac{2 (11 k+39)}{5 (k+5)^3} \,\,
 \hat{A}_{-} \pa \hat{B}_{-} \hat{G}_{12}
\nonu \\
&& -\frac{2 (7 k-37)}{15 (k+5)^3}
 \,\,\hat{A}_{-} \hat{B}_{-} \pa \hat{G}_{12}
-\frac{2 i \left(8740 k^4+85907 k^3+289448 k^2+366959 k+157094\right)}{5 (k+5)^3 (23 k+19) (35 k+47)} \,\,\pa^2 \hat{A}_{-} \hat{G}_{11}
\nonu \\
&& 
-\frac{2 i \left(22 k^2+277 k+438\right)}{15 (k+5)^3}
\,\,\pa \hat{A}_{-} \pa \hat{G}_{11}
-\frac{32 (7 k-31) (11 k+7)}{15 (k+5)^3 (23 k+19)} \hat{A}_{+} \hat{A}_{-} 
\pa \hat{G}_{21}
\nonu \\
& &+ 
\frac{4 i \left(10695 k^4+91592 k^3+417156 k^2+629500 k+272601\right)}{15 (k+5)^3 (23 k+19) (35 k+47)}
\,\,\hat{A}_{-} \pa^2 \hat{G}_{11}
\nonu \\
&& 
+ \frac{8 i}{5 (k+5)^2} \,\,\hat{A}_{-} \pa \hat{G}_{11} T^{(1)}
-\frac{12 i}{5 (k+5)^2} \,\, \hat{A}_{-} \hat{G}_{11} \pa T^{(1)}
-\frac{6 i}{(k+5)^2} \,\,\hat{A}_{-} \pa T^{(1)}  U^{(\frac{3}{2})} 
\nonu \\
&& + \frac{4 (5 k+28)}{5 (k+5)^3} \,\,
\pa \hat{A}_{+} \hat{A}_{-}  T_{+}^{(\frac{3}{2})}
+\frac{12 (k+4)}{(k+5)^3}\,\,  \hat{A}_{+} \pa \hat{A}_{-}  T_{+}^{(\frac{3}{2})}
+\frac{4 (35 k+204)}{15 (k+5)^3} \,\, 
\hat{A}_{+} \hat{A}_{-}  \pa T_{+}^{(\frac{3}{2})}
\nonu \\
&&-\frac{4 \left(267 k^2+263 k-272\right)}{5 (k+5)^3 (23 k+19)}
\pa \hat{A}_{+} \hat{A}_{-} \hat{G}_{21}
-\frac{4 (2 k-3)}{(k+5)^3} \hat{A}_{+} \pa \hat{A}_{-} \hat{G}_{21}
\nonu \\
&&  + \frac{2 i (5 k+31)}{(k+5)^2}\,\, \pa \hat{B}_3  W_{+}^{(\frac{5}{2})}
+ \frac{2 i \left(101 k^2+408 k+475\right)}{(k+5)^2 (23 k+19)}
 \,\,\hat{B}_3  \pa W_{+}^{(\frac{5}{2})}
-\frac{2 i \left(15 k^2+294 k+1004\right)}{5 (k+5)^3} 
\nonu \\
&& \times \pa^2 \hat{B}_3  T_{+}^{(\frac{3}{2})}
\nonu \\
&&-\frac{2 i \left(45 k^2+254 k-506\right)}{15 (k+5)^3}
 \,\,\pa \hat{B}_3  \pa T_{+}^{(\frac{3}{2})}
+ \frac{8 i (k+3) (3 k+23)}{3 (k+5)^3}
 \,\,\hat{B}_3  \pa^2 T_{+}^{(\frac{3}{2})}
\nonu \\
&& + \frac{4 (2 k-3)}{(k+5)^3}  \,\,\pa \hat{B}_3 \hat{B}_3  \hat{G}_{21}
+ \frac{32 (k+2)}{3 (k+5)^3} \,\,\hat{B}_3 \hat{B}_3  \pa \hat{G}_{21}
+ \frac{8 (2 k+15)}{3 (k+5)^3} \,\,\hat{B}_3 \hat{B}_3 \pa  T_{+}^{(\frac{3}{2})}
\nonu \\
&& 
+\frac{2 i \left(54 k^4+675 k^3-3842 k^2-49258 k-39615\right)}{15 (k+5)^3 (23 k+19)} \,\,\pa^2 \hat{B}_3 \hat{G}_{21}
\nonu \\
&& +  \frac{2 i \left(18 k^4+395 k^3-144 k^2+13674 k+10735\right)}{15 (k+5)^3 (23 k+19)} \,\,\pa \hat{B}_3 \pa \hat{G}_{21}
\nonu \\
&& -\frac{i \left(72 k^4+1312 k^3-3351 k^2-26910 k-26239\right)}{15 (k+5)^3 (23 k+19)}  \,\,\hat{B}_3 \pa^2 \hat{G}_{21}
\nonu \\
&& +\frac{2 i}{(k+5)^2}  \,\,\pa \hat{B}_3 \hat{G}_{21} T^{(1)}
+\frac{8 i}{(k+5)^2}  \,\,\hat{B}_3 \pa \hat{G}_{21} T^{(1)}
+\frac{10 i}{(k+5)^2}   \,\,\hat{B}_3 \hat{G}_{21} \pa T^{(1)}
\nonu \\
&& +
\frac{6 i \left(735 k^4+15022 k^3+78576 k^2+103858 k+30609\right)}{5 (k+5)^3 (7 k+3) (35 k+47)} \,\,\pa^2 \hat{B}_{-}  V^{(\frac{3}{2})} 
\nonu \\
&&
-\frac{2 i \left(k^2-131 k-842\right)}{15 (k+5)^3}
\,\, \pa  \hat{B}_{-}  \pa V^{(\frac{3}{2})} 
\nonu \\
&& -\frac{2 i \left(280 k^3+6751 k^2+24459 k+20136\right)}{15 (k+5)^3 (35 k+47)}
 \,\,\hat{B}_{-}  \pa^2 V^{(\frac{3}{2})} 
+ \frac{i (7 k+37)}{(k+5)^2} \,\,\pa \hat{B}_{-}  V^{(\frac{5}{2})}  
\nonu \\
&& + \frac{i (11 k-3)}{5 (k+5)^2} \,\, \hat{B}_{-}  \pa V^{(\frac{5}{2})}  
-\frac{8 (k+6)}{(k+5)^3} \,\,\pa \hat{B}_{-} \hat{B}_3  V^{(\frac{3}{2})} 
+ \frac{4 (9 k+152)}{5 (k+5)^3} \,\, \hat{B}_{-} \pa \hat{B}_3  V^{(\frac{3}{2})} 
\nonu \\
&& + \frac{16 (k+8)}{5 (k+5)^3}\,\,  \hat{B}_{-} \hat{B}_3 \pa V^{(\frac{3}{2})} 
+\frac{24 k}{5 (k+5)^3} \,\,\hat{B}_{-} \pa \hat{B}_3 \hat{G}_{22}
-\frac{16 k}{5 (k+5)^3} \,\,\hat{B}_{-} \hat{B}_3 \pa \hat{G}_{22}
\nonu \\
&& -\frac{2 i \left(33565 k^4+358075 k^3+1193848 k^2+1451883 k+567777\right)}{15 (k+5)^3 (23 k+19) (35 k+47)}\,\, \pa^2 \hat{B}_{-} \hat{G}_{22}
\nonu \\
&&-\frac{2 i \left(25 k^2+311 k+723\right)}{15 (k+5)^3}
 \,\,\pa \hat{B}_{-} \pa \hat{G}_{22}
\nonu \\
&&+
\frac{i \left(52080 k^4+477673 k^3+1386471 k^2+1619075 k+641877\right)}{15 (k+5)^3 (23 k+19) (35 k+47)}
 \,\,\hat{B}_{-} \pa^2 \hat{G}_{22}
\nonu \\
&& -\frac{8 i}{5 (k+5)^2} \,\,\hat{B}_{-} \pa \hat{G}_{22} T^{(1)}
+\frac{12 i}{5 (k+5)^2}  \,\,\hat{B}_{-} \hat{G}_{22} \pa T^{(1)}
-\frac{6 i}{(k+5)^2} \,\,\hat{B}_{-} \pa T^{(1)}   V^{(\frac{3}{2})}
\nonu \\
&& -\frac{4 (7 k+12)}{5 (k+5)^3}\,\, \pa \hat{B}_{+} \hat{B}_{-}  T_{+}^{(\frac{3}{2})}
-\frac{12 (k+4)}{(k+5)^3} \,\,\hat{B}_{+} \pa \hat{B}_{-}  T_{+}^{(\frac{3}{2})}
-\frac{4 (k-114)}{15 (k+5)^3} \,\,\hat{B}_{+} \hat{B}_{-}  \pa T_{+}^{(\frac{3}{2})}
\nonu \\
&& +\frac{2 \left(12 k^3-525 k^2-1966 k-1349\right)}{5 (k+5)^3 (23 k+19)}
\,\,\pa \hat{B}_{+} \hat{B}_{-} \hat{G}_{21}
-\frac{2 (7 k+27)}{(k+5)^3} \,\,\hat{B}_{+} \pa \hat{B}_{-} \hat{G}_{21}
\nonu \\
&& -\frac{4 \left(12 k^3+625 k^2+19 k-494\right)}{15 (k+5)^3 (23 k+19)}
\,\,\hat{B}_{+} \hat{B}_{-} \pa \hat{G}_{21}
-\frac{2 (3 k+16)}{(k+5)^2} \,\,\pa \hat{G}_{11}  V_{+}^{(2)}
-\frac{6 (k+4)}{(k+5)^2}  \,\,\hat{G}_{11}  \pa V_{+}^{(2)}
\nonu \\
&& +\frac{4 (2 k+9)}{(k+5)^2} \,\,\pa \hat{G}_{21} W^{(2)}
+ \frac{2 (3 k+10)}{(k+5)^2} \,\,\hat{G}_{21} \pa W^{(2)}
-\frac{2 (3 k+28)}{(k+5)^2}  \,\,\pa \hat{G}_{21} T^{(2)}
\nonu \\
&& -\frac{2 (2 k+17)}{(k+5)^2}  \,\,\hat{G}_{21} \pa T^{(2)}
\nonu \\
&& +
\frac{1
}{90 (k+5)^3 (7 k+3) (23 k+19) (35 k+47)}
(53410 k^6+1532853 k^5+21947810 k^4 \nonu \\
&& +96802830 k^3+
160741842 k^2+97220421 k+14908914) \pa^3 \hat{G}_{21}
\nonu \\
&& -\frac{4 \left(3 k^3+48 k^2+311 k-38\right)}{5 (k+5)^2 (23 k+19)}
\,\,\pa^2 \hat{G}_{21} T^{(1)}
+\frac{(6 k^3+201 k^2+1847 k+1024)}{5 (k+5)^2 (23 k+19)}
\,\,\pa \hat{G}_{21} \pa T^{(1)}
\nonu \\
&& + \frac{(54 k^3+539 k^2-5957 k-6654)}{15 (k+5)^2 (23 k+19)}
\,\,\hat{G}_{21} \pa^2 T^{(1)}
+ \frac{2 (3 k+16)}{(k+5)^2} \pa \hat{G}_{22}  U_{+}^{(2)}
+ \frac{6 (k+4)}{(k+5)^2} \hat{G}_{22}  \pa U_{+}^{(2)}
\nonu \\
&& -\frac{12 (k-3)}{(23 k+19)} \,\,T^{(1)} \pa  W_{+}^{(\frac{5}{2})}
-\frac{3 (3 k+14)}{(k+5)^2} \,\,\pa^2 T^{(1)}  T_{+}^{(\frac{3}{2})}
+ \frac{(3 k+16)}{(k+5)^2}  \,\,\pa T^{(1)}  \pa T_{+}^{(\frac{3}{2})}
\nonu \\
&& 
+ \frac{2}{(k+5)^2}  \,\,T^{(1)}  \pa^2 T_{+}^{(\frac{3}{2})}
+\frac{4 (3 k+14) \left(70 k^4-1895 k^3-8828 k^2-10515 k-1908\right)}{3 (k+5)^2 (7 k+3) (23 k+19) (35 k+47)} \,\,\pa
\hat{T}  T_{+}^{(\frac{3}{2})}
 \nonu \\
&& +\frac{4 \left(210 k^5-4285 k^4-23400 k^3+27093 k^2+113100 k+72270\right)}{3 (k+5)^2 (7 k+3) (23 k+19) (35 k+47)}
\,\,\hat{T}  \pa T_{+}^{(\frac{3}{2})}
\nonu \\
&& -\frac{2 \left(11172 k^5+221816 k^4+1078389 k^3+1968145 k^2+1347087 k+256959\right)}{3 (k+5)^2 (7 k+3) (23 k+19) (35 k+47)}
\,\,\pa \hat{T} \hat{G}_{21}
\nonu \\
&& \left.
-\frac{4 \left(3626 k^5+88711 k^4+433734 k^3+672005 k^2+309130 k+10110\right)}{3 (k+5)^2 (7 k+3) (23 k+19) (35 k+47)}
\,\,\hat{T} \pa \hat{G}_{21} \right](w).
\label{9halfthird}
\eea
The $(k-3)$ factor appears in the third term  of (\ref{9halfthird}).

Finally,
the first order singular term in (\ref{w5half-w3}) can be summarized by
\bea
& & \{ W_{-}^{(\frac{5}{2})} \, 
 W^{(3)} \}_{-1}(w)  = 
\left[ \frac{8 (k-3)}{(23 k+19)} \,\, \hat{T} {\bf P_{-}^{(\frac{5}{2})}}
+\frac{i (3 k+11)}{(k+5)^2} \,\, \pa \hat{A}_3 {\bf P_{-}^{(\frac{5}{2})}}
+\frac{i (3 k+11)}{(k+5)^2} \,\, \hat{A}_3 \pa {\bf P_{-}^{(\frac{5}{2})}}
\right. \nonu \\
&& -\frac{i (3 k+11)}{(k+5)^2}  \,\,\pa \hat{B}_3 {\bf P_{-}^{(\frac{5}{2})}}
-\frac{i (3 k+11)}{(k+5)^2}  \,\, \hat{B}_3 \pa {\bf P_{-}^{(\frac{5}{2})}}
\nonu \\
&& + \frac{(152 k^3-1459 k^2-12058 k-12879)}{30 (k+5)^2 (23 k+19)}
\,\,\pa^2 {\bf P_{-}^{(\frac{5}{2})}}  -\frac{(3 k+14)}{2 (k+5)}
\,\,\pa  {\bf S_{-}^{(\frac{7}{2})}}
\nonu \\
&& -\frac{2}{(k+5)} \,\,U^{(\frac{5}{2})} V_{-}^{(2)}
-\frac{2}{(k+5)}  \,\,U_{-}^{(2)}  V^{(\frac{5}{2})}
-\frac{4}{(k+5)^2} \,\,\hat{A}_3 \hat{A}_3  W_{-}^{(\frac{5}{2})}
\nonu \\
&& +\frac{8}{(k+5)^2} \,\,\hat{A}_3 \hat{B}_3  W_{-}^{(\frac{5}{2})}
-\frac{4}{(k+5)^2} \,\,\hat{A}_{+} \hat{A}_{-}  W_{-}^{(\frac{5}{2})}
-\frac{4}{(k+5)^2}  \,\,\hat{B}_3 \hat{B}_3  W_{-}^{(\frac{5}{2})}
\nonu \\
&& -\frac{4}{(k+5)^2} \,\, \hat{B}_{+} \hat{B}_{-}  W_{-}^{(\frac{5}{2})}
+ \frac{2}{(k+5)} \,\,\hat{G}_{12} W^{(3)}
-\frac{8 \left(k^2+4 k-7\right)}{(k+5) (23 k+19)}
\,\, \hat{T}  W_{-}^{(\frac{5}{2})}
\nonu \\
&& +
\frac{48 i (3 k+1)}{(k+5)^2 (23 k+19)}
\,\, \hat{T} \hat{A}_3 \hat{G}_{12}
+ \frac{160 i k}{(k+5)^2 (23 k+19)} \,\,\hat{T} \hat{B}_3 \hat{G}_{12}
\nonu \\
&& 
-\frac{(76 k^3-2751 k^2-15523 k-12108)}{15 (k+5)^2 (23 k+19)}
\,\,\pa^2  W_{-}^{(\frac{5}{2})} 
+ \frac{2 (5 k+31)}{3 (k+5)^2}  \,\,\pa U^{(\frac{3}{2})}   V_{-}^{(2)}
\nonu \\
&&  +
\frac{2 (3 k+13)}{(k+5)^2}  \,\, \pa U_{-}^{(2)} V^{(\frac{3}{2})} 
+ \frac{2 (7 k+25)}{3 (k+5)^2} \,\, U_{-}^{(2)} \pa V^{(\frac{3}{2})} 
-\frac{2 (7 k+34)}{(k+5)^2}  \,\, \pa T_{-}^{(\frac{3}{2})} T^{(2)}
\nonu \\
&& -\frac{2 (3 k+14)}{(k+5)^2} \,\, T_{-}^{(\frac{3}{2})} \pa T^{(2)}
+ \frac{16 (k-3)}{(k+5) (23 k+19)}
\,\, \hat{T} \hat{G}_{12} T^{(1)}
 +
\frac{2 (3 k+13)}{(k+5)^2}  \,\, U^{(\frac{3}{2})}   \pa V_{-}^{(2)}
\nonu \\
&& -\frac{
}{45 (k+5)^3 (7 k+3) (23 k+19) (35 k+47)}
(24570 k^6+84079 k^5-4527992 k^4 
\nonu \\
&& -25506514 k^3-
31019878 k^2-12380997 k+936252)
\,\, \pa^3  T_{-}^{(\frac{3}{2})} 
\nonu \\
&& + \frac{2 i (3 k+17)}{(k+5)^2}  \,\, \pa \hat{A}_3  W_{-}^{(\frac{5}{2})} 
+ \frac{2 i \left(15 k^2+68 k+157\right)}{(k+5)^2 (23 k+19)}
\,\, \hat{A}_3  \pa W_{-}^{(\frac{5}{2})} 
\nonu \\
&& -\frac{2 i \left(144 k^2+835 k+834\right)}{15 (k+5)^3} \pa^2
\hat{A}_3  T_{-}^{(\frac{3}{2})}
\nonu \\
&& -\frac{2 i \left(48 k^2+65 k-632\right)}{15 (k+5)^3}
\,\, \pa  \hat{A}_3  \pa T_{-}^{(\frac{3}{2})}
+\frac{4 i \left(48 k^2+365 k+763\right)}{15 (k+5)^3}
\,\, \hat{A}_3  \pa^2 T_{-}^{(\frac{3}{2})}
\nonu \\
&& + \frac{8 (3 k+14)}{(k+5)^3} \,\, \pa \hat{A}_3 \hat{A}_3 T_{-}^{(\frac{3}{2})}
+ \frac{52 (k+6)}{3 (k+5)^3} \,\, \hat{A}_3 \hat{A}_3 \pa T_{-}^{(\frac{3}{2})}
+\frac{4 (4 k-13)}{(k+5)^3}  \,\, \pa \hat{A}_3 \hat{A}_3  \hat{G}_{12}
\nonu \\
&& +
\frac{8 (2 k-29)}{3 (k+5)^3}  \,\, \hat{A}_3 \hat{A}_3  \pa \hat{G}_{12}
-\frac{8 (3 k+14)}{(k+5)^3} \,\, \pa \hat{A}_3 \hat{B}_3 T_{-}^{(\frac{3}{2})}
-\frac{8 (3 k+14)}{(k+5)^3}  \,\, \hat{A}_3 \pa \hat{B}_3 T_{-}^{(\frac{3}{2})}
\nonu \\ 
&& -\frac{104 (k+6)}{3 (k+5)^3}  \,\, \hat{A}_3 \hat{B}_3 \pa T_{-}^{(\frac{3}{2})}
-\frac{4 (k-4)}{(k+5)^3} \,\, \pa \hat{A}_3 \hat{B}_3  \hat{G}_{12}
-\frac{4 (k-4)}{(k+5)^3} \,\, \hat{A}_3 \pa \hat{B}_3  \hat{G}_{12}
\nonu \\
&&+
\frac{16 (k+20)}{3 (k+5)^3} \,\, \hat{A}_3 \hat{B}_3  \pa \hat{G}_{12}
-\frac{4 (5 k+19)}{(k+5)^3} \,\, \pa \hat{A}_3 \hat{B}_{+}  U^{(\frac{3}{2})} 
-\frac{4 (5 k+23)}{(k+5)^3} \,\, \hat{A}_3 \pa \hat{B}_{+}  U^{(\frac{3}{2})} 
\nonu \\
&& -\frac{4 (5 k+19)}{(k+5)^3} \,\, \hat{A}_3 \hat{B}_{+}  \pa U^{(\frac{3}{2})} 
-\frac{8 (3 k+14)}{(k+5)^3}   \,\, \pa \hat{A}_3 \hat{B}_{+} \hat{G}_{11}
-\frac{8 (3 k+14)}{(k+5)^3}  \,\, \hat{A}_3 \pa \hat{B}_{+} \hat{G}_{11}
\nonu \\
&&  -\frac{8 (3 k+14)}{(k+5)^3}  \,\, \hat{A}_3 \hat{B}_{+} \pa \hat{G}_{11}
+ \frac{8 i \left(495 k^3+5102 k^2+9229 k+5574\right)}{15 (k+5)^3 (23 k+19)}
\,\, \pa^2 \hat{A}_3 \hat{G}_{12}
\nonu \\
&& 
+\frac{2 i \left(660 k^3+8631 k^2+10372 k+4217\right)}{15 (k+5)^3 (23 k+19)}
\,\, \pa \hat{A}_3 \pa \hat{G}_{12}
\nonu \\
&&  
-\frac{i \left(2640 k^3+26669 k^2+75958 k+49593\right)}{15 (k+5)^3 (23 k+19)}
\,\,\hat{A}_3 \pa^2 \hat{G}_{12}
+ \frac{6 i}{(k+5)^2} \,\,\pa \hat{A}_3 \hat{G}_{12} T^{(1)}
\nonu \\
&& +\frac{8 i}{(k+5)^2}  \,\,\hat{A}_3 \pa \hat{G}_{12} T^{(1)}
+\frac{6 i}{(k+5)^2}  \,\,\hat{A}_3 \hat{G}_{12} \pa T^{(1)}
+ \frac{2 i \left(18 k^2+171 k+421\right)}{5 (k+5)^3} 
\,\,\pa^2 \hat{A}_{+}  V^{(\frac{3}{2})} 
\nonu \\
&& +\frac{2 i \left(8 k^2+201 k+751\right)}{15 (k+5)^3}
\,\,\pa \hat{A}_{+}  \pa V^{(\frac{3}{2})} 
-\frac{4 i \left(13 k^2+106 k+336\right)}{15 (k+5)^3}
\,\,\hat{A}_{+}  \pa^2 V^{(\frac{3}{2})} 
\nonu \\
&&-\frac{4 i (2 k+7)}{(k+5)^2} \,\,\pa \hat{A}_{+}  V^{(\frac{5}{2})} 
-\frac{2 i (2 k+9)}{(k+5)^2} \,\,\hat{A}_{+}  \pa V^{(\frac{5}{2})} 
-\frac{8 (2 k+13)}{(k+5)^3} \,\,\pa \hat{A}_{+} \hat{A}_3 V^{(\frac{3}{2})} 
\nonu \\
&&  -\frac{8 (2 k+11)}{(k+5)^3}\,\,  \hat{A}_{+} \pa \hat{A}_3 V^{(\frac{3}{2})} 
-\frac{8 (2 k+11)}{(k+5)^3} \,\, \hat{A}_{+} \hat{A}_3 \pa V^{(\frac{3}{2})} 
+ \frac{4 (4 k+17)}{(k+5)^3} \,\,\pa \hat{A}_{+} \hat{A}_{-} T_{-}^{(\frac{3}{2})}
\nonu \\
&& + \frac{4 (4 k+15)}{(k+5)^3} \,\, \hat{A}_{+} \pa \hat{A}_{-} T_{-}^{(\frac{3}{2})}
+ \frac{4 (16 k+87)}{3 (k+5)^3} \,\, \hat{A}_{+} \hat{A}_{-} \pa T_{-}^{(\frac{3}{2})}
+ \frac{8 (k-2)}{(k+5)^3} \,\,\pa \hat{A}_{+} \hat{A}_{-} \hat{G}_{12}
\nonu \\
&& +\frac{8 k}{(k+5)^3} \,\,\hat{A}_{+} \pa \hat{A}_{-} \hat{G}_{12}
+\frac{16 (k-10)}{3 (k+5)^3} \,\,\hat{A}_{+} \hat{A}_{-} \pa \hat{G}_{12}
+\frac{8 (2 k+13)}{(k+5)^3}  \,\,\pa \hat{A}_{+}  \hat{B}_3 V^{(\frac{3}{2})} 
\nonu \\
&&+ \frac{8 (2 k+11)}{(k+5)^3}  \,\,\hat{A}_{+} \pa \hat{B}_3 V^{(\frac{3}{2})} 
+ \frac{8 (2 k+11)}{(k+5)^3}  \,\,\hat{A}_{+}  \hat{B}_3 \pa V^{(\frac{3}{2})} 
-\frac{8 (3 k+14)}{(k+5)^3}  \,\,\pa \hat{A}_{+} \hat{B}_{3} \hat{G}_{22}
\nonu \\
&& -\frac{8 (3 k+14)}{(k+5)^3}  \,\,\hat{A}_{+} \pa \hat{B}_{3} \hat{G}_{22}
-\frac{8 (3 k+14)}{(k+5)^3}  \,\,\hat{A}_{+} \hat{B}_{3} \pa \hat{G}_{22}
-\frac{4 (3 k+14)}{(k+5)^3} \,\,\pa \hat{A}_{+} \hat{B}_{+} T_{+}^{(\frac{3}{2})} 
\nonu \\
&& -\frac{4 (3 k+14)}{(k+5)^3} \,\,\hat{A}_{+} \pa \hat{B}_{+} T_{+}^{(\frac{3}{2})} 
-\frac{4 (3 k+14)}{(k+5)^3} \,\,\hat{A}_{+} \hat{B}_{+} \pa T_{+}^{(\frac{3}{2})} 
-\frac{6}{(k+5)^2} \,\,\pa \hat{A}_{+} \hat{B}_{+} \hat{G}_{21}
\nonu \\
&& -\frac{6}{(k+5)^2} \,\,\hat{A}_{+} \pa \hat{B}_{+}  \hat{G}_{21} 
-\frac{2 (3 k+19)}{(k+5)^3} \,\,\hat{A}_{+} \hat{B}_{+} \pa \hat{G}_{21} 
-\frac{8 i \left(3 k^2+29 k+80\right)}{5 (k+5)^3} \,\,
\pa^2 \hat{A}_{+} \hat{G}_{22}
\nonu \\
&& -\frac{2 i \left(22 k^2+311 k+660\right)}{15 (k+5)^3}
\,\, \pa \hat{A}_{+} \pa \hat{G}_{22}
+ \frac{2 i \left(34 k^2+237 k+750\right)}{15 (k+5)^3}
\,\, \hat{A}_{+} \pa^2 \hat{G}_{22}
\nonu \\
&&  + \frac{2 i (7 k+25)}{(k+5)^2} \,\,\pa \hat{B}_3 W_{-}^{(\frac{5}{2})} 
+ \frac{2 i \left(55 k^2+416 k+513\right)}{(k+5)^2 (23 k+19)}
\,\,\hat{B}_3 \pa W_{-}^{(\frac{5}{2})} 
-\frac{2 i \left(3 k^2-16 k-160\right)}{5 (k+5)^3} \,\,\pa^2 \hat{B}_3  
T_{-}^{(\frac{3}{2})} 
\nonu \\
&& + \frac{2 i \left(7 k^2-24 k+370\right)}{15 (k+5)^3}
 \,\, \pa \hat{B}_3  
\pa T_{-}^{(\frac{3}{2})} 
-\frac{4 i \left(2 k^2+56 k+355\right)}{15 (k+5)^3}
 \,\,\hat{B}_3  
\pa^2 T_{-}^{(\frac{3}{2})} 
\nonu \\
&&  + \frac{8 (3 k+14)}{(k+5)^3} \,\,
\pa \hat{B}_3 \hat{B}_3  T_{-}^{(\frac{3}{2})} 
+ \frac{52 (k+6)}{3 (k+5)^3} \,\,\hat{B}_3 \hat{B}_3  \pa T_{-}^{(\frac{3}{2})} 
-\frac{4 (2 k-5)}{(k+5)^3} \,\,\pa \hat{B}_3 \hat{B}_3 \hat{G}_{12}
\nonu \\
&& -\frac{8 (4 k+11)}{3 (k+5)^3} \,\,\hat{B}_3 \hat{B}_3 \pa \hat{G}_{12}
+ \frac{2 i \left(54 k^4+18 k^3-779 k^2-20110 k-17727\right)}{
15 (k+5)^3 (23 k+19)} \,\,\pa^2 \hat{B}_3 \hat{G}_{12}
\nonu \\
&& + \frac{2 i \left(18 k^4+236 k^3+3657 k^2-720 k-4199\right)}{
15 (k+5)^3 (23 k+19)} \,\,\pa \hat{B}_3 \pa \hat{G}_{12}
\nonu \\
&& -\frac{i \left(72 k^4+484 k^3-1137 k^2-51150 k-48241\right)}{
15 (k+5)^3 (23 k+19)} \,\,\hat{B}_3 \pa^2 \hat{G}_{12}
\nonu \\
&& -\frac{6 i}{(k+5)^2} \,\,\pa \hat{B}_3 \hat{G}_{12} T^{(1)}
-\frac{8 i}{(k+5)^2}  \,\,\hat{B}_3 \pa \hat{G}_{12} T^{(1)}
-\frac{6 i}{(k+5)^2}  \,\,\hat{B}_3 \hat{G}_{12}  \pa T^{(1)}
\nonu \\
&& -\frac{2 i \left(27 k^2+319 k+939\right)}{15 (k+5)^3}
\,\,\pa^2 \hat{B}_{+} U^{(\frac{3}{2})} 
+ \frac{2 i \left(k^2-193 k-1118\right)}{15 (k+5)^3}
\,\,\pa \hat{B}_{+} \pa U^{(\frac{3}{2})} 
\nonu \\
&& + \frac{2 i \left(8 k^2-14 k-379\right)}{15 (k+5)^3}
\,\,\hat{B}_{+} \pa^2 U^{(\frac{3}{2})} 
+ \frac{7 i}{(k+5)}  \,\,\pa \hat{B}_{+} U^{(\frac{5}{2})} 
+ \frac{i (5 k+19)}{(k+5)^2}  \,\,\hat{B}_{+} \pa U^{(\frac{5}{2})} 
\nonu \\
&&  +\frac{4 (5 k+23)}{(k+5)^3} \,\,\pa \hat{B}_{+} \hat{B}_3  U^{(\frac{3}{2})}  
+ \frac{4 (5 k+19)}{(k+5)^3} \,\,\hat{B}_{+} \pa \hat{B}_3  U^{(\frac{3}{2})}  
+ \frac{4 (5 k+19)}{(k+5)^3} \,\,\hat{B}_{+} \hat{B}_3  \pa U^{(\frac{3}{2})}  
\nonu \\
&& + \frac{2 (5 k+19)}{(k+5)^3} \,\,
\pa \hat{B}_{+} \hat{B}_{-}  T_{-}^{(\frac{3}{2})} 
+ \frac{2 (5 k+23)}{(k+5)^3}  \,\,
\hat{B}_{+} \pa \hat{B}_{-}  T_{-}^{(\frac{3}{2})}  
+\frac{2 (23 k+129)}{3 (k+5)^3}  \,\,
\hat{B}_{+} \hat{B}_{-}  \pa T_{-}^{(\frac{3}{2})} 
\nonu \\
&& +\frac{4}{(k+5)^2}  \,\, \pa \hat{B}_{+} \hat{B}_{-}  \hat{G}_{12}
+\frac{4 (k+7)}{(k+5)^3}  \,\, \hat{B}_{+} \pa \hat{B}_{-}  \hat{G}_{12}
-\frac{2 (k+17)}{3 (k+5)^3}  \,\, \hat{B}_{+} \hat{B}_{-}  \pa \hat{G}_{12}
\nonu \\
&& -\frac{2 i \left(15 k^2+102 k+121\right)}{5 (k+5)^3} 
\,\, \pa^2 \hat{B}_{+} \hat{G}_{11}
-\frac{2 i \left(25 k^2+317 k+801\right)}{15 (k+5)^3}  
\,\, \pa \hat{B}_{+} \pa \hat{G}_{11}
\nonu \\
&& + \frac{i \left(80 k^2+373 k-441\right)}{15 (k+5)^3}
 \,\, \hat{B}_{+} \pa^2 \hat{G}_{11}
+ \frac{2 (13 k+42)}{3 (k+5)^2} \,\,\pa \hat{G}_{11} V_{-}^{(2)}
+ \frac{6 (k+4)}{(k+5)^2} \,\,\hat{G}_{11} \pa V_{-}^{(2)}
\nonu \\
&& -\frac{8 (k+4)}{(k+5)^2} \,\,\pa \hat{G}_{12} W^{(2)}
-\frac{6 (k+4)}{(k+5)^2}  \,\,\hat{G}_{12} \pa W^{(2)}
+ \frac{32}{(k+5)^2} \,\, \pa \hat{G}_{12} T^{(2)}
-\frac{2 (k-3)}{(k+5)^2} \,\, \hat{G}_{12} \pa T^{(2)}
\nonu \\
&& -\frac{1
}{45 (k+5)^3 (7 k+3) (23 k+19) (35 k+47)}
(26705 k^6+749924 k^5+9760926 k^4 \nonu \\
&& +45315454 k^3+
72253639 k^2+42974046 k+7156746)
\pa^3 \hat{G}_{12}
\nonu \\
&&  -\frac{6 (k-27) \left(2 k^2+17 k+19\right)}{5 (k+5)^2 (23 k+19)}
\,\,\pa^2 \hat{G}_{12} T^{(1)}
+ \frac{2 \left(3 k^3-3 k^2-565 k-647\right)}{5 (k+5)^2 (23 k+19)}
\,\,\pa \hat{G}_{12}  \pa T^{(1)}
\nonu \\
&& + \frac{2 \left(27 k^3-662 k^2-8095 k-6918\right)}{15 (k+5)^2 (23 k+19)}
 \,\,\hat{G}_{12} \pa^2 T^{(1)}
-\frac{2 (11 k+54)}{3 (k+5)^2} \,\,\pa \hat{G}_{22}  U_{-}^{(2)}
\nonu \\
&& -\frac{2 (3 k+14)}{(k+5)^2} \,\,\hat{G}_{22}  \pa U_{-}^{(2)}
-\frac{12 (k-3)}{(23 k+19)} \,\,T^{(1)} \pa W_{-}^{(\frac{5}{2})} 
+ \frac{3 (3 k+14)}{(k+5)^2} \,\,\pa^2 T^{(1)}  T_{-}^{(\frac{3}{2})}  
\nonu \\
&& +\frac{(k+4)}{(k+5)^2}   \,\,\pa T^{(1)}  \pa T_{-}^{(\frac{3}{2})}  
-\frac{2 (2 k+9)}{(k+5)^2}  \,\,T^{(1)}  \pa^2 T_{-}^{(\frac{3}{2})}  
\nonu \\
&& + 
\frac{4 (3 k+14) \left(70 k^4+3418 k^3+24403 k^2+24696 k+7497\right)}{3 (k+5)^2 (7 k+3) (23 k+19) (35 k+47)} \,\,\pa \hat{T}  T_{-}^{(\frac{3}{2})} 
\nonu \\
&& + \frac{4 \left(210 k^5+11654 k^4+155505 k^3+622926 k^2+671055 k+220014\right)}{3 (k+5)^2 (7 k+3) (23 k+19) (35 k+47)}
 \,\,\hat{T}  \pa T_{-}^{(\frac{3}{2})} 
\nonu \\
&& + \frac{2 \left(11172 k^5+213122 k^4+1009383 k^3+1715041 k^2+1176465 k+253881\right)}{3 (k+5)^2 (7 k+3) (23 k+19) (35 k+47)}
\,\, \pa \hat{T} \hat{G}_{12}
\nonu \\
&& \left.
+ \frac{8 \left(1813 k^5+42182 k^4+181503 k^3+222574 k^2+68246 k-7770\right)}{3 (k+5)^2 (7 k+3) (23 k+19) (35 k+47)}
\,\, \hat{T} \pa \hat{G}_{12} \right](w).
\label{lastexp}
\eea
The $(k-3)$ factor appears in the first term  in (\ref{lastexp}).
Of course, one can reexpress this first order term in terms of 
several descendant fields with correct numerical relative coefficients
coming from each higher singular terms plus several (quasi) primary fields as
done in previous many examples.

\end{document}